\documentclass{aastex631}

\usepackage{natbib}
\usepackage{graphicx}

\usepackage{graphics}
\usepackage{amssymb}
\usepackage{amsbsy}
\usepackage{amsfonts}
\usepackage{bm}
\usepackage{mathtools}
\usepackage{empheq}

\usepackage{amsmath}

\usepackage[makeroom]{cancel}
\usepackage{tikz}

\newcommand{\bhat}{\boldsymbol{\hat{b}}}
\newcommand{\bu}{\boldsymbol{u}}
\newcommand{\bV}{\boldsymbol{v}}
\newcommand{\bc}{\boldsymbol{c}}
\newcommand{\bE}{\boldsymbol{E}}
\newcommand{\bb}{\boldsymbol{B}}

\newcommand{\bp}{\bar{\bar{\boldsymbol{p}}}}

\newcommand{\bI}{\bar{\bar{\boldsymbol{I}}}}
\newcommand{\bW}{\bar{\bar{\boldsymbol{W}}}}
\newcommand{\bPi}{\bar{\bar{\boldsymbol{\Pi}}}}
\newcommand{\bQ}{\bar{\bar{\boldsymbol{Q}}}}

\newcommand{\bD}{\bar{\bar{\boldsymbol{D}}}}

\newcommand{\bH}{\bar{\bar{\boldsymbol{H}}}}
\newcommand{\bnu}{\bar{\nu}}
\newcommand{\bh}{\bar{\bar{\boldsymbol{h}}}}

\newcommand{\bg}{\boldsymbol{g}}

\newcommand{\hbV}{\hat{\boldsymbol{V}}}

\newcommand{\pr}{\partial}
\newcommand{\nn}{\nonumber}

\newcommand{\bx}{\boldsymbol{x}}
\newcommand{\bX}{\bar{\bar{\boldsymbol{X}}}}
\newcommand{\by}{\boldsymbol{y}}
\newcommand{\bC}{\boldsymbol{C}}

\newcommand{\bVV}{\boldsymbol{V}}

\newcommand{\bee}{\begin{eqnarray}}
\newcommand{\eee}{\end{eqnarray}}

\newcommand{\trace}{\textrm{Tr}}

\newcommand{\tbc}{\tilde{\boldsymbol{c}}}
\newcommand{\tc}{\tilde{c}}

\newcommand{\vecq}{\vec{\boldsymbol{q}}}
\newcommand{\vecX}{\vec{\boldsymbol{X}}}
\newcommand{\vecQ}{\vec{\boldsymbol{Q}}}

\newcommand{\erf}{\textrm{erf}}

\newcommand*\circled[1]{\tikz[baseline=(char.base)]{
    \node[shape=circle,draw,inner sep=1pt] (char) {#1};}}

\begin{document}
%\title{Generalized fluid models of the Braginskii type.\\
%   Part 2. The Boltzmann operator.}

\title{GENERALIZED FLUID MODELS OF THE BRAGINSKII TYPE.\\
   PART 2. THE BOLTZMANN OPERATOR.}

\author{P. Hunana}
\affiliation{Instituto de Astrof\'isica de Canarias (IAC), La Laguna, Tenerife, 38205, Spain; peter.hunana@gmail.com}
\affiliation{Universidad de La Laguna, La Laguna, Tenerife, 38206, Spain}

%============= older style ===============================
%\author{P. Hunana\altaffilmark{1,2}}

%\altaffiltext{1}{Instituto de Astrof\'isica de Canarias (IAC), La Laguna, Tenerife, 38205, Spain; peter.hunana@gmail.com}
%\altaffiltext{2}{Universidad de La Laguna, La Laguna, Tenerife, 38206, Spain}
%=========================================================

\begin{abstract}
  In our previous paper (Hunana et al. 2022) we have employed the Landau collisional operator together with
  the moment method of Grad and considered various generalizations of the Braginskii model,
  such as a multi-fluid formulation of the 21- and 22-moment models valid for general masses and temperatures,
  where all of the considered moments are described by their evolution equations
  (with fully non-linear left-hand-sides). Here we consider the same models, however, we employ
  the Boltzmann operator and calculate the collisional contributions via expressing them
  through the Chapman-Cowling collisional integrals. These ``integrals'' just represent a useful mathematical technique/notation introduced
   roughly  100 years ago,
  which (in the usual semi-linear approximation) allows one to postpone specifying the particular collisional process and finish all of the calculations
  with the Boltzmann operator. We thus consider multi-fluid 21- and 22-moment models which are valid
    for a large class of elastic collisional processes describable by the Boltzmann operator.
    Reduction into the 13-moment approximation recovers the models of Schunk and Burgers.
    We only focus on the particular cases of hard spheres, Coulomb collisions, purely repulsive inverse power force $|K|/r^\nu$ and
    attractive force $-|K|/r^\nu$ with repulsive rigid core (or potential $V(r)=\delta(r)- |c|/r^n$,
    so that the particles bounce from each other when they meet), but other cases can be found in the literature.  
  In the Appendix, we introduce the Boltzmann operator in a way suitable for newcomers and
  we discuss a surprisingly simple recipe how to calculate the collisional contributions with analytic software.
\end{abstract}

\tableofcontents

%\vspace{0.5cm}
%\noindent

\newpage
%==================================================================================================================
%==================================================================================================================
\section{Introduction}
%{\it - Note for the production office: Even though this is a single-author paper, it is written in a plural voice as ``we'' and 
%I strongly prefer it this way, so please do not change it. Thank you in advance.}\\

Even though kinetic plasma simulations are becoming increasingly common in recent years and are being employed to even model global astrophysical scales 
 (see e.g. \cite{Palmroth2023,Palmroth2018,Lapenta2022,Karimabadi2014} and references therein),
construction of fluid models from the kinetic Boltzmann equation is still of crucial importance for a very large area
of physical sciences, from the solar and astrophysical applications to laboratory studies of plasma fusion.  
It is worth noting that kinetic plasma simulations (particle-in-cell or Vlasov, fully kinetic or hybrid)
are typically focused on almost collisionless plasmas by modeling the
evolution of the Vlasov equation, with an assumption that the effects of collisions are subdominant,
but where various stabilization mechanisms (which can be viewed as heuristic collisions), often have to be added  to prevent numerical problems.
The Landau collisional operator contains an integral over the velocity space
(and the Boltzmann operator contains another integral over the solid angle) and simulating these collisional operators fully kinetically
would drastically increase the computational cost to a whole new level.
Additional complexity arises when collisions between numerous particle species need to be considered, such as if one wants to understand 
the evolution of minor abundances in the solar/stellar atmospheres and interiors
\citep{Asplund2021,Asplund2009,ChristensenDalsgaard2021,ChristensenDalsgaard2008,Paxton2018,Paxton2011,Michaud2015,Khomenko2014,Killie2007,Killie2004,Hansteen1997,Thoul1994,Michaud1993,VauclairVauclair1982,Noerdlinger1977},
in the Earth's and planetary ionospheres \citep{SchunkNagy2009,Schunk2004,Schunk1988,Schunk1975,Schunk1977},
or if one needs to model the evolution of plasma impurities at the edge (the scrape-off layer)
in a tokamak \citep{Makarov2023,Makarov2022,Raghunathan2022b,Raghunathan2022a,Raghunathan2021,Sytova2020,Sytova2018,WIESEN2015,ROZHANSKY2015,KUKUSHKIN2011}. Fluid models for plasmas are still relevant and in many areas fluid models will remain irreplaceable for a very long time.

%\subsection{Motivation}
This paper is a continuation of our previous paper \cite{Hunana2022} (Part 1), where the Landau collisional operator
has been used - the same operator as used by \cite{Braginskii1965,Braginskii1958}. 
The Landau collisional operator was developed by \cite{Landau1936,Landau1937} as a simplification
of the Boltzmann collisional operator, appropriate for cases when the collisional dynamics is
dominated by collisions with a small scattering angle (sometimes called the grazing collisions limit),
which is the case of Coulomb collisions between charged particles such as ions and electrons. Even though in Part 1 we have presented various generalizations
of the Braginskii model, such as a multi-fluid formulation for arbitrary masses and temperatures with the stress-tensors and heat fluxes 
described by their own evolution equations, our models remained valid only for fully ionized
weakly-coupled plasmas (with sufficiently large Coulomb logarithm $\ln\Lambda\gg 1$).
However, in many astrophysical as well as laboratory applications, one encounters partially ionized plasmas, 
where neutral particles are present and where several different collisional interactions need to be considered.
To be able to address at least some of these processes, here in Part 2 we employ the Boltzmann collisional operator.   
We only use the ``classical'' well-known Boltzmann operator (see eq. (\ref{eq:BoltzmannO})), which can describe only elastic collisions
(of non-rotating and non-vibrating particles/molecules, i.e. a monatomic species) and our models do not have the ionization and recombination processes.

Our initial motivation was to consider only one additional collisional process - the collisions of ideal hard spheres.
The hard sphere approximation is reasonable for modeling neutral-neutral collisions with sufficiently high temperatures exceeding roughly 1000 Kelvin and   
for the sake of simplicity of final models and numerical simulations, this approximation is sometimes used to model the
collisions between neutrals and charged particles as well.
Note that from an analytic perspective, the hard sphere approximation might look as surprisingly simple at first, because its differential cross-section
$\sigma_{ab}(g_{ab},\theta)=(r_a+r_b)^2/4$ is just a constant given by
the radii $r_a$, $r_b$ of the colliding spheres (independent of the relative velocity $\bg_{ab}=\bV_a-\bV_b$ of the colliding spheres and independent
of the scattering angle $\theta$),
so that the $\sigma_{ab}(g_{ab},\theta)$ can be immediately pulled outside of the Boltzmann operator. But in practice, this does not matter much at all,
and the collisional integrals for hard spheres
are still as difficult to calculate, as for the Coulomb collisions with the Rutherford differential cross-section
(and often the final results for hard spheres are actually more complicated).
Additionally, we feel that the collisions of ideal hard spheres are by far the most beautiful example for understanding
the effects of viscosity and thermal conductivity.
It is because hard spheres are easy to visualize (one can say that we are just studying a very large number of billiard balls),
so it feels very clarifying that during each collision the momentum and energy
is conserved exactly - and in spite of this, the entire system has the effects of viscosity and thermal conductivity
(as a consequence of the perturbation of the distribution function). Of course, the same is true for the Coulomb collisions, but there the nature of the
electrostatic interaction with the Coulomb logarithm makes the effects of viscosity and thermal conductivity much more blurry.
We therefore find it useful to consider the same 21- and 22-moment models as in Part 1, but this time for the collisions of ideal hard spheres.
Such models of hard spheres are very interesting even on their own, and some simple results are summarized below in Technical
Introduction \ref{sec:HardSphereP}, where the comparison of hard spheres with the Braginskii case of Coulomb collisions is discussed. Nevertheless,
perhaps more importantly,
coupling the evolution equations of these hard sphere models with the evolution equations of Coulomb collisions from Part 1,
offers a description of partially ionized plasmas - with the precision that matches or exceeds the Braginskii model.
As in Part 1 or in Braginskii, we use the restriction that the differences in bulk/drift velocities between species must be
sufficiently smaller than their thermal velocities, $|\bu_b-\bu_a|/\sqrt{v_{\textrm{th}a}^2+v_{\textrm{th}b}^2} \ll 1$. The only exception are Appendices
\ref{sec:Mom}, \ref{sec:Energy} and \ref{sec:Unrestricted},
where the pure Maxwellian distributions are considered (i.e. with no stress-tensors and no heat fluxes),
and the well-known momentum exchange rates
and energy exchange rates of hard spheres are reproduced with unrestricted drifts. 

Our second goal is to revisit the Coulomb collisions and by employing the Rutherford differential cross-section (see eq. (\ref{eq:buu})), to
rederive the Braginskii model and all of the results of Part 1 directly with the Boltzmann operator.
The calculations presented here in Part 2 with the Boltzmann operator are very different than those in Part 1 with the Landau operator,
so showing a complete analytic match of models for arbitrary masses and temperatures,
serves as an excellent verification tool that our models are formulated correctly. 
Additionally, with the Boltzmann operator, it is possible to capture corrections of the Coulomb logarithm,
where as an example the Landau operator yields $2\ln\Lambda$, whereas the Boltzmann operator yields more precise $\ln(\Lambda^2+1)$,
and these numbers can be kept in their un-approximated form, if the Coulomb logarithm is not sufficiently large.
Plasmas are usually separated to three broad categories of ``weakly-coupled plasmas'' (with $\ln\Lambda \ge 10$), ``moderately-coupled plasmas'' 
($2\le \ln\Lambda \le 10$) and ``strongly-coupled plasmas'' ($\ln\Lambda\le2$), so by considering corrections of the Coulomb logarithm,
one can extend the area of validity (with some limitations) also to moderately-coupled plasmas.
Such corrections of the Coulomb logarithm are already present in the models of \cite{ChapmanCowling1953} (p. 178),
\cite{Burgers1969} (p. 115), see also \cite{JiHeld2021}.
These corrections are obtained in a very easy way (see Appendix \ref{sec:AppendixA}),
where one considers integrals over the normalized impact parameters $x$ such as
$\int_0^\Lambda \frac{2x}{(1+x^2)}dx=\ln(\Lambda^2+1)$ (which corresponds to the upper cut-off at the Debye length $\lambda_D$ and no lower cut-off).
In contrast, the Landau operator has the $\ln\Lambda$ already in its definition, and by focusing on large $x$, 
the same integral is viewed as a simplified $\int_{x_{\textrm{min}}}^\Lambda \frac{2}{x}dx$, where 
it is necessary to introduce also some lower cut-off. For pure convenience, this lower cut-off is chosen to be $x_{\textrm{min}}=1$ 
(which corresponds to the impact parameter for 90-degree scattering), so that both integrals are the same for large $\Lambda$.
Our 21- and 22-moment models only need two additional integrals, given by eqs. (\ref{eq:A22x})-(\ref{eq:A23x}).  
Notably, in contrast to $\ln\Lambda$, expressions $\ln(\Lambda^2+1)$ together with (\ref{eq:A22x})-(\ref{eq:A23x})
do not become negative regardless of the encountered physical conditions. 
As a consequence, even though these simple corrections are not suitable to describe strongly-coupled plasmas (and their applicability to a full range of
moderately-coupled plasmas is also questionable),  
at least it is not possible to encounter the awkward situation that observational data/models yield negative Coulomb logarithms.
Moderately-coupled plasmas are encountered in laboratory experiments with laser produced plasmas to study
the inertial confinement fusion, see for example \cite{Adrian2022} and \cite{Lin2023}.
In astrophysics, corrections of the Coulomb logarithm are required to model the diffusion of helium and other heavy elements
in the solar/stellar interiors or in the envelopes of white dwarfs, see e.g. \citep{Michaud2015,Thoul1994,Paquette1986,IbenMacDonald1985} and references
therein. Interestingly, in some of these studies
more sophisticated corrections than ours are considered, where instead of the Coulomb potential $V(r)=q_a q_b /r$ with the cut-off at the
Debye length $\lambda_D$, one uses the Debye screening potential $V(r)=(q_a q_b /r) \exp(-r/\lambda_D)$ to calculate the collisional integrals 
and this potential was also used by \cite{Stanton2016,Dangola2008,Mason1967,Kihara1959} and \cite{Liboff1959}.
We do not use the Debye screened potential, but this directly brings us to our final goal.

Naturally, each time a new collisional process is considered, one does not want to start from scratch with a bare Boltzmann operator
and keep recalculating the underlying description of viscosity, thermal conductivity and diffusion. To prevent this, Chapman and Cowling developed
a very useful technique of expressing the final model through integrals $\Omega^{(l,j)}_{ab}$, now known as ``Chapman-Cowling integrals''. 
Essentially, one just takes an arbitrary/unspecified differential cross-section $\sigma_{ab}(g_{ab},\theta)$, and defines all the possible integrals over the relative
velocity $g_{ab}$ and the scattering angle $\theta$ that the model will need (for attractive forces, it is much better to integrate over the impact
parameter). In this way, the long process of obtaining the underlying fluid model is done only once, and one can consider a particular collisional
process a posteriori, by focusing at the calculation of $\Omega^{(l,j)}_{ab}$.
Our final goal therefore is to express our
21- and 22- moment models through the Chapman-Cowling integrals, for arbitrary masses and temperatures of species.
Importantly, reducing our models into the 13-moment approximation, yields the models of \cite{Schunk1975,Schunk1977} and \cite{Burgers1969}.  

After such a construction is done,  we want to of course provide some interesting Chapman-Cowling integrals for our model, and
not just the hard spheres and Coulomb collisions. In the literature, one can find a vast number of collisional processes that
  are considered with the Boltzmann operator, see for example \cite{ChapmanCowling1953} and \cite{Hirschfelder1954}.
  This is because in general the collisional forces between interacting atoms/molecules are not known and should be modeled quantum-mechanically. 
  For example, two particles can be attracted to each other at long distances and repel each other at short distances,
  which can be modeled by a general Lennard-Jones
force $F(r) = K_{ab}/r^\nu-K'_{ab}/r^{\nu '}$ (where positive $K_{ab}$ represents repulsion and positive $K_{ab}'$ represents attraction).
For neutral particles, the most studied combination is the repulsive force $\nu=13$ and the attractive force $\nu'=7$.  
Written with a potential $V(r)=4 \epsilon [ (\sigma/r)^{12} - (\sigma/r)^6]$ instead of a force, this is known as the Lennard-Jones 12-6 model. 
Another useful example is the Sutherland's model, where one prescribes repulsive force $\nu=\infty$ in the general Lennard-Jones model,
so that the model corresponds to hard spheres that are attracted to each other. The Sutherland's model
allows one to consider attractive forces between particles that have a finite radius, so that the particles bounce from each other once they meet.
Nevertheless, for our purposes these models are too complicated and we wanted to consider only the purely repulsive force
$F(r) =  |K_{ab}|/r^{\nu}$ and the purely attractive force $F(r) = - |K_{ab}|/r^{\nu}$. The repulsive case is quite easy, and it allows one
to join together the Coulomb collisions ($\nu=2$), the Maxwell molecules ($\nu=5$) and the hard spheres $(\nu=\infty)$. However, for attractive forces
steeper than $1/r^2$, there is a complication that particles spiral around each other, where for large impact parameters they reach some minimum distance
and separate, but for small impact parameters they spiral towards each other and hit each other.  
For the attractive Coulomb force, particles can only meet for the impact parameter $b_0=0$, but 
for steeper forces, there is a whole range of impact parameters when this happens, and one needs to specify what happens to the particles when they meet.
The integrals figured out by \cite{Eliason1956} used a ``transparent core'', where the particle trajectories just pass through each other.
We prefer the ``rigid core'' model considered by \cite{Kihara1960} (and references therein), where the particles bounce from each other. The rigid core model
is more realistic and it does not bring any additional complexity to the transparent core model. Essentially,  
the rigid core model represents a simplified Sutherland's model in the limit of infinitesimally small hard spheres. Its
attractive potential can be written as $V(r)=\delta(r) - |c|/r^n$, where $\delta(r)$ is delta function, and it is a very elegant solution
to the general problem of attractive forces, without introducing the complexity of finite particle sizes. In fact, it is rather
surprising that it is possible to integrate over all of the spiraling particle trajectories and create a fluid model out of it. 
We wanted to make sure that we are understanding these models correctly and in Section \ref{eq:AttractNU3}, we
reproduce the solution for the attractive case $\nu=3$ (or $n=2$) and in Section \ref{sec:AlnuA}, we briefly verify the numerical integrals
for the attractive Maxwell molecules $\nu=5$ and for the London force $\nu=7$. The case $\nu=5$ is especially important, because it
allows one to model the (non-resonant) ion-neutral collisions, where the attraction is caused by the ion polarizing the neutral.  
For other attractive cases, we simply adopt the numerical integrals of \cite{Higgins1968}. In Section \ref{sec:Alnu}, 
we also verified many numerical integrals for the purely repulsive forces. Our model is therefore ready to
be used with a wide variety of repulsive forces $r^{-\nu}$, as well as attractive forces $-r^{-\nu}$ with the rigid core, and for
other forces, one needs to provide the Chapman-Cowling integrals.

\newpage
%====================================================================================================
%====================================================================================================
\section{Technical introduction} \label{sec:Technical}
\subsection{Classification of models}
The classification of models obtained with the moment method of Grad was already addressed at great length in Part 1
 and is based on the expansion of the distribution function $f_a=f_a^{(0)}(1+\chi_a)$ around the Maxwellian 
 $f_a^{(0)} = \frac{n_a}{\pi^{3/2} v_{\textrm{th} a}^3} \exp(- c_a^2 /v_{\textrm{th} a}^2)$ in Hermite polynomials, see the Appendix B there, together with
 p. 34-35. The $\bc_a=\bV_a-\bu_a$ is the fluctuating/random velocity, with $\bu_a$ being the fluid/drift velocity and the thermal speed $v_{\textrm{th} a}=\sqrt{2T_a/m_a}$ 
 (we use the same notation as Braginskii, with the Boltzmann constant $k_{\rm{B}}=1$).  
Here we briefly repeat that the models describing strict Maxwellians (with no stress-tensors and no heat fluxes) are referred to
as 5-moment models, because only five fluid moments are present (one density, three velocities and one scalar pressure/temperature). 
The major models can be summarized as
 \begin{eqnarray}
&&  \textrm{5-moment}: \qquad \,\,
  \chi_a =  0;\nn\\    
&&  \textrm{13-moment}: \qquad
  \chi_a =  \bh^{(2)}_{a} :{\bH}^{(2)}_{a} +\vec{\boldsymbol{h}}^{(3)}_a \cdot \vec{\boldsymbol{H}}^{(3)}_{a};\nn\\   
&&  \textrm{21-moment}: \qquad
  \chi_a =  {\bh}^{(2)}_{a}: {\bH}^{(2)}_{a} +\vec{\boldsymbol{h}}^{(3)}_a \cdot \vec{\boldsymbol{H}}^{(3)}_{a} + {\bh}^{(4)}_{a}: {\bH}^{(4)}_{a}
  +\vec{\boldsymbol{h}}^{(5)}_a \cdot \vec{\boldsymbol{H}}^{(5)}_{a};\nn\\ 
&&  \textrm{22-moment}: \qquad
  \chi_a =  {\bh}^{(2)}_{a} :{\bH}^{(2)}_{a} +\vec{\boldsymbol{h}}^{(3)}_a \cdot \vec{\boldsymbol{H}}^{(3)}_{a} + h^{(4)}_a H^{(4)}_a + {\bh}^{(4)}_{a}: {\bH}^{(4)}_{a}
  +\vec{\boldsymbol{h}}^{(5)}_a \cdot \vec{\boldsymbol{H}}^{(5)}_{a}. \label{eq:TwoPart1}
 \end{eqnarray}
%=========================================
The big $H$ are (irreducible) Hermite polynomials and the small $h$ are Hermite moments. 
Matrices $\bh^{(2)}_{a}$ and $\bh^{(4)}_{a}$ can be viewed as stress-tensors and vectors $\vec{\boldsymbol{h}}^{(3)}_a$ and $\vec{\boldsymbol{h}}^{(5)}_a$
       can be viewed as heat fluxes.  Rewritten with fluid moments, the perturbation of the 22-moment model is given by (\ref{eq:beau2x}). 
Models with one usual stress-tensor $\bPi_a^{(2)}$ (which contains 5 independent components) and one usual heat flux vector $\vecq_a$
(which contains 3 independent components) are referred to as 13-moment models. These models were developed in great detail
by \cite{Burgers1969,Schunk1975,Schunk1977} and references therein, see also the book by \cite{SchunkNagy2009}. 
The model of \cite{Braginskii1958,Braginskii1965} (who used the Chapman-Enskog
expansions and not the method of Grad) can be viewed as a 21-moment model. 
Instead of formulation with Hermite moments, which are used for example in the models of
\cite{Balescu1988} and \cite{Zhdanov2002} (originally published in 1982), our final model is formulated in fluid moments, by
employing the ``stress-tensor'' $\bPi_a^{(4)}$ of the 4th-order fluid moment and the ``heat flux'' $\vecX^{(5)}_a$ of the 5th-order fluid moment.
We are using free wording, because $\bPi_a^{(4)}$ is not really a stress-tensor and $\vecX^{(5)}_a$ is not really a heat flux. 
The model of \cite{Braginskii1965} then can be interpreted as being constructed with  
two coupled stress-tensors $\bPi_a^{(2)}, \bPi^{(4)}_a$ and two coupled heat flux vectors $\vecq_a$, $\vecX^{(5)}_a$, which in the
highly-collisional/quasi-static approximation (by canceling the time-derivatives in the evolution equations for these quantities)
yields more precise stress-tensor $\bPi_a^{(2)}$ and more precise
heat flux $\vecq_a$ than the 13-moment moments. Our formulation of the Braginskii model through two stress-tensors and two heat flux vectors
seems to be very clarifying for newcomers to the subject and the formulation is starting to be appreciated in the academic environment as well. 
Note that in the semi-linear approximation the product $f_a f_b$ is approximated as $f_af_b=f_a^{(0)}f_b^{(0)}(1+\chi_a+\chi_b)$, with the
  $\chi_a\chi_b$ neglected, and the result is further simplified by expansions with small drifts (see Appendix \ref{eq:AppendixAC}),
  so in some works the $f_a^{(0)}$ is Maxwellian without drifts and the drift speed is considered as part of the perturbation.\\

Finally, the 22-moment model contains one more scalar quantity $h^{(4)}_a$, which is the (fully contracted) scalar perturbation of the 4th-order fluid moment
$\widetilde{X}^{(4)}_a =  m_a \int |\bc_a|^{4} (f_a -f_a^{(0)}) d^3v_a$.
This quantity describes the tail of a distribution function and by an analogy with a 1-dimensional
statistics, it can be viewed as an ``excess kurtosis''. For example, the positive $\widetilde{X}^{(4)}_a>0$ 
 means that the distribution
 is slimmer in the middle than Maxwellian (and usually with a higher peak), but that it has longer/heavier tails,
 i.e. that the tail contains more data than Maxwellian.
 Similarly, the negative  $\widetilde{X}^{(4)}_a<0$ means that the distribution is fatter in the middle than Maxwellian, but that it has shorter/lighter
 tails. In the highly-collisional/quasi-static approximation, the $\widetilde{X}_a^{(4)}$ is proportional to the divergence of the heat flux vectors and as
 a consequence, it has both the thermal part (proportional to $\nabla^2 T_a$ in the unmagnetized case) and the frictional part due to differences in
 drifts. Similarly to the heat flux vectors ($\sim\nabla T_a$), the scalar perturbations $\widetilde{X}_a^{(4)}$ can have
 either positive or negative values. Interestingly, as was shown in Part 1, 
 these scalars directly modify the energy exchange rates between species. 
 The scalar perturbations $\widetilde{X}_a^{(4)}$, together with the Chapman-Cowling integrals, are also considered
 in the models of \cite{Laguna2022}, \cite{Laguna2023},
who focus at the (1-Hermite) description for the electron species and by neglecting the viscosities, they consider 9-moment models. 
 We here focus only at the classical Boltzmann operator, whereas the last two references also consider generalized Boltzmann
operators, such as the Wang Chan-Uhlenbeck operator, allowing them to account for the inelastic collisions and the ionization processes.
The scalar perturbations are also considered with the so-called ``maximum entropy closures'', see for example
\cite{Levermore1996,Groth2009,Torrilhon2010,McDonaldTorrilhon2013,Bocelli2023,Bocelli2024} and references therein, where the
expansion of the distribution function is done differently to prevent the negativity of the $f_a$ (see our limitations Section \ref{sec:NEG}), but
where only the heuristic BGK (relaxation-type) collisional operator seems to be employable with this method.\\

The scalar perturbations of the 4th-order fluid moment are also frequently considered in the anisotropic fluid models
expanded around the bi-Maxwellian distribution function, to model the collisionless Landau damping phenomenon more precisely than it is possible with the heat flux
closures, see for example
\cite{HammettPerkins1990,Snyder1997,Snyder2001,Goswami2005,PassotSulem2007,PSH2012,SulemPassot2015,JosephDimits2016,HunanaPRL2018} and references therein.
In the general 3D geometry, these models contain two distinct scalar pressures $p_{\parallel a}, p_{\perp a}$ (temperatures $T_{\parallel a}, T_{\perp a}$)
along and across the magnetic field lines, and the 4th-order fluid moment contains three distinct scalar perturbations
$\widetilde{X}^{(4)}_{\parallel\parallel a}$, $\widetilde{X}^{(4)}_{\parallel\perp a}$ and $\widetilde{X}^{(4)}_{\perp\perp a}$ 
(often denoted as $\widetilde{r}_{\parallel\parallel a}$, $\widetilde{r}_{\parallel\perp a}$ and $\widetilde{r}_{\perp\perp a}$).
Importantly, the presence of temperature anisotropy in these models
allows one to consider effects which are encountered in kinetic plasma simulations of the Vlasov equation,
such as the firehose and mirror instabilities.     
The anisotropic temperatures were introduced into the fluid framework by  
Chew, Goldberger and Low \citep{Chew1956} (which is often abbreviated as CGL) and their model is also known as a ``collisionless MHD''.
 For an introductory guide to the CGL-type fluid models with anisotropic temperatures
 and Landau fluid models, see the two volume lecture notes of \cite{Hunana2019b,Hunana2019a}.
%===
\footnote{We note that the lecture notes of \cite{Hunana2019b,Hunana2019a}, as well as the last paragraph of \cite{HunanaPRL2018},
  contain an incorrect interpretation that Landau fluid closures are required to go beyond the 4th-order fluid moment in the hierarchy of moments.
   In reality, the much simpler Hermite closures (that we use right here) can be used as well.
    This was already addressed in \cite{Hunana2022}, see Section 8.6 ``Hermite closures'', together with Appendices B.8 and B.9.} 
%===
  With anisotropic temperatures, the collisional contributions become very complicated even without Landau damping, see for example
 \cite{ChoduraPohl1971,DemarsSchunk1979} and \cite{BarakatSchunk1982}.
We note that it is also possible to account for the Landau damping effect in the isotropic MHD-type framework considered here,
 see for example \cite{ChangCallen1992-1,ChangCallen1992-2,JiHeldJhang2013}, where in addition to the non-local parallel heat flux, one also obtains
non-local expressions for the parallel stress-tensor $\bPi^{(2)}_a:\bhat\bhat$ (see eqs. (12)-(13) in the last reference).  
Recently, even fluid closures which capture the cyclotron resonances and the associated cyclotron damping in the fluid framework
have been investigated by \cite{JikeiAmano2021,JikeiAmano2022}, see also the new development of \cite{ParkJi2024}.
Further discussion about the bi-Maxwellian expansions can be found in our limitations Section \ref{sec:MAX}.\\

As in Part 1, we here focus on the ``classical'' isotropic fluid models, with expansions around the Maxwellian distribution and without Landau damping.
 Of course, because the distribution function is expanded, these models still do contain anisotropic temperature fluctuations, 
  which can be shown easily by simply projecting the pressure tensor
  $\bp_a=p_a\bI+\bPi_a^{(2)}$ into $p_{\parallel a} = p_a + \bPi_a^{(2)}:\bhat\bhat$ and $p_{\perp a} = p_a - \bPi_a^{(2)}:\bhat\bhat/2$,
  meaning that the anisotropy
  $p_{\parallel a}-p_{\perp a}=(3/2)\bPi_{a}^{(2)}:\bhat\bhat$ actually represents the parallel viscosity. 
Our 21- and 22-moment models considered here are the closest to the 13-moment models
of \cite{Burgers1969,Schunk1975,Schunk1977} and \cite{SchunkNagy2009} and our models have the same properties as those models have:
1) All of the considered fluid moments are described by their own evolution equations, with fully nonlinear left-hand-sides.
2) Our models are formulated as multi-fluids and are valid for arbitrary temperatures $T_a$ \& $T_b$ and masses $m_a$ \& $m_b$ of all species.
3) Our models use the usual ``modern'' notation/definitions, where the fluid moments for species ``a'' are defined with respect to
the drift/bulk velocity $\bu_a$ of species ``a'', so that the random/fluctuating velocity is $\bc_a=\bV_a-\bu_a$. This is in contrast to
the ``older'' notation used for example in the early works of Chapman, Cowling, Enskog, Burnett and also in the model
of \cite{Zhdanov2002}, where the fluid moments are
defined with respect to the average velocity of all of the species $\langle \bu \rangle =\sum_a \rho_a \bu_a /\sum_a \rho_a $,
so that the random velocity is defined as $\bc_a=\bV_a - \langle \bu \rangle$
(and one defines the drift velocity for each species $\boldsymbol{w}_a=\bu_a-\langle \bu \rangle$).
As discussed already by Grad, the ``modern'' formulation becomes the natural choice if the differences in drifts become significant,
where it is more likely that the distribution function $f_a$ will become Maxwellian
with respect to its own velocity $\bu_a$ and not the average velocity $\langle \bu \rangle$, and this might be
further amplified when large temperature differences are considered. And finally the last property
4) The classical Boltzmann operator is considered and fluid models are expressed through
the ``Chapman-Cowling collisional integrals'' $\Omega_{ab}^{(l,j)}$
(defined by (\ref{eq:CCom0}) or (\ref{eq:CCo})),
where one integrates the differential cross-section $\sigma_{ab}(g,\theta)$ over both the scattering angle $\theta$
and the relative velocity $g$. 
 Our 21- and 22-moment models thus can be best viewed as a
generalized models of \cite{Burgers1969} and \cite{Schunk1975,Schunk1977}, where the only difference is that our models are developed to
higher orders in the fluid hierarchy, so that the 21-moment model matches the precision of the \cite{Braginskii1965} model. Freely speaking,
the 13-moment models can be viewed as ``1-Hermite'' (because one Hermite polynomial is used for the stress-tensors and heat fluxes),
whereas our 21-moment model can be viewed as ``2-Hermite'' (because two Hermite polynomials are used for the stress-tensors and heat fluxes).
Our 22-moment model is a 2-Hermite/1-Hermite hybrid, because the fully contracted scalars are described by only 1-Hermite polynomial.
 The possible improvement by the 23-moment model is discussed in our limitations Section \ref{sec:23mom}.\\  

%\newpage
Importantly, by employing the Boltzmann operator, the 21-moment model has been expressed through the Chapman-Cowling integrals
(for arbitrary temperatures and masses) also in the recent work of 
\cite{Raghunathan2021,Raghunathan2022b,Raghunathan2022a}. Unfortunately we did not verify equivalence, because their model is
formulated as a generalized model of \cite{Zhdanov2002} and as already noted in Part 1, we are puzzled by the
notation in the Zhdanov's model. (We did not verify equivalence for the case of Coulomb collisions with small temperature differences
of the ion species - that the Zhdanov's model considers - in Part 1 either.) We will try to verify equivalence or clarify the possible
differences with the above references in the near future.
With the Landau operator, collisional integrals for arbitrary temperatures and masses
(and even arbitrarily high-order N-Hermite expansions in the hierarchy of moments) are considered in the various papers of  
\cite{JiHeld2006,JiHeld2008,Ji2023} and references therein,
but we find their work to be quite difficult to follow and we were unable to verify equivalence with their general expressions for arbitrary
temperatures and masses either.
In Part 1, we have only verified equivalence with \cite{JiHeld2013} for the particular (2-Hermite) case of the Braginskii model of one ion-electron
plasma with small temperature differences, 
see for example our analytic electron Braginskii coefficients (56)-(60) there, which can be shown to be equivalent to the 2-Hermite formulation of
\cite{JiHeld2013} with their collisional matrices. The parallel (unmagnetized) electron coefficients are also identical to \cite{Simakov2014}  
and we were able to show analytic match with  \cite{Balescu1988} for these parallel coefficients as well (for the case of one ion species).
 Unfortunately, we were unable to establish analytic
match with the magnetized 21-moment transport coefficients of \cite{Balescu1988}, 
which is due to his rather ``obscure'' formulation for his final model, because from the perspective of the moment method of Grad,
if there is a match for the parallel (unmagnetized) coefficients, the collisional integrals on the right-hand-side of evolution equations were calculated correctly
and subsequently, there should be a match for the magnetized coefficients as well.
Here in Part 2, we additionally show equivalence with \cite{JiHeld2013} for the case of ``improved'' Braginskii ion species, where the
ion-electron collisions are retained (see Section \ref{sec:heavy}). 
Multi-fluid models for unmagnetized plasmas (with small temperature differences for the ion species) 
were considered also by \cite{Simakov2016a,Simakov2016b}, but we did not verify equivalence with their description either.\\

We note that even though our general evolution equations (both in Part 1 and Part 2) are valid for arbitrary temperatures $T_a$ and $T_b$
(and can be easily solved for a particular
case of interest, even if the temperature differences are vast), we prefer to write down quasi-static/highly-collisional solutions
only for the case with $T_a\simeq T_b$. The reason is that because when the ion temperature vastly exceeds the electron temperature,
expansions with mass-ratios will eventually break down. Here in Part 2, the situation is even more complicated, because one introduces the 
Chapman-Cowling integrals (whose values are technically undetermined for the general collisional case)
and the mass-ratio expansions with arbitrary temperatures might break down even easier than before.
Because we consider only the 2-Hermite approximation and impose the $T_a\simeq T_b$, the mass-ratio expansions in presence of
Chapman-Cowling integrals ``can still be kept under control'', by being guided by the Coulomb collisions and also
by the hard spheres. However, one can easily envision that for higher order N-Hermite schemes,
expansions with mass-ratios might become impossible in presence of general Chapman-Cowling integrals.
This might be especially true for the case of the heavyweight particles colliding with much lighter particles
(e.g. the ion-electron collisions), where the mass-ratios such as $\sqrt{m_e/m_i}=0.023$ are simply not large enough, because these mass-ratios
are multiplied by another large numbers coming from the collisional operators
(see later Section \ref{sec:heavy}).\\

For the case of the Landau collisional operator, very interesting discussions about the convergence of the transport coefficients
with high-order schemes can be found in \cite{JiHeld2013,Davies2021,Sadler2021,Simakov2022} and references therein.
In a recent study, \cite{Ji2023} also considers the fully contracted scalars (with up to 32 polynomials).
Additionally, it seems that unrestricted drifts with the Landau operator were considered by \cite{JiHeld2020}, see also \cite{Pfefferle2017},
and with the Boltzmann operator (by focusing at the Coulomb collisions) by \cite{JiHeld2021}.
The last reference shows a very interesting effect
that while for small drifts the Landau and Boltzmann operators yield the same results (for the Coulomb collisions),
for sufficiently large drifts the results start to differ and the simplified Landau operator becomes imprecise (see their Figures 1-4).
The differences are further amplified when the Coulomb logarithm is not sufficiently large.
 For the 5-moment models, unrestricted drifts with the Boltzmann operator (for a general collisional process)
  were considered by \cite{Draine1986}, see our Appendix \ref{sec:Unrestricted}, equations (\ref{eq:DraineRab3})-(\ref{eq:DraineQab3}).

%================================================================================
\subsection{How is our model formulated (``menu'' of our Collisional forces)}
  It is useful to summarize how to use our model, even if some definitions will be repeated in the next section.
   Each species ``a'' are described by their own evolution equations, which naturally have collisionless left-hand-sides,
   and collisional right-hand-sides, in the same fashion as the usual Boltzmann equation (\ref{eq:Boltz0}) is written. For the collisionless left-hand-sides, we
   consider the force of gravity and the Lorentz force and one can choose the desired level of complexity in 3 levels (or anything in between). By either choosing 
 the fully nonlinear evolution equations given in Section \ref{sec:EvolutionEQ}; or the semi-linear approximation given in Section \ref{sec:EvolutionEQ2} (where the
 coupling between stress-tensors and heat fluxes is retained); or one can select the simplest case of fully decoupled equations given in Section  \ref{sec:EvolutionEQ3}. 
 It might sound surprising that stress-tensors and heat fluxes are described by their own evolution equations, 
 but with the moment method of Grad, this is just a natural consequence of taking the Boltzmann equation, and integrating it without neglecting the time derivative
 of the distribution function.

 The collisional right-hand-sides of these equations are given in Section \ref{sec:Main} and are expressed through the
 Chapman-Cowling integrals
 \begin{eqnarray}
   \Omega_{ab}^{(l,j)} &\equiv&
  \frac{1}{2\sqrt{\pi}} \Big( \frac{1}{\beta_{ab}} \Big)^{2j+3} \int_0^\infty e^{-\frac{g^2}{\beta_{ab}^2}} g^{2j+3}
  \Big[  2\pi \int_0^{b_0^\textrm{max}} (1-\cos^l \theta) b_0 db_0 \Big] dg, \label{eq:CCom0}
 \end{eqnarray}
      where the product of two Maxwellians $f_a^{(0)}f_b^{(0)}$ is represented by its reduced expression $\exp(-g^2/\beta_{ab}^2)$ with
       parameter $\beta^2_{ab} = v_{\textrm{th} a}^2 + v_{\textrm{th} b}^2$ (see Appendix \ref{eq:AppendixAC}), and this expression is  
       integrated over all (positive) impact parameters $b_0$ and all relative velocities $\bg_{ab}=\bV_a-\bV_b$
       (where we dropped the species indices for $\bg_{ab}$ and $g=|\bg_{ab}|$). The definition (\ref{eq:CCom0}) is typically written with the maximum impact parameter
 $b_0^\textrm{max}=\infty$, but we find it useful to emphasize that for the Coulomb collisions one introduces cut-off at the Debye length $b_0^\textrm{max}=\lambda_D$,
and for the hard spheres the integral is calculated with $b_0^\textrm{max}=r_a+r_b$ (the sum of the sphere radii).
The indices ``$l$'' and ``$j$'' are integers, starting with the $l=1$ and $j=1$. 
A particular collisional process is given by prescribing the relation between the scattering angle $\theta$ and the impact parameter $b_0$. 
For example, for the hard spheres $\cos\theta = 2\hat{b}_0^2-1$, where the normalized impact parameter (with hat) $\hat{b}_0=b_0 /(r_a+r_b)$, 
and by simply using this expression, one can directly calculate the $\Omega_{ab}^{(l,j)}$ of hard spheres for any ``l'' and ``j''.  
We prefer to write the main definition (\ref{eq:CCom0}) with integration over
the impact parameter $b_0$ instead of integrating over the differential cross-section $\sigma_{ab}(g,\theta)$ with $d\theta$, because except of few very special cases,
the differential cross-section is never derived anyway, and one directly integrates over the $d b_0$ instead. Additionally, for attractive forces integration
over the $d \theta$ can be very confusing (see Section \ref{eq:AttractNU3}).  

The idea behind the definition (\ref{eq:CCom0}) might perhaps look blurry at first, but it just represents various integrals that one will encounter,
  when developing a fluid model. For example, the lowest-order integral  $\Omega_{ab}^{(1,1)}$ (i.e. with indices $l=1$ and $j=1$) defines the collisional frequencies
  \begin{equation}
  \nu_{ab} = \frac{16 \mu_{ab}}{3 m_a } n_b \Omega^{(1,1)}_{ab},
  \end{equation}
where $\mu_{ab}=m_a m_b/(m_a+m_b)$ is the reduced mass. The collisonal frequencies $\nu_{ab}$ for our forces are summarized in eq. (\ref{eq:Rsimple2x}).
For all higher-order ``l'' and ``j'' Chapman-Cowling integrals, we find it the best to simply normalize them with respect to the lowest-order integral
$\Omega_{ab}^{(1,1)}$, where we introduce notation
\begin{equation}
\Omega_{l,j} \equiv \frac{\Omega_{ab}^{(l,j)}}{\Omega_{ab}^{(1,1)}}, \label{eq:0}
\end{equation}
and the results of (\ref{eq:0}) are just pure (real, positive, dimensionless) numbers.
For the simple collisional forces considered here, these ratios do not depend on the temperature, and for example
\begin{eqnarray}
  \textrm{Coulomb collisions}\, (\ln\Lambda\gg 1): \qquad \Omega_{1,2} &=& 1; \qquad \Omega_{1,3}=2; \qquad \Omega_{2,2}=2; \qquad
  \Omega_{2,3} = 4;\qquad \Omega_{2,4} = 12;\label{eq:fun3}\\  
  \textrm{Hard spheres}: \qquad \Omega_{1,2} &=& 3; \qquad \Omega_{1,3}=12; \qquad \Omega_{2,2}=2; \qquad \Omega_{2,3} = 8; \qquad \Omega_{2,4} = 40.  \label{eq:fun1}
\end{eqnarray}

In the definition (\ref{eq:0}), we have eliminated the species indices ``ab'' (which the reader should put back anytime he/she misses them)
and we have moved the indices $l,j$ down, so that it is easy to write powers, where for example the $\Omega_{1,2}^2$ means $\Omega_{1,2}$ to the power of two.
The powers of the $\Omega_{l,j}$ are not needed for the collisional right-hand-sides of the evolution equations given in Section \ref{sec:Main}
(they are calculated in the semi-linear approximation),
but the powers are needed after one cancels the time-derivatives $d/dt$ and obtains the results in the quasi-static/highly-collisional
approximation. In later Sections, we will need to write down a lot of the $\Omega_{l,j}$ ratios and
because we consider only models where none of the indices ``l'' or ``j'' reach
the value of 10, we will further simplify the notation by removing the comma and we only write $\Omega_{lj}$. But for clarity, we keep the comma for now.
For the multi-fluid 13-moment models of \cite{Burgers1969} and \cite{Schunk1975,Schunk1977}, only four of these ratios are needed,
the $\Omega_{1,2}$; $\Omega_{1,3}$; $\Omega_{2,2}$ and $\Omega_{2,3}$.
After generalizing the \cite{Braginskii1965} model with the Chapman-Cowling integrals, it can be shown that
his ion viscosity and ion heat flux, as well as his electron viscosity, require just one more ratio, the $\Omega_{2,4}$.
The generalized Braginskii electron heat flux requires two more of these ratios, the $\Omega_{1,4}$ and $\Omega_{1,5}$.
Our multi-fluid 21- and 22- moment models for arbitrary masses and small temperature differences also require $\Omega_{3,3}$ and for arbitrary temperatures also the
$\Omega_{2,5}$, $\Omega_{3,4}$ and $\Omega_{3,5}$.

 We find it useful to summarize all collisional forces that we consider right here, for a general ``l'' and ``j''
  (which represents a ``menu'' of our collisional forces), given by
\begin{eqnarray}
  \textrm{Coulomb collisions}\, (\ln\Lambda\gg 1): \qquad \Omega_{l,j} &=&  (j-1)! \,l\,; \label{eq:Coulomb0}\\         
  \textrm{Coulomb collisions}: \qquad \Omega_{l,j} &=&  (j-1)! \frac{A_l(2)}{A_1(2)}\quad \textrm{where} \quad
  A_l(2) \equiv \int_0^{\Lambda} \Big[1-   \Big( \frac{x^2-1}{x^2+1} \Big)^l \,\Big] x dx; \label{eq:Coulomb1}\\
 \textrm{Hard spheres}:  \qquad   \Omega_{l,j} &=& (j+1)! \Big[\frac{1}{2} -\frac{1+(-1)^l}{4(l+1)} \Big] = \Bigg\{ \begin{array}{lr}
    (j+1)!/2;    &  l= \textrm{odd};\\
    (j+1)! l/[2(l+1)];  & \quad l= \textrm{even};
 \end{array}  \label{eq:Hard0}\\
  \textrm{Inverse power force}\, \pm 1/r^\nu: \qquad  \Omega_{l,j}
  &=&   \frac{\Gamma\big( j+2 - \frac{2}{\nu-1}\big)}{\Gamma\big( 3 - \frac{2}{\nu-1}\big)} \frac{A_l(\nu)}{A_1(\nu)}; \label{eq:Inv0}\\
   \textrm{Maxwell molecules}\, \pm 1/r^5: \qquad  \Omega_{l,j}
  &=&   \frac{1}{2^{j-1}} \frac{(2j+1)!!}{3} \frac{A_l(5)}{A_1(5)}. \label{eq:Inv0M}
\end{eqnarray}
     For the Coulomb collisions ($\nu=2$) in moderatelly-coupled plasmas (\ref{eq:Coulomb1}), the three required integrals
      $A_1(2)$, $A_2(2)$ and $A_3(2)$ that represent the corrections of the Coulomb logarithm are evaluated in eqs. (\ref{eq:A12x})-(\ref{eq:A23x}).
      The $\Gamma(x)$ in (\ref{eq:Inv0}) is the usual Gamma function.
      For all other forces, the required $\Omega_{l,j}$ are evaluated in eqs. (\ref{eq:beauty17x})-(\ref{eq:OmegaGI}). 
       Note that the repulsive and attractive forces in
      (\ref{eq:Inv0}) and (\ref{eq:Inv0M}) have the same form of $\Omega_{l,j}$, however, the difference is in the numerical integrals $A_l(\nu)$, 
       where for the repulsive forces the $A_l(\nu)$ numbers are given in Table \ref{Table:Anu},
       and for the attractive forces with rigid core in Table \ref{Table:Anu2}. \\

 Our general collisional contributions given in  Section \ref{sec:Main} may look complicated at first sight, but the coefficients that we
    freely call ``mass ratio coefficients'' only contain masses, temperatures and Chapman-Cowling integrals $\Omega_{l,j}$.
The great complexity of the model is caused by allowing each species to have arbitrary temperatures.  
When the temperature differences between species are small, the model drastically simplifies, see Section \ref{sec:Tsmall}.
Perhaps we could have moved the arbitrary temperatures into an Appendix, but we wanted to retain the structure of Part 1 as close as possible,
where the most general case with arbitrary temperatures is given first, and simplified only later. \\

To summarize, specifying a particular collisional process in our models consists of two
simple steps, where one chooses the ratios $\Omega_{l,j}$ from our current (rather limited, but not small) ``menu'' of collisional forces given by
(\ref{eq:Coulomb0})-(\ref{eq:Inv0M}) and pairs the choice with the collisional frequencies $\nu_{ab}$ given by (\ref{eq:Rsimple2x}).
Then, one can either numerically simulate these evolution equations, or one can find the solution for the stress-tensors and heat fluxes in
the highly-collisional/quasi-static approximation (by canceling the time-derivatives for these quantities).
We discuss few quasi-static solutions in Sections \ref{sec:ions} - \ref{sec:Kurtosis}, but we only
focus at the Braginskii case when only one collisional interaction is present, i.e. quasi-static solutions with two different
collisional interactions such as the ion-ion and ion-neutral collisions are not presented, and these will be discussed elsewhere.
The only small exception is Appendix \ref{sec:AppIN}, where the simpler 13-moment model is considered, just to show that the
  stress-tensor and heat flux of neutral particles become magnetized, if ions are present.

%===========================================================================================================================
%==========================================================================================================================

%\newpage
\subsection{Comparison of hard spheres and Coulomb collisions} \label{sec:HardSphereP}
            Throughout the text, we often compare the hard spheres with the \cite{Braginskii1965} model of Coulomb collisions, which we find useful
             to summarize right here. 
  For example, considering a simple gas consisting of only one species 
(or equivalently, when only the self-collisions ``a-a'' are retained and collisions with other particles are neglected),
the parallel viscosity $\eta_0^a$ and thermal conductivity $\kappa_\parallel^a$ can be compared as
\begin{eqnarray}
  \textrm{Coulomb collisions}: \qquad  \eta_0^a &=& \underbrace{\frac{1025}{1068}}_{0.960} \frac{p_a}{\nu_{aa}}; \qquad
  \kappa_\parallel^a = \underbrace{\frac{125}{32}}_{3.906} \frac{p_a}{\nu_{aa} m_a}; \nn\\  
  \textrm{Hard spheres}: \qquad \eta_0^a  &=& \underbrace{\frac{1025}{1212}}_{0.846} \frac{p_a}{\nu_{aa}}; \qquad
  \kappa_\parallel^a  = \underbrace{\frac{1125}{352}}_{3.196} \frac{p_a}{\nu_{aa} m_a}, \label{eq:NOW1}
\end{eqnarray}
where the values $0.96$ and $3.906$ are the famous Braginskii parallel coefficients for the ion species. Interestingly, all the values above 
can be deduced from the work of \cite{ChapmanCowling1953} (first publication was in 1939) in a fully analytic form and
correspond to their ``second approximation''
(p. 169 and 173, or see our eq. (\ref{eq:CC0}) and (\ref{eq:CC00})).
The comparison between the Coulomb collisions and hard spheres becomes more interesting by considering a population
of lightweight spheres, which in addition to self-collisions also collide with a population of much heavier spheres
(i.e. analogously to the Braginskii electron species), where the parallel viscosities are given by (\ref{eq:wow6})-(\ref{eq:wow7})
and the thermal conductivities by (\ref{eq:BragS10})-(\ref{eq:sleep3}). 
The comparison becomes even more interesting when the situation is reversed, now considering a population of very heavy spheres,
which in addition to self-collisions also collide with a population of much lighter spheres. This can be viewed as an ``improved'' Braginskii ion species,
where the ion-electron collisions are retained, and the parallel viscosities are given by (\ref{eq:wow33})-(\ref{eq:wow34}) and the thermal conductivities by
(\ref{eq:beau35HHHH})-(\ref{eq:beau35HHHHx}) (see the entertaining similarities of numerical factors).
It might sound surprising that collisions with particles that are 1836 times (or more) lighter can have any
significant effect on the ion viscosity and thermal conductivity, but the mass-ratios enter only as $\sqrt{m_e/m_i}$ and are multiplied by quite large numbers.
As an example, considering the proton-electron plasma, the Braginskii ion viscosity value 0.96 changes into 0.892 and the ion
thermal conductivity value 3.906 changes into 3.302,
which are quite significant differences of 8\% and 18\% (when divided by the smaller value).
The same values were also obtained in Part 1 in a more precise way without considering any exansions in mass-ratios (see eqs. (214) and (217) there). 
For the fully magnetized case, the ``improved Braginskii ion stress-tensor'' is then given by viscosities (\ref{eq:beau35HHH}) and
the ``improved Braginskii ion heat flux'' by thermal conductivities (\ref{eq:wow38}). For the magnetized proton-electron plasma
the differences then reach up to 43\%, which is obtained for the ion cross-conductivity
$\kappa_\times^i$ in the limit of weak magnetic field (where the $\kappa_\times^i$ is small). 
For the Coulomb collisions, such an improvement of the Braginskii model by retaining the ion-electron collisions
was considered before by \cite{JiHeld2013}, and in Section \ref{sec:JiHeld2013}
we show that our Coulomb results are equivalent to theirs (we only discuss quasi-static solutions with small temperature differences).
Interestingly, the same effect can be shown for the case of heavyweight hard spheres, where for example accounting for collisions with 
1836 times lighter spheres (that have the same number density and radius), yields in eq. (\ref{eq:NOW1})
the parallel viscosity value 0.800 and the thermal conductivity value 2.830, representing
differences of 6\% and 13\%.\\

 Because we already had the formulation of the entire Braginskii model through the Chapman-Cowling integrals,
  it felt slightly boring to only compare unmagnetized solutions of hard spheres and Coulomb collisions, and we wanted to somehow compare also the
  magnetized case. This is of course difficult to do, because one should consider proper coupling between neutral particles and charged particles,
  where the stress-tensors and heat fluxes of neutral particles become magnetized, such as briefly presented in Appendix \ref{sec:AppIN} for the
  simpler 13-moment model. To avoid this complexity, we probably went a bit too far and came up with an
  abstract idea/concept of generalized ``hard spheres'', which during the collisional
  encounter collide as hard spheres, but which otherwise feel the magnetic field (i.e. they have a non-zero cyclotron frequency). 
  Such a concept is difficult to justify, and any magnetized solutions marked as ``hard spheres'' should be viewed only as an academic curiosity, where only the
  parallel (unmagnetized) part of that solution is physically fully meaningful. We also use the same abstract idea for the collisions with
  an inverse power-law forces $\pm 1/r^\nu$, where during the collisional encounter this force is used, but otherwise the particles feel the
  magnetic field through the usual Lorentz force (which is present at the left-hand-side of the Boltzmann equation). Again, this is inconsistent,
  and one should consider neutral particles (with zero cyclotron frequency) that interact with charged particles. 
  As a consequence, in our quasi-static solutions only the parallel part is valid for any $\nu$ and the magnetized parts are valid only for $\nu=2$,
  nevertheless, these solutions are useful for double-checking the algebra and for deriving the generalization of the Braginskii model for moderately-coupled plasmas. 
  To summarize, our multi-fluid models as given by their evolution equations are formulated correctly, it is only the quasi-static solutions
  that we present in Sections \ref{sec:ions} - \ref{sec:Kurtosis} that are simplified too much in the magnetized case.

%================================================================================================
%================================================================================================
\subsection{Organization of the paper}

  In Section \ref{sec:TechIntro}, we state all of the required definitions, together with the evolution equations for the 22-moment model
  (the evolution equations are the same as in Section 7 of Part 1).

Section \ref{sec:Main} represents our main and most important section, where the general collisional contributions for the 22-moment model are
expressed through the Chapman-Cowling integrals, for arbitrary temperatures and masses. We also present a simplified model, where the
differences in temperatures between species are small. The collisional contributions might appear as slightly long (and perhaps boring) when seen for the first time,
 however, they represent our main results and the rest of the paper can be viewed only as application of these results
 for the particular simplified cases, serving as a verification tool for the results of Section \ref{sec:Main}. 
In Section \ref{sec:Particular}, we evaluate the collisional contributions for the particular cases of hard spheres and Coulomb collisions.
 
In Sections \ref{sec:ions}-\ref{sec:heavy}, we cancel the time-derivatives in the evolution equations for the stress-tensors and heat fluxes
and we discuss quasi-static solutions that are analogous to the Braginskii model, where
Section \ref{sec:ions} can be viewed as ``Braginskii ion species'', Section \ref{sec:electrons} as ``Braginskii electron species'' and
Section \ref{sec:heavy} as an ``improved Braginskii ion species'' (where the ion-electron collisions are retained).  
In Section \ref{sec:Kurtosis}, we discuss quasi-static solutions for the scalar $\widetilde{X}^{(4)}_a$.

In Section \ref{sec:Discussion}, we discuss various topics, such as 1) constants $A_l(\nu)$ for repulsive and attractive forces obtained by the numerical integration;
2) repulsive and attractive cube force $1/r^3$; 3) Maxwell molecules with force $1/r^5$ (where the attractive case represents ion-neutral collisions);
and we also discuss our limitations consisting of 4) the ideal equation of state; 5) possible improvement by the 23-moment model;
6) other forces/potentials that we did not consider;
7) the structure of the Braginskii model in the anisotropic (CGL) framework;
8) the negativity of the distribution function; and in Conclusions we also list numerical codes where the implementation of our models might be useful.

Appendix \ref{sec:AppendixA} represents our simple introduction into the Boltzmann operator, where we discuss the Coulomb logarithm and
also that the operator requires two distinct center-of-mass transformations, which we call ``simple'' and ``more advanced''.
In Appendix \ref{sec:Mom}, we calculate the momentum exchange rates (with unrestricted drifts) for the simple 5-moment models of
   hard spheres, Coulomb collisions and Maxwell molecules, and in Appendix \ref{sec:Energy}, we calculate the energy exchange rates $Q_{ab}$.
In Appendix \ref{sec:HardSpheresV}, we calculate the viscosity of hard spheres in the 1-Hermite approximation.
In Appendix \ref{sec:Integrals}, we discuss the integrals of the 22-moment model for a general collisional process,
which leads to a fully nonlinear system (\ref{eq:beau0})-(\ref{eq:beau7}).
We then discuss a recipe how these integrals are simplified in the semi-linear approximation and evaluated with an analytic software.
 In Appendix \ref{sec:Examples}, we show how to calculate some collisional integrals of Appendix \ref{sec:Integrals} by hand, where 
 we calculate the simplest (1-Hermite) self-collisional viscosity and thermal conductivity. 
 Also, in Appendix \ref{sec:Unrestricted}, we consider the 5-moment models with unrestricted drifts for a general collisional
 process.

\newpage
\section{Definitions and evolution equations} \label{sec:TechIntro}
Here we state few definitions and we also introduce the Boltzmann operator, together with the Chapman-Cowling integrals. 
%================================================
\subsection{Definition of fluid moments} \label{Sec:DefMoments}
As in Part 1, we use the traceless ``viscosity tensors'' and ``heat flux vectors''
\begin{eqnarray}
  \bPi^{(2)}_a &=& m_a \int \big(\bc_a\bc_a-\frac{\bI}{3}|\bc_a|^2\big) f_a d^3v_a; \qquad
  \bPi^{(4)}_a  = m_a \int \big(\bc_a\bc_a-\frac{\bI}{3}|\bc_a|^2\big)|\bc_a|^2 f_a d^3v_a; \nn\\    %\label{eq:Num31} 
\vecX^{(3)}_a &=& m_a \int \bc_a|\bc_a|^2 f_a d^3v_a = 2\vecq_a; \qquad \vecX^{(5)}_a = m_a \int \bc_a |\bc_a|^4 f_a d^3v_a, \label{eq:Num30}
\end{eqnarray}
together with the scalar
\begin{equation}
  \widetilde{X}^{(4)}_a =  m_a \int |\bc_a|^{4} (f_a -f_a^{(0)}) d^3v_a. \label{eq:Num30a}
\end{equation}
By using these fluid moments, the perturbation of the distribution function $f_a=f_0^{(0)}(1+\chi_a)$
for the 22-moment model then reads
\begin{eqnarray}
\chi_a &=& \chi_a^{\textrm{(visc)}} + \chi_a^{\textrm{(heat)}} + \chi_a^{\textrm{(scalar)}};\nn\\
\chi_a^{\textrm{(visc)}} &=& \frac{1}{2p_a} \big(\bPi_a^{(2)}:\tbc_a\tbc_a\big) +\frac{1}{28}\Big[
    \frac{\rho_a}{p_a^2} \bPi^{(4)}_a   -\frac{7}{p_a} \bPi_a^{(2)}\Big]:\tbc_a\tbc_a (\tc_a^2-7); \nn\\
%===
\chi_a^{\textrm{(heat)}} &=&  \frac{1}{5 p_a}\sqrt{\frac{m_a}{T_a}} (\vecq_a\cdot\tbc_a)(\tc_a^2-5) + \frac{1}{280 p_a}\sqrt{\frac{m_a}{T_a}}
\Big[\frac{\rho_a}{p_a} \vecX^{(5)}_a  -28 \vecq_a \Big]\cdot\tbc_a (\tc_a^4-14\tc_a^2+35); \nn\\
%===
\chi_a^{\textrm{(scalar)}} &=& \frac{1}{120}\frac{\rho_a}{p_a^2} \widetilde{X}^{(4)}_a (\tc_a^4-10\tc_a^2+15), \label{eq:beau2x}
\end{eqnarray}
with the normalized velocity (with tilde) $\tbc_a = \sqrt{m_a/T_a}\bc_a$. For a detailed discussion on how the expansions in Hermite polynomials
are performed, see Appendix B of \cite{Hunana2022}, see also Appendix of \cite{Balescu1988}.
It is noted that the only difference between the reducible and irreducible Hermite polynomials is
how these polynomials are initially defined/obtained, but up to a placement of the normalization constants, both polynomials are identical
and both polynomials yield the same perturbation (\ref{eq:beau2x}).   
The irreducible Hermite polynomials are defined through the Laguerre-Sonine polynomials (i.e. the same polynomials
that are used with the Chapman-Enskog method), while the reducible Hermite polynomials are obtained from their tensorial definition
(see also the summarizing Section 8.4, page 33 in Part 1). We prefer to work with the reducible Hermite polynomials, where no reference to
the Laguerre-Sonine polynomials has to be made and they feel as the ``natural'' choice when tensors beyond matrices are considered. 
The usefulness of the reducible Hermite polynomials can be further emphasized by considering Hermite 
expansions around an anisotropic bi-Maxwellian distribution function, which we do not discuss here (see also Section \ref{sec:MAX}, where
the possible extension of the Braginskii model to bi-Maxwellian CGL-type plasmas is briefly discussed).

%================================================
\subsection{Definition of collisional contributions}
By considering a general (for now unspecified) collisional
operator $C(f_a)=\sum_b C_{ab}(f_a,f_b)$, one defines the (tensorial) collisional contributions
\begin{eqnarray}
  && \boldsymbol{R}_a = m_a\int \bV_a C(f_a) d^3v_a; \qquad Q_a = \frac{m_a}{2}\int|\bc_a|^2 C(f_a)d^3v_a;\nn\\
  && \bQ^{(2)}_a = m_a \int \bc_a\bc_a C(f_a) d^3v_a; \qquad \bQ^{(3)}_a = m_a \int \bc_a\bc_a \bc_a C(f_a) d^3v_a; \nn\\
  && \bQ^{(4)}_a = m_a \int \bc_a\bc_a \bc_a \bc_a C(f_a) d^3v_a; \qquad \bQ^{(5)}_a = m_a \int \bc_a\bc_a \bc_a \bc_a \bc_a C(f_a) d^3v_a, \label{eq:Spec}
\end{eqnarray}
where for the last three it is useful to define simplified
\begin{eqnarray}
 \vecQ^{(3)}_a &=& \frac{1}{2} \trace \bQ^{(3)}_a = \frac{m_a}{2} \int \bc_a c_a^2 C(f_a) d^3v_a;  
 \qquad  \vecQ^{(5)}_a = \trace\trace \bQ^{(5)}_a = m_a \int \bc_a c_a^4 C(f_a) d^3v_a;\nn\\
 \bQ^{(4)*}_a &=& \trace \bQ^{(4)}_a = m_a \int \bc_a\bc_a c_a^2 C(f_a) d^3v_a; \qquad
 Q^{(4)}_a = \trace\bQ^{(4)*}_a = m_a \int c_a^4 C(f_a) d^3v_a,
\end{eqnarray}
so that we consider only vectors, matrices and scalars. The star on $\bQ^{(4)*}_a$ therefore represents trace. In contrast,
in our Appendices the star on the energy exchange rates $Q_{ab}^*$ represents a ``thermal part'' of $Q_{ab}$.

%\newpage
%============================================================================
\subsection{Introducing the Boltzmann operator}
Here we consider the well-known Boltzmann collisional operator, see e.g. the books by 
\cite{SchunkNagy2009,Burgers1969,ChapmanCowling1953}. 
For a reader who is not familiar with the Boltzmann operator, we highly recommend to
first read our brief Appendix \ref{sec:AppendixA} and come back here only later, because
the operator is introduced there with much better clarity than will be presented here.
 A reader familiar with the Chapman-Cowling integrals can skip the rest of this section, note our normalization (\ref{eq:norm})
  and continue with the evolution equations Section \ref{sec:EvolutionEQ}.

For any tensor $\bX_a$ (such as $\bc_a\bc_a$), the Boltzmann operator is integrated according to the recipe
\begin{equation}
\int \bX_a C_{ab}(f_a,f_b) d^3v_a = \iiint  g_{ab}  \sigma_{ab} (g_{ab},\theta)
f_a f_b \big[ \bX_a'-\bX_a \big] d\Omega d^3 v_a d^3 v_b, \label{eq:BoltzmannOOO}
\end{equation}
where $\bg_{ab}=\bV_a-\bV_b$ is the relative velocity with magnitude $g_{ab}=|\bg_{ab}|$.
We stop writing the species indices on $\bg_{ab}$. The primes represent quantities after
the collision and are related to the non-primed quantities before the collision by the conservation of momentum and energy.
Directly from the recipe (\ref{eq:BoltzmannOOO}), the required collisional contributions are then given by 
\begin{eqnarray}
  \boldsymbol{R}_{ab} &=& m_a 
  \iiint g \sigma_{ab}(g,\theta) f_a f_b \big[ \bV_a' - \bV_a\big] d\Omega d^3 v_a d^3 v_b;\nn\\
  %===
  Q_{ab} &=& \frac{m_a}{2} \iiint g \sigma_{ab}(g,\theta) f_a f_b \big[ c_a'^{2} -c_a^2 \big] d\Omega d^3 v_a d^3 v_b;\nn\\
  %===
 \bQ_{ab}^{(2)} &=&  m_a \iiint  g  \sigma_{ab} (g,\theta)
  f_a f_b \big[ \bc_a' \bc_a' - \bc_a \bc_a \big] d\Omega d^3 v_a d^3 v_b;\nn\\
  %=== 
  \vecQ^{(3)}_{ab} &=& \frac{m_a}{2} \iiint  g  \sigma_{ab} (g,\theta)
  f_a f_b \big[ \bc_a' c_a'^2 - \bc_a c_a^2 \big] d\Omega d^3 v_a d^3 v_b;\nn\\
  %===
  \bQ_{ab}^{(4)*} &=& m_a \iiint  g  \sigma_{ab} (g,\theta)
  f_a f_b \big[ \bc_a'\bc_a' c_a'^2 - \bc_a\bc_a c_a^2 \big] d\Omega d^3 v_a d^3 v_b;\nn\\
  %===
  Q_{ab}^{(4)} &=& m_a \iiint  g  \sigma_{ab} (g,\theta)
  f_a f_b \big[ c_a'^4 - c_a^4 \big] d\Omega d^3 v_a d^3 v_b;\nn\\
  %===
  \vecQ^{(5)}_{ab} &=& m_a \iiint  g  \sigma_{ab} (g,\theta)
  f_a f_b \big[ \bc_a' c_a'^4 - \bc_a c_a^4 \big] d\Omega d^3 v_a d^3 v_b. \label{eq:beauty10}
\end{eqnarray}
The $\sigma_{ab}(g,\theta)$ is the differential cross-section (sometimes denoted as $d\sigma/d\Omega$ instead), which specifies
the considered collisional process. Two simple examples are
\begin{eqnarray}
&&  \textrm{hard spheres:} \qquad \qquad \,\,\, \sigma_{ab}(g,\theta) = \frac{\alpha_0^2}{4}; \qquad \qquad \qquad \alpha_0= r_{ab} = r_a+r_b;\\
  &&  \textrm{Coulomb collisions:} \qquad \sigma_{ab}(g,\theta) = \frac{\alpha_0^2}{(1-\cos\theta)^2}; \qquad
  \alpha_0 = \frac{q_a q_b}{\mu_{ab} g^2}, \label{eq:Rutherford0}
\end{eqnarray}
where $r_{ab}$ is the sum of the sphere radii and $\mu_{ab}=m_a m_b/(m_a+m_b)$ is the reduced mass. The (\ref{eq:Rutherford0}) is the Rutherford
scattering cross-section and it is valid for both attractive and repulsive forces (positive and negative charges). 
The $\theta$ is the scattering angle and
the collisional contributions (\ref{eq:beauty10}) contain integral over the solid angle $d\Omega=\sin\theta d\theta d\phi$.
 Similar expressions as (\ref{eq:beauty10}), can be written by integrating over the impact parameter $b_0$, see the discussion
  in Appendix \ref{Sec:IntegratingB}, with the recipe (\ref{eq:BoltzmannOOB}).

When learning the Boltzmann operator for the first time, it is highly recommended to just use the above two cross-sections
and directly calculate at least the momentum exhange rates $\boldsymbol{R}_{ab}$ (see Appendix \ref{sec:Mom}) and the energy exchange rates $Q_{ab}$
(see Appendix \ref{sec:Energy})
for the 5-moment models, without introducing the Chapman-Cowling collisional integrals.
It is also beneficial to calculate the simplest viscosities with the 1-Hermite approximation (see Appendix \ref{sec:HardSpheresV}),  
or even consider the 2-Hermite approximation. Only after one is familiar with the Boltzmann operator,
one notices that the collisional integrals are actually very similar,
regardless if one considers the hard spheres, the Coulomb collisions or other interactions. Naturally, one starts to ask a question,
is it possible to calculate the collisional integrals (\ref{eq:beauty10}) only once, without specifying any particular collisional processes? 

Such a construction is indeed possible, by first defining the ``effective cross-sections''
of a general order ``$l$'' ($l$ is an integer, and we use the font mathbb\{Q\} to differentiate them from $Q_{ab}$), according to
\begin{eqnarray}
  \mathbb{Q}_{ab}^{(l)} (g) &=& \int \sigma_{ab} (g,\theta) (1-\cos^l \theta) d\Omega \nn\\
  &=& 2\pi \int_{\theta_{\textrm{min}}}^\pi \sigma_{ab} (g,\theta) (1-\cos^l \theta) \sin\theta d\theta; \qquad \textrm{(repulsion);}\nn\\
  &=& 2\pi \int_0^{b_0^\textrm{max}} (1-\cos^l \theta) b_0 db_0; \qquad \textrm{(attraction \& repulsion).}
  \label{eq:QabPicA} 
\end{eqnarray}
     We find it useful to differentiate between repulsive forces (where the scattering angle is positive and typically ranges from $0$ to $\pi$)
      and attractive forces (where the scattering angle is typically negative and for steeper forces than $1/r^2$ even reaches $\theta=-\infty$, 
      so it is better to integrate over the impact parameter $b_0$). 
The 5-moment models only need the first one $\mathbb{Q}_{ab}^{(1)}$, often called the ``momentum transfer cross-section''.
The 1-Hermite models (including those containing the fully contracted scalar) also
need the $\mathbb{Q}_{ab}^{(2)}$ and the multi-fluid 2-Hermite models also need the $\mathbb{Q}_{ab}^{(3)}$
(see e.g. the collisional integrals (\ref{eq:beau0})-(\ref{eq:beau7})).
We will see later that the simplified 2-Hermite case of \cite{Braginskii1958,Braginskii1965} is actually an exception, because considering
only self-collisions (for ions) and the case $m_e \ll m_b $ (for electrons) with small temperature differences, does not require the $\mathbb{Q}_{ab}^{(3)}$. 
Note that \cite{ChapmanCowling1953}, p. 157, define their effective cross-sections $\phi_{ab}^{(l)}$
without the factor of $2\pi$ and with the additional factor of $g$, so that $\mathbb{Q}_{ab}^{(l)}=(2\pi/g)\phi_{ab}^{(l)}$ holds.  For the case of hard spheres,
one immediatelly gets (with $\theta_{\textrm{min}}=0$ or $b_0^\textrm{max}=r_{ab}$)
\begin{equation}
{\textrm{Hard spheres:}} \qquad \mathbb{Q}_{ab}^{(1)} = \mathbb{Q}_{ab}^{(3)} = \pi r_{ab}^2; \qquad
  \mathbb{Q}_{ab}^{(2)} = \frac{2}{3}\pi r_{ab}^2; \qquad r_{ab}=r_a+r_b.
\end{equation}
The general $\mathbb{Q}_{ab}^{(n)}$ of hard spheres is also calculated easily and is given by (\ref{eq:HardPII}), showing that for odd ``l'' the
$\mathbb{Q}_{ab}^{(l)}$ is always equal to the geometrical cross-section $\pi r_{ab}^2$, and for even ``l'' it is slightly smaller.
Below we discuss the effective cross-sections for other particular cases.
 
%\newpage
%=======================================================================================================
%=======================================================================================================
\subsection{Effective cross-sections for particular cases}
  We consider the purely repulsive force $F(r) = + |K_{ab}| / r^\nu$, and the attractive force $F(r) = - |K_{ab}| / r^\nu$ with a ridig core repulsion
             (i.e. where the particles bounce from each other once they meet). With potentials, this can be written as
\begin{equation}
 \textrm{repulsion}: \qquad V(r)= + \frac{|K_{ab}|}{(\nu-1) r^{\nu-1}}; \qquad 
 \textrm{attraction}: \qquad V(r)= \delta(r) - \frac{|K_{ab}|}{(\nu-1) r^{\nu-1}}; \qquad \nu\ge 2. \label{eq:force}
\end{equation}
     For Coulomb collisions $\nu=2$, the delta function $\delta(r)$ does not influence the calculations at all, and the only difference is that
      now for the impact parameter $b_0=0$ an incoming electron bounces back from an ion as hard sphere, instead of going around the ion with an infinitely small loop,
      see Appendix \ref{sec:AppendixAC}.
By introducing the normalized impact parameter (with hat) $\hat{b}_0=b_0/\alpha_0$ (often denoted as $v_0$) and defining pure numbers
\begin{equation}
A_l(\nu) \equiv \int_0^{\hat{b}_0^{\textrm{max}}} (1-\cos^l \theta) \hat{b}_0 d \hat{b}_0, \label{eq:beauM1}
\end{equation}
the effective cross-sections (\ref{eq:QabPicA}) then can be written as
\begin{equation}
\begin{tabular}{lll}
  {\textrm{Hard spheres:}} & $\mathbb{Q}_{ab}^{(l)} = 2\pi \alpha_0^2 A_l(\infty)$; & $\alpha_0 = r_{ab}$; \\
  {\textrm{Coulomb collisions:}} &
  $\mathbb{Q}_{ab}^{(l)} = 2\pi \alpha_0^2 A_l(2)$; & $\alpha_0 = \frac{|q_a q_b|}{\mu_{ab} g^2}$; \\
  {\textrm{Inverse power:}} &
  $\mathbb{Q}_{ab}^{(l)} = 2\pi \alpha_0^2 A_l (\nu)$; & $\alpha_0= \Big(\frac{|K_{ab}|}{\mu_{ab} g^2}\Big)^{1/(\nu-1)}$; \\
  {\textrm{Maxwell molecules:}} &
  $\mathbb{Q}_{ab}^{(l)} = 2\pi \alpha_0^2 A_l (5)$; & $\alpha_0= \Big(\frac{|K_{ab}|}{\mu_{ab}}\Big)^{1/4} \frac{1}{\sqrt{g}}$. \label{eq:beauty15}
\end{tabular}
\end{equation}
 Note that from (\ref{eq:beauty15}), the hard sphere limit is obtained by $\lim_{\nu \to \infty} |K_{ab}|^{1/(\nu-1)}=r_{ab}$.

For the hard spheres and Coulomb collisions, the pure numbers (\ref{eq:beauM1}) can be calculated analytically by
\begin{eqnarray}
A_l(\infty) &\equiv& \int_0^{1} (1-\cos^l \theta) \hat{b}_0 d \hat{b}_0 ;\qquad \textrm{where}\qquad \cos\theta = 2\hat{b}_0^2-1;  \label{eq:beau31}\\
A_l(2) &\equiv& \int_0^{\Lambda} (1-\cos^l \theta) \hat{b}_0 d \hat{b}_0; \qquad \textrm{where}\qquad \cos\theta = \frac{\hat{b}_0^2-1}{\hat{b}_0^2+1}, \label{eq:beau32}
\end{eqnarray}
yielding the results
\begin{eqnarray}
&& \textrm{Hard spheres}:  \qquad  A_l(\infty) = \frac{1}{4}\int_{-1}^1 (1-x^l) dx 
 = \frac{1}{2} -\frac{1+(-1)^l}{4(l+1)} = \Bigg\{ \begin{array}{lr}
    1/2;    & \quad l= \text{odd};\\
    l /[2(l+1)];  & \quad l= \text{even};   
  \end{array} \label{eq:HardPII}\\
  && \textrm{Coulomb collisions}\quad (\ln\Lambda\gg 1): \qquad A_l(2) = 2l \ln\Lambda.
\end{eqnarray}
 If for Coulomb collisions the $\ln\Lambda$ is not very large, the exact integrals are given by
\begin{eqnarray}
  A_1(2) &=& \int_0^\Lambda \frac{2\hat{b}_0}{1+\hat{b}_0^2} d\hat{b}_0 = \ln(\Lambda^2+1); \label{eq:A12x}\\
  A_2(2) &=& \int_0^\Lambda \frac{4\hat{b}_0^3}{(1+\hat{b}_0^2)^2}d\hat{b}_0  
  = 2\ln(\Lambda^2+1)-2+\frac{2}{\Lambda^2+1};\label{eq:A22x}\\
  A_3(2) &=& \int_0^\Lambda \frac{(6\hat{b}_0^4+2)\hat{b}_0}{(1+\hat{b}_0^2)^3} d\hat{b}_0 =  3\ln(\Lambda^2+1) -4
  +\frac{6\Lambda^2+4}{(\Lambda^2+1)^2}. \label{eq:A23x}
\end{eqnarray}
 Note that the results (\ref{eq:A12x})-(\ref{eq:A23x}) do not become negative regardless of plasma conditions and for $\Lambda\to 0$ the expressions
  just converge to zero.
  
For Maxwell molecules ($\nu=5$), the constants have to be found by numerical integration, for repulsive forces see section \ref{sec:Alnu} and for
attractive forces see Section \ref{sec:AlnuA}.
As a summary, the effective cross-sections (\ref{eq:beauty15}) use the following pure numbers 
\begin{eqnarray}
&&  \textrm{Hard spheres}: \qquad\qquad\qquad\qquad A_1(\infty) = 1/2; \qquad \,A_2(\infty)=1/3; \qquad \, A_3(\infty)=1/2;\nn\\
&&  \textrm{Maxwell molecules (repuls.)}: \qquad A_1(5) = 0.422; \qquad A_2(5)=0.436; \qquad A_3(5)=0.585;\nn\\
&&  \textrm{Maxwell molecules (attrac.)}: \qquad A_1(5) = 0.781; \qquad A_2(5)=0.544; \qquad A_3(5)=0.902;\nn\\  
&&  \textrm{Coulomb collisions}: \qquad\qquad\qquad A_1(2) = 2\ln\Lambda; \qquad A_2(2)=4\ln\Lambda; \qquad A_3(2)=6\ln\Lambda.
\end{eqnarray}

%\newpage
%==========================================================================
\subsection{Chapman-Cowling collisional integrals}
Introducing the effective cross-sections (\ref{eq:QabPicA}) allows one to ``hide'' the particular collisional process, and integrate 
the collisional contributions (\ref{eq:beauty10}) over the solid angle $d\Omega$, where for example the 
momentum exhange rates become
\begin{eqnarray}
  \boldsymbol{R}_{ab} &=& -\mu_{ab} 
  \iint  f_a f_b  \, g \bg \mathbb{Q}_{ab}^{(1)} d^3 v_a d^3 v_b,
\end{eqnarray}
and the other collisional integerals are given by (\ref{eq:beau0})-(\ref{eq:beau7}). Importantly, in the semi-linear approximation,
it is possible to finish the calculations by expressing the results through 
the ``Chapman-Cowling collisional integrals'' 
\begin{equation}
  \Omega_{ab}^{(l,j)} \equiv
  \frac{1}{2\sqrt{\pi}} \Big( \frac{1}{\beta_{ab}} \Big)^{2j+3} \int_0^\infty e^{-\frac{g^2}{\beta_{ab}^2}} g^{2j+3} \mathbb{Q}_{ab}^{(l)}(g) dg, \label{eq:CCo}
\end{equation}
where ``l'' and ``j'' are integers and 
\begin{equation}
\beta^2_{ab} = v_{\textrm{th} a}^2 + v_{\textrm{th} b}^2 = \frac{2T_a}{m_a} + \frac{2T_b}{m_b} = \frac{2T_{ab}}{\mu_{ab}}. 
\end{equation}  
Our definition of $\Omega_{ab}^{(l,j)}$ is equal to \cite{SchunkNagy2009}, \cite{Schunk1977} and is also equivalent to the final definition
$\Omega_{ab}^{(l)}(j)$ of \cite{ChapmanCowling1953}, p. 157.

%\vspace{1cm}
% \begin{eqnarray}
%   \Omega_{ab}^{(l,j)} &\equiv&
%  \frac{1}{2\sqrt{\pi}} \Big( \frac{1}{\beta_{ab}} \Big)^{2j+3} \int_0^\infty e^{-\frac{g^2}{\beta_{ab}^2}} g^{2j+3}
%  \Big[  2\pi \int_0^{b_0^\textrm{max}} (1-\cos^l \theta) b_0 db_0 \Big] dg; \quad (\textrm{attract. \& repuls.})\label{eq:CCom}\\ 
%  \Omega_{ab}^{(l,j)} &=& \frac{1}{2\sqrt{\pi}} \Big( \frac{1}{\beta_{ab}} \Big)^{2j+3} \int_0^\infty e^{-\frac{g^2}{\beta_{ab}^2}} g^{2j+3}
%  \Big[ 2\pi \int_{\theta_{\textrm{min}}}^\pi \sigma_{ab} (g,\theta) (1-\cos^l \theta) \sin\theta d\theta \Big] dg, \quad (\textrm{repuls.})\label{eq:CComB}
% \end{eqnarray}
%\vspace{1cm}

For example, considering the simple 5-moment model, in the semi-linear approximation (for small drifts) yields the momentum exchange rates 
$\boldsymbol{R}_{ab} = (16/3) \mu_{ab} n_a n_b (\bu_b-\bu_a) \Omega^{(1,1)}_{ab}$, further yielding the following ``universal'' definition of the collisional frequency
\begin{equation}
\boldsymbol{R}_{ab} = m_a n_a \nu_{ab}(\bu_b-\bu_a); \qquad => \qquad \nu_{ab} = \frac{16}{3} \frac{\mu_{ab}}{m_a} n_b \Omega^{(1,1)}_{ab}. \label{eq:Rsimple}
\end{equation}
The result  (\ref{eq:Rsimple}) means that (in the semi-linear approximation) it is possible to calculate the $\boldsymbol{R}_{ab}$ only once
for all of the possible collisional processes and specify the particular collisional process only at the end. The particular examples are 
\begin{eqnarray}
&&  \textrm{Hard spheres:} \qquad \qquad \,\,\, \Omega_{ab}^{(1,1)} = \frac{\sqrt{\pi}}{2} r_{ab}^2 \beta_{ab}; \qquad r_{ab}=r_a+r_b\nn\\
&&  \textrm{Coulomb collisions:} \qquad \Omega_{ab}^{(1,1)} =  \frac{\sqrt{\pi}}{2} \frac{q_a^2 q_b^2}{\mu_{ab}^2\beta_{ab}^3} A_1 (2);\nn\\
&&  \textrm{Inverse power:} \qquad \qquad \Omega_{ab}^{(1,1)} =
  \frac{\sqrt{\pi}}{2}  \Big(\frac{|K_{ab}|}{\mu_{ab}}\Big)^{\frac{2}{\nu-1}} \beta_{ab}^{\frac{\nu-5}{\nu-1}}\, A_1(\nu)\Gamma\big( 3 - \frac{2}{\nu-1}\big);\nn\\
&&  \textrm{Maxwell molecules:} \qquad  \Omega_{ab}^{(1,1)} =
  \frac{\sqrt{\pi}}{2}  \Big(\frac{|K_{ab}|}{\mu_{ab}}\Big)^{\frac{1}{2}}  A_1(5)\Gamma\big( \frac{5}{2}\big),  \label{eq:Rsimple2}
\end{eqnarray}
the $\Gamma$ being the Gamma function
\begin{equation}
\int_0^\infty e^{-\frac{g^2}{\beta^2}} g^{\mu} dg = \frac{\beta^{\mu+1}}{2} \Gamma\Big(\frac{\mu+1}{2}\Big). 
\end{equation}

Similarly to the simple 5-moment model $\boldsymbol{R}_{ab}$ given by (\ref{eq:Rsimple}) (which defines the collisional frequency),
all of the collisional contributions (\ref{eq:beauty10}) of the 22-moment model
can be expressed through the Chapman-Cowling integrals (\ref{eq:CCo}), 
where one encounters (general ``l'' and ``j'')  $\Omega_{ab}^{(l,j)}$, with particular examples
\begin{eqnarray}
&&  \textrm{Hard spheres:} \qquad \qquad \,\,\, \Omega_{ab}^{(l,j)} = \frac{\sqrt{\pi}}{2} r_{ab}^2 \beta_{ab} A_l(\infty) (j+1)!;\nn\\
  &&  \textrm{Coulomb collisions:} \qquad \Omega_{ab}^{(l,j)}
  =  \frac{\sqrt{\pi}}{2} \frac{q_a^2 q_b^2}{\mu_{ab}^2 \beta_{ab}^3} A_l(2)(j-1)!;\nn\\
  &&  \textrm{Inverse power:} \qquad \qquad \Omega_{ab}^{(l,j)}
  =  \frac{\sqrt{\pi}}{2}  \Big(\frac{|K_{ab}|}{\mu_{ab}}\Big)^{\frac{2}{\nu-1}} \beta_{ab}^{\frac{\nu-5}{\nu-1}}\, A_l(\nu) \Gamma\big( j+2 - \frac{2}{\nu-1}\big);\nn\\
  &&  \textrm{Maxwell molecules:} \qquad \Omega_{ab}^{(l,j)}
  =  \frac{\sqrt{\pi}}{2}  \Big(\frac{|K_{ab}|}{\mu_{ab}}\Big)^{\frac{1}{2}} \, A_l(5) \Gamma\big( j+\frac{3}{2}\big).
\end{eqnarray}
After the collisional contributions are calculated, in the 13-moment models of \cite{Burgers1969,Schunk1977,SchunkNagy2009},
the following ratios are introduced
\begin{eqnarray}
  z_{ab} &=& 1-\frac{2}{5} \frac{\Omega_{ab}^{(1,2)}}{\Omega_{ab}^{(1,1)}}; \qquad
  z_{ab}' = \frac{5}{2} + \frac{2}{5} \frac{(\Omega_{ab}^{(1,3)}-5\Omega_{ab}^{(1,2)})}{\Omega_{ab}^{(1,1)}};\nn\\
  z_{ab}'' &=& \frac{\Omega_{ab}^{(2,2)}}{\Omega_{ab}^{(1,1)}}; \qquad
  z_{ab}''' = \frac{\Omega_{ab}^{(2,3)}}{\Omega_{ab}^{(1,1)}},
  \label{eq:zab}
\end{eqnarray}
where for example the $z_{ab}$ is the natural choice for the momentum exchange rates and the $z_{ab}'$ makes the heat flux description more concise.  
In the work of \cite{ChapmanCowling1953}, the following ratios are introduced (p. 164)
\begin{eqnarray}
  A = \frac{1}{5}\frac{\Omega_{ab}^{(2,2)}}{\Omega_{ab}^{(1,1)}}; \qquad B=\frac{5\Omega_{ab}^{(1,2)}-\Omega_{ab}^{(1,3)}}{5\Omega_{ab}^{(1,1)}}; \qquad
  C=\frac{2}{5} \frac{\Omega_{ab}^{(1,2)}}{\Omega_{ab}^{(1,1)}},  \label{eq:zab2}
\end{eqnarray}  
together with other ratios. Note that in \cite{ChapmanCowling1970} (third edition) the $C$ is redefined with an additional factor of $-1$, according to
 $C=2\Omega_{ab}^{(1,2)} /(5\Omega_{ab}^{(1,1)})-1$.
Introducing these ratios has a benefit of giving 
a more concise model at the end, however, for the multi-fluid 22-moment model we need more of these ratios and here we
are really not bothered by giving the most concise
formulation, all of the results will be just expressed through a ``mass ratio coefficients'', which for given
masses and temperatures are just pure numbers. 
Instead of defining the (\ref{eq:zab}) or (\ref{eq:zab2}), here we find the best to simply 
normalize all of the Chapman-Cowling integrals to $\Omega_{ab}^{(1,1)}$, by introducing simple notation
\begin{equation} \boxed{
\Omega_{l,j} = \frac{\Omega_{ab}^{(l,j)}}{\Omega_{ab}^{(1,1)}}, \label{eq:norm}}
\end{equation}
and the results are always pure numbers. The particular examples are
\begin{eqnarray}
&&  \textrm{Hard spheres:} \qquad \qquad \,\,\, \Omega_{l,j} = A_l(\infty) (j+1)!;\nn\\
&&  \textrm{Coulomb collisions:} \qquad \Omega_{l,j} =  \frac{A_l(2)}{A_1(2)}(j-1)!; \quad \textrm{and}\quad \ln\Lambda\gg 1: \quad \Omega_{l,j} = l (j-1)!\nn\\
&&  \textrm{Inverse power:} \qquad \qquad \Omega_{l,j}
  =  \frac{A_l(\nu)}{A_1(\nu)} \frac{\Gamma\big( j+2 - \frac{2}{\nu-1}\big)}{\Gamma\big( 3 - \frac{2}{\nu-1}\big)};\nn\\
&&  \textrm{Maxwell molecules:} \qquad \Omega_{l,j}
  =  \frac{A_l(5)}{A_1(5)} \frac{\Gamma\big( j+\frac{3}{2}\big)}{\Gamma\big( \frac{5}{2} \big)}
  = \frac{A_l(5)}{A_1(5)} \frac{1}{2^{j-1}} \frac{(2j+1)!!}{3}. \label{eq:best1}
\end{eqnarray}
For the Coulomb collisions, the normalized Chapman-Cowling integrals read (with $\ln\Lambda\gg 1$)
\begin{eqnarray}
  \textrm{Coulomb collisions:} \qquad
  \Omega_{1,2} &=& 1; \qquad \Omega_{1,3} = 2; \qquad \Omega_{1,4} = 6;  \qquad \Omega_{1,5} = 24; \nn\\
  \Omega_{2,2} &=& 2; \qquad \Omega_{2,3} = 4; \qquad \Omega_{2,4} = 12; \qquad \Omega_{2,5} = 48; \nn\\ 
  \Omega_{3,3} &=& 6; \qquad \Omega_{3,4} = 18;\qquad \Omega_{3,5} = 72, \label{eq:beauty17x}
\end{eqnarray}
where for the 13-moment models one only needs $\Omega_{1,2}$, $\Omega_{1,3}$, $\Omega_{2,2}$ and $\Omega_{2,3}$. 
For the hard spheres 
\begin{eqnarray}
  \textrm{hard spheres:} \qquad
  \Omega_{1,2} &=& 3; \qquad \Omega_{1,3} = 12; \qquad \Omega_{1,4} = 60; \qquad \Omega_{1,5} = 360; \nn\\
  \Omega_{2,2} &=& 2; \qquad \Omega_{2,3} = 8;  \qquad \Omega_{2,4} = 40; \qquad \Omega_{2,5} = 240; \nn\\
  \Omega_{3,3} &=& 12;\qquad \Omega_{3,4} = 60; \qquad \Omega_{3,5} = 360.   \label{eq:beauty17}
\end{eqnarray}
For the Maxwell molecules
\begin{eqnarray}
  \Omega_{l,2} = \frac{A_l(5)}{A_1(5)} \frac{5}{2}; \qquad \Omega_{l,3}=\frac{A_l(5)}{A_1(5)}\frac{35}{4};
  \qquad \Omega_{l,4}=\frac{A_l(5)}{A_1(5)}\frac{315}{8}; \qquad \Omega_{l,5} = \frac{A_l(5)}{A_1(5)} \frac{3465}{16}. \label{eq:Picus2}
\end{eqnarray}
By using the gamma function property
\begin{equation}
\Gamma\Big( n+\frac{p}{q}\Big) = \Gamma\Big(\frac{p}{q}\Big) \frac{1}{q^n} \prod_{k=1}^{n} (kq-q+p),
\end{equation}  
for the general inverse force one can also write
\begin{equation}
\textrm{Inverse force:} \qquad \quad\,\,\, \Omega_{l,j}
  =  \frac{A_l(\nu)}{A_1(\nu)} \frac{1}{(\nu-1)^{j-1}} \prod_{k=4}^{j+2} \big[ (k-1)\nu -(k+1) \big],
\end{equation}
further yielding the particular cases
\begin{eqnarray}
\Omega_{l,2} &=& \frac{A_l(\nu)}{A_1(\nu)} \frac{3\nu-5}{\nu-1}; \qquad
\Omega_{l,3} = \frac{A_l(\nu)}{A_1(\nu)} \frac{2(3\nu-5)(2\nu-3)}{(\nu-1)^2};\nn\\
\Omega_{l,4} &=& \frac{A_l(\nu)}{A_1(\nu)} \frac{2 (3 \nu-5) (2 \nu-3) (5 \nu-7)}{(\nu-1)^3}; \qquad
\Omega_{l,5} = \frac{A_l(\nu)}{A_1(\nu)} \frac{4 (3 \nu-5)(2\nu-3) (5\nu-7) (3 \nu-4)}{(\nu-1)^4}. \label{eq:OmegaGI}
\end{eqnarray}  
Our multi-fluid 22-moment model requires only eleven ratios of the Chapman-Cowling integrals $\Omega_{l,j}$ (see e.g. (\ref{eq:beauty17}))
where the numbers ``l'' and ``j'' never reach or exceed the value of 10, and in the rest
of the paper we use an abbreviated notation $\Omega_{lj}$.  

\newpage
%=============================================
\subsection{Collisional frequencies \texorpdfstring{$\nu_{ab}$}{}} \label{sec:Ratio}
It is useful to clarify the various collisional frequencies (defined also in (\ref{eq:Rsimple}))
\begin{equation}
  \nu_{ab} = \frac{16}{3} \frac{\mu_{ab}}{m_a} n_b \Omega^{(1,1)}_{ab}; \qquad
  \beta^2_{ab} = v_{\textrm{th} a}^2 + v_{\textrm{th} b}^2 = \frac{2T_a}{m_a} + \frac{2T_b}{m_b} = \frac{2T_{ab}}{\mu_{ab}}; \qquad
  T_{ab}=\frac{T_a m_b + T_b m_a}{m_a+m_b},\nn
\end{equation}
the $T_{ab}$ being the reduced temperature. The particular collisional frequencies read
%\begin{eqnarray}
%&&  \textrm{Hard spheres:} \qquad \qquad \,\,\,  \nu_{ab} = \frac{8}{3}\sqrt{\pi} \frac{\mu_{ab}}{m_a} n_b   r_{ab}^2 \beta_{ab};\nn\\
%  &&  \textrm{Coulomb collisions:} \qquad  \nu_{ab} = \frac{8}{3} \sqrt{\pi} \frac{\mu_{ab}}{m_a} n_b 
%  \Big(\frac{q_a q_b}{\mu_{ab}}\Big)^2 \frac{1}{\beta_{ab}^3} A_1 (2);\nn\\
%&&  \textrm{Inverse power:} \qquad \qquad  \nu_{ab} = \frac{8}{3} \sqrt{\pi} \frac{\mu_{ab}}{m_a} n_b 
%   \Big(\frac{K_{ab}}{\mu_{ab}}\Big)^{\frac{2}{\nu-1}} \beta_{ab}^{\frac{\nu-5}{\nu-1}}\, A_1(\nu)\Gamma\big( 3 - \frac{2}{\nu-1}\big);\nn\\
%&&  \textrm{Maxwell molecules:} \qquad   \nu_{ab} = \frac{8}{3} \sqrt{\pi} \frac{\mu_{ab}}{m_a} n_b 
%   \Big(\frac{K_{ab}}{\mu_{ab}}\Big)^{1/2}  A_1(5)\Gamma\big( \frac{5}{2}\big).  
%\end{eqnarray}
\begin{eqnarray}
&&  \textrm{Hard spheres:} \qquad \qquad \,\,\,  \nu_{ab} = \frac{8}{3}\sqrt{\pi} \frac{n_b}{m_a} \mu_{ab}  r_{ab}^2 \beta_{ab}; \qquad r_{ab}=r_a+r_b;\nn\\
  &&  \textrm{Coulomb collisions:} \qquad  \nu_{ab} = \frac{8}{3} \sqrt{\pi} \frac{n_b}{m_a \mu_{ab}} 
  (q_a q_b)^2 \beta_{ab}^{-3} A_1 (2); \quad \textrm{for}\, \ln\Lambda\gg 1: \, A_1(2)=2\ln\Lambda;\nn\\
&&  \textrm{Inverse power:} \qquad \qquad  \nu_{ab} = \frac{8}{3} \sqrt{\pi} \frac{n_b}{m_a} \mu_{ab}^{\frac{\nu-3}{\nu-1}} 
   |K_{ab}|^{\frac{2}{\nu-1}} \beta_{ab}^{\frac{\nu-5}{\nu-1}}\, A_1(\nu)\Gamma\big( 3 - \frac{2}{\nu-1}\big);\nn\\
&&  \textrm{Maxwell molecules:} \qquad   \nu_{ab} = \frac{8}{3} \sqrt{\pi} \frac{n_b}{m_a} \mu_{ab}^{1/2} 
   |K_{ab}|^{1/2}  A_1(5)\Gamma\big( \frac{5}{2}\big). \label{eq:Rsimple2x}
\end{eqnarray}
Note that $\Gamma(5/2) = (3/4)\sqrt{\pi}$ and 
\begin{equation*}
\mu_{ab}^{\frac{\nu-3}{\nu-1}} \beta_{ab}^{\frac{\nu-5}{\nu-1}} = \mu_{ab}^{1/2} (2T_{ab})^{\frac{(\nu-5)}{2(\nu-1)}}.
\end{equation*}
 The $\nu_{ab}$ of Maxwell molecules, where the ion polarizes the neutral (i.e. ion-neutral collisions), is given later by (\ref{eq:INnuab}).
Also note that in general $\nu_{ab} \neq \nu_{ba}$ (but one can still of course use the expression (\ref{eq:Rsimple2x}) and simply exchange
the indices to obtain the $\nu_{ba}$ expression) and instead the collisional frequencies are related by
the conservation of momentum 
\begin{equation}
m_a n_a \nu_{ab} = m_b n_b \nu_{ba}, \label{eq:cons}
\end{equation}
which is satisfied because $\Omega_{ab}^{(1,1)}=\Omega_{ba}^{(1,1)}$ is true. The expressions simplify for small temperature differences
(where the reduced temperature $T_{ab}\simeq T_a)$
\begin{eqnarray}
T_a\simeq T_b \qquad &&  \textrm{Hard spheres:} \qquad \qquad \,\,\,  \nu_{ab} = \frac{8}{3}\sqrt{\pi} n_b \frac{\mu_{ab}^{1/2}}{m_a}   r_{ab}^2 (2T_a)^{1/2};\nn\\
  &&  \textrm{Coulomb collisions:} \qquad  \nu_{ab} = \frac{8}{3} \sqrt{\pi} n_b \frac{\mu_{ab}^{1/2}}{m_a}
  (q_a q_b)^2 (2 T_a )^{-3/2} A_1 (2);\nn\\
&&  \textrm{Inverse power:} \qquad \qquad  \nu_{ab} = \frac{8}{3} \sqrt{\pi} n_b \frac{\mu_{ab}^{1/2}}{m_a} 
   |K_{ab}|^{\frac{2}{\nu-1}} (2T_a)^{\frac{\nu-5}{2(\nu-1)}}\, A_1(\nu)\Gamma\big( 3 - \frac{2}{\nu-1}\big);\nn\\
&&  \textrm{Maxwell molecules:} \qquad   \nu_{ab} = \frac{8}{3} \sqrt{\pi} n_b \frac{\mu_{ab}^{1/2}}{m_a} 
   |K_{ab}|^{1/2}  A_1(5)\Gamma\big( \frac{5}{2}\big), \label{eq:Rsimple2xx}
\end{eqnarray}
where the masses are represented by a universal factor of $\mu_{ab}^{1/2}/m_a$. For self-collisions this factor is just replaced by
$\mu_{aa}^{1/2}/m_a=1/\sqrt{2m_a}$ and we only write down the inverse power-law force
\begin{eqnarray}
\textrm{Inverse power:} \qquad \quad  \nu_{aa} &=& \frac{8}{3} \sqrt{\pi} n_a \frac{1}{\sqrt{2 m_a}} 
   |K_{aa}|^{\frac{2}{\nu-1}} (2T_a)^{\frac{\nu-5}{2(\nu-1)}}\, A_1(\nu)\Gamma\big( 3 - \frac{2}{\nu-1}\big); \label{eq:haha2}\\
   \frac{\nu_{aa}}{\nu_{ab}} &=& \frac{1}{\sqrt{2}} \Big| \frac{K_{aa}}{K_{ab}} \Big|^{\frac{2}{\nu-1}}
   \frac{n_a}{n_b} \Big(\frac{m_a+m_b}{m_b}\Big)^{1/2} \Big( \frac{T_a}{T_{ab}}\Big)^{\frac{(\nu-5)}{2(\nu-1)}}\label{eq:haha}.
\end{eqnarray}
 Note that the hard sphere limit is obtained by $\lim_{\nu \to \infty} |K_{aa}/K_{ab}|^{2/(\nu-1)}=(r_{aa}/r_{ab})^2$. Also note that the factor of $\sqrt{2}$ is present
in the ratio (\ref{eq:haha}). We will later consider further particular cases of
lightweight or heavyweight species ``a'', where for the small temperature differences $T_a\simeq T_b$ 
\begin{eqnarray}
m_a\ll m_b:  \qquad  \frac{\nu_{aa}}{\nu_{ab}} &=& \frac{1}{\sqrt{2}} \Big| \frac{K_{aa}}{K_{ab}} \Big|^{\frac{2}{\nu-1}} \frac{n_a}{n_b}; \nn\\
m_a \gg m_b: \qquad  \frac{\nu_{aa}}{\nu_{ab}} &=& \frac{1}{\sqrt{2}} \Big| \frac{K_{aa}}{K_{ab}} \Big|^{\frac{2}{\nu-1}} \frac{n_a}{n_b} \Big(\frac{m_a}{m_b}\Big)^{1/2}. \label{eq:happyA}
\end{eqnarray}
Using the last result and prescribing Coulomb collisions together with the charge neutrality
$n_e=Z_i n_i$ then yields the usual ratios  
$\nu_{ee}/\nu_{ei} = 1/(Z_i \sqrt{2})$ and $\nu_{ii}/\nu_{ie} = Z_i (m_i/m_e)^{1/2} /\sqrt{2}$, where the $\sqrt{2}$ is naturally present.
For a further discussion about the $\sqrt{2}$ factor see Section 8.2, p. 31 in \cite{Hunana2022} ``Collisional frequencies for ion-electron plasma''.

%As a clarifying example, let us use the repulsive inverse force, where by deriving the lowest
%Chapman-Cowling integral $\Omega_{ab}^{(1,1)}$, one obtains the collisional frequencies $\nu_{aa}$ (see later eq. (\ref{eq:haha2})).

%Also note that the (quasi-static) self-collisional viscosities and heat conductivities are proportional to $p_a/\nu_{aa}$,
%so from the temperature dependence in , both the viscosity $\eta_0^a$ and the thermal conductivity $\kappa_\parallel^a$
%have a temperature dependence as $T^s$, where the exponent   
%\begin{equation}
%s = 1-\frac{\nu-5}{2(\nu-1)} = \frac{1}{2}+\frac{2}{\nu-1} = \frac{\nu+3}{2(\nu-1)}. \label{eq:cool}
%\end{equation}
%For the Coulomb collisions the dependence is $T^{5/2}$, for the Maxwell molecules $T^{+1}$ and for the hard spheres $T^{1/2}$.  
%Interestingly, by measuring for example the viscosity dependence $T^s$, one can then deduce the power exponent $\nu$ for the
%intermolecular forces according to (\ref{eq:cool}), see e.g. Table 14, p. 223 in \cite{ChapmanCowling1953}, where the power ranges
%from $\nu=5$ for chlorine to $\nu=14.6$ for Helium (the numbers are modified in \cite{ChapmanCowling1970}).

%\newpage
%=================================================================================================================
%=================================================================================================================
\subsection{Evolution equations for 22-moment model} \label{sec:EvolutionEQ}
As in Part 1, the integration of the Boltzmann equation yields the usual ``basic'' evolution equations
\begin{eqnarray}
&& \frac{d_a}{dt} n_a + n_a \nabla\cdot\bu_a =0; \label{eq:Energy20}\\
&& \frac{d_a}{dt}\bu_a +\frac{1}{\rho_a}\nabla\cdot\bp_a -\boldsymbol{G} -\frac{eZ_a}{m_a}\Big( \bE+\frac{1}{c}\bu_a\times\bb\Big)
  = \frac{\boldsymbol{R}_a}{\rho_a}; \label{eq:Energy20x}\\
  && \frac{d_a }{dt}p_a + \frac{5}{3} p_a\nabla\cdot\bu_a
  +\frac{2}{3}\nabla\cdot\vec{\boldsymbol{q}}_a +\frac{2}{3}\bPi^{(2)}_a :(\nabla \bu_a)
  =\frac{2}{3} Q_a, \label{eq:Energy20xx}
\end{eqnarray}
which are accompanied by the following evolution equations for the higher-order moments (see Eqs. (9)-(12) of \cite{Hunana2022}  for the 21-moment model,
Eqs. (130)-(133) for the 22-moment model, or Eqs. (D13)-(D15) for a general ``n'') 
\begin{eqnarray}
&&  \frac{d_a}{dt}\bPi^{(2)}_a + \bPi^{(2)}_a \nabla\cdot\bu_a 
  +\Omega_a \big(\bhat\times \bPi^{(2)}_a \big)^S +\big( \bPi^{(2)}_a \cdot\nabla\bu_a\big)^S 
  -\frac{2}{3}\bI(\bPi^{(2)}_a:\nabla\bu_a) \nn\\
  && \quad +\frac{2}{5}\Big[(\nabla \vecq_a)^S -\frac{2}{3}\bI \nabla\cdot\vecq_a\Big]
 +p_a \bW_a = \bQ^{(2)}_a\,' \equiv \bQ^{(2)}_a -\frac{\bI}{3}\textrm{Tr}\bQ^{(2)}_a; \label{eq:Num1000}
\end{eqnarray}
%===
\begin{eqnarray} 
 && \frac{d_a}{d t}\vecq_a +\frac{7}{5}\vecq_a\nabla\cdot\bu_a  + \frac{7}{5}\vecq_a\cdot\nabla\bu_a +\frac{2}{5}(\nabla\bu_a)\cdot\vecq_a
  +\Omega_a\bhat\times\vecq_a + \frac{5}{2}p_a\nabla\Big(\frac{p_a}{\rho_a}\Big) \nn\\
 && \quad +\frac{1}{6}\nabla \widetilde{X}^{(4)}_a +\frac{1}{2}\nabla\cdot \bPi^{(4)}_a 
  -\frac{5}{2}\frac{p_a}{\rho_a}\nabla\cdot\bPi^{(2)}_a
  -\frac{1}{\rho_a}(\nabla\cdot\bp_a)\cdot\bPi^{(2)}_a \nn\\
  && \quad = \vec{\boldsymbol{Q}}^{(3)}_{a}\,' \equiv \vecQ^{(3)}_a   -\frac{5}{2}\frac{p_a}{\rho_a}\boldsymbol{R}_a
  -\frac{1}{\rho_a} \boldsymbol{R}_a\cdot\bPi^{(2)}_a; \label{eq:Thierry90}
\end{eqnarray}
%===
\begin{eqnarray}
  && \frac{d_a}{dt} \widetilde{X}^{(4)}_a +\nabla\cdot\vecX^{(5)}_a  -20 \frac{p_a}{\rho_a}\nabla\cdot\vecq_a
  +\frac{7}{3}\widetilde{X}^{(4)}_a(\nabla\cdot\bu_a) +4\big(\bPi^{(4)}_a -5\frac{p_a}{\rho_a}\bPi^{(2)}_a\big):\nabla\bu_a \nn\\
 && \quad -\frac{8}{\rho_a} (\nabla\cdot\bp_a)\cdot \vecq_a 
    = \widetilde{Q}^{(4)}_a\,' \equiv Q^{(4)}_a -20 \frac{p_a}{\rho_a}Q_a-\frac{8}{\rho_a} \boldsymbol{R}_a \cdot\vecq_a; \label{eq:Thierry92}
\end{eqnarray}
%===
\begin{eqnarray}
  && \frac{d_a}{dt} \bPi^{(4)}_a +\frac{1}{5}\Big[ (\nabla\vecX^{(5)}_a)^S-\frac{2}{3}\bI(\nabla\cdot\vecX^{(5)}_a)\Big]
  +\frac{9}{7}(\nabla\cdot\bu_a)\bPi^{(4)}_a +\frac{9}{7}(\bPi^{(4)}_a\cdot\nabla\bu_a)^S\nn\\
&&\quad  + \frac{2}{7}\big((\nabla\bu_a)\cdot\bPi^{(4)}_a\big)^S
  -\frac{22}{21}\bI (\bPi^{(4)}_a:\nabla\bu_a)
 -\, \frac{14}{5\rho_a} \Big[ \big((\nabla\cdot\bp_a)\vecq_a\big)^S -\frac{2}{3}\bI (\nabla\cdot\bp_a)\cdot\vecq_a\Big] \nn\\
 &&\quad +\Omega_a \big( \bhat\times \bPi^{(4)}_a \big)^S
 + \frac{7}{15}\big(15\frac{p_a^2}{\rho_a}+\widetilde{X}^{(4)}_a \big)\bW_a \nn\\
&&\quad = \bQ^{(4)}_a\,' \equiv \bQ^{(4)*}_a -\frac{\bI}{3}\trace \bQ^{(4)*}_a
 -\frac{14}{5\rho_a}\Big[ (\boldsymbol{R}_a\vecq_a)^S-\frac{2}{3}\bI (\boldsymbol{R}_a\cdot\vecq_a)\Big]; \label{eq:Thierry91}
\end{eqnarray}
%===
\begin{eqnarray}
&&  \frac{d_a}{d t}\vecX^{(5)}_a +\frac{1}{3}\nabla\widetilde{X}^{(6)}_a +\nabla\cdot \bPi^{(6)}_a\nn\\
  &&\quad  +\frac{9}{5}\vecX^{(5)}_a (\nabla\cdot\bu_a) + \frac{9}{5}\vecX^{(5)}_a\cdot\nabla\bu_a
  + \frac{4}{5}(\nabla\bu_a)\cdot\vecX^{(5)}_a +\Omega_a \bhat\times \vecX^{(5)}_a\nn\\
  &&\quad  +70 \frac{p_a^2}{\rho_a}\nabla\Big(\frac{p_a}{\rho_a}\Big) -35 \frac{p_a^2}{\rho_a^2} \nabla\cdot\bPi^{(2)}_a
  -\frac{7}{3\rho_a}\big(\nabla\cdot\bp_a\big)\widetilde{X}^{(4)}_a
  -\frac{4}{\rho_a} \big(\nabla\cdot\bp_a\big)\cdot \bPi^{(4)}_a \nn\\
  && \quad =\vecQ^{(5)}_a\,' \equiv \vecQ^{(5)}_a -35 \frac{p_a^2}{\rho_a^2}\boldsymbol{R}_a -\frac{7}{3\rho_a} \boldsymbol{R}_a \widetilde{X}^{(4)}_a
  - \frac{4}{\rho_a} \boldsymbol{R}_a\cdot\bPi^{(4)}_a. \label{eq:Thierry93}
\end{eqnarray}
As in Part 1, the last equation (\ref{eq:Thierry93}) is closed with the following closures for the stress-tensor $\bPi^{(6)}_a$ and the scalar
$\widetilde{X}^{(6)}_a$
\begin{eqnarray}
  \bPi^{(6)}_a  &=&  18 \frac{p_a}{\rho_a} \bPi^{(4)}_a -63 \frac{p_a^2}{\rho_a^2}\bPi^{(2)}_a; \qquad
 \widetilde{X}^{(6)}_a =  21 \frac{p_a}{\rho_a}\widetilde{X}^{(4)}_a.  \label{eq:Thierry70}
\end{eqnarray}
The reduction into the 21-moment model is obtained easily by neglecting the evolution equation (\ref{eq:Thierry92}) and
by simply prescribing $\widetilde{X}^{(4)}_a=0$ in the other evolution equations.  The left-hand-sides of the evolution equations
  contain the direction of the magnetic field $\bhat=\bb/|\bb|$, the cyclotron frequency of species
  $\Omega_a=q_a |\bb|/(m_a c)$ (which should not be confused with the Chapman-Cowling integrals),
  the rate-of-strain tensor $\bW_a=(\nabla\bu_a)^S-(2/3)\bI\nabla\cdot\bu_a$ and the symmetric operator ``S'', defined as $A_{ij}^S=A_{ij}+A_{ji}$.
The above evolution equations were obtained by integrating the Boltzmann equation
 \begin{equation}
  \frac{\pr f_a}{\pr t}+ \bV_a\cdot\nabla f_a + \Big[\boldsymbol{G}+\frac{eZ_a}{m_a}(\bE+\frac{1}{c}\bV_a\times\bb)\Big]\cdot\nabla_{v_a} f_a
  =C(f_a), \label{eq:Boltz0}
 \end{equation}
 see Appendices A, C, D in Part 1, together with Section 8.7 ``Inclusion of Gravity''.
 
%=========================================================================================================
%=========================================================================================================
\subsection{Coupled evolution equations (semi-linear approximation)} \label{sec:EvolutionEQ2}
In the semi-linear approximation, where many terms such as $(\nabla p_a) \vecq_a$, $\vecq_a (\nabla\cdot\bu_a)$, 
$\widetilde{X}^{(4)}_a \bW_a$, $\widetilde{X}^{(4)}_a \nabla T_a$ or $\boldsymbol{R}_a\cdot\vecq_a$ are neglected, which is equivalent
to prescribing zero large-scale gradients, the evolution equations become (we keep the full convective derivative $d_a/dt=\pr/\pr t+\bu_a\cdot\nabla$)
\begin{eqnarray}
&&  \frac{d_a}{dt}\bPi^{(2)}_a  +\Omega_a \big(\bhat\times \bPi^{(2)}_a \big)^S  +p_a \bW_a
   +\frac{2}{5}\Big[(\nabla \vecq_a)^S -\frac{2}{3}\bI \nabla\cdot\vecq_a\Big] = \bQ^{(2)}_a\,'; \label{eq:Num1000B}\\
%===  
   &&  \frac{d_a}{d t}\vecq_a + \Omega_a \bhat\times\vecq_a + \frac{5}{2}p_a \nabla \Big(\frac{p_a}{\rho_a}\Big)
  +\frac{1}{2}\nabla\cdot\bPi^{(4)}_a-\frac{5}{2}\frac{p_a}{\rho_a}\nabla\cdot\bPi^{(2)}_a  +\frac{1}{6}\nabla \widetilde{X}^{(4)}_a 
 = \vec{\boldsymbol{Q}}^{(3)}_{a}\,' ;\label{eq:Nomore110XX}\\
%===     
  && \frac{d_a}{dt} \widetilde{X}^{(4)}_a +\nabla\cdot\vecX^{(5)}_a  -20 \frac{p_a}{\rho_a}\nabla\cdot\vecq_a
     = \widetilde{Q}^{(4)}_a\,'; \label{eq:PPosled12X} \\
%===  
     && \frac{d_a}{dt} \bPi^{(4)}_a +\Omega_a \big( \bhat\times \bPi^{(4)}_a \big)^S  + 7 \frac{p_a^2}{\rho_a} \bW_a
     +\frac{1}{5}\Big[ (\nabla\vecX^{(5)}_a)^S-\frac{2}{3}\bI(\nabla\cdot\vecX^{(5)}_a)\Big] = \bQ^{(4)}_a\,';\label{eq:Nomore103X}\\
%===     
  &&  \frac{d_a}{d t}\vecX^{(5)}_a +\Omega_a \bhat\times \vecX^{(5)}_a  +70 \frac{p_a^2}{\rho_a}\nabla\Big(\frac{p_a}{\rho_a}\Big)
  +18\frac{p_a}{\rho_a}\nabla\cdot \bPi^{(4)}_a
  -98 \frac{p_a^2}{\rho_a^2}\nabla\cdot\bPi^{(2)}_a   + 7\frac{p_a}{\rho_a}\nabla \widetilde{X}^{(4)}_a = \vec{\boldsymbol{Q}}^{(5)}_{a}\,'. \label{eq:Nomore102X}
\end{eqnarray}
In these equations, the coupling between the stress-tensors and heat fluxes is retained.
Neglecting the scalars $\widetilde{X}^{(4)}_a$ and focusing
only at the 21-moment model, various explicit solutions with the coupled stress-tensors and heat fluxes (for the Coulomb collisions) can be found in Section 6 of Part 1,
where for simplicity only the unmagnetized solutions are given.

%=========================================================================================================
\subsection{Un-coupled evolution equations} \label{sec:EvolutionEQ3} 
To further simplify the system, one can de-couple the stress-tensors and heat fluxes. The evolution equations for the stress-tensors become
\begin{eqnarray}
  && \frac{d_a}{dt} \bPi^{(2)}_a  +\Omega_a \big(\bhat\times \bPi^{(2)}_a \big)^S + p_a \bW_a
  = \bQ_{a}^{(2)}\,' ; \label{eq:Posled21}\\
  && \frac{d_a}{dt} \bPi^{(4)}_a  +\Omega_a \big(\bhat\times \bPi^{(4)}_a \big)^S + 7 \frac{p_a^2}{\rho_a} \bW_a 
  =  \bQ^{(4)}_{a}\,', \label{eq:Energy22}
\end{eqnarray}
and the evolution equations for the heat fluxes read
\begin{eqnarray}
  &&  \frac{d_a}{d t}\vecq_a + \Omega_a \bhat\times\vecq_a + \frac{5}{2}p_a \nabla \Big(\frac{p_a}{\rho_a}\Big)
  = \vec{\boldsymbol{Q}}^{(3)}_{a}\,';\label{eq:Nomore110XXBB}\\
  %===
  &&  \frac{d_a}{d t}\vecX^{(5)}_a +\Omega_a \bhat\times \vecX^{(5)}_a  +70 \frac{p_a^2}{\rho_a}\nabla\Big(\frac{p_a}{\rho_a}\Big)
  = \vec{\boldsymbol{Q}}^{(5)}_{a}\,', \label{eq:Nomore102XBB} 
  %===
\end{eqnarray}
representing the 21-moment model, which eventually recovers the Braginskii model (in the quasi-static approximation for a plasma with only one ion species).   
As in Part 1, for the 22-moment model one additionally considers the evolution equation
\begin{equation}
 \frac{d_a}{dt} \widetilde{X}^{(4)}_a +\nabla\cdot\vecX^{(5)}_a  -20 \frac{p_a}{\rho_a}\nabla\cdot\vecq_a
     = \widetilde{Q}^{(4)}_a\,'.  \label{eq:PPosled12XB}
\end{equation}

We note that it is also possible to keep the stress-tensors (\ref{eq:Posled21})-(\ref{eq:Energy22}) separate as they are now, but
retain the contributions of scalars $\widetilde{X}^{(4)}_a$ into the heat flux equations 
\begin{eqnarray}
  &&  \frac{d_a}{d t}\vecq_a + \Omega_a \bhat\times\vecq_a + \frac{5}{2}p_a \nabla \Big(\frac{p_a}{\rho_a}\Big)
  +\frac{1}{6}\nabla \widetilde{X}^{(4)}_a = \vec{\boldsymbol{Q}}^{(3)}_{a}\,';\label{eq:Nomore110XXB}\\
  %===
  &&  \frac{d_a}{d t}\vecX^{(5)}_a +\Omega_a \bhat\times \vecX^{(5)}_a  +70 \frac{p_a^2}{\rho_a}\nabla\Big(\frac{p_a}{\rho_a}\Big)
  + 7\frac{p_a}{\rho_a}\nabla \widetilde{X}^{(4)}_a  = \vec{\boldsymbol{Q}}^{(5)}_{a}\,', \label{eq:Nomore102XB}
  %===
\end{eqnarray}
so that one considers system (\ref{eq:PPosled12XB})-(\ref{eq:Nomore102XB}) instead of (\ref{eq:Nomore110XXBB})-(\ref{eq:PPosled12XB}). 
As reported by \cite{Laguna2022,Laguna2023} for the 1-Hermite unmagnetized case (which in our notation should correspond to
eliminating the eq. (\ref{eq:Nomore102XB}) and prescribing $\vecX_a^{(5)}=28 (p_a/\rho_a) \vecq_a$ in (\ref{eq:PPosled12XB})), 
the quasi-static approximation then yields additional contributions to the heat fluxes, in
the form $\vecq_a \sim \nabla \widetilde{X}^{(4)}_a \sim \nabla \nabla^2 T_a$, which we neglect for simplicity.

\newpage
%==================================================================================================================
%==================================================================================================================
\section{Collisional contributions through Chapman-Cowling integrals} \label{sec:Main}
Here we write the collisional contributions by expressing them through the Chapman-Cowling integrals, for arbitrary temperatures and masses.
We keep the notation as close as possible to \cite{Hunana2022}, with the difference that here we keep the 1-Hermite and 2-Hermite contribution separate, 
so that the 22-moment model can be reduced to simpler models easily (this is further clarified below).  
%=============================================
\subsection{Momentum exchange rates  \texorpdfstring{$\boldsymbol{R}_{a}$}{R}}
The momentum exchange rates are given by
\begin{eqnarray}
  \boldsymbol{R}_a &=& \sum_{b\neq a} \nu_{ab} \Big\{ \rho_a (\bu_b-\bu_a)
  +\frac{\mu_{ab}}{T_{ab}} \Big(1-\frac{2}{5}\Omega_{12} \Big) \Big(\vecq_a -\frac{\rho_a}{\rho_b}\vecq_b \Big) \nn\\
  && - \Big(\frac{\mu_{ab}}{T_{ab}}\Big)^2 \Big( \frac{1}{8}- \frac{1}{10}\Omega_{12} + \frac{1}{70}\Omega_{13} \Big)
  \Big[\vecX^{(5)}_a -28\frac{p_a}{\rho_a}\vecq_a     - \frac{\rho_a}{\rho_b} \Big(\vecX^{(5)}_b -28\frac{p_b}{\rho_b}\vecq_b \Big) \Big] \Big\}, \label{eq:Final1}
\end{eqnarray}
where for later discussion of various collisional processes, it is beneficial to introduce coefficients
\begin{equation}
V_{ab (0)} = 1-\frac{2}{5}\Omega_{12}; \qquad V_{ab (3)} = \frac{1}{8}- \frac{1}{10}\Omega_{12} + \frac{1}{70}\Omega_{13}.\label{eq:Final11}
\end{equation}
Alternativelly, re-grouping the usual heat fluxes together
\begin{eqnarray}
  \boldsymbol{R}_a &=& \sum_{b\neq a} \nu_{ab} \Big\{ \rho_a (\bu_b-\bu_a)
  +\frac{\mu_{ab}}{T_{ab}} \Big[ V_{ab (1)}\vecq_a - V_{ab(2)}\frac{\rho_a}{\rho_b}\vecq_b \Big] \nn\\
  && - \Big(\frac{\mu_{ab}}{T_{ab}}\Big)^2 V_{ab (3)} 
  \Big[\vecX^{(5)}_a  - \frac{\rho_a}{\rho_b} \vecX^{(5)}_b \Big] \Big\}, \label{eq:Final1x}
\end{eqnarray}
yields coefficients 
\begin{eqnarray}
  V_{ab (1)} &=& 1-\frac{2}{5}\Omega_{12} + \frac{28 T_a m_b}{T_a m_b + T_b m_a} \Big( \frac{1}{8}- \frac{1}{10}\Omega_{12} + \frac{1}{70}\Omega_{13} \Big); \nn\\
  V_{ab (2)} &=& 1-\frac{2}{5}\Omega_{12} + \frac{28 T_b m_a}{T_a m_b + T_b m_a } \Big( \frac{1}{8}- \frac{1}{10}\Omega_{12} + \frac{1}{70}\Omega_{13} \Big). \label{eq:Vab12}
\end{eqnarray} 
For clarity, specifying the particular case of Coulomb collisions (by $\Omega_{12}=1$ and $\Omega_{13}=2$), yields
$V_{ab (0)}=+3/5$ and $V_{ab (3)}=+3/56$, recovering eqs. (15)-(16) of \cite{Hunana2022} (Part 1). Previously in Section 2.1 of Part 1,
all of the the collisional contributions were given in the form (\ref{eq:Final1x}), i.e. after
re-grouping the 1-Hermite and 2-Hermite contributions together (and the results before re-grouping can be found in Appendices of Part 1).
Here in Part 2, we prefer to keep the original form (\ref{eq:Final1}).  
The advantage is, that the reduction into the 1-Hermite approximation can be done easily (by prescribing
$\vecX^{(5)}_a =28\frac{p_a}{\rho_a}\vecq_a$ and $\vecX^{(5)}_b =28\frac{p_b}{\rho_b}\vecq_b$).
All the other collisional contributions will be given only in the form (\ref{eq:Final1}). We will also adopt the
free wording from Part 1 and call the ratios such as (\ref{eq:Vab12}) simply ``mass-ratio coefficients'',
even though they contain masses as well as temperatures and now also the dimensionless
Chapman-Cowling integrals.

%\newpage
%=============================================
\subsection{Energy exchange rates  \texorpdfstring{$Q_{a}$}{Q}}
The energy exchange rates of the 22-moment model are given by
%\begin{eqnarray}
%  Q_a &=& \sum_{b\neq a} 3\rho_a \nu_{ab} \frac{T_b-T_a}{m_a+m_b} 
%\end{eqnarray}
\begin{eqnarray}
 Q_{a} &=& \sum_{b\neq a} \frac{\rho_a \nu_{ab}}{(m_a+m_b)} \Big\{ 3(T_b-T_a)+ \hat{P}_{ab (1)}\frac{\rho_a}{n_a p_a} \widetilde{X}^{(4)}_a
 - \hat{P}_{ab (2)} \frac{\rho_b}{n_b p_b} \widetilde{X}^{(4)}_b  \Big\}, \label{eq:Thierry38}
\end{eqnarray}
with ``mass-ratio coefficients''  
%\begin{eqnarray}
% \hat{P}_{ab (1)} = \frac{3 T_a m_b (5 T_b m_b +4 T_b m_a -T_a m_b  )}{40 (T_a m_b +T_b m_a)^2};\qquad
%\hat{P}_{ab (2)} = \frac{3T_b m_a (5 T_a m_a +4 T_a m_b -T_b m_a  )}{40 (T_a m_b +T_b m_a)^2}. \label{eq:Thierry388}
%\end{eqnarray}
\begin{eqnarray}
  \hat{P}_{ab (1)} &=&  \frac{T_a m_b (7 T_b m_b +4 T_b m_a -3 T_a m_b  )}{8 (T_a m_b +T_b m_a)^2}
  - \Omega_{12} \frac{T_a m_b (7 T_b m_b  +2 T_b m_a -5 T_a m_b)}{10 (T_a m_b +T_b m_a)^2}\nn\\
  && \quad -  \Omega_{13} \frac{(T_a-T_b)T_a m_b^2}{10 (T_a m_b +T_b m_a)^2}; \nn\\
  \hat{P}_{ab (2)} &=&  \frac{T_b m_a (7 T_a m_a +4 T_a m_b -3 T_b m_a)}{8 (T_a m_b +T_b m_a)^2}
  - \Omega_{12} \frac{T_b m_a (7 T_a m_a +2 T_a m_b -5 T_b m_a)}{10 (T_a m_b +T_b m_a)^2} \nn\\
  && \quad + \Omega_{13} \frac{(T_a -T_b) T_b m_a^2  }{10 (T_a m_b +T_b m_a)^2}.
\end{eqnarray}
Terms proportional to $|\bu_b-\bu_a|^2$ are neglected in the semi-linear approximation and for a further discussion, see Section
    8.1 ``Energy Conservation'', p. 30 in Part 1 (and for the $Q_{ab}$ of 5-moment models, see also eq. (\ref{eq:Picus1}) here). 
In the formulation above, each Chapman-Cowling integral has its own mass-ratio coefficient (which for a given $T_a,T_b,m_a,m_b$ is a pure number).
This is the natural form how to write the results, because the above form comes directly from the calculations (where one naturally splits the results
to create the Chapman-Cowling integrals separately).
 Alternatively, the results can be re-arranged so that the Chapman-Cowling integrals are together, yielding 
\begin{eqnarray}
  \hat{P}_{ab (1)} &=&  \frac{T_a m_b}{2(T_a m_b +T_b m_a)^2}
  \Big[ T_b m_b \big(-\frac{7}{5}\Omega_{12} +\frac{1}{5}\Omega_{13} +\frac{7}{4} \big)
    +T_a m_b \big( \Omega_{12} - \frac{1}{5}\Omega_{13} - \frac{3}{4} \big) 
    - T_b m_a  \frac{2}{5}(\Omega_{12}- \frac{5}{2}) \Big];\nn\\
%===
  \hat{P}_{ab (2)} &=& \frac{T_b m_a}{2(T_b m_a +T_a m_b)^2}
  \Big[ T_a m_a \big(-\frac{7}{5}\Omega_{12} +\frac{1}{5}\Omega_{13} +\frac{7}{4} \big)
    +T_b m_a \big( \Omega_{12} - \frac{1}{5}\Omega_{13} - \frac{3}{4} \big) 
    - T_a m_b  \frac{2}{5}(\Omega_{12}- \frac{5}{2}) \Big]. \label{eq:P12}
\end{eqnarray}
%================================== 
As a quick double check, prescribing Coulomb collisions recovers the $\hat{P}_{ab (1)}, \hat{P}_{ab (2)}$ coefficients eq. (141) in Part 1.

%\newpage
%=============================================
\subsection{Stress tensor exchange rates  \texorpdfstring{$\bQ_{a}^{(2)}\,'$}{Q(2)'}}
The collisional exchange rates for the usual stress-tensor $\bPi_a^{(2)}$ are given by 
\begin{eqnarray}
  \bQ_{a}^{(2)}\,' &=& -\frac{3}{5}\nu_{aa} \Omega_{22} \bPi_a^{(2)}
  + \nu_{aa} \Big(\frac{3}{20} \Omega_{22} - \frac{3}{70} \Omega_{23} \Big)
  \Big( \frac{\rho_a}{p_a}\bPi^{(4)}_a- 7 \bPi^{(2)}_a\Big)    \nn\\
  && +\sum_{b\neq a} \frac{\rho_a \nu_{ab}}{m_a+m_b} \Big[ -K_{ab (1)} \frac{1}{n_a} \bPi^{(2)}_a +K_{ab (2)} \frac{1}{n_b} \bPi^{(2)}_b  \nn\\
   && +L_{ab (1)} \frac{1}{n_a} \Big( \frac{\rho_a}{p_a}\bPi^{(4)}_a- 7\bPi^{(2)}_a\Big)
   - L_{ab (2)} \frac{1}{n_b} \Big( \frac{\rho_b}{p_b}\bPi^{(4)}_b- 7\bPi^{(2)}_b\Big)  \Big], \label{eq:QabP}
\end{eqnarray}
with the 1-Hermite mass-ratio coefficients
\begin{eqnarray}
  K_{ab(1)} &=& \frac{2 T_b (m_a +m_b)}{(T_a m_b +T_b m_a)}
  - \frac{4 (T_b-T_a) m_b}{5 (T_a m_b +T_b m_a)} \Omega_{12} + \frac{ 3 m_b }{5 m_a} \Omega_{22};  \label{eq:Kab1T}\\
%===
  K_{ab(2)} &=& \frac{2 T_a (m_a +m_b)}{(T_a m_b +T_b m_a)}
  + \frac{4 (T_b-T_a) m_a }{5 (T_a m_b +T_b m_a)} \Omega_{12} - \frac{3}{5}\Omega_{22}. \label{eq:Kab2T}
\end{eqnarray}
It can be shown that this 1-Hermite viscosity model given by $K_{ab(1)}$; $K_{ab (2)}$
is equivalent to eq. (44) of \cite{Schunk1977}, or eq. (4.132a) of \cite{SchunkNagy2009}, obtained also by \cite{Burgers1969}.

The 2-Hermite coefficients read
%======
\begin{eqnarray}
  L_{ab(1)} &=& L_{a(11)} + L_{a(12)} +  L_{a(22)} +  L_{a(13)} +  L_{a(23)}; \label{eq:Lab1T0}\\
  L_{a(11)} &=& \frac{T_a T_b m_b  (m_a+m_b)}{(T_a m_b +T_b m_a)^2};\qquad
  L_{a(12)} = \Omega_{12} \frac{2 T_a m_b (T_a m_b -T_b m_a -2 T_b m_b) }{5(T_a m_b + T_b m_a)^2};\nn\\
  L_{a(22)} &=& \Omega_{22} \frac{3 T_a m_b^2 }{10(T_a m_b +T_b m_a)m_a};\qquad
  L_{a(13)} = \Omega_{13} \frac{4 T_a (T_b-T_a) m_b^2}{35 (T_a m_b +T_b m_a)^2};\nn\\
  L_{a(23)} &=& - \Omega_{23} \frac{3 T_a m_b^2 }{35 (T_a m_b +T_b m_a) m_a};\nn
\end{eqnarray}  
\begin{eqnarray}
  L_{ab(2)} &=& L_{b(11)} + L_{b(12)} +  L_{b(22)} +  L_{b(13)} +  L_{b(23)}; \label{eq:Lab2T0}\\
  L_{b(11)} &=& + \frac{T_a T_b m_a  (m_a + m_b)}{(T _a m_b +T_b m_a)^2}; \qquad
  L_{b(12)} =  - \Omega_{12} \frac{2 T_b m_a (2 T_a m_a +T_a m_b -T_b m_a)}{5(T_a m_b +T_b m_a)^2};\nn\\
  L_{b(22)} &=& - \Omega_{22} \frac{3 T_b m_a }{10(T_a m_b +T_b m_a)};\qquad
  L_{b(13)} = - \Omega_{13} \frac{4 T_b (T_b-T_a) m_a^2 }{35 (T_a m_b + T_b m_a)^2};\nn\\
  L_{b(23)} &=&  +\Omega_{23} \frac{3 T_b m_a }{35 (T_a m_b +T_b m_a)}.\nn
\end{eqnarray}

%\begin{eqnarray}
%  K_{ab(1)} &=& \frac{1}{5 m_a (T_a m_b +T_b m_a)} \Big\{ 10 T_b m_a^2
%  - 4 m_a m_b \Big[ (T_b-T_a) \Omega_{12} - T_b \big(\frac{3}{4} \Omega_{22}+\frac{5}{2} \big) \Big] 
%  +3  T_a m_b^2 \Omega_{22} \Big\};
%\end{eqnarray}  
%\begin{eqnarray}
%  L_{ab (1)} &=& -\, \frac{2 T_a m_b}{ 5(T_a m_b +T_b m_a)^2 m_a}
%  \Big\{ T_b m_a^2 (\Omega_{12}- \frac{5}{2}) - \frac{3}{4} T_a m_b^2 \big( \Omega_{22} -\frac{2}{7} \Omega_{23} \big) \nn\\
%&&  +m_a m_b \Big[ T_b \Big( \Omega_{12} - \frac{3}{4}\Omega_{22} + \frac{3}{14}\Omega_{23} -\frac{5}{2}  \Big)
%    +(T_b-T_a) \big( \Omega_{12} - \frac{2}{7} \Omega_{13}\big) \Big] \Big\};\nn\\
%===========
%  L_{ab(2)} &=& \frac{2 T_b m_a}{5 (T_a m_b + T_b m_a)^2} 
%   \Big\{ m_a \Big[ T_a \big( \frac{5}{2}-\Omega_{12}\big) 
%    + T_b \big( -\frac{3}{4}\Omega_{22} +\frac{3}{14}\Omega_{23} \big)
%    +(T_b-T_a) \big(\Omega_{12} - \frac{2}{7}\Omega_{13}\big) \Big]\nn\\
%   && -m_b T_a \big( \Omega_{12} + \frac{3}{4} \Omega_{22}
%   -\frac{3}{14} \Omega_{23} -\frac{5}{2} \big) \Big\}.
%\end{eqnarray}

Alternatively, the (\ref{eq:Lab1T0})-(\ref{eq:Lab2T0}) can be re-arranged into
\begin{eqnarray}
  L_{ab (1)} &=& -\, \frac{2 T_a m_b}{ 5(T_a m_b +T_b m_a)^2 m_a}
  \Big\{ T_b m_a^2 (\Omega_{12}- \frac{5}{2}) +  T_a m_b^2 \big( - \frac{3}{4} \Omega_{22} +\frac{3}{14} \Omega_{23} \big) \nn\\
&&  +m_a m_b \Big[ T_b \Big( \Omega_{12} -\frac{5}{2} - \frac{3}{4}\Omega_{22} + \frac{3}{14}\Omega_{23}   \Big)
    +(T_b-T_a) \big( \Omega_{12} - \frac{2}{7} \Omega_{13}\big) \Big] \Big\}; \label{eq:Lab1T}\\
%===========
  L_{ab(2)} &=& \frac{2 T_b m_a}{5 (T_a m_b + T_b m_a)^2} 
   \Big\{ m_a \Big[ - T_a \big( \Omega_{12} -\frac{5}{2} \big) 
    + T_b \big( -\frac{3}{4}\Omega_{22} +\frac{3}{14}\Omega_{23} \big)
    +(T_b-T_a) \big(\Omega_{12} - \frac{2}{7}\Omega_{13}\big) \Big]\nn\\
   && -m_b T_a \big( \Omega_{12} -\frac{5}{2} + \frac{3}{4} \Omega_{22}
   -\frac{3}{14} \Omega_{23}  \big) \Big\}.\label{eq:Lab2T}
\end{eqnarray}
As can be seen, in the re-arranged form (\ref{eq:Lab1T})-(\ref{eq:Lab2T}) the $L_{ab (1)}$ and $L_{ab (2)}$ coefficients are symmetric only partially,
and one keeps re-arranging them back and forth to show this partial symmetry. All higher-order 2-Hermite coefficients
for artibtrary temperatures will be given only in the form (\ref{eq:Lab1T0})-(\ref{eq:Lab2T0}) and a potential
user can rearrange these if needed. For small temperature
differences (see later Section \ref{sec:Tsmall}), all of the coefficients will be given in the re-arranged form  (\ref{eq:Lab1T})-(\ref{eq:Lab2T}).

Note that in the final collisional contributions of Part 1 given there by eqs. (22)-(23), a further re-arrangement is done 
for the $\bPi^{(2)}$ by introducing (hat) $\hat{K}_{ab (1)}=K_{ab (1)} +7L_{ab (1)}$ and $\hat{K}_{ab (2)}=K_{ab (2)} +7L_{ab (2)}$, and the (non-hat)
coefficients $K_{ab (1)}$ \& $K_{ab (2)}$ are given there by (L27). Now it it is easy to verify that prescribing Coulomb collisions indeed recovers
the equations of Part 1.

%\begin{eqnarray}
%L2 := -2/5 {  T_a}  \left( \Omega_{12}+3/4 \Omega_{22}-3/14 \Omega_{23}-5/2
% \right) {  m_a} {  T_b} {  m_b}+2/5  \left(  \left( 5/2-\Omega_{12} \right) {  T_a}+{  TM} \Omega_{12}+ \left( -3/4 \Omega22+3/14 
%\Omega_{23} \right) {  T_b}-2/7 {  TM} \Omega_{13} \right) {{  m_a}}^{2}{  T_b}
%\end{eqnarray}
  
%\newpage
%\subsubsection*{Equal temperatures}
%\begin{eqnarray}
% K_{ab(1)} &=& 2 + \frac{3}{5} \frac{m_b}{m_a}\Omega_{22};\qquad  K_{ab(2)} = 2 - \frac{3}{5}\Omega_{22};\nn\\
%  L_{ab(1)} &=& - \, \frac{2 m_b}{ 5 m_a (m_a +m_b)} \Big[
%  m_a(\Omega_{12}- \frac{5}{2}) - \frac{3}{4} m_b \big( \Omega_{22}- \frac{2}{7}\Omega_{23}\big) \Big] \nn\\
%L_{ab(2)} &=&  \frac{m_a}{70(m_a + m_b)} \big(-28 \Omega_{12} -21 \Omega_{22} +6 \Omega_{23}+70 \big); 
%\end{eqnarray}

%\subsubsection*{Self-collisions}
%\begin{eqnarray}
%  - K_{aa (1)} + K_{aa (2)} &=& -\frac{6}{5} \Omega_{22} ;\qquad
%   L_{aa (1)} - L_{aa (2)} = \frac{3}{10} \Omega_{22} - \frac{3}{35} \Omega_{23}.
%  L_{aa (1)} &=&  -\frac{1}{5}\Omega_{12} +\frac{3}{20}\Omega_{22} - \frac{3}{70} \Omega_{23} +\frac{1}{2} ; \nn\\
%  L_{aa (2)} &=& - \frac{1}{5}\Omega_{12} - \frac{3}{20}\Omega_{22} + \frac{3}{70}\Omega_{23} +\frac{1}{2}; \qquad 
%\end{eqnarray}

%\vspace{3cm}

%\newpage
%==========================================================================================
%==========================================================================================
\subsection{Higher-order stress tensor exchange rates  \texorpdfstring{$\bQ_{a}^{(4)}\,'$}{Q(4)'}}
The collisional exchange rates for the higher-order stress-tensor $\bPi_a^{(4)}$ are given by
%\begin{eqnarray}
%  \bQ_{a}^{(4)}\,' &=& -  \nu_{aa} \frac{3}{10} \big( 7\Omega_{22} +2\Omega_{23}\big) \frac{p_a}{\rho_a}\bPi_a^{(2)}
%  - \frac{1}{280}(7 \Omega_{22}+12 \Omega_{24})  \Big( \bPi^{(4)}_a- 7 \frac{p_a}{\rho_a} \bPi^{(2)}_a\Big) \nn\\
%&&   + \sum_{b\neq a} \nu_{ab} \Big[ - {M}_{ab (1)} \frac{p_a}{\rho_a} \bPi^{(2)}_a 
%    +{M}_{ab (2)} \frac{p_a^2}{\rho_a p_b} \bPi^{(2)}_b \nn\\
%    && \qquad -N_{ab (1)} \Big( \bPi^{(4)}_a- 7 \frac{p_a}{\rho_a} \bPi^{(2)}_a\Big)
%    - N_{ab (2)}\frac{p_a^2 \rho_b}{p_b^2\rho_a} \Big( \bPi^{(4)}_b- 7 \frac{p_b}{\rho_b} \bPi^{(2)}_b\Big) \Big],
%\end{eqnarray}
\begin{eqnarray}
  \bQ_{a}^{(4)}\,' &=& -  \nu_{aa} \big( \frac{21}{10}\Omega_{22} +\frac{3}{5}\Omega_{23}\big) \frac{p_a}{\rho_a}\bPi_a^{(2)}
  - \nu_{aa} \big( \frac{1}{40} \Omega_{22}+ \frac{3}{70} \Omega_{24} \big)  \Big( \bPi^{(4)}_a- 7 \frac{p_a}{\rho_a} \bPi^{(2)}_a\Big) \nn\\
&&   + \sum_{b\neq a} \nu_{ab} \Big[ - {M}_{ab (1)} \frac{p_a}{\rho_a} \bPi^{(2)}_a 
    +{M}_{ab (2)} \frac{p_a^2}{\rho_a p_b} \bPi^{(2)}_b \nn\\
    && \qquad -N_{ab (1)} \Big( \bPi^{(4)}_a- 7 \frac{p_a}{\rho_a} \bPi^{(2)}_a\Big)
    - N_{ab (2)}\frac{p_a^2 \rho_b}{p_b^2\rho_a} \Big( \bPi^{(4)}_b- 7 \frac{p_b}{\rho_b} \bPi^{(2)}_b\Big) \Big], \label{eq:Q4abP}
\end{eqnarray}
with  mass-ratio coefficients
\begin{eqnarray}
  {M}_{ab (1)} &=& M_{a(11)} + M_{a(12)} +  M_{a(22)} + M_{a(13)} + M_{a(23)} + M_{a(33)}; \label{eq:M1a}\\
  M_{a(11)} &=& \frac{14 T_b^2 m_a^2 (2 T_a m_a +T_a m_b -T_b m_a)}{T_a (m_a +m_b) (T_a m_b +T_b m_a)^2  };\nn\\
  M_{a(12)} &=& \Omega_{12} \frac{28 T_b m_a m_b (4 T_a^2 m_a^2 +T_a^2 m_a m_b +T_a^2 m_b^2 -7 T_a T_b m_a^2 +T_a T_b m_a m_b +4T_b^2 m_a^2)}{5 T_a (m_a+m_b)^2 (T_a m_b +T_b m_a)^2  };\nn\\
  M_{a(22)} &=& \Omega_{22} \frac{7 T_b m_a m_b (11 T_a m_a +3 T_a m_b -8 T_b m_a) }{5 T_a (m_a +m_b)^2 (T_a m_b +T_b m_a)};\nn\\
  M_{a(13)} &=& -\, \Omega_{13} \frac{4(T_b-T_a) m_a m_b^2 (4 T_a^2 m_a^2 +T_a^2 m_b^2 -8 T_a T_b m_a^2 +2 T_a T_b m_a m_b +5 T_b^2 m_a^2)}{5 T_a (m_a +m_b)^3 (T_a m_b +T_b m_a)^2 };\nn\\
  M_{a(23)} &=& \Omega_{23} \frac{2 m_b^2 (11 T_a^2 m_a^2 +3 T_a^2 m_b^2 -22 T_a T_b m_a^2 +6 T_a T_b m_a m_b +14 T_b^2 m_a^2)}{5 T_a (m_a +m_b)^3 (T_a m_b +T_b m_a)};\nn\\
  M_{a(33)} &=& -\,\Omega_{33} \frac{12 m_a m_b^2  (T_b-T_a)  }{5 T_a (m_a+ m_b)^3 }; \nn
\end{eqnarray}
%====================================
\begin{eqnarray}
  {M}_{ab (2)} &=& M_{b(11)} + M_{b(12)} +  M_{b(22)} + M_{b(13)} + M_{b(23)} + M_{b(33)}; \label{eq:M1b}\\
  M_{b(11)} &=& - \, \frac{14 T_b m_a^2 (T_a m_b -T_b m_a -2 T_b m_b) }{(m_a +m_b) (T_a m_b +T_b m_a)^2 } ;\nn\\
  M_{b(12)} &=& \Omega_{12} \frac{28 T_b m_a (3 T_a^2 m_a^2 m_b +T_a^2 m_b^3 -T_a T_b m_a^3 -7 T_a T_b m_a^2 m_b +2 T_a T_b m_a m_b^2 +T_b^2 m_a^3 +5 T_b^2 m_a^2 m_b)}{5 T_a (m_a+m_b)^2 (T_a m_b +T_b m_a)^2 };\nn\\
  M_{b(22)} &=& \Omega_{22} \frac{7 T_b  m_a^2 (8 T_a m_b -3 T_b m_a -11 T_b m_b) }{5 T_a (m_a +m_b)^2 (T_a m_b +T_b m_a) };\nn\\
  M_{b(13)} &=& \Omega_{13} \frac{4(T_b-T_a)T_b m_a^2 m_b (4 T_a^2 m_a^2 +T_a^2 m_b^2 -8 T_a T_b m_a^2 +2 T_a T_b m_a m_b +5 T_b^2 m_a^2)}{ 5 T_a^2 (m_a +m_b)^3 (T_a m_b +T_b m_a)^2};\nn\\
  M_{b(23)} &=&  -\, \Omega_{23}  \frac{2 T_b m_a m_b (11 T_a^2 m_a^2 +3 T_a^2 m_b^2 -22 T_a T_b m_a^2 +6 T_a T_b m_a m_b +14 T_b^2 m_a^2) }{5 T_a^2 (m_a +m_b)^3 (T_a m_b +T_b m_a)};\nn\\
  M_{b(33)} &=& \Omega_{33} \frac{12 (T_b-T_a)T_b m_a^2 m_b  }{5 (m_a + m_b)^3 T_a^2}; \nn
\end{eqnarray}
%===================================
\begin{eqnarray}
  {N}_{ab (1)} &=& N_{a(11)} + N_{a(12)} +  N_{a(22)} + N_{a(13)} + N_{a(23)} + N_{a(33)}  + N_{a(14)} + N_{a(24)} + N_{a(34)}; \label{eq:N1a}\\
  N_{a(11)} &=& -\, \frac{T_b^2 m_a^2 (14 T_a m_a m_b +7 T_a m_b^2 -4 T_b m_a^2 -11 T_b m_a m_b)}{(m_a +m_b)(T_a m_b + T_b m_a)^3 };\nn\\
%===  
  N_{a(12)} &=& -\, \Omega_{12} \frac{2 T_b  m_a m_b}{5 (m_a +m_b)^2 (T_a m_b +T_b m_a)^3 }  \Big( 28 T_a^2 m_a^2 m_b +7 T_a^2 m_a m_b^2 +7 T_a^2 m_b^3 -26 T_a T_b m_a^3 \nn\\
   && -82 T_a T_b m_a^2 m_b +19 T_b^2 m_a^3 +47 T_b^2 m_a^2 m_b \Big) ;\nn\\
  N_{a(22)} &=& -\, \Omega_{22} \frac{T_b  m_a m_b (77 T_a m_a m_b +21 T_a m_b^2 -22 T_b m_a^2 -78 T_b m_a m_b)}{10 (m_a+m_b)^2 (T_a m_b +T_b m_a)^2 };\nn\\
%===  
  N_{a(13)} &=& -\, \Omega_{13} \frac{2 m_a m_b^2 }{35 (m_a +m_b)^3 (T_a m_b +T_b m_a)^3 }
    \Big( 28 T_a^3 m_a^2 m_b +7 T_a^3 m_b^3 -80 T_a^2 T_b m_a^3 -178 T_a^2 T_b m_a^2 m_b \nn\\
    && -16 T_a^2 T_b m_a m_b^2
    -23 T_a^2 T_b m_b^3 +146 T_a T_b^2 m_a^3 +219 T_a T_b^2 m_a^2 m_b -32 T_a T_b^2 m_a m_b^2 -82 T_b^3 m_a^3 \nn\\
    && -117 T_b^3 m_a^2 m_b \Big);\nn\\
%===    
  N_{a(23)} &=& -\, \Omega_{23} \frac{m_b^2}{35 (m_a+m_b)^3 (T_a m_b +T_b m_a)^2 }  \Big( 77 T_a^2 m_a^2 m_b +21 T_a^2 m_b^3 -121 T_a T_b m_a^3 -296 T_a T_b m_a^2 m_b \nn\\
  &&   +21 T_a T_b m_a m_b^2 +100 T_b^2 m_a^3 +198 T_b^2 m_a^2 m_b \Big) ;\nn\\
%===  
  N_{a(33)} &=& -\, \Omega_{33} \frac{6 m_a m_b^2 (7 T_a m_b -2 T_b m_a -9 T_b m_b)}{35 (m_a +m_b)^3 (T_a m_b +T_b m_a)}; \nn\\
  N_{a(14)} &=& -\, \Omega_{14} \frac{4(T_b-T_a)m_a m_b^3 (4 T_a^2 m_a^2 +T_a^2 m_b^2 -8 T_a T_b m_a^2
    +2 T_a T_b m_a m_b +5 T_b^2 m_a^2)}{35(m_a+m_b)^3(T_a m_b +T_b m_a)^3 };\nn\\
  N_{a(24)} &=&  \Omega_{24} \frac{2 m_b^3 (11 T_a^2 m_a^2 +3 T_a^2 m_b^2 -22 T_a T_b m_a^2 +6 T_a T_b m_a m_b
    +14 T_b^2 m_a^2) }{ 35(m_a +m_b)^3 (T_a m_b +T_b m_a)^2 };\nn\\
  N_{a(34)} &=&  -\, \Omega_{34} \frac{12 m_a m_b^3 (T_b-T_a)  }{35 (m_a +m_b)^3 (T_a m_b +T_b m_a)};\nn
\end{eqnarray}
%===================================
\begin{eqnarray}
  {N}_{ab (2)} &=& N_{b(11)} + N_{b(12)} +  N_{b(22)} + N_{b(13)} + N_{b(23)} + N_{b(33)}  + N_{b(14)} + N_{b(24)} + N_{b(34)}; \label{eq:N1b}\\
  N_{b(11)} &=& - \, \frac{T_b^2 m_a^2  (11 T_a m_a m_b +4 T_a m_b^2 -7 T_b m_a^2 -14 T_b m_a m_b)}{(m_a +m_b) (T_a m_b +T_b m_a)^3 };\nn\\
  N_{b(12)} &=&  \Omega_{12} \frac{2 T_b^2 m_a^2}{5 T_a (m_a+m_b)^2 (T_a m_b +T_b m_a)^3} \Big( 40 T_a^2 m_a^2 m_b +19 T_a^2 m_a m_b^2
  +7 T_a^2 m_b^3 -14 T_a T_b m_a^3 -82 T_a T_b m_a^2 m_b \nn\\
  && \quad -12 T_a T_b m_a m_b^2 +7 T_b^2 m_a^3 +35 T_b^2 m_a^2 m_b \Big);\nn\\
  N_{b(22)} &=&  \Omega_{22} \frac{T_b^2 m_a^2 (78 T_a m_a m_b +22 T_a m_b^2 -21 T_b m_a^2 -77 T_b m_a m_b)}{10 T_a (m_a+m_b)^2 (T_a m_b +T_b m_a)^2 };\nn\\
%\end{eqnarray}
  %===
%\begin{eqnarray}
%===
  N_{b(13)} &=&  -\, \Omega_{13} \frac{2 T_b^2 m_a^2}{35 T_a^2 (m_a+m_b)^3 (T_a m_b +T_b m_a)^3 }\Big(
  94 T_a^3 m_a^3 m_b +66 T_a^3 m_a^2 m_b^2 +23 T_a^3 m_a m_b^3 +16 T_a^3 m_b^4  \nn\\
  && \quad -14 T_a^2 T_b m_a^4 -244 T_a^2 T_b m_a^3 m_b -100 T_a^2 T_b m_a^2 m_b^2 +25 T_a^2 T_b m_a m_b^3 \nn\\
  && \quad +14 T_a T_b^2 m_a^4 +201 T_a T_b^2 m_a^3 m_b +82 T_a T_b^2 m_a^2 m_b^2 -35 T_b^3 m_a^3 m_b \Big);\nn \\
%====  
  N_{b(23)} &=&  -\, \Omega_{23} \frac{T_b^2 m_a^2}{35 T_a^2 (m_a+m_b)^3 (T_a m_b+T_b m_a)^2 }
  \Big( 177 T_a^2 m_a^2 m_b +100 T_a^2 m_a m_b^2 +21 T_a^2 m_b^3 -21 T_a T_b m_a^3 \nn\\
  && \quad -296 T_a T_b m_a^2 m_b -79 T_a T_b m_a m_b^2 +98 T_b^2 m_a^2 m_b \Big) ;\nn\\
  N_{b(33)} &=&  -\, \Omega_{33} \frac{6 T_b^2 m_b m_a^2 (9 T_a m_a +2 T_a m_b -7 T_b m_a)}{35 T_a^2 (m_a +m_b)^3 (T_a m_b +T_b m_a) }; \nn\\
  N_{b(14)} &=&  -\, \Omega_{14} \frac{4 (T_b-T_a) T_b^2  m_a^3 m_b (4 T_a^2 m_a^2 +T_a^2 m_b^2 -8 T_a T_b m_a^2
    +2 T_a T_b m_a m_b +5 T_b^2 m_a^2)}{35 T_a^2 (m_a+m_b)^3 (T_a m_b+T_b m_a)^3 };\nn\\
  N_{b(24)} &=&  \Omega_{24} \frac{2 T_b^2 m_b m_a^2  (11 T_a^2 m_a^2 +3 T_a^2 m_b^2 -22 T_a T_b m_a^2
    +6 T_a T_b m_a m_b +14 T_b^2 m_a^2)}{35 T_a^2 (m_a+m_b)^3 (T_a m_b+T_b m_a)^2 };\nn\\
  N_{b(34)} &=&  -\, \Omega_{34} \frac{12 (T_b-T_a) T_b^2 m_a^3  m_b  }{35 T_a^2 (m_a+m_b)^3 (T_a m_b+T_b m_a) } .\nn
\end{eqnarray}
Obviously, the 2-Hermite viscosity model is much more complicated than the 1-Hermite model given by eqs. (\ref{eq:Kab1T})-(\ref{eq:Kab2T}).
Prescribing Coulomb collisions recovers eqs. (24)-(25) of Part 1, with the (non-hat) $M_{ab (1)}$ \& $M_{ab (2)}$ given by (L46) and further
re-arrangement (L51).

\subsection{Heat flux exchange rates  \texorpdfstring{$\vecQ^{(3)}_{a}\,'$}{Q(3)'}}
The exchange rates for the usual heat flux $\vecq_a$ are given by (note that the momentum exchange rates $\boldsymbol{R}_a$ enter and the
expressions are not further re-arranged)
\begin{eqnarray}
  \vecQ^{(3)}_{a}\,' &=& - \frac{2}{5}\nu_{aa}\Omega_{22}\vecq_a
  + \nu_{aa} \Big(\frac{1}{20}\Omega_{22}- \frac{1}{70}\Omega_{23}\Big) \Big(\frac{\rho_a}{p_a}\vecX^{(5)}_a -28 \vecq_a\Big)\nn\\
  && +\sum_{b\neq a} \nu_{ab}\Big\{  - D_{ab (1)} \vecq_a
  +  \frac{\rho_a}{\rho_b} D_{ab (2)} \vecq_b - p_a  U_{ab(1)} (\bu_b-\bu_a) \nn\\
  && + E_{ab (1)} \Big(\frac{\rho_a}{p_a}\vecX^{(5)}_a -28 \vecq_a\Big)
  + E_{ab (2)} \frac{\rho_a}{\rho_b} \Big(\frac{\rho_b}{p_b}\vecX^{(5)}_b -28 \vecq_b\Big) \Big\} 
   -\frac{5}{2} \frac{p_a}{\rho_a} \boldsymbol{R}_a, \label{eq:Q3P}
\end{eqnarray}  
with the 1-Hermite mass-ratios coefficients 
\begin{eqnarray}
D_{ab (1)} &=& D_{a (11)} + D_{a (12)} + D_{a (22)} + D_{a (13)} + D_{a (23)} ; \label{eq:HFSchunkD1}\\ 
D_{a (11)} &=& - \frac{m_a T_b (15 T_a m_a m_b +5 T_a m_b^2 -6 T_b m_a^2 -16 T_b m_a m_b)}{2(T_a m_b +T_b m_a)^2 (m_a+m_b)} ;\nn\\
D_{a (12)} &=& - \Omega_{12} \frac{m_b}{5 (T_a m_b +T_b m_a)^2 (m_a+m_b)^2} \Big( 15 T_a^2 m_a^2 m_b +5 T_a^2 m_b^3 -27 T_a T_b m_a^3 -62 T_a T_b m_a^2 m_b \nn\\
 && +5 T_a T_b m_a m_b^2 +22 T_b^2 m_a^3 +42 T_b^2 m_a^2 m_b \Big);\nn\\
%===
D_{a(22)} &=& - \Omega_{22} \frac{2  m_a m_b (5 T_a m_b -2 T_b m_a -7 T_b m_b)}{5 (T_a m_b +T_b m_a)(m_a+m_b)^2} ;\nn\\
D_{a(13)} &=& + \Omega_{13} \frac{2 m_b^2 (3 T_a^2 m_a^2 +T_a^2 m_b^2 -6 T_a T_b m_a^2 +2 T_a T_b m_a m_b +4 T_b^2 m_a^2)}{5 (T_a m_b +T_b m_a)^2 (m_a+m_b)^2} ;\nn\\
D_{a(23)} &=& - \Omega_{23} \frac{4 (T_b-T_a) m_a m_b^2 }{5 (T_a m_b +T_b m_a)(m_a +m_b)^2} ;\nn
\end{eqnarray}  
%=================
\begin{eqnarray}
D_{ab (2)} &=& D_{b (11)} + D_{b (12)} + D_{b (22)} + D_{b (13)} + D_{b (23)} ; \label{eq:HFSchunkD2}\\ 
D_{b (11)} &=& + \frac{T_a m_b (16 T_a m_a m_b +6 T_a m_b^2 -5 T_b m_a^2 -15 T_b m_a m_b)}{2 (T_a m_b + T_b m_a)^2 (m_a+ m_b)};\nn\\
D_{b (12)} &=& - \Omega_{12} \frac{m_b (37 T_a^2 m_a^2 m_b +22 T_a^2 m_a m_b^2 +5 T_a^2 m_b^3
  -5 T_a T_b m_a^3 -62 T_a T_b m_a^2 m_b -17 T_a T_b m_a m_b^2 +20 T_b^2 m_a^2 m_b) }{5 (T_a m_b +T_b m_a)^2 (m_a +m_b)^2};\nn\\
D_{b(22)} &=& - \Omega_{22} \frac{2 m_b^2 (7 T_a m_a +2 T_a m_b -5 T_b m_a)}{5 (m_a + m_b)^2 (T_a m_b +T_b m_a)};\nn\\
D_{b(13)} &=& + \Omega_{13} \frac{2 m_b^2 (3 T_a^2 m_a^2 +T_a^2 m_b^2 -6 T_a T_b m_a^2 +2 T_a T_b m_a m_b +4 T_b^2 m_a^2) }{5 (T_a m_b +T_b m_a)^2 (m_a +m_b)^2}
= D_{a(13)};\nn\\
D_{b(23)} &=& - \Omega_{23} \frac{4 (T_b-T_a) m_a m_b^2 }{5(T_a m_b+T_b m_a)(m_a +m_b)^2}= D_{a(23)};\nn
\end{eqnarray}
%================
\begin{eqnarray}
  U_{ab (1)} &=&  + \frac{5 m_a (2 T_a m_b -T_b m_a -3 T_b m_b)}{2 (T_a m_b +T_b m_a)(m_a+m_b)};\nn\\
            && - \Omega_{12} \frac{m_b (3 T_a^2 m_a^2 +T_a^2 m_b^2
    -6 T_a T_b m_a^2 +2 T_a T_b m_a m_b +4 T_b^2 m_a^2)}{T_a (m_a+m_b)^2 (T_a m_b +T_b m_a)};\nn\\
     && - \Omega_{22} \frac{2(T_a-T_b)m_a m_b }{T_a (m_a +m_b)^2}. \label{eq:HFSchunkU1}
\end{eqnarray}
Alternatively, these 1-Hermite coefficients can be re-arranged into 
%=========================================
\begin{eqnarray}
D_{ab (1)} &=& + \frac{1}{(m_a +m_b)^2 (T_a m_b +T_b m_a)^2} \Big\{
 T_a^2  m_b^4 \Big( \Omega_{12}- \frac{2}{5} \Omega_{13} \Big) \nn\\
&& +  T_a m_a m_b^3 \Big[ T_b \Big(\Omega_{12} - \frac{4}{5} \Omega_{13} - \frac{14}{5} \Omega_{22}
+\frac{4}{5}\Omega_{23} +\frac{5}{2}\Big)
 +2 T_a \Big( \Omega_{22}-\frac{2}{5}\Omega_{23} \Big) \Big] \nn\\
 && + 3 m_a^2 m_b^2 \Big[ \frac{14}{5} T_b^2\Big( \Omega_{12} - \frac{4}{21}\Omega_{13} - \frac{1}{3} \Omega_{22}
   + \frac{2}{21} \Omega_{23} -\frac{20}{21} \Big)  \nn\\
  && \qquad - \frac{62}{15} T_a T_b \Big( \Omega_{12} -\frac{6}{31} \Omega_{13} -\frac{3}{31} \Omega_{22}
  +\frac{2}{31} \Omega_{23} - \frac{25}{31} \Big) 
  +T_a^2 \Big( \Omega_{12} -\frac{2}{5} \Omega_{13} \Big) \Big] \nn\\
&& - \frac{1}{5} T_b m_a^3 m_b \Big[ - 22  T_b \Big( \Omega_{12} - \frac{2}{11}  \Omega_{22} - \frac{5}{2} \Big)
  + 27 T_a \Big( \Omega_{12}-\frac{25}{18} \Big) \Big]  -3 T_b^2 m_a^4 \Big\}; \label{eq:HFSchunkD1r}
\end{eqnarray}
%==================
\begin{eqnarray}
  D_{ab (2)} &=& -\frac{m_b }{(m_a +m_b)^2 (T_a m_b +T_b m_a)^2} 
   \Big\{ T_a^2 m_b^3 \Big(\Omega_{12} - \frac{2}{5} \Omega_{13}  +\frac{4}{5}\Omega_{22}-3 \Big) \nn\\
  &&  + \frac{22}{5} T_a  m_a m_b^2 \Big[
    T_a \Big(\Omega_{12}+\frac{7}{11}\Omega_{22}- \frac{2}{11} \Omega_{23}-\frac{5}{2} \Big)
    - \frac{17}{22} T_b \Big (\Omega_{12} + \frac{4}{17} \Omega_{13}
  + \frac{6}{17} \Omega_{22} - \frac{4}{17} \Omega_{23} - \frac{75}{34} \Big) \Big]  \nn\\
&&  + \frac{37}{5} m_a^2 m_b \Big[  T_a^2 \Big(\Omega_{12} - \frac{6}{37} \Omega_{13} - \frac{40}{37} \Big)
    - \frac{62}{37} T_b T_a \Big(\Omega_{12} - \frac{6}{31} \Omega_{13} -\frac{25}{31}
    - \frac{7}{31}\Omega_{22} + \frac{2}{31} \Omega_{23} \Big) \nn\\
&& + \frac{20}{37} T_b^2 \Big(\Omega_{12} - \frac{1}{2}\Omega_{22} + \frac{1}{5} \Omega_{23}  - \frac{2}{5} \Omega_{13} \Big) \Big] 
  -  T_a T_b m_a^3 \Big( \Omega_{12}-\frac{5}{2} \Big)  \Big\}; \label{eq:HFSchunkD2r}
\end{eqnarray}
%==================
%==================
\begin{eqnarray}
U_{ab (1)} &=& + \frac{1}{2 T_a (m_a +m_b)^2 (T_a m_b +T_b m_a)}
\Big\{ - 5 T_a T_b m_a^3 \nn\\
&& -6  m_a^2 m_b \Big[ T_a^2 \Big( \Omega_{12}-\frac{5}{3} \Big)
  -2  T_a T_b \Big( \Omega_{12} -\frac{1}{3}\Omega_{22} -\frac{5}{3}\Big)
  +\frac{4}{3} T_b^2 \Big( \Omega_{12} - \frac{1}{2}\Omega_{22} \Big)  \Big] \nn\\
&&  - 4 m_a m_b^2 T_a  \Big[ T_a \Big( \Omega_{22} - \frac{5}{2} \Big)  +T_b \Big( \Omega_{12} -\Omega_{22} + \frac{15}{4} \Big) \Big] 
-2 T_a^2 m_b^3 \Omega_{12}  \Big\}. \label{eq:HFSchunkU1r}
\end{eqnarray}
It can be shown that this 1-Hermite heat flux model is equivalent to eqs. (45)-(49) of \cite{Schunk1977}, or eqs. (4.132b)-(4.133d)
of \cite{SchunkNagy2009}, obtained also by \cite{Burgers1969}. \\

\noindent
The 2-Hermite coefficients read
\begin{eqnarray}
  {E}_{ab (1)} &=& E_{a(11)} + E_{a(12)} +  E_{a(22)} + E_{a(13)} + E_{a(23)}  + E_{a(14)} + E_{a(24)} ; \label{eq:E1}\\
  E_{a(11)} &=& -\, \frac{T_b T_a m_a m_b (15 T_a m_a m_b +5 T_a m_b^2 -12 T_b m_a^2 -22 T_b m_a m_b)}{16(m_a+m_b)(T_a m_b +T_b m_a)^3};\nn\\
  E_{a(12)} &=& - \Omega_{12} \frac{T_a m_b}{40 (T_a m_b +T_b m_a)^3 (m_a +m_b)^2}
  \Big(15 T_a^2 m_a^2 m_b^2 +5 T_a^2 m_b^4 -54 T_a T_b m_a^3 m_b  -94 T_a T_b m_a^2 m_b^2 \nn\\
  && +12 T_b^2 m_a^4 +68 T_b^2 m_a^3 m_b +76 T_b^2 m_a^2 m_b^2 \Big);\nn\\
%===  
  E_{a(22)} &=& - \Omega_{22} \frac{T_a m_a  m_b^2 (5 T_a m_b -4 T_b m_a -9 T_b m_b)}{20 (m_a+m_b)^2 (T_a m_b +T_b m_a)^2};\nn\\
  E_{a(13)} &=& + \Omega_{13} \frac{T_a m_b^2}{140(T_a m_b+T_b m_a)^3 (m_a+m_b)^2}
  \Big(42 T_a^2 m_a^2 m_b +14 T_a^2 m_b^3 -39 T_a T_b m_a^3  -128 T_a T_b m_a^2 m_b \nn\\
  && +23 T_a T_b m_a m_b^2 +34 T_b^2 m_a^3 +90 T_b^2 m_a^2 m_b \Big) ;\nn\\
%===  
  E_{a(23)} &=& + \Omega_{23} \frac{ T_a m_a m_b^2 (7 T_a m_b -2 T_b m_a -9 T_b m_b)}{35 (m_a+m_b)^2 (T_a m_b+T_b m_a)^2};\nn\\
  E_{a(14)} &=& -\, \Omega_{14} \frac{ T_a m_b^3 (3 T_a^2 m_a^2 +T_a^2 m_b^2 -6 T_a T_b m_a^2
    +2 T_a T_b m_a m_b +4 T_b^2 m_a^2)}{70 (T_a m_b +T_b m_a)^3 (m_a+m_b)^2};\nn\\
  E_{a(24)} &=& + \Omega_{24} \frac{(T_b-T_a)T_a m_a m_b^3 }{35(m_a+m_b)^2 (T_a m_b+T_b m_a)^2};\nn
\end{eqnarray}
%========================
\begin{eqnarray}
  {E}_{ab (2)} &=& E_{b(11)} + E_{b(12)} +  E_{b(22)} + E_{b(13)} + E_{b(23)} + E_{b(14)} + E_{b(24)}; \label{eq:E2}\\
  E_{b(11)} &=& -\, \frac{T_a T_b m_a m_b (22 T_a m_a m_b +12 T_a m_b^2 -5 T_b m_a^2 -15 T_b m_a m_b)}{16(T_a m_b +T_b m_a)^3 (m_a+m_b)};\nn\\
  E_{b(12)} &=& + \Omega_{12} \frac{ T_b m_a m_b } {40 (T_a m_b+T_b m_a)^3 (m_a+m_b)^2}
  \Big(71 T_a^2 m_a^2 m_b +68 T_a^2 m_a m_b^2 +17 T_a^2 m_b^3 -10 T_a T_b m_a^3 \nn\\
  &&  -94 T_a T_b m_a^2 m_b -44 T_a T_b m_a m_b^2 +20 T_b^2 m_a^2 m_b \Big) ;\nn\\
%===  
  E_{b(22)} &=& + \Omega_{22} \frac{  T_b m_a m_b^2 (9 T_a m_a +4 T_a m_b -5 T_b m_a)}{ 20 (T_a m_b+T_b m_a)^2 (m_a+m_b)^2} ;\nn\\
  E_{b(13)} &=& -\, \Omega_{13} \frac{T_b m_a m_b }{140 (T_a m_b +T_b m_a)^3 (m_a+m_b)^2}
  \Big(76 T_a^2 m_a^2 m_b +34 T_a^2 m_a m_b^2 +14 T_a^2 m_b^3 -5 T_a T_b m_a^3 \nn\\
   && -128 T_a T_b m_a^2 m_b -11 T_a T_b m_a m_b^2 +56 T_b^2 m_a^2 m_b \Big) ;\nn\\
%===  
  E_{b(23)} &=& - \, \Omega_{23} \frac{ T_b m_a m_b^2 (9 T_a m_a +2 T_a m_b -7 T_b m_a) }{35 (T_a m_b+T_b m_a)^2 (m_a+m_b)^2};\nn\\
  E_{b(14)} &=& + \Omega_{14} \frac{T_b m_a m_b^2 (3 T_a^2 m_a^2 +T_a^2 m_b^2 -6 T_a T_b m_a^2 +2 T_a T_b m_a m_b +4 T_b^2 m_a^2)  }{ 70 (T_a m_b+T_b m_a)^3 (m_a+m_b)^2};\nn\\
  E_{b(24)} &=& -\, \Omega_{24} \frac{T_b (T_b-T_a)  m_b^2 m_a^2}{35 (T_a m_b +T_b m_a)^2 (m_a+m_b)^2}.\nn
\end{eqnarray}

Prescribing Coulomb collisions recovers eqs. (18)-(19) of Part 1, with the (non-hat) coefficients
$D_{ab (1)}$; $D_{ab (2)}$; $U_{ab (1)}$  and $E_{ab (1)}$; $E_{ab (2)}$ given by (K45), with the further re-arrangement (K49).

%\newpage
%==============================================================================
%\subsubsection*{Equal temperatures}

\newpage
%========================================================================================================
%=======================================================================================================
\subsection{Higher-order heat flux exchange rates  \texorpdfstring{$\vecQ^{(5)}_{a}\,'$}{Q(5)'}}
The exchange rates for the higher-order heat flux $\vecX^{(5)}_a$
are given by (note that the momentum exchange rates $\boldsymbol{R}_a$ enter and the expressions are not further re-arranged)
\begin{eqnarray}
  \vecQ^{(5)}_{a}\,' &=& - \nu_{aa} \Big( \frac{8}{5}\Omega_{23}+ \frac{28}{5}\Omega_{22} \Big) \frac{p_a}{\rho_a}\vecq_a
  - \nu_{aa} \Big(\frac{2}{35}\Omega_{24} - \frac{3}{10}\Omega_{22} \Big) \Big(\vecX^{(5)}_a -28 \frac{p_a}{\rho_a} \vecq_a\Big)\nn\\
  && +\sum_{b\neq a} \nu_{ab}\Big\{  - F_{ab (1)} \frac{p_a}{\rho_a} \vecq_a
  -  \frac{p_a}{\rho_a} \frac{\rho_a}{\rho_b} F_{ab (2)} \vecq_b + \frac{p_a^2}{\rho_a}  U_{ab(2)} (\bu_b-\bu_a) \nn\\
  && - G_{ab (1)} \Big(\vecX^{(5)}_a -28 \frac{p_a}{\rho_a} \vecq_a\Big)
  + G_{ab (2)} \frac{p_a}{p_b} \Big(\vecX^{(5)}_b -28 \frac{p_b}{\rho_b}\vecq_b\Big) \Big\} 
   - 35 \frac{p_a^2}{\rho_a^2} \boldsymbol{R}_a, \label{eq:Qab5P}
\end{eqnarray}  
with mass-ratio coefficients
\begin{eqnarray}
  {F}_{ab (1)} &=& F_{a(11)} + F_{a(12)} +  F_{a(22)} + F_{a(13)} + F_{a(23)} + F_{a(33)}  + F_{a(14)} + F_{a(24)} + F_{a(34)}; \label{eq:F1}\\
  F_{a(11)} &=& -\, \frac{7 T_b^2 m_a^2(25 T_a^2 m_a m_b +5 T_a^2 m_b^2 -20 T_a T_b m_a^2 -32 T_a T_b m_a m_b + 8 T_b^2 m_a^2)}{T_a (T_a m_b +T_b m_a)^3 (m_a+m_b)} ;\nn\\
  F_{a(12)} &=&  -\, \Omega_{12} \frac{ 14 T_b m_b m_a}{5(T_a m_b+T_b m_a)^3 T_a (m_a+m_b)^3} \Big( 50 T_a^3 m_a^3 m_b +30 T_a^3 m_a^2 m_b^2 +30 T_a^3 m_a m_b^3
  +10 T_a^3 m_b^4 \nn\\
  && -85 T_a^2 T_b m_a^4 -259 T_a^2 T_b m_a^3 m_b -71 T_a^2 T_b m_a^2 m_b^2
  -17 T_a^2 T_b m_a m_b^3 +116 T_a T_b^2 m_a^4 +256 T_a  T_b^2 m_a^3 m_b \nn\\
  && +20 T_a T_b^2 m_a^2 m_b^2 -48 T_b^3 m_a^4 -88 T_b^3 m_a^3 m_b \Big);\nn\\
  F_{a(22)} &=&  -\, \Omega_{22}\frac{56 T_b m_b m_a^2 (10 T_a^2 m_a m_b +5 T_a^2 m_b^2 -8 T_a T_b m_a^2 -25 T_a T_b m_a m_b
    -7 T_a T_b m_b^2 +6 T_b^2 m_a^2 +11 T_b^2 m_a m_b) }{5 T_a (T_a m_b+T_b m_a)^2(m_a+m_b)^3};\nn\\
  F_{a(13)} &=& -\, \Omega_{13} \frac{4 m_b^2}{5 T_a (T_a m_b +T_b m_a)^3 (m_a+m_b)^4} \Big( 25 T_a^4 m_a^4 m_b +30 T_a^4 m_a^2 m_b^3
  +5 T_a^4 m_b^5 -110 T_a^3 T_b m_a^5 \nn\\
  && -252 T_a^3 T_b m_a^4 m_b -48 T_a^3 T_b m_a^3 m_b^2 -140 T_a^3 T_b m_a^2 m_b^3 +6 T_a^3 T_b m_a m_b^4 +288 T_a^2 T_b^2 m_a^5 +420 T_a^2 T_b^2 m_a^4 m_b \nn\\
  && -144 T_a^2 T_b^2 m_a^3 m_b^2 +84 T_a^2 T_b^2 m_a^2 m_b^3 -312 T_a T_b^3 m_a^5 -424 T_a T_b^3 m_a^4 m_b +128 T_a T_b^3 m_a^3 m_b^2 \nn\\
  && +120 T_b^4 m_a^5   +180 T_b^4 m_a^4 m_b \Big) ;\nn\\
  F_{a(23)} &=&  -\, \Omega_{23} \frac{16 m_b^2 m_a}{ 5 T_a (T_a m_b +T_b m_a)^2 (m_a+m_b)^4}\Big( 10 T_a^3 m_a^2 m_b +5 T_a^3 m_b^3 -26 T_a^2 T_b m_a^3
  -63 T_a^2 T_b m_a^2 m_b \nn\\
  && +T_a^2 T_b m_a m_b^2  -7 T_a^2 T_b m_b^3 +45 T_a T_b^2 m_a^3 +83 T_a T_b^2 m_a^2 m_b -7 T_a T_b^2 m_a m_b^2
  -21 T_b^3 m_a^3 -36 T_b^3 m_a^2 m_b \Big);\nn\\
  F_{a(33)} &=&  + \Omega_{33} \frac{16   (T_b-T_a) m_a^2 m_b^2 (5 T_a m_b -4 T_b m_a -9 T_b m_b) }{5 T_a (T_a m_b +T_b m_a) (m_a+m_b)^4};\nn\\
  F_{a(14)} &=& + \Omega_{14}  \frac{8 m_b^3 }{5 T_a (T_a m_b +T_b m_a)^3 (m_a+m_b)^4}
  \Big( T_a^2 m_a^2 +T_a^2 m_b^2 -2 T_a T_b m_a^2 +2 T_a T_b m_a m_b +2 T_b^2 m_a^2 \Big) \nn\\
  && \quad \times \Big(5 T_a^2 m_a^2 +T_a^2 m_b^2 -10 T_a T_b m_a^2 +2 T_a T_b m_a m_b +6 T_b^2 m_a^2 \Big);\nn\\
  F_{a(24)} &=& -\, \Omega_{24} \frac{32 (T_b- T_a)  m_b^3 m_a (2 T_a^2 m_a^2 +T_a^2 m_b^2 -4 T_a T_b m_a^2
    +2 T_a T_b m_a m_b +3 T_b^2 m_a^2)}{5 T_a (T_a m_b +T_b m_a)^2 (m_a+m_b)^4};\nn\\
  F_{a(34)} &=&  + \Omega_{34}\frac{32  (T_b -T_a)^2  m_a^2 m_b^3}{5 T_a (T_a m_b +T_b m_a) (m_a+m_b)^4};\nn
\end{eqnarray}
%=============================================
\newpage
%=============================================
\begin{eqnarray}
  {F}_{ab (2)} &=& F_{b(11)} + F_{b(12)} +  F_{b(22)} + F_{b(13)} + F_{b(23)} + F_{b(33)}  + F_{b(14)} + F_{b(24)} + F_{b(34)}; \label{eq:F2}\\
  F_{b(11)} &=&  +\frac{7 T_a m_a m_b  (8 T_a^2 m_b^2 -32 T_a T_b m_a m_b -20 T_a T_b m_b^2 +5 T_b^2 m_a^2 +25 T_b^2 m_a m_b) }{(T_a m_b +T_b m_a)^3 (m_a+m_b)};\nn\\
  F_{b(12)} &=&  -\, \Omega_{12} \frac{14 m_b}{5 (T_a m_b +T_b m_a)^3 (m_a+m_b)^3}
  \Big( 56 T_a^3 m_a^3 m_b^2 +36 T_a^3 m_a^2 m_b^3 +32 T_a^3 m_a m_b^4 +12 T_a^3 m_b^5 \nn\\
  && -74 T_a^2 T_b m_a^4 m_b -250 T_a^2 T_b m_a^3 m_b^2 -62 T_a^2 T_b m_a^2 m_b^3 -6 T_a^2 T_b m_a m_b^4 +5 T_a T_b^2 m_a^5 +119 T_a T_b^2 m_a^4 m_b \nn\\
  && +271 T_a T_b^2 m_a^3 m_b^2 +37 T_a T_b^2 m_a^2 m_b^3 -40 T_b^3 m_a^4 m_b -80 T_b^3 m_a^3 m_b^2 \Big);\nn\\
  F_{b(22)} &=&  -\, \Omega_{22} \frac{56  m_a m_b^2}{5 (T_a m_b +T_b m_a)^2 (m_a+m_b)^3}
  \Big(11 T_a^2 m_a m_b +6 T_a^2 m_b^2 -7 T_a T_b m_a^2 -25 T_a T_b m_a m_b -8 T_a T_b m_b^2 \nn\\
  && +5 T_b^2 m_a^2 +10 T_b^2 m_a m_b \Big);\nn\\
  F_{b(13)} &=&  + \Omega_{13} \frac{ 4 m_b^2 }{5 T_a (T_a m_b+T_b m_a)^3 (m_a+m_b)^4 }
  \Big( 93 T_a^4 m_a^4 m_b +68 T_a^4 m_a^3 m_b^2 +82 T_a^4 m_a^2 m_b^3 +52 T_a^4 m_a m_b^4 \nn\\
  && +5 T_a^4 m_b^5 -42 T_a^3 T_b m_a^5 -388 T_a^3 T_b m_a^4 m_b -96 T_a^3 T_b m_a^3 m_b^2 -36 T_a^3 T_b m_a^2 m_b^3 -46 T_a^3 T_b m_a m_b^4 \nn\\
  && +84 T_a^2 T_b^2 m_a^5 +576 T_a^2 T_b^2 m_a^4 m_b +60 T_a^2 T_b^2 m_a^3 m_b^2 -72 T_a^2 T_b^2 m_a^2 m_b^3 -56 T_a T_b^3 m_a^5 -392 T_a T_b^3 m_a^4 m_b \nn\\
  && -96 T_a T_b^3 m_a^3 m_b^2 +60 T_b^4 m_a^4 m_b \Big);\nn\\
  F_{b(23)} &=&  + \Omega_{23} \frac{16 m_b^2 }{5 T_a (T_a m_b +T_b m_a)^2 (m_a+m_b)^4 }
  \Big( 29 T_a^3 m_a^3 m_b +19 T_a^3 m_a^2 m_b^2 +7 T_a^3 m_a m_b^3 +2 T_a^3 m_b^4 -7 T_a^2 T_b m_a^4 \nn\\
  && -82 T_a^2 T_b m_a^3 m_b -31 T_a^2 T_b m_a^2 m_b^2 -T_a^2 T_b m_a m_b^3 +7 T_a T_b^2 m_a^4 +70 T_a T_b^2 m_a^3 m_b \nn\\
  && +18 T_a T_b^2 m_a^2 m_b^2 -15 T_b^3 m_a^3 m_b \Big)   ;\nn\\
  F_{b(33)} &=& -\, \Omega_{33} \frac{16 (T_b-T_a)  m_a  m_b^3 (9 T_a m_a +4 T_a m_b -5 T_b m_a)}{5 T_a (T_a m_b +T_b m_a)(m_a+m_b)^4 } ;\nn\\
  F_{b(14)} &=&  -\, \Omega_{14} \frac{8 m_b^3}{5 T_a (T_a m_b+T_b m_a)^3 (m_a+m_b)^4 } \Big( T_a^2 m_a^2 +T_a^2 m_b^2 -2 T_a T_b m_a^2 +2 T_a T_b m_a m_b
  +2 T_b^2 m_a^2 \Big) \nn\\
  && \quad \times \Big(5 T_a^2 m_a^2 +T_a^2 m_b^2 -10 T_a T_b m_a^2 +2 T_a T_b m_a m_b +6 T_b^2 m_a^2 \Big);\nn\\
  F_{b(24)} &=&  + \Omega_{24}\frac{ 32 (T_b-T_a) m_a  m_b^3 (2T_a^2 m_a^2 +T_a^2 m_b^2 -4 T_a T_b m_a^2
    +2 T_a T_b m_a m_b +3 T_b^2 m_a^2)}{5 T_a (T_a m_b+T_b m_a)^2 (m_a+m_b)^4 };\nn\\
  F_{b(34)} &=&  - \, \Omega_{34} \frac{32  (T_b-T_a)^2 m_a^2 m_b^3}{5  T_a (T_a m_b +T_b m_a) (m_a+m_b)^4};\nn
\end{eqnarray}

\newpage
%=============================================
%=============================================
\begin{eqnarray}
  {G}_{ab (1)} &=& G_{a(11)} + G_{a(12)} +  G_{a(22)} + G_{a(13)} + G_{a(23)} + G_{a(33)}  + G_{a(14)} + G_{a(24)} + G_{a(34)} \nn\\
  && \qquad + G_{a(15)} + G_{a(25)} + G_{a(35)}; \label{eq:G1}\\
  G_{a(11)} &=&  + \frac{T_b^2 m_a^2 (175 T_a^2 m_a m_b^2 +35 T_a^2 m_b^3 -280 T_a T_b m_a^2 m_b
    -308 T_a T_b m_a m_b^2 +40 T_b^2 m_a^3 +152 T_b^2 m_a^2 m_b)}{8 (T_a m_b+T_b m_a)^4 (m_a+m_b)};\nn\\
  G_{a(12)} &=&  + \Omega_{12} \frac{T_b m_a m_b }{10 (T_a m_b+T_b m_a)^4 (m_a+m_b)^3}
  \Big( 175 T_a^3 m_a^3 m_b^2 +105 T_a^3 m_a^2 m_b^3 +105 T_a^3 m_a m_b^4 +35 T_a^3 m_b^5 \nn\\
  && -595 T_a^2 T_b m_a^4 m_b -1393 T_a^2 T_b m_a^3 m_b^2 -497 T_a^2 T_b m_a^2 m_b^3 -119 T_a^2 T_b m_a m_b^4
  +220 T_a T_b^2 m_a^5 +1336 T_a T_b^2 m_a^4 m_b \nn\\
  && +1760 T_a T_b^2 m_a^3 m_b^2 +224 T_a T_b^2 m_a^2 m_b^3 -136 T_b^3 m_a^5 -608 T_b^3 m_a^4 m_b -612 T_b^3 m_a^3 m_b^2 \Big);\nn\\
  %====
  G_{a(22)} &=&  + \Omega_{22} \frac{T_b m_a^2 m_b}{5 (T_a m_b+T_b m_a)^3 (m_a+m_b)^3}
  \Big(70 T_a^2 m_a m_b^2 +35 T_a^2 m_b^3 -112 T_a T_b m_a^2 m_b -245 T_a T_b m_a m_b^2 \nn\\
  && -63 T_a T_b m_b^3 +16 T_b^2 m_a^3 +116 T_b^2 m_a^2 m_b +135 T_b^2 m_a m_b^2 \Big);\nn\\
  %====
  G_{a(13)} &=& + \Omega_{13} \frac{m_b^2}{70 (T_a m_b+T_b m_a)^4 (m_a+m_b)^4}
  \Big(175 T_a^4 m_a^4 m_b^2 +210 T_a^4 m_a^2 m_b^4 +35 T_a^4 m_b^6 -1540 T_a^3 T_b m_a^5 m_b \nn\\
  && -2828 T_a^3 T_b m_a^4 m_b^2 -1092 T_a^3 T_b m_a^3 m_b^3 -1540 T_a^3 T_b m_a^2 m_b^4 -56 T_a^3 T_b m_a m_b^5 +1255 T_a^2 T_b^2 m_a^6
  +7088 T_a^2 T_b^2 m_a^5 m_b \nn\\
  && +7218 T_a^2 T_b^2 m_a^4 m_b^2 -128 T_a^2 T_b^2 m_a^3 m_b^3 +1007 T_a^2 T_b^2 m_a^2 m_b^4 -1964 T_a T_b^3 m_a^6
  -8340 T_a T_b^3 m_a^5 m_b \nn\\
  && -6868 T_a T_b^3 m_a^4 m_b^2  +1188 T_a T_b^3 m_a^3 m_b^3 +960 T_b^4 m_a^6 +3600 T_b^4 m_a^5 m_b +3060 T_b^4 m_a^4 m_b^2 \Big) ;\nn\\
  %====
  G_{a(23)} &=&  + \Omega_{23} \frac{ 2 m_a m_b^2}{35 (T_a m_b +T_b m_a)^3 (m_a+m_b)^4}
  \Big(70 T_a^3 m_a^2 m_b^2 +35 T_a^3 m_b^4 -364 T_a^2 T_b m_a^3 m_b -672 T_a^2 T_b m_a^2 m_b^2  \nn\\
  && -56 T_a^2 T_b m_a m_b^3  -63 T_a^2 T_b m_b^4 +160 T_a T_b^2 m_a^4 +978 T_a T_b^2 m_a^3 m_b +1133 T_a T_b^2 m_a^2 m_b^2 \nn\\
  && -132 T_b^3 m_a^4 -558 T_b^3 m_a^3 m_b -531 T_b^3 m_a^2 m_b^2 \Big) ;\nn\\
  %====
  G_{a(33)} &=&  +\Omega_{33} \frac{2 m_a^2 m_b^2(35 T_a^2 m_b^2 -56 T_a T_b m_a m_b -126 T_a T_b m_b^2 +8 T_b^2 m_a^2
    +72 T_b^2 m_a m_b +99 T_b^2 m_b^2)}{35 (T_a m_b+T_b m_a)^2(m_a+m_b)^4};\nn\\
  %====
  G_{a(14)} &=&  -\, \Omega_{14} \frac{2 m_b^3}{35 (T_a m_b +T_b m_a)^4 (m_a+m_b)^4}
  \Big(35 T_a^4 m_a^4 m_b +42 T_a^4 m_a^2 m_b^3 +7 T_a^4 m_b^5 -75 T_a^3 T_b m_a^5  -236 T_a^3 T_b m_a^4 m_b \nn\\
  && +18 T_a^3 T_b m_a^3 m_b^2 -136 T_a^3 T_b m_a^2 m_b^3 +21 T_a^3 T_b m_a m_b^4
  +204 T_a^2 T_b^2 m_a^5 
  +408 T_a^2 T_b^2 m_a^4 m_b -192 T_a^2 T_b^2 m_a^3 m_b^2 \nn\\
  && +108 T_a^2 T_b^2 m_a^2 m_b^3 -228 T_a T_b^3 m_a^5 -404 T_a T_b^3 m_a^4 m_b
  +160 T_a T_b^3 m_a^3 m_b^2 +92 T_b^4 m_a^5 +176 T_b^4 m_a^4 m_b \Big);\nn\\
  %====
  G_{a(24)} &=& -\, \Omega_{24} \frac{4 m_a m_b^3 }{ 35 (T_a m_b +T_b m_a)^3 (m_a+m_b)^4}
  \Big( 28 T_a^3 m_a^2 m_b +14 T_a^3 m_b^3 -38 T_a^2 T_b m_a^3 -129 T_a^2 T_b m_a^2 m_b \nn\\
  && +17 T_a^2 T_b m_a m_b^2 -18 T_a^2 T_b m_b^3 +69 T_a T_b^2 m_a^3 +166 T_a T_b^2 m_a^2 m_b -29 T_a T_b^2 m_a m_b^2 \nn\\
  && -35 T_b^3 m_a^3 -77 T_b^3 m_a^2 m_b \Big);\nn\\
  G_{a(34)} &=&  + \Omega_{34} \frac{8 (T_b-T_a) m_a^2 m_b^3 (7 T_a m_b -4 T_b m_a -11 T_b m_b) }{35(T_a m_b+T_b m_a)^2 (m_a+m_b)^4};\nn\\
  G_{a(15)} &=&  + \Omega_{15} \frac{2  m_b^4 }{35 (T_a m_b +T_b m_a)^4 (m_a+m_b)^4}
  \Big(T_a^2 m_a^2 +T_a^2 m_b^2 -2 T_a T_b m_a^2 +2 T_a T_b m_a m_b +2 T_b^2 m_a^2 \Big) \nn\\
  && \quad \times \Big( 5 T_a^2 m_a^2 +T_a^2 m_b^2 -10 T_a T_b m_a^2 +2 T_a T_b m_a m_b +6 T_b^2 m_a^2 \Big)  ;\nn\\
  G_{a(25)} &=&  -\, \Omega_{25} \frac{ 8 (T_b-T_a) m_a m_b^4 (2 T_a^2 m_a^2 +T_a^2 m_b^2 -4 T_a T_b m_a^2
    +2 T_a T_b m_a m_b +3 T_b^2 m_a^2)}{35 (T_a m_b +T_b m_a)^3 (m_a+m_b)^4} ;\nn\\
  G_{a(35)} &=&  + \Omega_{35} \frac{8 (T_b-T_a)^2 m_b^4  m_a^2}{35 (T_a m_b +T_b m_a)^2 (m_a+m_b)^4} ;\nn
\end{eqnarray}
%=============================================
%=============================================

\newpage
\begin{eqnarray}
  {G}_{ab (2)} &=& G_{b(11)} + G_{b(12)} +  G_{b(22)} + G_{b(13)} + G_{b(23)} + G_{b(33)}  + G_{b(14)} + G_{b(24)} + G_{b(34)} \nn\\
  && \qquad + G_{b(15)} + G_{b(25)} + G_{b(35)}; \label{eq:G2}\\
  G_{b(11)} &=&  +\frac{m_a m_b T_a T_b(152 T_a^2 m_a m_b^2 +40 T_a^2 m_b^3 -308 T_a T_b m_a^2 m_b -280 T_a T_b m_a m_b^2
    +35 T_b^2 m_a^3 +175 T_b^2 m_a^2 m_b) }{8 (T_a m_b+T_b m_a)^4 (m_a+m_b)};\nn\\
%===  
  G_{b(12)} &=&  -\,\Omega_{12} \frac{m_a m_b  T_b}{10 (T_a m_b+T_b m_a)^4 (m_a+m_b)^3}
  \Big( 458 T_a^3 m_a^3 m_b^2 +524 T_a^3 m_a^2 m_b^3 +290 T_a^3 m_a m_b^4 +84 T_a^3 m_b^5 \nn\\
    && -497 T_a^2 T_b m_a^4 m_b -1823 T_a^2 T_b m_a^3 m_b^2 
    -1063 T_a^2 T_b m_a^2 m_b^3 -157 T_a^2 T_b m_a m_b^4 +35 T_a T_b^2 m_a^5 \nn\\
    && +623 T_a T_b^2 m_a^4 m_b  +1477 T_a T_b^2 m_a^3 m_b^2     +469 T_a T_b^2 m_a^2 m_b^3 -140 T_b^3 m_a^4 m_b -280 T_b^3 m_a^3 m_b^2 \Big); \nn\\
%===  
  G_{b(22)} &=&  -\, \Omega_{22} \frac{ m_a m_b^2 T_b}{5 (T_a m_b+T_b m_a)^3 (m_a+m_b)^3}
  \Big( 135 T_a^2 m_a^2 m_b+116 T_a^2 m_a m_b^2 +16 T_a^2 m_b^3 -63 T_a T_b m_a^3  \nn\\
  && -245 T_a T_b m_a^2 m_b   -112 T_a T_b m_a m_b^2 +35 T_b^2 m_a^3 +70 T_b^2 m_a^2 m_b \Big) ;\nn\\
%===  
  G_{b(13)} &=& + \Omega_{13} \frac{T_b m_a m_b}{70 (T_a m_b+T_b m_a)^4 (m_a+m_b)^4 T_a }
  \Big( 1871 T_a^4 m_a^4 m_b^2+2440 T_a^4 m_a^3 m_b^3+1898 T_a^4 m_a^2 m_b^4+1160 T_a^4 m_a m_b^5 \nn\\
  && +251 T_a^4 m_b^6-1064 T_a^3 T_b m_a^5 m_b-7648 T_a^3 T_b m_a^4 m_b^2-5736 T_a^3 T_b m_a^3 m_b^3-1184 T_a^3 T_b m_a^2 m_b^4-352 T_a^3 T_b m_a m_b^5 \nn\\
  && +35 T_a^2 T_b^2 m_a^6+1792 T_a^2 T_b^2 m_a^5 m_b+9402 T_a^2 T_b^2 m_a^4 m_b^2+5168 T_a^2 T_b^2 m_a^3 m_b^3+43 T_a^2 T_b^2 m_a^2 m_b^4 \nn\\
  && -784 T_a T_b^3 m_a^5 m_b  -4368 T_a T_b^3 m_a^4 m_b^2-1904 T_a T_b^3 m_a^3 m_b^3+420 T_b^4 m_a^4 m_b^2 \Big);\nn\\
%===  
  G_{b(23)} &=&  + \Omega_{23} \frac{2 T_b m_a m_b^2 }{35 (T_a m_b+T_b m_a)^3 (m_a+m_b)^4 T_a }
  \Big( 468 T_a^3 m_a^3 m_b+530 T_a^3 m_a^2 m_b^2+195 T_a^3 m_a m_b^3+28 T_a^3 m_b^4 \nn\\
  && -126 T_a^2 T_b m_a^4  -1154 T_a^2 T_b m_a^3 m_b-852 T_a^2 T_b m_a^2 m_b^2-139 T_a^2 T_b m_a m_b^3+98 T_a T_b^2 m_a^4  +735 T_a T_b^2 m_a^3 m_b \nn\\
  && +322 T_a T_b^2 m_a^2 m_b^2-105 T_b^3 m_a^3 m_b \Big);\nn\\
%===  
  G_{b(33)} &=&  +  \Omega_{33} \frac{2 T_b m_a m_b^3}{ 35 (T_a m_b+T_b m_a)^2 (m_a+m_b)^4 T_a}
  \Big( 99 T_a^2 m_a^2+72 T_a^2 m_a m_b+8 T_a^2 m_b^2-126 T_a T_b m_a^2 \nn\\
  && -56 T_a T_b m_a m_b+35 T_b^2 m_a^2 \Big) ;\nn\\
%===  
  G_{b(14)} &=& - \, \Omega_{14} \frac{2 T_b m_a m_b^2 }{35 (T_a m_b+T_b m_a)^4 (m_a+m_b)^4 T_a }
  \Big( 89 T_a^4 m_a^4 m_b+54 T_a^4 m_a^3 m_b^2+80 T_a^4 m_a^2 m_b^3+38 T_a^4 m_a m_b^4 \nn\\
  && +7 T_a^4 m_b^5-21 T_a^3 T_b m_a^5-344 T_a^3 T_b m_a^4 m_b-30 T_a^3 T_b m_a^3 m_b^2-60 T_a^3 T_b m_a^2 m_b^3  -17 T_a^3 T_b m_a m_b^4 \nn\\
  && +42 T_a^2 T_b^2 m_a^5+522 T_a^2 T_b^2 m_a^4 m_b-30 T_a^2 T_b^2 m_a^3 m_b^2-6 T_a^2 T_b^2 m_a^2 m_b^3  -28 T_a T_b^3 m_a^5-372 T_a T_b^3 m_a^4 m_b \nn\\
  && -8 T_a T_b^3 m_a^3 m_b^2+84 T_b^4 m_a^4 m_b \Big) ;\nn\\
  G_{b(24)} &=& - \, \Omega_{24} \frac{4 T_b m_a m_b^2}{35 (T_a m_b+T_b m_a)^3 (m_a+m_b)^4 T_a }
  \Big( 59 T_a^3 m_a^3 m_b+31 T_a^3 m_a^2 m_b^2+18 T_a^3 m_a m_b^3+4 T_a^3 m_b^4 \nn\\
  && -7 T_a^2 T_b m_a^4-160 T_a^2 T_b m_a^3 m_b-33 T_a^2 T_b m_a^2 m_b^2-6 T_a^2 T_b m_a m_b^3+7 T_a T_b^2 m_a^4 \nn\\
  && +147 T_a T_b^2 m_a^3 m_b+14 T_a T_b^2 m_a^2 m_b^2-42 T_b^3 m_a^3 m_b \Big) ;\nn\\
%===
  G_{b(34)} &=&  +\Omega_{34} \frac{8 (T_b-T_a)T_b m_a^2 m_b^3 (11 T_a m_a+4 T_a m_b-7 T_b m_a)  }{ 35 (T_a m_b+T_b m_a)^2 (m_a+m_b)^4 T_a };\nn\\
%===  
  G_{b(15)} &=&  + \Omega_{15} \frac{ 2 T_b m_a m_b^3 }{35 (T_a m_b+T_b m_a)^4 (m_a+m_b)^4 T_a }
 \Big(T_a^2 m_a^2+T_a^2 m_b^2-2 T_a T_b m_a^2+2 T_a T_b m_a m_b+2 T_b^2 m_a^2 \Big) \nn\\
 && \quad \times \Big( 5 T_a^2 m_a^2+T_a^2 m_b^2-10 T_a T_b m_a^2+2 T_a T_b m_a m_b+6 T_b^2 m_a^2 \Big);\nn\\
%=== 
 G_{b(25)} &=&  -\, \Omega_{25} \frac{8  (T_b-T_a) T_b m_a^2 m_b^3  (2 T_a^2 m_a^2+T_a^2 m_b^2-4 T_a T_b m_a^2+2 T_a T_b m_a m_b+3 T_b^2 m_a^2) }
 {35 (T_a m_b+T_b m_a)^3 (m_a+m_b)^4 T_a} ;\nn\\
  G_{b(35)} &=& + \Omega_{35} \frac{8  (T_b -T_a)^2 T_b m_a^3  m_b^3}{35 (T_a m_b+T_b m_a)^2 (m_a+m_b)^4 T_a};\nn
\end{eqnarray}
%=============================================
%=============================================
%Prescribing Coulomb collisions recovers mass-ratios (K61) of Part 1.

\begin{eqnarray}
  U_{ab (2)} &=& -\, \frac{35 m_a^2 T_b (4 T_a m_b -T_b m_a -5 T_b m_b)}{(m_a+m_b) (T_a m_b+T_b m_a)^2} \nn\\
  %===
  && -\, \Omega_{12} \frac{28 m_a m_b}{T_a (T_a m_b +T_b m_a)^2 (m_a+m_b)^3}
  \Big( 2 T_a^3 m_a^2 m_b +2 T_a^3 m_b^3 -3 T_a^2 T_b m_a^3 -9 T_a^2 T_b m_a^2 m_b \nn\\
  && \qquad +3 T_a^2 T_b m_a m_b^2 -3 T_a^2 T_b m_b^3 +6 T_a T_b^2 m_a^3 +12 T_a T_b^2 m_a^2 m_b  -6 T_a T_b^2 m_a m_b^2 \nn\\
  && \qquad -4 T_b^3 m_a^3 -8 T_b^3 m_a^2 m_b \Big)  \nn\\
  %===
  && + \Omega_{22} \frac{56(T_b-T_a) m_a^2 m_b (T_a m_b-T_b m_a -2 T_b m_b)}{T_a (T_a m_b+T_b m_a) (m_a+m_b)^3}\nn\\
  %===
  && + \Omega_{13} \frac{4 m_b^2}{T_a^2 (T_a m_b +T_b m_a)^2 (m_a+m_b)^4} \Big(5 T_a^2 m_a^2 +T_a^2 m_b^2 -10 T_a T_b m_a^2 +2 T_a T_b m_a m_b +6 T_b^2 m_a^2 \Big) \nn\\
  && \qquad \times \Big( T_a^2 m_a^2 +T_a^2 m_b^2 -2 T_a T_b m_a^2 +2 T_a T_b m_a m_b +2 T_b^2 m_a^2 \Big) \nn\\
  %===
  && - \Omega_{23}\frac{16(T_b-T_a) m_a  m_b^2 (2 T_a^2 m_a^2 +T_a^2 m_b^2 -4 T_a T_b m_a^2 +2 T_a T_b m_a m_b +3 T_b^2 m_a^2)}{T_a^2 (T_a m_b +T_b m_a) (m_a+m_b)^4}\nn\\
  %===
  && + \Omega_{33} \frac{16 (T_b-T_a)^2 m_a^2 m_b^2  }{T_a^2 (m_a+m_b)^4}.
\end{eqnarray}
Prescribing Coulomb collisions recovers eqs. (20)-(21) of Part 1, with the (non-hat) coefficients
$F_{ab (1)}$; $F_{ab (2)}$   and $G_{ab (1)}$; $G_{ab (2)}$; $U_{ab (2)}$ given by (K61), with the further re-arrangement (K64).

%Obviously, the 2-Hermite heat flux model given above is unfortunatelly much more complicated than the 1-Hermite model given
%by the mass-ratio coefficients (\ref{eq:HFSchunkD1})-(\ref{eq:HFSchunkU1}) or (\ref{eq:HFSchunkD1r})-(\ref{eq:HFSchunkU1r}). 

%\vspace{1cm}

%=============================================================================================================
%=============================================================================================================
\subsection{Scalar  exchange rates  \texorpdfstring{$\widetilde{Q}_{a}^{(4)}\,'$}{tilde Q(4)'}}
The exchange rates for the fully contracted scalar perturbation $\widetilde{X}^{(4)}_a$ are given by (note that the energy exchange rates $Q_a$ enter
and the expressions are not further re-arranged)
\begin{eqnarray}
  \widetilde{Q}_a^{(4)}\,' &=& - \nu_{aa} \frac{2}{5}\Omega_{22} \widetilde{X}^{(4)}_a
  + 3 \frac{p_a^2}{\rho_a} \sum_{b\neq a}  \nu_{ab} \Big\{ S_{ab (0)} \frac{(T_b-T_a)}{T_a}
  -S_{ab (1)} \frac{\rho_a}{p_a^2} \widetilde{X}^{(4)}_a -S_{ab (2)} \frac{\rho_b}{p_b^2} \widetilde{X}^{(4)}_b \Big\} \nn\\
  && -20 \frac{p_a}{\rho_a} Q_a; \label{eq:Q4exchange}
\end{eqnarray}
%\begin{eqnarray}
%  \widetilde{Q}_a^{(4)}\,' &=& - \nu_{aa} \frac{2}{5}\Omega_{22} \widetilde{X}^{(4)}_a
%  + 3  \sum_{b\neq a}  \nu_{ab} \Big\{ S_{ab (0)} \frac{p_a^2}{\rho_a} \frac{(T_b-T_a)}{T_a}
%  -S_{ab (1)} \widetilde{X}^{(4)}_a -S_{ab (2)} \frac{p_a^2}{\rho_a} \frac{\rho_b}{p_b^2} \widetilde{X}^{(4)}_b  \nn\\
%  && - S_{ab (3)} \Big( \frac{\rho_a}{p_a} \widetilde{X}^{(6)}_a -21 \widetilde{X}^{(4)}_a \Big)
%  - S_{ab (4)} \frac{p_a^2}{\rho_a} \frac{\rho_b}{p_b^2}  \Big( \frac{\rho_b}{p_b} \widetilde{X}^{(6)}_b -21 \widetilde{X}^{(4)}_b \Big)\Big\} \nn\\
%  && -20 \frac{p_a}{\rho_a} Q_a; \label{eq:Q4exchange0}
%\end{eqnarray}
with the 1-Hermite mass-ratio coefficients
\begin{eqnarray}
  S_{ab (0)} &=& + \frac{20 m_a^2 T_b}{(m_a+m_b) (T_a m_b+T_b m_a)} \nn\\
  && + \Omega_{12} \frac{8 m_a m_b (T_a^2 m_a^2+T_a^2 m_b^2 -2 T_a T_b m_a^2+2 T_a T_b m_a m_b+2 T_b^2 m_a^2) }{T_a (T_a m_b+T_b m_a) (m_a+m_b)^3} \nn\\
  && - \Omega_{22} \frac{ 8 (T_b-T_a) m_a^2  m_b}{T_a (m_a+m_b)^3};
\end{eqnarray}
%===========================
%===========================
\begin{eqnarray}
  {S}_{ab (1)} &=& S_{a(11)} + S_{a(12)} +  S_{a(22)} + S_{a(13)} + S_{a(23)}  + S_{a(14)} + S_{a(24)}; \label{eq:S1}\\
  S_{a(11)} &=& + \frac{T_b m_a^2 (15 T_a^2 m_b^2-40 T_a T_b m_a m_b-35 T_a T_b m_b^2+8 T_b^2 m_a^2+28 T_b^2 m_a m_b) }{6 (T_a m_b+T_b m_a)^3 (m_a+m_b)};\nn\\
%===  
  S_{a(12)} &=& +\Omega_{12} \frac{m_a m_b }{15 (m_a+m_b)^3 (T_a m_b+T_b m_a)^3} \Big(
         15 T_a^3 m_a^2 m_b^2+15 T_a^3 m_b^4-110 T_a^2 T_b m_a^3 m_b-205 T_a^2 T_b m_a^2 m_b^2 \nn\\
         && -40 T_a^2 T_b m_a m_b^3-35 T_a^2 T_b m_b^4+64 T_a T_b^2 m_a^4+318 T_a T_b^2 m_a^3 m_b+344 T_a T_b^2 m_a^2 m_b^2  -44 T_b^3 m_a^4 \nn\\
         && -168 T_b^3 m_a^3 m_b-154 T_b^3 m_a^2 m_b^2 \Big);\nn
\end{eqnarray}
%==================
\begin{eqnarray}
%===         
  S_{a(22)} &=& + \Omega_{22} \frac{ m_a^2 m_b (15 T_a^2 m_b^2-40 T_a T_b m_a m_b-70 T_a T_b m_b^2+8 T_b^2 m_a^2+56 T_b^2 m_a m_b+63 T_b^2 m_b^2)}
  {15 (m_a+m_b)^3 (T_a m_b+T_b m_a)^2 };\nn\\
%===  
  S_{a(13)} &=& - \,\Omega_{13} \frac{2  m_a m_b^2 }{15 (m_a+m_b)^3 (T_a m_b+T_b m_a)^3}
      \Big( 10 T_a^3 m_a^2 m_b+10 T_a^3 m_b^3-17 T_a^2 T_b m_a^3-52 T_a^2 T_b m_a^2 m_b \nn\\
      && +11 T_a^2 T_b m_a m_b^2-14 T_a^2 T_b m_b^3+29 T_a T_b^2 m_a^3+66 T_a T_b^2 m_a^2 m_b-23 T_a T_b^2 m_a m_b^2 \nn\\
      && -16 T_b^3 m_a^3-36 T_b^3 m_a^2 m_b \Big) ;\nn\\
%===      
  S_{a(23)} &=& + \Omega_{23}  \frac{4 (T_b-T_a)  m_a^2 m_b^2 (5 T_a m_b-4 T_b m_a-9 T_b m_b)}{15 (m_a+m_b)^3 (T_a m_b+T_b m_a)^2 };\nn\\
  S_{a(14)} &=& -\ \Omega_{14} \frac{4  (T_b-T_a) m_a m_b^3 (T_a^2 m_a^2+T_a^2 m_b^2-2 T_a T_b m_a^2+2 T_a T_b m_a m_b+2 T_b^2 m_a^2)  }{15 (m_a+m_b)^3 (T_a m_b+T_b m_a)^3};\nn\\
  S_{a(24)} &=& + \Omega_{24} \frac{4 (T_b-T_a)^2 m_a^2 m_b^3  }{15(m_a+m_b)^3 (T_a m_b+T_b m_a)^2};\nn
\end{eqnarray}
%=============================================================
%=============================================================
\begin{eqnarray}
  {S}_{ab (2)} &=& S_{b(11)} + S_{b(12)} +  S_{b(22)} + S_{b(13)} + S_{b(23)}  + S_{b(14)} + S_{b(24)} ; \label{eq:S2}\\
  S_{b(11)} &=& -\, \frac{  T_b^2 m_a^2 (28 T_a^2 m_a m_b+8 T_a^2 m_b^2-35 T_a T_b m_a^2-40 T_a T_b m_a m_b+15 T_b^2 m_a^2)}{6 T_a (T_a m_b+T_b m_a)^3 (m_a+m_b)};\nn\\
%===  
  S_{b(12)} &=& +\Omega_{12} \frac{T_b^2 m_a^2}{15 T_a^2 (m_a+m_b)^3 (T_a m_b+T_b m_a)^3}
  \Big( 119 T_a^3 m_a^3 m_b+148 T_a^3 m_a^2 m_b^2+79 T_a^3 m_a m_b^3+20 T_a^3 m_b^4 \nn\\
  && -70 T_a^2 T_b m_a^4-369 T_a^2 T_b m_a^3 m_b-248 T_a^2 T_b m_a^2 m_b^2-39 T_a^2 T_b m_a m_b^3 +50 T_a T_b^2 m_a^4 \nn\\
  && +240 T_a T_b^2 m_a^3 m_b+100 T_a T_b^2 m_a^2 m_b^2-30 T_b^3 m_a^3 m_b \Big)  ;\nn\\
%===         
  S_{b(22)} &=& + \Omega_{22} \frac{T_b^2 m_b m_a^2 (63 T_a^2 m_a^2+56 T_a^2 m_a m_b+8 T_a^2 m_b^2-70 T_a T_b m_a^2-40 T_a T_b m_a m_b+15 T_b^2 m_a^2) }
     {15 T_a^2 (m_a+m_b)^3 (T_a m_b+T_b m_a)^2};\nn\\
%===  
  S_{b(13)} &=& -\, \Omega_{13}   \frac{2 T_b^2 m_a^2}{15 T_a^2 (m_a+m_b)^3 (T_a m_b+T_b m_a)^3}
  \Big(22 T_a^3 m_a^3 m_b+12 T_a^3 m_a^2 m_b^2+14 T_a^3 m_a m_b^3+4 T_a^3 m_b^4 \nn\\
  && -5 T_a^2 T_b m_a^4-64 T_a^2 T_b m_a^3 m_b-T_a^2 T_b m_a^2 m_b^2-2 T_a^2 T_b m_a m_b^3+5 T_a T_b^2 m_a^4  +66 T_a T_b^2 m_a^3 m_b \nn\\
  && +T_a T_b^2 m_a^2 m_b^2-20 T_b^3 m_a^3 m_b \Big)   ;\nn\\
%===      
  S_{b(23)} &=& + \Omega_{23} \frac{4  (T_b-T_a) T_b^2  m_a^3 m_b (9 T_a m_a+4 T_a m_b-5 T_b m_a)}{15 T_a^2 (m_a+m_b)^3 (T_a m_b+T_b m_a)^2} ;\nn\\
%===  
  S_{b(14)} &=& -\, \Omega_{14}  \frac{4 (T_b-T_a) T_b^2 m_a^3 m_b (T_a^2 m_a^2+T_a^2 m_b^2-2 T_a T_b m_a^2+2 T_a T_b m_a m_b+2 T_b^2 m_a^2)  }
  {15 T_a^2 (m_a+m_b)^3 (T_a m_b+T_b m_a)^3};\nn\\
%===  
  S_{b(24)} &=& + \Omega_{24} \frac{4 (T_b -T_a)^2 T_b^2 m_a^4 m_b  }{15 T_a^2 (m_a+m_b)^3 (T_a m_b+T_b m_a)^2}.\nn
\end{eqnarray}
Prescribing Coulomb collisions recovers mass-ratio coefficients (M21) and (M23) in Part 1, with further re-arrangement (M26),
finally yielding eqs. (142)-(143) there.  \\

This concludes the description of the general multi-fluid 22-moment model, which is valid for arbitrary temperatures and masses.
A particular collisional process is obtained by simply specifying the
ratio of the Chapman-Cowling integrals $\Omega_{l,j}$, with examples given by (\ref{eq:best1}). For Coulomb collisions, the results
are for convenience summarized in Section \ref{sec:CoulombA} and for the hard spheres in Section \ref{sec:Hard}. 
The ratio of collisional frequencies $\nu_{aa}/\nu_{ab}$ is further discussed in Section \ref{sec:Ratio}. 
Even though the general model might appear quite complicated, all of the mass-ratio coefficients are just pure numbers.
Additionally, the model drastically simplifies by considering small temperature differences, which is addressed in the next Section. 

%Here we have considered the 22-moment model, which captures the possible heavy tails of distribution functions more accurately
%and also modifies the energy exchange rates. The 21-moment model is obtained by neglecting the scalar perturbations
%$\widetilde{X}^{(4)}_a$.

\newpage
%==================================================================================================================
\subsection{Collisional contributions for small temperature differences} \label{sec:Tsmall}
For small temperature differences, the momentum exchange rates (\ref{eq:Final1}) remain unchanged.
Alternatively, one can approximate $T_{ab}=T_a$, and in the second formulation (\ref{eq:Final1x}) the mass-ratio coefficients become
\begin{eqnarray}
  V_{ab (1)} &=& 1-\frac{2}{5}\Omega_{12} + \frac{28 m_b}{m_b + m_a} \Big( \frac{1}{8}- \frac{1}{10}\Omega_{12} + \frac{1}{70}\Omega_{13} \Big); \nn\\
  V_{ab (2)} &=& 1-\frac{2}{5}\Omega_{12} + \frac{28 m_a}{m_b + m_a} \Big( \frac{1}{8}- \frac{1}{10}\Omega_{12} + \frac{1}{70}\Omega_{13} \Big).
\end{eqnarray}
For the energy exchange rates $Q_a$ (\ref{eq:Thierry38}), the mass-ratios simplify into 
\begin{eqnarray}
 \hat{P}_{ab (1)} &=&  \frac{(5-2\Omega_{12})m_b}{10(m_a+m_b)};\qquad
 \hat{P}_{ab (2)} =   \frac{(5-2\Omega_{12})m_a}{10(m_a+m_b)}. 
\end{eqnarray}
%=========================
The exchange rates $\bQ_{a}^{(2)}\,'$ (\ref{eq:QabP}) become
\begin{eqnarray}
 K_{ab(1)} &=& 2 + \frac{3}{5} \frac{m_b}{m_a}\Omega_{22};\qquad  K_{ab(2)} = 2 - \frac{3}{5}\Omega_{22}; \label{eq:wow14}\\
 L_{ab(1)} &=& \frac{m_b}{ m_a (m_a +m_b)} \Big[ m_a \Big( 1 - \frac{2}{5} \Omega_{12} \Big)
   + \frac{3}{10} m_b \Big( \Omega_{22} - \frac{2}{7}\Omega_{23} \Big)  \Big]; \nn\\
L_{ab(2)} &=&  \frac{m_a}{(m_a + m_b)} \Big[ 1 - \frac{2}{5} \Omega_{12} - \frac{3}{10}\Omega_{22} + \frac{3}{35}\Omega_{23} \Big]. 
\end{eqnarray}
%=========================
The exchange rates $\bQ_{a}^{(4)}\,'$ (\ref{eq:Q4abP}) simplify into
\begin{eqnarray}
  {M}_{ab (1)} &=& \frac{1}{5 (m_a+m_b)^2} \Big[70 m_a^2 +28 m_a m_b (\Omega_{12} +\frac{3}{4} \Omega_{22}) +6 m_b^2 \Omega_{23}\Big];\nn\\
  {M}_{ab (2)} &=& \frac{1}{5 (m_a+m_b)^2} \Big[m_a^2 (70 -21 \Omega_{22})  +28 m_a m_b (\Omega_{12} - \frac{3}{14}\Omega_{23}) \Big];\\
%=====
  {N}_{ab (1)} &=& \frac{1}{(m_a+m_b)^3} \Big[   m_b^3 \Big(-\frac{3}{5}\Omega_{23} + \frac{6}{35}\Omega_{24} \Big) 
     - \frac{14}{5} m_a m_b^2 \Big( \Omega_{12}- \frac{16}{49} \Omega_{13} + \frac{3}{4} \Omega_{22} - \frac{3}{14}\Omega_{23} - \frac{6}{49} \Omega_{33} \Big) \nn\\
     &&   + \frac{14}{5} m_a^2 m_b \Big( \Omega_{12} + \frac{11}{14}\Omega_{22} - \frac{5}{2} \Big)  +4 m_a^3 \Big];\nn\\
  %====
  {N}_{ab (2)} &=& \frac{m_a^2}{(m_a+m_b)^3} \Big[ -\frac{14}{5}
    m_b \Big( - \Omega_{12} +\frac{16}{49} \Omega_{13} - \frac{11}{14}\Omega_{22} + \frac{3}{14}\Omega_{23}
    - \frac{3}{49}\Omega_{24} + \frac{6}{49}\Omega_{33} +\frac{10}{7} \Big) \nn\\
    && \quad -\frac{14}{5} m_a \Big( \Omega_{12} + \frac{3}{4}\Omega_{22} - \frac{3}{14} \Omega_{23} - \frac{5}{2} \Big) \Big]. \label{eq:wow15}
\end{eqnarray}
%=========================
The exchange rates $\vecQ^{(3)}_{a}\,'$ (\ref{eq:Q3P}) become
\begin{eqnarray}
  D_{ab (1)} &=& \frac{1}{(m_a+m_b)^2 } \Big[ m_b^2 \Big( - \Omega_{12}+ \frac{2}{5}\Omega_{13}\Big)
    +  m_a m_b \Big( \Omega_{12} + \frac{4}{5}\Omega_{22} - \frac{5}{2} \Big) +3 m_a^2 \Big]; \nn\\
%===
  D_{ab (2)} &=& \frac{ m_b}{ (m_a+m_b)^2} \Big[ m_b \Big(- \Omega_{12} + \frac{2}{5}\Omega_{13} - \frac{4}{5}\Omega_{22}+3 \Big) 
    + m_a (\Omega_{12}- \frac{5}{2} \Big)\Big];\nn\\
  U_{ab (1)} &=& -\, \frac{5 m_a + 2 m_b\Omega_{12} }{2(m_a+m_b)}; \label{eq:wow16}
\end{eqnarray}  
%\begin{eqnarray}
%  D_{ab (1)} &=& -\, \frac{1}{10(m_a+m_b)^2 } \Big[ m_b^2 \Big( 10 \Omega_{12} -4 \Omega_{13} \Big) 
%    -10 m_a m_b \Big(\Omega_{12} +\frac{4}{5}\Omega_{22} -\frac{5}{2}\Big) -30 m_a^2 \Big]; \nn\\
%===
%  D_{ab (2)} &=& -\, \frac{5 m_b}{2 (m_a+m_b)^2} \Big[ m_b \Big(- \frac{6}{5} +\frac{2}{5} \Omega_{12}
%    - \frac{4}{25} \Omega_{13} +\frac{8}{25} \Omega_{22} \Big) 
%    +m_a \Big(1 -\frac{2}{5} \Omega_{12} \Big) \Big];\nn\\
  %===%
%\end{eqnarray}
%================================
\begin{eqnarray}
  E_{ab (1)} &=& \frac{m_b}{ (m_a+m_b)^3}\Big[ m_b^2 \Big(-\frac{1}{8}\Omega_{12}+ \frac{1}{10}\Omega_{13}- \frac{1}{70}\Omega_{14} \Big)
    +m_a m_b \Big( -\frac{5}{16} +\frac{1}{4}\Omega_{12}- \frac{1}{28}\Omega_{13} + \frac{1}{5}\Omega_{22} - \frac{2}{35}\Omega_{23}  \Big) \nn\\
    && +m_a^2 \Big( \frac{3}{4} -\frac{3}{10}\Omega_{12}\Big) \Big];\nn\\
%===  
  E_{ab (2)} &=&  \frac{m_a m_b}{ (m_a+m_b)^3}\Big[ m_b \Big(- \frac{3}{4} +\frac{17}{40} \Omega_{12}  - \frac{1}{10}\Omega_{13}
    +\frac{1}{70}\Omega_{14} + \frac{1}{5}\Omega_{22} - \frac{2}{35}\Omega_{23}  \Big) \nn\\
   && + m_a \Big( \frac{5}{16} - \frac{1}{4}\Omega_{12} + \frac{1}{28}\Omega_{13} \Big) \Big].
\end{eqnarray}
%=================================
The exchange rates $\vecQ^{(5)}_{a}\,'$ (\ref{eq:Qab5P}) simplify into
\begin{eqnarray}
  F_{ab (1)} &=& \frac{1}{(m_a+m_b)^3} \Big[ m_b^3 \Big(-4 \Omega_{13} + \frac{8}{5} \Omega_{14} \Big)
    - 28 m_a m_b^2 \Big( \Omega_{12} -\frac{2}{5} \Omega_{13} - \frac{8}{35} \Omega_{23}\Big) \nn\\
    &&  +\frac{238}{5} m_a^2 m_b \Big( \Omega_{12} + \frac{8}{17} \Omega_{22} - \frac{25}{34} \Big)  + 84 m_a^3 \Big]; \nn\\
  F_{ab (2)} &=& \frac{1}{(m_a+m_b)^3} \Big[ m_b^3 \Big( - \frac{168}{5} \Omega_{12} + 4 \Omega_{13}
    - \frac{8}{5}\Omega_{14} + \frac{32}{5}\Omega_{23}\Big)  \nn\\
   && + 28 m_a m_b^2 \Big( \Omega_{12} - \frac{2}{5} \Omega_{13} + \frac{4}{5} \Omega_{22} -3 \Big) 
    - 14 m_a^2 m_b \Big( \Omega_{12} -\frac{5}{2}  \Big)  \Big];
\end{eqnarray}
%===================
%===================
\begin{eqnarray}
  G_{ab (1)} &=& \frac{1}{(m_a+m_b)^4} \Big[ m_b^4 \Big( \frac{1}{2}\Omega_{13}- \frac{2}{5}\Omega_{14} + \frac{2}{35}\Omega_{15} \Big) 
     + \frac{7}{2} m_a m_b^3 \Big( \Omega_{12} - \frac{4}{5}\Omega_{13} + \frac{4}{35}\Omega_{14}  - \frac{16}{35}\Omega_{23}
    + \frac{32}{245} \Omega_{24} \Big) \nn\\
    && +\frac{16}{35}  m_a^2 m_b^2 \Big( \Omega_{33} - \frac{833}{32} \Omega_{12}  + \frac{251}{32}\Omega_{13}
    - \frac{49}{4} \Omega_{22} + \frac{7}{2} \Omega_{23} + \frac{1225}{128} \Big) \nn\\
    && + \frac{42}{5} m_a^3 m_b \Big( \Omega_{12} + \frac{8}{21} \Omega_{22} - \frac{5}{2} \Big) +5 m_a^4 \Big]; \nn\\
%======  
  G_{ab (2)} &=& \frac{1}{(m_a+m_b)^4} \Big[
    - \frac{7}{2} m_a m_b^3 \Big(- \frac{32}{245}\Omega_{33} + \frac{12}{5}\Omega_{12}
    - \frac{251}{245}\Omega_{13}  + \frac{4}{35} \Omega_{14} - \frac{4}{245} \Omega_{15}
    + \frac{32}{35} \Omega_{22} - \frac{16}{35} \Omega_{23}  \nn\\
    && +  \frac{32}{245} \Omega_{24} - \frac{10}{7} \Big)  
      +\frac{119}{10} m_a^2 m_b^2 \Big( \Omega_{12} - \frac{4}{17} \Omega_{13} +   \frac{4}{119}\Omega_{14}
    + \frac{8}{17}\Omega_{22} - \frac{16}{119}\Omega_{23} - \frac{30}{17} \Big) \nn\\
   && - \frac{7}{2} m_a^3 m_b \Big( \Omega_{12}- \frac{1}{7}\Omega_{13} - \frac{5}{4} \Big) \Big];\label{eq:wow17m}
\end{eqnarray}
\begin{eqnarray}
U_{ab (2)} &=& \frac{ 4 \Omega_{13} m_b^2  +28 \Omega_{12} m_a m_b  +35 m_a^2}{(m_a+m_b)^2}. \label{eq:wow17}
\end{eqnarray}  
%====================
%====================
Finally, the exchange rates $\widetilde{Q}_a^{(4)}\,'$ (\ref{eq:Q4exchange}) become
\begin{eqnarray}
  S_{ab (0)} &=& \frac{4 m_a (2 \Omega_{12} m_b+5 m_a)}{(m_a+m_b)^2}; \nn\\
%===
  S_{ab (1)} &=& \frac{4 m_a}{3 (m_a+m_b)^3}  \Big[ m_a^2 + m_a m_b \Big(\Omega_{12}+ \frac{2}{5} \Omega_{22} - \frac{5}{2} \Big)
    -m_b^2 \Big( \Omega_{12} - \frac{2}{5} \Omega_{13}\Big) \Big] ;\nn\\
%===
  S_{ab (2)} &=&  \frac{4 m_a^2}{3 (m_a+m_b)^3} \Big[  m_b \Big( \Omega_{12} - \frac{2}{5}\Omega_{13} + \frac{2}{5}\Omega_{22} -1 \Big)
    - m_a \Big( \Omega_{12} - \frac{5}{2} \Big) \Big].\label{eq:Sab2-eq}
\end{eqnarray}
%=======
Obviously, for small temperature differences the formulation of the 22-moment model through
the Chapman-Cowling integrals is not overly-complicated and actually quite user-friendly.

%\newpage
%==================================================================================================================
\subsection{Collisional contributions for self-collisions (only double-check)}
The self-collisional contributions were already separated at the front of all the collisional contributions,
with the rest expressed as a $\sum_{b\neq a}$, and the following expressions are thus not needed. 
Nevertheless, for clarity and for the convenience of the reader, we provide these expressions as well.
For self-collisions, the $\bQ_{a}^{(2)}\,'$ exchange rates simplify into
\begin{eqnarray}
  {K}_{aa (1)} &=&  2 + \frac{3}{5} \Omega_{22};\qquad {K}_{aa (2)} =  2 - \frac{3}{5}\Omega_{22}; \nn\\
  {L}_{aa (1)} &=&  +\frac{1}{2} - \frac{1}{5}\Omega_{12} +\frac{3}{20} \Omega_{22} - \frac{3}{70}\Omega_{23};\nn\\
  {L}_{aa (2)} &=&  +\frac{1}{2} -\frac{1}{5}\Omega_{12}  - \frac{3}{20}\Omega_{22} + \frac{3}{70}\Omega_{23},
\end{eqnarray}
yielding self-collisional contributions
\begin{eqnarray}
  -K_{aa (1)} +  K_{aa (2)} = -\frac{6}{5} \Omega_{22}; \qquad
  L_{aa (1)} - L_{aa (2)} = \frac{3}{10}\Omega_{22} - \frac{3}{35}\Omega_{23}.
\end{eqnarray}
%=================================
For the $\bQ_{a}^{(4)}\,'$ exchange rates:
\begin{eqnarray}
  {M}_{aa (1)} &=& +\frac{7}{2} +\frac{7}{5} \Omega_{12} + \frac{21}{20}\Omega_{22}   + \frac{3}{10}\Omega_{23}; \nn\\
  {M}_{aa (2)} &=& +\frac{7}{2} +\frac{7}{5} \Omega_{12} - \frac{21}{20} \Omega_{22}  - \frac{3}{10}\Omega_{23}; \nn\\
  {N}_{aa (1)} &=& - \frac{3}{8}  + \frac{4}{35}\Omega_{13} + \frac{1}{80}\Omega_{22} + \frac{3}{140} \Omega_{24} + \frac{3}{70}\Omega_{33} ;\nn\\
  {N}_{aa (2)} &=& +\frac{3}{8} - \frac{4}{35}\Omega_{13} +\frac{1}{80}\Omega_{22} +\frac{3}{140} \Omega_{24} - \frac{3}{70}\Omega_{33},
\end{eqnarray}
yielding self-collisional contributions
\begin{eqnarray}
  -  M_{aa (1)} +  M_{aa (2)} = - \big( \frac{21}{10}\Omega_{22} +\frac{3}{5}\Omega_{23}\big); \qquad
  N_{aa (1)} +N_{aa (2)} = \frac{1}{40} \Omega_{22}+ \frac{3}{70} \Omega_{24}.
\end{eqnarray}
%=================================
For the  $\vecQ^{(3)}_{a}\,'$ exchange rates:
\begin{eqnarray}
  D_{aa (1)} &=& \frac{1}{8} + \frac{1}{5} \Omega_{22} +\frac{1}{10}\Omega_{13}; \qquad
  D_{aa (2)} = \frac{1}{8}  - \frac{1}{5}\Omega_{22}  +\frac{1}{10}\Omega_{13}; \nn\\
%=============
  E_{aa (1)} &=& - \frac{7}{320}\Omega_{12} +  \frac{9}{1120} \Omega_{13}  - \frac{1}{560} \Omega_{14}
  + \frac{1}{40}\Omega_{22}- \frac{1}{140}\Omega_{23} +\frac{7}{128};\nn\\
  E_{aa (2)} &=& + \frac{7}{320}\Omega_{12} -\frac{9}{1120} \Omega_{13} + \frac{1}{560}\Omega_{14} + \frac{1}{40}\Omega_{22}
  - \frac{1}{140}\Omega_{23} - \frac{7}{128}, 
\end{eqnarray}
yielding self-collisional contributions
\begin{eqnarray}
  -  D_{aa (1)} +  D_{aa (2)} = - \frac{2}{5}\Omega_{22}; \qquad
  E_{aa (1)} +E_{aa (2)} = \frac{1}{20}\Omega_{22}- \frac{1}{70}\Omega_{23}.
\end{eqnarray}
%=================================
For the  $\vecQ^{(5)}_{a}\,'$ exchange rates:
\begin{eqnarray}
  F_{aa (1)} &=& + \frac{49}{20}\Omega_{12} + \frac{9}{10}\Omega_{13} + \frac{1}{5}\Omega_{14}  + \frac{4}{5}\Omega_{23} + \frac{14}{5}\Omega_{22}
  + \frac{49}{8}; \nn\\
  F_{aa (2)} &=& - \frac{49}{20} \Omega_{12}- \frac{9}{10}\Omega_{13} - \frac{1}{5}\Omega_{14}
  + \frac{4}{5}\Omega_{23} + \frac{14}{5}\Omega_{22} - \frac{49}{8};\nn\\
%============
  G_{aa (1)} &=& + \frac{9}{112} \Omega_{13} + \frac{1}{280}\Omega_{15}+ \frac{1}{35}\Omega_{24}
  + \frac{1}{35}\Omega_{33} - \frac{3}{20}\Omega_{22} - \frac{93}{128};\nn\\
  G_{aa (2)} &=& + \frac{9}{112}\Omega_{13} + \frac{1}{280}\Omega_{15} - \frac{1}{35}\Omega_{24} +\frac{1}{35}\Omega_{33} 
  + \frac{3}{20}\Omega_{22}  - \frac{93}{128},
\end{eqnarray}
yielding self-collisional contributions
\begin{eqnarray}
  F_{aa (1)} + F_{aa (2)} &=& \frac{8}{5}\Omega_{23} + \frac{28}{5}\Omega_{22};\qquad
   - G_{aa (1)} + G_{aa (2)} = - \frac{2}{35} \Omega_{24} + \frac{3}{10}\Omega_{22}.
\end{eqnarray}
%=====================================
Finally, for the  $\widetilde{Q}_a^{(4)}\,'$ exchange rates:
\begin{eqnarray}
  S_{aa (1)} &=& -\frac{1}{4}  + \frac{1}{15}\Omega_{13} + \frac{1}{15}\Omega_{22};\qquad
  S_{aa (2)} = + \frac{1}{4} - \frac{1}{15}\Omega_{13} + \frac{1}{15}\Omega_{22},
\end{eqnarray}
yielding self-collisional contributions
\begin{equation}
 S_{aa (1)} +  S_{aa (2)} = \frac{2}{15}\Omega_{22}.
\end{equation}
This concludes the formulation of the multi-fluid 22-moment model through the Chapman-Cowling integrals. 
Below we briefly consider the particular cases of Coulomb collisions and hard spheres.

\newpage
%==================================================================================================================
%==================================================================================================================
\section{Collisional contributions for particular cases} \label{sec:Particular}
\subsection{Coulomb collisions (arbitrary temperatures and masses\texorpdfstring{, $\ln\Lambda\gg 1$}{})} \label{sec:CoulombA}
Because in the present formulation the 1-Hermite and 2-Hermite contributions were kept separately,
we often had to reference to various Appendices of Part 1. Considering the Coulomb collisions, it is beneficial to summarize the 
entire multi-fluid 22-moment model at one place, which we do right here. The results assume that the Coulomb logarithm is large, $\ln\Lambda\gg 1$.
If this is not necessarily true,
additional corrections for the $\ln\Lambda$ can be obtained from the previous general case by employing
Chapman-Cowling integrals (\ref{eq:best1}), which contain the coefficients $A_1(2)$, $A_2(2)$ and $A_3(2)$ given by (\ref{eq:A23x}).

The momentum exchange rates (\ref{eq:Final1})-(\ref{eq:Final11}) are given by the coefficients
\begin{eqnarray}
    V_{ab (0)} &=& +\frac{3}{5};\qquad  V_{ab (3)} = +\frac{3}{56}.
\end{eqnarray}
The energy exchange rates $Q_a$ (\ref{eq:Thierry38}) are given by
\begin{eqnarray}
  \hat{P}_{ab (1)} &=& \frac{3 T_a m_b (5 T_b m_b +4 T_b m_a -T_a m_b  )}{40 (T_a m_b +T_b m_a)^2};\qquad
  \hat{P}_{ab (2)} = \frac{3 T_b m_a (5 T_a m_a +4 T_a m_b -T_b m_a  )}{40 (T_a m_b +T_b m_a)^2}.
\end{eqnarray}
%=========================
The stress tensor $\bPi_a^{(2)}$ exchange rates $\bQ_{a}^{(2)}\,'$ (\ref{eq:QabP}) have coefficients
\begin{eqnarray}
%K_{ab (1)} &=& \frac{ 2 (2 T_a m_a m_b +3 T_a m_b^2 +5 T_b m_a^2 +6 T_b m_a m_b)}{5 (T_a m_b +T_b m_a) m_a};\nn\\
%K_{ab (2)} &=& \frac{2 (3 T_a m_a +2 T_a m_b -T_b m_a)}{5 (T_a m_b +T_b m_a)};\nn\\
%===
  K_{ab(1)} &=& \frac{2 T_b (m_a +m_b)}{(T_a m_b +T_b m_a)}
  - \frac{4 (T_b-T_a) m_b}{5 (T_a m_b +T_b m_a)}  + \frac{ 6 m_b }{5 m_a}; \\
%===
  K_{ab(2)} &=& \frac{2 T_a (m_a +m_b)}{(T_a m_b +T_b m_a)}
  + \frac{4 (T_b-T_a) m_a }{5 (T_a m_b +T_b m_a)} - \frac{6}{5};\nn\\
L_{ab (1)} &=& \frac{3 T_a m_b (2 T_a m_a m_b +3 T_a m_b^2 +7 T_b m_a^2 +8 T_b m_a m_b)}{35 (T_a m_b +T_b m_a)^2 m_a};\nn\\
L_{ab (2)} &=& \frac{ 3 m_a T_b (5 T_a m_a +4 T_a m_b -T_b m_a)}{35 (T_a m_b +T_b m_a)^2}. \label{eq:ViscGc}
\end{eqnarray}
The stress tensor $\bPi_a^{(4)}$ exchange rates $\bQ_{a}^{(4)}\,'$ (\ref{eq:Q4abP}) read
\begin{eqnarray}
  M_{ab (1)} &=& \frac{2}{5 T_a (T_a m_b +T_b m_a)^2 (m_b+m_a)} \Big(16 T_a^3 m_a m_b^2 +12 T_a^3 m_b^3 +56 T_a^2 T_b m_a^2 m_b +31 T_a^2 T_b m_a m_b^2 \nn\\
  && +70 T_a T_b^2 m_a^3  +14 T_a T_b^2 m_a^2 m_b -35 T_b^3 m_a^3 \Big) ;\nn\\
  M_{ab (2)} &=& -\, \frac{2 T_b m_a (9 T_a^2 m_a m_b -2 T_a^2 m_b^2 -21 T_a T_b m_a^2 -25 T_a T_b m_a m_b +7 T_b^2 m_a^2)}{5 (T_a m_b +T_b m_a)^2 T_a (m_b +m_a)};\nn\\
  N_{ab (1)} &=& -\, \frac{1}{35 (T_a m_b +T_b m_a)^3 (m_b +m_a)}\Big( 16 T_a^3 m_a m_b^3 +12 T_a^3 m_b^4 +72 T_a^2 T_b m_a^2 m_b^2 +21 T_a^2 T_b m_a m_b^3 \nn\\
  && +126 T_a T_b^2 m_a^3 m_b  -54 T_a T_b^2 m_a^2 m_b^2 -140 T_b^3 m_a^4 -273 T_b^3 m_a^3 m_b \Big);\nn\\
  N_{ab (2)} &=& -\, \frac{3 T_b^2 m_a^2 (35 T_a^2 m_a m_b +12 T_a^2 m_b^2 -35 T_a T_b m_a^2 -51 T_a T_b m_a m_b +7 T_b^2 m_a^2)}{35 (T_a m_b +T_b m_a)^3 T_a (m_b +m_a)}. \label{eq:Posled5}
\end{eqnarray}
The heat flux $\vecq_a$ exchange rates $\vecQ^{(3)}_{a}\,'$ (\ref{eq:Q3P}) read
\begin{eqnarray}
  U_{ab (1)} &=&  \frac{(4 T_a-11 T_b) m_a m_b-2 T_a m_b^2-5 T_b m_a^2}{2 (T_a m_b+T_b m_a) (m_b+m_a) };\nn\\
  D_{ab (1)} &=& -\, \frac{6 T_a^2 m_a m_b^2+2 T_a^2 m_b^3+21 T_a T_b m_a^2 m_b-5 T_a T_b m_a m_b^2-30 T_b^2 m_a^3-52 T_b^2 m_a^2 m_b}{10 (T_a m_b+T_b m_a)^2 (m_b+m_a)};\nn\\
  D_{ab (2)} &=& \frac{3 m_b T_a [(10 T_a -11 T_b) m_a m_b+4 T_a m_b^2-5 T_b m_a^2]}{10 (T_a m_b+T_b m_a)^2 (m_b+m_a)};\nn\\
  E_{ab (1)} &=& -\, \frac{3  T_a m_b [6 T_a^2 m_a m_b^2+2 T_a^2 m_b^3+27 T_a T_b m_a^2 m_b-11 T_a T_b m_a m_b^2-84 T_b^2 m_a^3-118 T_b^2 m_a^2 m_b]}{560 (T_a m_b+T_b m_a)^3 (m_b+m_a)};\nn\\
  E_{ab (2)} &=& -\, \frac{3  m_a m_b T_a T_b [16 T_a m_a m_b+10 T_a m_b^2-5 T_b m_a^2-11 T_b m_a m_b]}{112 (T_a m_b+T_b m_a)^3 (m_b+m_a)}.\nn\\
\end{eqnarray}
The heat flux $\vecX^{(5)}_a$ exchange rates $\vecQ^{(5)}_{a}\,'$ (\ref{eq:Qab5P}) become
\begin{eqnarray}
  U_{ab (2)} &=& -\, \frac{16 T_a^2 m_a m_b^2-8 T_a^2 m_b^3+56 T_a T_b m_a^2 m_b-52 T_a T_b m_a m_b^2-35 T_b^2 m_a^3-119 T_b^2 m_a^2 m_b}{(T_a m_b+T_b m_a)^2 (m_b+m_a)};\nn\\
  F_{ab (1)} &=& \big\{40 T_a^4 m_a m_b^3+8 T_a^4 m_b^4+180 T_a^3 T_b m_a^2 m_b^2+68 T_a^3 T_b m_a m_b^3+315 T_a^2 T_b^2 m_a^3 m_b
    +207 T_a^2 T_b^2 m_a^2 m_b^2 \nn\\
    && +700 T_a T_b^3 m_a^4+392 T_a T_b^3 m_a^3 m_b-280 T_b^4 m_a^4\big\} \big[5 (T_a m_b+T_b m_a)^3 (m_b+m_a) T_a\big]^{-1};\nn\\
    F_{ab (2)} &=& -\, \frac{3  T_a m_b \big[16 T_a^2 m_b^3+140 T_a T_b m_a^2 m_b+72 T_a T_b m_a m_b^2-35 T_b^2 m_a^3-119 T_b^2 m_a^2 m_b\big] }{5 (T_a m_b+T_b m_a)^3 (m_b+m_a)};\nn\\
    G_{ab (1)} &=&  -\, \big\{40 T_a^4 m_a m_b^4+8 T_a^4 m_b^5+220 T_a^3 T_b m_a^2 m_b^3+140 T_a^3 T_b m_a m_b^4 +495 T_a^2 T_b^2 m_a^3 m_b^2 \nn\\
    && +627 T_a^2 T_b^2 m_a^2 m_b^3+3640 T_a T_b^3 m_a^4 m_b+1916 T_a T_b^3 m_a^3 m_b^2-1400 T_b^4 m_a^5 \nn\\
    && -3304 T_b^4 m_a^4 m_b \big\} \big[280 (T_a m_b+T_b m_a)^4 (m_a+m_b)\big]^{-1};\nn\\
 G_{ab (2)} &=&  \frac{3  T_a T_b m_a^2 m_b\big[8 T_a^2 m_b^2-32 T_a T_b m_a m_b-28 T_a T_b m_b^2+5 T_b^2 m_a^2+17 T_b^2 m_a m_b\big]}{8 (T_a m_b+T_b m_a)^4 (m_a+m_b)}.\nn\\   
\end{eqnarray}
Finally, the scalar $\widetilde{X}^{(4)}_a$ exchange rates $\widetilde{Q}_a^{(4)}\,'$ (\ref{eq:Q4exchange}) become
\begin{eqnarray}
  S_{ab (0)} &=& \frac{4  m_a (2 T_a m_b +5 T_b m_a)}{(T_a m_b +T_b m_a) (m_b+m_a)};\nn\\
  S_{ab (1)} &=& -\, \frac{ m_a}{30 (T_a m_b +T_b m_a)^3 (m_b+m_a)} \Big( 2 T_a^3 m_b^3 +9 T_a^2 T_b m_a m_b^2 +6 T_a^2 T_b m_b^3 +72 T_a T_b^2 m_a^2 m_b
  +27 T_a T_b^2 m_a m_b^2 \nn\\
      &&  -40 T_b^3 m_a^3 -84 T_b^3 m_a^2 m_b \Big);\nn\\
  S_{ab (2)} &=& -\, \frac{T_b^2 m_a^3 (2 T_a^2 m_b -5 T_a T_b m_a -6 T_a T_b m_b +T_b^2 m_a)}{2T_a (T_a m_b +T_b m_a)^3  (m_b+m_a)}.
\end{eqnarray}

%=========================
%and for small temperature differences
%\begin{eqnarray}
%  {Z}_{ab (3)}^{\textrm{Coulomb}} &=& \frac{3  m_a (280 m_a^4-728 m_a^3 m_b-99 m_a^2 m_b^2-44 m_a m_b^3-8 m_b^4)}{140 (m_a+m_b)^5}; \nn\\
%  {Z}_{ab (4)}^{\textrm{Coulomb}} &=& -\, \frac{9 m_a^4 (5 m_a-4 m_b)}{4 (m_a+m_b)^5}.
%\end{eqnarray}

\newpage
%==================================================================================================================
\subsubsection{Coulomb collisions (small temperature differences)} \label{sec:Coulomb-eq}
Considering Coulomb collisions with small temperature differences, the model given in the previous section simplifies into 
\begin{eqnarray}
  \hat{P}_{ab (1)} &=& \frac{3 m_b}{10 (m_a+m_b)};\qquad  \hat{P}_{ab (2)} = \frac{3 m_a}{10 (m_a+m_b)};\nn\\
%===
  K_{ab(1)} &=& \frac{2(5 m_a+3 m_b)}{5 m_a}; \qquad  K_{ab(2)} = \frac{4}{5};\nn\\  
  L_{ab (1)} &=& \frac{3 m_b(7 m_a+3 m_b) }{35 m_a (m_a+m_b)};\qquad L_{ab (2)} = \frac{12 m_a}{35 (m_a+m_b)}; 
\end{eqnarray}
%The stress tensor $\bPi_a^{(4)}$ exchange rates $\bQ_{a}^{(4)}\,'$ (\ref{eq:Q4abP}) read
\begin{eqnarray}
  M_{ab (1)} &=&  \frac{2 (35 m_a^2+35 m_a m_b+12 m_b^2)}{5 (m_a+m_b)^2} ;\qquad M_{ab (2)} = \frac{4 m_a (7 m_a+m_b)}{5 (m_a+m_b)^2};\nn\\
%===
  N_{ab (1)} &=& \frac{140 m_a^3+7 m_a^2 m_b-25 m_a m_b^2-12 m_b^3}{35 (m_a+m_b)^3};\qquad   N_{ab (2)} = \frac{12 m_a^2 (7 m_a-3 m_b)}{35 (m_a+m_b)^3};
\end{eqnarray}
%The heat flux $\vecq_a$ exchange rates $\vecQ^{(3)}_{a}\,'$ (\ref{eq:Q3P}) read
\begin{eqnarray}
  U_{ab (1)} &=& -\, \frac{(5 m_a+2 m_b)}{2 (m_a+m_b)};\nn\\
  D_{ab (1)} &=& \frac{30 m_a^2+m_a m_b-2 m_b^2}{10 (m_a+m_b)^2} ;\qquad  D_{ab (2)} = -\, \frac{3 m_b (5 m_a-4 m_b)}{10 (m_a+m_b)^2} ;\nn\\
%===  
  E_{ab (1)} &=& \frac{3 m_b (84 m_a^2+7 m_a m_b-2 m_b^2) }{560 (m_a+m_b)^3};\qquad E_{ab (2)} = \frac{15 m_a m_b (m_a-2 m_b)}{112 (m_a+m_b)^3};
\end{eqnarray}
%The heat flux $\vecX^{(5)}_a$ exchange rates $\vecQ^{(5)}_{a}\,'$ (\ref{eq:Qab5P}) become
\begin{eqnarray}
&& U_{ab (2)} = \frac{35 m_a^2+28 m_a m_b+8 m_b^2}{(m_a+m_b)^2};\nn\\  
&& F_{ab (1)} = \frac{420 m_a^3+287 m_a^2 m_b+100 m_a m_b^2+8m_b^3}{5(m_a+m_b)^3};\nn\\  
&& F_{ab (2)} = \frac{3}{5}\frac{m_b(35 m_a^2-56m_a m_b -16m_b^2)}{(m_a+m_b)^3};\nn\\
&& G_{ab (1)} = \frac{1400 m_a^4 -1736 m_a^3 m_b -675 m_a^2 m_b^2 -172 m_a m_b^3-8m_b^4}{280 (m_a+m_b)^4};\nn\\
&& G_{ab (2)} = \frac{15}{8} \frac{m_a^2 m_b(m_a-4m_b)}{(m_a+m_b)^4};  
\end{eqnarray}
%Finally, the scalar $\widetilde{X}^{(4)}_a$ exchange rates $\widetilde{Q}_a^{(4)}\,'$ (\ref{eq:Q4exchange}) become
\begin{eqnarray}
  S_{ab (0)} &=& \frac{4 m_a (5 m_a+2 m_b)}{(m_a+m_b)^2};\nn\\
%===
  S_{ab (1)} &=& \frac{2 m_a (10 m_a^2-7 m_a m_b-2 m_b^2) }{15 (m_a+m_b)^3};\qquad
%===
  S_{ab (2)} = \frac{2 m_a^3}{(m_a+m_b)^3}.
\end{eqnarray}

\newpage
%==================================================================================================================
%==================================================================================================================
\subsection{Hard spheres collisions (arbitrary temperatures and masses)} \label{sec:Hard}
It is also beneficial to summarize the results for the collisions of hard spheres. 
The momentum exchange rates (\ref{eq:Final1})-(\ref{eq:Final11}) are given by the coefficients (note the opposite signs with respect to Coulomb
collisions)
\begin{eqnarray}
  V_{ab (0)} &=& -\,\frac{1}{5};\qquad  V_{ab (3)} = -\,\frac{1}{280}.
\end{eqnarray}
The energy exchange rates $Q_a$ (\ref{eq:Thierry38}) are given by
\begin{eqnarray}
  \hat{P}_{ab (1)} &=& -\, \frac{T_a  m_b (3 T_a m_b+4 T_b m_a+T_b m_b) }{40(T_a m_b+T_b m_a)^2};\qquad
  \hat{P}_{ab (2)} = -\, \frac{T_b m_a (T_a m_a+4 T_a m_b+3 T_b m_a) }{40 (T_a m_b+T_b m_a)^2}.
\end{eqnarray}
%=========================
The stress tensor $\bPi_a^{(2)}$ exchange rates $\bQ_{a}^{(2)}\,'$ (\ref{eq:QabP}) have coefficients
\begin{eqnarray}
%===
  K_{ab(1)} &=& \frac{2 T_b (m_a +m_b)}{(T_a m_b +T_b m_a)}
  - \frac{12 (T_b-T_a) m_b}{5 (T_a m_b +T_b m_a)}  + \frac{ 6 m_b }{5 m_a}; \\
%===
  K_{ab(2)} &=& \frac{2 T_a (m_a +m_b)}{(T_a m_b +T_b m_a)}
  + \frac{12 (T_b-T_a) m_a }{5 (T_a m_b +T_b m_a)} - \frac{6}{5};\nn\\  
L_{ab (1)} &=& -\, \frac{T_a m_b (6 T_a m_a m_b+3 T_a m_b^2+7 T_b m_a^2+4 T_b m_a m_b)}{35(T_a m_b+T_b m_a)^2 m_a};\nn\\
L_{ab (2)} &=& -\, \frac{T_b m_a (T_a m_a+4 T_a m_b+3 T_b m_a)}{35 (T_a m_b+T_b m_a)^2}.
\end{eqnarray}
The stress tensor $\bPi_a^{(4)}$ exchange rates $\bQ_{a}^{(4)}\,'$ (\ref{eq:Q4abP}) read
\begin{eqnarray}
  M_{ab (1)} &=& \frac{2}{5 (T_a m_b+T_b m_a)^2 (m_a+m_b)^3 T_a}
\Big( 96 T_a^3 m_a^3 m_b^2+88 T_a^3 m_a^2 m_b^3+96 T_a^3 m_a m_b^4+24 T_a^3 m_b^5 \nn\\
&& +168 T_a^2 T_b m_a^4 m_b+87 T_a^2 T_b m_a^3 m_b^2+198 T_a^2 T_b m_a^2 m_b^3+39 T_a^2 T_b m_a m_b^4+70 T_a T_b^2 m_a^5 \nn\\
&& -42 T_a T_b^2 m_a^4 m_b+138 T_a T_b^2 m_a^3 m_b^2+10 T_a T_b^2 m_a^2 m_b^3-35 T_b^3 m_a^5+42 T_b^3 m_a^4 m_b-3 T_b^3 m_a^3 m_b^2 \Big) ;\nn\\
%===
  M_{ab (2)} &=& -\, \frac{2 T_b m_a}{5 (T_a m_b+T_b m_a)^2 (m_a+m_b)^3 T_a^2}
\Big(5 T_a^3 m_a^3 m_b-24 T_a^3 m_a^2 m_b^2+33 T_a^3 m_a m_b^3-18 T_a^3 m_b^4 \nn\\
&& +7 T_a^2 T_b m_a^4-39 T_a^2 T_b m_a^3 m_b+93 T_a^2 T_b m_a^2 m_b^2-101 T_a^2 T_b m_a m_b^3-21 T_a T_b^2 m_a^4 \nn\\
&& +54 T_a T_b^2 m_a^3 m_b-165 T_a T_b^2 m_a^2 m_b^2-80 T_b^3 m_a^3 m_b \Big) ;\nn\\
%===
  N_{ab (1)} &=& \frac{1}{35 (T_a m_b+T_b m_a)^3 (m_a+m_b)^3} 
\Big( 288 T_a^3 m_a^3 m_b^3+264 T_a^3 m_a^2 m_b^4+288 T_a^3 m_a m_b^5+72 T_a^3 m_b^6 \nn\\
&& +744 T_a^2 T_b m_a^4 m_b^2+615 T_a^2 T_b m_a^3 m_b^3+774 T_a^2 T_b m_a^2 m_b^4+183 T_a^2 T_b m_a m_b^5+602 T_a T_b^2 m_a^5 m_b \nn\\
&& +390 T_a T_b^2 m_a^4 m_b^2+654 T_a T_b^2 m_a^3 m_b^3+146 T_a T_b^2 m_a^2 m_b^4+140 T_b^3 m_a^6+21 T_b^3 m_a^5 m_b \nn\\
&& +150 T_b^3 m_a^4 m_b^2+29 T_b^3 m_a^3 m_b^3 \Big);\nn\\
  N_{ab (2)} &=& -\, \frac{T_b^2 m_a^2}{35 T_a^2 (T_a m_b+T_b m_a)^3 (m_a+m_b)^3}
\Big(T_a^3 m_a^3 m_b+6 T_a^3 m_a^2 m_b^2-87 T_a^3 m_a m_b^3+148 T_a^3 m_b^4 \nn\\
&& +7 T_a^2 T_b m_a^4+45 T_a^2 T_b m_a^3 m_b-123 T_a^2 T_b m_a^2 m_b^2+559 T_a^2 T_b m_a m_b^3+21 T_a T_b^2 m_a^4 \nn\\
&& -54 T_a T_b^2 m_a^3 m_b+645 T_a T_b^2 m_a^2 m_b^2+240 T_b^3 m_a^3 m_b \Big).
\end{eqnarray}
The heat flux $\vecq_a$ exchange rates $\vecQ^{(3)}_{a}\,'$ (\ref{eq:Q3P}) read
\begin{eqnarray}
  U_{ab (1)} &=&  -\, \frac{(8 T_a^2 m_a^2 m_b-2 T_a^2 m_a m_b^2+6 T_a^2 m_b^3+5 T_a T_b m_a^3-8 T_a T_b m_a^2 m_b+19 T_a T_b m_a m_b^2+16 T_b^2 m_a^2 m_b)}
{2 T_a (T_a m_b+T_b m_a) (m_a+m_b)^2};\nn\\
  D_{ab (1)} &=&  \frac{3}{10 (T_a m_b+T_b m_a)^2 (m_a+m_b)^2}
\Big(18 T_a^2 m_a^2 m_b^2+8 T_a^2 m_a m_b^3+6 T_a^2 m_b^4+29 T_a T_b m_a^3 m_b \nn\\
&& +8 T_a T_b m_a^2 m_b^2+11 T_a T_b m_a m_b^3+10 T_b^2 m_a^4-2 T_b^2 m_a^3 m_b+4 T_b^2 m_a^2 m_b^2 \Big);\nn\\
%===
  D_{ab (2)} &=&  \frac{m_b (2 T_a^2 m_a^2 m_b-14 T_a^2 m_a m_b^2+32 T_a^2 m_b^3+5 T_a T_b m_a^3-8 T_a T_b m_a^2 m_b+83 T_a T_b m_a m_b^2+48 T_b^2 m_a^2 m_b)}
  {10 (T_a m_b+T_b m_a)^2 (m_a+m_b)^2};\nn\\
%\end{eqnarray}
%\begin{eqnarray}
%===  
  E_{ab (1)} &=&  -\, \frac{3 T_a m_b  }{560 (T_a m_b+T_b m_a)^3 (m_a+m_b)^2}
\Big(18 T_a^2 m_a^2 m_b^2+8 T_a^2 m_a m_b^3+6 T_a^2 m_b^4+43 T_a T_b m_a^3 m_b \nn\\
&& +24 T_a T_b m_a^2 m_b^2+13 T_a T_b m_a m_b^3+28 T_b^2 m_a^4+22 T_b^2 m_a^3 m_b+10 T_b^2 m_a^2 m_b^2 \Big) ;\nn\\
%===
E_{ab (2)} &=&  \frac{T_b m_b m_a}{560 (T_a m_b+T_b m_a)^3 (m_a+m_b)^2 }
\Big(4 T_a^2 m_a^2 m_b+26 T_a^2 m_a m_b^2+70 T_a^2 m_b^3-5 T_a T_b m_a^3 \nn\\
&& +8 T_a T_b m_a^2 m_b+109 T_a T_b m_a m_b^2+48 T_b^2 m_a^2 m_b \Big).
\end{eqnarray}
The heat flux $\vecX^{(5)}_a$ exchange rates $\vecQ^{(5)}_{a}\,'$ (\ref{eq:Qab5P}) become
\begin{eqnarray}
  U_{ab (2)} &=& \frac{1}{T_a^2 (T_a m_b+T_b m_a)^2 (m_a+m_b)^4}
\Big( 72 T_a^4 m_a^4 m_b^2-24 T_a^4 m_a^3 m_b^3+200 T_a^4 m_a^2 m_b^4-40 T_a^4 m_a m_b^5 \nn\\
&& +48 T_a^4 m_b^6+112 T_a^3 T_b m_a^5 m_b-116 T_a^3 T_b m_a^4 m_b^2+612 T_a^3 T_b m_a^3 m_b^3-380 T_a^3 T_b m_a^2 m_b^4 \nn\\
&& +316 T_a^3 T_b m_a m_b^5+35 T_a^2 T_b^2 m_a^6-112 T_a^2 T_b^2 m_a^5 m_b+606 T_a^2 T_b^2 m_a^4 m_b^2-712 T_a^2 T_b^2 m_a^3 m_b^3 \nn\\
&& +839 T_a^2 T_b^2 m_a^2 m_b^4+224 T_a T_b^3 m_a^5 m_b-352 T_a T_b^3 m_a^4 m_b^2+960 T_a T_b^3 m_a^3 m_b^3+384 T_b^4 m_a^4 m_b^2 \Big);\nn\\
%===
  F_{ab (1)} &=& \frac{1}{5 T_a (T_a m_b+T_b m_a)^3 (m_a+m_b)^4}
\Big(1200 T_a^4 m_a^4 m_b^3+1280 T_a^4 m_a^3 m_b^4+2400 T_a^4 m_a^2 m_b^5+640 T_a^4 m_a m_b^6 \nn\\
&& +240 T_a^4 m_b^7+3180 T_a^3 T_b m_a^5 m_b^2+2624 T_a^3 T_b m_a^4 m_b^3+6936 T_a^3 T_b m_a^3 m_b^4+640 T_a^3 T_b m_a^2 m_b^5 \nn\\
&& +828 T_a^3 T_b m_a m_b^6+2695 T_a^2 T_b^2 m_a^6 m_b+928 T_a^2 T_b^2 m_a^5 m_b^2+6966 T_a^2 T_b^2 m_a^4 m_b^3-1592 T_a^2 T_b^2 m_a^3 m_b^4 \nn\\
&& +1195 T_a^2 T_b^2 m_a^2 m_b^5+700 T_a T_b^3 m_a^7-756 T_a T_b^3 m_a^6 m_b+2844 T_a T_b^3 m_a^5 m_b^2-2524 T_a T_b^3 m_a^4 m_b^3 \nn\\
&& +856 T_a T_b^3 m_a^3 m_b^4-280 T_b^4 m_a^7+504 T_b^4 m_a^6 m_b-872 T_b^4 m_a^5 m_b^2+264 T_b^4 m_a^4 m_b^3\Big);\nn\\
%===
  F_{ab (2)} &=& -\, \frac{m_b}{5 T_a (T_a m_b+T_b m_a)^3 (m_a+m_b)^4 }
\Big(8 T_a^4 m_a^4 m_b^2-160 T_a^4 m_a^3 m_b^3+624 T_a^4 m_a^2 m_b^4-640 T_a^4 m_a m_b^5 \nn\\
&& +488 T_a^4 m_b^6+28 T_a^3 T_b m_a^5 m_b-332 T_a^3 T_b m_a^4 m_b^2+1812 T_a^3 T_b m_a^3 m_b^3-2468 T_a^3 T_b m_a^2 m_b^4 \nn\\
&& +3040 T_a^3 T_b m_a m_b^5+35 T_a^2 T_b^2 m_a^6-112 T_a^2 T_b^2 m_a^5 m_b+1950 T_a^2 T_b^2 m_a^4 m_b^2-2824 T_a^2 T_b^2 m_a^3 m_b^3 \nn\\
&& +6599 T_a^2 T_b^2 m_a^2 m_b^4+672 T_a T_b^3 m_a^5 m_b-1056 T_a T_b^3 m_a^4 m_b^2+5952 T_a T_b^3 m_a^3 m_b^3+1920 T_b^4 m_a^4 m_b^2 \Big);\nn\\
%\end{eqnarray}
%\begin{eqnarray}
%====
  G_{ab (1)} &=& \frac{1}{280 (T_a m_b+T_b m_a)^4 (m_a+m_b)^4}
\Big( 3600 T_a^4 m_a^4 m_b^4+3840 T_a^4 m_a^3 m_b^5+7200 T_a^4 m_a^2 m_b^6+1920 T_a^4 m_a m_b^7 \nn\\
&& +720 T_a^4 m_b^8+12780 T_a^3 T_b m_a^5 m_b^3+12864 T_a^3 T_b m_a^4 m_b^4+26136 T_a^3 T_b m_a^3 m_b^5+5760 T_a^3 T_b m_a^2 m_b^6 \nn\\
&& +2748 T_a^3 T_b m_a m_b^7+16385 T_a^2 T_b^2 m_a^6 m_b^2+14816 T_a^2 T_b^2 m_a^5 m_b^3+34650 T_a^2 T_b^2 m_a^4 m_b^4+5432 T_a^2 T_b^2 m_a^3 m_b^5 \nn\\
&& +3773 T_a^2 T_b^2 m_a^2 m_b^6+8680 T_a T_b^3 m_a^7 m_b+5980 T_a T_b^3 m_a^6 m_b^2+19332 T_a T_b^3 m_a^5 m_b^3+1204 T_a T_b^3 m_a^4 m_b^4 \nn\\
&& +2212 T_a T_b^3 m_a^3 m_b^5+1400 T_b^4 m_a^8-112 T_b^4 m_a^7 m_b+3168 T_b^4 m_a^6 m_b^2 
 -688 T_b^4 m_a^5 m_b^3+392 T_b^4 m_a^4 m_b^4 \Big);\nn
 %===
\end{eqnarray}
\begin{eqnarray}
  G_{ab (2)} &=& \frac{T_b m_a m_b}{ 280 (T_a m_b+T_b m_a)^4 (m_a+m_b)^4 T_a}
\Big(16 T_a^4 m_a^4 m_b^2+136 T_a^4 m_a^3 m_b^3+504 T_a^4 m_a^2 m_b^4-1736 T_a^4 m_a m_b^5 \nn\\
&& +3640 T_a^4 m_b^6+56 T_a^3 T_b m_a^5 m_b+548 T_a^3 T_b m_a^4 m_b^2+2364 T_a^3 T_b m_a^3 m_b^3-3892 T_a^3 T_b m_a^2 m_b^4 \nn\\
&& +17276 T_a^3 T_b m_a m_b^5-35 T_a^2 T_b^2 m_a^6+112 T_a^2 T_b^2 m_a^5 m_b+2082 T_a^2 T_b^2 m_a^4 m_b^2-3512 T_a^2 T_b^2 m_a^3 m_b^3 \nn\\
&& +29113 T_a^2 T_b^2 m_a^2 m_b^4+672 T_a T_b^3 m_a^5 m_b-1056 T_a T_b^3 m_a^4 m_b^2+21312 T_a T_b^3 m_a^3 m_b^3+5760 T_b^4 m_a^4 m_b^2\Big).\nn\\   
\end{eqnarray}
Finally, the scalar $\widetilde{X}^{(4)}_a$ exchange rates $\widetilde{Q}_a^{(4)}\,'$ (\ref{eq:Q4exchange}) become
\begin{eqnarray}
  S_{ab (0)} &=& \frac{4 m_a}{T_a (T_a m_b+T_b m_a) (m_a+m_b)^3}
\Big(6 T_a^2 m_a^2 m_b+4 T_a^2 m_a m_b^2+6 T_a^2 m_b^3+5 T_a T_b m_a^3 \nn\\
&& +2 T_a T_b m_a^2 m_b+13 T_a T_b m_a m_b^2+8 T_b^2 m_a^2 m_b \Big)  ;\nn\\
%===
  S_{ab (1)} &=& \frac{m_a}{30 (m_a+m_b)^3 (T_a m_b+T_b m_a)^3}
\Big(90 T_a^3 m_a^2 m_b^3+60 T_a^3 m_a m_b^4+90 T_a^3 m_b^5+231 T_a^2 T_b m_a^3 m_b^2 \nn\\
&& +132 T_a^2 T_b m_a^2 m_b^3+243 T_a^2 T_b m_a m_b^4-18 T_a^2 T_b m_b^5+184 T_a T_b^2 m_a^4 m_b+69 T_a T_b^2 m_a^3 m_b^2 \nn\\
&& +210 T_a T_b^2 m_a^2 m_b^3-35 T_a T_b^2 m_a m_b^4+40 T_b^3 m_a^5-12 T_b^3 m_a^4 m_b+48 T_b^3 m_a^3 m_b^2-20 T_b^3 m_a^2 m_b^3 \Big);\nn\\
%===
  S_{ab (2)} &=& -\, \frac{m_a^2 T_b^2}{30 T_a^2 (m_a+m_b)^3 (T_a m_b+T_b m_a)^3}
\Big( 2 T_a^3 m_a^3 m_b+12 T_a^3 m_a^2 m_b^2-30 T_a^3 m_a m_b^3+80 T_a^3 m_b^4 \nn\\
&& +5 T_a^2 T_b m_a^4+36 T_a^2 T_b m_a^3 m_b-39 T_a^2 T_b m_a^2 m_b^2+290 T_a^2 T_b m_a m_b^3+15 T_a T_b^2 m_a^4 \nn\\
&& -18 T_a T_b^2 m_a^3 m_b+327 T_a T_b^2 m_a^2 m_b^2+120 T_b^3 m_a^3 m_b \Big).
\end{eqnarray}
Note that the hard sphere mass-ratio coefficients are actually quite more complicated than the Coulomb coefficients.

\newpage
%==================================================================================================================
\subsubsection{Hard spheres collisions (small temperature differences)} \label{sec:Hard-eq}
The model significantly simplifies for small temperature differences, and it is given by 
%\begin{eqnarray}
%  V_{ab (0)} &=& -\,\frac{1}{5};\qquad  V_{ab (3)} = -\,\frac{1}{280};
%\end{eqnarray}
\begin{eqnarray}
  \hat{P}_{ab (1)} &=& - \frac{m_b}{10 (m_a+m_b)};\qquad  \hat{P}_{ab (2)} = -\, \frac{m_a}{10 (m_a+m_b)};\nn\\
%===
  K_{ab(1)} &=& \frac{2 (5 m_a+3 m_b)}{5 m_a}; \qquad  K_{ab(2)} = \frac{4}{5};\nn\\  
L_{ab (1)} &=& -\, \frac{(7 m_a+3 m_b) m_b}{35 m_a (m_a+m_b)} ;\qquad L_{ab (2)} = -\, \frac{4 m_a}{35 (m_a+m_b)}; 
\end{eqnarray}
%The stress tensor $\bPi_a^{(4)}$ exchange rates $\bQ_{a}^{(4)}\,'$ (\ref{eq:Q4abP}) read
\begin{eqnarray}
  M_{ab (1)} &=& \frac{2 (35 m_a^2+63 m_a m_b+24 m_b^2 )}{5 (m_a+m_b)^2 } ;\qquad M_{ab (2)} = \frac{4  m_a (7 m_a+9 m_b)}{5 (m_a+m_b)^2};\nn\\
%===
  N_{ab (1)} &=& \frac{(140 m_a^3+203 m_a^2 m_b+255 m_a m_b^2+72 m_b^3)}{35 (m_a+m_b)^3};\qquad   N_{ab (2)} = -\, \frac{ 4 m_a^2 (7 m_a+37 m_b) }{35 (m_a+m_b)^3};
\end{eqnarray}
%The heat flux $\vecq_a$ exchange rates $\vecQ^{(3)}_{a}\,'$ (\ref{eq:Q3P}) read
\begin{eqnarray}
  U_{ab (1)} &=& -\, \frac{(5 m_a+6 m_b)}{2 (m_a+m_b)};\nn\\
  D_{ab (1)} &=&  \frac{3 (10 m_a^2+7 m_a m_b+6 m_b^2)}{10 (m_a+m_b)^2};\qquad  D_{ab (2)} =  \frac{m_b (5 m_a+32 m_b) }{10 (m_a+m_b)^2 };\nn\\
%===  
  E_{ab (1)} &=&   -\, \frac{3 m_b (28 m_a^2+9 m_a m_b+6 m_b^2)}{560 (m_a+m_b)^3};\qquad E_{ab (2)} = -\, \frac{(m_a-14 m_b) m_a m_b}{112 (m_a+m_b)^3};
\end{eqnarray}
%The heat flux $\vecX^{(5)}_a$ exchange rates $\vecQ^{(5)}_{a}\,'$ (\ref{eq:Qab5P}) become
\begin{eqnarray}
  U_{ab (2)} &=& \frac{(35 m_a^2+84 m_a m_b+48 m_b^2)}{(m_a+m_b)^2};\nn\\
%===
  F_{ab (1)} &=& \frac{(420 m_a^3+763 m_a^2 m_b+508 m_a m_b^2+240 m_b^3)}{5 (m_a+m_b)^3};\nn\\
%===
  F_{ab (2)} &=& -\, \frac{m_b (35 m_a^2+448 m_a m_b+488 m_b^2)}{5 (m_a+m_b)^3};\nn\\
%\end{eqnarray}
%\begin{eqnarray}
%====
  G_{ab (1)} &=& \frac{(1400 m_a^4+2968 m_a^3 m_b+5261 m_a^2 m_b^2+1788 m_a m_b^3+720 m_b^4)}{280 (m_a+m_b)^4};\nn \\
 %===
  G_{ab (2)} &=& -\, \frac{m_a  m_b (m_a^2-28 m_a m_b-104 m_b^2)}{8 (m_a+m_b)^4};   
\end{eqnarray}
%Finally, the scalar $\widetilde{X}^{(4)}_a$ exchange rates $\widetilde{Q}_a^{(4)}\,'$ (\ref{eq:Q4exchange}) become
\begin{eqnarray}
  S_{ab (0)} &=& \frac{4 m_a (5 m_a+6 m_b)}{(m_a+m_b)^2}  ;\nn\\
%===
  S_{ab (1)} &=& \frac{2 m_a (10 m_a^2+13 m_a m_b+18 m_b^2)}{15 (m_a+m_b)^3};\qquad
%===
  S_{ab (2)} = -\, \frac{2 m_a^2 (m_a+4 m_b)}{3 (m_a+m_b)^3}.
\end{eqnarray}

\newpage
%==================================================================================================================
%==================================================================================================================
\section{Self-collisions (only one species)} \label{sec:ions}
\subsection{Viscosity-tensor  \texorpdfstring{$\bPi^{(2)}_a$}{Pi(2)} (self-collisions)}
 Here we consider a ``simple gas'', where only the self-collisions ``a-a'' are retained
and collisions with other species are neglected, analogously to the Braginskii ion species.
From the collisional contributions (\ref{eq:QabP}) and (\ref{eq:Q4abP}), the evolution equations for the stress-tensors read
\begin{eqnarray}
   \frac{d_a}{dt} \bPi^{(2)}_a  +\Omega_a \big(\bhat\times \bPi^{(2)}_a \big)^S + p_a \bW_a
  &=&  -\frac{3}{5}\nu_{aa} \Omega_{22} \bPi_a^{(2)}
  + \nu_{aa} \Big(\frac{3}{20} \Omega_{22} - \frac{3}{70} \Omega_{23} \Big)
  \Big( \frac{\rho_a}{p_a}\bPi^{(4)}_a- 7 \bPi^{(2)}_a\Big); \label{eq:Energy32}\\
   \frac{d_a}{dt} \bPi^{(4)}_a  +\Omega_a \big(\bhat\times \bPi^{(4)}_a \big)^S + 7 \frac{p_a^2}{\rho_a} \bW_a 
  &=&  -  \nu_{aa} \big( \frac{21}{10}\Omega_{22} +\frac{3}{5}\Omega_{23}\big) \frac{p_a}{\rho_a}\bPi_a^{(2)} \nn\\
 &&  - \nu_{aa} \big( \frac{1}{40} \Omega_{22}+ \frac{3}{70} \Omega_{24} \big)  \Big( \bPi^{(4)}_a- 7 \frac{p_a}{\rho_a} \bPi^{(2)}_a\Big). \label{eq:Energy31}
\end{eqnarray}
%===
As a quick double check, prescribing Coulomb collisions (with $\Omega_{22}=2$, $\Omega_{23}=4$, $\Omega_{24}=12$) recovers
equations (67)-(68) of Part 1. Neglecting the entire evolution equation  (\ref{eq:Energy31}) and also the last term of (\ref{eq:Energy32})
(with a closure $\bPi_a^{(4)}=7 (p_a/\rho_a)\bPi_a^{(2)}$),  yields the 1-Hermite approximation of \cite{Schunk1977,Schunk1975} and \cite{Burgers1969},
further discussed below in Section \ref{sec:1Hself}.  Here in the 2-Hermite approximation, the equations (\ref{eq:Energy32})-(\ref{eq:Energy31}) can be slightly re-arranged into
\begin{eqnarray}
   \frac{d_a}{dt} \bPi^{(2)}_a  +\Omega_a \big(\bhat\times \bPi^{(2)}_a \big)^S + p_a \bW_a
  &=&  - \nu_{aa} \Big(\frac{33}{20} \Omega_{22} -\frac{3}{10}\Omega_{23}\Big) \bPi_a^{(2)}
  + \nu_{aa} \Big(\frac{3}{20} \Omega_{22} - \frac{3}{70} \Omega_{23} \Big)
  \frac{\rho_a}{p_a}\bPi^{(4)}_a; \label{eq:Energy32B}\\
   \frac{d_a}{dt} \bPi^{(4)}_a  +\Omega_a \big(\bhat\times \bPi^{(4)}_a \big)^S + 7 \frac{p_a^2}{\rho_a} \bW_a 
  &=&  -  \nu_{aa} \Big( \frac{77}{40}\Omega_{22} +\frac{3}{5}\Omega_{23} - \frac{3}{10}\Omega_{24} \Big)\frac{p_a}{\rho_a}\bPi_a^{(2)} 
   - \nu_{aa} \Big( \frac{1}{40} \Omega_{22}+ \frac{3}{70} \Omega_{24} \Big)   \bPi^{(4)}_a, \nn
\end{eqnarray}
which can be compared with eqs. (69) of Part 1.
Then, prescribing the quasi-static/highly-collisional approximation
(by canceling the time-derivative $d_a/dt$) and solving the coupled system,
yields the stress-tensor $\bPi^{(2)}_a$ in the usual Braginskii form 
\begin{eqnarray}
  \bPi^{(2)}_a &=& -\eta_0^a \bW_0 -\eta_1^a\bW_1 -\eta_2^a\bW_2 +\eta_3^a\bW_3+\eta_4^a\bW_4; \label{eq:Energy081}\\
  \bW_0 &=& \frac{3}{2}\big(\bW_a:\bhat\bhat\big) \Big( \bhat\bhat-\frac{\bI}{3}\Big);\nn\\
  \bW_1 &=& \bI_\perp \cdot\bW_a\cdot\bI_\perp +\frac{1}{2}\big( \bW_a:\bhat\bhat\big) \bI_\perp;\qquad
  \bW_2 = \big( \bI_\perp\cdot\bW_a\cdot\bhat\bhat\big)^S;\nn\\
  \bW_3 &=& \frac{1}{2}\big( \bhat\times \bW_a\cdot\bI_\perp\big)^S;\qquad 
  \bW_4 = \big(\bhat\times\bW_a\cdot\bhat\bhat\big)^S, \nn
\end{eqnarray}
and now the viscosity coefficients read
\begin{eqnarray}
   \eta_0^a  &=&   \frac{5}{6} \frac{(301\Omega_{22}-84 \Omega_{23}+12 \Omega_{24})}{(77 \Omega_{22}^2
    +6 \Omega_{22}\Omega_{24}-6 \Omega_{23}^2)}  \frac{p_a}{\nu_{aa}}; \label{eq:beau033M} \\
  %===============
   \eta_2^a  &=& \frac{p_{a}}{\nu_{aa}\Delta}\, \Big\{
  \frac{3}{5}\Omega_{22} x^2
  +\frac{3}{196000} \big(301 \Omega_{22} -84 \Omega_{23}+ 12 \Omega_{24} \big) \big(77 \Omega_{22}^2 + 6 \Omega_{22}\Omega_{24} -6 \Omega_{23}^2 \big) \Big\}; \nn\\
   \eta_4^a  &=& \frac{p_{a}}{\nu_{aa} \Delta}\,
  \Big\{ x^3 + x \Big[\frac{2353}{1600}\Omega_{22}^2- \frac{33}{40}\Omega_{22}\Omega_{23}  + \frac{129}{1400}\Omega_{22}\Omega_{24} 
     +\frac{81}{700} \Omega_{23}^2 - \frac{9}{350} \Omega_{23} \Omega_{24} + \frac{9}{4900}\Omega_{24}^2 \Big]  \Big\}; \nn\\
  \Delta &=& x^4 + x^2 \Big[\frac{3433}{1600}\Omega_{22}^2 - \frac{201}{200} \Omega_{22}\Omega_{23} + \frac{129}{1400}\Omega_{22}\Omega_{24} 
     + \frac{99}{700}\Omega_{23}^2 - \frac{9}{350} \Omega_{23}\Omega_{24} + \frac{9}{4900} \Omega_{24}^2 \Big] \nn\\
  && \qquad + \Big(\frac{3}{700} \Big)^2 \Big[ 77 \Omega_{22}^2+6\Omega_{22}\Omega_{24} -6 \Omega_{23}^2 \Big]^2. \label{eq:beau033}
\end{eqnarray}
As before, the parameter which Braginskii uses to describe the strength of the magnetic field
(sometimes called the Hall parameter) $x=\Omega_a/\nu_{aa}$ is present
and the usual relations for viscosities hold as well,  $\eta_1^a(x)=\eta_2^a(2x)$; $\eta_3^a(x)=\eta_4^a(2x)$.

 Results (\ref{eq:Energy081})-(\ref{eq:beau033}) represent the Braginskii ion stress-tensor expressed through the Chapman-Cowling integrals, which enter through
  the ratios $\Omega_{l,j}$ and through the collisional frequencies $\nu_{aa}$. 
The parallel viscosity $\eta_0^a$ (\ref{eq:beau033M}) can be also used 
in the unmagnetized case (with solution $\bPi^{(2)}_a = -\eta_0^a \bW_a$) and it therefore has a general validity for a large class of self-collisional processes.
In contrast, the magnetized viscosities $\eta_1^a-\eta_4^a$ are valid only for Coulomb collisions and to obtain a more general result, one should obtain the quasi-static
approximation by considering two species of charged particles and neutral particles, where the results will naturally get more complicated
(see Appendix \ref{sec:AppIN}).

The Braginskii ion stress-tensor is recovered by prescribing Coulomb collisions ($\Omega_{22}=2$, $\Omega_{23}=4$ and $\Omega_{24}=12$) in (\ref{eq:beau033M})-(\ref{eq:beau033})
and by associating the $\nu_{aa}$ with the Coulomb collisional frequency, yielding
\begin{eqnarray}
  \textrm{Coulomb collisions}, (\ln\Lambda\gg 1): \qquad \eta_0^a  &=& \frac{1025}{1068}\, \frac{p_{a}}{\nu_{aa}}; \qquad \quad
   \Delta = x^4 + \frac{79321}{19600} x^2 + \Big(\frac{267}{175}\Big)^2; \nn\\
  \eta_2^a  &=& \frac{p_{a}}{\nu_{aa}\Delta}\, \Big( \frac{6}{5}x^2 + \frac{10947}{4900} \Big);\nn\\
  \eta_4^a  &=& \frac{p_{a}}{\nu_{aa}\Delta}\, \Big( x^3 + \frac{46561}{19600} x \Big),\label{eq:beau37}
\end{eqnarray}
recovering the analytic viscosities eq. (73) in Part 1, or the numerical viscosities eq. (4.44) in \cite{Braginskii1965}.  
This is a useful re-derivation of the Braginskii model directly through the Boltzmann operator.

Generalization to moderatelly-coupled plasmas reads
\begin{eqnarray}
  \textrm{Coulomb collisions}: \qquad \eta_0^a  &=& \frac{1025}{534} \frac{A_1(2)}{A_2(2)} \frac{p_a}{\nu_{aa}}; \qquad \quad
   \Delta = x^4 + \frac{79321}{78400} \Big(\frac{A_2(2)}{A_1(2)}\Big)^2 x^2 + \Big(\frac{267}{700}\Big)^2 \Big(\frac{A_2(2)}{A_1(2)}\Big)^4; \nn\\
  \eta_2^a  &=& \frac{p_{a}}{\nu_{aa}\Delta}\, \Big[ \frac{3}{5} \frac{A_2(2)}{A_1(2)} x^2 + \frac{10947}{39200}\Big(\frac{A_2(2)}{A_1(2)}\Big)^3  \Big];\nn\\
  \eta_4^a  &=& \frac{p_{a}}{\nu_{aa}\Delta}\, \Big[ x^3 + \frac{46561}{78400} \Big(\frac{A_2(2)}{A_1(2)}\Big)^2 x \Big],\label{eq:beau37M}
\end{eqnarray}
where as a reminder, the corrections of the Coulomb logarithm are given by
\begin{equation}
  A_1(2) = \ln(\Lambda^2+1);  \qquad A_2(2) =  2\ln(\Lambda^2+1)-2+\frac{2}{\Lambda^2+1}. \label{eq:A21U}
\end{equation}
The Braginskii case for weakly-coupled plasmas (\ref{eq:beau37}) is recovered by $A_2(2)/A_1(2)=2$. 
Note that the definition of the collisional frequency $\nu_{aa}$ contains the $A_1(2)$ coefficient as well, see eq. (\ref{eq:Rsimple2x}),  
so if the definition of $\nu_{aa}$ is used in (\ref{eq:beau37M}), the $A_1(2)$ coefficient cancels out and only the $A_2(2)$ coefficient remains.

Even better use of (\ref{eq:beau033}) would be to evaluate it with the Debye screened potential, but we 
do not provide the Chapman-Cowling integrals for this case (see Section \ref{sec:Other}, asymptotic limit for large temperatures can be found in \cite{Kihara1959}).
Obviously, expressing the entire magnetized Braginskii model through the Chapman-Cowling integrals is useful and physically meaningful,
even if one is interested only in the Coulomb collisions.

Now, we should continue by evaluating only the parallel viscosity for the inverse power-law force $F_{ab}=\pm |K_{ab}|/r^\nu$ (where the attractive force has a repulsive core).
Nevertheless, it feels slightly boring not to evaluate the magnetized viscosities as well, so we will evaluate them anyway, yielding
\begin{eqnarray}
  \eta_0^a  &=&  \frac{5}{6} \frac{(205\nu^2 -458 \nu+301)}{(101\nu-113)(3\nu-5)} \frac{A_1(\nu)}{A_2(\nu)} \frac{p_a}{\nu_{aa}}; \label{eq:beau38M}\\
  \eta_2^a  &=& \frac{p_{a}}{\nu_{aa}\Delta}\,
  \Big\{ \frac{3}{5} \frac{A_2(\nu)}{A_1(\nu)} \frac{(3\nu-5)}{(\nu-1)} x^2
  + \Big(\frac{A_2(\nu)}{A_1(\nu)} \Big)^3 \frac{3}{196000} \frac{(101\nu-113)(3 \nu-5)^3}{(\nu-1)^6} (205\nu^2-458\nu+301)  \Big\};\nn\\
  \eta_4^a  &=& \frac{p_{a}}{\nu_{aa}\Delta}\, \Big\{ x^3 + \frac{x}{78400} \Big(\frac{A_2(\nu)}{A_1(\nu)} \Big)^2
  \frac{(3\nu-5)^2}{(\nu-1)^6} \Big( 42529 \nu^4-193828 \nu^3 \nn\\
  && \qquad +356358 \nu^2-305956 \nu+103201 \Big)  \Big\};\nn\\
  \Delta &=& x^4
  +  \frac{x^2}{78400} \Big(\frac{A_2(\nu)}{A_1(\nu)} \Big)^2 \frac{(3 \nu-5)^2}{(\nu-1)^6} \Big( 71257 \nu^4 -312772 \nu^3 +548886 \nu^2 -449092 \nu+144025 \Big) \nn\\
 && + \Big(\frac{3}{700}\Big)^2 \Big(\frac{A_2(\nu)}{A_1(\nu)} \Big)^4 \, \frac{(101 \nu-113)^2 (3\nu-5)^4}{(\nu-1)^6}. \label{eq:beau38}
\end{eqnarray}
  Here, the parallel viscosity (\ref{eq:beau38M}) is valid for any power-law index $\nu$. 
   The magnetized viscosities are unfortunatelly valid only for $\nu=2$ (and to get more general results one should consider coupling between charged and neutral particles),
   nevertheless, these expressions are at least useful as an a posteriori double-check that the models are formulated correctly.

  Finally, the case of hard spheres is obtained by prescribing limit $\nu\to\infty$ together with $A_2(\infty)/A_1(\infty)=2/3$ in (\ref{eq:beau38M}),
   or equivalently, prescribing $\Omega_{22}=2$; $\Omega_{23}=8$; $\Omega_{24}=40$ in (\ref{eq:beau033M}). 
   Out of curiosity, let us consider the slightly academic case of generalized ``hard spheres'' discussed in Section \ref{sec:HardSphereP}
   (that have a non-zero cyclotron frequency and feel the magnetic field) and evaluate the magnetized viscosities as well, yielding
\begin{eqnarray}
  \textrm{``Hard spheres'':}\qquad \eta_0^a  &=& \frac{1025}{1212} \frac{p_{a}}{\nu_{aa}}; \qquad \quad
   \Delta = x^4 +\frac{71257}{19600}x^2 + \Big( \frac{303}{175} \Big)^2 ;\nn\\
  \eta_2^a  &=& \frac{p_{a}}{\nu_{aa}\Delta}\, \Big( \frac{6}{5}x^2 + \frac{12423}{4900} \Big);\nn\\
  \eta_4^a  &=& \frac{p_{a}}{\nu_{aa}\Delta}\, \Big( x^3+\frac{42529}{19600}x \Big). \label{eq:beau36}
\end{eqnarray}  
Note the perhaps surprising numerical similarities between the Coulomb collisions and hard spheres (\ref{eq:beau37})-(\ref{eq:beau36}).

%=========================================================
\subsubsection{Reduction into 1-Hermite approximation} \label{sec:1Hself}
In the 1-Hermite approximation, the stress-tensor evolves according to 
\begin{equation}
\frac{d_a}{dt} \bPi^{(2)}_a + \Omega_a \big(\bhat\times\bPi^{(2)}_a \big)^S +p_a \bW_a = -\frac{3}{5}\nu_{aa} \Omega_{22} \bPi_a^{(2)}. \label{eq:perfect4G}
\end{equation}
By applying the quasi-static approximation, the stress-tensor has the same form (\ref{eq:Energy081}), but now with the 1-Hermite viscosities
\begin{eqnarray}
  \big[ \eta_0^a \big]_1 &=&  \frac{5}{3 \Omega_{22} } \frac{p_a}{\nu_{aa}};\qquad
  \big[\eta_2^a \big]_1 = \frac{p_{a}}{\nu_{aa}}\,\frac{ 3\Omega_{22}/5}{x^2 + (3\Omega_{22}/5)^2}; \qquad
  \big[\eta_4^a \big]_1 = \frac{p_{a}}{\nu_{aa}}\, \frac{x}{x^2 + (3\Omega_{22}/5)^2}. \label{eq:pica33}
\end{eqnarray}
To emphasize that the results (\ref{eq:pica33}) represent viscosities in the simplified 1-Hermite approximation, we have added
the brackets $[\ldots]_1$ around the viscosity coefficients. Similarly, the previously given viscosities in the 2-Hermite approximation
(\ref{eq:beau033M})-(\ref{eq:beau36}) can be denoted by putting the brackets $[\ldots]_2$ around them. 
We use this notation only in the particular sub-sections, where the comparison to the 1-Hermite approximation is made and otherwise the 
2-Hermite designation $[\ldots]_2$ is ommited. 
An analogous notation is used by \cite{ChapmanCowling1953} to describe their ``first approximation'' and ``second approximation''.

Note that for both the Coulomb collisions and the hard spheres the $\Omega_{22}=2$, so the entire 1-Hermite stress-tensor (\ref{eq:pica33})
is identical for both cases (and only the collisional frequencies are different). 
Also note that in the limit of strong magnetic field ($x\gg 1$), the 2-Hermite perpendicular viscosities $\eta_2^a$ and
gyroviscosities $\eta_4^a$ (\ref{eq:beau033}) are identical to the 1-Hermite results (\ref{eq:pica33})
\begin{eqnarray}
  \textrm{Strong B-field:} \qquad \big[\eta_2^a \big]_2 &=& \big[\eta_2^a \big]_1 = \frac{3\Omega_{22}}{5} \frac{p_a \nu_{aa}}{\Omega_a^2} ; \qquad
   \big[\eta_4^a \big]_2 = \big[\eta_4^a \big]_1 = \frac{p_a}{\Omega_a},
\end{eqnarray}  
and only the parallel viscosities $\eta_0^a$ remain different.

%\subsubsection{Improvement of 2-Hermite approximation for parallel viscosities}
For the parallel viscosities, the improvement of the 2-Hermite
approximation with respect to the 1-Hermite approximation can be written in the following form (valid for any collisional process)
\begin{eqnarray}
\big[ \eta_0^a \big]_2 &=& \big[ \eta_0^a \big]_1
  \Big( 1+ \frac{3}{2} \frac{(7 \Omega_{22} -2 \Omega_{23})^2}{(77\Omega_{22}^2 +6 \Omega_{22} \Omega_{24} -6 \Omega_{23}^2)}\Big), \label{eq:RisY}
\end{eqnarray}
and the result does not contain the collisional frequency $\nu_{aa}$.
Evaluating the result (\ref{eq:RisY}) for our collisional forces then yields
\begin{eqnarray}
&&  \textrm{Coulomb collisions:} \qquad \big[ \eta_0^a \big]_2 = \big[ \eta_0^a \big]_1 \Big( 1+ \frac{27}{178} \Big) ;\nn\\  
&&  \textrm{Hard spheres:} \qquad \qquad \,\, \big[ \eta_0^a \big]_2 =  \big[ \eta_0^a \big]_1 \Big( 1+ \frac{3}{202}\Big);\nn\\
&&  \textrm{Inverse power:} \qquad \qquad \, \big[ \eta_0^a \big]_2 = 
  \big[ \eta_0^a \big]_1 \Big(1+  \frac{3(\nu-5)^2}{2(\nu-1)(101\nu-113)} \Big). \label{eq:CC0}
\end{eqnarray}
Interestingly, the equation (\ref{eq:CC0}) for the inverse power-law force can be found in \cite{ChapmanCowling1953}, p. 173,
implying that for the self-collisions, our 2-Hermite approximation is identical to the ``second approximation'' of the last reference.
%However, this is only true for the case of the self-collisions considered in this Section and the comparison with Chapman-Cowling
%is not clear if different species are considered,
%where our model should somewhat analogous to the ``third approximation'' of the last reference. 

An interested reader may plot the correction ratio given by (\ref{eq:CC0}) with respect to $\nu$. Considering only $\nu\ge 2$,
the correction ratio  is always non-negative and the largest correction (of $\sim 15\%$) is
indeed obtained for the case of the Coulomb collisions (i.e. the Braginskii case).
The correction ratio then sharply decreases (already for $\nu=3$ the correction is only 3/190) and 
becomes identicaly zero for the case of the Maxwell molecules ($\nu=5$) and then again slowly increases until the case of 
the hard spheres ($\nu=\infty$) is reached, with a small correction of only $\sim 1.5\%$.
Interestingly, the result (\ref{eq:CC0}) also implies that Chapman-Cowling essentially did know the 2-Hermite parallel ion viscosity much before Braginskii,
and in a fully analytic form.
The same conclusion will be reached for the parallel 2-Hermite thermal conductivity of the ion species, given later by (\ref{eq:CC00}).

%\newpage
\subsection{Higher-order viscosity-tensor  \texorpdfstring{$\bPi^{(4)}_a$}{Pi(4)} (self-collisions)} \label{sec:Pi4self}
The magnetized solution has a general form
\begin{equation} \label{eq:Energy88}
  \bPi^{(4)}_a = \frac{p_a}{\rho_a}\Big[-\eta_0^{a(4)} \bW_0 -\eta_1^{a(4)}\bW_1 -\eta_2^{a(4)}\bW_2 +\eta_3^{a(4)}\bW_3+\eta_4^{a(4)}\bW_4\Big],
\end{equation}
and for the unmagnetized case $\bPi^{(4)}_a = -\frac{p_a}{\rho_a}\eta_0^{a(4)} \bW_a$. 
The viscosities (of the 4th-order fluid moment) read
\begin{eqnarray}
  \eta_0^{a(4)} &=& \frac{p_a}{\nu_{aa}} \frac{35}{6} \frac{(385 \Omega_{22} -108 \Omega_{23}+12 \Omega_{24})}{(77\Omega_{22}^2 +6 \Omega_{22}\Omega_{24} -6 \Omega_{23}^2)};\nn\\
  \eta_2^{a(4)}  &=& \frac{p_{a}}{\nu_{aa} \Delta}\, \Big\{  \frac{3}{5} \Big( \frac{7}{2}\Omega_{22} + \Omega_{23} \Big)x^2
  + \frac{3}{28000}\big( 385 \Omega_{22} -108 \Omega_{23}+12\Omega_{24} \big) \big( 77 \Omega_{22}^2 +6 \Omega_{22}\Omega_{24} -6 \Omega_{23}^2 \big) \Big\};\nn\\
\eta_4^{a(4)}  &=& \frac{p_{a}}{\nu_{aa} \Delta}\, \Big\{ 7 x^3+ x \Big[ \frac{22099}{1600}\Omega_{22}^2- \frac{741}{100)}\Omega_{22}\Omega_{23} +\frac{147}{200}\Omega_{22}\Omega_{24} 
    +\frac{99}{100} \Omega_{23}^2 - \frac{36}{175}\Omega_{23}\Omega_{24}+\frac{9}{700}\Omega_{24}^2 \Big] \Big\};\nn\\
%===
 \Delta &=& x^4 + x^2 \Big[\frac{3433}{1600}\Omega_{22}^2 - \frac{201}{200} \Omega_{22}\Omega_{23} + \frac{129}{1400}\Omega_{22}\Omega_{24} 
     + \frac{99}{700}\Omega_{23}^2 - \frac{9}{350} \Omega_{23}\Omega_{24} + \frac{9}{4900} \Omega_{24}^2 \Big] \nn\\
  && \qquad + \Big(\frac{3}{700} \Big)^2 \Big[ 77 \Omega_{22}^2+6\Omega_{22}\Omega_{24} -6 \Omega_{23}^2 \Big]^2,
\end{eqnarray}
where the denominator $\Delta$ is the same as for the $\bPi^{(2)}_a$ in (\ref{eq:beau033}). Evaluation for the Coulomb collisions yields
\begin{eqnarray}
  \textrm{Coulomb collisions:}\qquad
  \eta_0^{a(4)} &=& \frac{8435}{1068} \frac{p_a}{\nu_{aa}}; \qquad \Delta = x^4 + (79321/19600) x^2 + (267/175)^2;  \nn\\
  \eta_2^{a(4)}  &=& \frac{p_{a}}{\nu_{aa}\Delta}\, \Big( (33/5)x^2+ (64347/3500)\Big);\nn\\
  \eta_4^{a(4)}  &=& \frac{p_{a}}{\nu_{aa}\Delta}\, \Big( 7x^3 +(59989/2800)x \Big),  \label{eq:cuc1}
\end{eqnarray}
recovering eq. (76) of Part 1. Evaluation for the generalized ``hard spheres'' yields
\begin{eqnarray}
  \textrm{``Hard spheres'':}\qquad
  \eta_0^{a(4)}  &=&  \frac{6755}{1212} \frac{p_a}{\nu_{aa}}; \qquad  \Delta = x^4 +(71257/19600)x^2 + (303/175)^2;   \nn\\
  \eta_2^{a(4)}  &=& \frac{p_{a}}{\nu_{aa}\Delta}\, \Big( 9x^2+(58479/3500) \Big);\nn\\
  \eta_4^{a(4)} &=& \frac{p_{a}}{\nu_{aa}\Delta}\, \Big( 7x^3+(38053/2800) x \Big).  \label{eq:cuc2} 
\end{eqnarray}
Note the numerical similarities between the (\ref{eq:cuc1}) and (\ref{eq:cuc2}). Finally, for the inverse power-law force
\begin{eqnarray}
  \textrm{Inverse force:}\qquad
  \eta_0^{a(4)}   &=&    \frac{35}{6} \frac{(193\nu^2-386\nu+241)}{(101\nu-113)(3\nu-5)} \frac{A_1(\nu)}{A_2(\nu)} \frac{p_a}{\nu_{aa}};\nn\\
  \eta_2^{a(4)}  &=& \frac{p_{a}}{\nu_{aa}\Delta}\,
  \Big\{
  \frac{A_2(\nu)}{A_1(\nu)} \frac{3}{10} \frac{(3\nu-5)(15\nu-19)}{(\nu-1)^2 } x^2 \nn\\
  && \qquad + \Big(\frac{A_2(\nu)}{A_1(\nu)}\Big)^3 \frac{3}{28000} \frac{(101\nu-113)(3\nu-5)^3}{(\nu-1)^6} (193\nu^2-386\nu+241) \Big\};\nn\\
  %=====
 \eta_4^{a(4)}  &=& \frac{p_{a}}{\nu_{aa}\Delta}\,
 \Big\{ 7x^3
 + x \Big(\frac{A_2(\nu)}{A_1(\nu)}\Big)^2 \frac{(3\nu-5)^2}{11200(\nu-1)^6} \Big( 38053\nu^4 -157444 \nu^3 \nn\\
 && \qquad +271182 \nu^2 -224548 \nu+75061 \Big) \Big\},
\end{eqnarray}
where the $\Delta$ is equal to (\ref{eq:beau38}). As a quick double-check, in the limit of weak magnetic field ($x\ll 1$) the
perpendicular viscosity $\eta_2^{a(4)}$ converges to the parallel viscosity $\eta_0^{a(4)}$, as it should.

%\newpage
%====================================================================================================
%====================================================================================================
\subsection{Heat flux  \texorpdfstring{$\vecq_a$}{q} (self-collisions)}
 For the heat flux vectors, the collisional contributions are given by (\ref{eq:Q3P}) and (\ref{eq:Qab5P}) and considering only self-collisions,
    the evolution equations read
\begin{eqnarray}
    \frac{d_a}{d t}\vecq_a + \Omega_a \bhat\times\vecq_a + \frac{5}{2}p_a \nabla \Big(\frac{p_a}{\rho_a}\Big) &=& - \frac{2}{5}\nu_{aa}\Omega_{22}\vecq_a
  + \nu_{aa} \Big(\frac{1}{20}\Omega_{22}- \frac{1}{70}\Omega_{23}\Big) \Big(\frac{\rho_a}{p_a}\vecX^{(5)}_a -28 \vecq_a\Big); \label{eq:Excite1}\\
  \frac{d_a}{d t}\vecX^{(5)}_a +\Omega_a\bhat\times\vecX^{(5)}_a +70\frac{p_a^2}{\rho_a}\nabla\Big(\frac{p_a}{\rho_a}\Big) 
  &=& - \nu_{aa} \Big( \frac{8}{5}\Omega_{23}+ \frac{28}{5}\Omega_{22} \Big) \frac{p_a}{\rho_a}\vecq_a \nn\\
&&  - \nu_{aa} \Big(\frac{2}{35}\Omega_{24} - \frac{3}{10}\Omega_{22} \Big) \Big(\vecX^{(5)}_a -28 \frac{p_a}{\rho_a} \vecq_a\Big).\label{eq:Excite2}
\end{eqnarray}
Neglecting the entire (\ref{eq:Excite2}) and also the last term of (\ref{eq:Excite1}), with a closure $\vecX^{(5)}_a=28 \frac{p_a}{\rho_a} \vecq_a$,
  yields the 1-Hermite approximation of \cite{Schunk1977,Schunk1975} and \cite{Burgers1969}, discussed further below in Section \ref{sec:1HermiteQa}. Here with the
  2-Hermite approximation, the equations can be slightly re-arranged into
\begin{eqnarray}
  \frac{d_a}{d t}\vecq_a + \Omega_a \bhat\times\vecq_a + \frac{5}{2}p_a \nabla \Big(\frac{p_a}{\rho_a}\Big) &=&
  - \nu_{aa} \Big(\frac{9}{5}\Omega_{22}-\frac{2}{5}\Omega_{23}\Big) \vecq_a
  + \nu_{aa} \Big(\frac{1}{20}\Omega_{22}- \frac{1}{70}\Omega_{23}\Big) \frac{\rho_a}{p_a}\vecX^{(5)}_a ; \label{eq:Excite3}\\
  \frac{d_a}{d t}\vecX^{(5)}_a +\Omega_a\bhat\times\vecX^{(5)}_a +70\frac{p_a^2}{\rho_a}\nabla\Big(\frac{p_a}{\rho_a}\Big) 
  &=& - \nu_{aa} \Big( \frac{8}{5}\Omega_{23} + 14\Omega_{22} - \frac{8}{5}\Omega_{24} \Big) \frac{p_a}{\rho_a}\vecq_a \nn\\
&&  - \nu_{aa} \Big(\frac{2}{35}\Omega_{24} - \frac{3}{10}\Omega_{22} \Big) \vecX^{(5)}_a.\label{eq:Excite4}
\end{eqnarray}
As a quick double-check, prescribing Coulomb collisions recovers eqs. (39)-(41) of Part 1.
Prescribing the quasi-static approximation (cancelling the $d_a/dt$) then yields the heat flux
\begin{equation} \label{eq:Braginskii_YES}
\vecq_a = -\kappa_\parallel^a \nabla_\parallel T_a - \kappa_\perp^a \nabla_\perp T_a + \kappa_\times^a \bhat\times\nabla T_a,
\end{equation}
with the thermal conductivities
\begin{eqnarray}
  \kappa_\parallel^a &=&  \frac{25 (77 \Omega_{22} - 28 \Omega_{23} +4 \Omega_{24})}{16 (7 \Omega_{22}^2 + \Omega_{22} \Omega_{24} - \Omega_{23}^2)} \, \frac{p_a}{\nu_{aa} m_a};
  \label{eq:Thierry51H}\\
\kappa_\perp^a &=& \frac{p_a}{\nu_{aa} m_a \Delta} \Big[ \Omega_{22} x^2+ \frac{1}{1225} \big(7 \Omega_{22}^2+\Omega_{22} \Omega_{24}-\Omega_{23}^2 \big)
  \big(77 \Omega_{22} -28 \Omega_{23} +4 \Omega_{24}  \big) \Big] ;\nn\\
\kappa_\times^a &=& \frac{p_a}{\nu_{aa} m_a \Delta} \, \Big[
  \frac{5}{2} x^3+ x \Big( \frac{149}{40}\Omega_{22}^2 - \frac{13}{5}\Omega_{22}\Omega_{23} + \frac{11}{35} \Omega_{22}\Omega_{24}
  + \frac{16}{35}\Omega_{23}^2 - \frac{4}{35} \Omega_{23}\Omega_{24} + \frac{2}{245}\Omega_{24}^2 \Big) \Big]; \nn\\
\Delta &=& x^4+ x^2 \Big[\frac{193}{100}\Omega_{22}^2- \frac{6}{5} \Omega_{22} \Omega_{23} + \frac{22}{175} \Omega_{22} \Omega_{24}
  + \frac{36}{175} \Omega_{23}^2 - \frac{8}{175} \Omega_{23}\Omega_{24} + \frac{4}{1225} \Omega_{24}^2 \Big] \nn\\
  && +\Big[\frac{4}{175} \big( 7 \Omega_{22}^2+\Omega_{22} \Omega_{24}-\Omega_{23}^2 \big) \Big]^2 ,
\label{eq:Thierry51}
\end{eqnarray}
where again $x=\Omega_a/\nu_{aa}$. The result (\ref{eq:Thierry51H}) represents parallel thermal conductivity for a general collisional process.
Prescribing Coulomb collisions ($\Omega_{22}=2$, $\Omega_{23}=4$, $\Omega_{24}=12$) yields
%=====
\begin{eqnarray}
\textrm{Coulomb collisions}, (\ln\Lambda\gg1): \qquad \kappa_\parallel^a &=&  \underbrace{\frac{125}{32}}_{3.906} \, \frac{p_a}{\nu_{aa} m_a}; \qquad
\kappa_\perp^a = \frac{p_a}{\nu_{aa} m_a \Delta} \Big[ 2 x^2 + \frac{648}{245} \Big] ; \label{eq:Thierry51B}\\
\kappa_\times^a &=& \frac{p_a}{\nu_{aa} m_a \Delta} \, \Big[   \frac{5}{2} x^3+  \frac{2277}{490} x \Big]; 
 \qquad  \Delta = x^4+  \frac{3313}{1225} x^2 +\Big(\frac{144}{175}\Big)^2, \nn
\end{eqnarray}
recovering eq. (43) of Part 1 and eq. (4.40) of \cite{Braginskii1965}.

Generalization to moderatelly-coupled plasmas reads
%=====
\begin{eqnarray}
\textrm{Coulomb collisions}: \qquad \kappa_\parallel^a &=&  \frac{125}{16} \frac{A_1(2)}{A_2(2)} \, \frac{p_a}{\nu_{aa} m_a}; \qquad
\kappa_\perp^a = \frac{p_a}{\nu_{aa} m_a \Delta} \Big[ \frac{A_2(2)}{A_1(2)} x^2  + \frac{81}{245}\Big(\frac{A_2(2)}{A_1(2)}\Big)^3 \Big] ; \label{eq:Thierry51MC}\\
\kappa_\times^a &=& \frac{p_a}{\nu_{aa} m_a \Delta} \, \Big[   \frac{5}{2} x^3+  \frac{2277}{1960}\Big(\frac{A_2(2)}{A_1(2)}\Big)^2  x \Big]; \nn\\
 \Delta &=& x^4+ \frac{3313}{4900} \Big(\frac{A_2(2)}{A_1(2)}\Big)^2 x^2 + \Big( \frac{36}{175} \Big)^2 \Big(\frac{A_2(2)}{A_1(2)}\Big)^4, \nn
\end{eqnarray}
where the corrections of the Coulomb logarithm are given by (\ref{eq:A21U}), and the limit of weakly-coupled plasmas $A_2(2)/A_1(2)=2$ recovers (\ref{eq:Thierry51B}).

Prescribing the generalized hard spheres
(the parallel conductivity is fully meaningful) yields
\begin{eqnarray}
\textrm{``Hard spheres''}: \qquad \kappa_\parallel^a &=&  \underbrace{\frac{1125}{352}}_{3.196} \, \frac{p_a}{\nu_{aa} m_a};\qquad 
\kappa_\perp^a = \frac{p_a}{\nu_{aa} m_a \Delta} \Big[ 2 x^2 + \frac{792}{245} \Big] ;  \label{eq:Thierry51BB}\\
\kappa_\times^a &=& \frac{p_a}{\nu_{aa} m_a \Delta} \, \Big[  \frac{5}{2} x^3+ \frac{2053}{490} x\Big]; 
 \qquad  \Delta = x^4+ \frac{573}{245} x^2  +\Big( \frac{176}{175} \Big)^2.\nn
\end{eqnarray}
%======

Finally, prescribing the inverse power-law force yields
\begin{eqnarray}
\kappa_\parallel^a &=&  \frac{25 (45 \nu^2-106 \nu+77)}{16 (11\nu-13) (3 \nu-5)} \frac{A_1(\nu)}{A_2(\nu)} \, \frac{p_a}{\nu_{aa} m_a};\nn\\
\kappa_\perp^a &=& \frac{p_a}{\nu_{aa} m_a \Delta} \Big[ \frac{A_2(\nu)}{A_1(\nu)} \frac{(3\nu-5)}{(\nu-1)} x^2
  + \Big( \frac{A_2(\nu)}{A_1(\nu)} \Big)^3  \frac{(45 \nu^2-106\nu+77) (11\nu-13) (3\nu-5)^3 }{1225 (\nu-1)^6 } \Big] ;\nn\\
\kappa_\times^a &=& \frac{p_a}{\nu_{aa} m_a \Delta} \, \Big[
  \frac{5}{2} x^3+ x \Big( \frac{A_2(\nu)}{A_1(\nu)} \Big)^2 \frac{(3 \nu-5)^2 (2053 \nu^4-9876 \nu^3 +19454 \nu^2-18004 \nu+6629)}{1960 (\nu-1)^6} \Big]; \nn\\
\Delta &=& x^4+ x^2 \Big( \frac{A_2(\nu)}{A_1(\nu)} \Big)^2 \frac{(3\nu-5)^2 (2865 \nu^4 -13348 \nu^3 +25446 \nu^2-22820 \nu+8113)}{4900 (\nu-1)^6} \nn\\
  && +\Big[ \Big( \frac{A_2(\nu)}{A_1(\nu)} \Big)^2 \, \frac{ 4 (11 \nu-13) (3\nu-5)^2}{175(\nu-1)^3} \Big]^2,
\label{eq:Thierry51A}
\end{eqnarray}
and prescribing $\nu=2$ and $\nu\to\infty$ of course recovers results (\ref{eq:Thierry51B})-(\ref{eq:Thierry51BB}).

%==========================
\subsubsection{Reduction into 1-Hermite approximation} \label{sec:1HermiteQa}
In the 1-Hermite approximation, the heat flux evolution equation reads
\begin{eqnarray}
    \frac{d_a}{d t}\vecq_a + \Omega_a \bhat\times\vecq_a + \frac{5}{2}p_a \nabla \Big(\frac{p_a}{\rho_a}\Big) &=& - \frac{2}{5}\nu_{aa}\Omega_{22}\vecq_a.\label{eq:Excite2P}
\end{eqnarray}
Cancelling the $d_a/dt$ then yields $\vecq_a$ in the same form (\ref{eq:Braginskii_YES}), but now with the 1-Hermite (self-collisional) thermal conductivities
\begin{eqnarray}
\big[ \kappa_\parallel^a \big]_1 &=&   \frac{25}{4\Omega_{22}}\, \frac{p_a}{\nu_{aa} m_a}; \qquad
\big[ \kappa_\perp^a \big]_1 = \frac{p_a}{\nu_{aa} m_a} \, \frac{ \Omega_{22}}{ x^2 + \big( 2 \Omega_{22}/5 \big)^2 };\qquad
\big[ \kappa_\times^a \big]_1 = \frac{p_a}{\nu_{aa} m_a } \, \frac{(5/2) x }{x^2 + \big( 2 \Omega_{22}/5 \big)^2}.
\label{eq:Thierry51P}
\end{eqnarray}
Note that only the $\Omega_{22}$ is again present and so for the Coulomb collisions and the generalized hard spheres the entire 1-Hermite heat flux has the same form
(and only the $\nu_{aa}$ are different). In the limit of strong magnetic field ($x\gg 1$), the 2-Hermite perpendicular and cross-conductivities (\ref{eq:Thierry51})
become identical to the 1-Hermite ones 
\begin{eqnarray}
  \textrm{Strong B-field:} \qquad \big[\kappa_\perp^a \big]_2 &=& \big[\kappa_\perp^a \big]_1 =  \Omega_{22} \frac{p_a \nu_{aa}}{m_a\Omega_a^2 }; \qquad
   \big[\kappa_\times^a \big]_2 = \big[\kappa_\times^a \big]_1 = \frac{5}{2}\frac{p_a}{m_a \Omega_a},
\end{eqnarray}  
and only the parallel conductivities $\kappa_\parallel^a$ remain different.

For the parallel conductivities, the difference between the 2-Hermite and 1-Hermite approximation can be written as
\begin{eqnarray}
  \big[ \kappa_\parallel^a \big]_2 &=& \big[ \kappa_\parallel^a \big]_1
  \Big( 1+ \frac{(7 \Omega_{2,2} -2 \Omega_{2,3})^2}{4 (7 \Omega_{2,2}^2 + \Omega_{2,2}\Omega_{2,4}-\Omega_{2,3}^2)}\Big),\label{eq:beau35b}
\end{eqnarray}
which again does not contain the collisional frequency $\nu_{aa}$. Evaluating (\ref{eq:beau35b}) for our collisional forces yields
\begin{eqnarray}
&&  \textrm{Coulomb collisions:} \qquad \big[ \kappa_\parallel^a \big]_2 = \big[ \kappa_\parallel^a \big]_1 \Big( 1+ \frac{1}{4}\Big) ;\nn\\  
&&  \textrm{Hard spheres:} \qquad \qquad \,\, \big[ \kappa_\parallel^a \big]_2 =  \big[ \kappa_\parallel^a \big]_1 \Big( 1+ \frac{1}{44} \Big);\nn\\
&&  \textrm{Inverse power:} \qquad \qquad \, \big[ \kappa_\parallel^a \big]_2 = \big[ \kappa_\parallel^a \big]_1
  \Big( 1+  \frac{(\nu-5)^2}{4(\nu-1)(11\nu-13)} \Big). \label{eq:CC00}
\end{eqnarray}
Equation (\ref{eq:CC00}) can be found in \cite{ChapmanCowling1953}, p. 173, and plotting the correction ratio with respect to $\nu$ yields
similar behavior than the viscosity (\ref{eq:CC0}). For $\nu\ge 2$, the correction ratio is the largest for the case of the
Coulomb collisions (with a correction of $\sim 25\%$), then the correction sharply decreases (it is only 1/40 for the case $\nu=3$), becomes
identically zero for $\nu=5$, and then slowly increases until the case of hard spheres is reached, with a small correction of only $\sim 2.3\%$.
Again, it is obvious that Chapman-Cowling knew the 2-Hermite parallel ion viscosity value much before Braginskii. 

Also note that for the 1-Hermite approximation, the ratio of the parallel thermal conductivity and viscosity
$\big[ \kappa_\parallel^a \big]_1 / \big[ \eta_0^a \big]_1 = 15/(4 m_a)$, meaning that the ratio is the same regardless of the considered collisional process.

%\newpage
%===================================================================================================
\subsection{Higher-order heat flux  \texorpdfstring{$\vecX^{(5)}_a$}{X(5)} (self-collisions)}
The solution for the heat flux $\vecX^{(5)}_a$ has a form
\begin{eqnarray}
  \vecX_a^{(5)} &=& \frac{p_a}{\rho_a} \Big[-\kappa_\parallel^{a(5)} \nabla_\parallel T_a - \kappa_\perp^{a(5)} \nabla_\perp T_a
    + \kappa_\times^{a(5)} \bhat\times\nabla T_a \Big], \label{eq:Thierry51M}
\end{eqnarray}
with the thermal conductivities (of the 5th-order fluid moment)
\begin{eqnarray}
\kappa_\parallel^{a(5)} &=&  \frac{175 (91 \Omega_{22}-32 \Omega_{23} +4 \Omega_{24})}{4 (7 \Omega_{22}^2+\Omega_{22} \Omega_{24} - \Omega_{23}^2)} \, \frac{p_a}{\nu_{aa} m_a};\nn\\
\kappa_\perp^{a(5)} &=& \frac{p_a}{\nu_{aa} m_a \Delta} \Big[ \big( 14 \Omega_{22}+4 \Omega_{23}\big) x^2
  +\frac{4}{175}\big( 7 \Omega_{22}^2 + \Omega_{22}\Omega_{24}-\Omega_{23}^2 \big) \big( 91 \Omega_{22} -32 \Omega_{23}+4 \Omega_{24} \big) \Big];\nn\\
\kappa_\times^{a(5)} &=& \frac{p_a}{\nu_{aa} m_a \Delta} \, \Big[
  70 x^3+ x \Big( \frac{1253}{10}\Omega_{22}^2 - \frac{422}{5}\Omega_{22}\Omega_{23} + \frac{48}{5}\Omega_{22}\Omega_{24}
  + \frac{72}{5}\Omega_{23}^2 - \frac{24}{7}\Omega_{23}\Omega_{24} + \frac{8}{35}\Omega_{24}^2 \Big) \Big]; \nn\\
\Delta &=& x^4+ x^2 \Big[\frac{193}{100}\Omega_{22}^2- \frac{6}{5} \Omega_{22} \Omega_{23} + \frac{22}{175} \Omega_{22} \Omega_{24}
  + \frac{36}{175} \Omega_{23}^2 - \frac{8}{175} \Omega_{23}\Omega_{24} + \frac{4}{1225} \Omega_{24}^2 \Big] \nn\\
  && +\Big[\frac{4}{175} \big( 7 \Omega_{22}^2+\Omega_{22} \Omega_{24}-\Omega_{23}^2 \big) \Big]^2 ,
\label{eq:Thierry51X5}
\end{eqnarray}
where the $\Delta$ is equal to (\ref{eq:Thierry51}). Prescribing the Coulomb collisions yields
\begin{eqnarray}
\textrm{Coulomb collisions}: \qquad \kappa_\parallel^{a(5)} &=&  \underbrace{\frac{2975}{24}}_{123.96} \, \frac{p_a}{\nu_{aa} m_a}; \qquad
\kappa_\perp^{a(5)} = \frac{p_a}{\nu_{aa} m_a \Delta} \Big[ 44 x^2 + (14688/175) \Big] ; \label{eq:Thierry51C}\\
\kappa_\times^{a (5)} &=& \frac{p_a}{\nu_{aa} m_a \Delta} \, \Big[   70 x^3+ x (1086/7) \Big]; 
 \qquad  \Delta = x^4+ x^2 (3313/1225)  +(144/175)^2, \nn
\end{eqnarray}
recovering eq. (46) of \cite{Hunana2022}. Prescribing the generalized hard spheres yields
\begin{eqnarray}
\textrm{``Hard spheres''}: \qquad \kappa_\parallel^{a(5)} &=&   \underbrace{\frac{7525}{88}}_{85.511} \, \frac{p_a}{\nu_{aa} m_a};\qquad 
\kappa_\perp^{a(5)} = \frac{p_a}{\nu_{aa} m_a \Delta} \Big[ 60 x^2 + (15136/175) \Big] ; \\
\kappa_\times^{a(5)} &=& \frac{p_a}{\nu_{aa} m_a \Delta} \, \Big[ 70  x^3+ x  (3814/35) \Big]; 
 \qquad  \Delta = x^4+ x^2 (573/245)  +(176/175)^2.\nn
\end{eqnarray}
Prescribing the inverse power-law force yields
\begin{eqnarray}
\kappa_\parallel^{a(5)} &=&  \frac{175 (43 \nu^2-94 \nu+67)}{4 (11 \nu-13) (3 \nu-5)} \frac{A_1(\nu)}{A_2(\nu)} \, \frac{p_a}{\nu_{aa} m_a};\nn\\
\kappa_\perp^{a(5)} &=& \frac{p_a}{\nu_{aa} m_a \Delta} \Big[ \frac{A_2(\nu)}{A_1(\nu)} \frac{2 (3\nu-5) (15 \nu-19)}{(\nu-1)^2} x^2
  + \Big( \frac{A_2(\nu)}{A_1(\nu)} \Big)^3 \frac{4 (43 \nu^2-94 \nu+67)(11 \nu-13) (3 \nu-5)^3 }{175 (\nu-1)^6 }  \Big] ;\nn\\
\kappa_\times^{a(5)} &=& \frac{p_a}{\nu_{aa} m_a \Delta} \, \Big[
  70 x^3+ x \Big( \frac{A_2(\nu)}{A_1(\nu)} \Big)^2 \frac{(3 \nu-5)^2 (1907 \nu^4-8676 \nu^3+16570 \nu^2-15124 \nu+5579)}{70 (\nu-1)^6 } \Big],
\label{eq:Thierry51AX5}
\end{eqnarray}
with the $\Delta$ equal to (\ref{eq:Thierry51A}).  
%=====

\newpage
%==================================================================================================================
%==================================================================================================================
\section{Case  \texorpdfstring{$m_a \ll m_b$}{ma<<mb} (lightweight particles such as electrons)} \label{sec:electrons}
Here we will assume $m_a \ll m_b$ and $T_a\simeq T_b$, which for the Coulomb collisions corresponds to the electron species of Braginskii.
To make this case easily distinguishable from the previous results, we will use a species index ``a=e'',
even though the results have a general validity for any particles ``e'' that have a small mass with respect to particles ``b'', and not just the electrons. 
For example, one can consider very light hard spheres, which collide with much heavier hard spheres and
as a reminder, we will sometimes write ``electrons''. 

This case is more complicated than the self-collisional case, because in addition to the ``e-e'' collisions, one also needs to take into account the 
``e-b'' collisions. As a consequence, in addition to the collisional frequency $\nu_{ee}$, the expressions will also contain
the collisional frequency $\nu_{eb}$. For general masses, the ratio of $\nu_{aa}/\nu_{ab}$ is given by (\ref{eq:haha}), 
which here simplifies into 
\begin{eqnarray}
  \textrm{General}\, (m_e\ll m_b): \qquad \frac{\nu_{ee}}{\nu_{eb}} &=& \frac{1}{2} \frac{n_e}{n_b} \frac{\Omega_{ee}^{(1,1)}}{\Omega_{eb}^{(1,1)}};\\
  \textrm{Coulomb collisions:} \qquad \frac{\nu_{ee}}{\nu_{eb}} &=& \frac{1}{\sqrt{2}} \frac{n_e}{Z_b^2 n_b} = \frac{1}{ Z_b \sqrt{2}}; \qquad
  (Z_b \,\,\textrm{is ion charge}); \label{eq:BigRit}\\   
  \textrm{Hard spheres:} \qquad \quad \,\, \frac{\nu_{ee}}{\nu_{eb}} &=& \frac{1}{\sqrt{2}} \frac{4 r_{e}^2}{(r_e+r_b)^2} \frac{n_e}{n_b};
  \qquad (r_e, r_b \,\, \textrm{are spheres radii})\label{eq:ritTT}\\
  \textrm{Inverse power:} \qquad \quad \,\, \frac{\nu_{ee}}{\nu_{eb}} &=& \frac{1}{\sqrt{2}} \Big( \frac{K_{ee}}{K_{eb}} \Big)^{\frac{2}{\nu-1}} \frac{n_e}{n_b},\label{eq:rit1}
\end{eqnarray}
where for the Coulomb collisions the charge-neutrality $n_e=Z_b n_b$ was assumed.
Note that no assumption was made about the hard sphere radius $r_e$ (i.e. it is not necesserily small with respect to $r_b$) and one can for example
consider the particular case of $r_e=r_b$ and $n_e=n_b$, with the ratio $\nu_{ee}/\nu_{eb}=1/\sqrt{2}$, where the same ratio is obtained for the Coulomb collisions with
the ion charge $Z_b=1$. Another simple example is the limit of the Lorentzian gas, where the ``e-e'' collisions become
insignificant and the ratio $\nu_{ee}/\nu_{eb}\ll 1$. For the Coulomb collisions, this corresponds to a large ion charge $Z_b\gg 1$
and for the hard spheres, this corresponds to for example $r_e\ll r_b$ and $n_e\simeq n_b$, or $r_e \simeq r_b$ and $n_e\ll n_b$. 
The Lorentzian limit is a bit academical (for example ions with $Z_b\gg 1$ are encountered much less frequently than the usual ions),
nevertheless, the limit is meaningful. It is just preferable to write it as $\nu_{ee}/\nu_{eb}\ll 1$ and not as $\nu_{ee}/\nu_{eb}\to 0$, 
because as the $\nu_{eb}$ increases, the final viscosities $\eta_0^e$ and thermal conductivities $\kappa_\parallel^e$ decrease towards zero.
We will of course derive the ``electron'' stress-tensors and heat fluxes for a general ratio $\nu_{ee}/\nu_{eb}$.

%==========================================
\subsection{Stress-tensors  \texorpdfstring{$\bPi^{(2)}_e$}{Pi(2)e} and  \texorpdfstring{$\bPi^{(4)}_e$}{Pi(4)e}}
Starting with the stress-tensors, the collisional contributions $\bQ_{e}^{(2)}\,'$ and $\bQ^{(4)}_{e}\,'$ given by (\ref{eq:QabP}) and (\ref{eq:Q4abP})
contain the mass-ratio coefficients for small temperature differences (\ref{eq:wow14})-(\ref{eq:wow15}), which for $m_e\ll m_b$ simplify into  
\begin{eqnarray}
  {K}_{eb (1)} &=& \frac{3}{5} \frac{m_b}{m_e}\Omega_{22}; \qquad {K}_{eb (2)} =2 - \frac{3}{5}\Omega_{22}; \nn\\
  L_{eb (1)} &=&  \frac{m_b}{m_e } \big( \frac{3}{10}\Omega_{22}- \frac{3}{35}\Omega_{23}\big);
  \qquad L_{eb (2)} = \frac{m_e}{70 m_b} \big(-28 \Omega_{12} -21 \Omega_{22} +6 \Omega_{23}+70 \big);\nn\\
  {M}_{eb (1)} &=& \frac{6}{5}\Omega_{23}; \qquad {M}_{eb (2)} = \frac{28 m_e}{5m_b} (\Omega_{12}-\frac{3}{14}\Omega_{23}); \nn\\
  N_{eb (1)} &=& -\frac{3}{5} \Omega_{23} + \frac{6}{35} \Omega_{24};\nn\\
  N_{eb (2)} &=& -\, \frac{14 m_e^2}{5 m_b^2} \Big( - \Omega_{12} +\frac{16}{49} \Omega_{13} - \frac{11}{14}\Omega_{22} + \frac{3}{14}\Omega_{23}
    - \frac{3}{49}\Omega_{24} + \frac{6}{49}\Omega_{33} +\frac{10}{7} \Big).
\end{eqnarray}
Note that the expansions with the small $m_e/m_b$ can be easily questioned, because the Chapman-Cowling
integrals $\Omega_{l,j}$ are technically undertermined at this stage and can be possibly large.
In such a case, the only correct approach is to solve a fully coupled system
for two species (see for example the Section 8.8, p. 38 ``Precision of $m_e/m_i$ expansions'' in Part 1, or Appendix N there). Nevertheless, one can be
guided by the electron equations of Part 1, and perform verification for the particular cases of interest a posteriori.

The evolution equations for the stress-tensors read
\begin{eqnarray}
  && \frac{d_e}{dt} \bPi^{(2)}_e  +\Omega_e \big(\bhat\times \bPi^{(2)}_e \big)^S + p_e \bW_e
  = \bQ_{e}^{(2)}\,' ; \nn\\
  && \frac{d_e}{dt} \bPi^{(4)}_e  +\Omega_e \big(\bhat\times \bPi^{(4)}_e \big)^S + 7 \frac{p_e^2}{\rho_e} \bW_e 
  =  \bQ^{(4)}_{e}\,', \label{eq:Posled31}
\end{eqnarray}
where the collisional contributions become
\begin{eqnarray}
  \bQ_{e}^{(2)}\,' &=& -\nu_{ee} \frac{3}{5} \Omega_{22} \bPi_e^{(2)}
  + \nu_{ee} \big(\frac{3}{20} \Omega_{22} - \frac{3}{70} \Omega_{23} \big)
  \Big( \frac{\rho_e}{p_e}\bPi^{(4)}_e- 7 \bPi^{(2)}_e\Big) \nn\\
  && -  \nu_{eb} \frac{3}{5} \Omega_{22} \bPi^{(2)}_e
  + \nu_{eb} \big( \frac{3}{10}\Omega_{22}- \frac{3}{35}\Omega_{23}\big) \Big( \frac{\rho_e}{p_e}\bPi^{(4)}_e- 7 \bPi^{(2)}_e\Big) ;\nn\\
%=====  
  \bQ_{e}^{(4)}\,' &=& -  \nu_{ee} \big( \frac{21}{10}\Omega_{22} +\frac{3}{5}\Omega_{23}\big) \frac{p_e}{\rho_e}\bPi_e^{(2)}
  - \nu_{ee} \big(\frac{1}{40} \Omega_{22}+ \frac{3}{70} \Omega_{24} \big)  \Big( \bPi^{(4)}_e- 7 \frac{p_e}{\rho_e} \bPi^{(2)}_e\Big) \nn\\
  && -  \nu_{eb} \frac{6}{5}\Omega_{23} \frac{p_e}{\rho_e} \bPi^{(2)}_e
  - \nu_{eb} \big(  -\frac{3}{5} \Omega_{23} + \frac{6}{35} \Omega_{24} \big) \Big( \bPi^{(4)}_e- 7 \frac{p_e}{\rho_e} \bPi^{(2)}_e\Big),
\end{eqnarray}
which can be further re-arranged into
\begin{eqnarray}
  \bQ_{e}^{(2)}\,' &=& - \Big[ \nu_{ee} \big(\frac{33}{20}\Omega_{22} -\frac{3}{10}\Omega_{23}\big) + \nu_{eb} \big(\frac{27}{10}\Omega_{22} -\frac{3}{5}\Omega_{23}\big)   \Big] \bPi_e^{(2)} \nn\\
 && + \Big[ \nu_{ee} \big(\frac{3}{20} \Omega_{22} - \frac{3}{70} \Omega_{23} \big) +\nu_{eb} \big( \frac{3}{10}\Omega_{22}- \frac{3}{35}\Omega_{23}\big) \Big] \frac{\rho_e}{p_e}\bPi^{(4)}_e;\nn\\
%=====  
  \bQ_{e}^{(4)}\,' &=& - \Big[ \nu_{ee} \big(\frac{77}{40}\Omega_{22} +\frac{3}{5}\Omega_{23}-\frac{3}{10}\Omega_{24}\big)
    + \nu_{eb} \big(\frac{27}{5}\Omega_{23} - \frac{6}{5}\Omega_{24}  \big)\Big] \frac{p_e}{\rho_e} \bPi_e^{(2)} \nn\\
  && + \Big[ - \nu_{ee} \big(\frac{1}{40} \Omega_{22}+ \frac{3}{70} \Omega_{24} \big) +\nu_{eb} \big(  \frac{3}{5} \Omega_{23} - \frac{6}{35} \Omega_{24} \big) \Big] \bPi^{(4)}_e.
  \label{eq:Posled31x}
\end{eqnarray}
As a quick double-check, prescribing the Coulomb collisions (with $\Omega_{22}=2$; $\Omega_{23}=4$; $\Omega_{24}=12$) recovers eq. (79) of Part 1.
The quasi-static solution with the general Chapman-Cowling integrals entering  the (\ref{eq:Posled31x}) can be slightly long to write down
and to clearly understand the solution, it is beneficial to introduce the following notation
\begin{eqnarray}
  \bQ_{e}^{(2)}\,' &=& -\nu_{eb} V_1 \bPi_e^{(2)} + \nu_{eb} V_2 \frac{\rho_e}{p_e}\bPi^{(4)}_e;\nn\\
  \bQ^{(4)}_{e}\,' &=& -\nu_{eb} V_3 \frac{p_e}{\rho_e} \bPi_e^{(2)} +\nu_{eb} V_4 \bPi^{(4)}_e,\label{eq:Posled31xx}
\end{eqnarray}
with the coefficients
\begin{eqnarray}
  V_1 &=&  \Big[ \frac{\nu_{ee}}{\nu_{eb}} \big(\frac{33}{20}\Omega_{22} -\frac{3}{10}\Omega_{23}\big) + \big(\frac{27}{10}\Omega_{22} -\frac{3}{5}\Omega_{23}\big)   \Big];\nn\\
  V_2 &=&  \Big[ \frac{\nu_{ee}}{\nu_{eb}} \big(\frac{3}{20} \Omega_{22} - \frac{3}{70} \Omega_{23} \big) + \big( \frac{3}{10}\Omega_{22}- \frac{3}{35}\Omega_{23}\big) \Big];\nn\\
  V_3 &=&  \Big[ \frac{\nu_{ee}}{\nu_{eb}} \big(\frac{77}{40}\Omega_{22} +\frac{3}{5}\Omega_{23}-\frac{3}{10}\Omega_{24}\big)
    +  \big(\frac{27}{5}\Omega_{23} - \frac{6}{5}\Omega_{24}  \big)\Big];\nn\\
  V_4 &=&  \Big[ - \frac{\nu_{ee}}{\nu_{eb}} \big(\frac{1}{40} \Omega_{22}+ \frac{3}{70} \Omega_{24} \big) + \big(  \frac{3}{5} \Omega_{23} - \frac{6}{35} \Omega_{24} \big) \Big].
  \label{eq:beau33mx}
\end{eqnarray}
The quasi-static solution of (\ref{eq:Posled31}) and (\ref{eq:Posled31xx}) then yields the stress-tensors in the usual form
\begin{eqnarray}
  \bPi^{(2)}_e &=& -\eta_0^e \bW_0 -\eta_1^e\bW_1 -\eta_2^e\bW_2 +\eta_3^e\bW_3+\eta_4^e\bW_4; \nn\\
  \bPi^{(4)}_e &=& \frac{p_e}{\rho_e}\Big[-\eta_0^{e(4)} \bW_0 -\eta_1^{e(4)}\bW_1 -\eta_2^{e(4)}\bW_2 +\eta_3^{e(4)}\bW_3+\eta_4^{e(4)}\bW_4\Big],
\end{eqnarray}
with the (2-Hermite) ``electron'' viscosities
\begin{eqnarray}
  \eta_0^e  &=& \frac{p_{e}}{\nu_{eb}} \frac{(7V_2-V_4)}{(V_2 V_3 - V_1 V_4)};\nn\\
  \eta_2^e  &=& \frac{p_{e}}{\nu_{eb}\Delta}\, \Big[ x^2 (V_1 -7 V_2) +(7 V_2 - V_4)(V_2 V_3 - V_1 V_4)\Big]; \nn\\
  \eta_4^e  &=& \frac{p_{e}}{\nu_{eb} \Delta}\, \Big[ x^3 + x (7 V_1 V_2 -V_2 V_3 -7 V_2 V_4 +V_4^2) \Big]; \nn\\
  \Delta &=& x^4 + x^2 (V_1^2 -2 V_2 V_3 +V_4^2) + (V_2 V_3 - V_1 V_4)^2 , \label{eq:beau33x}
\end{eqnarray}
and the ``electron'' viscosities of the stress-tensor $\bPi^{(4)}_e$
\begin{eqnarray}
  \eta_0^{e(4)}  &=& \frac{p_e}{\nu_{eb}} \frac{(7 V_1 - V_3)}{(V_2 V_3 - V_1 V_4)};\nn\\
  \eta_2^{e(4)}  &=& \frac{p_{e}}{\nu_{eb}\Delta}\, \Big[ x^2 (V_3 -7 V_4) +(7 V_1 - V_3)(V_2 V_3 - V_1 V_4)\Big]; \nn\\
  \eta_4^{e(4)}  &=& \frac{p_{e}}{\nu_{eb} \Delta}\, \Big[ 7 x^3+ x (7V_1^2 -V_1 V_3 -7 V_2 V_3 +V_3 V_4) \Big]; \nn\\
  \Delta &=& x^4 + x^2 (V_1^2 -2 V_2 V_3 +V_4^2) + (V_2 V_3 - V_1 V_4)^2, \label{eq:beau33xx}
\end{eqnarray}
where the $x=\Omega_e/\nu_{eb}$ and $\eta_1^e(x)=\eta_2^e(2x)$; $\eta_3^e(x)=\eta_4^e(2x)$. 
Results (\ref{eq:beau33x})-(\ref{eq:beau33xx}) together with the
coefficients (\ref{eq:beau33mx}) fully specify the final stress-tensors. Nevertheless, one might have a hope that some coefficients simplify (some indeed do and some
do not), so for clarity, for the usual stress-tensor $\bPi^{(2)}_e$ we also provide the fully explicit solution, where the parallel ``electron'' viscosity reads 
\begin{empheq}[box=\fbox]{align}
  \eta_0^e  &= \frac{p_{e}}{\nu_{eb} \Delta^*} \Big[\frac{\nu_{ee}}{\nu_{eb}} \Big( \frac{43}{40}\Omega_{22} - \frac{3}{10}\Omega_{23}
    + \frac{3}{70}\Omega_{24}\Big) + \frac{21}{10}\Omega_{22} -\frac{6}{5}\Omega_{23} + \frac{6}{35}\Omega_{24} \Big];\nn\\
  \Delta^* &\equiv (V_2 V_3 - V_1 V_4)=\Big(\frac{\nu_{ee}}{\nu_{eb}}\Big)^2 \Big( \frac{33}{100}\Omega_{22}^2 +\frac{9}{350}\Omega_{22}\Omega_{24} -\frac{9}{350}\Omega_{23}^2 \Big) \nn\\
  & \qquad + \frac{\nu_{ee}}{\nu_{eb}} \Big(  \frac{129}{200}\Omega_{22}^2 -\frac{9}{50}\Omega_{22}\Omega_{23} +\frac{9}{70}\Omega_{22}\Omega_{24}
  -\frac{18}{175} \Omega_{23}^2 \Big)
  + \frac{18}{175}\big( \Omega_{22}\Omega_{24} -\Omega_{23}^2\big), \label{eq:rit2}
\end{empheq}
and the perpendicular viscosities and gyroviscosities are given by
\begin{eqnarray}
  \eta_2^e  &=&  \frac{p_{e}}{\nu_{eb}\Delta}\, \Big\{ x^2 \frac{3}{5}\Omega_{22}\Big( \frac{\nu_{ee}}{\nu_{eb}} +1 \Big)
    + \Big[ \frac{\nu_{ee}}{\nu_{eb}} \Big( \frac{43}{40}\Omega_{22} - \frac{3}{10}\Omega_{23}
    + \frac{3}{70}\Omega_{24}\Big) + \frac{21}{10}\Omega_{22} -\frac{6}{5}\Omega_{23} + \frac{6}{35}\Omega_{24} \Big]\Delta^* \Big\};\nn\\
%===    
    \eta_4^e  &=& \frac{p_{e} x}{\nu_{eb} \Delta}\, \Big\{ x^2 +
    \Big(\frac{\nu_{ee}}{\nu_{eb}}\Big)^2 \Big[ \frac{2353}{1600}\Omega_{22}^2 -\frac{33}{40}\Omega_{22}\Omega_{23}
      +\frac{81}{700}\Omega_{23}^2 + \frac{129}{1400}\Omega_{22} \Omega_{24} - \frac{9}{350}\Omega_{23}\Omega_{24} + \frac{9}{4900} \Omega_{24}^2 \Big] \nn\\
    &&  \quad +\frac{\nu_{ee}}{\nu_{eb}}\Big[ \frac{18}{1225}\Omega_{24}^2 - \frac{36}{175}\Omega_{23}\Omega_{24}
      -\frac{114}{25}\Omega_{22}\Omega_{23} + \frac{144}{175}\Omega_{23}^2 + \frac{231}{40}\Omega_{22}^2 + \frac{96}{175}\Omega_{22}\Omega_{24} \Big] \nn\\
    &&  \quad + \frac{567}{100}\Omega_{22}^2 - \frac{144}{25}\Omega_{22} \Omega_{23} + \frac{54}{35}\Omega_{23}^2
    + \frac{18}{25} \Omega_{22}\Omega_{24} - \frac{72}{175} \Omega_{23}\Omega_{24} + \frac{36}{1225}\Omega_{24}^2 \Big\};\nn\\
    %===
\Delta &=& x^4 + x^2 \Big\{
 \Big(\frac{\nu_{ee}}{\nu_{eb}}\Big)^2 \Big[\frac{3433}{1600}\Omega_{22}^2- \frac{201}{200}\Omega_{22}\Omega_{23} + \frac{99}{700}\Omega_{23}^2
  + \frac{129}{1400}\Omega_{22}\Omega_{24} - \frac{9}{350}\Omega_{23}\Omega_{24} + \frac{9}{4900}\Omega_{24}^2 \Big]  \nn\\
 && \quad + \frac{\nu_{ee}}{\nu_{eb}} \Big[ \frac{1551}{200}\Omega_{22}^2 - \frac{132}{25}\Omega_{22}\Omega_{23}
   + \frac{162}{175}\Omega_{23}^2 + \frac{96}{175}\Omega_{22}\Omega_{24} - \frac{36}{175}\Omega_{23}\Omega_{24}
   +\frac{18}{1225}\Omega_{24}^2 \Big] \nn\\
 && \quad + \frac{729}{100}\Omega_{22}^2 - \frac{162}{25}\Omega_{22}\Omega_{23} + \frac{288}{175}\Omega_{23}^2
 + \frac{18}{25}\Omega_{22}\Omega_{24} - \frac{72}{175}\Omega_{23}\Omega_{24} + \frac{36}{1225}\Omega_{24}^2  \Big\} +  \Delta^{*2}. \label{eq:wow1}
\end{eqnarray}
 Result (\ref{eq:rit2}) represents the parallel electron viscosity of \cite{Braginskii1965}, here valid for a general class of collisional
  processes describable by the Boltzmann operator. Note that we assumed the same collisional process for the ``e-e'' and ``e-b'' collisions, and
  if these processes are different, the results will become more complicated.

Prescribing the Coulomb collisions in (\ref{eq:rit2})-(\ref{eq:wow1}) (with $\ln\Lambda\gg 1$ and $\Omega_{22}=2$; $\Omega_{23}=4$; $\Omega_{24}=12$) yields
\begin{eqnarray}
\textrm{Coulomb collisions}: \qquad  \eta_0^e  &=& \frac{p_{e}}{\nu_{eb}}\, \frac{\frac{\nu_{ee}}{\nu_{eb}} (41/28) + (51/35) }
  { [\big(\frac{\nu_{ee}}{\nu_{eb}}\big)^2  (267 /175)  + \frac{\nu_{ee}}{\nu_{eb}} (129 /50) +  (144 / 175) ]}; \label{eq:wow3}\\
  \eta_2^e  &=&  \frac{p_{e}}{\nu_{eb}\Delta}\, \Big\{ x^2 \frac{6}{5}\Big( \frac{\nu_{ee}}{\nu_{eb}} +1 \Big)
  + \Big[ \frac{\nu_{ee}}{\nu_{eb}} \frac{41}{28} + \frac{51}{35} \Big]
  \Big[\Big(\frac{\nu_{ee}}{\nu_{eb}}\Big)^2  \frac{267}{175}  + \frac{\nu_{ee}}{\nu_{eb}} \frac{129}{50} +  \frac{144}{175}\Big]  \Big\};\nn\\
%===    
    \eta_4^e  &=& \frac{p_{e} x}{\nu_{eb} \Delta}\, \Big\{ x^2 +
    \Big(\frac{\nu_{ee}}{\nu_{eb}}\Big)^2 \frac{46561}{19600}  +\frac{\nu_{ee}}{\nu_{eb}} \frac{12723}{2450}  + \frac{747}{245} \Big\};\nn\\
    %===
\Delta &=& x^4 + x^2 \Big[
\Big(\frac{\nu_{ee}}{\nu_{eb}}\Big)^2 \frac{79321}{19600} + \frac{\nu_{ee}}{\nu_{eb}} \frac{22047}{2450} +\frac{6633}{1225}  \Big]
+ \Big[ \Big(\frac{\nu_{ee}}{\nu_{eb}}\Big)^2  \frac{267}{175}  + \frac{\nu_{ee}}{\nu_{eb}} \frac{129}{50} +  \frac{144}{175} \Big]^2, \nn
\end{eqnarray}
where one needs to use the ratio of the collisional frequencies (\ref{eq:BigRit}), $\nu_{ee}/\nu_{eb} = 1 /(Z_b \sqrt{2})$.
The result (\ref{eq:wow3}) can be re-arranged into the eq. (82) in \cite{Hunana2022}  and
the result (\ref{eq:wow3}) represents the Braginskii electron viscosity for any ion charge $Z_i$.

%For clarity, we change the index $b\to i$, yielding
%\begin{eqnarray}
%\textrm{Coulomb collisions}: \qquad  \eta_0^e  &=& \frac{p_{e}}{\nu_{ei}}\, \frac{ (41 \sqrt{2} /56 Z_i) + (51/35) }
%  { [    (267 /350 Z_i^2)  +  (129 \sqrt{2}/100 Z_i) +  (144 / 175) ]}; \label{eq:wow4}\\
%  \eta_2^e  &=&  \frac{p_{e}}{\nu_{ei}\Delta}\, \Big\{ x^2 \frac{6}{5}\Big( \frac{\sqrt{2}}{2 Z_i }  +1 \Big)
%  + \Big[ \frac{41 \sqrt{2}}{56 Z_i } + \frac{51}{35} \Big]
%  \Big[ \frac{267}{350 Z_i^2}  + \frac{129 \sqrt{2}}{100 Z_i } +  \frac{144}{175}\Big]  \Big\};\nn\\
%===    
%    \eta_4^e  &=& \frac{p_{e} x}{\nu_{ei} \Delta}\, \Big\{ x^2 +
%    \frac{46561}{39200 Z_i^2}  + \frac{12723  \sqrt{2}}{4900 Z_i}  + \frac{747}{245} \Big\};\nn\\
%    %===
%\Delta &=& x^4 + x^2 \Big[
% \frac{79321}{39200 Z_i^2} + \frac{22047  \sqrt{2} }{4900 Z_i} +\frac{6633}{1225}  \Big]
%+ \Big[ \frac{267}{350 Z_i^2}  + \frac{129 \sqrt{2} }{100 Z_i } +  \frac{144}{175} \Big]^2. \nn
%\end{eqnarray}

Here we note that even though \cite{Braginskii1965} provides the electron viscosity only for the ion charge $Z_i=1$ in his eq. (4.45)   
(where the reported parallel electron viscosity 0.733 has a tiny rounding error and should be 0.731 instead), electron viscosities for
any $Z_i$ can be obtained from eq. (4.15) in his technical paper \cite{Braginskii1958}, by using the collisional matrices (A.14)
(where the first matrix has to be divided by $Z_i\sqrt{2}$). In fact, one can then obtain his model in a fully analytic form, for any $Z_i$,
and also verify that the 0.731 is the correct 2-Hermite result. As a side note, the rounding error is of course irrelevant,
but one might get very curious about it, because
from the work of \cite{JiHeld2013}, the 3-Hermite approximation indeed yields 0.733, so one might be wondering, if Braginskii used the 3-Hermite approximation
for the parallel electron viscosity. However, the perpendicular viscosities and gyroviscosities in eq. (4.45) of \cite{Braginskii1965} are provided in the
2-Hermite approximation and the collisional matrices for viscosities (A.14) in \cite{Braginskii1958}, do not go beyond the 2-Hermite approximation. Obviously, this is
only a rounding error, likely originating in eq. (4.16) of the last reference, as a result of $8.50/11.6=0.733$, which was evaluated several years later,
when writing the review paper.

It is interesting to evaluate the general results (\ref{eq:rit2})-(\ref{eq:wow1}) for other collisional forces
and for example, prescribing the hard spheres (with $\Omega_{22}=2$; $\Omega_{23}=8$; $\Omega_{24}=40$, the parallel viscosity is fully meaningfull) yields
\begin{eqnarray}
\textrm{``Hard spheres''}: \qquad  \eta_0^e  &=& \frac{p_{e}}{\nu_{eb}}\, \frac{\frac{\nu_{ee}}{\nu_{eb}} (41/28) + (51/35) }
  { [\big(\frac{\nu_{ee}}{\nu_{eb}}\big)^2  (303 /175)  + \frac{\nu_{ee}}{\nu_{eb}} (1191/350) + (288/175) ]}; \label{eq:wow5}\\
  \eta_2^e  &=&  \frac{p_{e}}{\nu_{eb}\Delta}\, \Big\{ x^2 \frac{6}{5}\Big( \frac{\nu_{ee}}{\nu_{eb}} +1 \Big)
  + \Big[ \frac{\nu_{ee}}{\nu_{eb}} \frac{41}{28} + \frac{51}{35} \Big]
  \Big[\Big(\frac{\nu_{ee}}{\nu_{eb}}\Big)^2  \frac{303}{175}  + \frac{\nu_{ee}}{\nu_{eb}} \frac{1191}{350} +  \frac{288}{175}\Big]  \Big\};\nn\\
%===    
    \eta_4^e  &=& \frac{p_{e} x}{\nu_{eb} \Delta}\, \Big\{ x^2 +
    \Big(\frac{\nu_{ee}}{\nu_{eb}}\Big)^2 \frac{42529}{19600}  +\frac{\nu_{ee}}{\nu_{eb}} \frac{10707}{2450}  + \frac{2727}{1225} \Big\};\nn\\
    %===
\Delta &=& x^4 + x^2 \Big[
\Big(\frac{\nu_{ee}}{\nu_{eb}}\Big)^2 \frac{71257}{19600} + \frac{\nu_{ee}}{\nu_{eb}} \frac{3603}{490} + \frac{4617}{1225} \Big]
+ \Big[ \Big(\frac{\nu_{ee}}{\nu_{eb}}\Big)^2  \frac{303}{175}  + \frac{\nu_{ee}}{\nu_{eb}} \frac{1191}{350} +  \frac{288}{175} \Big]^2, \nn
\end{eqnarray}
where the ratio of the collisional frequencies is given by (\ref{eq:ritTT}). Note the perhaps surprising numerical similarities between the
 (\ref{eq:wow3}) and (\ref{eq:wow5}) and the parallel viscosities can be also written as
\begin{empheq}[box=\fbox]{align}
  \textrm{Coulomb collisions}: \qquad  \eta_0^e   &=  \frac{1025 \frac{\nu_{ee}}{\nu_{eb}}  +1020}
         { [ 1068\big(\frac{\nu_{ee}}{\nu_{eb}}\big)^2 + 1806\frac{\nu_{ee}}{\nu_{eb}} + 576 ]} \, \frac{p_{e}}{\nu_{eb}}; \label{eq:wow6}\\
%===
  \textrm{Hard spheres}: \qquad  \eta_0^e  &=   \frac{1025 \frac{\nu_{ee}}{\nu_{eb}} +1020}
         {[ 1212 \big(\frac{\nu_{ee}}{\nu_{eb}}\big)^2 +2382 \frac{\nu_{ee}}{\nu_{eb}} +1152 ]} \, \frac{p_{e}}{\nu_{eb}}. \label{eq:wow7}
\end{empheq}
As a quick simple double-check, one can recover the self-collisional results by considering the limit of large $\nu_{ee}/\nu_{eb}\gg 1$.

\newpage
\subsubsection{``Electron'' viscosities for inverse power-law force}
For the inverse power-law force, the V-coefficients (\ref{eq:beau33mx}) are evaluated as 
\begin{eqnarray}
  V_1 &=&  + \,  \frac{A_2(\nu)}{A_1(\nu)} \frac{3 (3\nu-5)}{20 (\nu-1)^2}  \Big[ \frac{\nu_{ee}}{\nu_{eb}}(3\nu+1)+2\nu+6  \Big];\nn\\
  V_2 &=&  -\, \frac{A_2(\nu)}{A_1(\nu)} \frac{3 (3\nu-5) (\nu-5)}{140 (\nu-1)^2}\Big[ \frac{\nu_{ee}}{\nu_{eb}} +2 \Big];\nn\\
  V_3 &=&  -\, \frac{A_2(\nu)}{A_1(\nu)} \frac{(3\nu-5)}{40 (\nu-1)^3} \Big[ \frac{\nu_{ee}}{\nu_{eb}} (67\nu^2-302\nu+283) + 48 (2 \nu-3)(\nu-5)\Big];\nn\\
  V_4 &=&  -\, \frac{A_2(\nu)}{A_1(\nu)} \frac{(3\nu-5)}{280 (\nu-1)^3}\Big[ \frac{\nu_{ee}}{\nu_{eb}} (247 \nu^2-710 \nu+511) + 48 (2 \nu-3)(3 \nu-7)\Big],
  \label{eq:beau33mx2}
\end{eqnarray}
which fully define all of the viscosities (\ref{eq:beau33x})-(\ref{eq:beau33xx}). Nevertheless, one might have a hope that some coefficients simplify and for example
\begin{eqnarray}
7 V_2 -V_4 &=& \frac{A_2(\nu)}{A_1(\nu)} \frac{(3\nu-5)}{280(\nu-1)^3}
\Big[ \frac{\nu_{ee}}{\nu_{eb}}(205 \nu^2-458 \nu+301) +204 \nu^2-600 \nu+588  \Big];\nn\\
%===
  \Delta^* &\equiv& V_2 V_3 - V_1 V_4 = \Big(\frac{A_2(\nu)}{A_1(\nu)}\Big)^2 \frac{3 (3\nu-5)^2}{1400(\nu-1)^4} \Big[ \Big(\frac{\nu_{ee}}{\nu_{eb}}\Big)^2 2 (\nu-1)(101 \nu-113) \nn\\
    && \qquad  + \frac{\nu_{ee}}{\nu_{eb}}(397 \nu^2-938 \nu+589) + 96 (\nu-1)(2\nu-3) \Big];\nn\\
%===  
  V_1-7V_2 &=& \frac{A_2(\nu)}{A_1(\nu)} \frac{3 (3\nu-5)}{5(\nu-1)}\Big(\frac{\nu_{ee}}{\nu_{eb}}+1 \Big); \nn\\
  %===
  7 V_1 V_2 -V_2 V_3 -7 V_2 V_4 +V_4^2  &=& \Big(\frac{A_2(\nu)}{A_1(\nu)}\Big)^2 \frac{(3\nu-5)^2}{78400 (\nu-1)^6}
  \Big[ \Big(\frac{\nu_{ee}}{\nu_{eb}}\Big)^2(42529 \nu^4-193828\nu^3+356358\nu^2 -305956\nu \nn\\
    && +103201)  + \frac{\nu_{ee}}{\nu_{eb}} (85656 \nu^4-457056 \nu^3+1006224 \nu^2-1020768 \nu+404376)\nn\\
    && +43632 \nu^4-268992 \nu^3+692640 \nu^2-826560 \nu+396144 \Big]; \nn\\
  %===
  V_1^2-2 V_2 V_3+V_4^2 &=&  \Big(\frac{A_2(\nu)}{A_1(\nu)}\Big)^2 \frac{(3\nu-5)^2}{78400 (\nu-1)^6}
  \Big[ \Big(\frac{\nu_{ee}}{\nu_{eb}}\Big)^2 (71257 \nu^4-312772 \nu^3+548886 \nu^2-449092 \nu \nn\\
    && +144025)   +\frac{\nu_{ee}}{\nu_{eb}}(144120 \nu^4-707040 \nu^3+1437648 \nu^2-1367520 \nu+511224) \nn\\
    && +73872 \nu^4-406080 \nu^3+954720 \nu^2-1060416 \nu+474768 \Big], \label{eq:wow13}
\end{eqnarray}
and these expressions enter the viscosities (\ref{eq:beau33x}). Obviously, the factorizations simplify the final results only slightly and
if one wants a viscosity for a given $\nu$, it seems easier to first calculate the V-coefficients (\ref{eq:beau33mx2}) for that given $\nu$ and 
calculate the viscosities (\ref{eq:beau33x}) only afterwards. Nevertheless, for example the parallel viscosity reads
\begin{eqnarray}
  \eta_0^e  &=& \frac{p_{e}}{\nu_{eb}} \frac{A_1(\nu)}{A_2(\nu)} \frac{5(\nu-1)}{3(3\nu-5)}
  \Big[ \frac{\nu_{ee}}{\nu_{eb}}(205 \nu^2-458 \nu+301) +204 \nu^2-600 \nu+588  \Big] \nn\\
&&  \times  \Big[ \Big(\frac{\nu_{ee}}{\nu_{eb}}\Big)^2 2 (\nu-1)(101 \nu-113)   + \frac{\nu_{ee}}{\nu_{eb}}(397 \nu^2-938 \nu+589) + 96 (\nu-1)(2\nu-3) \Big]^{-1}. \label{eq:wow9}
\end{eqnarray}
The result (\ref{eq:wow9}) represents the Braginskii parallel ``electron'' viscosity for any force $K_{ab}/r^\nu$ 
and the cases (\ref{eq:wow6})-(\ref{eq:wow7}) are recovered by prescribing $\nu=2$ and $\nu\to \infty$. The self-collisional limit of (\ref{eq:wow9}) recovers the parallel
viscosity (\ref{eq:beau38}) and the same can be seen for the $\eta_2^e$ and $\eta_4^e$ given by (\ref{eq:wow13}).\\

\newpage
\subsubsection{Braginskii \texorpdfstring{($\nu=2$)}{} electron viscosities for moderately-coupled plasmas}
 Evaluation of the previous equations for $\nu=2$ yields a generalization of the Braginskii electron viscosities for moderately-coupled plasmas
\begin{eqnarray}
  \eta_0^e  &=& \frac{p_{e}}{\nu_{eb}} \frac{A_1(2)}{A_2(2)} \frac{5 (205  \frac{\nu_{ee}}{\nu_{eb}}+204)}{ 3 [178 \big(\frac{\nu_{ee}}{\nu_{eb}}\big)^2
    +301  \frac{\nu_{ee}}{\nu_{eb}}  +96 ] }; \label{eq:etaEmod}\\
  \eta_2^e  &=& \frac{p_{e}}{\nu_{eb}\Delta}\, \Big[ x^2  \frac{A_2(2)}{A_1(2)}\frac{3}{5} \Big(\frac{\nu_{ee}}{\nu_{eb}}+1 \Big)
    +  \Big(\frac{A_2(2)}{A_1(2)}\Big)^3  \frac{3}{392000} \Big(205 \frac{\nu_{ee}}{\nu_{eb}} +204 \Big)
    \Big(178 \big(\frac{\nu_{ee}}{\nu_{eb}}\big)^2 +301 \frac{\nu_{ee}}{\nu_{eb}} +96 \Big) \Big]; \nn\\
  \eta_4^e  &=& \frac{p_{e}}{\nu_{eb} \Delta}\, \Big[ x^3 + x \Big(\frac{A_2(2)}{A_1(2)}\Big)^2
    \frac{1}{78400}\Big( 46561 \big(\frac{\nu_{ee}}{\nu_{eb}}\big)^2   +101784 \frac{\nu_{ee}}{\nu_{eb}} +59760 \Big) \Big]; \nn\\
  \Delta &=& x^4 + x^2 \Big(\frac{A_2(2)}{A_1(2)}\Big)^2 \frac{1}{78400} \Big( 79321 \big(\frac{\nu_{ee}}{\nu_{eb}}\big)^2 +176376 \frac{\nu_{ee}}{\nu_{eb}}+106128 \Big) \nn\\
  && + \Big(\frac{A_2(2)}{A_1(2)}\Big)^4 \Big[\frac{3}{1400} \Big(178 \big(\frac{\nu_{ee}}{\nu_{eb}}\big)^2
    +301 \frac{\nu_{ee}}{\nu_{eb}}  +96 \Big)  \Big]^2, \nn
\end{eqnarray}
 where the $x=\Omega_e/\nu_{eb}$, the $\nu_{ee}/\nu_{eb} = 1 /(Z_b \sqrt{2})$ and the coefficients $A_l(2)$ are given by (\ref{eq:A12x})-(\ref{eq:A22x}).
  The limit of weakly-coupled plasmas $(\ln\Lambda\gg 1)$ is obtained by $A_2(2)/A_1(2)=2$, recovering eq. (82) of Part 1 (the last $Z_i$ there is missing the index ``i''),
  or here eq. (\ref{eq:wow3}).

%\newpage
%=================================================================================================================
\subsubsection{Reduction into 1-Hermite approximation}
In the simplified 1-Hermite approximation, the ``electron'' stress-tensor evolves according to 
\begin{eqnarray}
  && \frac{d_e}{dt} \bPi^{(2)}_e  +\Omega_e \big(\bhat\times \bPi^{(2)}_e \big)^S + p_e \bW_e
  = - \nu_{eb}(1+\frac{\nu_{ee}}{\nu_{eb}}) \frac{3}{5} \Omega_{22} \bPi_e^{(2)}, 
\end{eqnarray}
which in the quasi-static approximation yields the 1-Hermite viscosities
\begin{eqnarray}
 \big[ \eta_0^{e} \big]_1  &=& \frac{p_e}{\nu_{eb}} \frac{5}{3\Omega_{22}} \frac{1}{(1+\frac{\nu_{ee}}{\nu_{eb}})};\qquad
 \big[ \eta_2^{e} \big]_1  = \frac{p_{e}}{\nu_{eb} \Delta}\, (1+\frac{\nu_{ee}}{\nu_{eb}})\frac{3}{5} \Omega_{22}; \nn\\
 \big[ \eta_4^{e} \big]_1  &=& \frac{p_{e}}{\nu_{eb} \Delta}\, x; \qquad
           \Delta = x^2 + \Big( (1+\frac{\nu_{ee}}{\nu_{eb}})\frac{3}{5} \Omega_{22} \Big)^2. \label{eq:beau33xxx}
\end{eqnarray}
In the limit of strong magnetic field ($x\gg 1$), the 2-Hermite perpendicular viscosities and gyroviscosites (\ref{eq:wow1}) converge to
the 1-Hermite approximation 
\begin{equation}
  \textrm{Strong B-field}: \qquad  \eta_2^{e}   = \frac{p_{e} \nu_{eb}}{\Omega_e^2}\,\frac{3}{5} \Omega_{22} (1+\frac{\nu_{ee}}{\nu_{eb}}); \qquad
   \eta_4^{e}   = \frac{p_{e}}{\Omega_e},  
\end{equation}
and only the parallel viscosities $\eta_0^e$ remain different. For the inverse power-law force, the 1-Hermite parallel viscosity reads
\begin{equation}
  \textrm{Inverse power}: \qquad  \big[ \eta_0^e \big]_1  = \frac{p_e}{\nu_{eb}} \frac{A_1(\nu)}{A_2(\nu)} \frac{5(\nu-1)}{3(3\nu-5)} 
  \frac{1}{(\frac{\nu_{ee}}{\nu_{eb}}+1)},
\end{equation}
and the 2-Hermite result is given by (\ref{eq:wow9}).

%==============
\subsubsection{Improvement of the 2-Hermite approximation}
For a general collisional process, the improvement of the 2-Hermite approximation for the parallel viscosity can be written as
\begin{eqnarray}
  \big[ \eta_0^e \big]_2 &=&  \big[ \eta_0^e \big]_1 \Big\{ 1+  3 (7 \Omega_{22}-2 \Omega_{23})^2 \big(\frac{\nu_{ee}}{\nu_{eb}}+2 \big)^2 
      \Big[\Big(154 \Omega_{22}^2+12 \Omega_{22} \Omega_{24}-12 \Omega_{23}^2 \Big) \Big(\frac{\nu_{ee}}{\nu_{eb}}\Big)^2 \nn\\
  && \qquad   + \Big( 301 \Omega_{22}^2+\Omega_{22}(60 \Omega_{24}-84 \Omega_{23}) -48 \Omega_{23}^2 \Big) \frac{\nu_{ee}}{\nu_{eb}}
        +48 \Omega_{22}\Omega_{24}-48 \Omega_{23}^2  \Big]^{-1} \Big\},
\end{eqnarray}
which for the inverse power-law force becomes
\begin{equation}
\big[ \eta_0^e \big]_2 =  \big[ \eta_0^e \big]_1 \Big( 1+
\frac{3 (\nu-5)^2 (\frac{\nu_{ee}}{\nu_{eb}}+2)^2 }{2(\nu-1)(101\nu-113) (\frac{\nu_{ee}}{\nu_{eb}})^2+(397 \nu^2-938 \nu+589) \frac{\nu_{ee}}{\nu_{eb}} +96(\nu-1)(2\nu-3)} \Big),
\label{eq:wow10}
\end{equation}
and for the two particular cases
\begin{eqnarray}
  \textrm{Coulomb collisions}: \qquad  \big[ \eta_0^e \big]_2 &=&  \big[ \eta_0^e \big]_1
  \Big( 1+  \frac{27(\frac{\nu_{ee}}{\nu_{eb}} +2)^2}{178(\frac{\nu_{ee}}{\nu_{eb}})^2+301 \frac{\nu_{ee}}{\nu_{eb}}+96 } \Big);\\
   \textrm{Hard spheres}: \qquad  \big[ \eta_0^e \big]_2 &=&  \big[ \eta_0^e \big]_1
   \Big( 1+ \frac{3(\frac{\nu_{ee}}{\nu_{eb}} +2)^2}{202(\frac{\nu_{ee}}{\nu_{eb}})^2+397 \frac{\nu_{ee}}{\nu_{eb}}+192 }\Big).
\end{eqnarray}
Note that for the Maxwell molecules ($\nu=5$) the improvement of (\ref{eq:wow10}) is zero and in the self-collisional limit $\nu_{ee}/\nu_{eb}\gg 1$,
one recovers results (\ref{eq:RisY})-(\ref{eq:CC0}). The above results are useful, however, to get an idea about the improvement, one needs some concrete numbers.
First, we consider the particular case with the ratio of collisional frequences 
\begin{eqnarray}
  \frac{\nu_{ee}}{\nu_{eb}} = \frac{1}{\sqrt{2}}; \qquad \textrm{Coulomb collisions:}   \qquad \big[\eta_0^e \big]_2 &=& 0.731 \frac{p_e}{\nu_{eb}};
  \qquad \big[\eta_0^e \big]_2 = \big[\eta_0^e \big]_1 \big( 1+ 0.497 \big);  \nn\\
  \textrm{Hard spheres:} \qquad \big[ \eta_0^e \big]_2 &=&  0.507 \frac{p_e}{\nu_{eb}};
  \qquad \big[\eta_0^e \big]_2 = \big[\eta_0^e \big]_1 \big( 1+ 0.0383\big);\nn\\
  \textrm{Coulomb collisions \& Hard spheres:} \qquad 
\big[ \eta_0^e \big]_1 &=& 0.488 \frac{p_e}{\nu_{eb}}, \label{eq:wow11}
\end{eqnarray}
where the 0.731 is the Braginskii parallel electron viscosity (for the $Z_i=1$).
The result (\ref{eq:wow11}) shows that while for the Coulomb collisions the parallel electron viscosity is improved by almost 50\% by the 2-Hermite approximation,
for the hard spheres the improvement is less than 4\%. 

As a second example, we consider the Lorentz limit (which is the most extreme case), with the ratio of collisional frequencies 
\begin{eqnarray}
  \frac{\nu_{ee}}{\nu_{eb}} \ll 1; \qquad   
  \textrm{Coulomb collisions:} \qquad \big[\eta_0^e \big]_2 &=& \underbrace{(85/48)}_{1.77} \frac{p_e}{\nu_{eb}}; \qquad
  \big[ \eta_0^e \big]_2 = \big[ \eta_0^e \big]_1 \Big( 1+  \underbrace{(9/8)}_{1.125} \Big);\nn\\
  \textrm{Hard spheres:} \qquad \big[\eta_0^e \big]_2 &=& \underbrace{ (85/96)}_{0.885} \frac{p_e}{\nu_{eb}}; \qquad
  \big[ \eta_0^e \big]_2 = \big[ \eta_0^e \big]_1 \Big( 1+  \underbrace{(1/16)}_{0.0625} \Big);\nn\\
  \textrm{Coulomb collisions \& Hard spheres:} \qquad \big[\eta_0^e \big]_1 &=& \underbrace{ (5/6)}_{0.833} \frac{p_e}{\nu_{eb}}, \label{eq:wow12}
\end{eqnarray}
implying that while for the Coulomb collisions the correction of the 2-Hermite approximation is over 110\%, for the hard spheres it is
only 6\%.

\newpage
%===================================================================================================
%===================================================================================================
\subsection{Heat fluxes \texorpdfstring{$\vecq_e$}{qe}, \texorpdfstring{$\vecX^{(5)}_e$}{X(5)e} and momentum exchange rates \texorpdfstring{$\boldsymbol{R}_e$}{Re}}
For $m_e\ll m_b$, the mass-ratio coefficients for small temperature differences (\ref{eq:wow16})-(\ref{eq:wow17}) simplify into  
\begin{eqnarray}
  D_{eb (1)} &=& - \Omega_{12}+ \frac{2}{5}\Omega_{13}  ; \qquad
  D_{eb (2)} =  - \Omega_{12} + \frac{2}{5}\Omega_{13} - \frac{4}{5}\Omega_{22}+3;\nn\\
  U_{eb (1)} &=& -  \Omega_{12}; \qquad U_{eb (2)} = + 4 \Omega_{13};\nn\\
%===
  E_{eb (1)} &=& -\frac{1}{8}\Omega_{12}+ \frac{1}{10}\Omega_{13}- \frac{1}{70}\Omega_{14} ; \qquad
%===  
  E_{eb (2)} =  \frac{m_e}{ m_b} \Big(- \frac{3}{4} +\frac{17}{40} \Omega_{12}  - \frac{1}{10}\Omega_{13}
    +\frac{1}{70}\Omega_{14} + \frac{1}{5}\Omega_{22} - \frac{2}{35}\Omega_{23}  \Big); \nn\\
%\end{eqnarray}
%=======================
%\begin{eqnarray}
  F_{eb (1)} &=& -4 \Omega_{13} + \frac{8}{5} \Omega_{14}; \qquad
  F_{eb (2)} =  - \frac{168}{5} \Omega_{12} + 4 \Omega_{13}
    - \frac{8}{5}\Omega_{14} + \frac{32}{5}\Omega_{23}; \nn\\
%===
  G_{eb (1)} &=&  \frac{1}{2}\Omega_{13}- \frac{2}{5}\Omega_{14} + \frac{2}{35}\Omega_{15}; \nn\\
%===  
  G_{eb (2)} &=& - \frac{m_e}{m_b } 
     \frac{7}{2} \Big(- \frac{32}{245}\Omega_{33} + \frac{12}{5}\Omega_{12}
    - \frac{251}{245}\Omega_{13}  + \frac{4}{35} \Omega_{14} - \frac{4}{245} \Omega_{15}
    + \frac{32}{35} \Omega_{22} - \frac{16}{35} \Omega_{23}   +  \frac{32}{245} \Omega_{24} - \frac{10}{7} \Big).\nn
\end{eqnarray}
%=================
The collisional right-hand-sides become
\begin{eqnarray}
  \vecQ^{(3)}_{e}\,' &=& - \frac{2}{5}\nu_{ee}\Omega_{22}\vecq_e
  + \nu_{ee} \Big(\frac{1}{20}\Omega_{22}- \frac{1}{70}\Omega_{23}\Big) \Big(\frac{\rho_e}{p_e}\vecX^{(5)}_e -28 \vecq_e\Big)\nn\\
  && + \nu_{eb}\Big\{  + \Big( \Omega_{12} - \frac{2}{5}\Omega_{13} \Big) \vecq_e  + \Omega_{12} p_e   (\bu_b-\bu_e) \nn\\
  && + \Big( -\frac{1}{8}\Omega_{12}+ \frac{1}{10}\Omega_{13}- \frac{1}{70}\Omega_{14} \Big) \Big(\frac{\rho_e}{p_e}\vecX^{(5)}_e -28 \vecq_e\Big) \Big\} 
   -\frac{5}{2} \frac{p_e}{\rho_e} \boldsymbol{R}_e; \label{eq:Q3PE}
\end{eqnarray}
%======
\begin{eqnarray}
  \vecQ^{(5)}_{e}\,' &=& - \nu_{ee} \Big( \frac{8}{5}\Omega_{23}+ \frac{28}{5}\Omega_{22} \Big) \frac{p_e}{\rho_e}\vecq_e
  - \nu_{ee} \Big(\frac{2}{35}\Omega_{24} - \frac{3}{10}\Omega_{22} \Big) \Big(\vecX^{(5)}_e -28 \frac{p_e}{\rho_e} \vecq_e\Big)\nn\\
  && +\nu_{eb}\Big\{  + \Big( 4 \Omega_{13} - \frac{8}{5} \Omega_{14} \Big) \frac{p_e}{\rho_e} \vecq_e
   + 4 \Omega_{13} \frac{p_e^2}{\rho_e}   (\bu_b-\bu_e) \nn\\
  && - \Big( \frac{1}{2}\Omega_{13}- \frac{2}{5}\Omega_{14} + \frac{2}{35}\Omega_{15}  \Big) \Big(\vecX^{(5)}_e -28 \frac{p_e}{\rho_e} \vecq_e \Big)  \Big\} 
   - 35 \frac{p_e^2}{\rho_e^2} \boldsymbol{R}_e, \label{eq:Qab5PE}
\end{eqnarray}
%======
with the momentum exchange rates
\begin{eqnarray}
  \boldsymbol{R}_e &=& \nu_{eb} \Big\{ \rho_e (\bu_b-\bu_e)
  + \Big(1-\frac{2}{5}\Omega_{12} \Big) \frac{\rho_e}{p_e} \vecq_e  
   -  \Big( \frac{1}{8}- \frac{1}{10}\Omega_{12} + \frac{1}{70}\Omega_{13} \Big) \frac{\rho_e^2}{p_e^2}
  \Big(\vecX^{(5)}_e -28\frac{p_e}{\rho_e}\vecq_e \Big)\Big\}; \nn\\
%====
  \boldsymbol{R}_e &=& \nu_{eb} \Big\{ \rho_e (\bu_b-\bu_e)
  + \Big( \frac{9}{2}- \frac{16}{5}\Omega_{12} + \frac{2}{5} \Omega_{13} \Big) \frac{\rho_e}{p_e} \vecq_e  
   -  \Big( \frac{1}{8}- \frac{1}{10}\Omega_{12} + \frac{1}{70}\Omega_{13} \Big) \frac{\rho_e^2}{p_e^2}   \vecX^{(5)}_e \Big\},
  \label{eq:Final1E}
\end{eqnarray}
which can be re-arranged into 
\begin{eqnarray}
  \vecQ^{(3)}_{e}\,' &=& - \Big[ \nu_{ee} \Big(\frac{9}{5}\Omega_{22}-\frac{2}{5}\Omega_{23}\Big)
    + \nu_{eb} \Big(  \frac{45}{4} - \frac{25}{2}\Omega_{12} + \frac{21}{5}\Omega_{13} - \frac{2}{5}\Omega_{14}  \Big) \Big] \vecq_e \nn\\
  && + \Big[ \nu_{ee} \Big(\frac{1}{20}\Omega_{22}- \frac{1}{70}\Omega_{23}\Big)
  +\nu_{eb}  \Big(  \frac{5}{16} -\frac{3}{8}\Omega_{12}+  \frac{19}{140}\Omega_{13} - \frac{1}{70}\Omega_{14}  \Big)  \Big] \frac{\rho_e}{p_e}\vecX^{(5)}_e\nn\\
  &&   + \Big(\frac{5}{2}-\Omega_{12}\Big) \nu_{eb} p_e     (\bu_e-\bu_b); \label{eq:Q3PEE}
\end{eqnarray}
%==================
\begin{eqnarray}
  \vecQ^{(5)}_{e}\,' &=& - \Big[ \nu_{ee} \Big( \frac{8}{5}\Omega_{23} + 14\Omega_{22} - \frac{8}{5}\Omega_{24} \Big)
   +\nu_{eb} \Big( \frac{315}{2} -112 \Omega_{12} -4 \Omega_{13} + \frac{64}{5}\Omega_{14} - \frac{8}{5}\Omega_{15}  \Big) \Big]\frac{p_e}{\rho_e}\vecq_e \nn\\
 && - \Big[ \nu_{ee} \Big(\frac{2}{35}\Omega_{24} - \frac{3}{10}\Omega_{22} \Big)  
   +\nu_{eb} \Big( - \frac{35}{8} + \frac{7}{2}\Omega_{12} - \frac{2}{5}\Omega_{14} +\frac{2}{35}\Omega_{15}  \Big) \Big] \vecX^{(5)}_e  \nn\\
 &&  + \big( 35 - 4 \Omega_{13} \big) \nu_{eb} \frac{p_e^2}{\rho_e}  (\bu_e-\bu_b). \label{eq:Qab5PEE}
\end{eqnarray}
As a quick double-check, prescribing Coulomb collisions recovers eqs. (48)-(50) of \cite{Hunana2022},
where we have used the notation $\delta\bu=\bu_e-\bu_b$ (analogous to the Braginskii notation $\bu=\bu_e-\bu_i$).

The collisional contributions (\ref{eq:Q3PEE})-(\ref{eq:Qab5PEE}) enter the right-hand-side of the evolution equations for the heat flux vectors
\begin{eqnarray}
    \frac{d_e}{d t}\vecq_e + \Omega_e \bhat\times\vecq_e + \frac{5}{2}p_e \nabla \Big(\frac{p_e}{\rho_e}\Big) &=& \vecQ^{(3)}_{e}\,'; \nn\\
%=============    
  \frac{d_e}{d t}\vecX^{(5)}_e +\Omega_e\bhat\times\vecX^{(5)}_e +70\frac{p_e^2}{\rho_e}\nabla\Big(\frac{p_e}{\rho_e}\Big) 
  &=& \vecQ^{(5)}_{e}\,'.\label{eq:Excite2E}
\end{eqnarray}

To write the general solution in the quasi-static approximation, it is useful to introduce notation
\begin{eqnarray}
  \vecQ^{(3)}_{e}\,' &=& \nu_{eb} \Big[ - B_1 \vecq_e + B_2  \frac{\rho_e}{p_e}\vecX^{(5)}_e + B_5  p_e \delta\bu \Big];\nn\\
  \vecQ^{(5)}_{e}\,' &=& \nu_{eb} \Big[ - B_3 \frac{p_e}{\rho_e} \vecq_e - B_4  \vecX^{(5)}_e + B_6  \frac{p_e^2}{\rho_e} \delta\bu \Big];\nn\\
   \boldsymbol{R}_e &=& \nu_{eb} \Big[ -\rho_e \delta \bu +B_7 \frac{\rho_e}{p_e} \vecq_e -B_8 \frac{\rho_e^2}{p_e^2} \vecX^{(5)}_e  \Big], \label{eq:QeBrag}
\end{eqnarray}
with the following B-coefficients
(big B as Braginskii, who was the first to figure out all of the electron transport coefficents in the presence of magnetic field)
\begin{eqnarray}
B_1 &=& \Big[ \frac{\nu_{ee}}{\nu_{eb}} \Big(\frac{9}{5}\Omega_{22}-\frac{2}{5}\Omega_{23}\Big)
  +  \Big(  \frac{45}{4} - \frac{25}{2}\Omega_{12} + \frac{21}{5}\Omega_{13} - \frac{2}{5}\Omega_{14}  \Big) \Big];\nn\\
B_2 &=& \Big[ \frac{\nu_{ee}}{\nu_{eb}} \Big(\frac{1}{20}\Omega_{22}- \frac{1}{70}\Omega_{23}\Big)
  +  \Big(  \frac{5}{16} -\frac{3}{8}\Omega_{12}+  \frac{19}{140}\Omega_{13} - \frac{1}{70}\Omega_{14}  \Big)  \Big];\nn\\
B_3 &=& \Big[ \frac{\nu_{ee}}{\nu_{eb}} \Big( \frac{8}{5}\Omega_{23} + 14\Omega_{22} - \frac{8}{5}\Omega_{24} \Big)
  + \Big( \frac{315}{2} -112 \Omega_{12} -4 \Omega_{13} + \frac{64}{5}\Omega_{14} - \frac{8}{5}\Omega_{15}  \Big) \Big];\nn\\
B_4 &=& \Big[ \frac{\nu_{ee}}{\nu_{eb}} \Big(\frac{2}{35}\Omega_{24} - \frac{3}{10}\Omega_{22} \Big)  
  + \Big( - \frac{35}{8} + \frac{7}{2}\Omega_{12} - \frac{2}{5}\Omega_{14} +\frac{2}{35}\Omega_{15}  \Big) \Big];\nn\\
B_5 &=& \Big(\frac{5}{2}-\Omega_{12}\Big); \qquad B_6 = \big( 35 - 4 \Omega_{13} \big); \nn\\
B_7 &=&  1-\frac{2}{5}\Omega_{12}+28 B_8 = \Big( \frac{9}{2}- \frac{16}{5}\Omega_{12} + \frac{2}{5} \Omega_{13} \Big); \qquad
B_8 =  \Big( \frac{1}{8}- \frac{1}{10}\Omega_{12} + \frac{1}{70}\Omega_{13} \Big). \label{eq:Bcoeff}
\end{eqnarray}
The collisional contributions (\ref{eq:QeBrag})-(\ref{eq:Bcoeff}) enter the right-hand-sides of the evolution equations (\ref{eq:Excite2E}). Note that for 
the momentum exchange rates, the coefficients are related to the previously defined $B_5=(5/2) V_{ab(0)}$; $B_7=V_{eb (1)}$ and $B_8=V_{ab (3)}$, but
here to write down the solutions, we find it better to use the B-designation. 

%=====================================================================================
\subsubsection{Braginskii form for  \texorpdfstring{$\vecq_e$}{qe} (through Chapman-Cowling integrals)}
In the quasi-static approximation, the heat flux is split into the thermal and frictional part $\vecq_e=\vecq_e^T+\vecq_e^u$, where
by using the \cite{Braginskii1965} notation for the transport coefficients
\begin{eqnarray}
  \vecq_e^T &=& -\kappa_\parallel^e \nabla_\parallel T_e - \kappa_\perp^e \nabla_\perp T_e + \kappa_\times^e \bhat\times\nabla T_e;\nn\\
  %=====
&& \kappa_\parallel^e = \frac{p_e}{m_e \nu_{eb}}\gamma_0 ; \qquad   
\kappa_\perp^e =  \frac{p_e}{m_e \nu_{eb}} \frac{\gamma_1' x^2+\gamma_0'}{\triangle};
\qquad \kappa_\times^e =  \frac{p_e}{m_e \nu_{eb}}\frac{\gamma_1'' x^3+\gamma_0''x}{\triangle};
\qquad \triangle = x^4 + \delta_1 x^2 +\delta_0; \label{eq:QEnice}\\
%=====  
\vecq_e^u &=& \beta_0 p_e \delta\bu_\parallel + p_e \delta\bu_\perp \frac{\beta_1'x^2+\beta_0'}{\triangle} 
- p_e \bhat\times\delta\bu \frac{\beta_1''x^3+\beta_0''x}{\triangle}. \label{eq:Thierry63}
\end{eqnarray}
The thermal heat conductivities (\ref{eq:QEnice}) are given by
\begin{eqnarray}
  \gamma_0 &=& \frac{5 (28 B_2 +B_4)}{2 (B_1 B_4 +B_2 B_3)}; \qquad 
  \gamma_1' = \frac{5}{2} B_1 -70 B_2;\qquad
  \gamma_0' = \frac{5}{2}(28 B_2 +B_4)(B_1 B_4+B_2 B_3);\nn\\
  \gamma_1'' &=& \frac{5}{2};\qquad   \gamma_0'' = \frac{B_2}{2}(140 B_1 -5 B_3 +140 B_4)  +\frac{5}{2}B_4^2;\nn\\
  \delta_1 &=& B_1^2 -2 B_2 B_3 +B_4^2; \qquad \delta_0 = (B_1 B_4 +B_2 B_3)^2, \label{eq:HFgamma0}
\end{eqnarray}
where the B-coefficients (\ref{eq:Bcoeff}) contain the Chapman-Cowling integrals.  
The frictional heat flux (\ref{eq:Thierry63}) is given by  
\begin{eqnarray}
  \beta_0 &=& \frac{B_2 B_6 +B_4 B_5}{B_1 B_4 +B_2 B_3};\qquad
  \beta_1' = B_1 B_5 -B_2 B_6; \qquad \beta_0' = (B_2 B_6+B_4 B_5) (B_1 B_4 +B_2 B_3);\nn\\
  \beta_1'' &=& B_5; \qquad \beta_0'' = B_2(B_1 B_6 -B_3 B_5 +B_4 B_6) +B_4^2 B_5, \label{eq:HFbeta0}
\end{eqnarray}
with the same denomintor $\triangle$. 

%The Lorentz approximation is obtained easily by setting $\nu_{ee}/\nu_{eb}=0$.

%=====================================================================================
\subsubsection{Solution for  \texorpdfstring{$\vecX_e^{(5)}$}{X(5)e} (through Chapman-Cowling integrals)}
Analogously to $\vecq_e$, one writes the solution for the heat flux $\vecX_e^{(5)}$ split into the
thermal and frictional parts 
\begin{eqnarray}
  \vecX_e^{(5)T} &=& \frac{p_e}{\rho_e} \Big[-\kappa_\parallel^{e(5)} \nabla_\parallel T_e - \kappa_\perp^{e(5)} \nabla_\perp T_e
    + \kappa_\times^{e(5)} \bhat\times\nabla T_e \Big];\nn\\
%=======
&& \kappa_\parallel^{e(5)} = \frac{p_e}{ m_e \nu_{eb}}\gamma_0^{(5)}; \qquad   
\kappa_\perp^{e(5)} =  \frac{p_e}{ m_e \nu_{eb}} \frac{\gamma_1^{(5)'} x^2+\gamma_0^{(5)'}}{\triangle};
\qquad \kappa_\times^{e(5)} =  \frac{p_e}{ m_e \nu_{eb}}\frac{\gamma_1^{(5)''} x^3+\gamma_0^{(5)''} x}{\triangle}; \label{eq:Thierry61} \\
%=======  
\vecX_e^{(5)u} &=& \frac{p_e^2}{\rho_e} \Big[ \beta_0^{(5)}  \delta\bu_\parallel + \frac{\beta_1^{(5)'} x^2+\beta_0^{(5)'}}{\triangle} \delta\bu_\perp
-\frac{\beta_1^{(5)''}x^3+\beta_0^{(5)''}x}{\triangle} \bhat\times\delta\bu\Big], \label{eq:Thierry60}
\end{eqnarray}
where the $\triangle$ is the same as in (\ref{eq:QEnice}). The thermal transport coefficients read
\begin{eqnarray}
  \gamma_0^{(5)} &=& \frac{5 (28 B_1-B_3)}{2 (B_1 B_4+B_2 B_3)}; \qquad 
  \gamma_1^{(5)'} = \frac{5}{2} B_3+70 B_4;\qquad
  \gamma_0^{(5)'} = \frac{5}{2}(28 B_1-B_3)(B_1 B_4+B_2 B_3);\nn\\
  \gamma_1^{(5)''} &=& 70;\qquad   \gamma_0^{(5)''} = -\, \frac{B_3}{2}(5 B_1+140 B_2+5 B_4) +70 B_1^2,
\end{eqnarray}
and the frictional coefficients read
\begin{eqnarray}
  \beta_0^{(5)} &=& \frac{B_1 B_6 -B_3 B_5}{B_1 B_4+B_2 B_3};\qquad
  \beta_1^{(5)'} = B_3 B_5 + B_4 B_6; \qquad \beta_0^{(5)'} = (B_1 B_6 -B_3 B_5) (B_1 B_4 +B_2 B_3);\nn\\
  \beta_1^{(5)''} &=& B_6; \qquad \beta_0^{(5)''} = -\,  B_3 (B_1 B_5 +B_2 B_6 +B_4 B_5) +B_1^2 B_6,\label{eq:Thierry61P}
\end{eqnarray}
where the B-coefficients are given by (\ref{eq:Bcoeff}).

%=====================================================================================
\subsubsection{Momentum exchange rates \texorpdfstring{$\boldsymbol{R}_e$}{Re} (through Chapman-Cowling integrals)}
To finish the task of expressing all of the Braginskii transport coefficients through the Chapman-Cowling integrals,
one needs to substitute the $\vecq_e$ and $\vecX^{(5)}_e$ solutions given above into the momentum exchange rates $\boldsymbol{R}_e$ (\ref{eq:QeBrag})
and write the results in the Braginskii form, by also spllitting the $\boldsymbol{R}_e$ into the
frictional and thermal parts
\begin{eqnarray}
  \boldsymbol{R}_e^u &=& -\alpha_0 \rho_e\nu_{eb} \delta\bu_\parallel -\rho_e\nu_{eb}\delta\bu_\perp \Big( 1-\frac{\alpha_1' x^2+\alpha_0'}{\triangle}\Big)
  -\rho_e \nu_{eb}\bhat\times\delta \bu \frac{\alpha_1''x^3 +\alpha_0'' x}{\triangle}; \label{eq:RU}\\
  \boldsymbol{R}_e^T &=& -\beta_0 n_e \nabla_\parallel T_e-n_e \nabla_\perp T_e \frac{\beta_1' x^2+\beta_0'}{\triangle}
  +n_e \bhat\times\nabla T_e \frac{\beta_1'' x^3+\beta_0'' x}{\triangle}.\label{eq:RT}
\end{eqnarray}
The coefficients for the frictional part (\ref{eq:RU}) are then obtained by
\begin{eqnarray}
  \alpha_0 &=& 1-B_7\beta_0+ B_8\beta_0^{(5)};\qquad
  \alpha_1' = B_7\beta_1'- B_8\beta_1^{(5)'};\nn\\
  \alpha_0' &=& B_7\beta_0' - B_8\beta_0^{(5)'};\qquad
  \alpha_1'' = B_7\beta_1'' - B_8\beta_1^{(5)''};\qquad
  \alpha_0'' = B_7 \beta_0'' - B_8\beta_0^{(5)''},
\end{eqnarray}
and more explicitly as
\begin{eqnarray}
  \alpha_0 &=& 1-B_7 \frac{(B_2 B_6 +B_4 B_5)}{(B_1 B_4 +B_2 B_3)} + B_8\frac{(B_1 B_6 -B_3 B_5)}{(B_1 B_4+B_2 B_3)};\nn\\
  \alpha_1' &=& B_7 (B_1 B_5 -B_2 B_6) - B_8 (B_3 B_5 + B_4 B_6);\nn\\
  \alpha_0' &=& \big[ B_7 (B_2 B_6+B_4 B_5) - B_8 (B_1 B_6 -B_3 B_5)  \big]  (B_1 B_4 +B_2 B_3);\nn\\
  \alpha_1'' &=& B_7 B_5 - B_8 B_6; \nn\\
  \alpha_0'' &=& B_7 \big[ B_2(B_1 B_6 -B_3 B_5 +B_4 B_6) +B_4^2 B_5 \big] - B_8 \big[ -\,  B_3 (B_1 B_5 +B_2 B_6 +B_4 B_5) +B_1^2 B_6  \big], \label{eq:HFalpha0}
\end{eqnarray}
where the B-coefficients are given by (\ref{eq:Bcoeff}).

The coefficients for the thermal part (\ref{eq:RT}) (often called the thermal force)
were already written with the $\beta$-coefficients of the frictional heat flux (\ref{eq:Thierry63}), because
the system satisfies the Onsager symmetry, and it can be verified that the following relations are indeed true  
\begin{eqnarray}
    \beta_0 &=& B_7 \gamma_0 - B_8\gamma_0^{(5)};\qquad
  \beta_1' = B_7\gamma_1' - B_8 \gamma_1^{(5)'};\nn\\
  \beta_0' &=& B_7\gamma_0' - B_8\gamma_0^{(5)'}; \qquad
  \beta_1'' = B_7 \gamma_1'' - B_8 \gamma_1^{(5)''}; \qquad
  \beta_0'' = B_7 \gamma_0'' - B_8\gamma_0^{(5)''}.
\end{eqnarray}
  The Onsager symmetry - i.e. that the transport coefficients $\beta_0,\beta_1',\beta_0',\beta_1'',\beta_0''$ of the fricional heat flux (\ref{eq:Thierry63})
      are identical to the transport coefficients of the thermal force (\ref{eq:RT}), is actualy a useful double-check that our calculations are correct.

%===============================================================================
\subsubsection{Re-arranged Braginskii coefficients (through Chapman-Cowling integrals)}
At this stage, all of the Braginskii transport coefficients (the $\gamma$-coefficients for the thermal heat flux $\vecq_e^T$ (\ref{eq:HFgamma0}),
the $\beta$-coefficients for the frictional heat flux $\vecq_e^u$ (\ref{eq:HFbeta0}),
as well as the $\alpha$-coefficients for the momentum exchange rates (\ref{eq:HFalpha0})), were already expressed through
the Chapman-Cowling integrals fully explicitly, by using the B-coefficients (\ref{eq:Bcoeff}). One might naturally wonder, if a re-grouping of
these expressions might bring some simplifications, and here we re-group the results with respect to $\nu_{ee}/\nu_{eb}$.
For some coefficients, this re-grouping is partially beneficial, however for a few cases, the re-grouping yields expressions which are too long.   

Starting with the parallel heat conductivity, the re-grouping yields
\begin{empheq}[box=\fbox]{align}
  \kappa_\parallel^e &= \frac{5}{2 \Delta^*} \Big[ \frac{\nu_{ee}}{\nu_{eb}} \Big( \frac{11}{10} \Omega_{22}- \frac{2}{5}\Omega_{23}+ \frac{2}{35}\Omega_{24} \Big) 
    + \frac{35}{8}- 7 \Omega_{12} + \frac{19}{5}\Omega_{13} - \frac{4}{5}\Omega_{14} + \frac{2}{35}\Omega_{15} \Big]  \frac{p_e}{\nu_{eb} m_e};\nn\\
  \Delta^* & \equiv (B_1 B_4 +B_2 B_3) 
   = \frac{\nu_{ee}^2}{\nu_{eb}^2} \Big( \frac{4}{175}\Omega_{22}\Omega_{24}+\frac{4}{25}\Omega_{22}^2 - \frac{4}{175} \Omega_{23}^2 \Big) 
    + \frac{\nu_{ee}}{\nu_{eb}} \Big(  -\frac{4}{35} \Omega_{12}\Omega_{24} - \frac{4}{5} \Omega_{12}\Omega_{22} +\frac{4}{175} \Omega_{13} \Omega_{24} \nn\\
 &     + \frac{11}{25} \Omega_{13} \Omega_{22} 
    - \frac{4}{25} \Omega_{14} \Omega_{22} +\frac{4}{175} \Omega_{22}\Omega_{15}
  - \frac{2}{5} \Omega_{23}\Omega_{12} - \frac{8}{175} \Omega_{23}\Omega_{14} + \frac{48}{175} \Omega_{13} \Omega_{23} 
  + \frac{1}{7} \Omega_{24} + \Omega_{22} \Big) \nn\\
&    -\frac{7}{4} \Omega_{12}^2 + \frac{2}{5}\Omega_{12}\Omega_{14} - \frac{4}{35} \Omega_{12}\Omega_{15} + \Omega_{13}\Omega_{12} 
  + \frac{4}{35}\Omega_{13}\Omega_{14} + \frac{4}{175} \Omega_{13} \Omega_{15} - \frac{4}{175} \Omega_{14}^2 \nn\\
&  - \frac{19}{35} \Omega_{13}^2
  + \frac{1}{7} \Omega_{15}+ \frac{7}{4}\Omega_{13} - \Omega_{14}. \label{eq:Extreme0}
\end{empheq}
The result (\ref{eq:Extreme0}) can be viewed as the parallel electron thermal conductivity of \cite{Braginskii1965},
here generalized to a form valid for any lightweight particles $m_e\ll m_b$ that
the classical Boltzmann operator can describe. Note that we assumed the same collisional process for the ``e-e'' and ``e-b'' collisions,
and one could derive a more general result, where these two processes are different.
Prescribing the Coulomb collisions and the hard spheres (with the $\Omega_{l,j}$ given by (\ref{eq:beauty17x})-(\ref{eq:beauty17})) yields parallel thermal conductivities
\begin{empheq}[box=\fbox]{align}
  \textrm{Coulomb collisions}:\qquad   \kappa_\parallel^e  &= \frac{25}{4} 
  \frac{(360 \frac{\nu_{ee}}{\nu_{eb}}  +433 )}{(576 \big(\frac{\nu_{ee}}{\nu_{eb}} \big)^2  +1208 \frac{\nu_{ee}}{\nu_{eb}}  +217)} \,
  \frac{p_e}{\nu_{eb} m_e};\label{eq:BragS10} \\
%===
  \textrm{Hard spheres}:\qquad  \kappa_\parallel^e  &=
  \frac{25}{4}  \frac{(360 \frac{\nu_{ee}}{\nu_{eb}} +433 )}{( 704 \big(\frac{\nu_{ee}}{\nu_{eb}}\big)^2 +1944 \frac{\nu_{ee}}{\nu_{eb}} +1275 )} \, \frac{p_e}{\nu_{eb} m_e},
  \label{eq:sleep3}
\end{empheq}
where the collisional frequencies are given by (\ref{eq:BigRit})-(\ref{eq:ritTT}). As a partial double-check, taking the results (\ref{eq:Extreme0})-(\ref{eq:sleep3})
and considering the self-collisional limit $\nu_{ee}/\nu_{eb}\gg 1$,
recovers the self-collisional results (\ref{eq:Thierry51})-(\ref{eq:Thierry51BB}).
Prescribing $\nu_{ee}/\nu_{eb}=1/\sqrt{2}$ then recovers the famous Braginskii value of $\gamma_0=3.1616$ (valid for the ion charge $Z_i=1$) and for the
hard spheres $\gamma_0=1.4316$, also compared in (\ref{eq:wow20}) and together with other coefficients in Table \ref{eq:TableCH}.

Note that the re-grouping with $\nu_{ee}/\nu_{eb}$ did not simlify the final result (\ref{eq:Extreme0}) that much, and one might as well use the formulation
(\ref{eq:HFgamma0}) through the B-coefficients. Nevertheless, since the denominator $\Delta^*$ is already written down, the other two parallel coefficients are
%==========================
\begin{eqnarray}
  \beta_0 &=& \frac{1}{\Delta^*} \Big[ \frac{\nu_{ee}}{\nu_{eb}}
    \Big( \Omega_{22}- \frac{1}{5}\Omega_{13} \Omega_{22} - \frac{1}{2}\Omega_{23} + \frac{2}{35}\Omega_{13}\Omega_{23}
    + \frac{1}{7}\Omega_{24} - \frac{2}{35}\Omega_{12}\Omega_{24} + \frac{3}{10}\Omega_{12} \Omega_{22} \Big) \nn\\
   && +\frac{7}{2}\Omega_{13}+ \frac{3}{2}\Omega_{13}\Omega_{12} - \frac{19}{35}\Omega_{13}^2 - \frac{3}{2}\Omega_{14}
    +\frac{2}{35}\Omega_{13}\Omega_{14} - \frac{7}{2}\Omega_{12}^2 
     + \frac{2}{5}\Omega_{12}\Omega_{14} + \frac{1}{7}\Omega_{15}
    - \frac{2}{35}\Omega_{12}\Omega_{15} \Big]; \\
%\end{eqnarray}
%=================
%\begin{eqnarray}
  \alpha_0 &=& 1 - \frac{1}{\Delta^*} \Big[ \frac{\nu_{ee}}{\nu_{eb}}
    \Big(- \frac{8}{175}\Omega_{12}\Omega_{13} \Omega_{23} - \frac{4}{25}\Omega_{12}\Omega_{13} \Omega_{22}
    + \frac{4}{175}\Omega_{12}^2 \Omega_{24} + \frac{11}{25}\Omega_{12}^2 \Omega_{22} + \frac{4}{175}\Omega_{13}^2 \Omega_{22} \nn\\
    &&    + \frac{4}{25}\Omega_{23}\Omega_{12}^2- \frac{4}{35}\Omega_{12}\Omega_{24}- \frac{4}{5}\Omega_{12}\Omega_{22}
    - \frac{2}{5}\Omega_{23}\Omega_{12}+ \frac{4}{35}\Omega_{13}\Omega_{23}+ \frac{1}{7}\Omega_{24} + \Omega_{22} \Big) \nn\\
    &&  -\frac{8}{175} \Omega_{12}\Omega_{13}\Omega_{14} + \frac{4}{175}\Omega_{13}^3  + \frac{4}{175}\Omega_{12}^2 \Omega_{15}
    - \frac{7}{4}\Omega_{12}^2 +\frac{2}{5}\Omega_{12}\Omega_{14} - \frac{4}{35}\Omega_{12}\Omega_{15} \nn\\
 &&   +\Omega_{13} \Omega_{12} + \frac{4}{35}\Omega_{13}\Omega_{14} - \frac{19}{35}\Omega_{13}^2 +\frac{1}{7}\Omega_{15}
    + \frac{7}{4}\Omega_{13} - \Omega_{14} \Big].
\end{eqnarray}
For the thermal heat flux, the rest of the Braginskii $\gamma$-coefficients are given by
\begin{eqnarray}
  \gamma_1' &=& \frac{\nu_{ee}}{\nu_{eb}} \Omega_{22} + \frac{25}{4} -5 \Omega_{12}+\Omega_{13};\qquad   \gamma_0' = \gamma_0 \Delta^{*2};
  \qquad \gamma_1'' = \frac{5}{2}; \nn\\
  \gamma_0'' &=& \Big(\frac{\nu_{ee}}{\nu_{eb}}\Big)^2
  \Big( \frac{149}{40}\Omega_{22}^2- \frac{13}{5}\Omega_{23}\Omega_{22} + \frac{16}{35}\Omega_{23}^2 + \frac{11}{35}\Omega_{22} \Omega_{24}
  - \frac{4}{35}\Omega_{23}\Omega_{24}+ \frac{2}{245}\Omega_{24}^2 \Big) \nn\\ 
  &&  + \frac{\nu_{ee}}{\nu_{eb}} \Big( \frac{4}{245}\Omega_{24}\Omega_{15} -2 \Omega_{12}\Omega_{24} -49 \Omega_{12}\Omega_{22}
  + \frac{38}{35}\Omega_{13}\Omega_{24} + \frac{247}{10}\Omega_{13}\Omega_{22} - \frac{24}{5}\Omega_{14}\Omega_{22}
  + \frac{11}{35}\Omega_{22}\Omega_{15} \nn\\
  && \quad + 17 \Omega_{23} \Omega_{12} + \frac{12}{7}\Omega_{23}\Omega_{14}
  - \frac{304}{35}\Omega_{13}\Omega_{23} - \frac{8}{35}\Omega_{14}\Omega_{24} - \frac{4}{35}\Omega_{23}\Omega_{15}
  + \frac{5}{4}\Omega_{24} + \frac{525}{16}\Omega_{22} - \frac{45}{4}\Omega_{23} \Big) \nn\\
  &&   + \frac{9625}{128} - \frac{875}{4}\Omega_{12} + \frac{855}{8}\Omega_{13} -20\Omega_{14} + \frac{1295}{8}\Omega_{12}^2
  - \frac{323}{2}\Omega_{13}\Omega_{12} + 31 \Omega_{12} \Omega_{14} +\frac{1444}{35}\Omega_{13}^2 \nn\\
  && - \frac{114}{7} \Omega_{13}\Omega_{14} + \frac{58}{35}\Omega_{14}^2 + \frac{5}{4} \Omega_{15}
  -2 \Omega_{12}\Omega_{15}  +\frac{38}{35}\Omega_{13}\Omega_{15} - \frac{8}{35}\Omega_{14}\Omega_{15} + \frac{2}{245} \Omega_{15}^2,
\end{eqnarray}
together with
\begin{eqnarray}
  \delta_0 &=& \Delta^{*2}; \nn\\
  \delta_1 &=& \Big(\frac{\nu_{ee}}{\nu_{eb}}\Big)^2 \Big( \frac{193}{100}\Omega_{22}^2 -\frac{6}{5}\Omega_{23}\Omega_{22}
  +\frac{36}{175}\Omega_{23}^2 +\frac{22}{175}\Omega_{22}\Omega_{24}
  - \frac{8}{175}\Omega_{23} \Omega_{24} + \frac{4}{1225}\Omega_{24}^2 \Big) \nn\\
%===  
  && +\frac{\nu_{ee}}{\nu_{eb}} \Big( \frac{8}{1225}\Omega_{24}\Omega_{15} - \frac{4}{5}\Omega_{12}\Omega_{24}
  -\frac{127}{5}\Omega_{12}\Omega_{22} +\frac{76}{175}\Omega_{13}\Omega_{24}
  + \frac{293}{25}\Omega_{13}\Omega_{22} - \frac{52}{25}\Omega_{14}\Omega_{22}  +\frac{22}{175}\Omega_{22} \Omega_{15} \nn\\
  && \quad + 8\Omega_{23}\Omega_{12} +\frac{128}{175}\Omega_{23}\Omega_{14}
  - \frac{684}{175}\Omega_{13}\Omega_{23} - \frac{16}{175}\Omega_{14}\Omega_{24} - \frac{8}{175}\Omega_{23}\Omega_{15}
  + \frac{1}{2}\Omega_{24} + \frac{149}{8}\Omega_{22} - \frac{11}{2} \Omega_{23} \Big) \nn\\
%===  
  &&  + \frac{3025}{64} + \frac{4}{1225}\Omega_{15}^2  + \frac{169}{2}\Omega_{12}^2 + \frac{68}{5}\Omega_{12}\Omega_{14}
  - \frac{4}{5}\Omega_{12}\Omega_{15}  - \frac{388}{5}\Omega_{13}\Omega_{12}
  - \frac{1216}{175}\Omega_{13}\Omega_{14} \nn\\
  && + \frac{76}{175}\Omega_{13}\Omega_{15} + \frac{24}{35}\Omega_{14}^2
  + \frac{3277}{175}\Omega_{13}^2 - \frac{16}{175}\Omega_{14}\Omega_{15}
  + \frac{1}{2}\Omega_{15} - \frac{495}{4}\Omega_{12} + \frac{217}{4}\Omega_{13} -9 \Omega_{14}. \label{eq:PICAA3}
\end{eqnarray}
For the frictional heat flux, the magnetized Braginskii $\beta$-coefficients are given by 
\begin{eqnarray}
  \beta_1' &=& \frac{\nu_{ee}}{\nu_{eb}}\Big( \frac{11}{4}\Omega_{22}- \frac{9}{5}\Omega_{12}\Omega_{22}
  - \frac{1}{2}\Omega_{23} + \frac{2}{5}\Omega_{23}\Omega_{12} + \frac{1}{5}\Omega_{13} \Omega_{22} - \frac{2}{35}\Omega_{13} \Omega_{23} \Big) \nn\\
  &&  + \frac{275}{16} - \frac{235}{8}\Omega_{12} + \frac{25}{2}\Omega_{12}^2
  +7 \Omega_{13} - \frac{57}{10} \Omega_{13} \Omega_{12} - \frac{1}{2}\Omega_{14}
  + \frac{2}{5}\Omega_{12}\Omega_{14} + \frac{19}{35}\Omega_{13}^2 - \frac{2}{35}\Omega_{13}\Omega_{14};\nn\\
%=====  
  \beta_0' &=& \beta_0 \Delta^{*2}; \qquad \beta_1'' = \frac{5}{2}-\Omega_{12}; \nn\\
  %====
  \beta_0'' &=& \Big[ \frac{\nu_{ee}}{\nu_{eb}} \Big(\frac{1}{20}\Omega_{22}- \frac{1}{70}\Omega_{23}\Big)
    +  \Big(  \frac{5}{16} -\frac{3}{8}\Omega_{12}+  \frac{19}{140}\Omega_{13} - \frac{1}{70}\Omega_{14}  \Big)  \Big] \nn\\
  && \times \Big[ \frac{\nu_{ee}}{\nu_{eb}} \Big( \frac{35}{2}\Omega_{22} -6 \Omega_{13}\Omega_{22}
    -18 \Omega_{23}+\frac{8}{5}\Omega_{13} \Omega_{23} + \frac{8}{5}\Omega_{23} \Omega_{12}
    +14 \Omega_{12} \Omega_{22} +6 \Omega_{24} - \frac{8}{5}\Omega_{12} \Omega_{24} - \frac{8}{35}\Omega_{13}\Omega_{24} \Big) \nn\\
    &&  - \frac{1225}{8}+ \frac{259}{2}\Omega_{13} + \frac{245}{2}\Omega_{12} +32 \Omega_{13}\Omega_{12}
    - \frac{84}{5}\Omega_{13}^2 -60 \Omega_{14} + \frac{16}{5}\Omega_{13} \Omega_{14}
    -112 \Omega_{12}^2+ \frac{64}{5}\Omega_{12}\Omega_{14} \nn\\
    && \qquad + 6 \Omega_{15} - \frac{8}{5} \Omega_{12}\Omega_{15} - \frac{8}{35} \Omega_{13} \Omega_{15} \Big] \nn\\
  && + \Big[ \frac{\nu_{ee}}{\nu_{eb}} \Big(\frac{2}{35}\Omega_{24} - \frac{3}{10}\Omega_{22} \Big)  
    + \Big( - \frac{35}{8} + \frac{7}{2}\Omega_{12} - \frac{2}{5}\Omega_{14} +\frac{2}{35}\Omega_{15}  \Big) \Big]^2 
   \Big[\frac{5}{2}-\Omega_{12}\Big],
\end{eqnarray}
where the last expression for $\beta_0''$ was re-arranged only partially (otherwise further re-grouping with $\nu_{ee}/\nu_{eb}$ makes this expression too long).
Again, the heat flux expressions above did not simplify much and the original formulation by first calculating the B-coefficients (\ref{eq:Bcoeff})
for a given collisional process and then calculating the Braginskii coefficients (\ref{eq:HFgamma0})-(\ref{eq:HFbeta0}), seems to be more user-friendly.

For the momentum exchange rates, the first three $\alpha$-coefficients are given by 
\begin{eqnarray}
  \alpha_1' &=& \frac{\nu_{ee}}{\nu_{eb}} \Big( \frac{4}{1225} \Omega_{13}^2 \Omega_{24} - \frac{36}{25} \Omega_{23} \Omega_{12}^2 +\frac{4}{25} \Omega_{12}^2 \Omega_{24}
  + \frac{109}{25} \Omega_{12}^2 \Omega_{22} +\frac{11}{175} \Omega_{13}^2 \Omega_{22} -\frac{4}{175} \Omega_{13}^2 \Omega_{23} \nn\\
  && - \frac{26}{25} \Omega_{12} \Omega_{13} \Omega_{22}
  +\frac{64}{175} \Omega_{12} \Omega_{13} \Omega_{23} - \frac{8}{175} \Omega_{13} \Omega_{12} \Omega_{24} +\frac{2}{35} \Omega_{13} \Omega_{24}
  + \frac{3}{2} \Omega_{13} \Omega_{22} +4 \Omega_{23} \Omega_{12} \nn\\
  && - \frac{18}{35} \Omega_{13} \Omega_{23} - \frac{2}{5} \Omega_{12} \Omega_{24} -\frac{127}{10} \Omega_{12} \Omega_{22}
  + \frac{1}{4} \Omega_{24} + \frac{149}{16} \Omega_{22} - \frac{11}{4} \Omega_{23} \Big) \nn\\
%===
  &&   + \frac{3025}{64} - \frac{678}{175} \Omega_{12} \Omega_{13}^2  - \frac{64}{25} \Omega_{12}^2 \Omega_{14}  + \frac{4}{25}\Omega_{12}^2 \Omega_{15}
  + \frac{516}{25} \Omega_{13} \Omega_{12}^2   - \frac{8}{175} \Omega_{13}^2 \Omega_{14} \nn\\
 &&  + \frac{4}{1225}\Omega_{13}^2 \Omega_{15}  - \frac{144}{5} \Omega_{12}^3 + \frac{38}{175} \Omega_{13}^3 
  - \frac{8}{175} \Omega_{12} \Omega_{13} \Omega_{15} + \frac{24}{35} \Omega_{12} \Omega_{13} \Omega_{14}
  - \frac{32}{35} \Omega_{13} \Omega_{14} + \frac{2}{35} \Omega_{13} \Omega_{15} \nn\\
&&  + \frac{719}{140} \Omega_{13}^2 + \frac{419}{4} \Omega_{12}^2 + \frac{34}{5} \Omega_{12} \Omega_{14} - \frac{2}{5} \Omega_{12} \Omega_{15}
  - \frac{533}{10} \Omega_{13} \Omega_{12} + \frac{1}{4} \Omega_{15} - \frac{9}{2} \Omega_{14} - \frac{495}{4} \Omega_{12} +34 \Omega_{13};
\end{eqnarray} 
%======================== 
\begin{eqnarray}  
  \alpha_0' &=& \Delta^* \Big[  \frac{\nu_{ee}}{\nu_{eb}} \Big( \frac{4}{25}\Omega_{23} \Omega_{12}^2 + \frac{4}{175}\Omega_{12}^2 \Omega_{24}
    +\frac{11}{25}\Omega_{12}^2 \Omega_{22} + \frac{4}{175}\Omega_{13}^2 \Omega_{22} - \frac{8}{175}\Omega_{12} \Omega_{13}\Omega_{23}
    -\frac{4}{25}\Omega_{12} \Omega_{13} \Omega_{22} \nn\\
    && \qquad -\frac{4}{35}\Omega_{12} \Omega_{24} - \frac{4}{5}\Omega_{12}\Omega_{22}
    - \frac{2}{5}\Omega_{23} \Omega_{12} + \frac{4}{35}\Omega_{13} \Omega_{23}+ \frac{1}{7}\Omega_{24} + \Omega_{22} \Big) \nn\\
%===    
    &&    +\frac{4}{175} \Omega_{13}^3 - \frac{8}{175} \Omega_{12} \Omega_{13} \Omega_{14}+ \frac{4}{175} \Omega_{12}^2 \Omega_{15}
    - \frac{7}{4}\Omega_{12}^2 + \frac{2}{5} \Omega_{12} \Omega_{14} - \frac{4}{35} \Omega_{12} \Omega_{15} + \Omega_{13}\Omega_{12} \nn\\
    && + \frac{4}{35} \Omega_{13} \Omega_{14} - \frac{19}{35} \Omega_{13}^2 + \frac{1}{7}\Omega_{15}+ \frac{7}{4}\Omega_{13} - \Omega_{14} \Big]; \nn\\
  %====
\alpha_1'' &=& \frac{55}{8}-9 \Omega_{12} + \Omega_{13}+ \frac{16}{5}\Omega_{12}^2 - \frac{4}{5}\Omega_{13}\Omega_{12} + \frac{2}{35}\Omega_{13}^2.  
\end{eqnarray}
The last fourth coefficient $\alpha_0''$ given by (\ref{eq:HFalpha0}) becomes way too long after the re-grouping,
and calculation through the B-coefficients (\ref{eq:Bcoeff}) is much easier.

\newpage
%=============================================================================================================
%=======================================================================================
\subsubsection{Braginskii coefficients for inverse power-law force}
For the inverse power law force $F_{ab}=\pm |K_{ab}|/r^\nu$, the B-coefficients (\ref{eq:Bcoeff}) become
\begin{eqnarray}
B_1 &=&  \frac{\nu_{ee}}{\nu_{eb}} \frac{A_2(\nu)}{A_1(\nu)} \frac{(3\nu-5)(\nu+3)}{5 (\nu-1)^2}  + \frac{(3\nu^3+67\nu^2-191\nu+185)}{20(\nu-1)^3};\nn\\
B_2 &=&  - \, \frac{\nu_{ee}}{\nu_{eb}} \frac{A_2(\nu)}{A_1(\nu)} \frac{(3\nu-5) (\nu-5) }{140(\nu-1)^2}  - \frac{(\nu-5) (23 \nu^2-62 \nu+55)}{560 (\nu-1)^3};\nn\\
B_3 &=&  - \, \frac{\nu_{ee}}{\nu_{eb}} \frac{A_2(\nu)}{A_1(\nu)} \frac{2 (3 \nu-5) (29 \nu^2-122 \nu+109) }{5 (\nu-1)^3}
- \frac{(\nu-5) (15 \nu-19)  (23 \nu^2-62 \nu+55)}{10 (\nu-1)^4};\nn\\
B_4 &=&  \frac{\nu_{ee}}{\nu_{eb}}\frac{A_2(\nu)}{A_1(\nu)} \frac{ (3 \nu-5) (59 \nu^2-190 \nu+147)}{70 (\nu-1)^3}
+ \frac{(755 \nu^4 -5396 \nu^3 +13898 \nu^2 -16036 \nu+7035)}{280 (\nu-1)^4};\nn\\
B_5 &=& -\, \frac{(\nu-5)}{2 (\nu-1)} ; \qquad B_6 = -\, \frac{(\nu-5) (13 \nu-17) }{(\nu-1)^2}, \label{eq:Bcoeff-nu}
\end{eqnarray}
and the Braginskii $\gamma$-coefficients are obtained by (\ref{eq:HFgamma0}),
the $\beta$-coefficients by (\ref{eq:HFbeta0}) and the $\alpha$-coefficients by (\ref{eq:HFalpha0}).
The values of $A_l(\nu)$ are given in Table \ref{Table:Anu} and the
ratios of collisional frequencies $\nu_{ee}/\nu_{eb}$ are given by (\ref{eq:rit1}). Note that for the Maxwell molecules ($\nu=5$), the coefficients
$B_5$ and $B_6$ become zero and thus all of the $\beta$-coefficients (\ref{eq:HFbeta0}) and (\ref{eq:Thierry61P}) become zero as well,
eliminating the frictional heat fluxes $\vecq_e^u$ and $\vecX^{(5)u}_e$. Also, for the Maxwell molecules the coefficient $\alpha_0=1$ and all of the other $\alpha$-coefficients
(\ref{eq:HFalpha0}) become zero, yielding a simple (isotropic) momentum exchange rates $\boldsymbol{R}_e = - \rho_e\nu_{eb} \delta\bu$. 

Expressing the transport coefficients explicitly, the coefficient of the parallel thermal conductivity reads
\begin{empheq}[box=\fbox]{align}
  \gamma_0 &= \frac{5}{2 \Delta^*} \frac{1}{(\nu-1)^3}\Big[ \frac{\nu_{ee}}{\nu_{eb}} \frac{A_2(\nu)}{A_1(\nu)} \frac{(3\nu-5)(45\nu^2-106\nu+77)}{70}
    + \frac{(433 \nu^4-2596 \nu^3+6310 \nu^2-7076 \nu+3185)}{280(\nu-1)} \Big];\nn\\
  \Delta^* &= \frac{(3\nu-5)}{(\nu-1)^3}\Big[ \Big( \frac{\nu_{ee}}{\nu_{eb}} \frac{A_2(\nu)}{A_1(\nu)} \Big)^2 \frac{4(11\nu-13)(3\nu-5)}{175}
  + \frac{\nu_{ee}}{\nu_{eb}} \frac{A_2(\nu)}{A_1(\nu)}   \frac{(243 \nu^3-1077 \nu^2+1745 \nu-975)}{175(\nu-1)} \nn\\
  & \qquad + \frac{(425 \nu^4-2516\nu^3+6118\nu^2-7156\nu+3385)}{700(\nu-1)^2} \Big], \label{eq:Extreme1}
\end{empheq}
where the $(\nu-1)^3$ cancells out, but we kept this form to be able to use the $\Delta^*$ in the expressions below. 
The electron heat flux of \cite{Braginskii1965} is obtained by prescribing $\nu=2$. 
The other two parallel coefficients are given by
\begin{eqnarray}
  \beta_0 &=& \frac{1}{\Delta^*} \frac{(\nu-5)}{(\nu-1)^3} \frac{(3\nu-5)}{70}
  \Big[ -\, \frac{\nu_{ee}}{\nu_{eb}} \frac{A_2(\nu)}{A_1(\nu)} (23\nu-31)
    - \frac{(19 \nu^2-62\nu+59)}{(\nu-1)} \Big];\nn\\
  \alpha_0 &=& 1- \frac{(\nu-5)^2 (3\nu-5)}{175 (\nu-1)^4 \Delta^*}
  \Big[\frac{\nu_{ee}}{\nu_{eb}} \frac{A_2(\nu)}{A_1(\nu)} 2(6\nu-7)  + \frac{(41 \nu^2-122\nu+97)}{4 (\nu-1)}\Big].
\end{eqnarray}

The rest of coefficients for the thermal heat flux read
\begin{eqnarray}
  \gamma_1' &=&  \frac{1}{(\nu-1)} \Big[\frac{\nu_{ee}}{\nu_{eb}} \frac{A_2(\nu)}{A_1(\nu)} (3\nu-5)
  + \frac{(13\nu^2-42\nu+45)}{4(\nu-1)} \Big];\qquad   \gamma_0' = \gamma_0 \Delta^{*2};
  \qquad \gamma_1'' = \frac{5}{2}; \nn\\
  \gamma_0'' &=& \frac{1}{(\nu-1)^6} \Big[ \Big(\frac{\nu_{ee}}{\nu_{eb}} \frac{A_2(\nu)}{A_1(\nu)} \Big)^2
    \frac{(3\nu-5)^2 (2053 \nu^4-9876 \nu^3+19454 \nu^2-18004 \nu+6629)}{1960} \nn\\
    && + \frac{\nu_{ee}}{\nu_{eb}} \frac{A_2(\nu)}{A_1(\nu)}  \frac{ (3\nu-5) (20129 \nu^6-172182\nu^5+644463 \nu^4-1324148\nu^3+1570351\nu^2-1018262\nu+283745)}{3920 (\nu-1)} \nn\\
    && + \frac{1}{31360 (\nu-1)^2 } \Big( 202301 \nu^8-2505736 \nu^7+14021772 \nu^6-45784056 \nu^5+95060494 \nu^4 \nn\\
    && \qquad -128340920 \nu^3+110080076 \nu^2-54930120 \nu+12261725 \Big) \Big], 
\end{eqnarray}
together with
\begin{eqnarray}
  \delta_0 &=& \Delta^{*2};\nn\\
  \delta_1 &=& \frac{1}{(\nu-1)^6} \Big[ \Big(\frac{\nu_{ee}}{\nu_{eb}} \frac{A_2(\nu)}{A_1(\nu)} \Big)^2
    \frac{(3\nu-5)^2 (2865 \nu^4 -13348 \nu^3+25446 \nu^2-22820 \nu+8113)}{4900 } \nn\\
    && + \frac{\nu_{ee}}{\nu_{eb}} \frac{A_2(\nu)}{A_1(\nu)} \frac{ (3\nu-5) (30965 \nu^6-255342 \nu^5+924603 \nu^4-1840580 \nu^3+2115035 \nu^2-1328110 \nu+357525)}{9800 (\nu-1)} \nn\\
    && + \frac{1}{78400 (\nu-1)^2} \Big( 349609 \nu^8 -4149448 \nu^7 +22359612 \nu^6 -70510072 \nu^5 +141608310 \nu^4 \nn\\
    && \qquad -185034872 \nu^3 +153575612 \nu^2 -74100040 \nu+15966825 \Big) \Big]. 
\end{eqnarray}
%================
For the frictional heat flux the coefficients become
\begin{eqnarray}
  \beta_1' &=& \frac{(\nu-5)}{(\nu-1)^4} \Big[ -\, \frac{\nu_{ee}}{\nu_{eb}} \frac{A_2(\nu)}{A_1(\nu)} \frac{(27 \nu^2-54\nu+43) (3\nu-5) }{140}
    - \, \frac{(341 \nu^4-1796 \nu^3+4142 \nu^2-4516 \nu+2085)}{560 (\nu-1)} \Big]; \nn\\
  \beta_0' &=& \beta_0 \Delta^{*2}; \nn \qquad  \beta_1'' = -\, \frac{(\nu-5)}{2(\nu-1)};\\
  %=====
  \beta_0'' &=& \frac{(\nu-5)(3\nu-5)}{(\nu-1)^6} \Big[
    -\, \Big(\frac{\nu_{ee}}{\nu_{eb}} \frac{A_2(\nu)}{A_1(\nu)} \Big)^2 \frac{  (3\nu-5) (1063 \nu^3 -4029 \nu^2 +5365 \nu-2527) }{4900} \nn\\
    && - \, \frac{\nu_{ee}}{\nu_{eb}} \frac{A_2(\nu)}{A_1(\nu)} \frac{(21367 \nu^5-159667 \nu^4+492358 \nu^3-776006 \nu^2+624947 \nu-207095)}{19600 (\nu-1)} \nn\\
    && -\, \frac{(9193 \nu^6-87790 \nu^5+359407 \nu^4-801188 \nu^3+1025783 \nu^2-716174 \nu+214865)}{19600 (\nu-1)^2} \Big].
\end{eqnarray}
%===============
Finally for the momentum exchange rates the coefficients read
\begin{eqnarray}
  \alpha_1' &=& \frac{(\nu-5)^2}{(\nu-1)^7}\Big[ \frac{\nu_{ee}}{\nu_{eb}} \frac{A_2(\nu)}{A_1(\nu)} \frac{(3 \nu-5)(773 \nu^4-2748 \nu^3+5110 \nu^2-5196 \nu+2317)}{19600} \nn\\
    && + \frac{(9337 \nu^6-65070 \nu^5+218687 \nu^4 -430500 \nu^3+518439 \nu^2-356942 \nu+110145)}{78400 (\nu-1)} \Big];\nn\\
  \alpha_0' &=& \frac{(\nu-5)^2 (3 \nu-5)^2}{(\nu-1)^7}\Big[ \Big(\frac{\nu_{ee}}{\nu_{eb}} \frac{A_2(\nu)}{A_1(\nu)} \Big)^3
    \frac{8 (6 \nu-7)(11 \nu-13) (3 \nu-5) }{30625 } \nn\\
    && + \Big(\frac{\nu_{ee}}{\nu_{eb}} \frac{A_2(\nu)}{A_1(\nu)} \Big)^2 \frac{(4269 \nu^4-24206 \nu^3+53352 \nu^2-53178 \nu+19955)}{30625 (\nu-1)} \nn\\
    && + \frac{\nu_{ee}}{\nu_{eb}} \frac{A_2(\nu)}{A_1(\nu)} \frac{(15063 \nu^5-109945 \nu^4+335150 \nu^3-528858 \nu^2+429019 \nu-141965)}{ 122500 (\nu-1)^2}\nn\\
    && + \frac{ (41 \nu^2-122 \nu+97)(425 \nu^4-2516\nu^3+6118 \nu^2-7156 \nu+3385)}{490000 (\nu-1)^3 } \Big],
\end{eqnarray}
together with
\begin{eqnarray}
  \alpha_1'' &=& \frac{(\nu-5)^2 (29 \nu^2-50 \nu+37)}{280 (\nu-1)^4};\nn\\
  \alpha_0'' &=& \frac{(\nu-5)^2 (3 \nu-5)^2}{(\nu-1)^6}\Big[ \Big(\frac{\nu_{ee}}{\nu_{eb}} \frac{A_2(\nu)}{A_1(\nu)} \Big)^2     \frac{(557 \nu^2-1482 \nu+989)}{12250 } \nn\\
    && + \frac{\nu_{ee}}{\nu_{eb}} \frac{A_2(\nu)}{A_1(\nu)} \frac{(479 \nu^3-2197 \nu^2+3517 \nu-1927)}{6125 (\nu-1)} 
     + \frac{4 (53 \nu^4-347 \nu^3+885 \nu^2-1037 \nu+478)}{6125 (\nu-1)^2} \Big].
\end{eqnarray}
This concludes the formulation of the model for the inverse force. As previously, only the parallel coefficients $\alpha_0, \beta_0, \gamma_0$ are
  valid for any power-law index $\nu$ and unfortunatelly, the magnetized coefficients are valid only for $\nu=2$, nevertheless,
  we will use these equations in the next Section \ref{eq:ElectronMod} to obtain the Braginskii model for moderately-coupled plasmas.
 To obtain more general results, one needs to consider
coupling between charged particles and neutrals, where the results will of course become more complicated.

%=========================================
The Lorentz approximation is obtained easily from all of the coefficients and for example the parallel heat flux reads
\begin{equation}
  \frac{\nu_{ee}}{\nu_{eb}}\ll 1;\qquad
  \gamma_0 = \frac{25}{4} \frac{(\nu-1)(433 \nu^4-2596 \nu^3+6310 \nu^2-7076 \nu+3185)}{(3\nu-5)(425\nu^4-2516\nu^3+6118\nu^2-7156\nu+3385)}.
\end{equation}
Fixing the ratio of collisional frequencies into $\nu_{ee}/\nu_{eb}=1/\sqrt{2}$, the Braginskii parameters for various repulsive forces are compared
in Table \ref{eq:TableCH}.
%===========================================================================
\begin{table}[ht!]
\centering 
\begin{tabular}{| c | c | c | c |} 
  \hline
                 & $\alpha_0$  & $\beta_0$   &  $\gamma_0$ \\
  \hline
  $\nu=2$        & 0.513       &  0.711     &  3.1616 \\
  $\nu=3$        & 0.936       &  0.280     &  1.644 \\
  $\nu=5$        & 1           &  0         &  1.445 \\
  $\nu=13$       & 0.972       &  - 0.196    &  1.413  \\
  $\nu=\infty$   & 0.936       &  - 0.298    &  1.4316 \\
  \hline 
\end{tabular} 
\caption{Comparison of Braginskii parallel coefficients for the repulsive force $K_{ab}/r^\nu$, where the
  ratio of collisional frequencies is $\nu_{ee}/\nu_{eb}=1/\sqrt{2}$. For the Coulomb collisions ($\nu=2$), it does not matter that the ``e-i'' collisions
  are attractive, because the numbers are the same. For the other $\nu$ cases, both the ``e-e'' and ``e-b'' collisions are repulsive.
 One can create similar table for attractive forces, by using results of Section \ref{sec:AlnuA}.}
\label{eq:TableCH}
\end{table}

%\newpage
%=============================================================================================================
%=======================================================================================
\subsubsection{Braginskii \texorpdfstring{($\nu=2$)}{} electron coefficients for moderately-coupled plasmas} \label{eq:ElectronMod}
 Here we consider the Coulomb case $\nu=2$ and summarize the Braginskii electron coefficients for plasmas, where the Coulomb logarithm $\ln\Lambda$ is not necessarily large. 
For the brevity of expressions, we use a simple notation
\begin{equation}
A \equiv \frac{\nu_{ee}}{\nu_{ei}} \frac{A_2(2)}{A_1(2)} = \frac{1}{Z_i \sqrt{2}} \frac{A_2(2)}{A_1(2)},
\end{equation}
where as a reminder the $Z_i$ is the ion charge and  
\begin{equation*}
  A_1(2) = \ln(\Lambda^2+1);  \qquad A_2(2) =  2\ln(\Lambda^2+1)-2+\frac{2}{\Lambda^2+1}.
\end{equation*}
The parallel electron Braginskii coefficients are then given by
\begin{eqnarray}
  \alpha_0 = \frac{4 (16 +61 A+ 36 A^2)}{217 +604 A +144 A^2};\qquad
  \beta_0 = \frac{30 (11+15 A)}{217 +604 A +144 A^2};\qquad
  \gamma_0 = \frac{25(433+180 A)}{4 (217 +604 A +144 A^2)},
\end{eqnarray}
and the perpendicular coefficients read
\begin{eqnarray}
  \delta_0 &=& \Big(\frac{217 +604 A +144 A^2}{700}\Big)^2; \qquad
  \delta_1 = \frac{3313}{4900}A^2+\frac{41269}{9800}A+ \frac{586601}{78400};
\end{eqnarray}
\begin{eqnarray}
  \alpha_1' &=& \frac{24741}{19600}A + \frac{363033}{78400}; \qquad
  \alpha_0' = \frac{9 (217 +604 A +144 A^2)(40 A+17)}{490000}; \nn\\
  \alpha_1'' &=& \frac{477}{280};\qquad 
  \alpha_0'' = \frac{2277}{12250}A^2+ \frac{1359}{6125}A + \frac{576}{6125};
\end{eqnarray}
\begin{eqnarray}
  \beta_1' &=& \frac{129}{140}A + \frac{2127}{560}; \qquad
  \beta_0' = \frac{3 (217 +604 A +144 A^2)(15 A+11)}{49000}; \nn\\
  \beta_1'' &=& \frac{3}{2}; \qquad \beta_0'' = \frac{1773}{4900}A^2+\frac{20133}{19600}A+\frac{17187}{19600};
\end{eqnarray}
\begin{eqnarray}
  \gamma_1' &=& A+\frac{13}{4}; \qquad \gamma_0' = \frac{(217 +604 A +144 A^2)(180 A+433)}{78400};\nn\\
  \gamma_1'' &=& \frac{5}{2}; \qquad \gamma_0'' = \frac{2277}{1960}A^2+\frac{25281}{3920}A+\frac{320797}{31360}.
\end{eqnarray}
 The above expressions represent the Braginskii electron heat fluxes and momentum exchange rates for moderatelly-coupled plasmas. In the limit when the Coulomb logarithm becomes
large, the coefficient $A=\sqrt{2}/Z_i$ and one recovers expressions (56)-(60) of \cite{Hunana2022} for weakly-coupled plasmas.

%\newpage
%===========================================================================
\subsubsection{Reduction into 1-Hermite approximation}
In the 1-Hermite approximation, the heat flux evolves according to
\begin{equation}
  \frac{d_e}{d t}\vecq_e + \Omega_e \bhat\times\vecq_e + \frac{5}{2}p_e \nabla \Big(\frac{p_e}{\rho_e}\Big) = \nu_{eb} \Big[ - B_1 \vecq_e  + B_5  p_e \delta\bu \Big];
\end{equation}
with coefficients
\begin{equation}
  B_1 = \Big[ \frac{\nu_{ee}}{\nu_{eb}} \frac{2}{5}\Omega_{22} +\frac{5}{2} -2\Omega_{12} +\frac{2}{5}\Omega_{13} \Big]; \qquad
  B_5 = \big( \frac{5}{2} -\Omega_{12} \big).
\end{equation}
As a quick double-check, prescribing Coulomb collisions recovers eqs. (H40)-(H41) of Part 1. The solutions are
\begin{eqnarray}
  \vecq_e^T &=& -\kappa_\parallel^e \nabla_\parallel T_e - \kappa_\perp^e \nabla_\perp T_e + \kappa_\times^e \bhat\times\nabla T_e;\nn\\
  %=====
&& \kappa_\parallel^e = \frac{5}{2 B_1} \frac{p_e}{m_e \nu_{eb}}; \qquad   
\kappa_\perp^e =    \frac{5}{2} \frac{B_1}{(x^2+B_1^2)} \frac{p_e}{m_e \nu_{eb}};
\qquad \kappa_\times^e =  \frac{5}{2} \frac{x}{(x^2+B_1^2)}\frac{p_e}{m_e \nu_{eb}}; \label{eq:Thierry63AAm}\\
%=====  
\vecq_e^u &=&  \frac{B_5}{B_1} p_e \delta\bu_\parallel + p_e \delta\bu_\perp \frac{B_5 B_1}{(x^2+B_1^2)}  
- p_e \bhat\times\delta\bu \frac{B_5 x}{(x^2+B_1^2)}. \label{eq:Thierry63AA}
\end{eqnarray}
Note that for both the Coulomb collisions and hard spheres $B_1= (4/5)(\nu_{ee}/\nu_{eb}) + (13/10)$ and the entire thermal heat flux is the same. 
The 1-Hermite parallel thermal conductivity reads
\begin{eqnarray}
\big[ \gamma_0 \big]_1 &=& \frac{(25/4)}{ (\frac{\nu_{ee}}{\nu_{eb}} \Omega_{22} + (25/4) -5\Omega_{12} +\Omega_{13})}; \label{eq:qe1Herm} \nn\\
\big[ \gamma_0 \big]_1 &=& \frac{25}{4} \Big[ \frac{\nu_{ee}}{\nu_{eb}}\frac{A_2(\nu)}{A_1(\nu)}  \frac{(3\nu-5)}{(\nu-1)} + \frac{(13\nu^2-42\nu+45)}{4(\nu-1)^2} \Big]^{-1},
\end{eqnarray}
where the
dimensionless $\gamma_0$ coefficient is defined as $\kappa_\parallel^e=\gamma_0 p_e/(\nu_{eb} m_e)$.

%\newpage
%===============================================================
\subsubsection{Improvement of the 2-Hermite approximation}
For a general collisional process, the improvement of the 2-Hermite approximation for the parallel thermal conductivity reads
\begin{eqnarray}
  \big[ \kappa_\parallel^e \big]_2 &=& \big[ \kappa_\parallel^e \big]_1
  \Big\{1+
  \Big( (28 \Omega_{22}-8\Omega_{23})\frac{\nu_{ee}}{\nu_{eb}} -210\Omega_{12}+76\Omega_{13}-8 \Omega_{14}+175 \Big)^2 \nn\\
  && \times \Big[ \Big(448 \Omega_{22}^2+64 \Omega_{22}\Omega_{24}-64 \Omega_{23}^2 \Big) \Big(\frac{\nu_{ee}}{\nu_{eb}}\Big)^2
    +\Big( (1232 \Omega_{22}+768 \Omega_{23}+64 \Omega_{24}) \Omega_{13} \nn\\
  && \qquad  + (-2240 \Omega_{22}-1120 \Omega_{23}-320 \Omega_{24}) \Omega_{12} +(-448 \Omega_{14}+64 \Omega_{15}+2800) \Omega_{22} \nn\\
  && \qquad  -128 \Omega_{23}\Omega_{14}+400 \Omega_{24}\Big) \frac{\nu_{ee}}{\nu_{eb}} 
    -1520 \Omega_{13}^2 +(2800 \Omega_{12}+320 \Omega_{14}+64 \Omega_{15}+4900) \Omega_{13} \nn\\
  && \qquad  -4900 \Omega_{12}^2 +(1120 \Omega_{14}-320 \Omega_{15}) \Omega_{12}   -64 \Omega_{14}^2-2800 \Omega_{14}+400 \Omega_{15} \Big]^{-1} \Big\},
\end{eqnarray}
and for the inverse power-law force
\begin{eqnarray}
  \big[ \kappa_\parallel^e \big]_2 &=& \big[ \kappa_\parallel^e \big]_1
  \Big\{1+ (\nu-5)^2 \Big( 4(\nu-1)(3\nu-5)\frac{A_2(\nu)}{A_1(\nu)} \frac{\nu_{ee}}{\nu_{eb}}+23 \nu^2-62 \nu+55 \Big) \nn\\
  && \times \Big[ 16(11\nu-13) (3 \nu-5)(\nu-1)^2 \Big(\frac{A_2(\nu)}{A_1(\nu)} \frac{\nu_{ee}}{\nu_{eb}} \Big)^2
    +4 (\nu-1)(243 \nu^3-1077 \nu^2+1745 \nu-975) \frac{A_2(\nu)}{A_1(\nu)} \frac{\nu_{ee}}{\nu_{eb}}\nn\\
    && +425 \nu^4-2516 \nu^3+6118 \nu^2-7156 \nu+3385  \Big]^{-1}\Big\},
\end{eqnarray}
and for the two particular cases
\begin{eqnarray}
  \textrm{Coulomb collisions:}\qquad  \big[ \kappa_\parallel^e \big]_2 &=& \big[ \kappa_\parallel^e \big]_1
  \Big(1+  \frac{9 (8\frac{\nu_{ee}}{\nu_{eb}} +23)^2}{4 \big( 576 \big(\frac{\nu_{ee}}{\nu_{eb}} \big)^2  +1208 \frac{\nu_{ee}}{\nu_{eb}}  +217 \big) }   \Big); \label{eq:wow21}\\
%===
  \textrm{Hard spheres:}\qquad \big[ \kappa_\parallel^e \big]_2 &=& \big[ \kappa_\parallel^e \big]_1
  \Big(1+  \frac{ (8\frac{\nu_{ee}}{\nu_{eb}} +23)^2}{4 \big( 704 \big(\frac{\nu_{ee}}{\nu_{eb}}\big)^2 +1944 \frac{\nu_{ee}}{\nu_{eb}} +1275 \big) }  \Big).\label{eq:wow22}
\end{eqnarray}
To get a better sense about the improvement of (\ref{eq:wow21})-(\ref{eq:wow22}), it is again useful to consider two particular cases,
where in the first case the ratio of collisional frequencies is fixed as 
\begin{eqnarray}
  \frac{\nu_{ee}}{\nu_{eb}} = \frac{1}{\sqrt{2}}; \qquad \textrm{Coulomb collisions:}
  \qquad \big[\gamma_0 \big]_2 &=& \frac{25 (180\sqrt{2}+433)}{4 (505+604\sqrt{2})} = 3.1616;
  \qquad \big[\gamma_0 \big]_2 = \big[\gamma_0 \big]_1 \big( 1+ 1.3594 \big);  \nn\\
  \textrm{Hard spheres:} \qquad \big[ \gamma_0 \big]_2 &=& \frac{25 (180 \sqrt{2}+433)}{4 (1627+972 \sqrt{2})} = 1.4316;
  \qquad \big[\gamma_0 \big]_2 = \big[\gamma_0 \big]_1 \big( 1+ 0.06840 \big);\nn\\
  \textrm{Coulomb collisions \& Hard spheres:} \qquad 
\big[ \gamma_0 \big]_1 &=& \frac{25\sqrt{2}}{8  +13\sqrt{2} } = 1.34. \label{eq:wow20}
\end{eqnarray}  
Note the relatively large improvement of 136\% for the Coulomb collisions, whereas the improvement for the hard spheres is less than 7\%.
Finally, considering the extreme case of the Lorentz approximation yields
\begin{eqnarray}
  \frac{\nu_{ee}}{\nu_{eb}} \ll 1; \qquad \textrm{Coulomb collisions:} \qquad \big[\gamma_0 \big]_2 &=& \frac{10825}{868} = 12.47;
  \qquad \big[\gamma_0 \big]_2 = \big[\gamma_0 \big]_1 \big( 1+ 5.485 \big)\nn\\
  \textrm{Hard spheres:} \qquad \big[\gamma_0 \big]_2 &=& \frac{433}{204} = 2.12;
  \qquad \quad\,\, \big[\gamma_0 \big]_2 = \big[\gamma_0 \big]_1 \big( 1+ 0.104 \big)\nn\\
  \textrm{Coulomb collisions \& Hard spheres:} \qquad 
\big[ \gamma_0 \big]_1 &=& \frac{25}{13} = 1.923.   \label{eq:Fuf1}
\end{eqnarray}
Here the improvement for the Coulomb collisions is larger than 500\%, whereas the improvement for the hard spheres still remains
a marginal 10\%.

As noted previously, comparing results in the Lorentz approximation can be slightly confusing, because by using the collisional frequency $\nu_{eb}$
and considering full expressions yields
\begin{equation}
\textrm{Coulomb collisions:}\qquad  \kappa_\parallel^e = \frac{3}{4  \sqrt{2\pi}} \frac{ T_e^{5/2}}{ \sqrt{ m_e}  e^4 \ln\Lambda}  \, \frac{\gamma_0}{Z_b}; \
\qquad \,\,\,\, \big[\gamma_0 \big]_2 =  \frac{25}{4} \frac{ Z_b (433 Z_b+180 \sqrt{2})}{( 217 Z_b^2 +604  Z_b \sqrt{2}  +288 )}.
\end{equation}
So as the ion charge $Z_b$ increases, the parameter $\gamma_0$ converges to a constant value, but the entire conductivity $\kappa_\parallel^e$
actually decreases to zero. Similarly, for the hard spheres
(let us for simplicity use only the 1-Hermite approximation)
\begin{equation}
\textrm{Hard spheres:}\qquad  \kappa_\parallel^e = \frac{3}{8\sqrt{2\pi}} \frac{T_e^{1/2}}{\sqrt{m_e}} \, \frac{ n_e \gamma_0 }{n_b (r_{e}+r_b)^2}; \qquad 
\big[\gamma_0 \big]_1 = \frac{25}{\frac{32}{\sqrt{2}}\frac{r_e^2 n_e}{(r_{e}+r_b)^2 n_b} +13   },
\end{equation}  
and as the radius $r_b$ increases, the parameter $\gamma_0$ converges to a constant value, but the entire conductivity $\kappa_\parallel^e$
decreases to zero. 
%Nevertheless, the Lorentz approximation and the numbers in (\ref{eq:Fuf1}) are still meaningfull and should be just viewed as an asymptotic trend
%when the ``a-b'' collisions start to dominate. If one dislikes the Lorentz approximation, one can just calculate the results for some finite $Z_b$,
%see for example Table 10, p. 105 in \cite{Hunana2022} where the heat conductivity is compared.\\

\newpage
%==================================================================================================================
%==================================================================================================================
\section{Case  \texorpdfstring{$m_a \gg m_b$}{ma>>mb} (heavyweight particles such as ions)} \label{sec:heavy}
Here we assume $m_a \gg m_b$ and $T_a\simeq T_b$ and consider heavyweight particles ``a'', which in addition to self-collisions 
also collide with much lighter particles ``b''. For the Coulomb collisions,
this section can be viewed as an improved Braginskii ion species, where the ion-electron collisions are retained. 
In comparison to the self-collisional Section \ref{sec:ions}, the results will also contain corrections expressed through
the ratio of collisional frequencies
\begin{eqnarray}
  \textrm{Coulomb collisions}: \qquad  \frac{\nu_{ie}}{\nu_{ii}} &=& \frac{\sqrt{2}}{Z_i} \sqrt{\frac{m_e}{m_i}}; \label{eq:wow30}\\
  \textrm{Hard spheres}: \qquad \frac{\nu_{ab}}{\nu_{aa}} &=& \sqrt{2} \frac{(r_a+r_b)^2}{4r_a^2} \frac{n_b}{n_a}\sqrt{\frac{m_b}{m_a}}; \label{eq:wow30H}\\
  \textrm{Inverse power}: \qquad \frac{\nu_{ab}}{\nu_{aa}} &=& \sqrt{2} \Big( \frac{K_{ab}}{K_{aa}}\Big)^{\frac{2}{\nu-1}} \frac{n_b}{n_a} \sqrt{\frac{m_b}{m_a}},
\end{eqnarray}
where for the Coulomb collisions, for additional clarity we changed the indices to ions ``i'' and electrons ``e''. The above ratios might appear
to be small at first sight, where for example for the proton-electron plasma the $\sqrt{2m_e/m_p}=0.03300$, but as already reported by
\cite{JiHeld2013,JiHeld2015} for the Coulomb collisions, the final correction for the Braginskii
parallel ion viscosity is 8\% and for the parallel ion thermal conductivity is 18\%, which is not insignificant. Here we obtain the improved viscosities and
thermal conductivities through the Chapman-Cowling integrals. 
  
%===================================================
\subsection{Heavyweight viscosity}
Starting with the viscosity, the relevant mass-ratio coefficients (\ref{eq:wow14})-(\ref{eq:wow15}) for $m_a\gg m_b$ simplify into 
\begin{eqnarray}
 K_{ab(1)} &=& 2 ;\qquad  L_{ab(1)} = \frac{m_b}{ m_a} \Big( 1 - \frac{2}{5} \Omega_{12} \Big)\ll 1; \qquad
  {M}_{ab (1)} = \frac{70}{5}; \qquad {N}_{ab (1)} = 4, 
\end{eqnarray}
and the evolution equations read
\begin{eqnarray}
   \frac{d_a}{dt} \bPi^{(2)}_a  +\Omega_a \big(\bhat\times \bPi^{(2)}_a \big)^S + p_a \bW_a
  &=&  - \Big[ \frac{3}{5}\nu_{aa} \Omega_{22} +2 \nu_{ab} \Big]\bPi_a^{(2)} \nn\\
&&  + \nu_{aa} \Big(\frac{3}{20} \Omega_{22} - \frac{3}{70} \Omega_{23} \Big)  \Big( \frac{\rho_a}{p_a}\bPi^{(4)}_a- 7 \bPi^{(2)}_a\Big); \label{eq:Energy32H}\\
%=============  
   \frac{d_a}{dt} \bPi^{(4)}_a  +\Omega_a \big(\bhat\times \bPi^{(4)}_a \big)^S + 7 \frac{p_a^2}{\rho_a} \bW_a 
  &=&  -  \Big[ \nu_{aa} \big( \frac{21}{10}\Omega_{22} +\frac{3}{5}\Omega_{23}\big) +\frac{70}{5}\nu_{ab}  \Big]\frac{p_a}{\rho_a}\bPi_a^{(2)} \nn\\
  && - \Big[  \nu_{aa} ( \frac{1}{40} \Omega_{22}+\frac{3}{70} \Omega_{24}) + 4 \nu_{ab} \Big] \Big( \bPi^{(4)}_a- 7 \frac{p_a}{\rho_a} \bPi^{(2)}_a\Big), \label{eq:Energy31H}
\end{eqnarray}
which can be rewritten as
\begin{eqnarray}
   \frac{d_a}{dt} \bPi^{(2)}_a  +\Omega_a \big(\bhat\times \bPi^{(2)}_a \big)^S + p_a \bW_a
  &=&  - \nu_{aa} V_1 \bPi_a^{(2)}   + \nu_{aa} V_2   \frac{\rho_a}{p_a}\bPi^{(4)}_a ; \label{eq:Energy32HH}\\
%=============  
   \frac{d_a}{dt} \bPi^{(4)}_a  +\Omega_a \big(\bhat\times \bPi^{(4)}_a \big)^S + 7 \frac{p_a^2}{\rho_a} \bW_a 
  &=&  -  \nu_{aa} V_3 \frac{p_a}{\rho_a}\bPi_a^{(2)}  + \nu_{aa} V_4 \bPi^{(4)}_a, \label{eq:Energy31HH}
\end{eqnarray}
with the coefficients
\begin{eqnarray}
  V_1 &=& \Big[ \frac{33}{20}\Omega_{22} - \frac{3}{10}\Omega_{23}  +2 \frac{\nu_{ab}}{\nu_{aa}} \Big];\qquad
  V_2 = \Big[\frac{3}{20} \Omega_{22} - \frac{3}{70} \Omega_{23} \Big];\nn\\
  V_3 &=& \Big[ \frac{77}{40}\Omega_{22}+ \frac{3}{5}\Omega_{23} - \frac{3}{10}\Omega_{24} -14 \frac{\nu_{ab}}{\nu_{aa}}  \Big];\qquad
  V_4 = - \Big[    \frac{1}{40} \Omega_{22}+\frac{3}{70} \Omega_{24} + 4 \frac{\nu_{ab}}{\nu_{aa}} \Big].
\end{eqnarray}
For clarity, neglecting the corrections $\nu_{ab}/\nu_{aa}$ recovers the self-collisional system (\ref{eq:Energy32B}), which in the quasi-static
approximation yielded the self-collisional stress-tensor (\ref{eq:Energy081})-(\ref{eq:beau033}).  
Here the viscosities of the stress-tensor $\bPi_a^{(2)}$ are given by 
\begin{eqnarray}
  \eta_0^a  &=& \frac{p_{a}}{\nu_{aa}} \frac{(7V_2-V_4)}{(V_2 V_3 - V_1 V_4)}; \label{eq:beau33xH} \\
  \eta_2^a  &=& \frac{p_{a}}{\nu_{aa}\Delta}\, \Big[ x^2 (V_1 -7 V_2) +(7V_2-V_4)(V_2 V_3 - V_1 V_4)\Big]; \nn\\
  \eta_4^a  &=& \frac{p_{a}}{\nu_{aa} \Delta}\, \Big[ x^3 + x (7 V_1 V_2 -V_2 V_3 -7 V_2 V_4 +V_4^2) \Big]; \nn\\
  \Delta &=& x^4 + x^2 (V_1^2 -2 V_2 V_3 +V_4^2) + (V_2 V_3 - V_1 V_4)^2 , \nn 
\end{eqnarray}
where $x=\Omega_a/\nu_{aa}$ and re-grouping with the $\nu_{ab}/\nu_{aa}$ yields
\begin{eqnarray}
  7V_2 - V_4 &=& \frac{43}{40}\Omega_{22} - \frac{3}{10}\Omega_{23}+ \frac{3}{70}\Omega_{24} + 4 \frac{\nu_{ab}}{\nu_{aa}};\nn\\
  V_2 V_3 - V_1 V_4 &=&   \frac{33}{100}\Omega_{22}^2 + \frac{9}{350}\Omega_{22}\Omega_{24} - \frac{9}{350}\Omega_{23}^2
  +\frac{\nu_{ab}}{\nu_{aa}} \Big( \frac{91}{20}\Omega_{22} - \frac{3}{5}\Omega_{23} + \frac{3}{35}\Omega_{24}\Big)
  + 8 \Big(\frac{\nu_{ab}}{\nu_{aa}}\Big)^2;\nn\\
  %===
  V_1 -7V_2 &=& \frac{3}{5}\Omega_{22}+2 \frac{\nu_{ab}}{\nu_{aa}};\nn\\
  %===
  7 V_1 V_2 -V_2 V_3 -7 V_2 V_4 +V_4^2 &=& \frac{2353}{1600}\Omega_{22}^2 - \frac{33}{40}\Omega_{22}\Omega_{23} + \frac{129}{1400}\Omega_{22}\Omega_{24} + \frac{81}{700}\Omega_{23}^2 
  - \frac{9}{350}\Omega_{23}\Omega_{24} + \frac{9}{4900}\Omega_{24}^2 \nn\\
  && \quad + \frac{\nu_{ab}}{\nu_{aa}}\Big( \frac{43}{5}\Omega_{22}- \frac{12}{5}\Omega_{23}+ \frac{12}{35}\Omega_{24} \Big)
  + 16 \Big(\frac{\nu_{ab}}{\nu_{aa}}\Big)^2   ;\nn\\
  %===
  V_1^2 -2 V_2 V_3 +V_4^2 &=&  \frac{3433}{1600}\Omega_{22}^2 - \frac{201}{200}\Omega_{22}\Omega_{23} + \frac{129}{1400}\Omega_{22}\Omega_{24} + \frac{99}{700}\Omega_{23}^2
   - \frac{9}{350}\Omega_{23}\Omega_{24} + \frac{9}{4900}\Omega_{24}^2  \nn\\
  && \quad +\frac{\nu_{ab}}{\nu_{aa}} \Big( 11 \Omega_{22} - \frac{12}{5}\Omega_{23}+ \frac{12}{35}\Omega_{24} \Big)
  + 20 \Big(\frac{\nu_{ab}}{\nu_{aa}}\Big)^2.
\end{eqnarray}
The parallel viscosity can be slightly re-arranged into
\begin{eqnarray}
  \eta_0^a  &=&   \frac{5}{6} \frac{(301\Omega_{22}-84 \Omega_{23}+12 \Omega_{24} +1120 \frac{\nu_{ab}}{\nu_{aa}} )}
      {[77 \Omega_{22}^2   +6 \Omega_{22}\Omega_{24}-6 \Omega_{23}^2  + \frac{\nu_{ab}}{\nu_{aa}} (\frac{3185}{3} \Omega_{22}-140 \Omega_{23} +20 \Omega_{24} )
        + \frac{5600}{3} (\frac{\nu_{ab}}{\nu_{aa}})^2 ]}  \,\frac{p_a}{\nu_{aa}}. \label{eq:beau35H} 
\end{eqnarray}
The results (\ref{eq:beau33xH})-(\ref{eq:beau35H}) now represent the stress-tensor for heavyweight particles, where collisions with much lighter particles are retained.
Prescribing the Coulomb collisions (with $\ln\Lambda\gg 1$) yields
\begin{eqnarray}
\textrm{Coulomb collisions}:\qquad  \eta_0^i  &=&  \frac{p_i}{\nu_{ii}}\,  \frac{\frac{41}{28}+4 \frac{\nu_{ie}}{\nu_{ii}} }
       { \frac{267}{175} + \frac{541}{70}\frac{\nu_{ie}}{\nu_{ii}} +8 (\frac{\nu_{ie}}{\nu_{ii}})^2} ; \label{eq:beau35HHH} \\
%===       
  \eta_2^i  &=& \frac{p_{i}}{\nu_{ii}\Delta}\, \Big[ x^2 \Big(\frac{6}{5}+2 \frac{\nu_{ie}}{\nu_{ii}}\Big)
         +\Big( \frac{41}{28}+4 \frac{\nu_{ie}}{\nu_{ii}} \Big)\Big( \frac{267}{175} + \frac{541}{70}\frac{\nu_{ie}}{\nu_{ii}} +8 \big(\frac{\nu_{ie}}{\nu_{ii}}\big)^2 \Big)\Big]; \nn\\
%===
  \eta_4^i  &=& \frac{p_{i} x}{\nu_{ii} \Delta}\, \Big[ x^2 +   \frac{46561}{19600}  + \frac{82}{7}\frac{\nu_{ie}}{\nu_{ii}} +16 \big(\frac{\nu_{ie}}{\nu_{ii}} \big)^2  \Big]; \nn\\
  \Delta &=& x^4 + x^2 \Big[\frac{79321}{19600} +\frac{578}{35}\frac{\nu_{ie}}{\nu_{ii}} + 20 \big(\frac{\nu_{ie}}{\nu_{ii}}\big)^2 \Big]
  + \Big[\frac{267}{175} + \frac{541}{70}\frac{\nu_{ie}}{\nu_{ii}} +8 \big(\frac{\nu_{ie}}{\nu_{ii}}\big)^2 \Big]^2, \nn
\end{eqnarray}
where the ratio of collisional frequencies is given by (\ref{eq:wow30}). We will analyze the result further below, but first we find it entertaining to write down
the solutions for the generalized hard spheres (where the parallel viscosity is fully meaningfull), which reads
\begin{eqnarray}
\textrm{``Hard spheres''}:\qquad  \eta_0^a  &=&  \frac{p_a}{\nu_{aa}}\,  \frac{\frac{41}{28}+4 \frac{\nu_{ab}}{\nu_{aa}} }
       { \frac{303}{175} + \frac{541}{70}\frac{\nu_{ab}}{\nu_{aa}} +8 (\frac{\nu_{ab}}{\nu_{aa}})^2} ; \label{eq:beau35HHHx} \\
%===       
  \eta_2^a  &=& \frac{p_{a}}{\nu_{aa}\Delta}\, \Big[ x^2 \Big(\frac{6}{5}+2 \frac{\nu_{ab}}{\nu_{aa}}\Big)
         +\Big( \frac{41}{28}+4 \frac{\nu_{ab}}{\nu_{aa}} \Big)\Big( \frac{303}{175} + \frac{541}{70}\frac{\nu_{ab}}{\nu_{aa}} +8 \big(\frac{\nu_{ab}}{\nu_{aa}}\big)^2 \Big)\Big]; \nn\\
%===
  \eta_4^a  &=& \frac{p_{a} x}{\nu_{aa} \Delta}\, \Big[ x^2 + \frac{42529}{19600}  + \frac{82}{7}\frac{\nu_{ab}}{\nu_{aa}} +16 \big(\frac{\nu_{ab}}{\nu_{aa}} \big)^2  \Big]; \nn\\
  \Delta &=& x^4 + x^2 \Big[ \frac{71257}{19600} +\frac{578}{35}\frac{\nu_{ab}}{\nu_{aa}} + 20 \big(\frac{\nu_{ab}}{\nu_{aa}}\big)^2 \Big]
  + \Big[\frac{303}{175} + \frac{541}{70}\frac{\nu_{ab}}{\nu_{aa}} +8 \big(\frac{\nu_{ab}}{\nu_{aa}}\big)^2 \Big]^2, \nn
\end{eqnarray}
and the ratio of collisional frequencies is given by (\ref{eq:wow30H}). Note the many numerical similarities between the Coulomb collisions (\ref{eq:beau35HHH})
and the hard spheres (\ref{eq:beau35HHHx}) and the parallel viscosities can be also written as
\begin{empheq}[box=\fbox]{align}
\textrm{Coulomb collisions}:\qquad  \eta_0^i  &=   \frac{1025+2800 \frac{\nu_{ie}}{\nu_{ii}} }
       { 1068 + 5410\frac{\nu_{ie}}{\nu_{ii}} +5600 (\frac{\nu_{ie}}{\nu_{ii}})^2 } \, \frac{p_i}{\nu_{ii}}; \label{eq:wow33}\\
 %=== 
\textrm{Hard spheres}:\qquad   \eta_0^a  &=   \frac{1025+2800 \frac{\nu_{ab}}{\nu_{aa}} }
      { 1212 + 5410\frac{\nu_{ab}}{\nu_{aa}} +5600 (\frac{\nu_{ab}}{\nu_{aa}})^2 }  \,\frac{p_a}{\nu_{aa}}. \label{eq:wow34}
\end{empheq}
From the expressions (\ref{eq:wow33})-(\ref{eq:wow34}) it is obvious that even though the ratios $\nu_{ie}/\nu_{ii}$ might be small, they are multiplied by quite large
numbers. Note that because the ratio $\nu_{ie}/\nu_{ii}$ depends on the ion mass $m_i$, one can not just write numerical results for the ion charge $Z_i=1$
(as it was possible for the electron species, where the ratio $\nu_{ei}/\nu_{ee}=Z_i\sqrt{2}$ does not contain the ion mass) and to get some numerical
values, one needs to choose a particular case. Considering the proton-electron plasma yields 
\begin{eqnarray}
  \textrm{Coulomb collisions,}\,\,\frac{\nu_{ie}}{\nu_{ii}}=0.033 ; \qquad  \eta_0^i  &=&  0.892\,\frac{p_i}{\nu_{ii}}; \qquad
  \Delta = x^4 + 4.614 x^2   + 3.202; \label{eq:beau35HHHb} \\
%===       
  \eta_2^i  &=& \frac{p_{i}}{\nu_{ii}\Delta}\, \Big[ 1.266 x^2 +2.8565 \Big]; \nn\\
%===
  \eta_4^i  &=& \frac{p_{i} x}{\nu_{ii} \Delta}\, \Big[ x^2 +   2.780 \Big].\nn
\end{eqnarray}
Contrasting the result (\ref{eq:beau35HHHb}) with the self-collisional eq. (4.44) of \cite{Braginskii1965} or the eq. (74) of \cite{Hunana2022} reveals that
the numerical values are quite different, implying that the effects of the ion-electron collisions are not completely insignificant.
For example, the change of the parallel viscosity value from the 1025/1068=0.960 into the 0.892 represents a difference of almost 8\%
(when divided by the smaller value).   
The value of 0.892 is consistent with the eq. (217) of Part 1, there calculated even more precisely without any expansions in small mass-ratios. 
As another double-check, it can be shown that our Coulomb viscosities yield the same results as equations (89a)-(89c) of \cite{JiHeld2013}
(where we only consider $T_a\simeq T_b$), which is shown below in Section \ref{sec:JiHeld2013}. 

For completeness, considering the hard spheres with the same ratio of collisional frequencies as in (\ref{eq:beau35HHHb}), yields 
\begin{eqnarray}
  \textrm{``Hard spheres'',}\,\,\frac{\nu_{ab}}{\nu_{aa}}=0.033 ; \qquad  \eta_0^a  &=&  0.800\,\frac{p_a}{\nu_{aa}}; \qquad
  \Delta = x^4 + 4.202 x^2   + 3.981; \label{eq:beau35HHHc} \\
%===       
  \eta_2^a  &=& \frac{p_{a}}{\nu_{aa}\Delta}\, \Big[ 1.266 x^2 +3.1849 \Big]; \nn\\
%===
  \eta_4^a  &=& \frac{p_{a} x}{\nu_{aa} \Delta}\, \Big[ x^2 +   2.574 \Big],\nn
\end{eqnarray}
and for the parallel viscosity the change from the self-collisional value of 1025/1212=0.846 into the 0.800 represents a difference of 6\%.

%\newpage
%============================================================================
\subsection{Heavyweight thermal conductivity}
Continuing with the thermal conductivity, the relevant mass ratio coefficients (\ref{eq:wow16})-(\ref{eq:wow17m}) for $m_a\gg m_b$ simplify into
\begin{eqnarray}
D_{ab (1)} = 3; \qquad E_{ab (1)} \ll 1; \qquad F_{ab (1)} =84; \qquad G_{ab (1)} = 5,
\end{eqnarray}
and the evolution equations become
\begin{eqnarray}
  \frac{d_a}{d t}\vecq_a + \Omega_a \bhat\times\vecq_a + \frac{5}{2}p_a \nabla \Big(\frac{p_a}{\rho_a}\Big) &=&
  - \Big[ \frac{2}{5}\nu_{aa}\Omega_{22} +3 \nu_{ab}   \Big]\vecq_a \nn\\
  && + \nu_{aa} \Big(\frac{1}{20}\Omega_{22}- \frac{1}{70}\Omega_{23}\Big) \Big(\frac{\rho_a}{p_a}\vecX^{(5)}_a -28 \vecq_a\Big); \label{eq:Excite1H}\\
%===  
  \frac{d_a}{d t}\vecX^{(5)}_a +\Omega_a\bhat\times\vecX^{(5)}_a +70\frac{p_a^2}{\rho_a}\nabla\Big(\frac{p_a}{\rho_a}\Big) 
  &=& - \Big[ \nu_{aa} \Big( \frac{8}{5}\Omega_{23}+ \frac{28}{5}\Omega_{22} \Big) +84\nu_{ab}  \Big]\frac{p_a}{\rho_a}\vecq_a \nn\\
&&  - \Big[\nu_{aa} \Big(\frac{2}{35}\Omega_{24} - \frac{3}{10}\Omega_{22} \Big)+5\nu_{ab} \Big]\Big(\vecX^{(5)}_a -28 \frac{p_a}{\rho_a} \vecq_a\Big),\label{eq:Excite2H}
\end{eqnarray}
which can be rewritten as
%\begin{eqnarray}
%  \frac{d_a}{d t}\vecq_a + \Omega_a \bhat\times\vecq_a + \frac{5}{2}p_a \nabla \Big(\frac{p_a}{\rho_a}\Big) &=&
%  - \nu_{aa} \Big[ \frac{2}{5}\Omega_{22} +3 \frac{\nu_{ab}}{\nu_{aa}}   \Big] \vecq_a \nn\\
%  && + \nu_{aa} \Big(\frac{1}{20}\Omega_{22}- \frac{1}{70}\Omega_{23}\Big) \Big(\frac{\rho_a}{p_a}\vecX^{(5)}_a -28 \vecq_a\Big); \label{eq:Excite1HH}\\
%===  
%  \frac{d_a}{d t}\vecX^{(5)}_a +\Omega_a\bhat\times\vecX^{(5)}_a +70\frac{p_a^2}{\rho_a}\nabla\Big(\frac{p_a}{\rho_a}\Big) 
%  &=& - \nu_{aa} \Big[   \frac{8}{5}\Omega_{23}+ \frac{28}{5}\Omega_{22}  +84 \frac{\nu_{ab}}{\nu_{aa}}  \Big]\frac{p_a}{\rho_a}\vecq_a \nn\\
%&&  - \nu_{aa} \Big[ \frac{2}{35}\Omega_{24} - \frac{3}{10}\Omega_{22} +5\frac{\nu_{ab}}{\nu_{aa}} \Big]\Big(\vecX^{(5)}_a -28 \frac{p_a}{\rho_a} \vecq_a\Big),\label{eq:Excite2HH}
%\end{eqnarray}
\begin{eqnarray}
  \frac{d_a}{d t}\vecq_a + \Omega_a \bhat\times\vecq_a + \frac{5}{2}p_a \nabla \Big(\frac{p_a}{\rho_a}\Big) &=&
  - \nu_{aa} B_1 \vecq_a  + \nu_{aa} B_2 \frac{\rho_a}{p_a}\vecX^{(5)}_a; \label{eq:Excite1HHH}\\
%===  
  \frac{d_a}{d t}\vecX^{(5)}_a +\Omega_a\bhat\times\vecX^{(5)}_a +70\frac{p_a^2}{\rho_a}\nabla\Big(\frac{p_a}{\rho_a}\Big) 
  &=& - \nu_{aa} B_3 \frac{p_a}{\rho_a}\vecq_a   - \nu_{aa}  B_4 \vecX^{(5)}_a,\label{eq:Excite2HHH}
\end{eqnarray}
with the coefficients
\begin{eqnarray}
  B_1 &=&  \Big[ \frac{9}{5}\Omega_{22} -\frac{2}{5}\Omega_{23} +3 \frac{\nu_{ab}}{\nu_{aa}}   \Big] ;\qquad
  B_2 =  \Big[\frac{1}{20}\Omega_{22}- \frac{1}{70}\Omega_{23}\Big];\nn\\
  B_3 &=& \Big[   \frac{8}{5}\Omega_{23}+ 14\Omega_{22} -\frac{8}{5}\Omega_{24} -56 \frac{\nu_{ab}}{\nu_{aa}}  \Big];\qquad
  B_4 = \Big[ \frac{2}{35}\Omega_{24} - \frac{3}{10}\Omega_{22} +5\frac{\nu_{ab}}{\nu_{aa}} \Big]. \label{eq:pica7}
\end{eqnarray}
The quasi-static approximation then yields the thermal heat flux
\begin{eqnarray}
  \vecq_a &=& -\kappa_\parallel^a \nabla_\parallel T_a - \kappa_\perp^a \nabla_\perp T_a + \kappa_\times^a \bhat\times\nabla T_a; \label{eq:QEniceH}\\
  %=====
&& \kappa_\parallel^a = \frac{p_a}{m_a \nu_{aa}}\gamma_0 ; \qquad   
\kappa_\perp^a =  \frac{p_a}{m_a \nu_{aa}} \frac{\gamma_1' x^2+\gamma_0'}{\triangle};
\qquad \kappa_\times^a =  \frac{p_a}{m_a \nu_{aa}}\frac{\gamma_1'' x^3+\gamma_0''x}{\triangle};
\qquad \triangle = x^4 + \delta_1 x^2 +\delta_0, \nn
%=====  
\end{eqnarray}
with the transport coefficients (which should not be confused with the electron species)
\begin{eqnarray}
  \gamma_0 &=& \frac{5 (28 B_2 +B_4)}{2 (B_1 B_4 +B_2 B_3)}; \qquad 
  \gamma_1' = \frac{5}{2} B_1 -70 B_2;\qquad
  \gamma_0' = \frac{5}{2}(28 B_2 +B_4)(B_1 B_4+B_2 B_3);\nn\\
  \gamma_1'' &=& \frac{5}{2};\qquad   \gamma_0'' = \frac{B_2}{2}(140 B_1 -5 B_3 +140 B_4)  +\frac{5}{2}B_4^2;\nn\\
  \delta_1 &=& B_1^2 -2 B_2 B_3 +B_4^2; \qquad \delta_0 = (B_1 B_4 +B_2 B_3)^2, \label{eq:HFgamma0H}
\end{eqnarray}
and the thermal conductivities become
\begin{eqnarray}
\kappa_\parallel^a &=& \frac{p_a}{m_a \nu_{aa} \Delta^*} \Big( \frac{11}{4}\Omega_{22} - \Omega_{23}+ \frac{1}{7}\Omega_{24}+ \frac{25}{2}\frac{\nu_{ab}}{\nu_{aa}} \Big);\nn\\
%========
  \Delta^* &=&  (B_1 B_4 +B_2 B_3) = \frac{4}{175} \Big( 7 \Omega_{22}^2 +  \Omega_{24}\Omega_{22}  - \Omega_{23}^2 \Big)
  +\frac{\nu_{ab}}{\nu_{aa}} \Big( \frac{53}{10}\Omega_{22}- \frac{6}{5}\Omega_{23}+ \frac{6}{35}\Omega_{24} \Big) 
  +15 \big(\frac{\nu_{ab}}{\nu_{aa}} \big)^2;\nn\\
%========
  \kappa_\perp^a &=&  \frac{p_a}{m_a \nu_{aa} \triangle} \Big\{ x^2\Big(\Omega_{22}+ \frac{15}{2}\frac{\nu_{ab}}{\nu_{aa}}  \Big)
  +\Big( \frac{11}{4}\Omega_{22} - \Omega_{23}+ \frac{1}{7}\Omega_{24}+ \frac{25}{2}\frac{\nu_{ab}}{\nu_{aa}} \Big) \Delta^*  \Big\};\nn\\
%========
  \kappa_\times^a &=&  \frac{p_a x}{m_a \nu_{aa} \triangle} \Big\{ \frac{5}{2} x^2 +
\frac{149}{40}\Omega_{22}^2 - \frac{13}{5}\Omega_{22}\Omega_{23} + \frac{16}{35}\Omega_{23}^2 + \frac{11}{35}\Omega_{24}\Omega_{22}
  - \frac{4}{35} \Omega_{24}\Omega_{23}+ \frac{2}{245}\Omega_{24}^2 \nn\\
  && \quad + \frac{\nu_{ab}}{\nu_{aa}} \Big( \frac{55}{2}\Omega_{22}- 10 \Omega_{23} + \frac{10}{7}\Omega_{24} \Big) +\frac{125}{2}\big(\frac{\nu_{ab}}{\nu_{aa}} \big)^2 \Big\};\nn\\
%===
  \triangle &=&  x^4 + x^2 \Big[ \frac{193}{100}\Omega_{22}^2 - \frac{6}{5}\Omega_{22} \Omega_{23}
    + \frac{36}{175} \Omega_{23}^2 + \frac{22}{175}\Omega_{24} \Omega_{22} - \frac{8}{175}\Omega_{24}\Omega_{23}+ \frac{4}{1225} \Omega_{24}^2   \nn\\
  && \qquad  + \frac{\nu_{ab}}{\nu_{aa}}\Big( \frac{67}{5}\Omega_{22} - 4 \Omega_{23} + \frac{4}{7}\Omega_{24} \Big)  +34 \big(\frac{\nu_{ab}}{\nu_{aa}} \big)^2 \Big] + (\Delta^*)^2.
\end{eqnarray}
The parallel thermal conductivity can be also written as  
\begin{eqnarray}
  \kappa_\parallel^a  &=& \frac{25}{2} \frac{(77 \Omega_{22}-28 \Omega_{23}+4 \Omega_{24}+350 \frac{\nu_{ab}}{\nu_{aa}})}
        {\big[ 56 \Omega_{22}^2+8 \Omega_{22}\Omega_{24}-8 \Omega_{23}^2    
            + \frac{\nu_{ab}}{\nu_{aa}}\big(1855 \Omega_{22}-420 \Omega_{23}+60 \Omega_{24}\big) +5250 \big(\frac{\nu_{ab}}{\nu_{aa}} \big)^2    \big]} \,
        \frac{p_a}{\nu_{aa} m_a}. \label{eq:SUS1}
\end{eqnarray}

Prescribing the Coulomb collisions (with $\ln\Lambda\gg 1$) yields 
\begin{eqnarray}
\textrm{Coulomb collisions}: \qquad   
\kappa_\parallel^i &=& \frac{p_i}{m_i \nu_{ii}} \frac{ \frac{45}{14}+ \frac{25}{2} \frac{\nu_{ie}}{\nu_{ii}} }
      { \frac{144}{175}+ \frac{55}{7} \frac{\nu_{ie}}{\nu_{ii}}  +15 ( \frac{\nu_{ie}}{\nu_{ii}} )^2 } ; \label{eq:wow38}\\
      \kappa_\perp^i &=&   \frac{p_i}{m_i \nu_{ii} \triangle} \Big\{ x^2 \Big( 2+ \frac{15}{2}\frac{\nu_{ie}}{\nu_{ii}}  \Big)
      + \Big(\frac{45}{14}+ \frac{25}{2} \frac{\nu_{ie}}{\nu_{ii}} \Big)
      \Big[  \frac{144}{175}+ \frac{55}{7} \frac{\nu_{ie}}{\nu_{ii}}  +15 \big( \frac{\nu_{ie}}{\nu_{ii}} \big)^2   \Big]  \Big\};\nn\\
      %===
      \kappa_\times^i &=&  \frac{p_i x}{m_i \nu_{ii} \triangle} \Big\{ \frac{5}{2} x^2 + \frac{2277}{490}+\frac{225}{7}\frac{\nu_{ie}}{\nu_{ii}}
      +\frac{125}{2} \big( \frac{\nu_{ie}}{\nu_{ii}}\big)^2 \Big\};\nn\\
      %===
      \triangle &=&  x^4 + x^2 \Big[ \frac{3313}{1225} + \frac{618}{35}\frac{\nu_{ie}}{\nu_{ii}}  +34 \big(\frac{\nu_{ie}}{\nu_{ii}}  \big)^2 \Big]
      + \Big[ \frac{144}{175}+ \frac{55}{7} \frac{\nu_{ie}}{\nu_{ii}}  +15 ( \frac{\nu_{ie}}{\nu_{ii}} )^2  \Big]^2, \nn
\end{eqnarray}
with the ratio of collisional frequencies (\ref{eq:wow30}). For comparison, prescribing the generalized hard spheres (the parallel thermal
conductivity is fully meaningfull) yields
\begin{eqnarray}
\textrm{``Hard spheres''}: \qquad   
\kappa_\parallel^a &=& \frac{p_a}{m_a \nu_{aa}} \frac{ \frac{45}{14}+ \frac{25}{2} \frac{\nu_{ab}}{\nu_{aa}} }
      { \frac{176}{175}+ \frac{55}{7} \frac{\nu_{ab}}{\nu_{aa}}  +15 ( \frac{\nu_{ab}}{\nu_{aa}} )^2 } ; \label{eq:wow39}\\
      \kappa_\perp^a &=&   \frac{p_a}{m_a \nu_{aa} \triangle} \Big\{ x^2 \Big( 2+ \frac{15}{2}\frac{\nu_{ab}}{\nu_{aa}}  \Big)
      + \Big(\frac{45}{14}+ \frac{25}{2} \frac{\nu_{ab}}{\nu_{aa}} \Big)
      \Big[  \frac{176}{175}+ \frac{55}{7} \frac{\nu_{ab}}{\nu_{aa}}  +15 \big( \frac{\nu_{ab}}{\nu_{aa}} \big)^2   \Big]  \Big\};\nn\\
      %===
      \kappa_\times^a &=&  \frac{p_a x}{m_a \nu_{aa} \triangle} \Big\{ \frac{5}{2} x^2 + \frac{2053}{490} +\frac{225}{7}\frac{\nu_{ab}}{\nu_{aa}}
      +\frac{125}{2} \big( \frac{\nu_{ab}}{\nu_{aa}}\big)^2 \Big\};\nn\\
      %===
      \triangle &=&  x^4 + x^2 \Big[ \frac{573}{245} + \frac{618}{35}\frac{\nu_{ab}}{\nu_{aa}}  +34 \big(\frac{\nu_{ab}}{\nu_{aa}}  \big)^2 \Big]
      + \Big[ \frac{176}{175}+ \frac{55}{7} \frac{\nu_{ab}}{\nu_{aa}}  +15 ( \frac{\nu_{ab}}{\nu_{aa}} )^2  \Big]^2,\nn
\end{eqnarray}
with the ratio of collisional frequencies (\ref{eq:wow30H}). Note the many numerical similarities between the (\ref{eq:wow38})-(\ref{eq:wow39}) and
the parallel thermal conductivities can be also written as
\begin{empheq}[box=\fbox]{align}
 \textrm{Coulomb collisions}: \qquad \kappa_\parallel^i  &= \frac{1125+4375  \frac{\nu_{ie}}{\nu_{ii}}}
         { 288+2750  \frac{\nu_{ie}}{\nu_{ii}}   +5250 \big(\frac{\nu_{ie}}{\nu_{ii}} \big)^2 } \, \frac{p_i}{m_i \nu_{ii}}; \label{eq:beau35HHHH}\\
 \textrm{Hard spheres}: \qquad \kappa_\parallel^a  &= \frac{1125+4375  \frac{\nu_{ab}}{\nu_{aa}}}
         { 352+2750  \frac{\nu_{ab}}{\nu_{aa}}   +5250 \big(\frac{\nu_{ab}}{\nu_{aa}} \big)^2 } \, \frac{p_a}{m_a \nu_{aa}}. \label{eq:beau35HHHHx}    
\end{empheq}
Considering the proton-electron plasma with the ratio of collisional frequencies $\nu_{ie}/\nu_{ii}=0.033$, the Coulomb collisions then yield
in the magnetized case
\begin{eqnarray}
\textrm{Coulomb collisions,} \,\, \frac{\nu_{ie}}{\nu_{ii}}=0.033; \qquad     
\kappa_\parallel^i &=& 3.302 \frac{p_i}{m_i \nu_{ii}}; \qquad \triangle =  x^4 + 3.3242 x^2 +  1.2067; \label{eq:wow43}\\
\kappa_\perp^i &=&   \frac{p_i}{m_i \nu_{ii} \triangle} \Big[2.2475 x^2 + 3.9839 \Big];\nn\\
\kappa_\times^i &=&  \frac{p_i x}{m_i \nu_{ii} \triangle} \Big[ \frac{5}{2} x^2 + 5.7757 \Big].\nn
\end{eqnarray}
Contrasting these conductivities with the self-collisional eq. (4.40) of \cite{Braginskii1965} or eq. (44) in \cite{Hunana2022} reveals that some values  
are very different. For example, for the parallel conductivity the change from the self-collisional value of 1125/288 = 125/32 = 3.906 into the above value
3.302 represents a difference of 18\% (when divided by the smaller value).
The value 3.302 is consistent with the eq. (214) in Part 1, there calculated without any expansions in mass-ratios. 
From the numerical values (\ref{eq:wow43}), the largest correction is in the value $\delta_0=1.2067$,
where the self-collisional Braginskii result was $\delta_0=0.6771$. 
For example, considering the cross conductivity in the limit of weak magnetic field
($x\ll 1$), where the $\kappa_\times^i$ is small, yields
\begin{equation}
  \textrm{Weak B-field}:\quad \kappa_\times^i =  \frac{p_i x}{m_i \nu_{ii}} \underbrace{\frac{5.7757}{1.2067}}_{= 4.787};\qquad
  \big(\kappa_\times^i \big)^{\textrm{Self-coll}} =  \frac{p_i x}{m_i \nu_{ii}} \underbrace{\frac{4.6469}{0.6771}}_{= 6.863},
\end{equation}
which is a difference of 43\%. It is indeed counter-intuitive that collisions with particles that are 1836 times lighter can introduce such large differences.
For the Coulomb collisions, equivalence with \cite{JiHeld2013} (where we only consider $T_a\simeq T_b$) is shown below in Section \ref{sec:JiHeld2013}.

For completeness, considering the hard spheres with the same ratio of collisional frequencies as in (\ref{eq:wow43}) yields
\begin{eqnarray}
\textrm{``Hard spheres'',} \,\, \frac{\nu_{ab}}{\nu_{aa}}=0.033;\qquad    
\kappa_\parallel^a &=&  2.830 \frac{p_a}{m_a \nu_{aa}}; \qquad \triangle =  x^4 + 2.958 x^2 + 1.642; \label{eq:wow44}\\
\kappa_\perp^a &=&   \frac{p_a}{m_a \nu_{aa} \triangle} \Big[ 2.2475 x^2 + 4.647 \Big];\nn\\
\kappa_\times^a &=&  \frac{p_a x}{m_a \nu_{aa} \triangle} \Big[ \frac{5}{2} x^2 + 5.3186 \Big].\nn
\end{eqnarray}
The result can be contrasted with the self-collisional eq. (\ref{eq:Thierry51BB}), where the parallel conductivity of hard spheres was 3.196, which
represents difference of 13\%. For the $\kappa_\times^a$ in the limit of weak magnetic field, the difference is 28\%. \\

\newpage
%=========================================================================
\subsection{Comparison with Ji and Held 2013 (Coulomb collisions)} \label{sec:JiHeld2013}
As a double-check that our model is formulated correctly, it is useful to compare our results with \cite{JiHeld2013}.
As discussed already in the Introduction, even though our general model formulated through the evolution equations is valid for arbitrary temperature differences,
we prefer to write down quasi-static solutions only for the similar temperatures $T_a\simeq T_b$, so that the expansions with mass-ratios remain valid. 
For large temperature differences, especially if the ion temperature vastly exceeds the electron temperature, it is better to obtain the
quasi-static approximation numerically, without any expansions in mass-ratios.
That the case of large temperature differences is indeed not trivial, can be also seen from the discussion in \cite{JiHeld2015}. 
To compare our results with the former reference, we thus introduce the variable $\zeta=(1/Z_i)\sqrt{m_e/m_i}$, so that our $\nu_{ii}/\nu_{ie}=\sqrt{2}\zeta$.
Their choice of the ion collisional time differs from our/Braginskii choice
by $(1/\tau_{ii})^{\textrm{JH}}= \sqrt{2}/\tau_{ii} =\sqrt{2} \nu_{ii}$ (for a further discussion about this topic, see Section 8.2, p. 31 in \cite{Hunana2022}).  
So by using their variable $r=x/\sqrt{2}$ and by keeping our collisional frequencies, one can write the ion viscosities (\ref{eq:wow33}) and (\ref{eq:beau35HHH}) as
\begin{eqnarray}
 \eta_0^i  &=&    \frac{p_i}{\sqrt{2}\nu_{ii}}\,  \frac{\sqrt{2}\frac{1025}{1068}+\frac{1400}{267} \zeta }
      {  1 + \frac{2705}{534}\sqrt{2}\zeta +\frac{2800}{267} \zeta^2 }; \nn\\ 
\eta_2^i  &=& \frac{p_{i}}{\sqrt{2}\nu_{ii}\Delta}\, 4 \Big[  r^2 \Big(\frac{3}{5}\sqrt{2}+ 2 \zeta\Big)
         +\Big( \sqrt{2}\frac{1025}{1068}+\frac{1400}{267} \zeta \Big) \frac{1}{4}(\frac{267}{175})^2 \Big(1  + \frac{2705}{534} \sqrt{2}\zeta + \frac{2800}{267} \zeta^2 \Big)\Big]; \nn\\
%===
  \eta_4^i  &=& \frac{p_{i} r}{\sqrt{2}\nu_{ii} \Delta}\, 4\Big[ r^2 +   \frac{46561}{39200}  + \frac{41}{7}\sqrt{2} \zeta + 16 \zeta^2  \Big]; \nn\\
  \Delta &=& 4 \Big\{ r^4 +  r^2 \Big[\frac{79321}{39200} +\frac{289}{35} \sqrt{2} \zeta + 20 \zeta^2 \Big]
  + \frac{1}{4}(\frac{267}{175})^2 \Big[ 1+ \frac{2705}{534}\sqrt{2} \zeta + \frac{2800}{267}\zeta^2 \Big]^2 \Big\}, \nn
\end{eqnarray}
%=================
and numerical evaluation yields
\begin{eqnarray}
\eta_0^i &=& \frac{p_i}{\sqrt{2}\nu_{ii}}\, \frac{1.357+5.243\zeta}{1+7.164\zeta+10.487\zeta^2};\nn\\
\eta_2^i  &=& \frac{p_{i}}{\sqrt{2}\nu_{ii}\Delta}\, 4\Big[  r^2 \Big(\frac{3}{5}\sqrt{2} +2 \zeta\Big)
         +\Big( 1.357+5.243\zeta \Big)0.582 \Big( 1 + 7.164 \zeta + 10.487 \zeta^2 \Big)\Big]; \nn\\
%===
  \eta_4^i  &=& \frac{p_{i} r}{\sqrt{2}\nu_{ii} \Delta}\, 4\Big[  r^2 +   1.188  + 8.283 \zeta + 16 \zeta^2  \Big]; \nn\\
  \Delta &=& 4 \Big\{ r^4 +  r^2 \Big[2.023 + 11.677 \zeta + 20 \zeta^2 \Big]
  + 0.582 \Big[1 + 7.164 \zeta + 10.487 \zeta^2 \Big]^2 \Big\}, \nn
\end{eqnarray}
recovering equations (89a)-(89c) of \cite{JiHeld2013} (there is a small missprint in their $e\to i$). Similarly, one can write
the ion thermal conductivities (\ref{eq:beau35HHHH}) and (\ref{eq:wow38}) as
\begin{eqnarray}
\kappa_\parallel^i  &=& \frac{p_i}{\sqrt{2} m_i \nu_{ii}}\, \frac{\frac{125}{32}\sqrt{2}+ \frac{4375}{144} \zeta}
      { 1+\frac{1375}{144}  \sqrt{2}\zeta   +\frac{875}{24} \zeta^2 };\nn\\  
      \kappa_\perp^i &=&   \frac{p_i}{\sqrt{2} m_i \nu_{ii} \triangle} 4 \Big\{ r^2   \Big(\sqrt{2} + \frac{15}{2} \zeta  \Big)
      +  \Big( \frac{125}{32}\sqrt{2} + \frac{4375}{144} \zeta \Big)
      \frac{1}{4} \big(\frac{144}{175}\big)^2 \Big[ 1+ \frac{1375}{144} \sqrt{2} \zeta  + \frac{875}{24} \zeta^2    \Big]  \Big\};\nn\\
      %===
      \kappa_\times^i &=&  \frac{p_i r}{ \sqrt{2}m_i \nu_{ii} \triangle} 4 \Big\{ \frac{5}{2} r^2  + \frac{2277}{980}+\frac{225}{14} \sqrt{2} \zeta
      + \frac{125}{2} \zeta^2 \Big\};\nn\\
      %===
      \triangle &=&  4 \Big\{ r^4 +  r^2 \Big[ \frac{3313}{2450} + \frac{309}{35} \sqrt{2} \zeta  +34  \zeta^2 \Big]
      + \frac{1}{4} \big(\frac{144}{175}\big)^2 \Big[ 1+ \frac{1375}{144} \sqrt{2} \zeta  + \frac{875}{24} \zeta^2  \Big]^2 \Big\}, \nn
\end{eqnarray}
%===============================
and numerical evaluation yields
\begin{eqnarray}
\kappa_\parallel^i  &=& \frac{p_i}{\sqrt{2} m_i \nu_{ii}}\, \frac{ 5.524 + 30.382 \zeta}{ 1+13.504\zeta  + 36.458 \zeta^2 };\nn\\ 
      \kappa_\perp^i &=&   \frac{p_i}{\sqrt{2} m_i \nu_{ii} \triangle} 4 \Big\{ r^2   \Big(\sqrt{2} + \frac{15}{2} \zeta  \Big)
      +  \Big( 5.524 + 30.382 \zeta \Big)
      0.1693 \Big[ 1+13.504\zeta  + 36.458 \zeta^2  \Big]  \Big\};\nn\\
      %===
      \kappa_\times^i &=&  \frac{p_i r}{ \sqrt{2}m_i \nu_{ii} \triangle} 4 \Big\{ \frac{5}{2} r^2  + 2.323 + 22.728 \zeta
      + 62.500 \zeta^2 \Big\};\nn\\
      %===
      \triangle &=&  4 \Big\{ r^4 +  r^2 \Big[ 1.352 + 12.485 \zeta  +34  \zeta^2 \Big]
      +  0.1693 \Big[ 1+13.504\zeta  + 36.458 \zeta^2  \Big]^2 \Big\}, \nn
\end{eqnarray}
recovering equations (88a)-(88c) of \cite{JiHeld2013}. They also provide solutions in the 3-Hermite approximation.

%\newpage
%=================================================================================================
%=================================================================================================
\section{Scalar perturbation (excess-kurtosis)  \texorpdfstring{$\widetilde{X}^{(4)}_a$}{tilde X(4)}} \label{sec:Kurtosis}
Here we consider solutions for the scalar perturbations $\widetilde{X}^{(4)}_a$, which we separate into two cases of self-collisions 
and of lightweight particles $m_a\ll m_b$ (and we do not discuss corrections for the heavyweight particles $m_a\gg m_b$).
%==========================================================
\subsection{Scalar  \texorpdfstring{$\widetilde{X}^{(4)}_a$}{perturbation} for self-collisions}
At the semi-linear level, the evolution equation for the scalar perturbation reads (see eq. (\ref{eq:PPosled12XB}) with collisional contributions (\ref{eq:Q4exchange}))
\begin{eqnarray}
   \frac{d_a}{dt} \widetilde{X}^{(4)}_a +\nabla\cdot\vecX^{(5)}_a  -20 \frac{p_a}{\rho_a}\nabla\cdot\vecq_a
    &=& - \nu_{aa} \frac{2}{5}\Omega_{22} \widetilde{X}^{(4)}_a, \label{eq:Thierry92c}
\end{eqnarray}
which in the quasi-static approximation yields solution
\begin{eqnarray}
  \textrm{22-mom:} \qquad
  \widetilde{X}^{(4)}_a &=& -\, \frac{5}{2 \nu_{aa} \Omega_{22}} \Big( \nabla\cdot \vecX^{(5)}_a
  - 20 \frac{p_a}{\rho_a} \nabla\cdot \vecq_a \Big),
\end{eqnarray}
where the (2-Hermite) heat fluxes $\vecq_a$ and $\vecX^{(5)}_a$ are given by (\ref{eq:Thierry51}) and (\ref{eq:Thierry51X5}).
As noted below eq. (\ref{eq:PPosled12XB}), we neglected the contributions of $\widetilde{X}^{(4)}_a$ in the heat flux evolution equations, resulting
in suppression of terms such as $\vecq_a\sim \nabla \nabla^2 T_a$ (in the unmagnetized case). Considering the magnetized case, at the semi-linear level one can simplify
\begin{eqnarray}
  \nabla\cdot \vecX^{(5)}_a = -\, \frac{p_a}{\rho_a} \big( \kappa_\parallel^{a(5)} \nabla_\parallel^2 T_a + \kappa_\perp^{a(5)} \nabla_\perp^2 T_a \big); \qquad
  \nabla\cdot \vecq_a = -\, \big(\kappa_\parallel^a \nabla_\parallel^2 T_a + \kappa_\perp^a \nabla_\perp^2 T_a \big), 
\end{eqnarray}
and the general solution for the $\widetilde{X}^{(4)}_a$ then has a form
\begin{eqnarray}
  \widetilde{X}^{(4)}_a &=& + \frac{p_a}{\nu_{aa} \rho_a} \Big( \kappa_\parallel^{a (4)} \nabla_\parallel^2 T_a + \kappa_\perp^{a(4)} \nabla_\perp^2 T_a \Big); \label{eq:X4form}\\
%===  %
 {\kappa}_\parallel^{a (4)} &=&  \frac{125 (63 \Omega_{22}-21 \Omega_{23}+2 \Omega_{24}) }
         {2 \Omega_{22} (7 \Omega_{22}^2+ \Omega_{22} \Omega_{24}-\Omega_{23}^2) }\, \frac{p_a}{\nu_{aa} m_a}; \nn\\
%===
         {\kappa}_\perp^{a (4)} &=& \frac{p_a}{\nu_{aa} m_a \Delta} \Big[ \frac{5(2\Omega_{23}-3 \Omega_{22})}{\Omega_{22}} x^2
          + \frac{8 (7 \Omega_{22}^2+\Omega_{22}\Omega_{24}-\Omega_{23}^2)(63 \Omega_{22}-21 \Omega_{23}+2 \Omega_{24})}{245 \Omega_{22}} \Big] ;\nn\\
\Delta &=& x^4+ x^2 \Big[\frac{193}{100}\Omega_{22}^2- \frac{6}{5} \Omega_{22} \Omega_{23} + \frac{22}{175} \Omega_{22} \Omega_{24}
  + \frac{36}{175} \Omega_{23}^2 - \frac{8}{175} \Omega_{23}\Omega_{24} + \frac{4}{1225} \Omega_{24}^2 \Big] \nn\\
  && +\Big[\frac{4}{175} \big( 7 \Omega_{22}^2+\Omega_{22} \Omega_{24}-\Omega_{23}^2 \big) \Big]^2 , \label{eq:Thierry51S}
\end{eqnarray}
where the Hall parameter $x=\Omega_a/\nu_{aa}$. The $\kappa_\parallel^{a (4)}$ and $\kappa_\perp^{a (4)}$ can be viewed as the thermal conductivities
of the 4th-order fluid moment and can be also written as 
\begin{equation}
\kappa_\parallel^{a (4)} = \frac{5}{2 \Omega_{22}} \big( \kappa_\parallel^{a (5)} -20 \kappa_\parallel^a \big); \qquad
   \kappa_\perp^{a (4)} = \frac{5}{2 \Omega_{22}} \big( \kappa_\perp^{a (5)} -20 \kappa_\perp^a \big).
\end{equation}
Evaluating (\ref{eq:Thierry51S}) for the case of the Coulomb collisions (with $\ln\Lambda\gg 1$) yields
\begin{eqnarray}
\textrm{Coulomb collisions}:\qquad {\kappa}_\parallel^{a (4)} &=&  \underbrace{\frac{1375}{24}}_{57.29}\, \frac{p_a}{\nu_{aa} m_a}; \nn\\
%===
       {\kappa}_\perp^{a (4)} &=& \frac{p_a}{\nu_{aa} m_a \Delta} \Big[ 5 x^2  + \frac{9504}{245} \Big] ;\qquad
       \Delta = x^4+  \frac{3313}{1225}x^2  +\Big(\frac{144}{175} \Big)^2 , \label{eq:X4Coulomb}
\end{eqnarray}
recovering equations (149)-(150) of \cite{Hunana2022}. For comparison, evaluating (\ref{eq:Thierry51S}) for the case of the
generalized hard spheres yields (the parallel conductivity is fully meaningful)
\begin{eqnarray}
\textrm{``Hard spheres''}:\qquad {\kappa}_\parallel^{a (4)} &=&  \underbrace{\frac{2375}{88}}_{26.99} \, \frac{p_a}{\nu_{aa} m_a}; \nn\\
%===
       {\kappa}_\perp^{a (4)} &=& \frac{p_a}{\nu_{aa} m_a \Delta} \Big[ 25 x^2  + \frac{6688}{245} \Big] ;\qquad
       \Delta = x^4+  \frac{573}{245} x^2  +\Big( \frac{176}{175} \Big)^2. \label{eq:X4Hard}
\end{eqnarray}
Finally, evaluating (\ref{eq:Thierry51S}) for the inverse power-law force yields
\begin{eqnarray}
  \textrm{Inverse power}:\qquad {\kappa}_\parallel^{a (4)} &=&  \frac{125 (19 \nu^2-32 \nu+21)(\nu-1)}{2 (11\nu-13) (3 \nu-5)^2}
  \Big(\frac{A_1(\nu)}{A_2(\nu)}\Big)^2 \, \frac{p_a}{\nu_{aa} m_a}; \nn\\
%===
      {\kappa}_\perp^{a (4)} &=& \frac{p_a}{\nu_{aa} m_a \Delta} \Big[ \frac{5(5\nu-9)}{(\nu-1)} x^2
        + \frac{8(3\nu-5)^2 (11\nu-13)(19 \nu^2-32\nu+21)}{245(\nu-1)^5} \Big(\frac{A_2(\nu)}{A_1(\nu)}\Big)^2 \Big] ;\nn\\
      \Delta &=& x^4+  \frac{(3\nu-5)^2 (2865 \nu^4-13348 \nu^3+25446 \nu^2-22820 \nu+8113)}{4900 (\nu-1)^6} \Big(\frac{A_2(\nu)}{A_1(\nu)}\Big)^2 x^2  \nn\\
      && +\Big[ \frac{4 (11 \nu-13) (3 \nu-5)^2}{175 (\nu-1)^3} \Big(\frac{A_2(\nu)}{A_1(\nu)}\Big)^2 \Big]^2 .
\end{eqnarray}

As discussed already in the Introduction, the above 22-moment model represents a 2-Hermite and 1-Hermite hybrid model, because the heat fluxes are described by
two Hermite polynomials (and are analogous to the Braginskii precision), whereas the scalar perturbations are described by one Hermite polynomial (see also the
limitations Section \ref{sec:23mom}, where the 23-moment model is briefly discussed).

%==========================================================
\subsubsection{Reduction into 14-moment model}
It is useful to briefly explore the influence of the 2-Hermite heat fluxes, by reducing the 22-moment model into the 14-moment model, where the heat fluxes
are also described by only 1-Hermite polynomial. By prescribing closure $\vecX^{(5)}_a = 28 (p_a/\rho_a)\vecq_a$, the evolution equation (\ref{eq:Thierry92c}) in the
1-Hermite approximation reads
\begin{eqnarray}
   \frac{d_a}{dt} \widetilde{X}^{(4)}_a +8 \frac{p_a}{\rho_a} \nabla\cdot \big[ \vecq_a \big]_1
    &=& - \nu_{aa} \frac{2}{5}\Omega_{22} \widetilde{X}^{(4)}_a, \label{eq:Thierry92cB}
\end{eqnarray}
where the 1-Hermite heat flux $\big[\vecq_a\big]_1$ is given by (\ref{eq:Thierry51P}).
The quasi-static approximation then yields the solution
\begin{eqnarray}
  \textrm{14-mom:} \qquad
  \widetilde{X}^{(4)}_a &=& -\, \frac{20}{  \Omega_{22}} \frac{p_a}{\nu_{aa} \rho_a} \nabla\cdot \big[\vecq_a \big]_1,
\end{eqnarray}
which at the semi-linear level simplifies into the same form as (\ref{eq:X4form})
\begin{equation}
  \widetilde{X}^{(4)}_a = + \frac{p_a}{\nu_{aa} \rho_a} \Big( \kappa_\parallel^{a (4)} \nabla_\parallel^2 T_a + \kappa_\perp^{a(4)} \nabla_\perp^2 T_a \Big),
\end{equation}
but now the 1-Hermite thermal conductivities (of the 4th-order fluid moment) read
\begin{eqnarray}
\big[ \kappa_\parallel^{a (4)} \big]_1 &=&   \frac{125}{\Omega_{22}^2}\, \frac{p_a}{\nu_{aa} m_a}; \qquad
\big[ \kappa_\perp^{a (4)} \big]_1 = \frac{p_a}{\nu_{aa} m_a} \, \frac{ 20 }{ x^2 + \big( 2 \Omega_{22}/5 \big)^2 }.
\label{eq:Thierry51PX}
\end{eqnarray}
Note that for both the Coulomb collisions and the hard spheres the (\ref{eq:Thierry51PX}) has the same form (because $\Omega_{22}=2$)
and for example the parallel value reads $125/4=31.250$, which is quite different from the Coulomb value of $57.29$ given by (\ref{eq:X4Coulomb}),
representing a correction of 83\%. In contrast, the hard sphere value of $26.99$ given by (\ref{eq:X4Hard}) is much closer to the 1-Hermite result
(with a negative correction of 14\%).

Thus, already from the self-collisional case it is possible to conclude that the scalar perturbations $\widetilde{X}^{(4)}_a$ are far more sensitive to the
choice of the Hermite approximation than the heat fluxes or stress-tensors and it seems that it is necessary to consider the 23-moment model 
(see Section \ref{sec:23mom}) or possibly beyond to obtain a more reliable $\widetilde{X}^{(4)}_a$ values (the convergence was studied by \cite{Ji2023}, but
we were unable to deduce the converged $\widetilde{X}^{(4)}_a$ value from their work).

\newpage
%=======================================================================================
%=======================================================================================
\subsection{Scalar  \texorpdfstring{$\widetilde{X}^{(4)}_e$}{perturbation} for lightweight particles  \texorpdfstring{$m_e\ll m_b$}{me<<mb}}
Considering the case $m_e\ll m_b$ with temperatures $T_e\simeq T_b$, the mass-ratio coefficients $S_{ab}$ (\ref{eq:Sab2-eq})
entering the collisional exchange rates $\widetilde{Q}^{(4)}_a\,'$ (\ref{eq:Q4exchange})
are small in comparison to the self-collisions (of the order of $m_e/m_b$ or smaller).  
The evolution equation for the scalar $\widetilde{X}^{(4)}_e$ thus has the same form as (\ref{eq:Thierry92c}) 
\begin{eqnarray}
   \frac{d_e}{dt} \widetilde{X}^{(4)}_e +\nabla\cdot\vecX^{(5)}_e  -20 \frac{p_e}{\rho_e}\nabla\cdot\vecq_e
    &=& - \nu_{ee} \frac{2}{5}\Omega_{22} \widetilde{X}^{(4)}_e, \label{eq:Thierry92cV}
\end{eqnarray}
with the quasi-static solution
\begin{eqnarray}
  \textrm{22-mom:} \qquad
  \widetilde{X}^{(4)}_e &=& -\, \frac{5}{2 \nu_{ee} \Omega_{22}} \Big( \nabla\cdot \vecX^{(5)}_e
  - 20 \frac{p_e}{\rho_e} \nabla\cdot \vecq_e \Big), \label{eq:PICC1}
\end{eqnarray}
but now the (2-Hermite) heat fluxes $\vecq_e$ and $\vecX^{(5)}_e$ for the lightweight particles are given by
(\ref{eq:QEnice})-(\ref{eq:Thierry61P}) with the B-coefficients (\ref{eq:Bcoeff}), which contain the Chapman-Cowling integrals. 
For the particular case of Coulomb collisions, the solution (\ref{eq:PICC1}) recovers eq. (154) of \cite{Hunana2022}. We here directly simplify
the (\ref{eq:PICC1}) by further applying the semi-linear approximation,
with the thermal and frictional parts ($\widetilde{X}^{(4)}_e= \widetilde{X}^{(4)T}_e + \widetilde{X}^{(4)u}_e$ and $\vecX^{(5)}_e= \vecX^{(5) T}_e+\vecX^{(5) u}_e$ and
$\vecq_e=\vecq_e^T+\vecq_e^u$) 
\begin{eqnarray}
  \nabla\cdot \vecX^{(5) T}_e &=& -\, \frac{p_e}{\rho_e} \big( \kappa_\parallel^{e(5)} \nabla_\parallel^2 T_e + \kappa_\perp^{e(5)} \nabla_\perp^2 T_e \big); \nn\\
  \nabla\cdot \vecX^{(5) u}_e &=& \frac{p_e^2}{\rho_e} \Big[ \beta_0^{(5)}  \nabla\cdot \delta\bu_\parallel + \frac{\beta_1^{(5)'} x^2+\beta_0^{(5)'}}{\triangle} \nabla\cdot \delta\bu_\perp
-\frac{\beta_1^{(5)''}x^3+\beta_0^{(5)''}x}{\triangle} \nabla\cdot (\bhat\times\delta\bu)\Big]; \nn\\
  \nabla\cdot \vecq_e^T &=& -\, \big(\kappa_\parallel^e \nabla_\parallel^2 T_e + \kappa_\perp^e \nabla_\perp^2 T_e \big);\nn\\
   \nabla\cdot \vecq_e^u &=&  \beta_0 p_e \nabla\cdot \delta\bu_\parallel + p_e  \frac{\beta_1'x^2+\beta_0'}{\triangle} \nabla\cdot \delta\bu_\perp
   - p_e  \frac{\beta_1''x^3+\beta_0''x}{\triangle} \nabla\cdot(\bhat\times\delta\bu);\nn\\
   \triangle &=& x^4 + \delta_1 x^2 +\delta_0,
\end{eqnarray}
where the $\delta \bu=\bu_e-\bu_b$, the $x=\Omega_e/\nu_{eb}$ and all the other coefficients given by (\ref{eq:QEnice})-(\ref{eq:Thierry61P}).
The thermal part of (\ref{eq:PICC1}) then becomes (changing from $\nu_{ee}$ to $\nu_{eb}$ to make easy comparison with Part 1)
\begin{eqnarray}
 \widetilde{X}^{(4)T}_e &=& + \frac{p_e}{\nu_{eb}\rho_e} \Big(  \kappa_\parallel^{e(4)} \nabla_\parallel^2 T_e
 +  \kappa_\perp^{e(4)} \nabla_\perp^2 T_e \Big); \nn\\
 \kappa_\parallel^{e(4)} &=& \frac{5}{2 \Omega_{22}} \frac{\nu_{eb}}{\nu_{ee}}(\kappa_\parallel^{e(5)}-20 \kappa_\parallel^{e}); \qquad
 \kappa_\perp^{e(4)} = \frac{5}{2 \Omega_{22}} \frac{\nu_{eb}}{\nu_{ee}} (\kappa_\perp^{e(5)}-20 \kappa_\perp^{e}),
\end{eqnarray}
where the thermal conductivities (of the 4th-order fluid moment) are analogous to eq. (157) of Part 1. 
It is useful to introduce the $\gamma$-coefficients (of the 4th-order fluid moment) and write the thermal conductivities as
\begin{eqnarray}
  \kappa_\parallel^{e(4)} &=& \frac{p_e}{m_e \nu_{eb}} \gamma_0^{(4)}; \qquad
  \kappa_\perp^{e(4)} =  \frac{p_e}{ m_e \nu_{eb}} \frac{\gamma_1^{(4)'} x^2+\gamma_0^{(4)'}}{\triangle};\label{eq:kappaEscalar}\\
  \gamma_0^{(4)} &=& \frac{5}{2 \Omega_{22}} \frac{\nu_{eb}}{\nu_{ee}} \big(\gamma_0^{(5)}-20\gamma_0 \big);
  \qquad \gamma_1^{(4)'} = \frac{5}{2 \Omega_{22}} \frac{\nu_{eb}}{\nu_{ee}} \big(\gamma_1^{(5)'}-20\gamma_1'\big); \qquad
  \gamma_0^{(4)'} =  \frac{5}{2 \Omega_{22}} \frac{\nu_{eb}}{\nu_{ee}} \big(\gamma_0^{(5)'}-20\gamma_0'\big),
\end{eqnarray}
which are analogous to equations (159) and (162) of Part 1. These $\gamma$-coefficients then can be expressed through the Chapman-Cowling integrals explicitly, where the
parallel coefficient reads
\begin{eqnarray}
  \gamma_0^{(4)} &=& \frac{5}{2 \Omega_{22} \triangle^*} \frac{\nu_{eb}}{\nu_{ee}} \Big[ \frac{\nu_{ee}}{\nu_{eb}}\Big( 36 \Omega_{22}-12 \Omega_{23}+\frac{8}{7} \Omega_{24} \Big) +175-245 \Omega_{12}+114 \Omega_{13}-20 \Omega_{14}+ \frac{8}{7}\Omega_{15} \Big]; \label{eq:PICAA}\\
  %====
  \Delta^* &\equiv& (B_1 B_4 +B_2 B_3) 
   = \frac{\nu_{ee}^2}{\nu_{eb}^2} \Big( \frac{4}{175}\Omega_{22}\Omega_{24}+\frac{4}{25}\Omega_{22}^2 - \frac{4}{175} \Omega_{23}^2 \Big) 
    + \frac{\nu_{ee}}{\nu_{eb}} \Big(  -\frac{4}{35} \Omega_{12}\Omega_{24} - \frac{4}{5} \Omega_{12}\Omega_{22} +\frac{4}{175} \Omega_{13} \Omega_{24} \nn\\
 &&     + \frac{11}{25} \Omega_{13} \Omega_{22} 
    - \frac{4}{25} \Omega_{14} \Omega_{22} +\frac{4}{175} \Omega_{22}\Omega_{15}
  - \frac{2}{5} \Omega_{23}\Omega_{12} - \frac{8}{175} \Omega_{23}\Omega_{14} + \frac{48}{175} \Omega_{13} \Omega_{23} 
  + \frac{1}{7} \Omega_{24} + \Omega_{22} \Big) \nn\\
&&    -\frac{7}{4} \Omega_{12}^2 + \frac{2}{5}\Omega_{12}\Omega_{14} - \frac{4}{35} \Omega_{12}\Omega_{15} + \Omega_{13}\Omega_{12} 
  + \frac{4}{35}\Omega_{13}\Omega_{14} + \frac{4}{175} \Omega_{13} \Omega_{15} - \frac{4}{175} \Omega_{14}^2 \nn\\
&&  - \frac{19}{35} \Omega_{13}^2
  + \frac{1}{7} \Omega_{15}+ \frac{7}{4}\Omega_{13} - \Omega_{14}, \label{eq:PICAA5}
\end{eqnarray}
and the perpendicular coefficients are given by
\begin{eqnarray}
  \gamma_1^{(4)'} &=& \frac{5}{2 \Omega_{22}} \frac{\nu_{eb}}{\nu_{ee}}\Big[
    \frac{\nu_{ee}}{\nu_{eb}}\big( 4 \Omega_{23}-6 \Omega_{22}\big) - \frac{75}{2}+65\Omega_{12}-30 \Omega_{13}+4 \Omega_{14} \Big];\\
  \gamma_0^{(4)'} &=& \frac{5}{2 \Omega_{22}} \frac{\nu_{eb}}{\nu_{ee}} \triangle^*
  \Big[ \frac{\nu_{ee}}{\nu_{eb}}\Big( 36\Omega_{22}-12 \Omega_{23}+ \frac{8}{7}\Omega_{24} \Big) +175 - 245 \Omega_{12}+114 \Omega_{13}-20 \Omega_{14}+ \frac{8}{7}\Omega_{15} \Big]
  = \gamma_0^{(4)} \triangle^{*2}.
  \label{eq:PICAA2}
\end{eqnarray}
As a summary, the thermal part reads
\begin{equation}
\widetilde{X}^{(4)T}_e = + \frac{p_e^2}{\nu_{eb}^2 \rho_e m_e} \Big(  \gamma_0^{(4)} \nabla_\parallel^2 T_e
 +  \frac{\gamma_1^{(4)'} x^2+\gamma_0^{(4)'}}{\triangle} \nabla_\perp^2 T_e \Big); \qquad   \triangle = x^4 + \delta_1 x^2 +\delta_0, \label{eq:PICAA4}
\end{equation}
with the $\gamma$-coefficients given by (\ref{eq:PICAA})-(\ref{eq:PICAA2}) and the $\delta$-coefficients $\delta_1$ and $\delta_0$ given by (\ref{eq:PICAA3}),
which fully expresses the thermal part through the Chapman-Cowling integrals.
As a double-check, prescribing Coulomb collisions recovers the $\gamma$-coefficients (164) of Part 1. Also, in the limit of zero magnetic field
the result (\ref{eq:PICAA4}) simplifies into an isotropic $\widetilde{X}^{(4)T}_e = (p_e^2 /\nu_{eb}^2 \rho_e m_e)  \gamma_0^{(4)} \nabla^2 T_e$. 
For comparison, the parallel coefficients can be also written as
\begin{eqnarray}
  \textrm{Coulomb collisions}: \qquad  \gamma_0^{(4)} &=& \frac{\nu_{eb}}{\nu_{ee}}
  \frac{250 (132\frac{\nu_{ee}}{\nu_{eb}} +229)}{(576 \frac{\nu_{ee}^2}{\nu_{eb}^2}+1208 \frac{\nu_{ee}}{\nu_{eb}}+217)}; \\
  \textrm{Hard spheres}: \qquad  \gamma_0^{(4)} &=& \frac{\nu_{eb}}{\nu_{ee}}
  \frac{1000 (19 \frac{\nu_{ee}}{\nu_{eb}}+17)}{(704 \frac{\nu_{ee}^2}{\nu_{eb}^2}+1944 \frac{\nu_{ee}}{\nu_{eb}}+1275)},
\end{eqnarray}
and to have some numerical values, for the particular case of
\begin{equation}
  \frac{\nu_{ee}}{\nu_{eb}}=\frac{1}{\sqrt{2}}; \quad \textrm{Coulomb collisions}: \gamma_0^{(4)}=83.847; \qquad
  \textrm{Hard spheres}: \gamma_0^{(4)}=14.339. \label{eq:PICAA9}
\end{equation}  
%which for $\nu_{ee}/\nu_{eb}=1/\sqrt{2}$ yields values $\gamma_0^{(4)}=83.847$ and $\gamma_0^{(4)}=14.339$, respectivelly.   

A similar construction can be done for the frictional part $\widetilde{X}^{(4)u}_e$, by using the $\beta$-coefficients (\ref{eq:HFbeta0}) and (\ref{eq:Thierry61P})
and defining
\begin{eqnarray}
 \widetilde{X}^{(4)u}_e &=& -\,\frac{p_e^2}{\nu_{eb}\rho_e}
  \Big[ \beta_0^{(4)} \nabla\cdot\delta\bu_\parallel + \frac{\beta_1^{(4)'} x^2+\beta_0^{(4)'}}{\triangle} \nabla\cdot\delta\bu_\perp
    -\frac{\beta_1^{(4)''}x^3+\beta_0^{(4)''}x}{\triangle} \nabla\cdot(\bhat\times\delta\bu)\Big];\label{eq:PICAA6}\\
  %===
    \beta_0^{(4)} &=& \frac{5}{2\Omega_{22}} \frac{\nu_{eb}}{\nu_{ee}}\big(\beta_0^{(5)}-20\beta_0\big); \qquad
  \beta_1^{(4)'} = \frac{5}{2\Omega_{22}} \frac{\nu_{eb}}{\nu_{ee}}\big(\beta_1^{(5)'}-20\beta_1'\big);
  \qquad \beta_0^{(4)'} = \frac{5}{2\Omega_{22}} \frac{\nu_{eb}}{\nu_{ee}} \big(\beta_0^{(5)'}-20\beta_0'\big);\nn\\
   \beta_1^{(4)''} &=& \frac{5}{2\Omega_{22}} \frac{\nu_{eb}}{\nu_{ee}}\big(\beta_1^{(5)''}-20\beta_1''\big);\qquad
  \beta_0^{(4)''} = \frac{5}{2\Omega_{22}} \frac{\nu_{eb}}{\nu_{ee}}\big(\beta_0^{(5)''}-20\beta_0''\big),
\end{eqnarray}
which is analogous to the equations (158) and (168) of Part 1. All of the $\beta$-coefficients are then expressed through the Chapman-Cowling
integrals easily, where for example the parallel coefficient reads
\begin{eqnarray}
  \beta_0^{(4)} &=& \frac{5}{2\Omega_{22} \Delta^*} \frac{\nu_{eb}}{\nu_{ee}} \Big[
    \frac{\nu_{ee}}{\nu_{eb}} \Big(8 \Omega_{22}- \frac{16}{5}\Omega_{13}\Omega_{22}-8 \Omega_{23}+ \frac{16}{35}\Omega_{13}\Omega_{23}
    +\frac{8}{5}\Omega_{23}\Omega_{12}+ 8\Omega_{12}\Omega_{22}+\frac{8}{7}\Omega_{24}-\frac{16}{35}\Omega_{12}\Omega_{24} \Big)\nn\\ 
    &&    - 42 \Omega_{12}^2+ \frac{24}{5} \Omega_{12} \Omega_{14}- \frac{16}{35}\Omega_{12}\Omega_{15}
    +16 \Omega_{13}\Omega_{12}+ \frac{16}{35}\Omega_{13}\Omega_{14} - \frac{208}{35}\Omega_{13}^2 +\frac{8}{7}\Omega_{15}+42\Omega_{13}-16\Omega_{14} \Big], \label{eq:PICAA7}
\end{eqnarray}
with the $\Delta^*$ given by (\ref{eq:PICAA5}). Evaluating the (\ref{eq:PICAA7}) then yields
\begin{eqnarray}
  \textrm{Coulomb collisions}: \qquad  \beta_0^{(4)} &=& \frac{\nu_{eb}}{\nu_{ee}} \frac{150 (32\frac{\nu_{ee}}{\nu_{eb}} +29)}
         {(576 \frac{\nu_{ee}^2}{\nu_{eb}^2}+1208 \frac{\nu_{ee}}{\nu_{eb}}+217)}; \\
         \textrm{Hard spheres}: \qquad  \beta_0^{(4)} &=& -\, \frac{\nu_{eb}}{\nu_{ee}} \frac{50 (64 \frac{\nu_{ee}}{\nu_{eb}}+51)}
                {(704 \frac{\nu_{ee}^2}{\nu_{eb}^2}+1944 \frac{\nu_{ee}}{\nu_{eb}} +1275)},
\end{eqnarray}
and to have some numerical values, for the particular case of 
\begin{equation}
  \frac{\nu_{ee}}{\nu_{eb}}=\frac{1}{\sqrt{2}}; \quad \textrm{Coulomb collisions}: \beta_0^{(4)}=8.06; \qquad
  \textrm{Hard spheres}: \beta_0^{(4)}=-2.27. \label{eq:PICAA10}
\end{equation}
In the limit of zero magnetic field, the solution (\ref{eq:PICAA6}) simplifies into an isotropic
$\widetilde{X}^{(4)u}_e = - (p_e^2/\nu_{eb}\rho_e) \beta_0^{(4)} \nabla\cdot\delta\bu$.

%====================================
\subsubsection{Reduction into 14-moment model}
Prescribing closure $\vecX^{(5)}_e = 28 (p_e/\rho_e)\vecq_e$ in equations (\ref{eq:Thierry92cV})-(\ref{eq:PICC1})
yields a quasi-static solution
\begin{eqnarray}
  \textrm{14-mom:} \qquad
  \widetilde{X}^{(4)}_e &=&  -\, \frac{p_e}{\nu_{ee} \rho_e} \frac{20}{\Omega_{22}} \nabla\cdot \big[\vecq_e\big]_1,  \label{eq:S1H1e}
\end{eqnarray}
where the 1-Hermite heat flux $\big[\vecq_e\big]_1$ is given by (\ref{eq:Thierry63AAm})-(\ref{eq:Thierry63AA}).
Further applying the semi-linear approximation yields the thermal and frictional parts
\begin{eqnarray}
  \nabla\cdot \big[\vecq_e^T\big]_1 &=& -\kappa_\parallel^e \nabla^2_\parallel T_e - \kappa_\perp^e \nabla^2_\perp T_e; \qquad
 \kappa_\parallel^e = \frac{5}{2 B_1} \frac{p_e}{m_e \nu_{eb}}; \qquad   
\kappa_\perp^e =    \frac{5}{2} \frac{B_1}{(x^2+B_1^2)} \frac{p_e}{m_e \nu_{eb}}; \label{eq:Thierry63AAm2}\\
%=====  
\nabla\cdot \big[\vecq_e^u \big]_1 &=&  \frac{B_5}{B_1} p_e \nabla\cdot\delta\bu_\parallel + p_e  \frac{B_5 B_1}{(x^2+B_1^2)} \nabla\cdot\delta\bu_\perp 
- p_e  \frac{B_5 x}{(x^2+B_1^2)} \nabla\cdot(\bhat\times\delta\bu); \nn\\
%===
  B_1 &=& \Big[ \frac{\nu_{ee}}{\nu_{eb}} \frac{2}{5}\Omega_{22} +\frac{5}{2} -2\Omega_{12} +\frac{2}{5}\Omega_{13} \Big]; \qquad
  B_5 = \big( \frac{5}{2} -\Omega_{12} \big), \label{eq:PICAA8}
\end{eqnarray}
and the thermal part of (\ref{eq:S1H1e}) then can be written as (changing to $\nu_{eb}$ similarly to (\ref{eq:PICAA4}))
\begin{eqnarray}
  \widetilde{X}^{(4)T}_e &=& + \frac{p_e^2}{\nu_{eb}^2 \rho_e m_e} \Big( \gamma_0^{(4)}\nabla_\parallel^2 T_e
  + \frac{50}{\Omega_{22}} \frac{\nu_{eb}}{\nu_{ee}}  \frac{B_1}{(x^2+B_1^2)} \nabla_\perp^2 T_e \Big);   \qquad
  \gamma_0^{(4)} = \frac{50}{\Omega_{22} B_1} \frac{\nu_{eb}}{\nu_{ee}},
\end{eqnarray}
and the frictional part of (\ref{eq:S1H1e}) reads (similarly to (\ref{eq:PICAA6}))
\begin{eqnarray}
 \widetilde{X}^{(4)u}_e &=& -\,\frac{p_e^2}{\nu_{eb}\rho_e}
  \Big[ \beta_0^{(4)} \nabla\cdot\delta\bu_\parallel +  \frac{20}{\Omega_{22}} \frac{\nu_{eb}}{\nu_{ee}} \frac{B_5 B_1}{(x^2+B_1^2)} \nabla\cdot\delta\bu_\perp
    -  \frac{20}{\Omega_{22}} \frac{\nu_{eb}}{\nu_{ee}}\frac{B_5 x}{(x^2+B_1^2)} \nabla\cdot(\bhat\times\delta\bu)\Big]; \nn\\
  \beta_0^{(4)} &=& \frac{20}{\Omega_{22}} \frac{\nu_{eb}}{\nu_{ee}}\frac{B_5}{B_1},
\end{eqnarray}
with the $B_1$ and $B_5$ given by (\ref{eq:PICAA8}). The evaluation of the parallel coefficients yields (emphasizing the 1-Hermite approximation with brackets $[\ldots]_1$ )
\begin{eqnarray}
&&  \textrm{Coulomb collisions and Hard spheres}: \qquad \big[\gamma_0^{(4)}\big]_1 = \frac{\nu_{eb}}{\nu_{ee}} \frac{250}{(8 \frac{\nu_{ee}}{\nu_{eb}}+13) };\\
&&  \textrm{Coulomb collisions}: \quad \big[\beta_0^{(4)}\big]_1 = \frac{\nu_{eb}}{\nu_{ee}} \frac{150}{(8 \frac{\nu_{ee}}{\nu_{eb}}+13)}; \qquad
  \textrm{Hard spheres}: \quad \big[\beta_0^{(4)}\big]_1 = -\, \frac{\nu_{eb}}{\nu_{ee}} \frac{50}{(8 \frac{\nu_{ee}}{\nu_{eb}}+13)},  
\end{eqnarray}
and to have some numerical values, for the particular case of
\begin{eqnarray}
 \frac{\nu_{ee}}{\nu_{eb}}=\frac{1}{\sqrt{2}}; \quad && \textrm{Coulomb collisions and Hard spheres}: \big[\gamma_0^{(4)}\big]_1 =18.95; \\
 && \textrm{Coulomb collisions}: \big[\beta_0^{(4)}\big]_1 = 11.37; \qquad \textrm{Hard spheres}: \big[\beta_0^{(4)}\big]_1 = -3.79.
\end{eqnarray}
Comparing the numerical values with the previously obtained (\ref{eq:PICAA9}) and (\ref{eq:PICAA10}) reveals that for the Coulomb collisions,
the 2-Hermite heat fluxes in the 22-moment model have a huge influence at the thermal coefficient $\gamma_0^{(4)}$,
where the 14-moment value of $18.95$ increases over 4.4 times into the value of $83.847$. It seems that the 23-moment model yields a further drastic increase,
see the limitations Section \ref{sec:23mom} and eq. (\ref{eq:pica3}).

\newpage
%================================================================================================================
%================================================================================================================
\section{Discussion and Conclusions} \label{sec:Discussion}
Here we discuss various topics that we find of interest.

%======================================================================
\subsection{Numerical constants  \texorpdfstring{$A_{l}(\nu)$}{Al(nu)} for repulsive forces} \label{sec:Alnu}
Detailed discussions on how to solve the scattering process for a central force can be found in various books,
which we are not interested in repeating here, see for example \cite{ChapmanCowling1953}, p. 170, or \cite{SchunkNagy2009}, p. 77
(we use the notation of the last reference where the scattering angle is $\theta$ and not $\chi$). 
  The numerical values $A_1(\nu)$ and $A_2(\nu)$ for repulsive forces which enter the Chapman-Cowling integrals
  are given in the first reference in Table 3, p. 172 and in the second reference on p. 103 and
also in \cite{Hirschfelder1954}, p. 548. 
However, a small complication arises because our multi-fluid 21- and 22-moment models also require the $A_3(\nu)$ numbers, which are not given by the
above references and we have to provide these numbers, to make our model usable.
(Note that the Chapman-Cowling table contains a 3rd column with numbers $A$, but this number is unrelated to $A_3(\nu)$ and defined as (\ref{eq:zab2})).
Also, the above tables can be quite confusing for a newcomer,
because the tables only consider $\nu \ge 5$ and at the first sight it is not clear, if all of the forces below $\nu<5$ require the Coulomb logarithm cut-off,
or if the ``trouble'' starts only exactly at the case $\nu=2$ (the latter is true, and no cut-off is required for $\nu=2.1$).
We thus created a new Table \ref{Table:Anu}, which contains the required $A_3(\nu)$ numbers and for clarity, we also included the $\nu=3$ and $\nu=4$ cases.
 Later on, we found the cases $\nu=3$ and $\nu=4$, calculated also for the $A_3(\nu)$ in Table 1 of \cite{Kihara1960}
and more precise results can be found in Table 2 of \cite{Higgins1968}.  

Let us summarize a quick recipe how the numbers $A_l(\nu)$ are calculated.
 For any interaction potential $V(r)$, the distance of the closest approach is calculated by solving the equation
\begin{equation}
1-\frac{b_0^2}{r^2} -\frac{2V(r)}{\mu_{ab}g_{ab}^2} =0,
\end{equation}
     where $b_0$ is the impact parameter, $\mu_{ab}$ the reduced mass and $\bg_{ab}=\bV_a-\bV_b$ the relative velocity.
      The repulsive force $F_{ab}=K_{ab}/r^\nu$ (where $K_{ab}$ is positive) corresponds
      to potential $V(r)=\frac{K_{ab}}{(\nu-1)r^{\nu-1}}$ and it 
is useful to introduce normalization parameter $\alpha_0 =(\frac{K_{ab}}{\mu_{ab} g_{ab}^2})^{1/(\nu-1)}$. Then, by 
adopting the notation of Chapman-Cowling with the dimensionless quantity $v=b_0/r$  and the normalized impact parameter $v_0=b_0/\alpha_0$
(which is equivalent to our $\hat{b}_0$), the recipe consists of finding the $v$ value which satisfies the equation
\begin{equation}
1-v^2-\frac{2}{\nu-1} \Big(\frac{v}{v_0}\Big)^{\nu-1} =0, \label{eq:v00}
\end{equation}  
and denoting the real positive solution as $v_{00}$. Then, the relation between the scattering angle $\theta$ and the
normalized impact parameter $v_0$ is calculated according to
\begin{equation}
\theta = \pi-2 \Phi; \qquad \Phi=\int_0^{v_{00}} \frac{1}{\sqrt{1-v^2-\frac{2}{\nu-1} (v/v_0)^{\nu-1} }} dv. \label{eq:ritP}
\end{equation}  
For example, for the Coulomb collisions ($\nu=2$) one obtains $v_{00}=(-1 + \sqrt{v_0^2+1})/v_0$, leading to the
relation $\theta=2\arcsin (1/\sqrt{v_0^2+1})$, which can be rewritten as $\cos\theta=(v_0^2+1)/(v_0^2-1)$ or equivalently
as $\tan(\theta/2)=1/v_0=\alpha_0/b_0= q_a q_b /(\mu_{ab} g^2 b_0)$.
\footnote{For attractive forces, the recipe is modified by changing the signs in front of $2/(\nu-1)$ in (\ref{eq:v00})-(\ref{eq:ritP}) to plus signs and replacing
$K_{ab}$ with $|K_{ab}|$.  
For the Coulomb collisions one obtains $v_{00}=(1 + \sqrt{v_0^2+1})/v_0$, leading to the
relation $\theta=-2\arcsin (1/\sqrt{v_0^2+1})$, so that the scattering angle $\theta$ is negative, however the $\cos\theta=(v_0^2+1)/(v_0^2-1)$ is the same, because
$\cos(-x)=\cos(x)$. One can also write $\tan(\theta/2) = - 1/v_0 =  -|q_a q_b|/(\mu_{ab} g^2 b_0) = + q_a q_b /(\mu_{ab} g^2 b_0)$, which is again
the same result as for the repulsive case.}
The repulsive case $\nu=3$ is addressed in the next section. Unfortunately for a general $\nu$, the relation (\ref{eq:ritP}) can not be
obtained in primitive functions and the relationship between the scattering angle $\theta$ and the normalized impact parameter $v_0$ is only 
numerical, i.e. one prescribes some concrete $v_0$ and numerically obtains the corresponding $\theta$.
Note that without solving the relation (\ref{eq:ritP}), one can not write the differential cross-section $\sigma_{ab}(g,\theta)$ either.
Nevertheless, the effective cross-sections $\mathbb{Q}_{ab}^{(l)}$ integrate over all the possible normalized
impact parameters $v_0$, and it is possible to put all of the numerical factors inside of the constants
\begin{equation}
  A_l(\nu)=\int_0^\infty (1-\cos^l\theta) v_0 dv_0, \label{eq:pupu1}
\end{equation}
which are pure numbers. Here, we have already used for the upper boundary the usual $\infty$, because the cases
of hard spheres (with the cut-off $v_0^{\textrm{max}}=1$) and Coulomb collisions (with the cut-off $v_0^{\textrm{max}}=\Lambda$)
were already addressed, and all the other cases (for $\nu>2$) do not require a cut-off. By plugging the (\ref{eq:ritP}) into (\ref{eq:pupu1}), 
for a given $\nu$ and $l$, we need to numerically integrate 
\begin{equation}
A_l(\nu) = \int_0^\infty \Big\{ 1-(-1)^l \cos^l \Big[ \int_0^{v_{00}} \frac{2}{\sqrt{1-v^2-\frac{2}{\nu-1} (v/v_0)^{\nu-1} }} dv \Big] \Big\} v_0 dv_0. \label{eq:picc}
\end{equation}
To solve the (\ref{eq:picc}), for the inner integral we have used the built-in numerical integration in the Maple software. For the outer integral,
we have written a very primitive ``midpoint'' quadrature numerical routine, without re-scaling the $\int_0^\infty$ integrals 
and some of the last digits given in Table \ref{Table:Anu} might be slightly imprecise. The case $\nu=3$ is precise and given by (\ref{eq:A3coeff}).
The table should be ideally re-calculated with a more sophisticated numerical quadrature, nevertheless, the Table can be used with confidence.
\footnote{ Notably, our  $A_3(3)=1.4252$ given by the semi-analytic  (\ref{eq:A3coeff}),
  slightly differs from the $A_3(3)=1.4272=0.7136 \times 2$ value given by \cite{Kihara1960} and cited also by \cite{Higgins1968}
  (their value should have been $0.7126$). Otherwise (perhaps surprisingly), our numerical results are consistent with \cite{Higgins1968},
  implying that the purely repulsive case is easy to integrate and that our precision can be improved easily.}

%================================
\begin{table}[ht!]
\centering 
\begin{tabular}{| c | c | c | c |} 
  \hline
            & $A_1(\nu)$       & $A_2(\nu)$     &  $A_3(\nu)$ \\
  \hline
  $\nu=2$      & $2\ln\Lambda$   & $4\ln\Lambda$    & $6\ln\Lambda$  \\
  $\nu=3$      & 0.7952        & 1.0557         & 1.4252 \\
  $\nu=4$      & 0.494         & 0.561          & 0.750 \\
  \hline
  $\nu=5$      & 0.422         & 0.436          & 0.585 \\
  $\nu=6$      & 0.396         & 0.384          & 0.519 \\
  $\nu=7$      & 0.385         & 0.357          & 0.486 \\
  $\nu=8$      & 0.382         & 0.341          & 0.467 \\
  $\nu=9$      & 0.381         & 0.330          & 0.456 \\
  $\nu=10$     & 0.382         & 0.324          & 0.449 \\
  $\nu=11$     & 0.383         & 0.319          & 0.444 \\
  $\nu=13$     & 0.388         & 0.313          & 0.440 \\
  $\nu=15$     & 0.393         & 0.310          & 0.438 \\
  $\nu=21$     & 0.407         & 0.307         & 0.440 \\
  $\nu=25$     & 0.414         & 0.307         & 0.443 \\
  $\nu=51*$    & 0.443         & 0.311          & 0.458 \\
  $\nu=\infty$ & 1/2           & 1/3            & 1/2 \\
  \hline 
\end{tabular} 
\caption{Values $A_l(\nu)$ for {\it repulsive} inter-particle force $1/r^\nu$, as a numerical solution of eq. (\ref{eq:picc}). From the cases given,
  note that the $A_1(\nu)$ reaches a minimum around $\nu=9$ and
  the $A_3(\nu)$ around $\nu=15$. However, frustratingly, the $A_2(\nu)$ still did not reach minimum at $\nu=25$, so we were very pleased to discover that
    \cite{Higgins1968} also provide the case $\nu=51$ (their $n=50$, marked with star because we did not verify it), where the $A_2(\nu)$ minimum
    is finally visible. It is quite fascinating that one needs to go to such steep forces to recover the hard sphere limit
    (and the $\nu=51$ values are still not close).
    It could be interesting to figure out, what value of $\nu$ is required to recover two decimal digits of hard spheres.}
\label{Table:Anu}
\end{table}
%================================

 Note that we use the $A_l(\nu)$ numbers of \cite{ChapmanCowling1953}, by considering the force $1/r^\nu$.  
In many papers, the potential $1/r^n$ is considered instead, so our $\nu=n+1$. There are additional differences in normalizations, and many 
 papers  use the $A^{(l)}(n)$ numbers of \cite{Hirschfelder1954}, p. 548, and these numbers are related by 
\begin{equation} \boxed{
A_l(\nu) = A^{(l)}(n)^{\textrm{Hirschfelder}} \times 2^{2/n}, \quad \textrm{where}\quad \nu=n+1. \label{eq:Import}}
\end{equation}

\newpage
%======================================================================
\subsection{Collisions with repulsive inverse cube force  \texorpdfstring{$1/r^3$}{1/r^3}}
The case with the repulsive force $F_{ab}=K_{ab}/r^3$ can be treated semi-analytically, because the solution of (\ref{eq:v00}) reads
$v_{00}=v_0/\sqrt{v_0^2+1}$, which yields the relation between the scattering angle $\theta$ and the normalized impact parameter $v_0=\hat{b}_0=b_0/\alpha_0$,
in the form
\begin{equation}
\theta = \pi - \frac{\pi v_0}{\sqrt{v_0^2+1}}; \quad \textrm{or} \quad v_0^2 = \frac{(\pi-\theta)^2}{\theta (2\pi-\theta)}, 
\end{equation}
further yielding the differential cross-section
\begin{equation}
 \sigma_{ab}(g,\theta) = \frac{\alpha_0^2}{(\sin \theta)} \, \frac{\pi^2 (\pi-\theta)}{\theta^2 (2\pi-\theta)^2}; \qquad
  \alpha_0 = \frac{1}{g} \Big( \frac{K_{ab}}{\mu_{ab}} \Big)^{1/2}. \label{eq:sigma3}
\end{equation}
Note that similarly to Coulomb collisions, the  $\sigma_{ab}(g,\theta)$ becomes divergent at $\theta=0$. However, here no cut-off for $v_0^{\textrm{max}}$
or $\theta_{\textrm{min}}$ is
required and the effective cross-sections are well-defined for $v_0^{\textrm{max}}=\infty$ and $\theta_{\textrm{min}}=0$.
The numerical constants $A_l(3)$ need to be evaluated as
\begin{eqnarray}
  A_l (3) &=& \int_0^\infty \Big[ 1 - (-1)^l \cos^l \Big(\frac{\pi v_0}{\sqrt{v_0^2+1}} \Big) \Big] v_0 d v_0 =
   \int_0^\pi \frac{\pi^2 (\pi-\theta)}{\theta^2 (2\pi-\theta)^2} \big( 1-\cos^l\theta \big) d\theta, \label{eq:Al3}
\end{eqnarray}
 where one can choose to integrate over the impact parameter or the scattering angle.
One can directly evaluate (\ref{eq:Al3}) numerically, or alternatively, introduce the so-called
``sine integrals'' $\textrm{Si}(x) \equiv \int_0^x [(\sin t)/t] dt$. Regardless of the choice, the $A_l(3)$ values are one of the few exceptions, because
they can be easily evaluated with any precision, as
\begin{eqnarray}
\textrm{Repulsion}:\qquad  A_1 (3) &=& \frac{\pi}{2} \Big[ \textrm{Si}(\pi)-\frac{1}{2}\textrm{Si}(2\pi) \Big] -1 = 0.795202;\nn\\
  A_2 (3) &=& \frac{\pi}{2} \Big[ \textrm{Si}(2\pi)-\frac{1}{2}\textrm{Si}(4\pi) \Big] = 1.055687; \nn\\
  A_3 (3) &=& \frac{3\pi}{8} \Big[ \textrm{Si}(\pi)+\textrm{Si}(3\pi) - \frac{1}{2}\textrm{Si}(2\pi)
    - \frac{1}{2}\textrm{Si}(6\pi) \Big] -1 = 1.425238. \label{eq:A3coeff}
\end{eqnarray}  
     The first two results were first calculated by \cite{Eliason1956} (they use potential $1/r^n$ and
      their numbers must be multiplied by 2, see conversion (\ref{eq:Import})).

%======================================================================
\subsection{Collisions with attractive inverse cube force  \texorpdfstring{$1/r^3$}{1/r^3} (and repulsive core)} \label{eq:AttractNU3}
 We follow \cite{Kihara1960} and \cite{Eliason1956}.
  Let us first consider the particular case of the attractive force $F_{ab}= - |K_{ab}|/r^3$ and 
  only in the next section to consider the attractive force $F_{ab}= - |K_{ab}|/r^\nu$ for a general $\nu$. As discussed already in the introduction, it is important
  to specify what happens to particles when they meet and we prefer the ``rigid core'' model considered by \cite{Kihara1960} (and references therein),  
  where the particles represent infinitesimally small hard spheres and the potential can be written as $V(r)=\delta(r)- |K_{ab}| /(2r^{2})$.
  This delta function influences the calculations
only through specifying what happens to particles that meet, and the rigid core model is given by the usual relation $\theta = \pi-2 \Phi$.
In contrast, the transparent core model considered by \cite{Eliason1956} (where particles pass through each other) is given by $\theta = -2 \Phi$.
This assumption does not enter the calculations until eq. (\ref{eq:ritP33}).
For the particular case $\nu=3$, both models actually yield the same results, but for steeper $\nu$ the models start to differ. 

By using the variable $v=b_0/r$ and the normalized impact parameter $v_0=b_0/\alpha_0$ 
where $\alpha_0 =(\frac{|K_{ab}|}{\mu_{ab} g_{ab}^2})^{1/2}$, the equation representing the distance of the closest approach reads
\begin{equation*}
1-v^2 +\Big(\frac{v}{v_0}\Big)^{2} =0, \label{eq:v00BB}
\end{equation*}
 and the solution for $v$ becomes $v_{00}=v_0/\sqrt{v_0^2-1}$.
Obviously, the solution is well-defined only for $v_0>1$ and there is a critical value $v_0^{\textrm{crit}}=1$, below which the solution becomes imaginary. 
This is because for small normalized impact parameters $v_0<1$, the particles actually hit each other, so the distance of the closest approach is zero.
To calculate the collisional integrals, the solutions have to be split into two distinct cathegories. Starting with the case $v_0>1$,
one proceeds similarly as before, and calculates the relationship between the scattering angle $\theta$ and $v_0$ as
\begin{equation}
  v_0> 1; \qquad \theta = \pi-2 \int_0^{v_{00}} \frac{1}{\sqrt{1-v^2 +(v/v_0)^{2} }} dv = \pi - \frac{\pi v_0}{\sqrt{v_0^2-1}}. \label{eq:ritP3}
\end{equation}
 For large impact parameters $v_0\gg 1$, the scattering angle of course approaches $\theta=0$. It is useful to numerically
  explore the (\ref{eq:ritP3}) for few values of $v_0$. For example, large $v_0=10$ yields small $\theta=-0.008$.
But as the $v_0$ decreases towards the critical value, choosing $v_0=1.1547$ yields $\theta=-\pi$,
the $v_0=1.06066$ yields $\theta=-2\pi$, the $v_0=1.0328$ yields $\theta=-3\pi$, the $v_0=1.0206$ yields $\theta=-4\pi$ and so on.  
So as the $v_0$ approaches the critical value, particles spiral around each other an increasing number of times, before they separate. 
Finally, the $v_0^{\textrm{crit}}=1$ yields $\theta^{\textrm{crit}}=-\infty$, meaning that the particles keep orbiting each other. The scattering angle therefore ranges
from $-\infty$ to $0$. Note that one could write the differential cross-section in an analogous form to (\ref{eq:sigma3}), but now the $\sin\theta$ would create 
strong oscillations as $v_0$ approaches the critical value.  
The integrals can be calculated as  
\begin{eqnarray}
v_0 > 1; \qquad  A_l (3) &=& \int_1^\infty \Big[ 1 - (-1)^l \cos^l \Big(\frac{\pi v_0}{\sqrt{v_0^2-1}} \Big) \Big] v_0 d v_0 =
   \int_{-\infty}^0 \frac{\pi^2 (\pi-\theta)}{\theta^2 (2\pi-\theta)^2} \big( 1-\cos^l\theta \big) d\theta, \label{eq:Al3P}
\end{eqnarray}
 where one can choose to integrate over the impact parameter, or over the scattering angle.
The resuls can be again written in a semi-analytic form by using the sine integrals
\begin{eqnarray}
  v_0> 1; \qquad  A_1 (3) &=& \frac{\pi}{4} \textrm{Si}(2\pi) = 1.11381; \qquad
  A_2(3) = \frac{\pi}{4} \textrm{Si}(4\pi)  = 1.17194; \nn\\
   A_3(3) &=& \frac{3\pi}{16}\big[ \textrm{Si}(2\pi)+\textrm{Si}(6 \pi)\big] = 1.72956, \label{eq:Int1}
\end{eqnarray}
 and can be easily evaluated to any precission.

\begin{figure*}[!htpb]
  $$\includegraphics[width=0.4\linewidth]{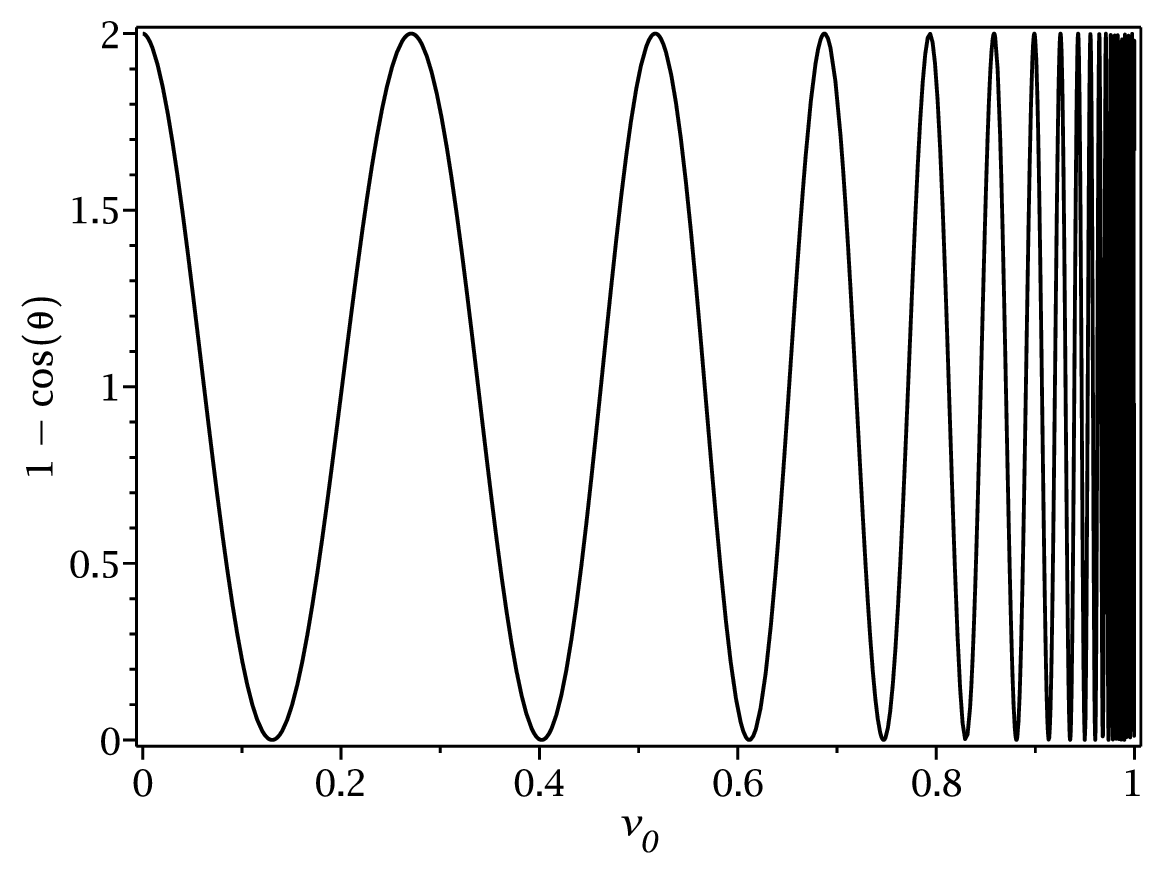}\hspace{0.02\textwidth}\includegraphics[width=0.4\linewidth]{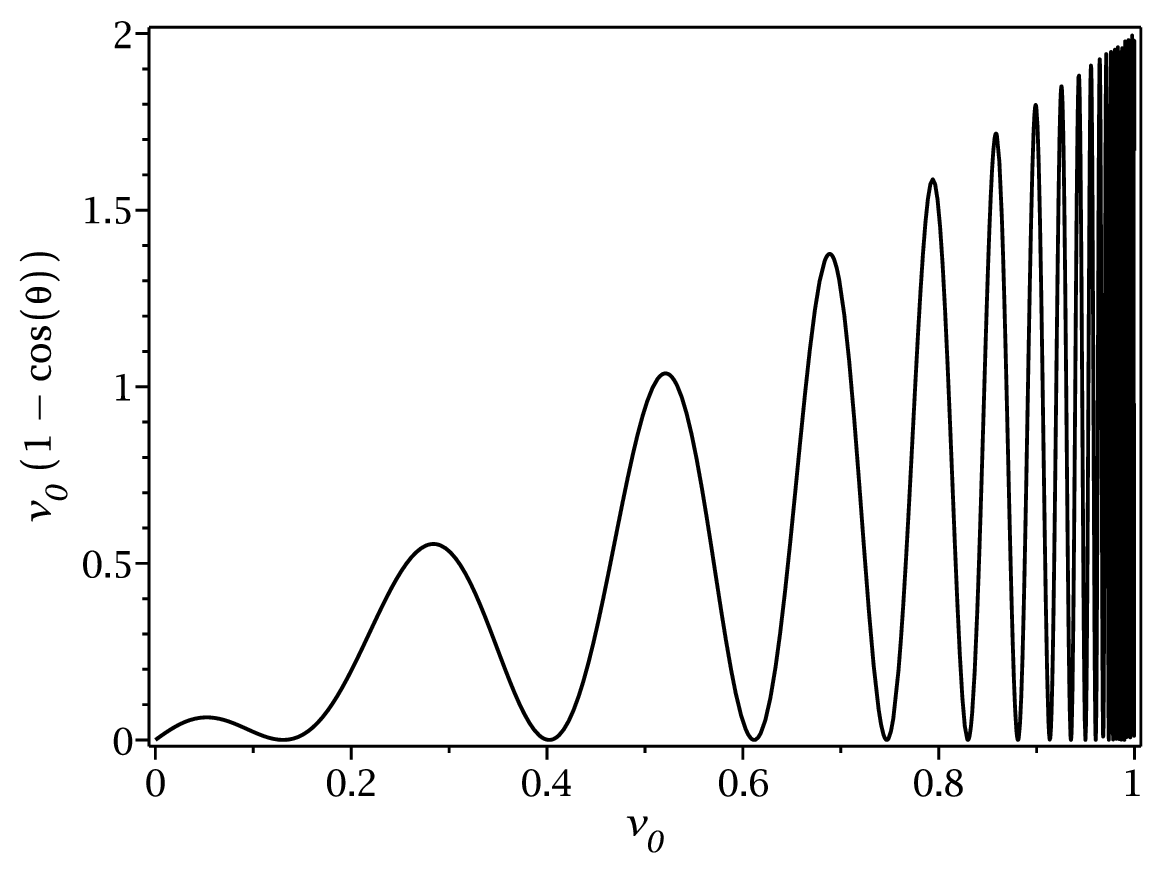}$$  
  \caption{Left panel: Function $1-\cos\theta$ representing the rigid core model (\ref{eq:ritP33}),
    with the chosen value $v=10000$. For larger $v$-values the function just becomes more oscillatory,
    but the average is obviously $1$. Right panel: Function $(1-\cos\theta) v_0$ representing integral (\ref{eq:Int2}), plotted for the same $v=10000$.
    As $v$ increases, the area under the curve converges to $1/2$.} \label{fig:1}
\end{figure*}

Now for the second part with $v_0<1$. Since there is no real $v_{00}$ solution and the distance of the closest approach is zero, the upper integration boundary
in (\ref{eq:ritP3}) is $v=\infty$, and the relation between the scattering angle $\theta$ and $v_0$ has a form
\begin{equation}
  v_0<1; \qquad \theta = \pi-2 \Phi; \qquad \Phi= \int_0^{\infty} \frac{1}{\sqrt{1-v^2 +(v/v_0)^{2} }} dv
  = \frac{v_0}{\sqrt{1-v_0^2}} \lim_{v\to\infty} \textrm{arcsinh} \Big( \frac{v}{v_0} \sqrt{1-v_0^2} \Big).\label{eq:ritP33}
\end{equation}
 Instead of the inverse hyperbolic sine, one can also use $\textrm{arcsinh}(x)=\ln (x+\sqrt{1+x^2})$.
      Note that the result (\ref{eq:ritP33}) technically diverges for $v\to\infty$ (meaning as the particles approach each other at $r\to 0$).
Nevertheless, one can still continue the calculations, because the result (\ref{eq:ritP33}) enters the next integral as $\cos\theta$, and therefore large $v$ just yields a 
function $\cos\theta$ that is rapidly oscilating. 
It is very useful to prescribe some large value of $v$ in (\ref{eq:ritP33}) and simply plot the functions $1-\cos^l\theta$, which for $v=10000$ and
$l=1$ we plot in Figure \ref{fig:1}.
 It certainly should be possible to figure out the integral (\ref{eq:Int2}) in a mathematically more appealing way,
  but simply from Figure \ref{fig:1}, it is obvious that
(for $l=1$) the integral must be equal to $1/2$. We have verified the result by numerical integration. 
One can easily plot similar figures for the $l=2$ and $l=3$ cases and as a result, one obtains 
\begin{eqnarray}
  v_0 < 1; \qquad  A_l (3) &=& \int_0^1 \Big[ 1 - \cos^l \theta \Big] v_0 d v_0 ; \label{eq:Int2}\\
  A_1(3) &=& \frac{1}{2}; \qquad A_2(3)=\frac{1}{4}; \qquad A_3(3)=\frac{1}{2}. \nn
\end{eqnarray}
     These contributions therefore come from particles that spiral toward each other, collide as infinitesimally small hard spheres, and spiral away
      from each other afterwards. Interestingly, the same
      results are obtained if one lets the particles pass through each other, by considering the tranparent core model
      with $\theta = -2 \Phi$ in (\ref{eq:ritP33}),
      because the plots look similar to Figure \ref{fig:1} (the curves are just ``symmetrical'').   

Adding the results (\ref{eq:Int1}) and (\ref{eq:Int2}) together then yields the final numbers
\begin{eqnarray}
\textrm{Attraction}:\qquad   A_1 (3) &=& \frac{\pi}{4} \textrm{Si}(2\pi) +\frac{1}{2}= 1.61381; \nn\\
  A_2(3) &=& \frac{\pi}{4} \textrm{Si}(4\pi)+\frac{1}{4}  = 1.42194; \nn\\
   A_3(3) &=& \frac{3\pi}{16}\big[ \textrm{Si}(2\pi)+\textrm{Si}(6 \pi)\big] +\frac{1}{2}= 2.22956. \label{eq:Int12}
\end{eqnarray}
 These results are of course different than for the repulsive case (\ref{eq:A3coeff}).

%\newpage
%============================================================================================================
\subsection{Numerical constants  \texorpdfstring{$A_{l}(\nu)$}{Al(nu)} for attractive forces (with repulsive core)} \label{sec:AlnuA}
  Here we consider the general attractive force $F_{ab}= - |K_{ab}|/r^\nu$ with the repulsive rigid core,
  or the potential $V(r)=\delta(r)-\frac{|K_{ab}|}{(\nu-1)r^{\nu-1}}$.
The normalized impact parameter $v_0=b_0/\alpha_0$ is defined with $\alpha_0 =(\frac{|K_{ab}|}{\mu_{ab} g_{ab}^2})^{1/(\nu-1)}$ and the equation representing the distance of
the closest approach becomes
\begin{equation}
1-v^2 +\frac{2}{\nu-1} \Big(\frac{v}{v_0}\Big)^{\nu-1} =0. \label{eq:v00B}
\end{equation}
 The calculations then proceed in the same fashion as discussed for the case $\nu=3$ in the previous section.
  First, one needs to find the critical $v_0$ value, below which eq. (\ref{eq:v00B}) does not have any real positive solution (and all solutions are either negative,
  or imaginary numbers). At first look, one would guess that for a general $\nu$, this has to be done numerically. Nevertheless, as shown by
\cite{Eliason1956}, this critical value can be actually found analytically, and is given by a very simple relation (we write both cases $\nu=n+1$)
\begin{equation}
v_0^{\textrm{crit}} = \Big( \frac{\nu-1}{\nu-3} \Big)^{\frac{\nu-3}{2(\nu-1)}} = \Big( \frac{n}{n-2} \Big)^{\frac{n-2}{2n}}. \label{eq:v0crit}
\end{equation}
 This result is obtained by realizing that at some criticial $v_0$ the eq. (\ref{eq:v00B}) should have a double root,
  so the equation is supplemented with its derivative with respect to $v$, and solving the coupled system yields $v^2=n/(n-2)$ and $v_0^{\textrm{crit}}=v^{(n-2)/n}$.
  The relationship between the scattering angle $\theta$
  and the normalized impact parameter $v_0$ is therefore easily split into two cathegories
\begin{equation}
  v_0>v_0^{\textrm{crit}}; \qquad \theta = \pi-2 \Phi; \qquad \Phi= \int_0^{v_{00}} \frac{1}{\sqrt{1-v^2 +\frac{2}{\nu-1} (v / v_0)^{\nu-1} }} dv; \label{eq:uu1}
\end{equation}
\begin{equation}
  v_0<v_0^{\textrm{crit}}; \qquad \theta = \pi-2 \Phi; \qquad \Phi= \int_0^{\infty} \frac{1}{\sqrt{1-v^2 +\frac{2}{\nu-1} (v / v_0)^{\nu-1} }} dv. \label{eq:uu2}
\end{equation}
     The first range $v_0>v_0^{\textrm{crit}}$ represents particles that do not hit each other, where
      one solves the eq. (\ref{eq:v00B}) and denotes its {\it smallest} real positive solution as $v_{00}$.
      The critical value $v_0=v_0^{\textrm{crit}}$ represents particles that keep orbiting. 
      The second range $v_0<v_0^{\textrm{crit}}$ represents particles that hit each other, where for the rigid core model $\theta = \pi-2 \Phi$,
      implying $\cos\theta=-\cos(2\Phi)$. 
   One can consider the transparent core model by using $\theta = -2 \Phi$ in (\ref{eq:uu2}), implying $\cos\theta=+\cos(2\Phi)$.
   We prefer the rigid core model and the numerical integrals are calculated as
\begin{eqnarray}
  A_l(\nu) &=& \int^\infty_{v_0^{\textrm{crit}}}
  \Big\{ 1-(-1)^l \cos^l \Big[ \int_0^{v_{00}} \frac{2}{\sqrt{1-v^2+\frac{2}{\nu-1} (v/v_0)^{\nu-1} }} dv \Big] \Big\} v_0 dv_0 \nn\\
  && +\int_0 ^{v_0^{\textrm{crit}}}
  \Big\{ 1-(-1)^l \cos^l \Big[ \int_0^{\infty} \frac{2}{\sqrt{1-v^2+\frac{2}{\nu-1} (v/v_0)^{\nu-1} }} dv \Big] \Big\} v_0 dv_0. \label{eq:piccI}
\end{eqnarray}
      Below in Table \ref{Table:Anu2}, we provide results calculated by \cite{Higgins1968}, where we also added the Coulomb case $\nu=2$ for
      visual reference. Notably, \cite{Kihara1960} also figured out the 
      quite ``head-spinning'' case of attraction force $\nu=\infty$ (with rigid core repulsion) and rather surprisingly,
      the integral $A_l(\infty)$ is the same as for pure repulsion.

%================================
\begin{table}[ht!]
\centering 
\begin{tabular}{| c | c | c | c |} 
  \hline
            & $A_1(\nu)$       & $A_2(\nu)$     &  $A_3(\nu)$ \\
  \hline
  $\nu=2$      & $2\ln\Lambda$   & $4\ln\Lambda$    & $6\ln\Lambda$  \\
  $\nu=3$      & 1.61381         & 1.42194          &  2.22956\\
  $\nu=4$      & 1.0177         & 0.7358           &  1.224\\
  \hline
  $\nu=5$      & 0.7811   &  0.5439   &  0.9018 \\
  $\nu=6$      & 0.6361   &  0.4588   &  0.7310\\
  $\nu=7$      & 0.5472   &  0.4128   &  0.6297\\
  $\nu=51$     & 0.4408   &  0.3099   &  0.4561\\
  $\nu=\infty$ & 1/2      &  1/3      &  1/2 \\
  \hline 
\end{tabular} 
\caption{Values $A_l(\nu)$ for {\it attractive} inter-particle force $-1/r^\nu$ with repulsive rigid core, as a numerical solution of eq. (\ref{eq:piccI}).
  The case $\nu=3$ is precise and given by the semi-analytic (\ref{eq:Int12}).
  The other values were calculated by \cite{Higgins1968} (they provide longer table), see also \cite{Kihara1960}.
  We multiplied their numbers by $2^{2/n}$, see conversion (\ref{eq:Import}).
  We have briefly verified the cases $\nu=5$ (the ion-neutral collisions), $\nu=7$ (the London force) and also $\nu=4,6$, but only to two decimal digits,
  because our numerical routine is too simple. }
\label{Table:Anu2}
\end{table}
%================================

\newpage
%================================================================================
\subsection{Maxwell molecules (collisions with force  \texorpdfstring{$1/r^5$}{1/r^5})}
 Let us discuss the case with $\nu=5$ in more detail.
The case is special, because one does not need to perform any expansions of the distribution function and
it is possible to calculate the collisional integrals with the Boltzmann operator for a general unspecified $f_a$, see 
Appendix \ref{sec:MaxwellR} for the momentum exchange rates and Appendix \ref{sec:MaxwellQ} for the energy exchange rates. 
As a consequence, after one prescribes the Maxwell molecules in our general model, for some equations the model collapses to the basic 5-moment model
and for some other equations, it collapses to a 1-Hermite approximation. This behavior is  natural. 
For example, considering the momentum exchange rates, evaluation for the power-law force $1/r^\nu$ yields  
\begin{eqnarray}
  \boldsymbol{R}_a &=& \sum_{b\neq a} \nu_{ab} \Big\{ \rho_a (\bu_b-\bu_a)
  +\frac{\mu_{ab}}{T_{ab}} \frac{(5-\nu)}{5(\nu-1)} \Big(\vecq_a -\frac{\rho_a}{\rho_b}\vecq_b \Big) \nn\\
  && - \Big(\frac{\mu_{ab}}{T_{ab}}\Big)^2  \frac{(5-\nu)(\nu+3)}{280(\nu-1)^2}
  \Big[\vecX^{(5)}_a -28\frac{p_a}{\rho_a}\vecq_a     - \frac{\rho_a}{\rho_b} \Big(\vecX^{(5)}_b -28\frac{p_b}{\rho_b}\vecq_b \Big) \Big] \Big\}, \label{eq:Final1z}
\end{eqnarray}
so prescribing $\nu=5$ eliminates the contributions from both the 1-Hermite and 2-Hermite heat fluxes.
The same is true for the energy exchange rates $Q_a$ (\ref{eq:Thierry38}), where the coefficients 
$\hat{P}_{ab (1)}$ and $\hat{P}_{ab (2)}$ read
\begin{eqnarray}
\hat{P}_{ab (1)} &=& -\, \frac{(\nu-5) T_a m_b }{40 (\nu-1)^2 (T_a m_b+T_b m_a)^2} \big(3 T_a m_b \nu +4 T_b m_a \nu +T_b m_b \nu  -7 T_a m_b-4 T_b m_a+3 T_b m_b \big);\nn\\
\hat{P}_{ab (2)} &=& -\, \frac{(\nu-5) T_b m_a }{40 (\nu-1)^2 (T_a m_b+T_b m_a)^2} \big(3 T_b m_a \nu +4 T_a m_b \nu +T_a m_a \nu -7 T_b m_a -4 T_a m_b +3 T_a m_a  ),
\end{eqnarray}
and become identically zero for $\nu=5$. Considering for example the self-collisional viscosities and thermal conductivities in a quasi-static
approximation (where additionally, the stress-tensors and heat fluxes are de-coupled),
the description collapses into a 1-Hermite approximation,
see for example equations (\ref{eq:CC0})-(\ref{eq:CC00}), where the collapse can be traced back into the evolution equations   
\begin{eqnarray}
   \frac{d_a}{dt} \bPi^{(2)}_a  +\Omega_a \big(\bhat\times \bPi^{(2)}_a \big)^S + p_a \bW_a
  &=&  - \nu_{aa} \frac{3}{5} \frac{A_2(\nu)}{A_1(\nu)} \frac{(3\nu-5)}{(\nu-1)} \bPi_a^{(2)} \nn\\
  && \quad + \nu_{aa}  \frac{A_2(\nu)}{A_1(\nu)} \frac{3 (3\nu-5)(5-\nu)}{140(\nu-1)^2}
  \Big( \frac{\rho_a}{p_a}\bPi^{(4)}_a- 7 \bPi^{(2)}_a\Big); \label{eq:Energy32xx}\\
   \frac{d_a}{dt} \bPi^{(4)}_a  +\Omega_a \big(\bhat\times \bPi^{(4)}_a \big)^S + 7 \frac{p_a^2}{\rho_a} \bW_a 
  &=&  -  \nu_{aa} \frac{A_2(\nu)}{A_1(\nu)} \frac{3 (3 \nu-5) (15\nu-19)}{10 (\nu-1)^2} \frac{p_a}{\rho_a}\bPi_a^{(2)} \nn\\
   && \quad - \nu_{aa} \frac{A_2(\nu)}{A_1(\nu)} \frac{(3\nu-5) (247\nu^2-710\nu+511)}{280(\nu-1)^3}
   \Big( \bPi^{(4)}_a- 7 \frac{p_a}{\rho_a} \bPi^{(2)}_a\Big). \label{eq:Energy31xx}
\end{eqnarray}
The last term of (\ref{eq:Energy32xx}) becomes zero for $\nu=5$, so the higher-order stress-tensor $\bPi_a^{(4)}$
does not change the value of the $\bPi^{(2)}_a$ anymore and the result is as 1-Hermite. In the general evolution equations for arbitrary temperatures, 
it can be shown that the following coefficients become identically zero for $\nu=5$: the $L_{ab (1)}$ and $L_{ab (2)}$ in the stress-tensor contributions 
$\bQ_{a}^{(2)}\,'$ (\ref{eq:QabP}); and the $E_{ab (1)}$ and $E_{ab (2)}$ in the heat flux contributions $\vecQ^{(3)}_{a}\,'$ (\ref{eq:Q3P}).
This is not a problem of our specific model and such a behavior is unavoidable, because to put it simply,
the Maxwell molecules do not want their distribution function to be expanded.
In the models of \cite{Schunk1977,SchunkNagy2009}, the collisional contributions between the Maxwell molecules are
expressed with their own right-hand-sides, which are fully non-linear.
We did not make such calculations since it is not really clear, if it is beneficial to have in a multi-fluid model
one fully non-linear collisional interaction, when all the other ones are only semi-linear. In our model, the  
interaction between Maxwell molecules is thus described as any other force $1/r^\nu$.  The case is often used
to describe the non-resonant interactions between ions and neutrals, where the ion polarizes the neutral and creates a dipole,
 with the resulting long range attraction force $1/r^5$ and a short range repulsion.

Considering only a single-species gas consisting of Maxwell molecules, the 2-Hermite approximation becomes slightly awkward,
if the stress-tensors and heat fluxes are de-coupled. For a gas of pure Maxwell molecules, one should retain the coupling between the stress-tensors and heat fluxes
on the left-hand-sides of evolution equations and consider at least the coupled system (\ref{eq:Num1000B})-(\ref{eq:Nomore102X})
or the fully non-linear system (\ref{eq:Num1000})-(\ref{eq:Thierry70}),
so that the hierarchy of evolution equations remains coupled and so that it brings additional information with respect to the 1-Hermite scheme.

Importantly, there is no problem at all, if a two-fluid model consisting of ions (i) and neutrals (n) is considered, where the
usual stress-tensors of ions $\bPi^{(2)}_i$ and neutrals $\bPi^{(2)}_n$ can be described by the following evolution equations     
\begin{eqnarray}
  &&   \frac{d_i}{dt} \bPi^{(2)}_i  +\Omega_i \big(\bhat\times \bPi^{(2)}_i \big)^S + p_i \bW_i =  \bQ_{i}^{(2)}\,' \nn\\
  && \quad = -\frac{3}{5}\nu_{ii} \Omega_{22}^{ii} \bPi_i^{(2)}
  + \nu_{ii} \Big(\frac{3}{20} \Omega_{22}^{ii} - \frac{3}{70} \Omega_{23}^{ii} \Big)
  \Big( \frac{\rho_i}{p_i}\bPi^{(4)}_i- 7 \bPi^{(2)}_i\Big)    \nn\\
  && \quad +\frac{\rho_i \nu_{in}}{m_i+m_n} \Big[ -K_{in (1)} \frac{1}{n_i} \bPi^{(2)}_i +K_{in (2)} \frac{1}{n_n} \bPi^{(2)}_n  \nn\\
   && \quad +L_{in (1)} \frac{1}{n_i} \Big( \frac{\rho_i}{p_i}\bPi^{(4)}_i- 7\bPi^{(2)}_i\Big)
   - L_{in (2)} \frac{1}{n_n} \Big( \frac{\rho_n}{p_n}\bPi^{(4)}_n- 7\bPi^{(2)}_n\Big)  \Big]; \label{eq:Xfun1}
\end{eqnarray}
%===============
\begin{eqnarray}
 &&  \frac{d_n}{dt} \bPi^{(2)}_n  + p_n \bW_n =  \bQ_{n}^{(2)}\,' \nn\\
   && \quad = -\frac{3}{5}\nu_{nn} \Omega_{22}^{nn} \bPi_n^{(2)}
  + \nu_{nn} \Big(\frac{3}{20} \Omega_{22}^{nn} - \frac{3}{70} \Omega_{23}^{nn} \Big)
  \Big( \frac{\rho_n}{p_n}\bPi^{(4)}_n- 7 \bPi^{(2)}_n\Big)    \nn\\
  && \quad +\frac{\rho_n \nu_{ni}}{m_i+m_n} \Big[ -K_{ni (1)} \frac{1}{n_n} \bPi^{(2)}_n +K_{ni (2)} \frac{1}{n_i} \bPi^{(2)}_i  \nn\\
   && \quad +L_{ni (1)} \frac{1}{n_n} \Big( \frac{\rho_n}{p_n}\bPi^{(4)}_n- 7\bPi^{(2)}_n\Big)
   - L_{ni (2)} \frac{1}{n_i} \Big( \frac{\rho_i}{p_i}\bPi^{(4)}_i- 7\bPi^{(2)}_i\Big)  \Big], \label{eq:Xfun2}
\end{eqnarray}
and which are coupled to the evolution equations for the $\bPi^{(4)}_i$ and $\bPi^{(4)}_n$. Note that we introduced back the species indices ``ab'' on the
ratios of the Chapman-Cowling integrals (which as noted before has to be done, to differentiate between the various collisional processes).
For example, for the self-collisions of ions, one prescribes the Coulomb interaction, with $\Omega_{22}^{ii}=2$ and $\Omega_{23}^{ii}=4$.
For the self-collisions of neutrals, one prescribes the hard sphere interaction, with $\Omega_{22}^{nn}=2$ and $\Omega_{23}^{nn}=8$.
For the collisions between ions and neutrals,
one can prescribe the Maxwell molecule interaction, where considering the small temperature differences for simplicity 
\begin{eqnarray}
  &&  K_{in(1)} = 2 + \frac{3}{5} \frac{m_n}{m_i}\Omega_{22}^{in};\qquad  K_{in(2)} = 2 - \frac{3}{5}\Omega_{22}^{in};
  \qquad \Omega_{22}^{in}=\frac{5}{2} \frac{A_2(5)}{A_1(5)}=\frac{5}{2} \times \frac{0.5439}{0.7811} = 1.741;\nn\\
&&  K_{ni(1)} = 2 + \frac{3}{5} \frac{m_i}{m_n}\Omega_{22}^{in};\qquad  K_{ni(2)} = 2 - \frac{3}{5}\Omega_{22}^{in},   
\end{eqnarray}
and all of the $L_{in (1)}=L_{in (2)}=L_{ni (1)}=L_{ni (2)}=0$. Note that it does not matter that the L-coefficients are zero, because the
self-collisions keep the coupling to the higher-order stress-tensors $\bPi^{(4)}_i$ and $\bPi^{(4)}_n$ and the system is well-defined.\\

 Finally, it is useful to clarify the collisional frequency, which for the force $F=\pm |K_{ab}|/r^5$ is given by (see eq. (\ref{eq:Rsimple2x}))
\begin{equation}
 \nu_{ab} = 2 \pi n_b \frac{\mu_{ab}^{1/2}}{m_a} |K_{ab}|^{1/2}  A_1(5).
\end{equation}
In the models of \cite{Schunk1975,Schunk1977}, see also \cite{SchunkNagy2009}, p. 90, 
the ion polarizes the neutral and the neutral becomes a dipole with attractive potential $V(r)= -\gamma_n e^2/(2r^4)$, where the
$\gamma_n$ is the neutral polarizability (given by the table on the same page). Because a general attractive force $F=-K_{ab}/r^\nu$ corresponds
to potential $V=-\frac{K_{ab}}{(\nu-1) r^{\nu-1}}$, which for $\nu=5$ means $V=- K_{ab} /(4 r^{4})$, further implying that the $K_{ab}=2\gamma_n e^2$. 
The ion-neutral collisional frequency then can be rewritten as
\begin{equation}
 \nu_{in} = \underbrace{2 \sqrt{2} A_1(5)}_{2.210}  \frac{\pi n_n m_n}{m_i+m_n} \Big(\frac{\gamma_n e^2}{\mu_{in}} \Big)^{1/2}, \label{eq:INnuab}
\end{equation}
where we used the $A_1(5)=0.7811$ from our Table \ref{Table:Anu2}. The proportionality constant of 2.21 agrees with the eq.
(4.88) of the last reference, see also \cite{Dalgarno1958} and references therein (the attraction case $1/r^5$ with rigid core repulsion
was first calculated by Langevin in 1905).

\newpage
%================================================================================
\subsection{Limitations of our approach}
%===
\subsubsection{Ideal equation of state} \label{sec:Ideal}
We note that the use of the ideal equation of state might seem contradictory at first, because
in the statistical mechanics concerning systems in
equilibrium (a gas enclosed in a box without gradients), the ideal equation of state is often interpreted as an ideal gas -
where the particles do not interact with each other (and only collide with the box).
One introduces expansion in virial coefficients $p/(k_{\rm{B}} T)=n+B_2(T)n^2+B_3(T)n^3+\cdots$,
where the second virial coefficient $B_2(T)$ is associated with the binary collisions,
the third virial coefficient $B_3(T)$ with the three-body collisions and so on. Prescribing any collisions then automatically represents the non-ideal behavior,
where for example for the hard spheres
with the diameter $\sigma$ the second virial coefficient $B_2= (2\pi/3)\sigma^3$ and the $B_3=(5/8) (B_2)^2$.
It is also possible to calculate the virial coefficients for other collisional forces, such as the repulsive power-law force that we consider,
see for example \cite{Hirschfelder1954}, p. 157.    
In contrast, with the Boltzmann operator, the non-ideal
behavior is obtained by considering non-equilibrium systems with perturbations around the $f_a^{(0)}$, where the collisions
yield the effects of viscosity and thermal conductivity, but the classical Boltzmann operator does not modify the ideal equation of state.
One can thus use two very different methods to determine the collisional forces from the experimental data, by either
1) measuring the (equilibrium) virial coefficients; or 2) measuring the (non-equilibrium) coefficients, such as the viscosity.
It is quite remarkable that these two very different methods can yield similar results for the collisional forces in some gases, see for example the Table on
p. 1110 in the last reference, where the fits for the Lennard-Jones 12-6 model are given.
The apparent controversy between these two approaches can be resolved by considering a more general Boltzmann operator appropriate for dense gases,
which takes into account the restricted space that the particles of finite
volume occupy. However, it seems that in practice such a generalized operator is typically considered only for the particular case of hard spheres,
where (by still retaining only the binary collisions) this Boltzmann operator finally yields
a pressure tensor which contains the $B_2$ virial coefficient of hard spheres, see p. 645 in \cite{Hirschfelder1954}, or
p. 284 in \cite{ChapmanCowling1953}. For other interaction potentials, perhaps the restricted space must be taken into account
quantum-mechanically.

%=============================================================================
\subsubsection{Possible improvement by the 23-moment model} \label{sec:23mom}
 As already noted in the Technical Introduction \ref{sec:Technical},
our 22-moment model represents a 2-Hermite/1-Hermite hybrid, because our stress-tensors and heat fluxes are described
 by two Hermite polynomials, whereas the fully contracted scalars are described by one Hermite polynomial.
 Our motivation for such a model in Part 1 was that we just wanted to study various generalizations of
the Braginskii 21-moment model, where as an interesting additional complication,
we added the scalars in their simplest possible form, represented by only one additional moment.    
However, from the perspective of high-order convergence studies, it is indeed more natural to consider fluid models, where the scalar perturbations are described by
the same number of polynomials, as the stress-tensors and heat fluxes are. 
In this case, it is appropriate to modify the equation (2) of Part 1 and expand the distribution function in the irreducible Hermite polynomials $H$, according to
\begin{equation}
f_a = f_a^{(0)} (1+\chi_a); \qquad \chi_a = \sum_{n=1}^{N}
  \Big[ h^{(2n)}_{ij}  H^{(2n)}_{ij} +{h}^{(2n+1)}_i {H}^{(2n+1)}_i + h^{(2n+2)} H^{(2n+2)} \Big], 
\end{equation}
where the difference is the last term for the scalar perturbations, which before contained $h^{(2n)} H^{(2n)}$. 
This new formulation has a benefit of eliminating the scalar $h^{(2)}=0$ automatically, and this formulation yields models where all quantities
are described by the same number of Hermite polynomials. Cutting the series at some chosen
``$N$'' now represents a $(5+9N)$-moment model and for example $N=1$ yields a 14-moment model, $N=2$ yields a 23-moment model, and 
$N=32$ considerd by \cite{Ji2023} yields a 293-moment model.
For the 23-moment model, the scalar $\widetilde{X}^{(4)}_a$ is coupled to another scalar $\widetilde{X}^{(6)}_a =  m_a \int |\bc_a|^{6} (f_a -f_a^{(0)}) d^3v_a$, 
and the perturbation of the 22-moment model (\ref{eq:beau2x}) becomes 
\begin{eqnarray}
\chi_a^{\textrm{(scalar)}} &=& \frac{1}{120}\frac{\rho_a}{p_a^2} \widetilde{X}^{(4)}_a (\tc_a^4-10\tc_a^2+15)
+\frac{1}{5040} \frac{\rho_a}{p_a^2}\Big[ \frac{\rho_a}{p_a}\widetilde{X}^{(6)}_a - 21 \widetilde{X}^{(4)}_a\Big]
\big( \tc_a^6-21\tc_a^4+105\tc_a^2-105\big). \label{eq:beau2xP}
\end{eqnarray}
We have actually calculated the 23-moment model and initially we had an intention to present this model here in Part 2. 
However, we have concluded that at least for the arbitrary temperatures, the collisional contributions with the Chapman-Cowling integrals are just too long
to be presented (it was very surprising, how much complexity this one additional moment brings), and the model brought additional complications.
For arbitrary temperatures, the model also requires the Chapman-Cowling
integrals $\Omega_{1,6}$, $\Omega_{2,6}$ and $\Omega_{3,6}$ (which are nevertheless straightforward to calculate,
and which are not needed if the temperature differences are small). Notably, we have encountered an unexpected behavior, where for self-collisions the perpendicular
heat conductivity (of the 4th-order fluid moment) $\hat{\kappa}_\perp^{a (4)}$ changed its sign in front of the $x^2$ term in the numerator (so that in the limit
of strong magnetic field, the conductivity became negative).
The minus sign does not imply that the result is necessarily incorrect, because the scalar $\widetilde{X}^{(4)}_a$
represents excess kurtosis, which can be positive or negative. Nevertheless, the change of sign with respect to the 22-model was unexpected
and we became uncertain, if our calculations are correct. We did not spent sufficient time to verify our calculations and
we expect that the following numbers are incorrect. We nevertheless provide them, as a motivation that the 23-moment model is expected to have quite different solutions
for the scalars $\tilde{X}^{(4)}_a$ than the 22-moment model. For the case of
self-collisions, we obtained $\hat{\kappa}_\parallel^{a (4)} =   47875/528 = 90.672$, which contrasts with the value
$1375/24 = 57.29$ of the 22-moment model, see eq. (\ref{eq:X4Coulomb}), and with the value $125/4 = 31.25$ of the 14-moment model. 
For the electron species with $Z_i=1$, we obtained 
\begin{eqnarray}
   \textrm{23-mom:} \qquad \gamma_0^{(4)} &=& 139.49; \qquad  \beta_0^{(4)} = 10.26;\nn\\
  %===
  \textrm{22-mom:} \qquad  \gamma_0^{(4)} &=& 83.85; \qquad \,\,\, \beta_0^{(4)} = 8.06;\nn\\
  %===
  \textrm{14-mom:} \qquad \gamma_0^{(4)} &=& 18.95; \qquad \,\,\, \beta_0^{(4)} = 11.37. \label{eq:pica3}
\end{eqnarray}
Again, the numbers for the 23-moment model might be incorrect, nevertheless, the large differences in the thermal conductivities $\gamma_0^{(4)}$ imply
that the 23-moment model might be the right multi-fluid model worth considering, and not our 22-moment model. 
In the recent paper of \cite{Ji2023}, the scalar perturbations are
considered to high-orders, but we were unable to make a comparison with our results.

%\newpage
\subsubsection{Other collisional interaction forces/potentials} \label{sec:Other}
 In the literature, one can find the Chapman-Cowling integrals calculated for many other collisional processes.
One particular case of the general Lennard-Jones model $F(r) = K/r^\nu-K'/r^{\nu '}$ is the Sutherland's model with $\nu=\infty$, which corresponds to hard spheres that are
attracted to each other. If the attraction of the spheres is weak $K'\ll K$, the model can be treated in a similar fashion as the inverse power-law force,  
where in the Chapman-Cowling integrals it is possible to separate the temperature and introduce
numerical integrals similar to the $A_l(\nu)$, which are independent of the temperature, see \cite{ChapmanCowling1953}, p. 180.
However, if the attraction force is not weak, one looses the
ability to separate the temperature from the numerical integrals.  One can therefore find various tables in the literature, where the Chapman-Cowling integrals
are tabulated with respect to a (normalized) temperature.  
For example, the Lennard-Jones 12-6 model $V(r)=4 \epsilon [ (\sigma/r)^{12} - (\sigma/r)^6]$ is often used to describe gases for temperatures below 1000 Kelvin and  
tabulated Chapman-Cowling integrals can be found in Appendices of \cite{Hirschfelder1954}, p. 1126.
In our notation, the ratios $\Omega_{1,2}$, $\Omega_{1,3}$, $\Omega_{2,2}$, $\Omega_{2,3}$, $\Omega_{2,4}$, $\Omega_{2,5}$
are given (together with the $\Omega_{2,6}$ and $\Omega_{4,4}$ that we do not need). By using these results, one can therefore
obtain the Braginskii ion viscosity, ion heat flux and electron viscosity, calculated for the Lennard-Jones 12-6 model. 
Additional integrals for this model can be found in \cite{Saxena1956}, where also the $\Omega_{1,4}$, $\Omega_{1,5}$ are given (together with the $\Omega_{4,3}$
that we do not need),
which is sufficient to recover the Braginskii electron heat flux. The last reference also argues that it is fine to just approximate
the $\Omega_{3,3}\simeq \Omega_{2,3}$, which then specifies our entire model for small temperature differences
(and only the $\Omega_{3,4}$ and $\Omega_{3,5}$ are further needed for arbitrary temperatures).  Obviously, our model could be potentially 
used with the Lennard-Jones 12-6 model if more effort is made, and here we considered only its simplification $V(r)=\delta(r) - 4\epsilon\sigma/r^6$.     
As a side note, which Chapman-Cowling integrals are needed by a given fluid model, is a usefull guide for judging the model's
complexity and which effects are included/excluded. \\

For temperatures higher than 1000 Kelvin, in addition to the repulsive power-law potential $V(r)=V_0/r^n$ that we use,
the repulsive exponential potential $V(r)=V_0 \exp(-r/\rho)$ is often considered in the literature as well,
see for example \cite{Monchick1959,Higgins1968}. Notably, all of the Chapman-Cowling integrals that our model needs are tabulated,
 so after re-formulation to our notation, the exponential potential could be added
to our ``menu'' of collisional forces (\ref{eq:Coulomb0})-(\ref{eq:Inv0M}) relatively easily. 
Importantly, instead of tabulation with respect to temperature, in more recent works one can find the
Chapman-Cowling integrals fitted with some approximant, so that a value for any temperature can be used.
See for example \cite{Capitelli2000} (Tables A1-A5), where the two temperature regions below and above 1000 Kelvin described by the Lennard-Jones 12-6 model
and the repulsive exponential model are fitted together, yielding the Chapman-Cowling integrals which are valid from 50 Kelvin to 100,000 Kelvin.
Nevertheless, only some Chapman-Cowling integrals that we need are given. 

 Especially interesting is the Debye screening potential $V(r)=\pm (V_0/r) \exp(-r/\lambda_D)$, which allows one
to address the ``artificial'' Coulomb logarithm cut-off. This potential was considered for example by
\cite{Stanton2016,Dangola2008,Paquette1986,Mason1967,Liboff1959,Kihara1959} (and references therein),
where the last reference provides analytic $\Omega^{(l,j)}_{ab}$ in the limit of
large temperatures (large Coulomb logarithm), see his eq. (4.5). Using this analytic result
in our quasi-static expressions would constitute a generalization of the Braginskii model to the Debye screened potential.
Nevertheless, we need to study this case in much better detail and the discussion is postponed to future venues.   
Additionally, judging from the work of \cite{Liboff1959}, for large $\ln\Lambda$ the Debye screened potential seem to yield only 
small differences in comparison to the usual artificial cut-off (the claimed differences are only 2\% for the diffusion
and only 0.5\% for the coefficients of viscosity and thermal conductivity). 
In other words, for large $\ln\Lambda$ the usual cut-off at the Debye length is actually a very reasonable approximation, allowing one to avoid
the complexity of the Debye screened potential. This is however not true if the $\ln\Lambda$ is not large, and the Debye screened potential
is often used to describe the diffusion of elements in solar/stellar interiors and around white dwarfs. Here our quasi-static expressions
presented in Section \ref{sec:electrons} (``electrons'') and Section \ref{sec:heavy} (``improved ions'') show the limitation that we
have assumed that the cases of repulsion and attraction have the same Chapman-Cowling integrals, which is not true for the
Debye screened potential in moderately-coupled plasmas, and these expressions have to be revisited (nevertheless, this is easy to do
from our general formulation).

%\newpage
\subsubsection{Extending the Braginskii model into anisotropic (CGL) framework} \label{sec:MAX}
As already discussed in Section 8.9.1, p. 39 of \cite{Hunana2022}, one of the major limitations of our Braginskii-type models in a weakly-collisional
regime is the neglection of the possible temperature anisotropy of the equilibrium distribution function $f_a^{(0)}$, around which the models are expanded. 
Here we want to briefly discuss, how an extension of the Braginskii model into the anisotropic framework pioneered by \cite{Chew1956} would look like. 
Such a model obviously needs to incorporate the ``stress-tensors'' (matrices) coming from the 4th-order fluid moment $X^{(4)}_{ijkl}$ and the ``heat fluxes'' (vectors)
coming from the 5th-order fluid moment $X^{(5)}_{ijklm}$, however, the anisotropic CGL-type decomposition of these tensors with respect to magnetic field lines
is much more complicated than the isotropic MHD-type decomposition. It can be shown that in addition to the usual CGL stress-tensor (coming from the 2nd-order fluid moment)
\begin{eqnarray}
\bPi_a^{(2)\textrm{CGL}} = m_a \int \Big( \bc_a\bc_a -c_{\parallel a}^2 \bhat\bhat -\frac{c_{\perp a}^2}{2} \bI_\perp \Big) f_a d^3v_a, \label{eq:PiCGL}
\end{eqnarray}
it is necessary to consider two independent stress-tensors of the 4th-order fluid moment
\begin{eqnarray}
 && \bPi^{\parallel (4)}_a = m_a \int \Big[ \bc_a \bc_a -c_{\parallel a}^2\bhat\bhat -\frac{c_{\perp a}^2}{2} \bI_\perp\Big] c_{\parallel a}^2 f_a d^3v_a;\nn\\
 && \bPi^{\perp(4)}_a = \frac{m_a}{2} \int \Big[ \bc_a \bc_a -c_{\parallel a}^2\bhat\bhat
    -\frac{c_{\perp a}^2}{2} \bI_\perp\Big] c_{\perp a}^2 f_a d^3v_a, \label{eq:happy15}
\end{eqnarray}
where each stress-tensor has 4 independent components. Similarly, in addition to the two usual CGL heat flux vectors (coming from the 3rd-order fluid moment)
\begin{eqnarray}
&& \boldsymbol{S}^\parallel_a = m_a\int c_{\parallel a}^2 \bc_a f_a d^3v_a;\qquad
 \boldsymbol{S}^\perp_a = \frac{m_a}{2}\int c_{\perp a}^2 \bc_a f_a d^3v_a, \label{eq:Sdef}
\end{eqnarray}
it is necessary to consider three independent heat flux vectors of the 5th-order fluid moment
\begin{eqnarray}
  \boldsymbol{S}^{\parallel\parallel (5)}_a &=& m_a\int c_{\parallel a}^4 \bc_a f_a d^3v_a;\qquad
  \boldsymbol{S}^{\parallel\perp (5)}_a = \frac{m_a}{2}\int c_{\parallel a}^2 c_{\perp a}^2 \bc_a f_a d^3v_a;\qquad
  \boldsymbol{S}^{\perp\perp (5)}_a = \frac{m_a}{4}\int c_{\perp a}^4 \bc_a f_a d^3v_a, \label{eq:S5def}
\end{eqnarray}
where each vector has 3 independent components. The basic CGL model has 6 independent components (1 density, 3 velocities and 2 scalar pressures $p_\parallel$ and $p_\perp$).
Incorporating the usual CGL stress-tensor (\ref{eq:PiCGL}) and heat flux vectors (\ref{eq:Sdef}) then represents a 16-moment model. Considering expansions
around a bi-Maxwellian $f_a^{(0)}$, the collisional contributions
for the 16-moment model with the Landau collisional operator were calculated by \cite{ChoduraPohl1971} and with the Boltzmann operator by \cite{DemarsSchunk1979,BarakatSchunk1982}. 
Incorporating also the stress-tensors (\ref{eq:happy15}) and heat fluxes (\ref{eq:S5def}) then yields a 33-moment model and
  incorporating the 3 scalar perturbations 
\begin{eqnarray}
  \widetilde{X}^{(4)}_{\parallel\parallel a} =  m_a\int c_{\parallel a}^4 (f_a-f_a^{(0)}) d^3v_a;\qquad
   \widetilde{X}^{(4)}_{\parallel\perp a} = \frac{m_a}{2}\int c_{\parallel a}^2 c_{\perp a}^2 (f_a-f_a^{(0)}) d^3v_a;\qquad
  \widetilde{X}^{(4)}_{\perp\perp a} = \frac{m_a}{4}\int c_{\perp a}^4 (f_a-f_a^{(0)}) d^3v_a, \nn
\end{eqnarray}
  yields a 36-moment model. We note that these scalar perturbations seem to be required to capture the 2-Hermite 
  Braginskii parallel viscosity (which is absent in the CGL stress-tensors (\ref{eq:happy15})).
  However, more comments on the minimal model capable of reproducing this parallel viscosity cannot be provided without
  performing explicit calculations, which will be addressed elsewhere. (I.e., it is not clear whether it would be possible to
  construct a 35-moment model with only two scalars $\widetilde{X}^{(4)}_{\parallel\parallel a}+2\widetilde{X}^{(4)}_{\parallel\perp a}$ and
  $\widetilde{X}^{(4)}_{\parallel\perp a}+2\widetilde{X}^{(4)}_{\perp\perp a}$).

%We have made a significant effort and obtained the fully non-linear evolution equations of this anisotropic 36-moment model,
%together with the anisotropic Hermite expansions around a bi-Maxwellian $f_a^{(0)}$ and we have constructed the underlying perturbation 
%of the distribution function $f_a=f_a^{(0)}(1+\chi_a)$ (without calculating the collisional integrals),
%however, the results are too technical to report here and will be addressed elsewhere. 

%\newpage
%=============================================================================
\subsubsection{Negativity of the distribution function} \label{sec:NEG}
Here we need to address again the possible negativity of the distribution function, previously addressed in Part 1 in   
 Section 8.9.5, p. 44 ``Comments on the positivity of the perturbed distribution function'' and we suggest that a reader reads
 that section first, before returning here. The discussion there has some good points, however, 
 we now have a much simpler view and we want to clearly state that the negativity of the distribution function is not a possibility,
 but a certainty. Before, we were under the impression that there is some threshold how large the fluid moments - such as the
 heat flux - can become and if the heat flux is kept sufficiently small, we thought that the distribution function remains positive. 
 This is however not true, which can be easily seen from the 1-Hermite heat flux perturbation
 \begin{equation}
f_a = f_a^{(0)} \big[ 1+ \frac{m_a}{p_a T_a}(\vecq_a\cdot\bc_a)\big( c_a^2 \frac{m_a}{5T_a} - 1 \big) \big].\nn
\end{equation}
 The heat flux can be positive or negative, so by making it negative, regardless how small the $\vecq_a$ is chosen to be,
 there is always a sufficiently large velocity $c_a$, where the distribution function becomes negative. 
 Basically, no more discussion is required and for example the criticism of \cite{Scudder2021} and \cite{Cranmer2021} (and references therein)
 about the $f_a<0$ is correct. The problem are the polynomials and that the velocity $c_a$ is unrestricted.  
 Perhaps, as one goes sufficiently high in the fluid hierarchy, the region where the $f_a<0$ occurs might be moved to higher velocity values
 and employing relativistic effects might help (or perhaps not). It is useful to note that this is not a problem specific to our model,
 or to the method of Grad, and the method of Chapman-Enskog expansions has the same issue (and in spite of this, these methods explain
 the experimentally measured gas viscosities and thermal conductivities with excellent accuracy).
 In fact, the issue with the negativity of the distribution with these methods have been known for quite some time
 and it is the motivation behind the so-called ``maximum entropy closures'', see for example
 \cite{Levermore1996,Groth2009,Torrilhon2010,McDonaldTorrilhon2013,Bocelli2023,Bocelli2024} and references therein, where for example the last reference
 defines the 14-moment model as
\begin{equation}
 f_a = \exp\big( \alpha_0+ \alpha_i v_i + \alpha_{ij} v_i v_j + \alpha_i^{(3)}v_i v^2 + \alpha^{(4)} v^4 \big), \nn
\end{equation}  
which is understood easily as somewhat analogous to the expansion in the 14-moment model of Grad, but importantly, now the expansion is up in the exponential,
which ensures the positivity of $f_a$.
%===
The method seems promising, even though the method has its own problems, such as the complicated relation of the $\alpha$-moments
to fluid moments and it seems that only
the heuristic BGK operator is typically employed. For clarity, we note that in those references the 21-moment model is defined as
\begin{equation}
  f_a = \exp\big( \alpha_0+ \alpha_i v_i + \alpha_{ij} v_i v_j + \alpha_{ijk}^{(3)} v_iv_jv_k + \alpha^{(4)} v^4 \big), \nn 
\end{equation}
i.e. corresponding with the method of Grad to a model, where the full heat flux tensor $q_{ijk}$ is present.
In contrast, our 22-moment model only contains (the heat flux vector) $v_iv^2$, but we go higher in the hierarchy of moments and
we also consider the $v^2 v_i v_j$ and $v_i v^4$ and our 21-moment is obtained by $\alpha^{(4)}=0$.

\newpage
%====================================================================================================================================
\subsection{Conclusions} \label{sec:Conclusions}
We have considered the classical Boltzmann operator and by keeping the restriction of small drifts between species, we have
expressed the 21- and 22-moment models through the Chapman-Cowling integrals, for arbitrary temperatures and masses of species. 
Our models are valid for a large class of collisional processes that the classical Boltzmann operator
can describe, even though we have discussed only Coulomb collisions, hard spheres, Maxwell molecules, purely repulsive force $|K|/r^\nu$,
and attractive force $-|K|/r^\nu$ with rigid repulsive core. 
Our models are best described as an improved multi-fluid 13-moment models of \cite{Burgers1969,Schunk1975,Schunk1977,SchunkNagy2009}, where
the precision of our 21-moment model is equivalent to the precision of the \cite{Braginskii1965} model.
Even though already discussed in Part 1, it is useful to clarify again that Braginskii calculated
the stress-tensors and heat fluxes only for the case of fully ionized plasma, comprising only one species of ions and electrons,
where he used the Landau collisional operator. His review paper \cite{Braginskii1965} contains Section 7 about multi-component plasmas, but there no stress-tensors
or heat fluxes were calculated. Nevertheless, in that section Braginskii actually uses the Boltzmann operator, where he cites the work of \cite{ChapmanCowling1953} and
in his short Appendix he calculates the momentum exchange rates $\boldsymbol{R}_{ab}$ and the energy exchange rates $Q_{ab}$ for the basic 5-moment models with
arbitrary masses (and small temperature differences and small drifts). 
In fact, even though he does not refer to it that way, Braginskii actually uses the Chapman-Cowling integral $\Omega_{ab}^{(1,1)}$ valid
for an arbitrary differential cross-section,
where the $\sigma_{ab}'$ given by his eq. (A6) is the momentum transfer cross-section $\sigma_{ab}'=\mathbb{Q}_{ab}^{(1)}(g)$ and taking his eq. (A5)
and using our notation yields relations
\begin{equation}
    \alpha_{ab}' = \frac{8}{3 \sqrt{\pi}} \Big( \frac{\mu_{ab}}{2 T} \Big)^{5/2} \int g^5 \mathbb{Q}_{ab}^{(1)} e^{-\mu_{ab}^2 g^2 /(2T)} dg
    = \frac{16}{3} \Omega_{ab}^{(1,1)} = \nu_{ab} \frac{m_a}{\mu_{ab} n_b}, \label{eq:BragPIC}
\end{equation}
together with his $\alpha_{ab} = m_a n_a \nu_{ab}$. So his $\alpha_{ab}'$ is the Chapman-Cowling integral $\Omega_{ab}^{(1,1)}$, where he just decided to
omit the factor of $16/3$, so that the relation of his $\alpha_{ab}'$ to collisional frequencies $\nu_{ab}$ does not contain any numerical factor.
That the above relation is indeed true, can be also seen by taking his eqs. (7.5) and (7.6), where the
$\alpha_{ab}'$  for the hard spheres and Coulomb collisions is given, and comparing these to our Chapman-Cowling integrals (\ref{eq:Rsimple2}).
To conclude, eq. (\ref{eq:BragPIC}) shows that the \cite{Braginskii1965} model uses the Chapman-Cowling integrals as well, even though only for the basic 5-moment models.
As a consequence, the Braginskii model is often cited in various papers considering solar partially ionized plasmas,
even if stress-tensors and heat fluxes are neglected in those particular papers, which can be confusing at first from a perspective of fully ionized plasma literature.  
 Here in Part 2, we have essentially generalized the Braginskii model to multi-fluid partially ionized plasmas - albeit only for elastic collisions
and without the ionization process (which are limitations of the Boltzmann operator).\\

%{\it Note for the production office - if possible, please keep the first letters of the numerical codes in the paragraphs that follow
%  with the bold font (I think it is very clarifying, because otherwise the names of codes are sometimes very confusing).}\\

Our models could be potentially useful for a large number of numerical codes in various areas, where a multi-fluid description needs to be
considered. For example, the equations of \cite{Burgers1969} were used by \cite{Thoul1994} (see also \cite{Thoul2007}) to describe the diffusion of elements in
the solar interior and their description is used in the MESA code ({\bf M}odules for {\bf E}xperiments in {\bf S}tellar {\bf A}strophysics)
- \cite{Paxton2011,Paxton2015,Paxton2018}.
There is also the numerical routine of \cite{Thoul2013}, which calculates the diffusion of elements in stars. We unfortunately do not explicitly discuss
the diffusion (which obviously would be beneficial to address in the future, see e.g. Section E.6, p. 85 in Part 1, where the BGK operator is used),
but we believe that our model is sufficiently
comprehensible and since our model reduces to the simpler Burger's model, obtaining the required coefficients should be relatively easy.
The diffusion formulation of \cite{Michaud1993}, which uses the equations of Burgers, is also implemented
in the ASTEC code (the {\bf A}arhus {\bf ST}ellar {\bf E}volution {\bf C}ode)
\cite{ChristensenDalsgaard2008}.
In the ionospheric physics, there is a system of various codes called GAIM ({\bf G}lobal {\bf A}ssimilation of {\bf I}onospheric {\bf M}easurements),
summarized by \cite{Schunk2004}.

In the solar community, there is a large number of other codes, which currently do not solve the Burgers-Schunk equations and
focus on other non-ideal MHD effects, such as the radiative transfer. Nevertheless, many of these codes can be viewed as ``multi-purpose MHD codes''
and even though some were originally developed to study the photosphere, these codes are being extended
to also study the solar chromosphere and corona and maybe some of these codes could benefit from our multi-fluid description.
Examples include the MANCHA3D code ({\bf M}ultifluid {\bf A}dvanced {\bf N}on-ideal MHD {\bf C}ode
for {\bf H}igh resolution simulations in {\bf A}strophysics 3D - \cite{Modestov2024,Popescu2019,Khomenko2018,Felipe2010,Khomenko2006};
the MURaM code (The {\bf M}ax-Planck-Institute for Aeronomy/ {\bf U}niversity of Chicago
{\bf Ra}diation {\bf M}agneto-hydrodynamics code ) - \cite{Vogler2005,Rempel2009,Rempel2017};
the MPI-AMRVAC code ({\bf M}essage {\bf P}assing {\bf I}nterface-{\bf A}daptive {\bf M}esh {\bf R}efinement {\bf V}ersatile {\bf A}dvection {\bf C}ode)
- \cite{Keppens2023,Keppens2012}; 
the Bifrost code (means a rainbow bridge in Norse mythology) - \cite{Gudiksen2011}; or the Pencil code - \cite{PencilCode2021}.

For the applications to the plasma fusion, there is a large number of codes for modeling the plasma impurities in the scrape-off layer (SOL) in a tokamak. 
Here the complexity of our model should not be a problem, because in a number of these applications,
the multi-fluid model of \cite{Zhdanov2002} (originally published in 1982) has been already implemented, which is a 21-moment model.
It is again noted that the 21-moment model of
Zhdanov, has been recently expressed through the Chapman-Cowling integrals in the papers of
\cite{Raghunathan2022b,Raghunathan2022a,Raghunathan2021}. We did not verify equivalence here in Part 2, 
nor in Part 1 with the simplified model of Zhdanov, because we are confused by the formulation of the Zhdanov's model.
The model of Zhdanov is surely very useful and we do not want to criticize it.
Some differences, such as the definition of fluid moments with respect to the bulk/drift velocity 
$\bu_a$ and not the average velocity $\langle \bu\rangle$, were already discussed in Section \ref{sec:Technical}.
Importantly, our 22-moment model contains the fully contracted scalars $\widetilde{X}^{(4)}_a$, which the 21-moment models of
Zhdanov and Raghunathan do not have. As already shown in Part 1 for the case of the Coulomb collisions, as well as here in Part 2
for the general collisional interaction, these scalars enter the energy exchange rates between species and it might be interesting to see to what degree
the introduction of these scalars modifies the predictions of fluid codes used for plasma fusion.

Examples of such codes include: the SOLBS code ({\bf S}crape-{\bf O}ff {\bf L}ayer {\bf P}lasma {\bf S}imulator) - \cite{Schneider2006,KUKUSHKIN2011},
which is a coupling of a plasma fluid code B2 with the neutral kinetic Monte-Carlo code EIRENE. Or the updated
SOLBS-ITER code package - \cite{WIESEN2015}, where the plasma fluid code B2.5 is used,
which was selected in year 2015 by the ITER organization to be the principal simulation tool for the
scrape-off layer of the future ITER machine (International Thermonuclear Experimental Reactor).
An interesting reading about the various Braginskii-Zhdanov implementations
can be found in \cite{Sytova2020,Sytova2018,ROZHANSKY2015,Makarov2023,Makarov2022}.
The Zhdanov's closure has also been implemented into the SOLEDGE3X code
\cite{Bufferand2022}. Other notable fluid codes used by the plasma fusion community are the UEDGE
code \cite{ROGNLIEN2002}; the TOKAM3X code \citep{Tamain2016,Tamain2010}; the JEREK code \citep{Korving2023}; or the
 BOUT++ ({\bf BOU}ndary {\bf T}urbulence) framework \citep{Dudson2009,Dudson2015}, from which other codes are developed, such as the 
the HERMES-3 code \citep{DUDSON2024}.

Of course, there is a large number of other MHD-type numerical codes which do not have names,
where the implementation of our models might be useful. Also,
our multi-fluid models may find some limited applications for the description of the solar wind,
see e.g. reviews by \cite{Marsch2006,BrunoCarbone2013,Verscharen2019} and
references therein, even though in this case the kinetic description seems to be more appropriate.

\section{Acknowledgments}
This work was supported by the European Research Council in the frame
of the Consolidating Grant ERC-2017-CoG771310-PI2FA ``Partial Ionisation: Two-Fluid Approach'', led by Elena Khomenko.
PH expresses his greatest gratitude to Elena Khomenko, Thierry Passot, Manuel Collados, Gary Paul Zank, Anna Tenerani,
David Mart\'inez-G\'omez and Mart\'in Manuel G\'omez M\'iguez  
for supporting this research in various ways and for the invaluable discussions about this subject. 
This manuscript can be also found at arxiv.org/abs/2406.00467.

\subsection{Notable missprints in \cite{Hunana2022}}
In eq. (82), the last $Z_i$ is missing index i. In Table 2, p. 35, the last $\tc_a$ is missing index a.
The top of eq. (B130) should read $\tilde{h}^{(4)}=(\rho/p^2) \widetilde{X}^{(4)}$ and the power is missing on $p^2$
(previous equations such as (B48) or (202) are correct). The first sentence of Section G.2 (p. 92) should read
``according to (F6)'' and not (F8). In Table 8, p. 104, the number $0.50{\bf 8}$ is missing the bold font on $8$. 
Sentence at the bottom of p. 128 should read ``introduced in Appendix K.3'' and not C.3. 

%==================================================================================================================
%==================================================================================================================
\newpage
\appendix
%====================================================================================================================
%====================================================================================================================
\section{The Boltzmann operator} \label{sec:AppendixA}
\subsection{Basic properties}
We follow \cite{SchunkNagy2009,Schunk1977,Burgers1969,Tanenbaum1967,ChapmanCowling1953}. The Boltzmannn collisional operator reads
\begin{equation}
  C_{ab} = \iint g_{ab}  \sigma_{ab} (g_{ab},\theta)
  \big[ f_a'f_b' - f_a f_b \big] d\Omega d^3 v_b, \label{eq:BoltzmannO}
\end{equation}
where $\bg_{ab}=\bV_a-\bV_b$ is the relative velocity with magnitude $g_{ab}=|\bg_{ab}|$.
The $\bV_a,\bV_b$ are velocities before the collision, the $\bV_a',\bV_b'$ are velocities after the
collision, the $f_a f_b= f_a(\bV_a) f_b(\bV_b)$ and the $f_a' f_b'= f_a(\bV_a') f_b(\bV_b')$. The primed and non-primed velocities are related through
the usual conservations of momentum and energy
\begin{eqnarray}
&&  m_a \bV_a + m_b \bV_b = m_a \bV_a' + m_b \bV_b';\nn\\
&&  m_a v_a^2 + m_b v_b^2 = m_a v_a'^2 + m_b v_b'^2,
\end{eqnarray}  
where the center-of-mass velocity $\boldsymbol{V}_c = (m_a \bV_a+m_b\bV_b)/(m_a+m_b)$ is unchanged by the collision, 
and the required transformations
from the domain $(\bV_a,\bV_b)$ into $(\boldsymbol{V}_c,\bg_{ab})$ read   
\begin{eqnarray}
  \bV_a = \boldsymbol{V}_c + \frac{m_b}{m_a+m_b} \bg_{ab}; \qquad
  \bV_b = \boldsymbol{V}_c - \frac{m_a}{m_a+m_b} \bg_{ab};\nn\\
  \bV_a' = \boldsymbol{V}_c + \frac{m_b}{m_a+m_b} \bg'_{ab}; \qquad
  \bV_b' = \boldsymbol{V}_c - \frac{m_a}{m_a+m_b} \bg'_{ab}. \label{eq:CMSimple}
\end{eqnarray}
The relative velocity after the collision $\bg_{ab}'=\bV_a'-\bV_b'$ has
a different direction than $\bg_{ab}$, but from the conservation of momentum and energy it is straightforward to show, that its
magnitude stays constant, $g_{ab}=g_{ab}'$.  
Introducing spherical co-ordinates with orthogonal unit vectors
$\hat{\boldsymbol{e}}_1, \hat{\boldsymbol{e}}_2,\hat{\boldsymbol{e}}_3$ where the direction of vector
$\boldsymbol{g}_{ab}$ forms the axis $\hat{\boldsymbol{e}}_3=\bg_{ab}/g_{ab}$,
the relative velocity after the collision then can be written as
\begin{eqnarray}
  \bg_{ab}' &=& g_{ab} \big[ \sin\theta\cos\phi \hat{\boldsymbol{e}}_1
  + \sin\theta\sin\phi \hat{\boldsymbol{e}}_2 + \cos\theta \hat{\boldsymbol{e}}_3 \big]; \nn\\
  \bg_{ab}'-\bg_{ab} &=& g_{ab} \big[ \sin\theta\cos\phi \hat{\boldsymbol{e}}_1
  + \sin\theta\sin\phi \hat{\boldsymbol{e}}_2 - (1-\cos\theta)\hat{\boldsymbol{e}}_3 \big], \label{eq:nice}
\end{eqnarray}
where the last expression represents the change of the relative velocity during the collision. 
The Boltzmann operator contains an integral over
the solid angle $d\Omega = \sin\theta d\theta d\phi$, where $\theta$ is called the scattering angle.
During calculations, the integral over the $d\Omega$ is typically calculated first. For clarity, simply integrating over (\ref{eq:nice}) yields
\begin{eqnarray}
&&  \int_0^{2\pi} \bg_{ab}' d\phi = 2\pi g_{ab} \cos\theta \hat{\boldsymbol{e}}_3 = 2\pi \bg_{ab}\cos\theta; \qquad
  \int \bg_{ab}' d\Omega = 0; \nn\\
&&    \int_0^{2\pi} (\bg_{ab}'-\bg_{ab}) d\phi = -2\pi\bg_{ab} (1-\cos\theta);\qquad
  \int (\bg_{ab}'-\bg_{ab}) d\Omega = -4\pi \bg_{ab}, \label{eq:Integ0}
\end{eqnarray}
i.e., the non-zero results are in the direction of $\bg_{ab}$.

The $\sigma_{ab} (g_{ab},\theta)$ in (\ref{eq:BoltzmannO})
is the differential cross-section (sometimes denoted as $d\sigma/d\Omega$ instead),
and it is defined through the impact parameter $b_0$ as 
\begin{equation}
  \sigma_{ab} (g_{ab},\theta )= \frac{b_0}{\sin\theta}
  \Big| \frac{db_0}{d\theta} \Big|, \label{eq:sigmaABdef}
\end{equation}
meaning that it can be obtained once the relation between $\theta$ and $b_0$ is determined for a considered interaction.
 (The impact parameter is denoted as $b_0$ to distinguish it from the species index ``b''.)

%========================================
\subsubsection{Hard spheres collisions}
For example, for
hard spheres of radii $r_a$ and $r_b$, the relation between $\theta$ and $b_0$ is obtained purely geometrically as
(see e.g. \cite{ChapmanCowling1953}, Figure 5, p. 59)
\begin{equation}
  \cos \Big( \frac{\theta}{2} \Big) = \hat{b}_0; \qquad \hat{b}_0=\frac{b_0}{r_a+r_b};
  \qquad => \qquad \sigma_{ab} (g_{ab},\theta) = \frac{(r_a+r_b)^2}{4}, \label{eq:simple3}
\end{equation}
where the $\hat{b}_0$ (with hat) can be viewed as a normalized (dimensionless) impact parameter.
The relationship between $\theta$ and $\hat{b}_0$ can be also 
written as
\begin{equation}
\cos\theta = 2\cos^2 \big(\frac{\theta}{2}\big) -1; \qquad => \qquad \cos\theta = 2\hat{b}_0^2-1.
\end{equation}  
For hard sphere collisions the 
scattering angle $\theta$ ranges from 0 to $\pi$, and the normalized impact parameter $\hat{b}_0$ ranges from 0 to 1
(for larger impact parameter the spheres do not collide).

%========================================
\subsubsection{Coulomb collisions (Coulomb logarithm)} \label{sec:AppendixAC}
For Coulomb collisions, the relation between $\theta$ and $b_0$ can be
shown to be (see e.g. \cite{SchunkNagy2009}, eqs. 4.37 and 4.51, or \cite{Burgers1969}, p. 114-115, and many other books)
\begin{eqnarray}
&&  \tan \Big( \frac{\theta}{2} \Big) = \frac{1}{\hat{b}_0}; \qquad \hat{b}_0=\frac{b_0}{\alpha_0};
  \qquad \alpha_0 = \frac{q_a q_b}{\mu_{ab} g_{ab}^2}; \qquad => \nn\\
&&  \sigma_{ab} (g_{ab},\theta) = \frac{\alpha_0^2}{4\sin^4(\theta/2)} = \frac{\alpha_0^2}{(1-\cos\theta)^2}, \label{eq:buu}
\end{eqnarray}
which is the famous Rutherford scattering cross-section. For Coulomb collisions, the normalization parameter (distance) $\alpha_0$ has a nice interpretation
  of representing the impact parameter for 90-degree scattering. Writing the $\alpha_0$ in relation (\ref{eq:buu}) without the absolute value on charges has an 
advantage that the relation is valid for both cases of repulsion and attraction. The impact parameter $b_0$ is by definition always positive, so   
for repulsion ($q_a q_b>0$) the scattering angle $\theta$ is positive, and for attraction ($q_a q_b <0$) the $\theta$ is negative.
The relationship between the $\theta$ and $\hat{b}_0$ can be also written as
\begin{eqnarray}
\textrm{repulsion}: \quad \sin \Big( \frac{\theta}{2} \Big) = + \frac{1}{\sqrt{1+\hat{b}_0^2}}; \qquad \qquad 
\textrm{attraction}: \quad \sin \Big( \frac{\theta}{2} \Big) = - \frac{1}{\sqrt{1+\hat{b}_0^2}}, \label{eq:tyt}
\end{eqnarray}
and also as
\begin{equation}
  \cos\theta = \frac{1-\tan^2(\frac{\theta}{2})}{1+\tan^2(\frac{\theta}{2})}; \qquad => \qquad
  \cos\theta = \frac{\hat{b}_0^2-1}{\hat{b}_0^2+1}. \label{eq:simple}
\end{equation}
 Even though the relations are of course symmetrical in $\theta$, we find it useful to write few values for a given normalized $\hat{b}_0$, see the table below,
  starting with large impact parameters and continuing to small ones.
%====
\begin{table}[ht!]
\centering 
\begin{tabular}{| c | c | c | c | c | c | c |} 
  \hline
                                             & $|\hat{b}_0|=100$       & $|\hat{b}_0|=10$     & $|\hat{b}_0|=1$  & $|\hat{b}_0|=0.1$ & $|\hat{b}_0|=0.01$  & $|\hat{b}_0|=0$ \\
  \hline
  Repulsion (with $\hat{b}_0\ge 0$)      & $\theta=1^\circ$   & $\theta=11^\circ$    & $\theta=90^\circ$  & $\theta=169^\circ$ & $\theta=179^\circ$ & $\theta=180^\circ$ \\
  Attraction (with $\hat{b}_0\le 0$)     & $\theta=-1^\circ$   & $\theta=-11^\circ$    & $\theta=-90^\circ$  & $\theta=-169^\circ$ & $\theta=-179^\circ$ & $\theta=-180^\circ$\\
  \hline 
\end{tabular}
%\caption{Coulomb collisions, scattering angle $\theta$ for a given normalized impact parameter $\hat{b}_0$.}
\end{table}
%====

\noindent
  Note that the attraction case for small impact parameters might seem counter-intuitive. It is important to recognize that for the cases of repulsion and
  attraction particles follow two different trajectories, representing two branches of a hyperbola. Since the relations are written in the center-of-mass reference
  frame, it is good to envision an incoming electron that approaches stationary ion. Such electron goes {\it around} an ion with an orbit analogous to a
  hyperbolic comet going around the Sun and small impact parameters indeed yield an electron that is deflected backwards. (Speeking freely, the attraction case can
  be very confusing, because if you shoot an electron towards an ion - the electron comes directly back at you!).   
 For both cases, the eccentricity of the hyperbola is the same $\epsilon=\sqrt{1+(b_0/\alpha_0)^2}$, but the distance of the closest approach written
 as $r_{\textrm{min}}=\alpha_0+\sqrt{b_0^2 +\alpha_0^2}$ is different.   
 For the repulsion case ($\alpha_0>0$), direct impact with $b_0=0$ yields $r_{\textrm{min}}=2\alpha_0$
 (which is understood easily by equating the kinetic energy $\mu_{ab} g^2_{ab}/2$ to the potential energy $q_aq_b/r_{\textrm{min}}$).
 In contrast, for the attraction case ($\alpha_0<0$), as the impact parameter decreases $b_0\to 0$, the distance of the closest approach (the perihelion) 
 $r_{\textrm{min}}\to 0$, because the Coulomb potential does not account for finite particle sizes. In the simplified picture given by eqs. (\ref{eq:buu}),
 a direct $b_0=0$ collision of an incoming electron with an ion is therefore viewed as an extreme limit of the hyperbolic orbit, where
 the electron is scattered backwards by going around the ion with an infinitly small loop, even though in reality they should of course hit each other.
 A natural fix is to introduce a rigid/hard sphere core, by considering the Sutherland's model, and to avoid the complexity of finite particle sizes,
 one can replace the Coulomb potential with $V(r)=\delta(r)+q_a q_b/r$, where $\delta(r)$ is delta function.
 But then, nothing is changed and for $b_0=0$ the electron is again scattered backwards, this time by hitting the ion as hard sphere.
 Integration over the scattering angles is therefore not affected, the repulsion and attraction cases are symmetrical for Coulomb collisions, and one can safely evaluate 
 the collisional integrals with the minimum impact parameter $b_0^{\textrm{min}}=0$.

 However, as is well-known, for Coulomb collisions the Boltzmann operator  yields collisional integrals which are technically infinite, because
the differential cross-section (\ref{eq:buu}) is too strongly divergent at $\theta=0$.  
Arguing by the Debye screening, the maximum impact parameter is then chosen to be the Debye length, $b_0^{\textrm{max}}=\lambda_D$,
which needs to be related to some minimum scattering angle $\theta_{\textrm{min}}$.
Estimating that since for pure Maxwellians with zero relative drifts one can calculate the average
 $\langle|\bV_a-\bV_b|^2\rangle = 3T_{ab}/\mu_{ab}$,
where $T_{ab}=(m_a T_b+m_b T_a)/(m_a+m_b)$ is the reduced temperature (see the integral (\ref{eq:NiceX})),
then yields the maximum normalized impact parameter
\begin{equation}
\hat{b}_0^{\textrm{max}} =  \frac{\lambda_D}{|\alpha_0|} = \frac{3T_{ab}}{|q_a q_b|} \lambda_D  \equiv \Lambda, \label{eq:lambdaWeird}
\end{equation}
  and $\Lambda$ is typically very large.
      For multi-species plasmas, the definition of the Debye length reads $\lambda_D = ( 4\pi \sum_s  n_s q_s^2 /T_s )^{-1/2}$
      (one can also define $\lambda_{D s}=\sqrt{T_s/(4\pi q_s^2 n_s )}$ for each species, with the summation $\lambda_D^{-2} = \sum_s \lambda_{D s}^{-2}$).  
Note that (\ref{eq:lambdaWeird}) is only a rough estimate, and because the final contributions will come through
$\ln\Lambda$, various simplifications of (\ref{eq:lambdaWeird}) are employed in the literature, where
the differences in temperatures and charges are often ignored. For example, considering equal temperatures and only charges $q_s=\pm e$ with
the Debye length $\lambda_D = \sqrt{T/(4\pi e^2 \sum_s n_s )}$ then yields   
\begin{equation}
  \Lambda =\frac{3 T}{e^2} \lambda_D = \frac{3 T^{3/2}}{e^3 \sqrt{4 \pi n}}= 12 \pi \lambda_D^3 n = 9\big( \frac{4}{3} \pi \lambda_D^3 \big) n; \qquad
  n=\sum_s n_s, \label{eq:DebyE}
\end{equation}
  i.e. $12\pi$ times the number of particles in the Debye cube ($\lambda_D^3$), or, $9$ times the number of particles in the Debye sphere.
   Alternativelly, one can simply take the electron Debye length $\lambda_{De}$ in the estimation of (\ref{eq:lambdaWeird}),
   so that in (\ref{eq:DebyE}) the $n$ is replaced by the electron density $n_e$  (which for $n_e=n_i$ creates only a difference of
  $\ln\sqrt{2}=0.35$ for the final $\ln\Lambda$ value). 
Essentially, when calculating Coulomb collisions, one hides the uncertainty of calculations inside of the
Coulomb logarithm, and keeps (or tries to keep) the rest of the calculations in a rigorous form. 
Importantly, the cut-off (\ref{eq:lambdaWeird}) allows one to define $\theta_{\textrm{min}}$, which then
yields the following crucial integral
\begin{equation}
  \theta_{\textrm{min}} = 2 \arctan \frac{1}{\Lambda}; \qquad => \qquad
  \int_{\theta_{\textrm{min}}}^\pi \frac{\sin\theta}{1-\cos\theta}d\theta = \ln(\Lambda^2+1) \simeq 2 \ln \Lambda, \label{eq:crucial}
\end{equation}
which is calculated easily by a substitution $x=\cos\theta$ and by using $\cos\theta_{\textrm{min}}=(\Lambda^2-1)/(\Lambda^2+1)$, see
eq. (\ref{eq:simple}). 
Integral (\ref{eq:crucial}) appears when calculating the momentum exchange rates $\boldsymbol{R}_{ab}$ for Coulomb collisions
(see Section \ref{sec:Coulomb}), where one encounters the following integral (which is useful to calculate right here)
\begin{equation}
  \textrm{Coulomb:}\qquad \int \sigma_{ab}(g_{ab},\theta) \big[ \bg_{ab}' - \bg_{ab} \big] d\Omega
  = - 2\pi \Big( \frac{q_a q_b}{\mu_{ab} g_{ab}^2} \Big)^2 \bg_{ab} \ln(\Lambda^2+1),
  \label{eq:CoulombA}
\end{equation}
and which would be unbounded if the cut-off (\ref{eq:lambdaWeird}) is not applied. 

 Alternatively, the same integrals (\ref{eq:crucial})-(\ref{eq:CoulombA}) can be calculated by switching the integration from $d\theta$ into the
  integration over $d\hat{b}_0$, according to
\begin{equation} 
 \int_0^\Lambda \frac{2\hat{b}_0}{1+\hat{b}_0^2} d\hat{b}_0 = \ln(\Lambda^2+1).  \label{eq:crucialB}
\end{equation}
%==============
 (One can use substitution (\ref{eq:simple}) directly in (\ref{eq:crucial}), or use the $\sigma_{ab}(g_{ab},\theta)$ definition (\ref{eq:sigmaABdef}) in  (\ref{eq:CoulombA})).

      Note that if one chooses to focus only on large impact parameters (as is typically done with the Landau collisional operator) and 
   for example neglects the number 1 in the denominator of (\ref{eq:crucialB}), subsequently creates a problem at the
   lower integration boundary, where it is necessary to introduce a cut-off at some $\hat{b}_0^{\textrm{min}}>0$. Nevertheless, knowing
   that the correct integral (\ref{eq:crucialB}) is equal to $2\ln\Lambda$ for large $\Lambda$, the problem is overcomed easily by a trick of
   choosing the cut-off at the normalized $\hat{b}_0^{\textrm{min}}=1$ (which corresponds to $\theta_{\textrm{max}}=\pi/2$) and which yields the same result at the end.
   With the simplified Landau operator, the Coulomb logarithm is then typically interpreted as coming from the integral
\begin{equation*}      
 \int_1^\Lambda \frac{d\hat{b}_0}{\hat{b}_0}   =\int^{\lambda_D}_{b_0^{\textrm{min}}} \frac{d b_0}{b_0} = \ln \frac{\lambda_D}{b_0^{\textrm{min}}} =\ln\Lambda,
\end{equation*}  
 with the lower cut-off $b_0^{\textrm{min}}=|\alpha_0|=|q_a q_b|/(3T_{ab})$ corresponding to 90-degree scattering.
In contrast, in the full integral (\ref{eq:crucialB}) the $\alpha_0$ does not represent a minimum distance cut-off, but simply represents a normalization distance
which is present in the Coulomb scattering relations (\ref{eq:buu}) completely naturally, and which then enters the maximum distance cut-off as the normalized $\lambda_D/|\alpha_0|$.
%==============

%\vspace{1cm}
%========================================
\subsubsection{Integrating over the Boltzmann operator} \label{Sec:IntegratingB}
During calculations, one often works with the fluctuating (random) velocity $\bc_a=\bV_a-\bu_a$, where $\bu_a$ is the bulk (fluid, drift) velocity.
The collisional contributions are calculated by multiplying the Boltzmann operator (\ref{eq:BoltzmannO}) by some tensor $\bX_a$, such
as $\bc_a\bc_a$ and integrated over $d^3v_a$.
As is well documented in various books, instead of working with the Boltzmann operator (\ref{eq:BoltzmannO}),
it is actually easier to calculate (see e.g. Appendix G  of \cite{SchunkNagy2009} or their eq. 4.60; p. 40 of \cite{Burgers1969}, or
p. 62-66 of \cite{ChapmanCowling1953})
\begin{equation} \boxed{
\int \bX_a C_{ab} d^3v_a = \iiint  g_{ab}  \sigma_{ab} (g_{ab},\theta)
  f_a f_b \big[ \bX_a'-\bX_a \big] d\Omega d^3 v_a d^3 v_b, }\label{eq:BoltzmannOO}
\end{equation}
where $\bX_a'=\bX_a(\bV_a')$ is the tensor after the collision, such as $\bc_a'\bc_a'$. Note that the bulk flow velocity $\bu_a$ does not change
during the collision, i.e. $\bu_a'=\bu_a$.
One first calculates the integral over the $d\Omega$ with the center-of-mass transformation (\ref{eq:CMSimple}) and
afterwards integrates over the $d^3v_a d^3v_b$,
where a more sophisticated center-of-mass transformation is needed, which is addressed in the next section.

 We note that often, calculations with the Boltzmann operator are presented with repulsive forces in mind,
  by integrating over the solid angle $d\Omega=\sin\theta d\theta d\phi$ in a spherical geometry, where the scattering angle $\theta$ is naturally positive and
  ranges from $0$ to $\pi$ (unless a cut-off at $\theta_{\textrm{min}}$ is required) and the $\phi$ ranges from $0$ to $2\pi$. 
  However, for attractive forces this can be very confusing, because the scattering angle $\theta$ is negative, and for steeper forces than $1/r^2$, the
  particles even spiral around each other and one needs to consider integrals over $d\theta$ with integral boundaries from $-\infty$ to $0$ (see the example
  in Section \ref{eq:AttractNU3}).
  For the attractive forces, it is better to get rid off the differential cross-section $\sigma_{ab} (g_{ab},\theta)$ and the integration over the
  solid angle $d\Omega$ from the beginning, and instead rewrite the Boltzmann operator (\ref{eq:BoltzmannOB}) back into its ``old fashioned'' form 
\begin{equation}
  C_{ab} = \iiint g_{ab}    \big[ f_a'f_b' - f_a f_b \big]  b_0 db_0 d\phi d^3 v_b, \label{eq:BoltzmannOB}
\end{equation}
     where one integrates over the positive impact parameter $b_0$ from $0$ to $\infty$ (unless a cut-off at $b_0^{\textrm{max}}$ is required)
and the angle $\phi$ from $0$ to $2\pi$. Then, the recipe (\ref{eq:BoltzmannOO}) is modified into
\begin{equation} \boxed{ 
\int \bX_a C_{ab} d^3v_a = \iiiint  g_{ab}  
  f_a f_b \big[ \bX_a'-\bX_a \big] b_0 db_0 d\phi d^3 v_a d^3 v_b.} \label{eq:BoltzmannOOB}
\end{equation}
     Additionally, even for repulsive forces $1/r^\nu$ the relationship between the $\theta$ and $b_0$ often can not be
      expressed in elementary functions (and is only numerical), so one does not want to be bothered by deriving
      the differential cross-section (\ref{eq:sigmaABdef}), when integration over the $db_0$ is readily available. Basically, the relationship between
      the $\theta$ and $b_0$ is more fundamental than the differential cross-section, and we actually wish we wrote the entire paper with recipe
      (\ref{eq:BoltzmannOOB}) instead of (\ref{eq:BoltzmannOO}). We will continue by using (\ref{eq:BoltzmannOO}), but as can be verified,
      all of the final results are valid for attractive as well as repulsive forces, and
    when in doubt, just return to (\ref{eq:BoltzmannOOB}) and repeat the calculations.\\

\newpage
%================================================
\subsection{Center-of-mass velocity transformation for Maxwellian product $f_a f_b$} \label{eq:AppendixAC}
The Boltzmann operator (\ref{eq:BoltzmannO}) or the recipe (\ref{eq:BoltzmannOO}) contain a product $f_a f_b$.
Considering colliding Maxwellians $f_a^{(0)} = \frac{n_a}{\pi^{3/2} v_{\textrm{th} a}^3} \exp(- c_a^2/v_{\textrm{th} a}^2)$
and $f_b^{(0)} = \frac{n_b}{\pi^{3/2} v_{\textrm{th} b}^3} \exp(- c_b^2/v_{\textrm{th} b}^2)$
with different temperatures $T_a, T_b$ and drifts $\bu_a, \bu_b$,
to be able to integrate over the $d^3v_a d^3v_b$, requires a more sophisticated transformation 
than (\ref{eq:CMSimple}).  For brevity, we often stop writing the species indices on the relative velocity $\bg_{ab}$ and 
we also define the difference in bulk (drift) velocities, according to 
\begin{equation}
\bg=\bV_a-\bV_b ; \qquad \bu=\bu_b-\bu_a. \label{eq:VU}
\end{equation}  
(Here we adopted the choice of \cite{SchunkNagy2009}, for example \cite{Burgers1969} defines $\bg$ with an opposite sign, see his eq. (8.2),
also in comparison to Appendix G.3 of \cite{Hunana2022} now $\bg=-\bx$.)
The required transformations from the velocity space $(\bc_a,\bc_b)$ into $(\boldsymbol{C}^*,\bg)$ are given by
\begin{eqnarray}
  \bc_a &=& \boldsymbol{C}^* +\frac{v_{\textrm{th} a}^2}{v_{\textrm{th} a}^2 + v_{\textrm{th} b}^2}(\bg+\bu);\nn\\
  \bc_b &=& \boldsymbol{C}^* -\frac{v_{\textrm{th} b}^2}{v_{\textrm{th} a}^2 + v_{\textrm{th} b}^2} (\bg+\bu), \label{eq:Pic}
\end{eqnarray}
which corresponds to defining the ``center-of-mass velocity'' as
\begin{eqnarray}
  \boldsymbol{C}^* = \frac{m_a \bc_a+m_b \bc_b}{m_a+m_b} +\frac{m_am_b}{(m_a+m_b)} \, \frac{T_b-T_a}{(m_b T_a +m_a T_b)}(\bg+\bu), \label{eq:uglyX}
\end{eqnarray}
and which transforms the product of two Maxwellians
\begin{equation}
  f_a^{(0)} f_b^{(0)} = \frac{n_a n_b}{\pi^3 v_{\textrm{th} a}^3v_{\textrm{th} b}^3} \exp\Big( -\frac{|\bc_a|^2}{v_{\textrm{th} a}^2}
  -\frac{|\bc_b|^2}{v_{\textrm{th} b}^2}\Big), 
\end{equation}
into
\begin{eqnarray}
  f_a^{(0)} f_b^{(0)} &=& \frac{n_a n_b}{\pi^3 {\alpha}^3 \beta^3}\exp\Big( -\frac{|\boldsymbol{C}^*|^2}{{\alpha}^2}
  -\frac{|\bg+\bu|^2}{\beta^2}\Big), \label{eq:fafb}
\end{eqnarray}
with new thermal speeds
\begin{equation}
  {\alpha}^2 = \frac{v_{\textrm{th} a}^2 v_{\textrm{th} b}^2}{v_{\textrm{th} a}^2 + v_{\textrm{th} b}^2};
  \qquad \beta^2 = v_{\textrm{th} a}^2+v_{\textrm{th} b}^2. 
\end{equation}  
As can be verified by calculating the Jacobian, $d^3c_a d^3c_b = d^3 C^* d^3 g$.

Considering only small drifts $u/\beta\ll 1$, the last term in (\ref{eq:fafb}) can be expanded into
(simply by using $|\bg+\bu|^2=g^2+2\bg\cdot\bu+u^2$ and expanding $\exp(-x)=1-x+x^2/2$) 
\begin{equation}
\exp\Big(-\frac{|\bg+\bu|^2}{\beta^2}\Big) = e^{-\frac{g^2}{\beta^2}}
  \Big( 1-2\frac{\bg\cdot\bu}{\beta^2} -\frac{u^2}{\beta^2} +2\frac{(\bg\cdot\bu)^2}{\beta^4}+\cdots \Big), \label{eq:Martin}
\end{equation}  
where terms of higher order than $u^2$ are neglected. In the semi-linear approximation, one also neglects the $u^2$ terms and
the product (\ref{eq:fafb}) then becomes
\begin{eqnarray}
  f_a^{(0)} f_b^{(0)} &=& \frac{n_a n_b}{\pi^3 {\alpha}^3 \beta^3}
  e^{ -\frac{C^{*2}}{{\alpha}^2}} e^{-\frac{g^2}{\beta^2}}
  \big( 1-2\frac{\bg\cdot\bu}{\beta^2} \big). \label{eq:fafbS}
\end{eqnarray}

When perturbations around Maxwellians are considered with distribution functions expanded as $f_a=f_a^{(0)}(1+\chi_a)$,
the product $f_a f_b$ in the semi-linear approximation reads
\begin{eqnarray}
  f_a f_b &=& \frac{n_a n_b}{\pi^3 {\alpha}^3 \beta^3}
  e^{ -\frac{C^{*2}}{{\alpha}^2}} e^{-\frac{g^2}{\beta^2}}
  \big( 1-2\frac{\bg\cdot\bu}{\beta^2} +\chi_a + \chi_b \big). \label{eq:fafbS2}
\end{eqnarray}

Also note that for pure Maxwellians with zero relative drifts ($\bu_a=\bu_b$)
\begin{eqnarray}
\bu=0: \qquad && \frac{1}{n_a n_b} \iint |\bV_a-\bV_b|^2 f_a^{(0)} f_b^{(0)} d^3v_a d^3v_b = \frac{1}{\pi^3 {\alpha}^3 \beta^3}
  \iint g^2 e^{ -\frac{C^{*2}}{{\alpha}^2}} e^{-\frac{g^2}{\beta^2}} d^3C^* d^3g \nn\\  
&& = \frac{1}{\pi^{3/2} \beta^3}
  \int g^2 e^{-\frac{g^2}{\beta^2}} d^3g =\frac{3}{2}\beta^2 = 3 \frac{T_{ab}}{\mu_{ab}}, \label{eq:NiceX}
\end{eqnarray}  
which is the integral used in the estimation of the Coulomb logarithm (\ref{eq:lambdaWeird}). 

\newpage
%================================================
\subsection{Summary of center-of-mass transformations (``simple'' vs. ``more advanced'')}
The Boltzmann operator can be very confusing at first, because of the various transformations that
are being used during the calculations, which we summarize right here.
In many instances (for higher-order moments than the momentum exchange rates $\boldsymbol{R}_{ab}$),
we will see that instead of the center-of-mass velocity $\boldsymbol{V}_c$, the ``natural language'' of the
Boltzmann operator is actually the modified velocity (introducing hat)
\begin{equation}
\hat{\bVV}_c \equiv \bVV_c - \bu_a.
\end{equation}
Then, by using this velocity, it is important to emphasize that the Boltzmann operator requires two distinct 
center-of-mass transformations, which for easy reference we will call ``simple'' and ``more advanced''.

1) The ``simple'' center-of-mass transformation (\ref{eq:CMSimple}) (which relates the velocities before and after the collision)
\begin{eqnarray}
  \bV_a &=& \bVV_c + \frac{\mu_{ab}}{m_a} \bg; \qquad
  \bV_a' = \bVV_c + \frac{\mu_{ab}}{m_a} \bg'; \nn\\ 
  \bc_a &=& \hat{\bVV}_c + \frac{\mu_{ab}}{m_a} \bg; \qquad
  \bc_a' = \hat{\bVV}_c + \frac{\mu_{ab}}{m_a} \bg', \label{eq:CMSimple0}
\end{eqnarray}
and which is used in the recipe (\ref{eq:BoltzmannOO}), with a subsequent integration over the solid angle $d\Omega$. Because the recipe 
(\ref{eq:BoltzmannOO}) does not contain $\bX_b$, transformations for $\bV_b$ and $\bc_b$ are not needed.

2) The ``more advanced'' center-of-mass transformation (\ref{eq:Pic}) for the Maxwellian product $f_a f_b$ 
\begin{eqnarray}
  \bc_a &=& \bC^* +\frac{v_{\textrm{th} a}^2}{\beta^2}(\bg+\bu);\qquad
  \bc_b = \bC^* -\frac{v_{\textrm{th} b}^2}{\beta^2} (\bg+\bu); \nn\\
\hat{\bVV}_c &=&
   \bC^* - \frac{2}{\beta^2}\frac{(T_b-T_a)}{(m_a+m_b)}(\bg+\bu) +\frac{\mu_{ab}}{m_a}\bu,  \label{eq:hatV}
\end{eqnarray}
where $\beta^2 = v_{\textrm{th} a}^2+v_{\textrm{th} b}^2$, which is used to transfer everything
into the space $(\bC^*,\bg)$, and integrate over the $d^3 C^* d^3 g$. Note that for equal temperatures
$v_{\textrm{th} a}^2 /\beta^2 = \mu_{ab}/m_a$ (and $v_{\textrm{th} b}^2 /\beta^2 = \mu_{ab}/m_b$) and so the velocities $\bc_a$ in the two transformations become equal. 
It is indeed the difference in the temperature of species, which makes the calculations sometimes very complicated, especially when
perturbations of the distribution function $\chi_a+\chi_b$ are considered. 
In this case, it is often the best practice to calculate the collisional contributions with equal temperatures first,
and only then repeat the same calculation with arbitrary temperatures.

\newpage
%=================================================
\section{Momentum exchange rates for 5-moment models} \label{sec:Mom}
\setcounter{equation}{0}
Here we consider only strict Maxwellians $f_a^{(0)} f_b^{(0)}$ (i.e. the 5-moment models) and calculate the momentum exchange rates for the hard sphere collisions,
Coulomb collisions and Maxwell molecules collisions.
%=============
\subsection{Hard spheres collisions (small drifts)}
Let us define the total radius $r_{ab}=r_a+r_b$ of the spheres. 
The momentum exchange rates $\boldsymbol{R}_{ab}$ are given by
\begin{equation}
  \boldsymbol{R}_{ab} = m_a \int \bV_a C_{ab} d^3v_a = m_a \frac{r_{ab}^2}{4}
  \iiint g_{ab} f_a^{(0)} f_b^{(0)} \big[ \bV_a' - \bV_a\big] d\Omega d^3 v_a d^3 v_b.
\end{equation}
One first calculates the integral over the $d\Omega$, by transforming the $ \bV_a' - \bV_a$
with (\ref{eq:CMSimple}) into
\begin{equation}
\bV_a'-\bV_a = \frac{m_b}{m_a+m_b} (\bg_{ab}'-\bg_{ab}), \label{eq:VT}
\end{equation}
which by using the $d\Omega$ integral (\ref{eq:Integ0}) yields
\begin{equation}
  \boldsymbol{R}_{ab} = - \pi r_{ab}^2 \mu_{ab} 
  \iint g_{ab} \bg_{ab} f_a^{(0)} f_b^{(0)} d^3 v_a d^3 v_b. \label{eq:RabEx}
\end{equation}

The exact integral will be calculated in the next section, and here we first focus on small drifts  
where the product $f_a^{(0)} f_b^{(0)}$ is approximated by (\ref{eq:fafbS}). Let us again drop the
species indices on $\bg_{ab}$, implying
\begin{equation}
  \boldsymbol{R}_{ab} = - r_{ab}^2 \mu_{ab}  \frac{n_a n_b}{\pi^2 {\alpha}^3 \beta^3}
  \iint  e^{ -\frac{C^{*2}}{{\alpha}^2}} e^{-\frac{g^2}{\beta^2}} g \bg
   \big( 1-2\frac{\bg\cdot\bu}{\beta^2} \big) d^3 C^* d^3 g,
\end{equation}
and the integral over $d^3C^*$ can be calculated easily, yielding
\begin{equation}
  \boldsymbol{R}_{ab} = - r_{ab}^2 \mu_{ab}  \frac{n_a n_b}{\pi^{1/2} \beta^3}
  \int e^{-\frac{g^2}{\beta^2}} g \bg
   \big( 1-2\frac{\bg\cdot\bu}{\beta^2} \big) d^3 g. \label{eq:RabC}
\end{equation}
Now finally the integration over the g-space. It is useful to rotate the spherical coordinates, and consider
unit vectors $\hat{\boldsymbol{e}}_1^*, \hat{\boldsymbol{e}}_2^*,\hat{\boldsymbol{e}}_3^*$, where now
the direction of vector $\bu$ forms the axis $\hat{\boldsymbol{e}}_3^* = \bu/u$, and the relative velocity $\bg$
reads
\begin{equation}
 \bg = g \big( \sin\theta^* \cos\phi^* \hat{\boldsymbol{e}}_1^*
  + \sin\theta^* \sin\phi^* \hat{\boldsymbol{e}}_2^* + \cos\theta^* \hat{\boldsymbol{e}}_3^* \big). \label{eq:refF}
\end{equation}
Then for example $\int \bg d\phi^* = 2\pi g \cos\theta^* \bu/u$
(i.e. the result is in the direction of $\bu$) and further integrating $\int \bg d\Omega^*=0$, meaning
that the first term in (\ref{eq:RabC}) is zero. The second term in (\ref{eq:RabC}) can be calculated by using
$\bg\cdot\bu= g u \cos\theta^*$ and $d^3g= g^2\sin\theta^* dg d\theta^* d \phi^*$, yielding integral
\begin{eqnarray}
\int e^{-\frac{g^2}{\beta^2}} g \bg
(\bg\cdot\bu ) d^3 g &=& u \int e^{-\frac{g^2}{\beta^2}} g^4 \bg \cos\theta^* \sin\theta^* dg d\theta^* d \phi^*
= 2\pi \bu \int e^{-\frac{g^2}{\beta^2}} g^5 \cos\theta^{*2} \sin\theta^* dg d\theta^* \nn\\
&=& \frac{4}{3} \pi \bu \int_0^\infty e^{-\frac{g^2}{\beta^2}} g^5 dg = \frac{4}{3} \pi \bu \beta^6. \label{eq:IntNice}
\end{eqnarray}
Alternatively, the same integral can be calculated by first pulling the dot product $\bu\cdot$ out of the integral and calculating
\begin{eqnarray}
  \bu\cdot \Big[ \int e^{-\frac{g^2}{\beta^2}} g \bg \bg d^3 g \Big] &=& \bu\cdot \Big[ \frac{4\pi}{3} \bI \int_0^\infty e^{-\frac{g^2}{\beta^2}} g^5 dg \Big]
  = \bu\cdot \Big[ \frac{4\pi}{3} \bI \beta^6 \Big],
\end{eqnarray}
where $\bI$ is the unit matrix and $\bu\cdot\bI=\bu$ (see also a more general integral (\ref{eq:formula1})). 
The (\ref{eq:RabC}) then becomes 
\begin{equation}
  \boldsymbol{R}_{ab} = + \frac{8}{3} \sqrt{\pi} r_{ab}^2 \mu_{ab}  n_a n_b \beta \bu,
  \label{eq:RabCC}
\end{equation}
which finally defines the collisional frequency $\nu_{ab}$ for hard spheres interractions, according to
\begin{equation} \boxed{
  \boldsymbol{R}_{ab} = m_a n_a \nu_{ab}(\bu_b-\bu_a); \qquad => \qquad
  \nu_{ab} = \frac{8}{3} \sqrt{\pi} (r_a+r_b)^2 \frac{m_b n_b}{m_a+m_b} \sqrt{ v_{\textrm{th} a}^2+v_{\textrm{th} b}^2 }.} \label{eq:nuab}
\end{equation}
The result agrees for example with eq. (C4) of \cite{Schunk1977}, where
\begin{equation}
2\frac{T_{ab}}{\mu_{ab}} = \frac{2T_a}{m_a}+\frac{2T_b}{m_b} = v_{\textrm{th} a}^2+v_{\textrm{th} b}^2= \beta^2.
\end{equation}

\newpage
%==================================================================================
%==================================================================================
\subsection{Hard spheres collisions (unrestricted drifts)} \label{eq:HSMomUnrestr}
Here the product $f_a^{(0)}f_b^{(0)}$ is given by the
non-expanded (\ref{eq:fafb}), and the momentum exchange rates are given by 
\begin{equation}
  \boldsymbol{R}_{ab} = - r_{ab}^2 \mu_{ab} \frac{n_a n_b}{\sqrt{\pi} \beta^3} 
  \int g \bg e^{ - \frac{|\bg+\bu|^2}{\beta^2} }   d^3 g. \label{eq:niceA}
\end{equation}
The integral is calculated by introducing the same reference frame as before (\ref{eq:refF}),
where $\hat{\boldsymbol{e}}_3^* = \bu/u$. In that reference frame $|\bg+\bu|^2=g^2+u^2+2gu\cos\theta^*$ and
the integration over $d\phi^*$ can be carried out
\begin{equation}
  \int g \bg e^{ - \frac{|\bg+\bu|^2}{\beta^2} }   d^3 g
  = 2\pi \frac{\bu}{u} \int_0^\infty \int_0^\pi g^4 e^{ - \frac{|\bg+\bu|^2}{\beta^2} } \cos\theta^* \sin\theta^* dg d\theta^*. \label{eq:gntHS}
\end{equation}
It is beneficial to introduce constant $\epsilon= u/\beta$ (as a general value which is not necessarily small)
and change the integration into new variables $z,s$ defined as (where substitution (\ref{eq:zsB}) is used as a middle step)
\begin{equation}
z=\frac{g}{\beta}+s; \qquad s=\epsilon\cos\theta^*; \qquad => \qquad \frac{|\bg+\bu|^2}{\beta^2} = z^2-s^2+\epsilon^2, \label{eq:zs}
\end{equation}
and the integral calculates
\begin{eqnarray}
  \int g \bg e^{ - \frac{|\bg+\bu|^2}{\beta^2} }  d^3 g &=& 2\pi \bu \frac{\beta^4}{\epsilon^3} e^{-\epsilon^2}
  \int_{-\epsilon}^\epsilon ds \int_s^\infty dz e^{-z^2+s^2} s(z-s)^4 \nn\\
  &=& -\pi \bu \beta^4 \Big[
    e^{-\epsilon^2} \Big( 1+\frac{2}{\epsilon^2}\Big) +\sqrt{\pi}\Big( \epsilon+\frac{1}{\epsilon}-\frac{1}{4\epsilon^3}\Big)
    \erf(\epsilon) \Big]. \label{eq:Rect}
\end{eqnarray}
To obtain the integral, it is necessary to first integrate over $dz$ 
\begin{equation}
  \int_s^\infty e^{-z^2} (z-s)^4  dz = -\, \frac{e^{-s^2}}{2}\big( s^3+\frac{5}{2}s \big)
  + \frac{\sqrt{\pi}}{2} \big(1-\erf(s)\big)  \big( s^4+3s^2+\frac{3}{4}\big),
\end{equation}
and then over $ds$.
In the limit of small $\epsilon$ the expression inside of rectangle brackets in (\ref{eq:Rect}) is equal to 8/3. Multiplying the bracket by
$3/8$ and defining $\Phi_{ab}$ (which now for small $\epsilon$ is equal to 1) then yields
the final momentum exchange rates
\begin{empheq}[box=\fbox]{align}
&  \boldsymbol{R}_{ab} = m_a n_a \nu_{ab}(\bu_b-\bu_a)\Phi_{ab}; \label{eq:RhardX}\\
& \Phi_{ab} = \frac{3}{8}\Big[
    e^{-\epsilon^2} \Big( 1+\frac{1}{2\epsilon^2}\Big) +\sqrt{\pi}\Big( \epsilon+\frac{1}{\epsilon}-\frac{1}{4\epsilon^3}\Big)
    \erf(\epsilon) \Big]; \qquad
\epsilon=\frac{|\bu_b-\bu_a|}{\sqrt{v_{\textrm{th} a}^2 +v_{\textrm{th} b}^2}},  
\end{empheq}
with the collisional frequency (\ref{eq:nuab}). The result agrees with eq. B3 of \cite{Schunk1977}, with p. 97-98 of
\cite{SchunkNagy2009}, eq. 15.14 of \cite{Burgers1969}  and also eq. 3.10 of \cite{Draine1986}
  (there is a missprint with the momentum transfer cross-section missing, his $\tilde\sigma$, which for the case of hard spheres is $\pi r_{ab}^2$). 
%The integration technique used to obtain the result
%(\ref{eq:Rect}) is thus completely analogous to the calculations with the Landau operator (for Coulomb collisions),
%the difference being that for hard spheres the integral (\ref{eq:Rect}) contains $g\bg$, whereas for Coulomb collisions
%the integral contains $\bg/g^3$. 
  
For completeness, instead of substitution (\ref{eq:zs}) with variable $z$, it is possible to introduce perhaps more intuitive variable $x$
\begin{equation}
x=\frac{g}{\beta}; \qquad s=\epsilon\cos\theta^*; \qquad => \qquad  \frac{|\bg+\bu|^2}{\beta^2} = x^2+2xs+\epsilon^2, \label{eq:zsB}
\end{equation}
and calculate the (\ref{eq:gntHS}) according to
\begin{eqnarray}
  \int g \bg e^{ - \frac{|\bg+\bu|^2}{\beta^2} }  d^3 g &=& 2\pi \bu \frac{\beta^4}{\epsilon^3} e^{-\epsilon^2}
  \int_{0}^\infty dx x^4 e^{-x^2} \int_{-\epsilon}^\epsilon ds e^{-2xs} s,
\end{eqnarray}
where one can use
\begin{eqnarray}
  \int_{-\epsilon}^\epsilon e^{-2xs} s ds &=& \frac{1}{4x^2}\Big( (1-2x\epsilon)e^{+2x\epsilon} - (1+2x\epsilon)e^{-2x\epsilon} \Big) \nn\\
  &=& \frac{1}{2x^2} \Big( \sinh(2x\epsilon)-2x\epsilon\cosh(2x\epsilon)\Big),\label{eq:DraineINT}
\end{eqnarray}
and then integrate over the $dx$, yielding the same result (\ref{eq:Rect}).

\newpage
%============================================================================
%============================================================================
\subsection{Coulomb collisions (unrestricted drifts)} \label{sec:Coulomb}
We start from a completely general equation for the momentum exchange rates
\begin{equation}
  \boldsymbol{R}_{ab} = m_a \int \bV_a C_{ab} d^3v_a = \mu_{ab}
  \iiint g_{ab} \sigma_{ab}(g_{ab},\theta) f_a f_b \big[ \bg_{ab}' - \bg_{ab} \big] d\Omega d^3 v_a d^3 v_b, \label{eq:nice2}
\end{equation}
where the $[\bV_a'-\bV_a]$ was just transformed into the center of mass velocities with (\ref{eq:CMSimple}) as before.
Then, considering Coulomb collisions, the integration over the $d\Omega$ is achieved by the already pre-calculated (\ref{eq:CoulombA}),
directly yielding 
\begin{equation}
  \boldsymbol{R}_{ab} = - 2\pi \ln(\Lambda^2+1) \frac{q_a^2 q_b^2 }{\mu_{ab}}  
  \iint \frac{\bg_{ab}}{g^3_{ab}} f_a f_b d^3 v_a d^3 v_b. \label{eq:nice3}
\end{equation}
Prescribing $f_a f_b$ to be Maxwellians with unrestricted drifts (\ref{eq:fafb}) and integrating over $d^3C^*$ yields
\begin{equation}
  \boldsymbol{R}_{ab} = - 2\pi \ln(\Lambda^2+1) \frac{q_a^2 q_b^2 }{\mu_{ab}} \frac{n_a n_b}{\pi^{3/2} \beta^3}  
  \int \frac{\bg}{g^3} e^{ -\frac{|\bg+\bu|^2}{\beta^2} }  d^3 g, \label{eq:nice5}
\end{equation}
where $\beta^2 = v_{\textrm{th} a}^2+v_{\textrm{th} b}^2$ and $\bu=\bu_b-\bu_a$. (The same integral with  $\bg=-\bx$ is calculated
for example in \cite{Hunana2022}, eq. G45, but for clarity we will calculate the integral again.)

Note the similarity with the integral (\ref{eq:Rect}),
and the integral is calculated with the same technique, first by integrating over the $d\phi^*$
\begin{equation}
  \int \frac{\bg}{g^3} e^{ - \frac{|\bg+\bu|^2}{\beta^2} }   d^3 g
  = 2\pi \frac{\bu}{u} \int_0^\infty \int_0^\pi e^{ - \frac{|\bg+\bu|^2}{\beta^2} } \cos\theta^* \sin\theta^* dg d\theta^*, 
\end{equation}
and then defining $\epsilon= u/\beta$ and changing into variables $z,s$ with (\ref{eq:zs}), yielding 
\begin{eqnarray}
  \int \frac{\bg}{g^3}   e^{ - \frac{|\bg+\bu|^2}{\beta^2} }  d^3 g &=& 2\pi \bu \frac{e^{-\epsilon^2}}{\epsilon^3} 
  \int_{-\epsilon}^\epsilon ds \int_s^\infty dz e^{-z^2+s^2} s \nn\\
  &=& - \pi \bu \Big[  \frac{\sqrt{\pi}}{\epsilon^3}\erf(\epsilon) -\frac{2e^{-\epsilon^2}}{\epsilon^2} \Big], \label{eq:Rect2}
\end{eqnarray}
where it is necessary to first integrate over $dz$
\begin{equation}
  \int_s^\infty e^{-z^2} dz = \frac{\sqrt{\pi}}{2}\big[ 1-\erf(s) \big],
\end{equation}
and then over $ds$. In the limit of small $\epsilon$ the expression inside of rectangle brackets in (\ref{eq:Rect2}) is equal to 4/3.
Multiplying the bracket by
$3/4$ and defining $\Phi_{ab}$ (which now for small $\epsilon$ is equal to 1) then yields the final result for unrestricted drifts
\begin{equation} \boxed{
   \boldsymbol{R}_{ab} = m_a n_a \nu_{ab}(\bu_b-\bu_a)\Phi_{ab}; \qquad 
\Phi_{ab} = \frac{3}{4} \Big[  \frac{\sqrt{\pi}}{\epsilon^3}\erf(\epsilon) -\frac{2e^{-\epsilon^2}}{\epsilon^2} \Big]; \qquad
\epsilon=\frac{|\bu_b-\bu_a|}{\sqrt{v_{\textrm{th} a}^2 +v_{\textrm{th} b}^2}}, \label{eq:RCoulombX}}
\end{equation}
recovering eq. (B1) of \cite{Schunk1977} or (26.4) of \cite{Burgers1969}, 
with the collisional frequency 
\begin{equation} \boxed{
    \nu_{ab} = \frac{8}{3} \ln(\Lambda^2+1) \sqrt{\pi} \frac{n_b q_a^2 q_b^2}{(v_{\textrm{th} a}^2+v_{\textrm{th} b}^2)^{3/2} m_a^2 } \big(1+\frac{m_a}{m_b}\big),}
  \label{eq:nuabC}
\end{equation}
where one can approximate $\ln(\Lambda^2+1)\simeq 2\ln\Lambda$.

\newpage
%============================================================================
%============================================================================
\subsection{Maxwell molecules collisions} \label{sec:MaxwellR}
We again start from a completely general equation for the momentum exchange rates (\ref{eq:nice2})
\begin{equation}
  \boldsymbol{R}_{ab} = \mu_{ab}
  \iiint g_{ab} \sigma_{ab}(g_{ab},\theta) f_a f_b \big[ \bg_{ab}' - \bg_{ab} \big] d\Omega d^3 v_a d^3 v_b.\nn
\end{equation}
The wording ``Maxwell molecules'' does not mean that $f_a f_b$ are necessarily Maxwellian distribution functions. 
Instead, the wording means any collisional process, where the differential cross-section $\sigma_{ab} \sim 1/g_{ab}$,
so that the product $g_{ab} \sigma_{ab}$ is independent of $g_{ab}$. One might consider the special case $g_{ab} \sigma_{ab}=\textrm{const.}$
  (which can be immediatelly pulled outside of the Boltzmann operator),
but it is much better to consider a general class of Maxwell molecules, where the product
$g_{ab} \sigma_{ab} = \mathcal{F}(\theta)$ is some function of $\theta$. Actually, because an exact form of the $\sigma_{ab}$ is not given,
it makes sense for a moment to consider all collisional processes, for any $\sigma_{ab}$, by employing the
definition of the momentum transfer cross-section
\begin{equation}
 \mathbb{Q}_{ab}^{(1)} (g_{ab})= \int  \sigma_{ab}(g_{ab},\theta)(1-\cos\theta)d\Omega. \label{eq:MTCS}
\end{equation}
Because integration over the $d\phi$ part of the solid angle yields $\int_0^{2\pi} (\bg_{ab}'-\bg_{ab}) d\phi = -2\pi\bg_{ab} (1-\cos\theta)$,
a completely general momentum exchange rates (for any collisional process) are then given by
\begin{equation}
 \boldsymbol{R}_{ab} = -\mu_{ab}   \iint g_{ab} \bg_{ab} \mathbb{Q}_{ab}^{(1)}  f_a f_b    d^3 v_a d^3 v_b.
\end{equation}
Now, because for Maxwell molecules the product $g_{ab} \mathbb{Q}_{ab}^{(1)}$ is independent of $g_{ab}$, it can be pulled
out of the integrals, yielding
\begin{equation}
  \boldsymbol{R}_{ab} =  - \Big[g_{ab}  \mathbb{Q}_{ab}^{(1)} \Big] \mu_{ab} \iint \bg_{ab} f_a f_b  d^3 v_a d^3 v_b.
\end{equation}
Importantly, for Maxwell molecules, to calculate the integrals over the velocity space, one does not have to assume that $f_a f_b$ are Maxwellian.
Instead, one can consider general (unspecified) $f_a f_b$ and just use the definition of fluid moments $\int f_a d^3v_a=n_a$ and 
$\int f_a \bV_a d^3v_a=n_a \bu_a$, meaning
\begin{equation}
\iint f_a f_b [\bV_a-\bV_b] d^3 v_a d^3 v_b = n_a n_b (\bu_a-\bu_b).
\end{equation}
Thus, for collisions of Maxwell molecules, the momentum exchange rates for general $f_a f_b$
(even with unrestricted drifts) are given by
\begin{equation} \boxed{
  \boldsymbol{R}_{ab} = m_a n_a \nu_{ab}(\bu_b-\bu_a); \qquad \nu_{ab} = \frac{n_b m_b}{m_a+m_b} \Big[ g_{ab} \mathbb{Q}_{ab}^{(1)} \Big],} \label{eq:Maxnuab}
\end{equation}
where the $g_{ab} \mathbb{Q}_{ab}^{(1)}$ is independent of the relative velocity $g_{ab}$. The result agrees with (4.83) of \cite{SchunkNagy2009}.

Maxwell molecules are interesting, because it is the simplest possible case, much simpler
than the case of constant $\sigma_{ab}$ for the hard spheres.  
The case was considered by Maxwell already in year 1866, who noticed that the determination of $f_a f_b$ is not
necessary to evaluate the collisional integrals.
 We recommend reading section ``Historical summary'' p. 380 in \cite{ChapmanCowling1953}
(which was eliminated in \cite{ChapmanCowling1970}).
It can be shown that the case of Maxwell molecules corresponds to interraction force $F=\pm |K_{ab}|/r^5$,
where the momentum transfer cross-section is given by (see eq. (\ref{eq:beauty15}))
\begin{equation}
  \mathbb{Q}_{ab}^{(1)} = \frac{2\pi}{g_{ab}} \frac{|K_{ab}|^{1/2}}{\mu_{ab}^{1/2}} A_1 (5); \qquad => \qquad
  \nu_{ab} = 2\pi n_b \frac{\mu_{ab}^{1/2}}{m_a} |K_{ab}|^{1/2} A_1(5),
\end{equation}
and the $A_1(5)$ represents a numerical integral (\ref{eq:beauM1}), where for the repulsive force $A_1(5)=0.422$ and for the attractive force
  (with rigid repulsive core) $A_1(5)=0.781$. The attractive case corresponds to (non-resonant) collisions between ions and neutrals,
  where the ion polarizes the neutral, see also the collisional frequency (\ref{eq:INnuab}).

\newpage
%==================================================================================
%==================================================================================
\section{Energy exchange rates for 5-moment models} \label{sec:Energy}
\setcounter{equation}{0}
By using the recipe (\ref{eq:BoltzmannOO}), a completely general energy exchange rates are given by
\begin{equation}
  Q_{ab} = \frac{m_a}{2} \int c_a^2 C_{ab} d^3 v_a  =\frac{m_a}{2} \iiint g_{ab} \sigma_{ab} f_a f_b \big[ c_a'^{2} -c_a^2 \big] d\Omega d^3 v_a d^3 v_b.
  \label{eq:GenQab}
\end{equation}
First, one needs to use the simple center-of-mass transformation (\ref{eq:CMSimple}), which relates the quantities before and after the collision by
\begin{eqnarray}
  c_a^2 &=& ( \bVV_c - \bu_a)^2 + 2 \frac{m_b}{m_a+m_b}( \bVV_c - \bu_a)\cdot\bg_{ab} + \Big(\frac{m_b}{m_a+m_b}\Big)^2 g^2_{ab};\nn\\
  c_a'^2 &=& ( \bVV_c - \bu_a)^2 + 2 \frac{m_b}{m_a+m_b}( \bVV_c - \bu_a)\cdot\bg_{ab}' + \Big(\frac{m_b}{m_a+m_b}\Big)^2 g'^2_{ab},
\end{eqnarray}
and so 
\begin{eqnarray}
  c_a'^2-c_a^2 &=& 2 \frac{m_b}{m_a+m_b}( \bVV_c - \bu_a)\cdot ( \bg_{ab}' -\bg_{ab} ), \label{eq:C2s}
\end{eqnarray}
yielding a general energy exchange rates
\begin{equation}
  Q_{ab} = \mu_{ab} \iiint g_{ab} \sigma_{ab} f_a f_b \big[ ( \bVV_c - \bu_a)\cdot ( \bg_{ab}' -\bg_{ab} )  \big] d\Omega d^3 v_a d^3 v_b.
  \label{eq:GenQab2}
\end{equation}
The $(\bVV_c - \bu_a)$ stays constant during the collision and it is not affected by the $d\Omega$ integral. The integral over the solid angle
is thus calculated in the same way as previously for the momentum exchange rates $\boldsymbol{R}_{ab}$ (which were defined by (\ref{eq:nice2})),
there is no difference, the result of the $d\Omega$ integral is just multiplied by $\bVV_c - \bu_a$.
Actually, one might be tempted to pull out the $\bu_a$ out of the entire integral (\ref{eq:GenQab2})
to immediatelly claim a relation to the general $\boldsymbol{R}_{ab}$. However, this is not
advisable, and from the definition of $\bVV_c$ it is better to write
\begin{equation}
\hat{\bVV}_c \equiv \bVV_c - \bu_a = \bVV_c^* +\frac{m_b}{m_a+m_b} \bu; \qquad \bVV_c^* \equiv \frac{m_a\bc_a+m_b \bc_b}{m_a+m_b}, \label{eq:VcStar}
\end{equation}
with $\bu=\bu_b-\bu_a$, and pull the $\bu$ out of the (\ref{eq:GenQab2}) instead, yielding a general recipe
\begin{empheq}[box=\fbox]{align}
  Q_{ab} &=  Q_{ab}^* + \frac{m_b}{m_a+m_b}(\bu_b-\bu_a)\cdot  \boldsymbol{R}_{ab};\nn\\
  Q_{ab}^* &= \mu_{ab} \iiint g_{ab} \sigma_{ab} f_a f_b \big[ \bVV_c^* \cdot ( \bg_{ab}' -\bg_{ab} )  \big] d\Omega d^3 v_a d^3 v_b;\nn\\
  Q_{ab} &= \mu_{ab} \iiint g_{ab} \sigma_{ab} f_a f_b \big[ \hat{\bVV_c} \cdot ( \bg_{ab}' -\bg_{ab} )  \big] d\Omega d^3 v_a d^3 v_b.
  \label{eq:GenQab3}
\end{empheq}
Recipe (\ref{eq:GenQab3}) saves a lot of time when calculating the $Q_{ab}$, because the $\boldsymbol{R}_{ab}$ was already calculated and one can
focus only on the $Q_{ab}^*$. Additionally, the general conservation of energy $Q_{ab}+Q_{ba}=(\bu_b-\bu_a)\cdot\boldsymbol{R}_{ab}$ and the conservation of
momentum $\boldsymbol{R}_{ab}+\boldsymbol{R}_{ba}=0$ imply that $Q_{ab}^*+Q_{ba}^*=0$, which the recipe satisfies. 

Later on (after the integral over the $d\Omega$ is calculated), for Maxwellian $f_a f_b$ one uses the more complicated
center-of-mass transformation  (\ref{eq:uglyX}) (which relates only the non-primed quantities before the collision),
where the velocity $\bVV_c^*$ transforms as
\begin{equation}
\bVV_c^* = \bC^* - \frac{2}{\beta^2}\frac{(T_b-T_a)}{(m_a+m_b)}(\bg_{ab}+\bu), \label{eq:VcC}
\end{equation}
and as a reminder
\begin{equation}
\frac{2}{\beta^2} = \frac{2}{v_{\textrm{th} a}^2+v_{\textrm{th} b}^2} = \frac{m_a m_b}{(m_b T_a + m_a T_b)}.
\end{equation}
The term $Q_{ab}^*$ in the recipe (\ref{eq:GenQab3}) is thus typically associated with the temperature differences $T_b-T_a$, and it is sometimes 
called the ``thermal part'' of $Q_{ab}$. Below, we will calculate the $Q_{ab}^*$ with unrestricted drifts $\bu$ for both the hard spheres
and Coulomb collisions.
%where the velocity difference $\bu$ enters the final results through the parameter $\epsilon=u/\beta$.
%Nevertheless, 
%if one is interested only in small drifts ($\epsilon\ll 1$) which are proportional to only $T_b-T_a$, a quick (a bit of cheating) shortcut is to neglect
%the $\bu$ in (\ref{eq:VcC}).

\newpage
\subsubsection*{Total energy exchange rates}
Note that the above $Q_{ab}$ enters the right-hand-side of the pressure equation $d_a p_a/dt +\cdots = (2/3) Q_{a}$. Sometimes, one instead considers
evolution equation for the total energy (kinetic plus internal, which can be defined as a fluid moment)
\begin{equation}
E^{\textrm{tot}} = \frac{m_a}{2}\int v_a^2 f_a d^3v_a = \frac{3}{2}p_a + \frac{\rho_a}{2}u_a^2,
\end{equation}
in which case its evolution equation has on the right-hand-side
\begin{eqnarray}
  Q_{ab}^{\textrm{tot}} &\equiv& \frac{m_a}{2}\int v_a^2 C_{ab} d^3v_a 
  = \frac{m_a}{2} \iiint g_{ab} \sigma_{ab} f_a f_b \big[ v_a'^{2} -v_a^2 \big] d\Omega d^3 v_a d^3 v_b \nn\\
  &=& \mu_{ab} \iiint g_{ab} \sigma_{ab} f_a f_b \big[ \bVV_c \cdot (\bg_{ab}'-\bg_{ab}) \big] d\Omega d^3 v_a d^3 v_b, \label{eq:QabTot}
\end{eqnarray}
i.e., where the $\bu_a$ was now indeed taken out of (\ref{eq:GenQab2}) and useful relations also are
\begin{equation}
Q_{ab}^{\textrm{tot}} = Q_{ab} + \bu_a\cdot \boldsymbol{R}_{ab} = Q_{ab}^* + \frac{m_a\bu_a+m_b\bu_b}{m_a+m_b}\cdot \boldsymbol{R}_{ab}. \label{eq:QabTot2}
\end{equation}

\newpage
%===========================
\subsection{Hard spheres (unrestricted drifts)} \label{eq:HSQabUnrest}
One starts with the general recipe (\ref{eq:GenQab3}), where the velocities $\hat{\bVV}_c$ and $\bVV_c^*$ are
defined in (\ref{eq:VcStar}), and both velocities are unchanged during the collision (because the $\bVV_c$, $\bu_a$ and $\bu_b$ are unchanged). 
Then, for the hard spheres, the integral over the solid angle is calculated simply with (\ref{eq:Integ0}) as 
\begin{equation}
\int \sigma_{ab} \big[ \bVV_c^* \cdot ( \bg_{ab}' -\bg_{ab} )  \big] d\Omega = -\pi r_{ab}^2 [\bVV_c^* \cdot \bg_{ab}],
\end{equation}
yielding energy exchange rates
\begin{eqnarray}
 Q_{ab}   &=& -\pi r_{ab}^2 \mu_{ab} \iint g_{ab} f_a f_b \big[ \hat{\bVV_c} \cdot \bg_{ab} \big] d^3 v_a d^3 v_b;\nn\\ 
 Q_{ab}^* &=& -\pi r_{ab}^2 \mu_{ab} \iint g_{ab} f_a f_b \big[ \bVV_c^* \cdot \bg_{ab} \big] d^3 v_a d^3 v_b. \label{eq:QabHardS}
\end{eqnarray}
The velocity $\bVV_c^*$ is transformed with (\ref{eq:VcC}) and the Maxwellian product $f_af_b$ is given by (\ref{eq:fafb}), and so
(we again stop writing species indices on $\bg_{ab}$)
\begin{equation}
  Q_{ab}^* = -\pi r_{ab}^2 \mu_{ab} \frac{n_a n_b}{\pi^3 {\alpha}^3 \beta^3}
  \iint  e^{-\frac{C^{*2}}{{\alpha}^2}} e^{-\frac{|\bg+\bu|^2}{\beta^2}} g  \bg \cdot
  \Big[  \bC^* - \frac{2}{\beta^2}\frac{(T_b-T_a)}{(m_a+m_b)}(\bg+\bu) \Big] d^3 C^* d^3 g,
\end{equation}
and the integral over the $d^3C^*$ can be carried out trivially, yielding
\begin{equation}
  Q_{ab}^* =  + r_{ab}^2 \mu_{ab} \frac{n_a n_b}{\pi^{1/2}  \beta^5} \frac{2 (T_b-T_a)}{(m_a+m_b)}
  \int  e^{-\frac{|\bg+\bu|^2}{\beta^2}} g  \bg \cdot
    (\bg+\bu)  d^3 g, \label{eq:HSPic2}
\end{equation}
and by employing the collisional frequency $\nu_{ab}$ for hard spheres (\ref{eq:nuab})
\begin{equation}
  Q_{ab}^* =     \frac{m_a n_a \nu_{ab} (T_b-T_a) }{(m_a+m_b)}  \frac{3}{4\pi \beta^6}
  \int  e^{-\frac{|\bg+\bu|^2}{\beta^2}} g  (g^2+\bg\cdot\bu)  d^3 g.
\end{equation}
To calculate the integral, one chooses that the direction of $\bu$ forms the axis
$\hat{\boldsymbol{e}}_3=\bu/u$, and the direction of $\bg$ is given by (\ref{eq:refF}), 
where $\bg\cdot\bu=gu\cos\theta^*$ and $|\bg+\bu|^2=g^2+2gu\cos\theta^*+u^2$, which allows one to integrate over the $d\phi$ by hand
\begin{equation}
  \int  e^{-\frac{|\bg+\bu|^2}{\beta^2}} g  (g^2+\bg\cdot\bu)  d^3 g = 2\pi \int e^{-(\frac{g^2}{\beta^2}+2\frac{g}{\beta}\epsilon\cos\theta^*+\epsilon^2)}
  g^3 (g^2 +g\epsilon\beta \cos\theta^*) \sin\theta^* d\theta^* dg,
\end{equation}
where the parameter $\epsilon=u/\beta$ was introduced, and one can just calculate the rest with an analytic software. 
Alternatively, by employing the substitutions (\ref{eq:zs})
\begin{eqnarray}
  \int  e^{-\frac{|\bg+\bu|^2}{\beta^2}} g  (g^2+\bg\cdot\bu)  d^3 g &=& 2\pi \beta^6 \frac{e^{-\epsilon^2}}{\epsilon}
  \int_{-\epsilon}^{\epsilon} ds \int_s^\infty dz e^{-z^2+s^2}  z(z-s)^4 ;\nn\\
  &=& 2\pi \beta^6 \Big[ e^{-\epsilon^2} +\sqrt{\pi}\Big( \epsilon+\frac{1}{2\epsilon}\Big)\erf(\epsilon)\Big]\equiv 4\pi \beta^6 \Psi_{ab},\label{eq:Picc}
\end{eqnarray}
where one first integrates over the $dz$
\begin{equation}
\int_s^\infty e^{-z^2} z(z-s)^4 dz = (s^2+1)e^{-s^2} +\sqrt{\pi}\big(\erf(s)-1\big)\big(s^3+\frac{3}{2}s\big), 
\end{equation}  
and then over the $ds$. For small $\epsilon=u/\beta$, the expression inside of rectangle brackets in (\ref{eq:Picc}) is equal to 2,
and multiplying the bracket by a factor of $1/2$ defines the $\Psi_{ab}$ (which is now equal to 1 for small $\epsilon$).
For hard spheres, the final energy exhange rates are then given by
\begin{empheq}[box=\fbox]{align}
&  Q_{ab} = \frac{m_a n_a \nu_{ab}}{m_a+m_b} \Big[ 3(T_b-T_a)\Psi_{ab} + m_b |\bu_b-\bu_a|^2 \Phi_{ab} \Big]; \qquad
  \Psi_{ab} = \frac{1}{2} \Big[ e^{-\epsilon^2} +\sqrt{\pi}\Big( \epsilon+\frac{1}{2\epsilon}\Big)\erf(\epsilon)\Big];\nn\\
& \qquad \qquad \Phi_{ab} = \frac{3}{8}\Big[
    e^{-\epsilon^2} \Big( 1+\frac{1}{2\epsilon^2}\Big) +\sqrt{\pi}\Big( \epsilon+\frac{1}{\epsilon}-\frac{1}{4\epsilon^3}\Big)
    \erf(\epsilon) \Big]; \qquad
\epsilon=\frac{|\bu_b-\bu_a|}{\sqrt{v_{\textrm{th} a}^2 +v_{\textrm{th} b}^2}}, \label{eq:PsiExact}
\end{empheq}
recovering e.g. \cite{Schunk1977} and \cite{Burgers1969}. 
For small $\epsilon\ll 1$ the scalar $\Phi_{ab}= 1+\epsilon^2/5$ and $\Psi_{ab}= 1+\epsilon^2/3$.
For large $\epsilon\gg 1$ the scalar $\Phi_{ab}= (3\sqrt{\pi}/8)\epsilon$ and $\Psi_{ab}= (\sqrt{\pi}/2)\epsilon$, and
both keep increasing linearly with $\epsilon$ (more-less after $\epsilon>2$).
If one wants a simple approximation of these expressions valid for all $\epsilon$ values, 
for example the function $\sqrt{1+\alpha \epsilon^2}$ has the power series expansion
$1+(\alpha/2)\epsilon^2$ and the asymptotic series expansion $\sqrt{\alpha}\epsilon$, i.e. exactly the same as both $\Phi_{ab}$ and $\Psi_{ab}$.  
If one chooses to match (only) the asymptotic expansion, then yields the following approximations
\begin{equation}
  \Phi_{ab}^{\textrm{Approx.}} = \sqrt{1+\pi \Big(\frac{3\epsilon}{8}\Big)^2 }; \qquad
  \Psi_{ab}^{\textrm{Approx.}} = \sqrt{1+\frac{\pi}{4} \epsilon^2}, \label{eq:PsiApprox}
\end{equation}
which nicely clarifies how the solutions behave with $\epsilon$.

Curiously, the same scalar $\Psi_{ab}$ (\ref{eq:PsiExact})
also comes out from the following integral (where $\bx$ can be anything)
\begin{eqnarray}
  \int  e^{-\frac{|\bx+\bu|^2}{\beta^2}} x  d^3 x &=& 2\pi \beta^4 \frac{e^{-\epsilon^2}}{\epsilon}
  \int_{-\epsilon}^{\epsilon} ds \int_s^\infty dz e^{-z^2+s^2}  (z-s)^3 ;\nn\\
  &=& 2\pi \beta^4 \Psi_{ab} (\epsilon); \qquad \epsilon=u/\beta.
\end{eqnarray}
Such integrals are obtained already at the kinetic level (as a right hand side of the Boltzmann equation) when using the hard sphere collisions to
estimate the charge-exchange frequency
\begin{equation}
  \int |\bV_a - \bV_b| f_a d^3v_a = \frac{n_a}{\pi^{3/2}v_{\textrm{th} a}^3} \int |\bV_a - \bV_b| e^{-\frac{|\bV_a-\bu_a|^2}{v_{\textrm{th} a}^2}} d^3v_a
  =   n_a v_{\textrm{th} a} \frac{2}{\sqrt{\pi}} \Psi_{ab}(\epsilon); \qquad \epsilon=\frac{|\bV_b-\bu_a|}{v_{\textrm{th} a}}, \label{eq:beau55}
\end{equation}
see for example eqs. (A1)-(A3) in \cite{PaulsZank1995}, and their approximant $\Psi_{ab}^{\textrm{Approx.}}$
(\ref{eq:PsiApprox}) is often used in the Heliospheric community to model the charge-exchange. The same
integral also appears for
\begin{equation}
  \int |\bV_a - \bV_b| f_a f_b d^3v_a d^3v_b = \frac{n_an_b}{\pi^{3/2} \beta^3} \int g e^{-\frac{|\bg+\bu|^2}{\beta^2}} d^3g =
  n_an_b \beta  \frac{2}{\sqrt{\pi}}\Psi_{ab} (\epsilon); \qquad \epsilon=u/\beta.
\end{equation}  

%\vspace{1cm}
%============================================
\subsubsection*{Total energy exchange rates}
The $\Psi_{ab}$ also appears when considering the total energy exchange rates (\ref{eq:QabTot}),
where after the integration over the solid angle (here written for a general constant cross-section, where for the hard sphere $\sigma_{ab}=r_{ab}^2/4$)
\begin{equation}
 Q_{ab}^{\textrm{tot}} = -4\pi \sigma_{ab} \mu_{ab} \iint g_{ab} f_a f_b \big[ \bVV_c \cdot \bg_{ab} \big] d^3 v_a d^3 v_b,
\end{equation}
and sometimes expressions are written by going back to the velocities $\bV_a$ \& $\bV_b$, by using
\begin{equation}
\mu_{ab} \bVV_c \cdot\bg_{ab} = \frac{m_a m_b}{(m_a+m_b)^2} \big[ m_a v_a^2 + (m_b-m_a)\bV_a\cdot\bV_b - m_b v_b^2 \big],
\end{equation}  
which for equal masses (applicable for the charge-exchange) is sometimes written as
\begin{equation}
m_a=m_b:\qquad Q_{ab}^{\textrm{tot}} = +\pi \sigma_{ab} m_a \iint f_a f_b |\bV_a-\bV_b|  \big( v_b^2-v_a^2 \big) d^3 v_a d^3 v_b,
\end{equation}
where the results (\ref{eq:QabTot2}), (\ref{eq:PsiExact}) can be used.

%If one wants to match also the power series expansion, one can consider
%for example 
%\begin{equation}
%  \Phi_{ab}^{\textrm{APR}} = \frac{1+\frac{8}{15\sqrt{\pi}}\epsilon + \frac{1}{5}\epsilon^2}{1+\frac{8}{15\sqrt{\pi}}\epsilon}; \qquad
%  \Psi_{ab}^{\textrm{APR}} = \frac{1+\frac{2}{3\sqrt{\pi}}\epsilon + \frac{1}{3}\epsilon^2}{1+\frac{2}{3\sqrt{\pi}}\epsilon},
%\end{equation}
%or higher order Pad\'e approximants. 

\newpage
%===========================
\subsection{Coulomb collisions (unrestricted drifts)}
Starting with the $Q_{ab}$ given by (\ref{eq:GenQab}) or the recipe (\ref{eq:GenQab3}), for Coulomb collisions the integral over the solid angle
was already calculated in (\ref{eq:CoulombA}), and here it is just multiplied by $\bVV_c^*$ (defined by (\ref{eq:VcStar})),
yielding the thermal energy exchange rates
\begin{equation}
  Q_{ab}^* = - 2\pi \ln(\Lambda^2+1)  \frac{q_a^2 q_b^2}{\mu_{ab}}  \iint \frac{f_a f_b}{ g_{ab}^3}  \big[ \bVV_c^* \cdot\bg_{ab} \big]  d^3 v_a d^3 v_b.
\end{equation}
The Maxwellian product $f_a f_b$ is given by (\ref{eq:fafb}), the $\bVV_c^*$ is transformed by (\ref{eq:VcC}),
and the thermal energy exchange rates thus become (we stop writing species indices for $\bg_{ab}$)
\begin{equation}
  Q_{ab}^* = - 2\pi \ln(\Lambda^2+1)  \frac{q_a^2 q_b^2}{\mu_{ab}} \frac{n_a n_b}{\pi^3 {\alpha}^3 \beta^3}
  \iint e^{-\frac{C^{*2}}{{\alpha}^2}} e^{-\frac{|\bg+\bu|^2}{\beta^2}}\frac{\bg}{ g^3}
  \cdot \big[  \bC^* - \frac{2}{\beta^2}\frac{(T_b-T_a)}{(m_a+m_b)}(\bg+\bu) \big]  d^3 C^* d^3 g,
\end{equation}
where the integral over the $d^3C^*$ can be carried out trivially, yielding
\begin{equation}
  Q_{ab}^* = + 4 \ln(\Lambda^2+1)  \frac{q_a^2 q_b^2}{\mu_{ab}} \frac{n_a n_b}{\pi^{1/2} \beta^5} \frac{(T_b-T_a)}{(m_a+m_b)}
  \int e^{-\frac{|\bg+\bu|^2}{\beta^2}}\frac{\bg}{ g^3}
  \cdot (\bg+\bu) d^3 g,
\end{equation}
and by using the Coulomb collisional frequency $\nu_{ab}$ (\ref{eq:nuabC}), then
\begin{equation}
  Q_{ab}^* = \frac{m_a n_a \nu_{ab}(T_b-T_a)}{m_a+m_b} \frac{3}{2\pi\beta^2}
  \int e^{-\frac{|\bg+\bu|^2}{\beta^2}}\frac{\bg}{ g^3}
  \cdot (\bg+\bu) d^3 g.
\end{equation}
As before, by choosing the axis $\hat{\boldsymbol{e}}_3=\bu/u$ and by employing substitutions (\ref{eq:zs}), the required integral calculates
\begin{eqnarray}
  \int e^{ - \frac{|\bg+\bu|^2}{\beta^2} } \frac{(g^2+\bg\cdot\bu)}{g^3} d^3 g &=&
  2\pi \int e^{ - \frac{|\bg+\bu|^2}{\beta^2} } \big( g+u\cos\theta^*) \sin\theta^* dg d\theta^* \nn\\
  &=&  2\pi \beta^2 \frac{e^{-\epsilon^2}}{\epsilon}
  \int_{-\epsilon}^{\epsilon} ds \int_s^\infty dz e^{-z^2+s^2}  z 
  = 2\pi \beta^2 e^{-\epsilon^2},  
\end{eqnarray}
where first integrating over $\int_s^\infty e^{-z^2}z dz = e^{-s^2}/2$ and then over $\int_{-\epsilon}^\epsilon ds = 2\epsilon$. 
For Coulomb collisions, the final energy exhange rates are then given by
\begin{empheq}[box=\fbox]{align}
&  Q_{ab} = \frac{m_a n_a \nu_{ab}}{m_a+m_b} \Big[ 3(T_b-T_a)\Psi_{ab} + m_b |\bu_b-\bu_a|^2 \Phi_{ab} \Big]; \qquad
  \Psi_{ab} = e^{-\epsilon^2};\nn\\
& \qquad \qquad \Phi_{ab} = \frac{3}{4} \Big[  \frac{\sqrt{\pi}}{\epsilon^3}\erf(\epsilon) -\frac{2e^{-\epsilon^2}}{\epsilon^2} \Big]; \qquad
\epsilon=\frac{|\bu_b-\bu_a|}{\sqrt{v_{\textrm{th} a}^2 +v_{\textrm{th} b}^2}}, \label{eq:NiceA}
\end{empheq}
recovering e.g. \cite{Schunk1977} and \cite{Burgers1969}. 
For small $\epsilon\ll 1$ the scalar $\Phi_{ab}= 1-(3/5)\epsilon^2$ and for large $\epsilon\gg 1$ the scalar
$\Phi_{ab}= 3\sqrt{\pi}/(4\epsilon^3)$  (for the scalar $\Psi_{ab} = e^{-\epsilon^2}$ the asymptotic expansion does not exist).
The scalar $\Phi_{ab}$ is rarely approximated, however, if one wants something simple and quite precise, we propose the following approximant
\begin{equation}
   \Phi_{ab}^{\textrm{Approx.}} = \frac{1}{\sqrt{1+\frac{6}{5}\epsilon^2+\frac{16}{9\pi}\epsilon^6}}, \label{eq:NiceApprox}
\end{equation}
%  \Phi_{ab}^{\textrm{APR}} = \frac{1}{1+\frac{3}{5}\epsilon^2+\frac{4}{3\sqrt{\pi}}\epsilon^3}; \qquad
which nicely clarifies how the $\Phi_{ab}$ behaves for small $\epsilon$ and large $\epsilon$ values (it has a correct power series and asymptotic series behavior). 
Note that the momentum exchange rates $R_{ab}\sim \epsilon \Phi_{ab}$
and the frictional part for the energy exchange rates $Q_{ab}(u)\sim \epsilon^2 \Phi_{ab}$. Both have a maximum at some $\epsilon$, and then go to zero
for higher $\epsilon$ values, which represents the runaway effect, see e.g. Figure 6, p. 95 in \cite{Hunana2022}. For example,
the momentum exchange rates reach a maximum at $\epsilon=0.97$, with a value $\epsilon\Phi_{ab}=0.57$. For the approximated (\ref{eq:NiceApprox}),
the maximum is reached at $\epsilon=0.98$, with a value $\epsilon\Phi_{ab}=0.60$, implying the approximant (\ref{eq:NiceApprox}) is capturing the runaway
effect quite precisely (Excercise : plot the two curves $\epsilon\Phi_{ab}$ with (\ref{eq:NiceA}) and (\ref{eq:NiceApprox})).

\newpage
%=====================================================================================
\subsection{Maxwell molecules} \label{sec:MaxwellQ}
Starting with the general recipe (\ref{eq:GenQab3})
\begin{equation}
 Q_{ab}^* = \mu_{ab} \iiint g_{ab} \sigma_{ab} f_a f_b \big[ \bVV_c^* \cdot ( \bg_{ab}' -\bg_{ab} )  \big] d\Omega d^3 v_a d^3 v_b,\nn
\end{equation}
one integrates over the $d\phi$ with (\ref{eq:Integ0}) and it is again the best for a moment to consider a general
 collisional process, where by employing the momentum transfer cross-section (\ref{eq:MTCS}) the $Q_{ab}^*$ becomes
\begin{equation}
 Q_{ab}^* = - \mu_{ab} \iint g_{ab}  \mathbb{Q}_{ab}^{(1)} f_a f_b \big[ \bVV_c^* \cdot \bg_{ab}  \big] d^3 v_a d^3 v_b.
\end{equation}
For the case of Maxwell molecules, the product $g_{ab}  \mathbb{Q}_{ab}^{(1)}$ is independent of $g_{ab}$ and can be pulled outside
of the integrals and by employing the collisional frequency  (\ref{eq:Maxnuab}) yields
\begin{equation}
  Q_{ab}^* = - \frac{m_an_a \nu_{ab}}{n_a n_b} \iint  f_a f_b \big[ \bVV_c^* \cdot \bg_{ab}  \big] d^3 v_a d^3 v_b.
\end{equation}
Here one assumes general $f_a f_b$, and by simply using the definition of $\bVV_c^*$ (eq. (\ref{eq:VcStar}))
and $\bg_{ab}=\bV_a-\bV_b$, one calculates integral
\begin{equation}
  Q_{ab}^* = - \frac{m_an_a \nu_{ab}}{(m_a+m_b)}\frac{1}{n_a n_b} \iint  f_a f_b (m_a\bc_a+m_b\bc_b) \cdot (\bV_a-\bV_b) d^3 v_a d^3 v_b,
\end{equation}
by employing only definitions of general fluid moments, such as 
$3p_a = m_a \int c_a^2 f_a d^3 v_a$ and $\int f_a \bc_a d^3v_a=0$, yielding
\begin{equation}
  Q_{ab}^* = 3\frac{m_an_a \nu_{ab}}{(m_a+m_b)} \Big[ \frac{p_b}{n_b} - \frac{p_a}{n_a}\Big].
\end{equation}
The final energy exchange rates for the Maxwell molecules thus become
\begin{equation} \boxed{
  Q_{ab} = \frac{m_a n_a \nu_{ab}}{m_a+m_b} \Big[ 3(T_b-T_a) + m_b |\bu_b-\bu_a|^2  \Big].} \label{eq:MMQAB}
\end{equation}

\newpage
%=========================================================================================
%=========================================================================================
\section{Hard spheres viscosity (1-Hermite)} \label{sec:HardSpheresV}
\setcounter{equation}{0}
The pressure tensor is defined as $\bp_a=m_a \int \bc_a\bc_a f_a d^3v_a$ and therefore its evolution equation contains the
following collisional contributions  
\begin{equation} 
\bQ_{ab}^{(2)} \equiv m_a \int  \bc_a \bc_a C_{ab} d^3v_a = m_a \iiint  g_{ab}  \sigma_{ab} (g_{ab},\theta)
  f_a f_b \big[ \bc_a' \bc_a' - \bc_a \bc_a \big] d\Omega d^3 v_a d^3 v_b. \label{eq:Viscos}
\end{equation}
We stop writing the species indices on $\bg_{ab}$. One first employs the ``simple'' center-of-mass transformation (\ref{eq:CMSimple0}), where the 
modified velocity (with hat) $\hat{\bVV}_c \equiv \bVV_c - \bu_a$ is present  
\begin{equation}
\bc_a = \bV_a - \bu_a = \hat{\bVV}_c + \frac{\mu_{ab}}{m_a}\bg; \qquad
  \bc_a' = \hat{\bVV}_c + \frac{\mu_{ab}}{m_a} \bg', \nn
\end{equation}
directly yielding matrices
\begin{eqnarray}
&&  \bc_a \bc_a = \hat{\bVV}_c \hat{\bVV}_c +\frac{\mu_{ab}}{m_a}\big( \bg \hat{\bVV}_c + \hat{\bVV}_c \bg \big) + \frac{\mu_{ab}^2}{m_a^2}\bg\bg;\nn\\
&&  \bc_a' \bc_a' = \hat{\bVV}_c \hat{\bVV}_c +\frac{\mu_{ab}}{m_a}\big( \bg' \hat{\bVV}_c + \hat{\bVV}_c \bg' \big) + \frac{\mu_{ab}^2}{m_a^2}\bg'\bg',  
\end{eqnarray}
and so the required expression which enters the (\ref{eq:Viscos}) reads
\begin{eqnarray}
  &&  m_a (\bc_a' \bc_a' -\bc_a\bc_a) = \mu_{ab} \Big( (\bg'-\bg) \hat{\bVV}_c + \hat{\bVV}_c (\bg'-\bg) \Big)
  + \frac{\mu_{ab}^2}{m_a} \big(\bg'\bg'-\bg\bg \big).  \label{eq:ViscosP}
\end{eqnarray}
To get familiar with the integrals, it is useful to first consider the case of the hard spheres, where the
differential cross-section $\sigma_{ab} (g_{ab},\theta)=r_{ab}^2/4$ can be pulled out of (\ref{eq:Viscos}). 

For the hard spheres, the first term of (\ref{eq:ViscosP}) is integrated over the $d\Omega$ by the simple (\ref{eq:Integ0}) and the last term
of (\ref{eq:ViscosP}) is integrated according to
\begin{eqnarray}
&&  \int \bg'\bg' d\Omega = \frac{4\pi}{3} \bI g^2; \qquad \int \bg\bg d\Omega = 4\pi\bg\bg; \nn\\
&& \int (\bg'\bg'-\bg\bg ) d\Omega = 4\pi \Big( \frac{\bI}{3}g^2 - \bg\bg \Big), \label{eq:Integ0A}
\end{eqnarray}
yielding the solid angle integral
\begin{eqnarray}
  &&  m_a \int (\bc_a' \bc_a' -\bc_a\bc_a)d\Omega = -4\pi \mu_{ab} \Big( \bg \hat{\bVV}_c + \hat{\bVV}_c \bg \Big)
  +  4\pi\frac{\mu_{ab}^2}{m_a} \Big( \frac{\bI}{3}g^2 - \bg\bg \Big),   
\end{eqnarray}
and so the collisional contributions (\ref{eq:Viscos}) for the hard spheres are then given by
\begin{equation} 
{\bQ}_{ab}^{(2)} = - \pi r_{ab}^2 \mu_{ab}  \iint   
  f_a f_b \, g\Big[ \bg \hat{\bVV}_c + \hat{\bVV}_c \bg 
  -  \frac{\mu_{ab}}{m_a} \Big( \frac{\bI}{3}g^2 - \bg\bg \Big) \Big]  d^3 v_a d^3 v_b. \label{eq:Viscos2}
\end{equation}
As a quick double check, calculating $Q_{ab}=(1/2)\trace \bQ_{ab}^{(2)}$ recovers the energy exchange rates (\ref{eq:QabHardS}).
Instead of working with the collisional contributions for the pressure tensor (\ref{eq:Viscos2}), it is also possible to directly define traceless  
collisional contributions for the stress-tensor
\begin{eqnarray} 
  \bQ_{ab}^{(2)}\,' &=& \bQ_{ab}^{(2)} - \frac{\bI}{3}\trace \bQ_{ab}^{(2)} \nn\\
  &=& - \pi r_{ab}^2 \mu_{ab}  \iint   
  f_a f_b \, g\Big[ \bg \hat{\bVV}_c + \hat{\bVV}_c \bg -\frac{2}{3}\bI (\bg\cdot \hat{\bVV}_c )
  -  \frac{\mu_{ab}}{m_a} \Big( \frac{\bI}{3}g^2 - \bg\bg \Big) \Big]  d^3 v_a d^3 v_b, \label{eq:Viscos3}
\end{eqnarray}
but we will work with the (\ref{eq:Viscos2}). In the semi-linear approximation, it can be shown that there will be no contributions
from the drifts $\bu=\bu_b-\bu_a$ at the end,
so for clarity of the shown calculations, we will neglect the drifts from the beginning.
Then by using the perturbed Maxwellians (\ref{eq:fafbS2}) and by employing the hard sphere collisional frequency $\nu_{ab}$ (\ref{eq:nuab}),
the collisional contributions are given by
\begin{equation} 
{\bQ}_{ab}^{(2)} = -  \frac{3}{8} \frac{n_a m_a \nu_{ab}}{\pi^{5/2} \alpha^3 \beta^4}  \iint   
  e^{ -\frac{C^{*2}}{{\alpha}^2}} e^{-\frac{g^2}{\beta^2}}
  \big( 1 +\chi_a + \chi_b \big)  \, g\Big[ \bg \hat{\bVV}_c + \hat{\bVV}_c \bg 
  -  \frac{\mu_{ab}}{m_a} \Big( \frac{\bI}{3}g^2 - \bg\bg \Big) \Big]  d^3 v_a d^3 v_b. \label{eq:Viscos2z}
\end{equation}
The integrals need to be calculated by the ``more advanced'' center-of-mass transformation (\ref{eq:hatV}), by moving everything
into the space $(\bC^*,\bg)$, where for zero drifts 
\begin{eqnarray}
  \hat{\bVV}_c &=& \bC^* - \frac{2}{\beta^2}\frac{(T_b-T_a)}{(m_a+m_b)}\bg. \label{eq:VcCC3}
\end{eqnarray}

%==============================================================
\subsection{Pressure tensor contributions from strict Maxwellians}
It is useful to separate the calculations, by first considering only strict Maxwellians (with zero perturbations $\chi_a$ and $\chi_b$),
where the (\ref{eq:Viscos2z}) calculates
\begin{eqnarray} 
{\bQ}_{ab}^{(2)} &=& -  \frac{3}{8} \frac{n_a m_a \nu_{ab}}{\pi^{5/2} \alpha^3 \beta^4}  \iint   
  e^{ -\frac{C^{*2}}{{\alpha}^2}} e^{-\frac{g^2}{\beta^2}}
    \, g\Big[ \bg \hat{\bVV}_c + \hat{\bVV}_c \bg 
      -  \cancel{\frac{\mu_{ab}}{m_a} \Big( \frac{\bI}{3}g^2 - \bg\bg \Big)} \Big]  d^3 C^* d^3 g \nn\\
    &=& -  \frac{3}{8} \frac{n_a m_a \nu_{ab}}{\pi^{5/2} \alpha^3 \beta^4}  \iint   
  e^{ -\frac{C^{*2}}{{\alpha}^2}} e^{-\frac{g^2}{\beta^2}}
  \, g\Big[ \bg \Big( \cancel{\bC^*} - \frac{2}{\beta^2}\frac{(T_b-T_a)}{(m_a+m_b)}\bg \Big)
    + \Big( \cancel{\bC^*} - \frac{2}{\beta^2}\frac{(T_b-T_a)}{(m_a+m_b)}\bg \Big) \bg \Big]  d^3 C^* d^3 g \nn\\
  &=&  -  \frac{3}{8} \frac{n_a m_a \nu_{ab}}{\pi^{5/2} \alpha^3 \beta^4}  \iint   
  e^{ -\frac{C^{*2}}{{\alpha}^2}} e^{-\frac{g^2}{\beta^2}}
  \, g\Big[ - \frac{4}{\beta^2}\frac{(T_b-T_a)}{(m_a+m_b)}\bg \bg \Big]  d^3 C^* d^3 g \nn\\
  &=& +  \frac{3}{2} \frac{(T_b-T_a)}{(m_a+m_b)} \frac{n_a m_a \nu_{ab}}{\pi^{5/2} \alpha^3 \beta^6}  \iint   
  e^{ -\frac{C^{*2}}{{\alpha}^2}} e^{-\frac{g^2}{\beta^2}}
  \, g\Big[ \bg \bg \Big]  d^3 C^* d^3 g \nn\\
  &=& +  \frac{3}{2} \frac{(T_b-T_a)}{(m_a+m_b)} \frac{n_a m_a \nu_{ab}}{\pi \beta^6}  \int e^{-\frac{g^2}{\beta^2}}
  \, g\Big[ \bg \bg \Big]  d^3 g
  =  +  \frac{3}{2} \frac{(T_b-T_a)}{(m_a+m_b)} \frac{n_a m_a \nu_{ab}}{\pi \beta^6} \frac{4\pi}{3}\bI \int_0^\infty e^{-\frac{g^2}{\beta^2}}
  \, g^5 d g\nn\\
  &=& +  2 \bI \frac{(T_b-T_a)}{(m_a+m_b)} n_a m_a \nu_{ab}.
    \label{eq:Viscos2z2}
\end{eqnarray}
As a quick double check, $Q_{ab}=(1/2)\trace \bQ_{ab}^{(2)} = 3(T_b-T_a)n_a m_a \nu_{ab}/(m_a+m_b)$, as it should. 
The result (\ref{eq:Viscos2z2}) is valid for arbitrary temperature differences, even though in the next section we will only
consider small temperature differences. 

%\newpage
%=========================================================================================
\subsection{Viscosity for arbitrary masses $m_a$ and $m_b$ (and small temperature differences)} \label{sec:10momEqT}
Here we go back to (\ref{eq:Viscos2z}) and now calculate the contributions coming from the perturbations
$\chi_a$ and $\chi_b$, which are given by (see for example Appendix B of \cite{Hunana2022})
\begin{equation}
\chi_a = \frac{m_a}{2T_a p_a} \bPi^{(2)}_a:\bc_a\bc_a; \qquad \chi_b = \frac{m_b}{2T_b p_b} \bPi^{(2)}_b:\bc_b\bc_b. \label{eq:Pert}
\end{equation}
We consider only small temperature differences, and the term (\ref{eq:Viscos2z2}) coming from the strict Maxwellians was already
calculated. For the rest of the calculations, the velocity (\ref{eq:VcCC3}) can be simplified
into $ \hat{\bVV}_c = \bC^*$, so that here we need to calculate
\begin{eqnarray} 
  \bQ_{ab}^{(2)} &=& -  \frac{3}{8} \frac{n_a m_a \nu_{ab}}{\pi^{5/2} \alpha^3 \beta^4} \iint   
 e^{ -\frac{C^{*2}}{{\alpha}^2}} e^{-\frac{g^2}{\beta^2}}
  \big( \cancel{1} +\chi_a + \chi_b \big)  \, g\Big[ \bg \bC^* + \bC^* \bg 
  -  \frac{\mu_{ab}}{m_a} \Big( \frac{\bI}{3}g^2 - \bg\bg \Big) \Big]  d^3 C^* d^3g, \label{eq:Viscos5}
\end{eqnarray}
where the strict Maxwellian term was scratched because it was already calculated (it is non-zero) and is given by (\ref{eq:Viscos2z2}).
For small temperature differences, the perturbations (\ref{eq:Pert}) are transformed with 
\begin{equation}
\bc_a = \bC^* +\frac{\mu_{ab}}{m_a}\bg; \qquad \bc_b = \bC^* -\frac{\mu_{ab}}{m_b}\bg,
\end{equation}
yielding transformed perturbations
\begin{eqnarray}
  \chi_a &=& \frac{m_a}{2T_a p_a} \bPi^{(2)}_a:\Big[ \bC^* \bC^* + \frac{\mu_{ab}}{m_a} \big( \bC^* \bg + \bg \bC^*\big)
    + \frac{\mu_{ab}^2}{m_a^2}\bg\bg \Big] ; \nn\\
  \chi_b &=& \frac{m_b}{2T_b p_b} \bPi^{(2)}_b:\Big[ \bC^* \bC^* - \frac{\mu_{ab}}{m_b} \big( \bC^* \bg + \bg \bC^*\big)
    + \frac{\mu_{ab}^2}{m_b^2}\bg\bg \Big], \label{eq:chab}
\end{eqnarray}
which enter the  (\ref{eq:Viscos5}). We calculate the $\chi_a$ and $\chi_b$ contributions separately, term by term.
Importantly, if one wants to calculate only self-collisions,
$\chi_a \neq \chi_b$, and one needs to use
\begin{equation} 
\textrm{self-collisions:} \qquad \chi_a + \chi_b = \frac{m_a}{T_a p_a} \bPi^{(2)}_a:\Big[ \bC^* \bC^* + \frac{1}{4}\bg\bg \Big], \label{eq:SelfPI}
\end{equation}
because the middle terms in (\ref{eq:chab}) cancel out.
%==============================================================
\subsubsection*{Useful integrals}
To calculate the integrals, one can use the following scheme (see e.g. page 115 of \cite{Hunana2022}) by assuming a well-behaved (non-singular) scalar function $f(y)$
such as polynomials
\begin{eqnarray}
  \int \by\by f(y) e^{- y^2/\alpha^2} d^3y &=& \bI \frac{4\pi}{3} \int_0^\infty y^4 f(y) e^{-y^2/\alpha^2}dy;
  \label{eq:formula1}\\
  \int \by(\bPi_a^{(2)}\cdot\by) f(y) e^{-y^2/\alpha^2} d^3y &=& \bPi_a^{(2)} \frac{4\pi}{3} \int_0^\infty y^4 f(y) e^{-y^2/\alpha^2} dy;\\
 \bPi_a^{(2)}: \int \by\by f(y) e^{-y^2/\alpha^2} d^3y &=& 0, 
\end{eqnarray}
together with
\begin{eqnarray}
  \bPi_a^{(2)}: \int \by\by\by\by f(y) e^{-y^2/\alpha^2} d^3y &=&
   \bPi_a^{(2)} \frac{8\pi}{15}\int_0^\infty y^6 f(y) e^{-y^2/\alpha^2} dy. \label{eq:formula2}
\end{eqnarray}
For example $\int e^{-C^2/\alpha^2}C_i C_j d^3C = \delta_{ij}\pi^{3/2}\alpha^5/2$. Similarly, for any well-behaved scalar function $f(C,g)$
\begin{eqnarray}
&&  \iint e^{ -\frac{C^{2}}{{\alpha}^2}} e^{-\frac{g^2}{\beta^2}} \bPi_a^{(2)} : (\bC\bg+\bg\bC)(\bC\bg+\bg\bC) f(C,g) d^3C d^3g \nn\\
&&  = \Big(\frac{4\pi}{3} \Big)^2 4 \bPi_a^{(2)} \int_0^\infty \int_0^\infty e^{ -\frac{C^{2}}{{\alpha}^2}} e^{-\frac{g^2}{\beta^2}} C^4 g^4 f(C,g) dC dg.
\end{eqnarray}

%\newpage
%==============================================================
\subsubsection*{Species ``a''}
Let us stop writing the star on $\bC^*$. The $\chi_a$ consists of 3 terms,
\begin{eqnarray}
  \chi_a &=& \frac{m_a}{2T_a p_a} \bPi^{(2)}_a:\Big[ \underbrace{\bC \bC}_{\circled{1}}
    + \underbrace{\frac{\mu_{ab}}{m_a} \big( \bC \bg + \bg \bC\big)}_{\circled{2}}
    + \underbrace{\frac{\mu_{ab}^2}{m_a^2}\bg\bg}_{\circled{3}} \Big],\nn
\end{eqnarray}
and we need to calculate the collisional contributions (\ref{eq:Viscos5}).  
The first term calculates
\begin{eqnarray}
&&  \circled{1} = \iint  e^{ -\frac{C^{2}}{{\alpha}^2}} e^{-\frac{g^2}{\beta^2}}
  \bPi^{(2)}_a: \Big[ \bC \bC   \Big]  \, g\Big[ \cancel{\bg \bC} + \cancel{\bC \bg} 
    -  \frac{\mu_{ab}}{m_a} \Big( \frac{\bI}{3}g^2 - \bg\bg \Big) \Big]  d^3 C d^3g  =0.
\end{eqnarray}
The second term calculates
\begin{eqnarray}
&&  \circled{2} = \frac{\mu_{ab}}{m_a} \iint  e^{ -\frac{C^{2}}{{\alpha}^2}} e^{-\frac{g^2}{\beta^2}}
  \bPi^{(2)}_a: \Big[ \bC \bg + \bg \bC    \Big]  \, g\Big[ \bg \bC + \bC \bg 
    -  \cancel{\frac{\mu_{ab}}{m_a} \Big( \frac{\bI}{3}g^2 - \bg\bg \Big)} \Big]  d^3 C d^3g  \nn\\
  && = \frac{\mu_{ab}}{m_a} \Big(\frac{4\pi}{3}\Big)^2  4 \bPi^{(2)}_a  \int_0^\infty e^{ -\frac{C^{2}}{{\alpha}^2}} C^4 dC
  \int_0^\infty e^{-\frac{g^2}{\beta^2}} g^5 dg  
    = \frac{\mu_{ab}}{m_a} \Big(\frac{4\pi}{3}\Big)^2  4 \bPi^{(2)}_a \frac{3}{8}\sqrt{\pi} \alpha^5 \beta^6\nn\\
 && =  \frac{8}{3} \alpha^5 \beta^6 \pi^{5/2} \frac{\mu_{ab}}{m_a}   \bPi^{(2)}_a. \label{eq:second}
\end{eqnarray}
The third term calculates
\begin{eqnarray}
&& \circled{3} =\frac{\mu_{ab}^2}{m_a^2} \iint  e^{ -\frac{C^{2}}{{\alpha}^2}} e^{-\frac{g^2}{\beta^2}}
  \bPi^{(2)}_a: \Big[ \bg \bg \Big]  \, g\Big[ \cancel{\bg \bC} + \cancel{\bC \bg }
    -  \frac{\mu_{ab}}{m_a} \Big( \frac{\bI}{3}g^2 - \bg\bg \Big) \Big]  d^3 C d^3g \nn\\
&& = - \alpha^3 \pi^{3/2} \frac{\mu_{ab}^3}{m_a^3} \int  e^{-\frac{g^2}{\beta^2}}
  \bPi^{(2)}_a: \Big[ \bg \bg \Big]  \, g  \Big( \frac{\bI}{3}g^2 - \bg\bg \Big)  d^3g  \nn\\
&& = + \alpha^3 \pi^{3/2} \frac{\mu_{ab}^3}{m_a^3} \int  e^{-\frac{g^2}{\beta^2}}
  \bPi^{(2)}_a: \Big[ \bg \bg \Big] \,  \bg\bg g  d^3g  \nn\\
&& = + \alpha^3 \pi^{3/2} \frac{\mu_{ab}^3}{m_a^3} \bPi^{(2)}_a \frac{8\pi}{15} \int_0^\infty  e^{-\frac{g^2}{\beta^2}}
  g^7  dg
%  =  + \alpha^3 \pi^{3/2} \frac{\mu_{ab}^3}{m_a^3} \bPi^{(2)}_a \frac{8\pi}{15} 3\beta^8 \nn\\
 = +\frac{8}{5} \alpha^3 \beta^8 \pi^{5/2} \frac{\mu_{ab}^3}{m_a^3} \bPi^{(2)}_a. \label{eq:third}
\end{eqnarray}
Then the total contributions from species ``a'' read
\begin{eqnarray} 
  \bQ_{ab}^{(2)}(\chi_a) &=& -  \frac{3}{8} \frac{n_a m_a \nu_{ab}}{\pi^{5/2} \alpha^3 \beta^4} \frac{m_a}{2T_a p_a}
  \big[ \frac{8}{3} \alpha^5 \beta^6 \pi^{5/2} \frac{\mu_{ab}}{m_a} + \frac{8}{5} \alpha^3 \beta^8 \pi^{5/2} \frac{\mu_{ab}^3}{m_a^3} \Big]\bPi^{(2)}_a\nn\\
  &=&  -    \nu_{ab} \frac{n_a m_a }{2T_a p_a} \mu_{ab}
  \big[ \alpha^2 \beta^2  + \frac{3}{5} \beta^4  \frac{\mu_{ab}^2}{m_a^2} \Big]\bPi^{(2)}_a. \label{eq:perfect1}
\end{eqnarray}

%\newpage
%===============================================================
\subsubsection*{Species ``b''}
The $\chi_b$ also consists of 3 terms,
\begin{eqnarray}
  \chi_b &=& \frac{m_b}{2T_b p_b} \bPi^{(2)}_b:\Big[ \underbrace{\bC \bC}_{\circled{1}}
    - \underbrace{\frac{\mu_{ab}}{m_b} \big( \bC \bg + \bg \bC\big)}_{\circled{2}}
    + \underbrace{\frac{\mu_{ab}^2}{m_b^2}\bg\bg}_{\circled{3}} \Big].\nn
\end{eqnarray}
As previously, the integration over the first term yields zero. The second and third terms calculate
(just by looking at the integrals (\ref{eq:second}) and (\ref{eq:third}))
\begin{equation}
  \circled{2} = - \frac{8}{3} \alpha^5 \beta^6 \pi^{5/2} \frac{\mu_{ab}}{m_b}   \bPi^{(2)}_b; \qquad
  \circled{3} = + \frac{8}{5} \alpha^3 \beta^8 \pi^{5/2} \frac{\mu_{ab}^3}{m_b^2 m_a} \bPi^{(2)}_b,
\end{equation}
yielding a total contributions from species ``b''
\begin{eqnarray} 
  \bQ_{ab}^{(2)}(\chi_b)
%  &=&
%  -  \frac{3}{8} \frac{n_a m_a \nu_{ab}}{\pi^{5/2} \alpha^3 \beta^4} \frac{m_b}{2T_b p_b}
%  \Big[ - \frac{8}{3} \alpha^5 \beta^6 \pi^{5/2} \frac{\mu_{ab}}{m_b} + \frac{8}{5} \alpha^3 \beta^8 \pi^{5/2} \frac{\mu_{ab}^3}{m_b^2 m_a}  \Big] \bPi^{(2)}_b \nn\\
  &=& -   \nu_{ab} \frac{n_a m_a}{2T_b p_b} \mu_{ab}
  \Big[ - \alpha^2 \beta^2  + \frac{3}{5} \beta^4  \frac{\mu_{ab}^2}{m_b m_a}  \Big] \bPi^{(2)}_b. \label{eq:perfect2}
\end{eqnarray}

%===============================================================
\subsubsection*{Final result for small temperature differences}
Results (\ref{eq:perfect1}) and (\ref{eq:perfect2}) were kept in their more general form with constants $\alpha$ and $\beta$
(just in case we want to use them later for re-calculation with arbitrary temperatures),
and here for small temperature differences the results are simplified with
\begin{eqnarray}
\alpha^2 = \frac{2T_a}{m_a+m_b}; \qquad \beta^2= \frac{2T_a}{\mu_{ab}}; \qquad \alpha^2\beta^2= \frac{4T_a^2}{m_a m_b}.
\end{eqnarray}
Adding the strict Maxwellian result (\ref{eq:Viscos2z2}) together with (\ref{eq:perfect1}) and (\ref{eq:perfect2}) then
yields the final collisional contributions for the pressure tensor
\begin{eqnarray} 
  \bQ_{ab}^{(2)} &=&  2 (T_b-T_a) \bI \frac{n_a m_a \nu_{ab}}{(m_a+m_b)}  -  \nu_{ab}   
  \frac{2(5m_a + 3m_b)}{5 (m_a+m_b)} \bPi^{(2)}_a +\frac{4}{5}   \nu_{ab} \frac{n_a }{ n_b} \frac{m_a}{(m_a+m_b)} \bPi^{(2)}_b\nn\\
  &=& \frac{2 \nu_{ab} m_a }{m_a+m_b} \Big[  (T_b-T_a) \bI n_a  -     
   \Big(1+\frac{3}{5}\frac{m_b}{m_a}  \Big)\bPi^{(2)}_a +\frac{2}{5}   \frac{n_a }{ n_b} \bPi^{(2)}_b \Big].
\end{eqnarray}
Introducing summation over all of the
``b'' species and separating the self-collisions up front, then yields the final stress-tensor contributions for the hard spheres
\begin{equation} \boxed{ 
  \bQ_{a}^{(2)}\,' =  -  \frac{6}{5}\nu_{aa}\bPi^{(2)}_a +\sum_{b\neq a} \frac{2\nu_{ab} m_a}{m_a+m_b}\Big[   
  -\Big(1+\frac{3}{5}\frac{m_b}{m_a}  \Big) \bPi^{(2)}_a +\frac{2}{5}  \frac{n_a }{ n_b} \bPi^{(2)}_b\Big].} \label{eq:perfect3}
\end{equation}
The hard sphere result (\ref{eq:perfect3})
is consistent with the general result for any collisional process (\ref{eq:QabP}), which in the 1-Hermite approximation simplifies into
\begin{eqnarray}
  \bQ_{a}^{(2)}\,' &=& -\frac{3}{5}\nu_{aa} \Omega_{22} \bPi_a^{(2)}
   +\sum_{b\neq a} \frac{\rho_a \nu_{ab}}{m_a+m_b} \Big[ -K_{ab (1)} \frac{1}{n_a} \bPi^{(2)}_a +K_{ab (2)} \frac{1}{n_b} \bPi^{(2)}_b \Big], 
\end{eqnarray}
where for the small temperature differences
\begin{equation}
K_{ab(1)} = 2 + \frac{3}{5} \frac{m_b}{m_a}\Omega_{22};\qquad  K_{ab(2)} = 2 - \frac{3}{5}\Omega_{22},\nn
\end{equation}
and for the hard spheres $\Omega_{22}=2$. Note that for the Coulomb collisions $\Omega_{22}=2$ as well, so the equation (\ref{eq:perfect3}) actually 
remains the same also for the Coulomb collisions
(see also eq. (J27) in \cite{Hunana2022}, there obtained with the Landau operator, through the Rosenbluth potentials).

Result (\ref{eq:perfect3}) enters the right-hand-side of the evolution equation for the viscosity tensor $\bPi^{(2)}_a$ 
(here written in a simplified form already at the semi-linear level)
\begin{equation}
\frac{d_a}{dt} \bPi^{(2)}_a + \cancel{\Omega_a \big(\bhat\times\bPi^{(2)}_a \big)^S} +p_a \bW_a = \bQ_{a}^{(2)'}, \label{eq:perfect4}
\end{equation}
where for the hard spheres the cyclotron frequency $\Omega_a=0$. The viscosity of a general gas (approximated as hard spheres) with
N-species present, is thus given by N coupled equations (\ref{eq:perfect4}), with the right-hand-side (\ref{eq:perfect3}).
For the particular case of self-collisions, the quasi-static approximation
finally yields the (1-Hermite) viscosity of hard spheres
\begin{equation} \boxed{
    \textrm{self-collisions:}\qquad \bPi^{(2)}_a = - \eta_a \bW_a; \qquad \eta_a = \frac{5}{6} \frac{p_a}{\nu_{aa}}
    = \frac{5}{16} \frac{\sqrt{\pi} (T_a m_a)^{1/2}}{ \pi (2r_{a})^2},  \label{eq:PiHardSelf}   }
\end{equation}
where  $\pi (2r_a)^2=\sigma_{\textrm{tot}}$ is the total cross-section. The result agrees with \cite{ChapmanCowling1953}, page 169
(in their notation $\sigma=r_{ab}=2r_a$) and with \cite{Schunk1975}, eq. (4.14b). In the 2-Hermite approximation,
the coefficient $5/6$ of hard spheres changes into $1025/1212$, see eq. (\ref{eq:beau36}).\\

The self-collisional result (\ref{eq:PiHardSelf}) coming from the first term of (\ref{eq:perfect3}) can be of course derived in a much more straightforward manner,
by considering self-collisions from the beginning and by plugging the perturbations (\ref{eq:SelfPI}) into (\ref{eq:Viscos5}), 
yielding a simple integral for the hard spheres
\begin{eqnarray}  
&&  \iint  e^{ -\frac{C^{2}}{{\alpha}^2}} e^{-\frac{g^2}{\beta^2}}
   \bPi^{(2)}_a:\Big[ \bC \bC + \frac{1}{4}\bg\bg \Big]  \, g\Big[ \cancel{\bg \bC} + \cancel{\bC \bg} 
     -  \frac{1}{2} \Big( \frac{\bI}{3}g^2 - \bg\bg \Big) \Big]  d^3 C d^3g \nn\\
&& =   -  \frac{1}{2} \iint  e^{ -\frac{C^{2}}{{\alpha}^2}} e^{-\frac{g^2}{\beta^2}}
   \bPi^{(2)}_a:\Big[ \cancel{\bC \bC} + \frac{1}{4}\bg\bg \Big]  \, g  \Big( \cancel{\frac{\bI}{3}g^2} - \bg\bg \Big)  d^3 C d^3g \nn\\
&& =   +  \frac{1}{8} \iint  e^{ -\frac{C^{2}}{{\alpha}^2}} e^{-\frac{g^2}{\beta^2}}
   \bPi^{(2)}_a: \bg\bg \bg\bg g  d^3 C d^3g \nn\\
   && =  +  \frac{1}{8}  \bPi^{(2)}_a (4\pi) \int_0^\infty e^{ -\frac{C^{2}}{{\alpha}^2}} C^2 dC  \Big(\frac{8\pi}{15}\Big) \int_0^\infty e^{-\frac{g^2}{\beta^2}} g^7 dg\nn\\
 && =  +  \frac{1}{5}  \bPi^{(2)}_a \pi^{5/2} \alpha^3 \beta^8,  \label{eq:beauty11}
\end{eqnarray}
and further yielding the self-collisional exchanges
\begin{equation}
\bQ^{(2)}_{a}\,' = -\frac{6}{5} \nu_{aa} \bPi^{(2)}_a.
\end{equation}
The viscosity of hard spheres is very clarifying (perhaps because one can easily envision a large number of billiard balls),
where the calculation nicely shows that even though during each collision the momentum
and energy is conserved exactly, the entire system is still viscous, as a consequence of the perturbation of the distribution function.
The same is true for the Coulomb collisions (and other collisional processes with the Boltzmann operator), but there the nature of the
electrostatic interaction and the required Coulomb logarithm cut-off can make the viscosity effect perhaps less clear.

\newpage
%==================================================================================================================
\section{Calculation of general collisional integrals} \label{sec:Integrals}
\setcounter{equation}{0}
One starts with the collisional integrals (\ref{eq:beauty10}), and transform these with the ``simple'' center-of-mass transformation
(\ref{eq:CMSimple0}), where one 
introduces the center-of-mass velocity $\boldsymbol{V}_c$, together with the modified center-of-mass velocity $\hbV_c$
(with hat)
\begin{equation}
\boldsymbol{V}_c \equiv \frac{m_a \bV_a+m_b\bV_b}{m_a+m_b}; \qquad \hbV_c \equiv \bVV_c - \bu_a; 
\end{equation}
and the quantities before and after the collision are related by 
\begin{equation}
  \bc_a = \hbV_c + \frac{\mu_{ab}}{m_a} \bg; \qquad  \bc_a' = \hbV_c + \frac{\mu_{ab}}{m_a} \bg'; \qquad
  g'=g. \label{eq:CMSimple0a}
\end{equation}
Directly from (\ref{eq:CMSimple0a}) one calculates expressions such as
\begin{eqnarray}
  c_a'^{2} &=& \hat{V}_c^2+ 2\frac{\mu_{ab}}{m_a} \hat{\bVV}_c\cdot\bg' +\frac{\mu_{ab}^2}{m_a^2} g^2;\qquad
  c_a^2 = \hat{V}_c^2+ 2\frac{\mu_{ab}}{m_a} \hat{\bVV}_c\cdot\bg +\frac{\mu_{ab}^2}{m_a^2} g^2;\\
  \bc_a' \bc_a' &=& \hat{\bVV}_c \hat{\bVV}_c +\frac{\mu_{ab}}{m_a}\big( \bg' \hat{\bVV}_c + \hat{\bVV}_c \bg' \big) + \frac{\mu_{ab}^2}{m_a^2}\bg'\bg';\qquad
  \bc_a \bc_a = \hat{\bVV}_c \hat{\bVV}_c +\frac{\mu_{ab}}{m_a}\big( \bg \hat{\bVV}_c + \hat{\bVV}_c \bg \big) + \frac{\mu_{ab}^2}{m_a^2}\bg\bg,\nn 
\end{eqnarray}
and subtracting them yields
\begin{eqnarray}
   m_a \big[ \bV_a'-\bV_a \big] &=& \mu_{ab} ( \bg'-\bg ); \nn\\
   m_a \big[ c_a'^{2} -c_a^2 \big] &=& 2\mu_{ab} \hbV_c \cdot ( \bg'-\bg );\nn\\
%===
    m_a \big[\bc_a' \bc_a' -\bc_a\bc_a \big] &=& \mu_{ab} \Big[  (\bg'-\bg) \hbV_c + \hbV_c (\bg'-\bg) 
  + \frac{\mu_{ab}}{m_a} \big(\bg'\bg'-\bg\bg \big)\Big]; \nn\\  
%===
  m_a \big[ \bc_a' c_a'^2 - \bc_a c_a^2 \big] &=& \mu_{ab} \Big[ 2 \hbV_c \hbV_c\cdot(\bg'-\bg)
  + \Big(\hat{V}_c^2 + \frac{\mu_{ab}^2}{m_a^2} g^2 \Big)   (\bg'-\bg) 
  + 2 \frac{\mu_{ab}}{m_a}  \hbV_c\cdot (\bg'\bg'-\bg\bg ) \Big]. \label{eq:INT1}
\end{eqnarray}
Expressions (\ref{eq:INT1}) are needed to derive the 13-moment models of \cite{SchunkNagy2009}, \cite{Schunk1977} and \cite{Burgers1969}.
Here for the 22-moment model, we additionally need 
\begin{eqnarray}
 m_a\big[ c_a'^{2}\bc_a' \bc_a' -  c_a^{2}\bc_a \bc_a \big]
  &=& \mu_{ab} \Big\{ \Big[ \hat{V}_c^2 +\frac{\mu_{ab}^2}{m_a^2} g^2 \Big]
  \Big[   (\bg'-\bg) \hbV_c + \hbV_c (\bg'-\bg)  + \frac{\mu_{ab}}{m_a}\big(\bg'\bg'-\bg\bg\big) \Big]\nn\\
  && + 2 \hbV_c \hbV_c \hbV_c\cdot(\bg'-\bg) 
   + 2\frac{\mu_{ab}}{m_a} \Big[ \hbV_c \cdot (\bg'\bg'-\bg\bg)\hbV_c +  \hbV_c \hbV_c \cdot (\bg'\bg'-\bg\bg) \Big]\nn\\
  && + 2\frac{\mu_{ab}^2}{m_a^2} \hbV_c\cdot (\bg'\bg'\bg'-\bg\bg\bg) \Big\}; \\
 %\end{eqnarray}
%=====================   
%\begin{eqnarray}
m_a\big[ c_a'^{4} -  c_a^{4} \big]
&=& \mu_{ab} \Big\{ 4 \Big( \hat{V}_c^2  +  \frac{\mu_{ab}^2}{m_a^2} g^2  \Big) \hbV_c\cdot(\bg'-\bg)
+ 4\frac{\mu_{ab}}{m_a} \hbV_c\hbV_c :(\bg'\bg'-\bg\bg) \Big\}; \\
%\end{eqnarray}
%===
%\begin{eqnarray}
  m_a \big[ \bc_a' c_a'^4 - \bc_a c_a^4 \big] &=& \mu_{ab} \Big\{
   \Big( \hat{V}_c^2  + \frac{\mu_{ab}^2}{m_a^2} g^2\Big)^2 (\bg'-\bg)
   +4  \Big( \hat{V}_c^2  +  \frac{\mu_{ab}^2}{m_a^2} g^2  \Big) \Big[ \hbV_c  \hbV_c\cdot(\bg'-\bg)
   +  \frac{\mu_{ab}}{m_a} \hbV_c\cdot(\bg'\bg'-\bg\bg) \Big]\nn\\
  &&   +4\frac{\mu_{ab}}{m_a} \hbV_c \hbV_c\hbV_c: (\bg'\bg' -\bg\bg)
  + 4\frac{\mu_{ab}^2}{m_a^2} \hbV_c\hbV_c: (\bg'\bg'\bg' -\bg\bg\bg) \Big\}. \label{eq:INT2}
\end{eqnarray}
%=====
Note that for example $(\hbV_c\cdot\bg')^2 = \hbV_c\hbV_c:\bg'\bg'$. 
Expressions (\ref{eq:INT1})-(\ref{eq:INT2}) enter the collisional integrals (\ref{eq:beauty10}), where one needs to integrate over the
solid angle $d\Omega=\sin\theta d\theta d\phi$, 
by introducing the effective cross-sections $\mathbb{Q}_{ab}^{(l)}$ (\ref{eq:QabPicA}). One first integrates over the angle $d\phi$, by employing integrals 
\begin{eqnarray}
 \int_0^{2\pi} (\bg'-\bg) d\phi &=& -2 \pi (1-\cos\theta)\bg;\\ 
 \int_0^{2\pi} (\bg'\bg'-\bg\bg) d\phi &=& 3 \pi (1-\cos^2\theta) \Big(  \frac{\bI}{3}g^2 - \bg\bg \Big); \label{eq:INT3}\\
%===
  \int_0^{2\pi} (\bg'\bg'\bg'-\bg\bg\bg) d\phi &=&
  \pi g^2 \Big[ (1-\cos^3\theta) - (1-\cos\theta) \Big] \big[\bI \bg \big]^S \nn\\
 && \quad -\pi \Big[ 5(1-\cos^3\theta)-3(1-\cos\theta) \Big] \bg\bg\bg, \label{eq:INT4}
\end{eqnarray}
%===================================
and then over the scattering angle $d\theta$, yielding simple recipes
\begin{eqnarray}
  \int \sigma_{ab} (g,\theta) \big[\bg'-\bg \big] d\Omega &=& -\bg \mathbb{Q}_{ab}^{(1)};\nn\\
%==
  \int \sigma_{ab} (g,\theta) \big[\bg'\bg'-\bg \bg \big] d\Omega
  &=& \frac{3}{2} \Big(  \frac{\bI}{3}g^2 - \bg\bg \Big) \mathbb{Q}_{ab}^{(2)};\nn\\
%==
  \int \sigma_{ab}(g,\theta) \big[ \bg'\bg'\bg'-\bg\bg\bg \big] d\Omega &=&
  \frac{g^2}{2} \Big( \mathbb{Q}_{ab}^{(3)} - \mathbb{Q}_{ab}^{(1)} \Big) \big[\bI \bg \big]^S 
  - \frac{1}{2} \Big( 5 \mathbb{Q}_{ab}^{(3)} - 3 \mathbb{Q}_{ab}^{(1)} \Big) \bg\bg\bg. \label{eq:INT5}
\end{eqnarray}
A few clarifying notes about the $d\phi$ integration. The (\ref{eq:INT3}) is obtained easily by
\begin{equation}
 \int_0^{2\pi} \bg'\bg' d\phi = \pi g^2 \sin^2\theta \bI + \pi g^2 (3\cos^2\theta-1)\hat{\boldsymbol{e}}_3 \hat{\boldsymbol{e}}_3; \qquad
\int_0^{2\pi} \bg\bg d\phi= 2\pi g^2 \hat{\boldsymbol{e}}_3 \hat{\boldsymbol{e}}_3.\nn
\end{equation}
The (\ref{eq:INT4}) is obtained by 
\begin{eqnarray}
   \int_0^{2\pi} \bg'\bg'\bg' d\phi &=& \pi g^3 \sin^2\theta \cos\theta \big[\bI \hat{\boldsymbol{e}}_3 \big]^S
  + \pi g^3 \big[ 5\cos^3\theta-3\cos\theta\big]\hat{\boldsymbol{e}}_3\hat{\boldsymbol{e}}_3\hat{\boldsymbol{e}}_3;\nn\\
  \int_0^{2\pi} \bg\bg\bg d\phi &=& 2\pi g^3 \hat{\boldsymbol{e}}_3\hat{\boldsymbol{e}}_3\hat{\boldsymbol{e}}_3;\nn\\
  \int_0^{2\pi} (\bg'\bg'\bg'-\bg\bg\bg) d\phi &=&
  \pi g^3 \sin^2\theta \cos\theta \big[\bI \hat{\boldsymbol{e}}_3 \big]^S
  + \pi g^3 \Big[ 5\cos^3\theta-3\cos\theta  -2   \Big]\hat{\boldsymbol{e}}_3\hat{\boldsymbol{e}}_3\hat{\boldsymbol{e}}_3,
\end{eqnarray}
and by using identities $\sin^2\theta \cos\theta = (1-\cos^3\theta) - (1-\cos\theta)$ and also
$5\cos^3\theta-3\cos\theta-2 = -5(1-\cos^3\theta)+3(1-\cos\theta)$. 
%\begin{eqnarray}
% \sin^2\theta \cos\theta &=& (1-\cos^3\theta) - (1-\cos\theta);\nn\\
% 5\cos^3\theta-3\cos\theta-2 &=& -5(1-\cos^3\theta)+3(1-\cos\theta).\nn
%\end{eqnarray}
Then, by using the simple recipes (\ref{eq:INT5}), the collisional integrals become   
\begin{empheq}[box=\fbox]{align}
  \boldsymbol{R}_{ab} &= -\mu_{ab} 
  \iint  f_a f_b  \, g \bg \mathbb{Q}_{ab}^{(1)} d^3 v_a d^3 v_b;\label{eq:beau0}\\
  %===
  Q_{ab} &= - \mu_{ab} \iint  f_a f_b \, g (\hbV_c \cdot \bg) \mathbb{Q}_{ab}^{(1)}  d^3 v_a d^3 v_b;\label{eq:beau0Pm}\\
  %===
 \bQ_{ab}^{(2)} &=   - \mu_{ab} \iint   
  f_a f_b \, g \Big[  \Big(\bg \hbV_c + \hbV_c \bg\Big)\mathbb{Q}_{ab}^{(1)} 
  - \frac{3}{2} \frac{\mu_{ab}}{m_a}  \Big(  \frac{\bI}{3}g^2 - \bg\bg \Big) \mathbb{Q}_{ab}^{(2)} \Big]  d^3 v_a d^3 v_b;\label{eq:beau0P}\\
  %=== 
  \vecQ^{(3)}_{ab} &=  - \frac{\mu_{ab}}{2} \iint   
  f_a f_b \, g \Big\{ \Big[  2 \hbV_c \hbV_c\cdot \bg +\Big(\hat{V}_c^2  + \frac{\mu_{ab}^2}{m_a^2} g^2 \Big)\bg \Big] \mathbb{Q}_{ab}^{(1)} 
  - 3 \frac{\mu_{ab}}{m_a}    \Big(  \frac{\hbV_c}{3}g^2 - \hbV_c\cdot\bg\bg \Big) \mathbb{Q}_{ab}^{(2)} \Big\}  d^3 v_a d^3 v_b.  \label{eq:beau4}
  %===
\end{empheq}
Expressions (\ref{eq:beau0})-(\ref{eq:beau4}) are equivalent to eqs. (4.79a)-(4.79d), p. 88 of \cite{SchunkNagy2009}. 
The rest of the collisional integrals that we needed for the 22-moment model are given by
\begin{empheq}[box=\fbox]{align}
    \bQ_{ab}^{(4)*}   &=
  - \mu_{ab} \iint f_a f_b \, g \Big\{ \Big( \hat{V}_c^2 +\frac{\mu_{ab}^2}{m_a^2} g^2 \Big)
 \Big[  \Big( \bg \hbV_c + \hbV_c \bg \Big)\mathbb{Q}_{ab}^{(1)}
   - \frac{3}{2} \frac{\mu_{ab}}{m_a}  \Big(  \frac{\bI}{3}g^2 - \bg\bg \Big) \mathbb{Q}_{ab}^{(2)} \Big]\nn\\
  &  + 2 \hbV_c \hbV_c (\hbV_c\cdot\bg) \mathbb{Q}_{ab}^{(1)} 
 - 3\frac{\mu_{ab}}{m_a} \Big[   \frac{2}{3} \hbV_c \hbV_c g^2 - (\hbV_c \cdot\bg) \big(\bg\hbV_c+\hbV_c\bg\big) \Big] \mathbb{Q}_{ab}^{(2)} \nn\\
  &  - \frac{\mu_{ab}^2}{m_a^2}  g^2  \Big( \mathbb{Q}_{ab}^{(3)} - \mathbb{Q}_{ab}^{(1)} \Big) \Big[ \hbV_c \bg + \bg \hbV_c +\bI (\hbV_c\cdot\bg) \Big] 
 + \frac{\mu_{ab}^2}{m_a^2}\Big( 5 \mathbb{Q}_{ab}^{(3)} - 3 \mathbb{Q}_{ab}^{(1)} \Big) \bg\bg (\hbV_c\cdot\bg)  \Big\} d^3 v_a d^3 v_b;  \label{eq:beau5}\\
%===
  Q_{ab}^{(4)}   &=
  - \mu_{ab} \iint f_a f_b \, g \Big\{ 4 \Big( \hat{V}_c^2 +\frac{\mu_{ab}^2}{m_a^2} g^2 \Big)
    (\hbV_c \cdot \bg )\mathbb{Q}_{ab}^{(1)} 
   - 6\frac{\mu_{ab}}{m_a}  \Big(  \frac{V_c^2}{3}g^2 - (\hbV_c \cdot\bg)^2 \Big)  \mathbb{Q}_{ab}^{(2)} \Big\} d^3 v_a d^3 v_b;  \label{eq:beau51} \\
%===
  \vecQ^{(5)}_{ab} &= - \mu_{ab} \iint f_a f_b \, g \Big\{
  \Big( \hat{V}_c^2  + \frac{\mu_{ab}^2}{m_a^2} g^2\Big)^2 \bg \mathbb{Q}_{ab}^{(1)}
  -6\frac{\mu_{ab}}{m_a} \hbV_c  \Big(  \frac{\hat{V}_c^2}{3}g^2 - (\hbV_c\cdot\bg)^2 \Big) \mathbb{Q}_{ab}^{(2)}\nn\\
&   +4  \Big( \hat{V}_c^2  +  \frac{\mu_{ab}^2}{m_a^2} g^2  \Big) \Big[ \hbV_c  (\hbV_c\cdot\bg) \mathbb{Q}_{ab}^{(1)}
   -  \frac{3}{2} \frac{\mu_{ab}}{m_a}   \Big(  \frac{\hbV_c}{3}g^2 - (\hbV_c\cdot\bg)\bg \Big) \mathbb{Q}_{ab}^{(2)} \Big]
     \nn\\
&  - 2 \frac{\mu_{ab}^2}{m_a^2}  g^2  \Big( \mathbb{Q}_{ab}^{(3)} - \mathbb{Q}_{ab}^{(1)} \Big) \Big[ \hat{V}_c^2 \bg + 2\hbV_c(\hbV_c\cdot \bg) \Big]
  + 2\frac{\mu_{ab}^2}{m_a^2} \Big( 5 \mathbb{Q}_{ab}^{(3)} - 3 \mathbb{Q}_{ab}^{(1)} \Big) (\hbV_c\cdot\bg)^2 \bg   \Big\} d^3 v_a d^3 v_b.  \label{eq:beau7}
\end{empheq}

The collisional integrals (\ref{eq:beau0})-(\ref{eq:beau7}) were derived without assuming any specific distribution functions $f_a f_b$.
Here we consider the 22-moment model, with the distribution function $f_a=f_a^{(0)}(1+\chi_a)$, where the perturbation $\chi_a$ is given by (\ref{eq:beau2x}).
Similarly, the $f_b=f_b^{(0)}(1+\chi_b)$ with the perturbation $\chi_b$ obtained by replacing $a\to b$ in (\ref{eq:beau2x}).
The collisional integrals (\ref{eq:beau0})-(\ref{eq:beau7}) need to be calculated by the ``more advanced'' center-of-mass
transformation (see Appendix \ref{sec:AppendixA}), by moving everything into the space $(\bC^*,\bg)$ 
\begin{eqnarray}
  \bc_a &=& \bC^* +\frac{v_{\textrm{th} a}^2}{\beta^2}(\bg+\bu);\qquad
  \bc_b = \bC^* -\frac{v_{\textrm{th} b}^2}{\beta^2} (\bg+\bu); \nn\\
\hat{\bVV}_c &=&
   \bC^* - \frac{2}{\beta^2}\frac{(T_b-T_a)}{(m_a+m_b)}(\bg+\bu) +\frac{\mu_{ab}}{m_a}\bu,  \label{eq:hatVx}
\end{eqnarray}
where $\bu=\bu_b-\bu_a$ is the difference in drifts/bulk velocities. The product $f_a f_b$ becomes
\begin{equation}
  f_a^{(0)} f_b^{(0)} = \frac{n_a n_b}{\pi^3 {\alpha}^3 \beta^3}\exp\Big( -\frac{|\boldsymbol{C}^*|^2}{{\alpha}^2}
  -\frac{|\bg+\bu|^2}{\beta^2}\Big); \qquad f_a f_b = f_a^{(0)} f_b^{(0)} (1+\chi_a+\chi_b + \chi_a \chi_b),
  %\label{eq:fafbx}
\end{equation}
with the new thermal speeds
\begin{equation}
  {\alpha}^2 = \frac{v_{\textrm{th} a}^2 v_{\textrm{th} b}^2}{v_{\textrm{th} a}^2 + v_{\textrm{th} b}^2};
  \qquad \beta^2 = v_{\textrm{th} a}^2+v_{\textrm{th} b}^2, 
\end{equation}
and one integrates over the $d^3 v_a d^3 v_b = d^3C^* d^3g$. Everything is fully non-linear at this stage
and if the integrals (\ref{eq:beau0})-(\ref{eq:beau7}) were indeed calculated, one would
obtain a fully non-linear 22-moment model.

%=================================================================================================
\subsection{Semi-linear approximation}
In practice, one proceeds with the semi-linear approximation, where terms such as $u^2$, $\bu\cdot\bPi^{(2)}$, $\bu\cdot\vecq$ or $\vecq\cdot\bPi^{(2)}$
are neglected, while expressions such as $p/\rho$ are retained and the product $f_a f_b$ is given by
\begin{equation}
  f_a f_b = \frac{n_a n_b}{\pi^3 {\alpha}^3 \beta^3}
  e^{ -\frac{C^{*2}}{{\alpha}^2}} e^{-\frac{g^2}{\beta^2}}
  \big( 1-2\frac{\bg\cdot\bu}{\beta^2} +\chi_a + \chi_b \big). \label{eq:fafbS5}
\end{equation}
The collisional integrals (\ref{eq:beau0})-(\ref{eq:beau7}) then can be calculated in two well-defined steps.
\begin{itemize}
\item[1)] One neglects the perturbations $\chi_a+ \chi_b$ and focuses only on the contributions from the drifts $\bu=\bu_b-\bu_a$
  (and also from the temperature differences $T_b-T_a$), by prescribing  
  \begin{eqnarray}
  f_a f_b &=& \frac{n_a n_b}{\pi^3 {\alpha}^3 \beta^3}
  e^{ -\frac{C^{*2}}{{\alpha}^2}} e^{-\frac{g^2}{\beta^2}} \big( 1-2\frac{\bg\cdot\bu}{\beta^2} \big);\nn\\
  \hat{\bVV}_c &=& \bC^* - \frac{2}{\beta^2}\frac{(T_b-T_a)}{(m_a+m_b)}(\bg+\bu) +\frac{\mu_{ab}}{m_a}\bu,  \label{eq:Split1}
  \end{eqnarray}
  where the transformations for $\bc_a$, $\bc_b$ are not needed anymore. During calculations, the collisional integrals are further linearized
  in $\bu$ and one can show that the $\bu$ contributions appear only in $\boldsymbol{R}_{ab}$, $\vecQ^{(3)}_{ab}$ and
  $\vecQ^{(5)}_{ab}$. In contrast, the scalar equations $Q_{ab}, Q_{ab}^{(4)}$ as well as
  the matrix equations $\bQ_{ab}^{(2)}$, $\bQ_{ab}^{(4)*}$ contain no semi-linear $\bu$ contributions, 
  and only contain contributions from the temperatures $T_b-T_a$.
  %Additionally, by considering the traceless
  %$\bQ^{(2)}_a\,' = \bQ^{(2)}_a -(\bI/3)\trace \bQ^{(2)}_a$ and  $\bQ^{(4)}_a\,' = \bQ^{(4)*}_a - (\bI/3)\trace \bQ^{(4)*}_a$, the contributions from the
  %$T_b-T_a$ are eliminated as well.  
%==  
\item[2)] One neglects the drifts $\bu$ and focuses only on the contributions from the $\chi_a + \chi_b$, by prescribing
  \begin{eqnarray}
  f_a f_b &=& \frac{n_a n_b}{\pi^3 {\alpha}^3 \beta^3}
  e^{ -\frac{C^{*2}}{{\alpha}^2}} e^{-\frac{g^2}{\beta^2}}  \big( \chi_a + \chi_b \big);\nn\\
  \bc_a &=& \bC^* +\frac{v_{\textrm{th} a}^2}{\beta^2}\bg;\qquad
  \bc_b = \bC^* -\frac{v_{\textrm{th} b}^2}{\beta^2}\bg; \nn\\
\hat{\bVV}_c &=&  \bC^* - \frac{2}{\beta^2}\frac{(T_b-T_a)}{(m_a+m_b)}\bg.  \label{eq:Split2}
  \end{eqnarray}
Then one can show that the $\boldsymbol{R}_{ab}$, $\vecQ^{(3)}_{ab}$ and
$\vecQ^{(5)}_{ab}$ only contain contributions from the heat fluxes $\chi^{\textrm{(heat)}}$. Also, the scalar equations
$Q_{ab}$, $Q_{ab}^{(4)}$ only contain contributions from the scalars $\chi^{\textrm{(scalar)}}$. Finally, 
considering the traceless $\bQ^{(2)}_a\,' = \bQ^{(2)}_a -(\bI/3)\trace \bQ^{(2)}_a$ and
$\bQ^{(4)}_a\,' = \bQ^{(4)*}_a - (\bI/3)\trace \bQ^{(4)*}_a$, these equations only contain contributions from the viscosities $\chi^{\textrm{(visc.)}}$.
%\end{itemize}

\noindent
For example, by using the $\bc_a$, $\bc_b$ transformations (\ref{eq:Split2}), the 1-Hermite viscosity perturbations are transformed as
\begin{eqnarray}
  \chi_a^{(2)} &=& \frac{m_a}{2T_a p_a} \bPi^{(2)}_a:\Big[ \bC^* \bC^* + \frac{v_{\textrm{th} a}^2}{\beta^2} \big( \bC^* \bg + \bg \bC^*\big)
    + \frac{v_{\textrm{th} a}^4}{\beta^4}\bg\bg \Big] ; \nn\\
  \chi_b^{(2)} &=& \frac{m_b}{2T_b p_b} \bPi^{(2)}_b:\Big[ \bC^* \bC^* - \frac{v_{\textrm{th} b}^2}{\beta^2} \big( \bC^* \bg + \bg \bC^*\big)
    + \frac{v_{\textrm{th} b}^4}{\beta^4}\bg\bg \Big], \label{eq:chabB}
\end{eqnarray}
and the 1-Hermite heat flux perturbations as
\begin{eqnarray}
  \chi_a^{(3)}  &=& \frac{1}{5} \frac{m_a}{T_a p_a} \boldsymbol{q}_a \cdot \Big(\bC^* + \frac{v_{\textrm{th} a}^2}{\beta^2} \bg \Big)
  \Big[ \frac{m_a}{T_a} \Big( C^{*2} +2 \frac{v_{\textrm{th} a}^2}{\beta^2} \bC^*\cdot\bg 
    + \frac{v_{\textrm{th} a}^4}{\beta^4} g^2 \Big) -5 \Big]; \nn\\
%=======
\chi_b^{(3)} &=& \frac{1}{5} \frac{m_b}{T_b p_b} \boldsymbol{q}_b \cdot \Big( \bC^* - \frac{v_{\textrm{th} b}^2}{\beta^2} \bg  \Big)
\Big[ \frac{m_b}{T_b} \Big( C^{*2} -2 \frac{v_{\textrm{th} b}^2}{\beta^2} \bC^*\cdot\bg 
    + \frac{v_{\textrm{th} b}^4}{\beta^4} g^2 \Big) -5 \Big]. \label{eq:HFchiab} 
\end{eqnarray}

If one is not interested in the arbitrary temperatures, the step 2) can be hugely simplified by considering small
temperature differences, or only self-collisions
  \begin{eqnarray}
&& T_b=T_a: \qquad \quad  \bc_a = \bC^* +\frac{\mu_{ab}}{m_a}\bg;\qquad
  \bc_b = \bC^* -\frac{\mu_{ab}}{m_b}\bg; \qquad \hat{\bVV}_c =  \bC^*; \label{eq:Split3c}\\   
&& \textrm{self-collisions:} \qquad \bc_a = \bC^* +\frac{1}{2}\bg;\qquad
  \bc_b = \bC^* -\frac{1}{2}\bg; \qquad \quad \hat{\bVV}_c =  \bC^*. \label{eq:Split3}
  \end{eqnarray}
  Importantly, for self-collisions $\chi_a\neq \chi_b$ (!), or in other words $\chi_a+\chi_b\neq 2\chi_a$ (an error which is very
  easy to make). 
\end{itemize}

%=============================================================================
\subsection{Semi-automatic integration of the collisional integrals}
If in the previous step 2) one considers only the small temperature differences (\ref{eq:Split3c}), the collisional integrals of
1-Hermite moments are actually not
overly complicated to calculate by hand and this is especially true if only the self-collisions (\ref{eq:Split3}) are considered.
When learning the Boltzmann operator for the first time, it is highly recommended to recover at least
parts of the 1-Hermite models by hand (and as stated previously, it is recommended to initially ignore the Chapman-Cowling integrals
and directly consider the hard spheres and
Coulomb collisions from the beginning). Perhaps only then one can clearly see the logic (and the beauty) behind the ``semi-automatic'' procedure that
we discuss here.

When calculating the integrals over the $d^3C^* d^3 g$ by hand, one keeps rotating the spherical co-ordinate system back and forth by choosing the
appropriate direction of the axis $\hat{\boldsymbol{e}}_3$. However, this is not necessary 
and it is possible to define two (unrelated) co-ordinate systems with two vectors 
\begin{eqnarray}
\bC^* &=& C^* \big[ \sin\theta^*\cos\phi^*, \sin\theta^*\sin\phi^*, \cos\theta^* \big]; \nn\\
\bg &=& g \big[ \sin\theta\cos\phi, \sin\theta\sin\phi, \cos\theta \big], \label{eq:best}
\end{eqnarray}
and integrate over the $d^3C^*=C^{*2}\sin\theta^* dC^* d\theta^* d\phi^*$ and
$d^3g = g^2 \sin\theta dg d\theta d\phi$. This extremelly simple trick allows one to use analytic software such as
Maple or Mathematica and calculate the collisional integrals easily, by simply
performing six successive one-dimensional integrals with the command ``int'' (and ignoring all the advanced features
that these programs offer). For the scalar equations $Q_{ab}$, $Q_{ab}^{(4)}$ nothing more is required, because
all of the quantities such as $\hbV_c\cdot\bg$ are scalars.
The vector equations $\boldsymbol{R}_{ab}$, $\vecQ^{(3)}_{ab}$, $\vecQ^{(5)}_{ab}$ contain other vectors,
such as the $\bu$ and $\vecq_a$. It is of course possible to use a general directions for these vectors,
by writing $\bu=[u_x,u_y,u_z]$. Nevertheless, by performing the integrals by hand, one learns that the final result of the
integration is always proportional to the entire vector $\bu$ or $\vecq_a$, and for the fastest calculations one can just
choose for example $\bu=[0,0,u]$ or $\boldsymbol{q}_a=[0,0,q_a]$ and obtain the same result (which in addition to some
computational speedup, has even a greater benefit of reducing the eye-strain when looking at the complicated expressions at a screen). 

For the matrix equations $\bQ^{(2)}_a$, $\bQ^{(4)*}_a$ the situation is more complicated, but only marginally.
For example, the Maple command ``int'' does not directly integrate over each component of a matrix and one needs to use the
command map(int,expr,$\phi=0..2\pi$) instead. In addition to matrices such as $\bg \hbV_c$, the matrix equations contain other
matrices, such as the $\bPi^{(2)}_a$. It is of course possible to consider all of the components of the matrix $\bPi^{(2)}_a$, 
but by performing the integrals by hand, one again learns that the final result of the integration is always proportional to
the entire matrix $\bPi_a^{(2)}$. Now, before drastically reducing the calculation to just one component of the $\bPi^{(2)}_a$,
it is indeed re-assuring to consider a small middle-step, by specifying that the matrix $\bPi_a^{(2)}$
has a traceless diagonal form such as $\bhat\bhat - \bI/3$, i.e. a diagonal matrix $[-1/3,-1/3,+2/3]$,
and verify that the result of the integration is the same diagional matrix.

%==================================
\subsection{Relation to the Fokker-Planck operator}
     It is worth noting that the corrections of the Coulomb logarithm such as $\ln(\Lambda^2+1)$ can be
      also derived by considering a more general class of Fokker-Planck operators,  
    where the dynamical friction vector $\boldsymbol{A}_{ab}$ and the diffusion tensor $\bD_{ab}$ (a matrix)
    are expressed through a differential cross-section $\sigma_{ab}(g_{ab},\theta)$, according to
\begin{eqnarray}
 C_{ab} &=& -\, \frac{\pr}{\pr \bV_a }\cdot ( \boldsymbol{A}_{ab} f_a) +\frac{1}{2} \frac{\pr}{\pr \bV_a } \frac{\pr}{\pr \bV_a } :(\bD_{ab} f_a); \label{eq:FP}\\
  \boldsymbol{A}_{ab} &=& \iint  g_{ab} \sigma_{ab}(g_{ab},\theta) f_b \big[\bV_a'-\bV_a \big] d\Omega d^3 v_b;\nn\\
 \bD_{ab} &=& \iint  g_{ab} \sigma_{ab}(g_{ab},\theta) f_b \big[(\bV_a'-\bV_a)(\bV_a'-\bV_a)\big] d\Omega d^3 v_b. \nn
\end{eqnarray}
     This operator was considered for example by \cite{Tanenbaum1967}, p. 283. Fokker-Planck operators are often derived from the
      Boltzmann operator by Taylor expanding the distribution functions $f_a'$ around $f_a$
      in velocities $\triangle \bV_a=\bV_a'-\bV_a$, which is a very tedious procedure,
      and in the last reference the (\ref{eq:FP}) is derived by a trick of using the integration recipe (\ref{eq:BoltzmannOO}) and
      expanding the $X(\bV_a)$ in $\triangle \bV_a$ instead. Here in Part 2, we did not spend much time exploring this operator,
      because as noted already in the last reference, the Fokker-Planck operator is actually as difficult
      to integrate as the full Boltzmann operator.
      This can be easily seen by looking at the Fokker-Planck collisional contributions in Appendix F of Part 1
\begin{eqnarray}
  \boldsymbol{R}_{ab} &=& m_a \int f_a \boldsymbol{A}_{ab} d^3v_a; \label{eq:FP1}\\
  Q_{ab} &=& m_a \int f_a \Big[ \boldsymbol{A}_{ab}\cdot\bc_a + \frac{1}{2} \textrm{Tr} \bD_{ab} \Big] d^3v_a; \label{eq:FP2}\\
  \bQ_{ab}^{(2)} &=& m_a \int f_a \Big[ \big(\boldsymbol{A}_{ab}\bc_a\big)^S  +\frac{1}{2} \big(\bD_{ab}\big)^S \Big]d^3v_a;\\
  \vecQ^{(3)}_{ab} &=& \frac{m_a}{2} \int f_a \Big[ 2(\boldsymbol{A}_{ab}\cdot\bc_a)\bc_a + \boldsymbol{A}_{ab}|\bc_a|^2 
    + (\textrm{Tr}\bD_{ab})\bc_a +2 \bD_{ab}\cdot\bc_a\Big] d^3v_a, \label{eq:Sch1}
\end{eqnarray}  
  together with eqs. (F12)-(F14) for $n=2$ there, and comparing them with the Boltzmann contributions
   given here by (\ref{eq:beauty10}). More importantly, one will always keep guessing at what order the Fokker-Planck operator 
   starts to fail, where for example the operator yields corrections of the Coulomb logarithm $A_1(2)$ and $A_2(2)$,
   but not the $A_3(2)$ given by (\ref{eq:A23x}) (which will be somehow approximated, perhaps).  
   Nevertheless, if one is interested only in simple fluid models, the dynamical friction vector and the diffusion tensor can be
   actually expressed through the effective cross-sections, according to
\begin{eqnarray}
  \boldsymbol{A}_{ab} &=& -\, \frac{\mu_{ab}}{m_a} \int g\bg f_b \mathbb{Q}_{ab}^{(1)} d^3 v_b;\nn\\
  \bD_{ab} &=& \frac{\mu_{ab}^2}{m_a^2} \int g f_b
  \Big[ 2\bg\bg \mathbb{Q}_{ab}^{(1)}  +\frac{3}{2} \big( \frac{\bI}{3}g^2 -\bg\bg \big)\mathbb{Q}_{ab}^{(2)}  \Big] d^3v_b, 
\end{eqnarray}   
and by using these expressions yields that the $\boldsymbol{R}_{ab}$, $Q_{ab}$ and even $\bQ_{ab}^{(2)}$ are identical to 
the Boltzmann expressions (\ref{eq:beau0})-(\ref{eq:beau0P}). The heat flux contributions $\vecQ^{(3)}_{ab}$ seem to be different,
but we did not spent much time with it, because again, why to work with some Taylor expanded function, when working with a non-expanded
function is not more complicated (and we would say actually even easier).

\newpage
%=========================================================================================================
%=========================================================================================================
\section{Examples of calculations for general collisional processes} \label{sec:Examples}
\setcounter{equation}{0}
Let us show few examples how to calculate the general collisional integrals with the Boltzmann operator by hand. As a reminder, here we mostly
use the semi-linear approximation. The exception is Appendix \ref{sec:Unrestricted}, where unrestricted drifts for the 5-moment
  models are considered and
  we also discuss the $|\bu_b-\bu_a|^2$ contributions to $Q_{ab}$.
%======================================
\subsection{Simplest momentum exchange rates (5-moment model)}
Starting with the momentum exchange rates (\ref{eq:beau0}) and considering the strict Maxwellians
(with perturbations $\chi_a$ and $\chi_b$ being zero), the momentum exchange rates for small drifts $\bu=\bu_b-\bu_a$ calculate
\begin{eqnarray}
\boldsymbol{R}_{ab} &=& -\mu_{ab} 
\iint  f_a f_b  \, g \bg \mathbb{Q}_{ab}^{(1)} d^3 v_a d^3 v_b;\nn\\
 &=& -\mu_{ab} 
\frac{n_a n_b}{\pi^3 {\alpha}^3 \beta^3} \iint
  e^{ -\frac{C^{*2}}{{\alpha}^2}} e^{-\frac{g^2}{\beta^2}}
  \big( \cancel{1} -2\frac{\bg\cdot\bu}{\beta^2} \big)   \, g \bg \mathbb{Q}_{ab}^{(1)} d^3 C^* d^3 g \nn\\
  &=& +2 \mu_{ab} 
\frac{n_a n_b}{\pi^3 {\alpha}^3 \beta^5} \bu\cdot \iint
  e^{ -\frac{C^{*2}}{{\alpha}^2}} e^{-\frac{g^2}{\beta^2}}
  \bg\bg  g \mathbb{Q}_{ab}^{(1)} d^3 C^* d^3 g \nn\\
  &=& +2 \mu_{ab} 
\frac{n_a n_b}{\pi^{3/2} \beta^5} \bu\cdot \int
  e^{-\frac{g^2}{\beta^2}}
  \bg\bg g  \mathbb{Q}_{ab}^{(1)} d^3 g \nn\\
  &=& +2 \mu_{ab} 
\frac{n_a n_b}{\pi^{3/2} \beta^5} \bu\cdot \bI \frac{4\pi}{3} \int_0^\infty
  e^{-\frac{g^2}{\beta^2}}
  g^5  \mathbb{Q}_{ab}^{(1)} d g \nn\\
&=& +\frac{16}{3} \mu_{ab} 
  n_a n_b \bu  \Big[ \frac{1}{2 \pi^{1/2} \beta^5}\int_0^\infty
  e^{-\frac{g^2}{\beta^2}}
  g^5  \mathbb{Q}_{ab}^{(1)} d g \Big] =  \frac{16}{3} \mu_{ab}   n_a n_b \bu \Omega_{ab}^{(1,1)},
\end{eqnarray}
where the definition of the lowest-level Chapman-Cowling integral (\ref{eq:CCo}) was used in the last step. 
So by chosing to define the momentum exchange rates through the collisional frequency $\nu_{ab}$ as
$\boldsymbol{R}_{ab}=m_a n_a \nu_{ab} \bu$, then yields the definition
\begin{equation}
\nu_{ab} \equiv \frac{16}{3} \frac{\mu_{ab}}{m_a} n_b \Omega_{ab}^{(1,1)}. \label{eq:nuabDEF}
\end{equation}

%======================================
\subsection{Simplest energy exchange rates (5-moment model)}
Instead of calculating the energy exchange rates $Q_{ab}$ given by (\ref{eq:beau0Pm}), let us calculate the more general pressure tensor contributions
$\bQ_{ab}^{(2)}$ given by (\ref{eq:beau0P}) and then do the trace. In the semi-linear approximation, the contributions from the drifts $\bu$
will be zero at the end, so let us neglect them from the beginning and the (\ref{eq:beau0P}) then calculates 
\begin{eqnarray}
 \bQ_{ab}^{(2)} &=&   - \mu_{ab} \frac{n_a n_b}{\pi^3 {\alpha}^3 \beta^3} \iint
  e^{ -\frac{C^{*2}}{{\alpha}^2}} e^{-\frac{g^2}{\beta^2}} 
   g \Big[  \Big(\bg \hbV_c + \hbV_c \bg\Big)\mathbb{Q}_{ab}^{(1)} 
     - \cancel{\frac{3}{2} \frac{\mu_{ab}}{m_a}  \Big(  \frac{\bI}{3}g^2 - \bg\bg \Big) \mathbb{Q}_{ab}^{(2)} }\Big]  d^3 C^* d^3 g \nn\\
   %===
   &=&   - \mu_{ab} \frac{n_a n_b}{\pi^3 {\alpha}^3 \beta^3} \iint
  e^{ -\frac{C^{*2}}{{\alpha}^2}} e^{-\frac{g^2}{\beta^2}} 
  g \Big[  \bg \Big(  \cancel{\bC^*} - \frac{2}{\beta^2}\frac{(T_b-T_a)}{(m_a+m_b)}\bg \Big) \nn\\
    && \qquad + \Big( \cancel{\bC^*} - \frac{2}{\beta^2}\frac{(T_b-T_a)}{(m_a+m_b)}\bg \Big) \bg \Big]
  \mathbb{Q}_{ab}^{(1)}   d^3 C^* d^3 g \nn\\
   %===
   &=&   + 4 \frac{(T_b-T_a)}{(m_a+m_b)}\mu_{ab} \frac{n_a n_b}{\pi^{3/2} \beta^5} \int
  e^{-\frac{g^2}{\beta^2}}  g \bg\bg  \mathbb{Q}_{ab}^{(1)} d^3 g
  = + 4 \frac{(T_b-T_a)}{(m_a+m_b)}\mu_{ab} \frac{n_a n_b}{\pi^{3/2} \beta^5} \frac{4\pi}{3} \bI \int_0^\infty
  e^{-\frac{g^2}{\beta^2}}  g^5 \mathbb{Q}_{ab}^{(1)} d g \nn\\
  &=& + \frac{32}{3} \frac{(T_b-T_a)}{(m_a+m_b)}\mu_{ab} n_a n_b \bI \Big[ \frac{1}{2 \pi^{1/2} \beta^5}  \int_0^\infty
  e^{-\frac{g^2}{\beta^2}}  g^5 \mathbb{Q}_{ab}^{(1)} d g \Big] = \frac{32}{3} \frac{(T_b-T_a)}{(m_a+m_b)}\mu_{ab} n_a n_b \bI \Omega_{ab}^{(1,1)},
\end{eqnarray}
which by using the collisional frequency (\ref{eq:nuabDEF}) can be re-written as
\begin{equation}
  \bQ_{ab}^{(2)} = 2 \bI \frac{(T_b-T_a)}{(m_a+m_b)} m_a n_a \nu_{ab}; \qquad
  Q_{ab}=\frac{1}{2} \trace \bQ_{ab}^{(2)} = 3 \frac{(T_b-T_a)}{m_a+m_b} m_a n_a \nu_{ab}. \label{eq:QabFin}
\end{equation}
The result (\ref{eq:QabFin}) is now valid for a general collisional process (and arbitrary temperature differences),
and it matches the result of the hard spheres, Coulomb collisions and Maxwell molecules. If one wants to keep the
  $|\bu_b-\bu_a|^2$ term in $Q_{ab}$, one can just use the first line of general recipe (\ref{eq:GenQab3}), directly yielding
\begin{equation} 
  Q_{ab} = \frac{m_a n_a \nu_{ab}}{m_a+m_b} \Big[ 3(T_b-T_a) + m_b |\bu_b-\bu_a|^2  \Big], \label{eq:QabGenA}
\end{equation}
which however introduces restriction that the temperature differences $T_b-T_a$ are small (because otherwise, by doing proper expansions in velocities $u^2$,
the term $T_b-T_a$ would yield $u^2$ contributions, see eq. (\ref{eq:Picus1X})).
The particular case of Maxwell molecules is an
exception, where the same $Q_{ab}$ was obtained in (\ref{eq:MMQAB}) for arbitrary temperatures and drifts.

%======================================
\subsection{Momentum and energy exchange rates with unrestricted drifts (5-moment model)} \label{sec:Unrestricted}
Still considering only the 5-moment models, it is useful to clarify the $R_{ab}$ and $Q_{ab}$
for a general collisional process with unrestricted drifts $\bu=\bu_b-\bu_a$.
By starting with the (\ref{eq:beau0})-(\ref{eq:beau0Pm}) and integrating over $d^3C^*$, it is easy to show that 
\begin{eqnarray}
  \boldsymbol{R}_{ab} &=& -\mu_{ab} \frac{n_a n_b}{\pi^{3/2} \beta^3} \int g \bg e^{-\frac{|\bg+\bu|^2}{\beta^2}} \mathbb{Q}_{ab}^{(1)} d^3g;
  \qquad  Q_{ab} =  Q_{ab}^* + \frac{m_b}{m_a+m_b} \bu\cdot  \boldsymbol{R}_{ab}; \\
  Q_{ab}^* &=& \mu_{ab} \frac{n_a n_b}{\pi^{3/2}} \frac{2}{\beta^5} \frac{(T_b-T_a)}{m_a+m_b}
  \int g (g^2+\bu\cdot\bg) e^{-\frac{|\bg+\bu|^2}{\beta^2}} \mathbb{Q}_{ab}^{(1)}d^3g,
\end{eqnarray}
which as a quick double-check recovers the hard spheres expressions (\ref{eq:niceA}) and (\ref{eq:HSPic2})
(where the $\mathbb{Q}_{ab}^{(1)}=\pi r_{ab}^2$). Then as before, one introduces variable $\epsilon=u/\beta$ and by using
either substitution (\ref{eq:zsB}) with $x=g/\beta$ and $s=\epsilon\cos\theta^*$, or substitution (\ref{eq:zs}) with $z=\frac{g}{\beta}+s$,
it is easy to show that the momentum exchange rates are given by (for convenience, we write results for both substitutions, because
for some cases one is easier to calculate than the other) 
\begin{eqnarray}
  \boldsymbol{R}_{ab} &=& - (\bu_b-\bu_a) \mu_{ab} \frac{n_a n_b}{\sqrt{\pi}} 2\beta \frac{e^{-\epsilon^2}}{\epsilon^3}
  \int_{-\epsilon}^{\epsilon} ds \int_0^\infty dx\, x^4 s e^{-x^2 -2xs} \mathbb{Q}_{ab}^{(1)} (g=\beta x) \label{eq:DraineRab}\\
  &=& - (\bu_b-\bu_a) \mu_{ab} \frac{n_a n_b}{\sqrt{\pi}} 2\beta \frac{e^{-\epsilon^2}}{\epsilon^3}
  \int_{-\epsilon}^{\epsilon} ds \int_s^\infty dz\, (z-s)^4 s e^{-z^2 +s^2} \mathbb{Q}_{ab}^{(1)} (g=\beta (z-s)). \label{eq:DraineRab2}
\end{eqnarray}
In (\ref{eq:DraineRab}) one can choose if to first integrate over the $ds$ or $dx$ and in (\ref{eq:DraineRab2}) one has to first
integrate over the $dz$. Similarly, it is easy to show that
\begin{eqnarray}
    Q_{ab}^* &=& \frac{(T_b-T_a)}{m_a+m_b} \mu_{ab} \frac{n_a n_b}{\sqrt{\pi}}   4\beta
    \frac{e^{-\epsilon^2}}{\epsilon} \int_{-\epsilon}^{\epsilon} ds \int_0^\infty dx\, x^4 (x+s) e^{-x^2 -2xs} \mathbb{Q}_{ab}^{(1)} (g=\beta x) \label{eq:DraineQab}\\
    &=& \frac{(T_b-T_a)}{m_a+m_b} \mu_{ab} \frac{n_a n_b}{\sqrt{\pi}}   4\beta
    \frac{e^{-\epsilon^2}}{\epsilon} \int_{-\epsilon}^{\epsilon} ds \int_s^\infty dz\, (z-s)^4 z e^{-z^2 +s^2} \mathbb{Q}_{ab}^{(1)} (g=\beta (z-s)).
\end{eqnarray}
After specifying particular collisional process with $\mathbb{Q}_{ab}^{(1)} (g)$, we find it the best to just calculate the above double
integrals with analytic software. Nevertheless, expressions (\ref{eq:DraineRab}) and (\ref{eq:DraineQab}) can be integrated over the $ds$
by (\ref{eq:DraineINT}), together with $\int_{-\epsilon}^\epsilon x e^{-2xs} ds = \sinh(2x\epsilon)$ so that in (\ref{eq:DraineQab})
\begin{eqnarray}
  \int_{-\epsilon}^\epsilon (x+s) e^{-2xs}  ds &=& \sinh(2x\epsilon)   +\frac{1}{2x^2} \Big( \sinh(2x\epsilon)-2x\epsilon\cosh(2x\epsilon)\Big),
\end{eqnarray}
finally yielding
\begin{empheq}[box=\fbox]{align}
  \boldsymbol{R}_{ab} &= (\bu_b-\bu_a) \mu_{ab} \frac{n_a n_b}{\sqrt{\pi}} \beta \frac{e^{-\epsilon^2}}{\epsilon^3}
  \int_0^\infty dx\, x^2 e^{-x^2} \Big( 2x\epsilon\cosh(2x\epsilon) -\sinh(2x\epsilon) \Big)
  \mathbb{Q}_{ab}^{(1)} (g=\beta x); \label{eq:DraineRab3}\\
  %===
   Q_{ab}^* &= \frac{(T_b-T_a)}{m_a+m_b} \mu_{ab} \frac{n_a n_b}{\sqrt{\pi}}   4\beta
   \frac{e^{-\epsilon^2}}{\epsilon} \int_0^\infty dx\, x^4 e^{-x^2}
   \Big[ \sinh(2x\epsilon)   -\frac{1}{2x^2} \Big( 2x\epsilon\cosh(2x\epsilon) -\sinh(2x\epsilon) \Big) \Big]
   \mathbb{Q}_{ab}^{(1)} (g=\beta x).
\end{empheq}
For clarity, the full $Q_{ab}$ reads
\begin{eqnarray}
Q_{ab} &=& \frac{(T_b-T_a)}{m_a+m_b} \mu_{ab} \frac{n_a n_b}{\sqrt{\pi}}   4\beta
   \frac{e^{-\epsilon^2}}{\epsilon} \int_0^\infty dx\, x^4 e^{-x^2}
   \Big[ \sinh(2x\epsilon)   -\frac{1}{2x^2} \Big( 2x\epsilon\cosh(2x\epsilon) -\sinh(2x\epsilon) \Big) \Big]
   \mathbb{Q}_{ab}^{(1)} (g=\beta x) \nn\\
 && + \frac{m_b}{m_a+m_b} \mu_{ab}\frac{n_a n_b}{\sqrt{\pi}} \beta^3 \frac{e^{-\epsilon^2}}{\epsilon}
  \int_0^\infty dx\, x^2 e^{-x^2} \Big( 2x\epsilon\cosh(2x\epsilon) -\sinh(2x\epsilon) \Big)
  \mathbb{Q}_{ab}^{(1)} (g=\beta x). \label{eq:DraineQab3}
\end{eqnarray}
This $\boldsymbol{R}_{ab}$ (\ref{eq:DraineRab3}) is directly equivalent to eq. (3.1) of \cite{Draine1986} and it can be shown that
 the $Q_{ab}$ (\ref{eq:DraineQab3}) is equivalent to his eq. (3.7).
Prescribing hard spheres (with $\mathbb{Q}_{ab}^{(1)}=\pi r_{ab}^2$) recovers previous results with unrestricted drifts
(\ref{eq:RhardX}) and (\ref{eq:PsiExact}), prescribing Coulomb collisions (with eq. (\ref{eq:beauty15}))
recovers results (\ref{eq:RCoulombX}) and (\ref{eq:NiceA}),
and prescribing Maxwell molecules recovers results (\ref{eq:Maxnuab}) and (\ref{eq:MMQAB}). Note that for a collisional force $r^{-\nu}$, the
$\mathbb{Q}_{ab}^{(1)}(g)\sim g^{-\frac{4}{\nu-1}}$.

As another double-check of (\ref{eq:DraineRab3})-(\ref{eq:DraineQab3}), it is useful to consider small drifts,
where expansions with small $\epsilon\ll 1$ yield
\begin{eqnarray}
  \frac{e^{-\epsilon^2}}{\epsilon^3} \Big(2x\epsilon\cosh(2x\epsilon) -\sinh(2x\epsilon)\Big) &=&
  \frac{8}{3} x^3 \Big[ 1+\Big(\frac{2}{5} x^2 -1\Big)\epsilon^2 + \ldots \Big];\\
  \frac{e^{-\epsilon^2}}{\epsilon} \Big[\sinh(2x\epsilon)   -\frac{1}{2x^2} \Big( 2x\epsilon\cosh(2x\epsilon) -\sinh(2x\epsilon) \Big) \Big]
  &=& 2x\Big[ 1 +\Big( \frac{2}{3}x^2-\frac{5}{3}\Big)\epsilon^2 +\ldots \Big],\nn
\end{eqnarray}
and by neglecting contributions of higher-orders than $|\bu_b-\bu_a|^2$ (the $\boldsymbol{R}_{ab}$ already contains $\bu_b-\bu_a$ up front, so the
 $(\bu_b-\bu_a)u^2$ is neglected), further implies that for small drifts
\begin{eqnarray}
  \boldsymbol{R}_{ab} &=& (\bu_b-\bu_a) \mu_{ab} \frac{n_a n_b}{\sqrt{\pi}} \frac{8}{3} \beta 
  \int_0^\infty dx\, x^5 e^{-x^2}  \mathbb{Q}_{ab}^{(1)} (g=\beta x); \label{eq:DraineRab5}\\
   Q_{ab}^* &=& \frac{(T_b-T_a)}{m_a+m_b} \mu_{ab} \frac{n_a n_b}{\sqrt{\pi}}   8\beta
   \int_0^\infty dx\, x^5 e^{-x^2}
   \Big[ 1 +\Big( \frac{2}{3}x^2-\frac{5}{3}\Big)\epsilon^2 \Big]
   \mathbb{Q}_{ab}^{(1)} (g=\beta x).\label{eq:DraineQab5}
\end{eqnarray}
The second term in (\ref{eq:DraineQab5}) proportional to $(T_b-T_a)|\bu_b-\bu_a|^2$ must be retained, if in the full $Q_{ab}$
 one wants to keep the $|\bu_b-\bu_a|^2$ contributions and also retain the validity for arbitrary temperature differences. 

Now, one can employ the technique with Chapman-Cowling integrals, where from the definition (\ref{eq:CCo})
\begin{eqnarray}
\Omega_{ab}^{(l,j)} &=& \frac{\beta}{2\sqrt{\pi}} \int_0^\infty dx x^{2j+3} e^{-x^2} \mathbb{Q}_{ab}^{(l)} (g=\beta x),
\end{eqnarray}
and so only the $\Omega_{ab}^{(1,1)}$ and $\Omega_{ab}^{(1,2)}$ are present in (\ref{eq:DraineQab5}).
Then, by using the collisional frequency $\nu_{ab}$ (\ref{eq:nuabDEF})
and employing our notation $\Omega_{1,2}=\Omega_{ab}^{(1,2)}/\Omega_{ab}^{(1,1)}$, directly yields
%\begin{eqnarray}
%R_{ab} &=& (\bu_b-\bu_a) m_a n_a \nu_{ab};\nn\\
%Q_{ab}^* &=& 3 (T_b-T_a) \frac{n_a m_a \nu_{ab}}{m_a+m_b} \big( 1 + \Upsilon_{ab} \epsilon^2 \big);
%\qquad \Upsilon_{ab}= \frac{2}{3} \big(\Omega_{1,2}-\frac{5}{2}\big);
%\qquad \epsilon^2=\frac{|\bu_b-\bu_a|^2}{v_{\textrm{th} a}^2 +v_{\textrm{th} b}^2};\nn\\
%Q_{ab} &=& \frac{n_a m_a \nu_{ab}}{m_a+m_b}\Big[ 3 (T_b-T_a) \big( 1 +\Upsilon_{ab} \epsilon^2 \big)
%+ m_b |\bu_b-\bu_a|^2 \Big], \label{eq:Picus1}
%\end{eqnarray}
\begin{eqnarray}
R_{ab} &=& (\bu_b-\bu_a) m_a n_a \nu_{ab};\nn\\
Q_{ab}^* &=& 3 (T_b-T_a) \frac{n_a m_a \nu_{ab}}{m_a+m_b} \Big( 1 + \Upsilon_{ab} \frac{|\bu_b-\bu_a|^2}{v_{\textrm{th} a}^2 +v_{\textrm{th} b}^2} \Big);
\qquad \Upsilon_{ab}= \frac{2}{3} \big(\Omega_{1,2}-\frac{5}{2}\big);\nn\\
Q_{ab} &=& \frac{n_a m_a \nu_{ab}}{m_a+m_b}\Big[ 3 (T_b-T_a) \Big( 1 +\Upsilon_{ab} \frac{|\bu_b-\bu_a|^2}{v_{\textrm{th} a}^2 +v_{\textrm{th} b}^2}\Big)
+ m_b |\bu_b-\bu_a|^2 \Big], \label{eq:Picus1}
\end{eqnarray}
where we have used symbol $\Upsilon_{ab}$ (Upsilon) to differentiate between various collisional processes, with examples
\begin{eqnarray}
&&  \textrm{Hard spheres:} \qquad \qquad \,\,\, \Omega_{1,2}=3; \qquad \qquad\, \Upsilon_{ab} = 1/3;\nn\\
  &&  \textrm{Coulomb collisions:} \qquad \Omega_{1,2}=1; \qquad \qquad \, \Upsilon_{ab} = -1 ;\nn\\
  &&  \textrm{Inverse power:} \qquad \qquad \Omega_{1,2}=\frac{3\nu-5}{\nu-1};\qquad \Upsilon_{ab} = \frac{\nu-5}{3(\nu-1)};\nn\\
  &&  \textrm{Maxwell molecules:} \qquad \Omega_{1,2}=5/2; \qquad \quad \,\,\,\Upsilon_{ab}  =  0.
\end{eqnarray}
Result (\ref{eq:Picus1}) is valid for arbitrary temperature differences and small drifts. 
  Notably, prescribing Coulomb collisions recovers eq. (174) or (G33) of Part 1. 
Importantly, the differences in temperature modify the $|\bu_b-\bu_a|^2$ contributions, according to 
\begin{equation}
  Q_{ab} = \frac{n_a m_a \nu_{ab}}{m_a+m_b}\Big[ 3 (T_b-T_a)
    + m_b |\bu_b-\bu_a|^2 \Big( 1+\Upsilon_{ab} \frac{ 3 m_a (T_b-T_a)}{2(T_a m_b+T_b m_a)}  \Big)\Big]. \label{eq:Picus1X}
\end{equation}
Also note that to obtain (\ref{eq:Picus1}) or (\ref{eq:Picus1X}), the route through the unrestricted drifts is not necessary,
and instead one can expand the $f_a f_b$ in small drifts from the beginning by (\ref{eq:Martin}) and retain the $u^2$ contributions. 
Prescribing small temperature differences in (\ref{eq:Picus1}) or (\ref{eq:Picus1X})
recovers the usual $Q_{ab}$ given by (\ref{eq:QabGenA}).

\newpage
%======================================
\subsection{Simplest viscosity (10-moment model, self-collisions)}
Starting again with the $\bQ_{ab}^{(2)}$ given by (\ref{eq:beau0P}), let us now calculate the simplest self-collisional (1-Hermite) viscosities.
The calculation proceeds in a similar manner as the already calculated viscosity of hard spheres in
Section \ref{sec:10momEqT}, where one employs the perturbations of the distribution function $\chi_a$, $\chi_b$ given by (\ref{eq:Pert}), which are
transformed with the self-collisional center-of-mass transformations $\bc_a = \bC^* +\frac{1}{2}\bg$ and $\bc_b = \bC^* -\frac{1}{2}\bg$, yielding 
\begin{equation*} 
\chi_a + \chi_b = \frac{m_a}{T_a p_a} \bPi^{(2)}_a:\Big[ \bC^* \bC^* + \frac{1}{4}\bg\bg \Big]. 
\end{equation*}
Additionally, for self-collisions
\begin{equation}
\mu_{aa}=m_a/2; \quad \alpha^2=T_a/m_a; \quad \beta^2=4T_a/m_a; \quad \textrm{and} \quad \hat{\bVV}_c =\bC^*, \label{eq:picaa}
\end{equation}
and so the (\ref{eq:beau0P}) calculates
\begin{eqnarray}
 \bQ_{aa}^{(2)} (\chi) &=&   - \frac{m_a}{2} \frac{n_a n_a}{\pi^3 {\alpha}^3 \beta^3} \iint
 e^{ -\frac{C^{*2}}{{\alpha}^2}} e^{-\frac{g^2}{\beta^2}}  \nn\\
 && \quad \times \frac{m_a}{T_a p_a} \bPi^{(2)}_a:\Big[ \bC^* \bC^* + \frac{1}{4}\bg\bg \Big] 
   g \Big[  \Big(\cancel{\bg \bC^*} + \cancel{\bC^* \bg} \Big)\mathbb{Q}_{aa}^{(1)} 
     - \frac{3}{4} \Big(  \frac{\bI}{3}g^2 - \bg\bg \Big) \mathbb{Q}_{aa}^{(2)} \Big]  d^3 C^* d^3 g \nn\\
   &=& + \frac{3}{8} \frac{m_a^2 n_a^2}{T_a p_a} \frac{1}{\pi^3 {\alpha}^3 \beta^3} \iint
 e^{ -\frac{C^{*2}}{{\alpha}^2}} e^{-\frac{g^2}{\beta^2}} 
 \bPi^{(2)}_a:\Big[ \cancel{\bC^* \bC^*} + \frac{1}{4}\bg\bg \Big] 
 g   \Big(  \cancel{\frac{\bI}{3}g^2} - \bg\bg \Big) \mathbb{Q}_{aa}^{(2)}   d^3 C^* d^3 g \nn\\
%  &=& - \frac{3}{32} \frac{m_a^2 n_a^2}{T_a p_a} \frac{1}{\pi^3 {\alpha}^3 \beta^3}  \bPi^{(2)}_a: \iint
% e^{ -\frac{C^{*2}}{{\alpha}^2}} e^{-\frac{g^2}{\beta^2}} 
%\bg\bg \bg\bg g \mathbb{Q}_{aa}^{(2)}   d^3 C^* d^3 g \nn\\
 %===
  &=& - \frac{3}{32} \frac{m_a^2 n_a^2}{T_a p_a} \frac{1}{\pi^{3/2} \beta^3} \bPi^{(2)}_a: \int
  e^{-\frac{g^2}{\beta^2}} 
  \bg\bg \bg\bg g \mathbb{Q}_{aa}^{(2)}   d^3 g \nn\\
  &=& - \frac{3}{32} \frac{m_a^2 n_a^2}{T_a p_a} \frac{1}{\pi^{3/2} \beta^3} \bPi^{(2)}_a \frac{8\pi}{15} \int_0^\infty
  e^{-\frac{g^2}{\beta^2}} g^7 \mathbb{Q}_{aa}^{(2)}   d g
  = -\frac{1}{10} \frac{m_a^2 n_a^2}{T_a p_a}\beta^4 \bPi^{(2)}_a \Big[ \frac{1}{2 \pi^{1/2} \beta^7} \int_0^\infty
    e^{-\frac{g^2}{\beta^2}} g^7 \mathbb{Q}_{aa}^{(2)}   d g \Big] \nn\\
  &=& -\frac{8}{5} n_a \bPi^{(2)}_a \Omega_{aa}^{(2,2)},
\end{eqnarray}
where in the last step the Chapman-Cowling integral $\Omega_{aa}^{(2,2)}$ defined by (\ref{eq:CCo}) was employed.  
The projection $\bQ^{(2)}_{aa}\,' = \bQ^{(2)}_{aa} -(\bI/3)\trace \bQ^{(2)}_{aa}$ does not change the result, because the $\bPi^{(2)}_a$ is traceless.
Only one species are present (so that $\bQ^{(2)}_{aa}\,'=\bQ^{(2)}_{a}\,'$) and using the collisional frequency $\nu_{aa}=(8/3)n_a \Omega_{aa}^{(1,1)}$
then yields the final collisional contribution 
\begin{equation} 
 \bQ_{a}^{(2)}\,' = -\frac{3}{5} \frac{\Omega_{aa}^{(2,2)}}{\Omega_{aa}^{(1,1)}} \nu_{aa} \bPi^{(2)}_a =  -\frac{3}{5} \Omega_{22} \nu_{aa} \bPi^{(2)}_a, \label{eq:ViscG}
\end{equation}
where the abbreviated $\Omega_{22}=\Omega_{2,2}$. 
The result (\ref{eq:ViscG}) enters the right-hand-side of the evolution equation for the stress-tensor, where in the quasi-static approximation
an equation in a form 
\begin{equation}
\cancel{\frac{d_a}{dt} \bPi^{(2)}_a} + \Omega_a \big(\bhat\times\bPi^{(2)}_a \big)^S +p_a \bW_a = - \bnu_a \bPi_a^{(2)}, \label{eq:perfect4Gb}
\end{equation}
has a solution (see details in Appendix E.4 of Part 1)
\begin{eqnarray}
  \bPi^{(2)}_a &=& -\eta_0^a \bW_0 -\eta_1^a\bW_1 -\eta_2^a\bW_2 +\eta_3^a\bW_3+\eta_4^a\bW_4;\nn\\ 
  \eta_0^a &=& \frac{p_a}{\bnu_a}; \quad \eta_1^a = \frac{p_a\bnu_a}{4\Omega_a^2+\bnu_a^2};
  \quad \eta_2^a = \frac{p_a\bnu_a}{\Omega_a^2+\bnu_a^2};
\quad \eta_3^a = \frac{2p_a\Omega_a}{4\Omega_a^2+\bnu_a^2}; \quad \eta_4^a = \frac{p_a\Omega_a}{\Omega_a^2+\bnu_a^2},
\end{eqnarray}
which for our case with $\bnu_a =  (3/5)\Omega_{22} \nu_{aa}$ yields the 1-Hermite viscosities (adding a designation $[\ldots]_1$)
\begin{eqnarray}
  \big[ \eta_0^a \big]_1 &=&  \frac{5}{3 \Omega_{22} } \frac{p_a}{\nu_{aa}}; \label{eq:pica33GGM}\\
  \big[\eta_1^a \big]_1 &=& \frac{p_{a} \nu_{aa}(3\Omega_{22}/5)}{(2\Omega_a)^2 + \nu_{aa}^2(3\Omega_{22}/5)^2}; \qquad
  \big[\eta_2^a \big]_1 = \frac{p_{a} \nu_{aa}(3\Omega_{22}/5)}{\Omega_a^2 + \nu_{aa}^2(3\Omega_{22}/5)^2}; \nn\\
  \big[\eta_3^a \big]_1 &=& \frac{2 p_a \Omega_a}{(2\Omega_a)^2 + \nu_{aa}^2 (3\Omega_{22}/5)^2};\qquad
  \big[\eta_4^a \big]_1 = \frac{p_a \Omega_a}{\Omega_a^2 + \nu_{aa}^2 (3\Omega_{22}/5)^2}. \label{eq:pica33GG}
\end{eqnarray}
 The parallel viscosity (\ref{eq:pica33GGM}) is valid for a general self-collisional process describable by the Boltzmann operator
 (because one can consider unmagnetized case with the solution $\bPi^{(2)}_a = -\eta_0^a \bW_a$).
  In contrast, the magnetized viscosities $\eta_1^a-\eta_4^a$ are valid only for Coulomb collisions (because the Lorentz force is present at the left-hand-side
  of (\ref{eq:perfect4Gb}), and to get more general solutions, one needs to consider coupling between neutrals and ions, see Appendix
  \ref{sec:AppIN}). 
By using the parameter $x=\Omega_a/\nu_{aa}$ that Braginskii uses (which describes
the strength of the magnetic field, sometimes also called the Hall parameter), the viscosities can be also written as (given also by (\ref{eq:pica33}))
\begin{eqnarray}
  \big[ \eta_0^a \big]_1 &=&  \frac{5}{3 \Omega_{22} } \frac{p_a}{\nu_{aa}};\nn\\
  \big[\eta_1^a \big]_1 &=& \frac{p_{a}}{\nu_{aa}}\,\frac{ 3\Omega_{22}/5}{(2x)^2 + (3\Omega_{22}/5)^2};\qquad 
  \big[\eta_2^a \big]_1 = \frac{p_{a}}{\nu_{aa}}\,\frac{ 3\Omega_{22}/5}{x^2 + (3\Omega_{22}/5)^2}; \nn\\
  \big[\eta_3^a \big]_1 &=& \frac{p_{a}}{\nu_{aa}}\, \frac{2x}{(2x)^2 + (3\Omega_{22}/5)^2}; \qquad
  \big[\eta_4^a \big]_1 = \frac{p_{a}}{\nu_{aa}}\, \frac{x}{x^2 + (3\Omega_{22}/5)^2}. \label{eq:pica33again}
\end{eqnarray}
Note that in the 1-Hermite approximation, the only Chapman-Cowling integral which enters the self-collisional viscosities (\ref{eq:pica33again}) is the $\Omega_{22}$,
where for example for the Coulomb collisions with the large Coulomb logarithm $\ln\Lambda\gg 1$ (as well as for the hard spheres), the $\Omega_{22}=2$.
For the collisional force $K/r^\nu$, the $\Omega_{22} = \frac{A_2(\nu)}{A_1(\nu)} \frac{3\nu-5}{\nu-1}$, with the constants $A_l(\nu)$ given in Table \ref{Table:Anu} and the collisional
frequencies $\nu_{aa}$ given by (\ref{eq:haha2}).
In the more precise 2-Hermite approximation, the self-collisional viscosities are given by (\ref{eq:beau033}), and the Chapman-Cowling integrals $\Omega_{23}$ and
$\Omega_{24}$ enter as well.

Also note that as a function of $x$, the relations $\eta_1^a(x)=\eta_2^a(2x)$ and $\eta_3^a(x)=\eta_4^a(2x)$ always hold (regardless of the level of the Hermite approximation,
or if one uses the Boltzmann or the heuristic BGK operator), so typically only the viscosities $\eta_0^a$, $\eta_2^a$ and $\eta_4^a$ are written down, because one can easily deduce
the $\eta_1^a$ and $\eta_3^a$ by simply replacing the $x\to 2x$ in the expressions for the $\eta_2^a$ and $\eta_4^a$.
The parallel viscosity $\eta_0^a$ is sometimes omitted as well, because its value
can be easily deduced from the $\eta_2^a$ by prescribing zero magnetic field $x=0$. One can obtain the same structure of the (1-Hermite) viscosity coefficients with the
very simple BGK operator, see e.g. \cite{Kaufman1960} or eq. (E14) in \cite{Hunana2022}. As already noted in the last reference (p. 77), in the
work of \cite{HelanderSigmar2002} (p. 86) and also \cite{ZankBook2014} (p. 164), the BGK viscosity coefficient $\eta_4$ is erroneously related to the $\eta_3$ by $\eta_3=2\eta_4$,
which is a valid relation only in the limit of weak magnetic field (small $x$).

%\newpage
%======================================
\subsection{Simplest thermal conductivity (8-moment model, self-collisions)}
Starting with the heat flux exchange rates $\vecQ^{(3)}_{ab}$ given by (\ref{eq:beau4}) and considering self-collisions (where the (\ref{eq:picaa}) applies),
we need to calculate
\begin{eqnarray}
  \vecQ^{(3)}_{aa} &=&  - \frac{m_a}{4} \frac{n_a^2}{\pi^3 {\alpha}^3 \beta^3} \iint
 e^{ -\frac{C^{*2}}{{\alpha}^2}} e^{-\frac{g^2}{\beta^2}}  (\cancel{1}+\chi_a+\chi_b ) \nn\\    
 && \times \, g \Big\{ \Big[  2 \bC^*\bC^*\cdot \bg +\Big(C^{*2}  + \frac{1}{4} g^2 \Big)\bg \Big] \mathbb{Q}_{aa}^{(1)} 
  -  \frac{3}{2}    \Big(  \frac{\bC^*}{3}g^2 - \bC^*\cdot\bg\bg \Big) \mathbb{Q}_{aa}^{(2)} \Big\}  d^3 C^* d^3 g,  \label{eq:beau4x}
\end{eqnarray}
where the strictly Maxwellian term was already scratched, because it yields zero. 
The heat flux perturbations of the distribution function are given by   
\begin{eqnarray}
  \chi_a^{(3)}  &=& \frac{1}{5} \frac{m_a}{T_a p_a} \boldsymbol{q}_a \cdot \bc_a \Big( \frac{m_a}{T_a}c_a^2 -5 \Big); \qquad
   \chi_b^{(3)}  = \frac{1}{5} \frac{m_b}{T_b p_b} \boldsymbol{q}_b \cdot \bc_b \Big( \frac{m_b}{T_b}c_b^2 -5 \Big),
\end{eqnarray}
and with the self-collisional transformations $\bc_a = \bC^* +\frac{1}{2}\bg$ and $\bc_b = \bC^* -\frac{1}{2}\bg$ 
they simplify into (again note that $\chi_a+\chi_b\neq 2\chi_a$)
\begin{equation}
  \textrm{self-collisions}: \qquad \chi_a^{(3)}+\chi_b^{(3)} =   \frac{m_a}{5 p_a T_a} \boldsymbol{q}_a \cdot
  \Big\{ 2\bC^* \Big[ \frac{m_a}{T_a}\big( C^{*2} +\frac{g^2}{4}\big) -5 \Big]
  + \frac{m_a}{T_a} \bg (\bC^*\cdot\bg) \Big\}. \label{eq:chi3self}   
\end{equation}
Heat flux perturbations (\ref{eq:chi3self}) enter the expression (\ref{eq:beau4x}), where in the first step several integrals cancel out
\begin{eqnarray}
  \vecQ^{(3)}_{aa} &=&  - \frac{m_a}{4} \frac{n_a^2}{\pi^3 {\alpha}^3 \beta^3} \frac{m_a}{5 p_a T_a} \boldsymbol{q}_a \cdot \iint
 e^{ -\frac{C^{*2}}{{\alpha}^2}} e^{-\frac{g^2}{\beta^2}} \Big\{ \cancel{2\bC^* \Big[ \frac{m_a}{T_a}\big( C^{*2} +\frac{g^2}{4}\big) -5 \Big] }
  + \frac{m_a}{T_a} \bg (\bC^*\cdot\bg) \Big\} \nn\\    
 && \times   \, g \Big\{ \Big[  \cancel{2 \bC^*\bC^*\cdot \bg} + \cancel{\Big(C^{*2}  + \frac{1}{4} g^2 \Big)\bg } \Big] \mathbb{Q}_{aa}^{(1)} 
  -  \frac{3}{2}    \Big(  \frac{\bC^*}{3}g^2 - \bC^*\cdot\bg\bg \Big) \mathbb{Q}_{aa}^{(2)} \Big\}  d^3 C^* d^3 g\nn\\
%===
 &=&  +\frac{3}{40} \frac{n_a^2}{\pi^3 {\alpha}^3 \beta^3} \frac{m_a^3}{p_a T_a^2} \boldsymbol{q}_a \cdot \iint
 e^{ -\frac{C^{*2}}{{\alpha}^2}} e^{-\frac{g^2}{\beta^2}} \bg (\bC^*\cdot\bg)  \, g   
      \Big(  \frac{\bC^*}{3}g^2 - \bC^*\cdot\bg\bg \Big) \mathbb{Q}_{aa}^{(2)}   d^3 C^* d^3 g,  \label{eq:beau4xxx}
\end{eqnarray}
where for example 
\begin{equation}
  \iint   e^{ -\frac{C^{*2}}{{\alpha}^2}} e^{-\frac{g^2}{\beta^2}} \bC^*  \Big[ \frac{m_a}{T_a}\big( C^{*2} +\frac{g^2}{4}\big) -5 \Big]
  g \Big(  \frac{\bC^*}{3}g^2 - \bC^*\cdot\bg \bg \Big) \mathbb{Q}_{aa}^{(2)} d^3 C^* d^3 g = 0.
\end{equation}
The rest of the (\ref{eq:beau4xxx}) calculates
\begin{eqnarray}
  \vecQ^{(3)}_{aa} &=& +\frac{3}{40} \frac{n_a^2}{\pi^3 {\alpha}^3 \beta^3} \frac{m_a^3}{p_a T_a^2} \boldsymbol{q}_a \cdot
  \Big[ -\frac{8\pi}{9} \int_0^\infty e^{ -\frac{C^{*2}}{{\alpha}^2}} C^{*4} dC^* \frac{4\pi}{3} \bI \int_0^\infty e^{-\frac{g^2}{\beta^2}} g^7 \mathbb{Q}_{aa}^{(2)} dg \Big] \nn\\
  &=& +\frac{3}{40} \frac{n_a^2}{\pi^3 {\alpha}^3 \beta^3} \frac{m_a^3}{p_a T_a^2} \boldsymbol{q}_a \cdot
  \Big[ -\frac{8\pi^3 \alpha^5\beta^7}{9}  \bI \frac{1}{2\pi^{1/2} \beta^7} \int_0^\infty e^{-\frac{g^2}{\beta^2}} g^7 \mathbb{Q}_{aa}^{(2)} dg \Big] \nn\\
  &=& -\frac{16}{15} n_a \boldsymbol{q}_a \Omega^{(2,2)}_{aa},
\end{eqnarray}
and by using the collisional frequency $\nu_{aa}=(8/3)n_a \Omega_{aa}^{(1,1)}$, the final result reads
\begin{equation}
  \vecQ^{(3)}_{a}\,' = -\frac{2}{5}  \frac{\Omega^{(2,2)}_{aa}}{\Omega^{(1,1)}_{aa}} \nu_{aa} \vecq_a = -\frac{2}{5} \Omega_{22} \nu_{aa} \vecq_a. \label{eq:Qpica}
\end{equation}
The result (\ref{eq:Qpica}) enters the right-hand-side of the evolution equation for the heat flux, where in the quasi-static approximation
an equation in a form 
\begin{equation}
    \cancel{\frac{d_a}{d t}\vecq_a} + \Omega_a \bhat\times\vecq_a + \frac{5}{2} \frac{p_a}{m_a} \nabla T_a = - \bnu_a\vecq_a.\label{eq:Excite2Pa}
\end{equation}
has a solution
\begin{eqnarray} 
\vecq_a &=& -\kappa_\parallel^a \nabla_\parallel T_a - \kappa_\perp^a \nabla_\perp T_a + \kappa_\times^a \bhat\times\nabla T_a; \nn\\
 \kappa_\parallel^a &=& \frac{5}{2}\frac{p_a}{\bnu_a m_a}; \qquad
  \kappa_\perp^a = \frac{5}{2}\frac{p_a}{m_a}\frac{\bnu_a}{(\Omega_a^2+\bnu_a^2)}; \qquad
  \kappa_\times^a = \frac{5}{2}\frac{p_a}{m_a}\frac{\Omega_a}{(\Omega_a^2+\bnu_a^2)},\label{eq:HF_BGK2X}
\end{eqnarray}
which for our case with $\bnu_a=(2/5) \Omega_{22} \nu_{aa}$ yields the 1-Hermite thermal conductivities (adding a designation $[\ldots]_1$)
\begin{equation}
 \big[\kappa_\parallel^a \big]_1 = \frac{25}{4 \Omega_{22}}\frac{p_a}{\nu_{aa} m_a}; \qquad
 \big[\kappa_\perp^a \big]_1 = \frac{p_a}{m_a}\,\frac{\Omega_{22}\nu_{aa}}{\Omega_a^2+\nu_{aa}^2 (2\Omega_{22}/5)^2}; \qquad
  \big[\kappa_\times^a \big]_1= \frac{5}{2}\frac{p_a}{m_a}\,\frac{\Omega_a}{\Omega_a^2+\nu_{aa}^2 (2\Omega_{22}/5)^2}. \label{eq:wow2}
\end{equation}
     The parallel thermal conductivity $\kappa_\parallel^a$ is valid for a general self-collisional process (because one can consider the unmagnetized case
      with solution $\vecq_a = -\kappa_\parallel^a \nabla T_a$) and the magnetized conductivities are valid only for Coulomb collisions. 
      By using the parameter $x=\Omega_a/\nu_{aa}$, the results can be also written as (\ref{eq:Thierry51P}).

\newpage
%==========
\subsection{Coupling between ions and neutrals (1-Hermite)} \label{sec:AppIN}
\subsubsection*{Ion-neutral viscosity}
  Let us again consider only the 1-Hermite approximation and write the evolution equations for the ion (i) and neutral (n)
  stress-tensors with a general coefficients $V_1, V_2, V_3, V_4$ as
\begin{eqnarray}
     \frac{d_i}{dt} \bPi^{(2)}_i  +\Omega_i \big(\bhat\times \bPi^{(2)}_i \big)^S + p_i \bW_i &=&  
    -  V_1 \bPi_i^{(2)} +  V_2  \bPi^{(2)}_n; \nn\\
   \frac{d_n}{dt} \bPi^{(2)}_n  + p_n \bW_n &=&  
    +  V_3 \bPi_i^{(2)} -  V_4 \bPi^{(2)}_n. \label{eq:Xfun2B}
\end{eqnarray}
In the quasi-static approximation, the ion stress-tensor $\bPi^{(2)}_i$ then contains the rate-of-strain tensors of both ions $\bW^i$ and neutrals $\bW^n$
(we moved the species indices up), with components 
\begin{eqnarray}
  \bPi^{(2)}_i &=& - \frac{V_4}{D} p_i \bW_0^i - \frac{V_4 D }{\Delta^*} p_i\bW_1^i -  \frac{V_4 D }{\Delta} p_i\bW_2^i
  + \frac{2 \Omega_i V_4^2 }{\Delta^*} p_i\bW_3^i
  + \frac{\Omega_i V_4^2 }{\Delta} p_i \bW_4^i \nn\\
  && - \frac{V_2}{D} p_n \bW_0^n - \frac{V_2 D }{\Delta^*} p_n \bW_1^n - \frac{V_2 D }{\Delta} p_n \bW_2^n
  + \frac{2\Omega_i V_2 V_4 }{\Delta^*} p_n \bW_3^n+ \frac{\Omega_i V_2 V_4 }{\Delta} p_n \bW_4^n,
\end{eqnarray}
where $\bW_0-\bW_4$ are the usual Braginskii matrices (\ref{eq:Energy081}) and we have introduced notation
\begin{equation}
  D = V_1 V_4-V_2 V_3; \qquad \Delta = D^2 +\Omega_i^2 V_4^2; \qquad \Delta^* = D^2 +4 \Omega_i^2 V_4^2. \label{eq:lost}
\end{equation}
Similarly, the stress-tensor for neutrals $\bPi^{(2)}_n$ becomes magnetized and contains the rate-of-strain tensors of both species
\begin{eqnarray}
  \bPi^{(2)}_n &=& -\frac{V_1}{D} p_n \bW_0^n -\frac{(V_1 D +4\Omega_i^2 V_4)}{\Delta^*} p_n \bW_1^n
  - \frac{(V_1 D +\Omega_i^2 V_4)}{\Delta} p_n \bW_2^n +\frac{2\Omega_i V_2 V_3}{\Delta^*} p_n \bW_3^n
  + \frac{\Omega_i V_2 V_3}{\Delta} p_n\bW_4^n \nn\\
  && - \frac{V_3}{D} p_i \bW_0^i -\frac{V_3  D }{\Delta^*} p_i\bW_1^i - \frac{V_3  D }{\Delta} p_i \bW_2^i +\frac{2 \Omega_i V_3 V_4 }{\Delta^*} p_i \bW_3^i
  + \frac{\Omega_i V_3 V_4 }{\Delta} p_i \bW_4^i,
\end{eqnarray}
with the same notation (\ref{eq:lost}). Considering that ion-ion collisions are Coulomb and
both the ion-neutral and neutral-neutral collisions are hard spheres, the V-coefficients for small temperature differences read
\begin{eqnarray}
  V_1 &=& \frac{6}{5} \nu_{ii} + \nu_{in}\frac{m_i}{m_i+m_n}\Big(2+\frac{6}{5}\frac{m_n}{m_i}\Big);\qquad
  V_2 = \nu_{in} \frac{4}{5} \frac{m_i}{(m_i+m_n)}\, \frac{n_i}{n_n};\nn\\
  V_3 &=& \nu_{ni}\frac{4}{5} \frac{m_n}{(m_i+m_n)}\, \frac{n_n}{n_i} ; \qquad
  V_4 = \frac{6}{5} \nu_{nn} + \nu_{ni}\frac{m_n}{m_i+m_n}\Big(2+\frac{6}{5}\frac{m_i}{m_n}\Big),
\end{eqnarray}
and the collisional frequencies are related by $\rho_i \nu_{in}=\rho_n \nu_{ni}$ and (see Section \ref{sec:Ratio})
\begin{eqnarray}
\frac{\nu_{nn}}{\nu_{ni}} &=& \frac{1}{\sqrt{2}} \Big( \frac{r_{nn}}{r_{in}}\Big)^2 \Big(\frac{m_n+m_i}{m_i}\Big)^{1/2} \frac{n_n}{n_i}; \qquad  
\frac{\nu_{nn}}{\nu_{in}} = \frac{1}{\sqrt{2}} \Big( \frac{r_{nn}}{r_{in}}\Big)^2 \Big(\frac{m_i}{m_n}\Big)^{1/2} \Big( \frac{m_i+m_n}{m_n} \Big)^{1/2};\nn\\
\frac{\nu_{ii}}{\nu_{nn}} &=& \frac{n_i}{n_n} \sqrt{\frac{m_n}{m_i}} \frac{q_i^4 \ln\Lambda}{r_{nn}^2} \frac{1}{2T^2}. \label{eq:CollFreq}
\end{eqnarray}

\newpage
\subsubsection*{Ion-neutral thermal conductivity}
Again considering only the 1-Hermite approximation, the evolution equations for heat fluxes are written with a general B-coefficients as
\begin{eqnarray}
  \frac{d_i}{d t}\vecq_i + \Omega_i \bhat\times\vecq_i + \frac{5}{2} \frac{p_i}{m_i} \nabla T_i &=&
  -  B_1 \vecq_i  + B_2 \vecq_n;\\
%===  
  \frac{d_n}{d t}\vecq_n + \frac{5}{2} \frac{p_n}{m_n} \nabla T_n &=& 
   + B_3 \vecq_i   - B_4 \vecq_n,
\end{eqnarray}
where here for simplicity we neglected the differences in drifts $\bu_b-\bu_a$. The quasi-static solution then yields the ion heat flux
\begin{eqnarray}
  \vecq_i &=& - \frac{B_4}{D} \Big(\frac{5}{2} \frac{p_i}{m_i}\Big) \nabla_\parallel T_i - \frac{B_4 D}{\Delta} \Big(\frac{5}{2} \frac{p_i}{m_i}\Big) \nabla_\perp T_i
  + \frac{B_4^2 \Omega_i}{\Delta} \Big(\frac{5}{2} \frac{p_i}{m_i}\Big) \bhat\times \nabla T_i \nn\\
  && - \frac{B_2}{D} \Big(\frac{5}{2} \frac{p_n}{m_n}\Big) \nabla_\parallel T_n - \frac{B_2 D}{\Delta} \Big(\frac{5}{2} \frac{p_n}{m_n}\Big) \nabla_\perp T_n
  + \frac{B_2 B_4 \Omega_i}{\Delta} \Big(\frac{5}{2} \frac{p_n}{m_n}\Big) \bhat\times \nabla T_n,
\end{eqnarray}
where we have introduced notation
\begin{equation}
  D = B_1 B_4-B_2 B_3; \qquad \Delta = D^2 +\Omega_i^2 B_4^2. \label{eq:lost2}
\end{equation}
The heat flux for the neutral particles becomes magnetized and reads
\begin{eqnarray}
  \vecq_n &=& - \frac{B_1}{D} \Big(\frac{5}{2} \frac{p_n}{m_n}\Big) \nabla_\parallel T_n - \frac{B_1 D + B_4 \Omega_i^2}{\Delta} \Big(\frac{5}{2} \frac{p_n}{m_n}\Big) \nabla_\perp T_n
  + \frac{B_2 B_3 \Omega_i}{\Delta} \Big(\frac{5}{2} \frac{p_n}{m_n}\Big) \bhat\times \nabla T_n \nn\\
  && - \frac{B_3}{D} \Big(\frac{5}{2} \frac{p_i}{m_i}\Big) \nabla_\parallel T_i - \frac{B_3 D}{\Delta} \Big(\frac{5}{2} \frac{p_i}{m_i}\Big) \nabla_\perp T_i
  + \frac{B_3 B_4 \Omega_i}{\Delta} \Big(\frac{5}{2} \frac{p_i}{m_i}\Big) \bhat\times \nabla T_i,
\end{eqnarray}
with the same notation (\ref{eq:lost2}). Considering that ion-ion collisions are Coulomb and
both the ion-neutral and neutral-neutral collisions are hard spheres, the B-coefficients for small temperature differences are given by
\begin{eqnarray}
  B_1 &=& \frac{4}{5}\nu_{ii} +\nu_{in} \frac{3 (10 m_i^2+7 m_i m_n+6 m_n^2)}{10 (m_i+m_n)^2}; \qquad
  B_2 = \nu_{in}\frac{\rho_i}{\rho_n} \frac{m_n (5 m_i+32 m_n) }{10 (m_i+m_n)^2 }; \nn\\
  B_3 &=& \nu_{ni}\frac{\rho_n}{\rho_i} \frac{m_i (5 m_n+32 m_i) }{10 (m_i+m_n)^2 }; \qquad
  B_4 = \frac{4}{5}\nu_{nn} +\nu_{ni} \frac{3 (10 m_n^2+7 m_n m_i+6 m_i^2)}{10 (m_i+m_n)^2},
\end{eqnarray}
and the collisional frequencies are related by (\ref{eq:CollFreq}).

\newpage
\bibliographystyle{jpp}
\bibliography{hunana_mhd}

\end{document}